\documentclass[phd,lfcs,twoside, singlespacing, logo, listintoc, timesfont]{infthesis}

\usepackage[OT1]{fontenc}
\usepackage{cmbright}
\usepackage[most]{tcolorbox}
\usepackage{graphicx, color, graphpap}      
\usepackage{amsmath}
\usepackage{enumitem}
\usepackage{amssymb}
\usepackage{thmtools, thm-restate}

\usepackage{amsthm}

\usepackage{pstricks}
\usepackage{float}
\usepackage[colorlinks=true, citecolor=blue, linkcolor=cyan, pagebackref]{hyperref}
\usepackage[T1]{fontenc}
\usepackage{bbm}
\usepackage{dsfont}
\usepackage[linesnumbered,ruled,vlined]{algorithm2e}
\SetKwInput{kwInit}{Init}
\usepackage{mathtools}
\DeclarePairedDelimiter\ceil{\lceil}{\rceil}

\usepackage{tikz}
\usepackage{graphicx}
\usetikzlibrary{positioning, scopes,svg.path,shapes.geometric,shadows}
\usepackage{graphicx}
\usepackage{xcolor}
\usepackage[caption=false]{subfig}
\usepackage{xfrac}
\usepackage[ruled,vlined]{algorithm2e}
\usepackage{siunitx}
\usepackage{qcircuit}
\usepackage{pifont} 
\usepackage{tabularx}
\usepackage{wrapfig}
\usepackage{url}

\usepackage{breakurl}
\usepackage{relsize}
\MHInternalSyntaxOn
\def\MT_extended_eqref:n #1{
  \protected@write\@auxout{}
  {\string\MT@newlabel{#1}}
  \textup{\let\df@label\@empty\MT_prev_tagform:n {\ref{#1}}}
}
\MHInternalSyntaxOff
\makeatletter
\newtheorem*{rep@theorem}{\rep@title}
\newcommand{\newreptheorem}[2]{%
\newenvironment{rep#1}[1]{%
 \def\rep@title{#2 \ref{##1}}%
 \begin{rep@theorem}}%
 {\end{rep@theorem}}}
\makeatother
\newreptheorem{theorem}{Theorem}
\newreptheorem{lemma}{Lemma}
\usepackage{lettrine}

\makeatletter
\renewcommand{\@chapapp}{}
\newenvironment{chapquote}[2][2em]
  {\setlength{\@tempdima}{#1}%
   \def\chapquote@author{#2}%
   \parshape 1 \@tempdima \dimexpr\textwidth-2\@tempdima\relax%
   \itshape}
  {\par\normalfont\hfill--\ \chapquote@author\hspace*{\@tempdima}\par\bigskip}
\makeatother

\DeclareUnicodeCharacter{2212}{-}

\newcommand{\corrref}[1]{\hyperref[#1]{Corollary~\ref{#1}}}
\newcommand{\defref}[1]{\hyperref[#1]{Definition~\ref{#1}}}
\newcommand{\secref}[1]{\hyperref[#1]{Sec.~\ref{#1}}}
\newcommand{\chapref}[1]{\hyperref[#1]{Chapter~\ref{#1}}}
\newcommand{\appref}[1]{\hyperref[#1]{Appendix~\ref{#1}}}
\newcommand{\thmref}[1]{\hyperref[#1]{Theorem~\ref{#1}}}
\newcommand{\lemref}[1]{\hyperref[#1]{Lemma~\ref{#1}}}
\newcommand{\figref}[1]{\hyperref[#1]{Fig.~\ref{#1}}}
\renewcommand{\eqref}[1]{\hyperref[#1]{Eq. (\ref{#1})}}
\newcommand{\algref}[1]{\hyperref[#1]{Algorithm~\ref{#1}}}
\newcommand{\attref}[1]{\hyperref[#1]{Attack~\ref{#1}}}

\newcommand{\braket}[2]{\langle #1 \hspace{1pt} | \hspace{1pt} #2 \rangle}
\newcommand{\ket}[1]{| #1 \rangle}
\newcommand{\bra}[1]{\langle #1 |}

\newcommand{\ketbra}[2]{| \hspace{1pt} #1 \rangle \langle #2 \hspace{1pt} |}

\newcommand{\norm}[2][]{#1| \! #1| #2 #1| \! #1|}

\newcommand\crule[3][black]{\textcolor{#1}{\rule{#2}{#3}}}

\newcommand{\sq}{\textsf{sq}}

\renewcommand{\geq}{\geqslant}
\renewcommand{\leq}{\leqslant}

\renewcommand{\vec}[1]{\boldsymbol{#1}}  

\newcommand{\xbs}{\boldsymbol{x}}
\newcommand{\ybs}{\boldsymbol{y}}
\newcommand{\zbs}{\boldsymbol{z}}
\newcommand{\vbs}{\boldsymbol{v}}

\newcommand{\rhox}{\rho_{\xbs}}
\newcommand{\rhoxi}{\rho_{\xbs_i}}
\newcommand{\rhotildex}{\tilde{\rho}_{\xbs}}
\newcommand{\rhotildexi}{\tilde{\rho}_{\xbs_i}}
\newcommand{\pzero}{p_{\xbs}}

\newcommand{\sigmaxi}{\sigma_{\xbs_i}}

\newcommand{\yhat}{\hat{y}}

\DeclareMathOperator{\TV}{\textsf{TV}}
\DeclareMathOperator{\OT}{\textsf{OT}}
\DeclareMathOperator{\W}{\textsf{W}}

\DeclareMathOperator{\KL}{\textsf{KL}}
\DeclareMathOperator{\Hsf}{\textsf{H}}
\DeclareMathOperator{\XE}{\textsf{XE}}
\DeclareMathOperator{\SH}{\textsf{SHD}}
\DeclareMathOperator{\SD}{\textsf{SD}}
\DeclareMathOperator{\MMD}{\textsf{MMD}}
\DeclareMathOperator{\BA}{\textsf{BA}}
\DeclareMathOperator{\HS}{\textsf{HS}}

\newcommand{\revneg}{\mathrel{\reflectbox{\rotatebox[origin=c]{180}{$\neg$}}}}
\newcommand{\IBM}{\mathsf{QCIBM}}
\newcommand{\QCIBM}{\mathsf{QCIBM}}

\DeclareMathOperator{\GEN}{\mathsf{GEN}}
\DeclareMathOperator{\EVAL}{\mathsf{EVAL}}

\newcommand{\paramtheta}{\boldsymbol{\theta}}
\DeclareMathOperator{\aspennine}{\text{\computerfont{Aspen}\computerfont{-9}}}
\DeclareMathOperator{\aspeneight}{\text{\computerfont{Aspen}\computerfont{-8}}}
\DeclareMathOperator{\aspenseven}{\text{\computerfont{Aspen}\computerfont{-7}}}
\DeclareMathOperator{\aspenfour}{\text{\computerfont{Aspen}\computerfont{-4}}}

\DeclareMathOperator{\Lbs}{\textsf{L}}
\DeclareMathOperator{\Gbs}{\textsf{G}}
\DeclareMathOperator{\Cbs}{\textsf{C}}

\newcommand{\dataset}{\{(\xbs_i, y_i)\}_{i=1}^M}

\newcommand{\E}{\mathcal{E}}
\newcommand{\depo}{\E_p^{\text{depo}}}          
\newcommand{\depon}{\E_p^{\text{GD}}}           

\newcommand{\flip}{\E_p^{\text{BF}}}            
\newcommand{\deph}{\E_p^{\text{deph}}}          
\newcommand{\paul}{\E_{\mathbf{p}}^{\text{P}}}  
\newcommand{\damp}{\E_p^{\text{AD}}}            

\newcommand{\Ansatze}{\text{ans\"{a}tze}}
\newcommand{\Ansatz}{\text{ansatz}}
\newcommand{\ansatze}{\text{ans\"{a}tze}}

\newtheoremstyle{example}{\topsep}{\topsep}%
{}
{}
{\bfseries}
{:}
{   }
{\thmname{#1}\thmnumber{ #2}}
\theoremstyle{example}
\newtheorem{theorem}{Theorem}
\newtheorem{lemma}{Lemma}
\newtheorem{corollary}{Corollary}

\theoremstyle{definition}
\newtheorem{definition}{Definition}

\newcommand{\DAE}{\mathsf{DAE}}
\newcommand{\GQE}{\mathsf{GQE}}
\newcommand{\GAE}{\mathsf{GAE}}

\newcommand{\SDAE}{\mathsf{SDAE}}
\newcommand{\QAOA}{\mathsf{QAOA}}
\newcommand{\VQE}{\mathsf{VQE}}
\newcommand{\VQA}{\mathsf{VQA}}

\newcommand{\QELA}{\mathsf{QELA}}
\newcommand{\IQP}{\mathsf{IQP}}
\newcommand{\BQP}{\mathsf{BQP}}
\newcommand{\BPP}{\mathsf{BPP}}
\newcommand{\C}{\mathsf{C}}
\newcommand{\BosonSampling}{\mathsf{BosonSampling}}
\newcommand{\poly}{\mathsf{poly}}

\newcommand{\Pee}{\mathsf{P}}
\newcommand{\NP}{\mathsf{NP}}
\newcommand{\RCS}{\mathsf{RCS}}

\newcommand{\XG}{\mathsf{X}}
\newcommand{\HG}{\mathsf{H}}

\newcommand{\ZG}{\mathsf{Z}}
\newcommand{\YG}{\mathsf{Y}}
\newcommand{\TG}{\mathsf{T}}
\newcommand{\SG}{\mathsf{S}}
\newcommand{\UG}{\mathsf{U}}
\newcommand{\VG}{\mathsf{V}}

\newcommand{\RX}{\mathsf{R}_x}
\newcommand{\RZ}{\mathsf{R}_z}
\newcommand{\RY}{\mathsf{R}_y}
\newcommand{\CNOT}{\mathsf{CNOT}}
\newcommand{\CZ}{\mathsf{CZ}}
\newcommand{\CX}{\mathsf{CX}}

\newcommand{\CU}{\mathsf{CU}}

\newcommand{\SWAP}{\mathsf{SWAP}}
\newcommand{\XY}{\mathsf{XY}}

\newcommand{\irm}{\mathrm{i}}
\newcommand{\erm}{\mathrm{e}}

\DeclareMathOperator{\VQC}{\textsf{VarQlone}}
\DeclareMathOperator{\VarQlone}{\textsf{VarQlone}}

\newtheorem*{theorem*}{Theorem}

\def\orcid#1{\kern -0.4em\href{https://orcid.org/#1}{\includegraphics[keepaspectratio,width=0.7em]{orcid_logo.pdf}}} 

\DeclareMathOperator*{\argmax}{arg\,max}
\DeclareMathOperator*{\argmin}{arg\,min}

\renewcommand{\>}{\rangle}
\newcommand{\<}{\langle}

\newcommand{\opt}{\mathrm{opt}}

\newcommand{\hilb}{\mathcal{H}}
\newcommand{\X}{\mathcal{X}}
\newcommand{\Y}{\mathcal{Y}}

\renewcommand{\epsilon}{\varepsilon}

\newcommand{\tr}{\mathsf{Tr}}
\newcommand{\Tr}{\mathsf{Tr}}

\definecolor{ForestGreen}{RGB}{34, 139, 34}
\definecolor{DeepSkyBlue}{RGB}{0, 191,255}
\definecolor{Lavender}{RGB}{230, 230, 250}

\long\def\ca#1\cb{} 
\apptocmd{\sloppy}{\hbadness 10000\relax}{}{}
\newcommand{\computerfont}[1]{{\fontfamily{cmtt}\selectfont #1}}

\def\orcid#1{\kern -0.4em\href{https://orcid.org/#1}{\includegraphics[keepaspectratio,width=0.7em]{images/orcid_logo.pdf}}}

\newcounter{attack}

\newcommand{\sbline}{\\[.5\normalbaselineskip]}

\newcommand{\threeqqvm}{%
    \begin{tikzpicture}[transform canvas={scale=0.09}]
\draw[fill=gray] (0,0) circle (20pt);
\draw[fill=gray] (4,0) circle (20pt);
\draw[fill=gray] (2,3) circle (20pt);
\draw[gray, thick] (0,0) -- (4,0) -- (0,0) -- (2,3) -- (4,0) -- (2,3);
\end{tikzpicture}
}

\newcommand{\fourqqvm}{%
\begin{tikzpicture}[transform canvas={scale=0.08}]
\draw[fill=gray] (0,0) circle (20pt);
\draw[fill=gray] (4,0) circle (20pt);
\draw[fill=gray] (0,4) circle (20pt);
\draw[fill=gray] (4,4) circle (20pt);
\draw[gray, thick] (0,0) -- (4,0) -- (4,4) -- (0,4) -- (0,0) -- (4,4) -- (4,0) -- (0,4) ;
\end{tikzpicture}
}

\newtcolorbox{mybox}[3][]
{
  colframe = #2!25,
  colback  = #2!10,
  coltitle = #2!20!black,  
  title    = {#3},
  #1,
}

\newtcolorbox{defbox}{colback=cyan!5!white,colframe=blue!75!black}
\newtcolorbox{thmbox}{colback=violet!5!white,colframe=violet!75!black}
\newtcolorbox{corrbox}{colback=orange!5!white,colframe=orange!75!black}
\newtcolorbox{lembox}{colback=green!5!white,colframe=green!75!black}

\DefineNamedColor{named}{GreenYellow}   {cmyk}{0.15,0,0.69,0}
\DefineNamedColor{named}{Yellow}        {cmyk}{0,0,1,0}
\DefineNamedColor{named}{Goldenrod}     {cmyk}{0,0.10,0.84,0}
\DefineNamedColor{named}{Dandelion}     {cmyk}{0,0.29,0.84,0}
\DefineNamedColor{named}{Apricot}       {cmyk}{0,0.32,0.52,0}
\DefineNamedColor{named}{Peach}         {cmyk}{0,0.50,0.70,0}
\DefineNamedColor{named}{Melon}         {cmyk}{0,0.46,0.50,0}
\DefineNamedColor{named}{YellowOrange}  {cmyk}{0,0.42,1,0}
\DefineNamedColor{named}{Orange}        {cmyk}{0,0.61,0.87,0}
\DefineNamedColor{named}{BurntOrange}   {cmyk}{0,0.51,1,0}
\DefineNamedColor{named}{Bittersweet}   {cmyk}{0,0.75,1,0.24}
\DefineNamedColor{named}{RedOrange}     {cmyk}{0,0.77,0.87,0}
\DefineNamedColor{named}{Mahogany}      {cmyk}{0,0.85,0.87,0.35}
\DefineNamedColor{named}{Maroon}        {cmyk}{0,0.87,0.68,0.32}
\DefineNamedColor{named}{BrickRed}      {cmyk}{0,0.89,0.94,0.28}
\DefineNamedColor{named}{Red}           {cmyk}{0,1,1,0}
\DefineNamedColor{named}{OrangeRed}     {cmyk}{0,1,0.50,0}
\DefineNamedColor{named}{RubineRed}     {cmyk}{0,1,0.13,0}
\DefineNamedColor{named}{WildStrawberry}{cmyk}{0,0.96,0.39,0}
\DefineNamedColor{named}{Salmon}        {cmyk}{0,0.53,0.38,0}
\DefineNamedColor{named}{CarnationPink} {cmyk}{0,0.63,0,0}
\DefineNamedColor{named}{Magenta}       {cmyk}{0,1,0,0}
\DefineNamedColor{named}{VioletRed}     {cmyk}{0,0.81,0,0}
\DefineNamedColor{named}{Rhodamine}     {cmyk}{0,0.82,0,0}
\DefineNamedColor{named}{Mulberry}      {cmyk}{0.34,0.90,0,0.02}
\DefineNamedColor{named}{RedViolet}     {cmyk}{0.07,0.90,0,0.34}
\DefineNamedColor{named}{Fuchsia}       {cmyk}{0.47,0.91,0,0.08}
\DefineNamedColor{named}{Lavender}      {cmyk}{0,0.48,0,0}
\DefineNamedColor{named}{Thistle}       {cmyk}{0.12,0.59,0,0}
\DefineNamedColor{named}{Orchid}        {cmyk}{0.32,0.64,0,0}
\DefineNamedColor{named}{DarkOrchid}    {cmyk}{0.40,0.80,0.20,0}
\DefineNamedColor{named}{Purple}        {cmyk}{0.45,0.86,0,0}
\DefineNamedColor{named}{Plum}          {cmyk}{0.50,1,0,0}
\DefineNamedColor{named}{Violet}        {cmyk}{0.79,0.88,0,0}
\DefineNamedColor{named}{RoyalPurple}   {cmyk}{0.75,0.90,0,0}
\DefineNamedColor{named}{BlueViolet}    {cmyk}{0.86,0.91,0,0.04}
\DefineNamedColor{named}{Periwinkle}    {cmyk}{0.57,0.55,0,0}
\DefineNamedColor{named}{CadetBlue}     {cmyk}{0.62,0.57,0.23,0}
\DefineNamedColor{named}{CornflowerBlue}{cmyk}{0.65,0.13,0,0}
\DefineNamedColor{named}{MidnightBlue}  {cmyk}{0.98,0.13,0,0.43}
\DefineNamedColor{named}{NavyBlue}      {cmyk}{0.94,0.54,0,0}
\DefineNamedColor{named}{RoyalBlue}     {cmyk}{1,0.50,0,0}
\DefineNamedColor{named}{Blue}          {cmyk}{1,1,0,0}
\DefineNamedColor{named}{Cerulean}      {cmyk}{0.94,0.11,0,0}
\DefineNamedColor{named}{Cyan}          {cmyk}{1,0,0,0}
\DefineNamedColor{named}{ProcessBlue}   {cmyk}{0.96,0,0,0}
\DefineNamedColor{named}{SkyBlue}       {cmyk}{0.62,0,0.12,0}
\DefineNamedColor{named}{Turquoise}     {cmyk}{0.85,0,0.20,0}
\DefineNamedColor{named}{TealBlue}      {cmyk}{0.86,0,0.34,0.02}
\DefineNamedColor{named}{Aquamarine}    {cmyk}{0.82,0,0.30,0}
\DefineNamedColor{named}{BlueGreen}     {cmyk}{0.85,0,0.33,0}
\DefineNamedColor{named}{Emerald}       {cmyk}{1,0,0.50,0}
\DefineNamedColor{named}{JungleGreen}   {cmyk}{0.99,0,0.52,0}
\DefineNamedColor{named}{SeaGreen}      {cmyk}{0.69,0,0.50,0}
\DefineNamedColor{named}{Green}         {cmyk}{1,0,1,0}
\DefineNamedColor{named}{ForestGreen}   {cmyk}{0.91,0,0.88,0.12}
\DefineNamedColor{named}{PineGreen}     {cmyk}{0.92,0,0.59,0.25}
\DefineNamedColor{named}{LimeGreen}     {cmyk}{0.50,0,1,0}
\DefineNamedColor{named}{YellowGreen}   {cmyk}{0.44,0,0.74,0}
\DefineNamedColor{named}{SpringGreen}   {cmyk}{0.26,0,0.76,0}
\DefineNamedColor{named}{OliveGreen}    {cmyk}{0.64,0,0.95,0.40}
\DefineNamedColor{named}{RawSienna}     {cmyk}{0,0.72,1,0.45}
\DefineNamedColor{named}{Sepia}         {cmyk}{0,0.83,1,0.70}
\DefineNamedColor{named}{Brown}         {cmyk}{0,0.81,1,0.60}
\DefineNamedColor{named}{Tan}           {cmyk}{0.14,0.42,0.56,0}
\DefineNamedColor{named}{Gray}          {cmyk}{0,0,0,0.50}
\DefineNamedColor{named}{Black}         {cmyk}{0,0,0,1}
\DefineNamedColor{named}{White}         {cmyk}{0,0,0,0}

\title{Machine learning applications for noisy intermediate-scale quantum computers}
\author{Brian Coyle}

\abstract{
Quantum machine learning (QML) has proven to be a fruitful area in which to search for potential applications of quantum computers. This is particularly true for those available in the near term, so called noisy intermediate-scale quantum (NISQ) devices. In this Thesis, we develop and study QML algorithms in three application areas. We focus our attention towards heuristic algorithms of a variational (meaning hybrid quantum-classical) nature, using parameterised quantum circuits as the underlying quantum machine learning model. The variational nature of these models makes them especially suited for NISQ computers. We order these applications in terms of the increasing complexity of the data presented to them.

Firstly, we study a variational quantum classifier in supervised machine learning, and focus on how (classical) data, feature vectors, may be encoded in such models in a way that is robust to the inherent noise on NISQ computers. We provide a framework for studying the robustness of these classification models, prove theoretical results relative to some common noise channels, and demonstrate extensive numerical results reinforcing these findings.

Secondly, we move to a variational generative model called the Born machine, where the data becomes a (classical or quantum) probability distribution. Now, the problem falls into the category of unsupervised machine learning. Here, we develop new training methods for the Born machine which outperform the previous state of the art, discuss the possibility of quantum advantage in generative modelling, and perform a systematic comparison of the Born machine relative to a classical competitor, the restricted Boltzmann machine. We also demonstrate the largest scale implementation (28 qubits) of such a model on real quantum hardware to date, using the Rigetti superconducting platform.

Finally, for our third QML application, the data becomes purely quantum in nature. We focus on the problem of approximately cloning quantum states, an important primitive in the foundations of quantum mechanics. For this, we develop a variational quantum algorithm which can learn to clone such states, and show how this algorithm may be used to improve quantum cloning fidelities on NISQ hardware. Interestingly, this application can be viewed as either supervised or unsupervised in nature. Furthermore, we demonstrate how this algorithm is useful in discovering novel implementable attacks on quantum cryptographic protocols, focusing on quantum coin flipping and key distribution as examples. For the algorithm, we derive differentiable cost functions, prove theoretical guarantees such as faithfulness, and incorporate state of the art methods such as quantum architecture search. 
}

\begin{document}

\begin{preliminary}

\maketitle

\begin{acknowledgements}

First and foremost, I want to thank my supervisor, Elham Kashefi for unending support and the opportunity to take my research in directions almost perpendicular to my original intended path. I would also like to thank my extensive team of co-supervisors, Vincent Danos, Tony Kennedy and Ajitha Rajan, for keeping me on the rails for the last 4 years. Of particular importance are my PhD examiners, Raul Garcia-Patron Sanchez and Iordanis Kerenidis, for agreeing to give me a PhD at the end of the road.

Edinburgh is a fantastic city, but made even more so by the people there who made the PhD what it was. The journey began with those in the CDT in Pervasive Parallelism, from the Informatics forum to the Bayes centre and back; Bruce Collie, Jack Turner, Pablo Andr\'{e}s-Martínez (a pleasant surprise at having someone else studying quantum computing in a classical computer science doctoral training centre), Mattia Bradascio, Nicolai Oswald, Martin Kristien, Maximiliana Behnke, Aleksandr Maramzin and Margus Lind. 

Next, of particular importance, the quantum team - First, the quomrade: Daniel Mills, who knows how much he contributed. Matty Hoban, my first quantum mentor and Andru Gheorghiu, who gave me my first lesson on complexity theory (among others), and generously donated the title of my first PhD paper.  Petros Wallden, the ever present rock of the quantum group. Ellen Derbyshire, who the journey started with. Mina Doosti, the renaissance woman, Alex Cojocaru the master cryptographer. The postdocs who taught me so much about life, the universe and everything; Niraj Kumar, Atul Mantri, Rawad Mehzer, Mahshid Delavar Theodoros Kapourniotis and Kaushik Chakraborty. Finally, those in the larger quantum group at the University of Edinburgh: James Mills, Chris Heunen, Myrto Arapinis, Ieva \v{C}epait\.{e}, Nuiok Dicaire,  Meisam Tarabkhah, Nishant Jain, Parth Padia, Lakshika Rathi, Patric Fulop and Jonas Landman.

Our group was fortunate enough to span the wisdom of two countries. The second half of this wisdom belonged to those in Paris, where I was fortunate enough to spend 3 months at the start of my PhD and interact with many amazing academics including: Pierre-Emmmanuel Emariau, Dominik Leichtle, Ulysse Chabaud, Armando Angrisani, L\'{e}o Colisson, Luka Music, Shane Mansfield, Tom Douce, Anu Unnikrishnan,  Marc Kaplan, Rhea Parekh, Natansh Mathur, Shraddha Singh, Mathieu Bozzio, Fred Groshans, Harold Ollivier, Damian Markham, Eleni Diamanti.

Then, my detour into quantum industry where I learned from fantastic researchers like Stasja Stanisic, Lana Mineh, Charles Derby, Raul Santos, Joel Klassen and Ashley Montanaro at Phasecraft, and Mattia Fiorentini, Michael Lubasch, Matthias Rosenkranz, David Amaro, Kirill Plekhanov, Carlo Modica, Chiara Leadbeater and Louis Sharrock at Cambridge Quantum. 

Thanks to my collaborators and friends, Ryan LaRose, Max Henderson, Justin Chan (who taught me how to software engineer) Alexei Kondratyev, Graham Enos, Mark Hodson, and Marco Paini.

A special thanks goes to those who read and gave feedback on my Thesis; Dan Mills, Ellen Derbyshire, Atul Mantri, Elham Kashefi and especially Marcello Benedetti, a quantum machine learning pioneer who I was fortunate enough to collaborate with on two papers while at CQC.

To those friends and family outside of academia, Kyle, Conor, Mark, Alan, Fionn, Kieran, Stephen, Duncan, Carmel, Marie, Laura, Ann \& Isabel and all of the Coyles too abundant to name. To my Scottish family; Sandra \& Pat, Mark \& Dot, Kenny \& Ann, Lucy, Fern \& Michael, Ronan, Amy \& Greg, and last but not least; Dudley the dog.

Finally, the one who deserves the most thanks is Kathryn, without whose endless support, none of this would have been possible.

To finish, the coronavirus pandemic from 2019-present, an entity which does not deserve thanks, but certainly deserves acknowledgement. 

\begin{figure}[h!]
    \centering
\begin{tikzpicture}[
  mytree/.style={scale=0.04, rotate=180, draw=green!60!black, thick, line join=round, inner color=green!60!yellow, outer color=green!50!black}]
\filldraw[fill=brown!50!black!,draw=black,thick] (-0.85,-0.6) rectangle (-0.75,-0.85);
  {[mytree]
  \shadedraw svg "M355,430
    q90,10 105,-85 30,0 50,-30 20,30 50,30 50,-20 100,0 10,88 105,85
    -45,90 -205,25 Q400,520 355,430";
  \shadedraw svg "M380,325
    q83,10 105,-80 25,0 35,-30 20,25 40,30 20,-10 35,-25 20,20 40,25
    25,90 105,82 -15,50 -120,15 -30,-2 -60,12 -30,0 -52,-28
    C490,370 380,360 380,325"  ;
  \shadedraw svg "M435,225
    q65,-8 90,-70 35,40 70,0 25,60 90,70 -30,52 -90,5 -36,48 -73,-3
    C520,254 445,265 435,225 " ;
  \shadedraw svg "M470,139
    q50,5 90,-80 50,90 90,80 -30,30 -50,20 -40,45 -78,0
    Q500,170 470,139";
    (-1.75,-2)
  }
\end{tikzpicture}
\end{figure}

\end{acknowledgements}

\standarddeclaration

\dedication{\begin{large}To my parents.\end{large}}

\newpage

\section*{Lay summary} \label{sec:lay_summary}

\begin{small}
In modern times, almost everyone on the planet has access to some form of `classical' computer. For most, this will be a simple smartphone, but what we refer to as classical computers also encompass everything up to the largest supercomputers on the planet. The computational capacities of these two extremes are vastly different, but they all obey the same laws under the hood. In contrast, quantum computers allow us to access a fundamentally different computational paradigm, by manipulating \emph{quantum} information directly. Given this difference, we believe that there exists an unbridgeable gap between what quantum and classical computers can do. Given this, in the long run we have a handful of quantum algorithms which can capitalise on this distinction and have real impact on problems which we cannot hope to solve using purely classical means.

However, actually building and scaling quantum computers is a significant engineering challenge, albeit one towards which tremendous progress is being made. As such, rather than being a clear cut advantage between classical and quantum devices promised by the theory, we currently have a rat race between the most powerful of both examples competing with each other. This is primarily due to the physical imperfections in the quantum devices and their small sizes, which classical computers can exploit to effectively nullify any theoretical advantages.

Nonetheless, small and error-prone quantum computers do exist and in the coming years they will only improve and grow in size. Therefore, the question arises; what should we do with them? This Thesis attempts to address exactly this question and looks to the field of machine learning to find answers. Machine learning is the ability of computers to learn for themselves without being explicitly programmed (``intelligent machines''), and using it we can, for example, recognise patterns in data invisible to the human eye. In its own right, machine learning is ubiquitous in our lives, for example, the recommendation system algorithms which underpin many social networks are machine learning based in nature. Given this ubiquity, it is not surprising that any potential for quantum computing having an impact in this area has created excitement.

In this vein, the Thesis investigates and develops three potential machine learning-based models and algorithms, suitable as applications for the quantum computers available now. The first is the problem of classifying data using a quantum model. In this application, we study the effect of quantum errors on such models, and whether clever model design could be used to make the models more stable against such errors.

The second, is using a quantum model to generate synthetic data. Here, we discuss questions of whether such models may provide and advantage over classical counterparts. We also provide new methods to make the quantum model learn better and more efficiently and run large scale experiments on a real quantum computer.

Finally, the third application has no obvious classical counterpart, in that we show how to train a quantum computer to learn how to clone, or make copies of, quantum states. We show an application of the algorithm we develop in the field of quantum cryptography, by demonstrating how the algorithm can learn to attack certain quantum protocols. Finally, we discuss the possibility of using the algorithm to discover things about the foundations of quantum mechanics.
\end{small}

\newpage

\section*{Publications and manuscripts}
%
%
During the period of time in which the work of this Thesis was completed I have been a part of the following articles:
\subsection*{Included in this Thesis}
The contents of this thesis are based on the following publications and one unpublished manuscript:

\begin{enumerate}
    \item \textbf{Robust Data Encodings for Quantum Classifiers,~\cite{larose_robust_2020}}\\ Ryan LaRose and \textbf{\underline{Brian Coyle}}.\\
    Publication: \href{https://link.aps.org/doi/10.1103/PhysRevA.102.032420}{\textit{Physical Review A 102, 032420 (2020)}}. \\
    Preprint: \href{https://arxiv.org/abs/2003.01695}{ArXiv: 2003.01695}.
    \item \textbf{The Born Supremacy: Quantum Advantage and \\
    Training of an Ising Born Machine.~\cite{coyle_born_2020}} \\
    \textbf{\underline{Brian Coyle}}, Daniel Mills, Vincent Danos and Elham Kashefi. \\
    Publication: \href{https://www.nature.com/articles/s41534-020-00288-9}{\textit{npj Quantum Information 6, 60 (2020)}}.\\
    Preprint: \href{https://arxiv.org/abs/1904.02214}{ArXiv: 1904.02214}.
    \item \textbf{Quantum versus Classical Generative Modelling in Finance.~\cite{coyle_quantum_2021}} \\
    \textbf{\underline{Brian Coyle}}, Maxwell Henderson, Justin Chan Jin Le, Niraj Kumar, Marco Paini and Elham Kashefi. \\
    Publication: \href{https://iopscience.iop.org/article/10.1088/2058-9565/abd3db/meta}{\textit{Quantum Science and Technology, Volume 6, Number 2 (2021)}}.\\
    Preprint: \href{https://arxiv.org/abs/2008.00691}{ArXiv: 2008.00691}.
    \item \textbf{Variational Quantum Cloning: Improving Practicality\\
    for Quantum Cryptanalysis.~\cite{coyle_variational_2020, coyle_progress_2022}} \\
    \textbf{\underline{Brian Coyle}}, Mina Doosti, Elham Kashefi and Niraj Kumar. \\
    Publication: (Progress toward practical quantum cryptanalysis by variational quantum cloning) \href{https://link.aps.org/doi/10.1103/PhysRevA.105.042604}{\textit{Physical Review A 105, 042604 (2022).}}\\
    Preprint: \href{https://arxiv.org/abs/2012.11424}{ArXiv: 2012.11424}.
\end{enumerate}

\newpage
\subsection*{Excluded from this Thesis}
I also coauthored the following publications which are excluded from this Thesis:
\begin{enumerate}
\setcounter{enumi}{4}
    \item \textbf{Certified Randomness From Steering Using Sequential Measurements.~\cite{coyle_certified_2019}} \\
    \textbf{\underline{Brian Coyle}}, Elham Kashefi and Matty Hoban. \\
    Publication: \href{https://doi.org/10.3390/cryptography3040027}{\textit{Cryptography 2019, 3(4), 27 (2019)}}.\\
    Preprint: \href{https://arxiv.org/abs/2008.00705}{ArXiv: 2008.00705}.
    \item \textbf{A Continuous Variable Born Machine.~\cite{cepaite_continuous_2020}} \\
    Ieva \v{C}epait\.{e}, \textbf{\underline{Brian Coyle}} and Elham Kashefi. \\
    Publication: \href{https://link.springer.com/article/10.1007/s42484-022-00063-3}{Quantum Machine Intelligence, 4(6),  (2022)}.\\
    Preprint: \href{https://arxiv.org/abs/2011.00904}{ArXiv: 2011.00904}.
    \item \textbf{Graph neural network initialisation of quantum approximate optimisation.~\cite{jain_graph_2021}} \\
    Nishant Jain, \textbf{\underline{Brian Coyle}}, Elham Kashefi and Niraj Kumar. \\
    Preprint: \href{https://arxiv.org/abs/2111.03016}{ArXiv: 2111.03016}.
    \item \textbf{Variational inference with a quantum computer.~\cite{benedetti_variational_2021}}\\ 
    Marcello Benedetti,  \textbf{\underline{Brian Coyle}}, Mattia Fiorentini, Michael Lubasch, Matthias Rosenkranz.\\
    Publication: \href{https://link.aps.org/doi/10.1103/PhysRevApplied.16.044057}{Phys. Rev. Applied 16, 044057}\\
    Preprint: \href{https://arxiv.org/abs/2103.06720}{ArXiv: 2103.06720}.
    \item \textbf{$f$-divergences and cost function locality in generative modelling with quantum circuits.~\cite{leadbeater_f-divergences_2021}} \\
    Chiara Leadbeater, Louis Sharrock, \textbf{\underline{Brian Coyle}} and Marcello Benedetti.\\
    Publication: \href{https://www.mdpi.com/1099-4300/23/10/1281}{Entropy 2021, 23(10), 1281}.\\
    Preprint: \href{https://arxiv.org/abs/2110.04253}{ArXiv: 2110.04253}.
\end{enumerate}

{
\hypersetup{linkcolor=blue}
\tableofcontents
}
\listoffigures

\end{preliminary}



\chapter{Introduction \& background} \label{chap:introduction}

\begin{chapquote}{Doctor Who, series 1, episode 13}
    \textbf{The Doctor:} ``. . . It's a bit dodgy, this process. You never know what you're going to end up with.''
\end{chapquote}

\section[\texorpdfstring{\color{black}}{} Introduction]{Introduction}

Quantum machine learning (QML) is nascent, and has the potential to dramatically impact the lives of every human in ways unbeknownst to them. \indent This is perhaps not surprising, given the nature of its parent fields. Both machine learning (ML) and quantum computing (QC) have the potential, individually, to reach into and impact our lives in myriad ways. On one hand, for \emph{classical} machine learning, this potential is partially realised, and ML is now commonplace in the products we consume and the services provided to us. This is due to many reasons, including the abundance of `big data' given to machine learning algorithms, and the availability of `big compute' in the development of specialised hardware for performing intensive machine learning calculations, like tensor/graphics processing units (TPUs/GPUs). On the other hand, we have quantum technologies (including, but not limited to, quantum computing, quantum information, quantum cryptography and quantum sensing and metrology), whose potential is (to date) mostly unrealised, but theoretically strong.

One key variable in modern times is the existence (and rapid development) of \emph{small scale} quantum computers. These devices utilise many features of quantum mechanics which are desirable but, due to their small size, are burdened by the unwelcome aspects also. One primary example is the destruction of quantum information via outside interaction, which is difficult to eliminate. Fortunately, once we have quantum computers of a suitable scale, these undesired `decoherences' can be corrected via mechanisms from quantum error correction theory. At the time of writing, however, large enough `self-correcting' quantum computers are at least (optimistically) $10$ years away. In the meantime, we would like to use the small systems we have for any interesting purpose whatsoever. 

As we have already hinted at, an area which has emerged as a promising source of applications is in machine learning. However, one may still ask the very pertinent question why should we have any reason to expect quantum computers to help machine learning? Of course it is understandable why one may want them to; machine learning is everywhere and is already (and will continue to) changing our world, and the way we interact with it in many ways. If quantum computers can aid or accelerate machine learning solutions, one could imagine a plethora of scientific and business use-cases, with real world impact. 

One commonly expressed answer is that both, in many cases, can be reduced a core element of multiplying large matrices; performing linear algebra in high dimensional spaces. Such operations are crucial for machine learning, hence the development of specialised (classical) hardware to do just these operations very efficiently. Quantum computers do these operations very naturally. Such an argument is not dissimilar to the original motivation for quantum computers, proposed by Feynman and others, in the simulation of complex physical systems. Indeed, tools from `quantum simulation' provided some of the early proofs of `exponential speedups' that quantum computers could (potentially) deliver to machine learning problems. These proofs kicked off the field of quantum machine learning proper. However, in contrast with quantum simulation, it becomes significantly trickier to be sure of a guaranteed advantage in machine learning using quantum techniques. In many cases, it is very likely that classical machine learning may perform (almost) equally as well, and in these cases, one would be very justified to ask why should we bother building an expensive new technology? Indeed, recent `dequantisations' of quantum algorithms have done exactly this, and reduced dramatically our hope for large speedups in ML problems, except perhaps in edge cases. 

However, fortunately for this Thesis, the nature and flavour of research quantum machine learning has drastically shifted the last few years, coinciding with the simultaneous development of the small quantum computers mentioned above. With direct access to quantum hardware the question somewhat changed from `How can quantum computers \emph{deliver speedups} to machine learning problems?' to `How can we \emph{use} the quantum computers \emph{we have now} for machine learning problems?'. Such a mindset change has delivered an explosion of engagement and excitement to a field which was previously dominated almost exclusively by provable computer science. Now, anyone with an internet connection may conduct quantum machine learning research. Of course, the theoretical aspects of quantum machine learning, and the development of provable algorithms is still as important as ever. In terms of motivation and content, this Thesis sits somewhere in the intersection. In particular, we aim to bring provable guarantees and theoretical justifications to a more experimental \emph{plug-and-play} approach.

In order to set the scene for this Thesis, and to hopefully aid reading, we provide two introductory viewpoints. The first, is for machine learning practitioners to grasp some useful intuition about what quantum technologies have to offer. The second is for the reverse scenario; to give quantum scientists a brief glimpse into how quantum technologies may impact machine learning.

\section[\texorpdfstring{\color{black}}{} Quantum computing for machine learning]{Quantum computing for machine learning} \label{sec:qc_for_ml}

Modern computers and information technology are embedded in almost every aspect of our daily lives. Given this ubiquity, it is difficult to look back in time and imagine an era in which computers were no more than highly specialised pieces of equipment, used only for tedious calculations and research purposes. Even Thomas Watson, chairman of IBM, famously said in $1943$; ``\emph{I think there's a world market for maybe five computers}''. It is not hard, however, to observe the modern development of quantum computers and draw parallels between the two eras. Similar to the clunky computers of the mid-1900s, quantum devices of the current day are only accessible via a few large industrial bodies, fill entire rooms in some cases, lack large computational capacity and are used primarily for research purposes. Unlike the pioneers of the past, modern `quantum engineers' have the luxury of a precursor, and can use this information to guide and accelerate development of quantum technologies. Unlike Watson, we almost have the reverse problem; not imagining the scope of potential use cases for the novel technology, but actually tempering the hype surrounding it, with claims\footnote{Whether these are true or not is unknown.} that quantum computing will revolutionise every aspect of our lives, from finance to engineering to medicine. As such, the cautious quantum algorithm developer must not only design the algorithm, but also provide evidence that \emph{no classical algorithm could achieve the same thing}; a so-called `quantum advantage'. This is part of the reason why quantum algorithm development is difficult.

A common misconception is that this advantage\footnote{Here we use the term `advantage' loosely to mean `quantum doing \textbf{something} better than classical in \textbf{some} capacity'. In fact, the nature and manifestation of quantum advantage is a subtle, deep and sometimes controversial point~\cite{ronnow_defining_2014, arute_quantum_2019, cho_ibm_2019}. We solidify what we mean by quantum advantage later at the relevant points in this Thesis.} is due to the fact that quantum computers can `\emph{try all solutions in parallel}' (since the fundamental building block of quantum logic, the qubit, can exist in a superposition of all possible states simultaneously), and hence arrive at the problem solution exponentially faster than is possible using purely `classical' logic. In reality, superposition is only one aspect of quantum advantage (other sources include interference, non-locality, and contextuality) and, in practice, careful manipulation\footnote{Both in theory and in practice.} of quantum systems is required to realise such advantages and build quantum algorithms. The famous algorithms that kindled early interest in quantum computation, such as Shor's celebrated factoring algorithm, and Grover's algorithm to accelerate search, do exactly this - using the natural abilities of quantum computers to deal with complex problems.

It is also commonly stated that quantum computers are a natural solution to the demise of Moore's law, as the rapid decrease of available real estate on integrated circuits precipitates the emergence of quantum effects, which are intentionally suppressed by chip manufacturers. It is perhaps less likely that quantum computers will `replace' modern classical computers, but instead will become another piece of specialised hardware used for specific problems only. It is not likely we will have quantum mobile phones any time in the (near) future. With this perspective in mind, it is more natural to fit quantum computers into the current machine learning ecosphere, supplementing the specialised computing devices that we use currently for machine learning tasks, such as GPUs.

Looking at the similar development track between classical and quantum computational hardware, one may also draw parallels between early machine learning, and the development of its quantum counterpart. As mentioned above, much of the success of modern machine learning and deep learning is due to the access to \emph{hardware}. Similarly, access to (albeit small scale) quantum computers has changed the manner in which much of quantum machine learning research is conducted. The physical demonstrations of `quantum computational supremacy' beginning in 2019 indicated that we are now at a transition period, as fully programmable devices exist which can perform tasks out of the reach of even the largest supercomputer on the planet. These tasks are not yet \emph{useful} in any practical sense, but are specifically designed to play to the natural strengths of the devices on which they are implemented. As mentioned above, access to these devices enables a new, more experimental type of quantum machine learning research. In many cases, we have to sacrifice provable guarantees for the algorithms we run on near term quantum devices, since they are primarily heuristic in nature (not very dissimilar to modern machine learning in fact), but in return we gain extreme flexibility in algorithm design and we can simply try them and see.

\section[\texorpdfstring{\color{black}}{} Machine learning for quantum computing]{Machine learning for quantum computing} \label{sec:ml_for_qc}

According to Arthur Samuel, a pioneer credited with the popularisation of the name ``machine learning'', the field of machine learning is the ``\emph{field of study that gives computers the ability to learn without being explicitly programmed}''. This is perhaps a slight misconception, since modern computers, despite being incredibly successful at playing complex games such as chess, or Go\footnote{\href{https://deepmind.com/research/case-studies/alphago-the-story-so-far}{AlphaGo: The story so far}.}, are not considered to be `intelligent', or even learn in the same manner that humans do. In a more practical definition, machine learning models and algorithms reproduce and importantly, \emph{generalise} from observations, or data. The human brain has a remarkable capacity for learning and generalisation, and ML algorithms have only been able to emulate this by focusing on very specialised scenarios, with limited cross-domain applicability.

Nevertheless, machine learning has the potential to be (and already is) extremely impactful in our lives. For example, aiding doctors in reducing false positive and negative cancer diagnoses\footnote{\href{https://www.nature.com/articles/d41586-020-00847-2}{How AI is improving cancer diagnostics}.}, or reducing road traffic accidents with autonomous vehicles by removing a major cause of accidents; human error\footnote{\href{https://www.zdnet.com/article/how-autonomous-vehicles-could-save-over-350k-lives-in-the-us-and-millions-worldwide/}{How autonomous vehicles could save over 350K lives in the US and millions worldwide}.}.

Just as machine learning is ubiquitous in our daily lives, it has also become an extremely useful tool in many aspects of quantum science and technology. For example, it is used to calibrate and stabilise quantum experiments, which was an instrumental piece in the quantum computational supremacy experiment discussed in the previous section. It is has also been used as a tool for foundational research. For example, in sifting through large amounts of data produced by the large hadron collider in CERN, looking for patterns in which new particles many be lurking. Machine learning has also been used successfully in the representation of quantum states~\cite{carleo_solving_2017} or even in the discovery of new experiments entirely~\cite{krenn_automated_2016}.

While all of these are certainly impactful and exciting applications, there are a number of problems with modern machine and deep learning. The first is the extreme expense required to train huge models, for example the recent demonstration of the impressive natural language processing model, GPT-3\footnote{GPT-3 stands for the third iteration of a `generative pre-trained transformer'~\href{https://arxiv.org/abs/2005.14165}{Language Models are Few-Shot Learners}.}, reportedly cost an estimated $\$12$ million dollars to train. Secondly, deep learning is very hungry for high-quality data, the lack of which in many situations can lead to poor results. Manually collecting and labelling the data required for supervised learning is an expensive an time consuming task, having to be done manually by humans in many cases. It can also be very difficult to \emph{interpret} modern machine learning models, in particular large neural networks, which arrive at problem solutions via the complex interaction between their billions of parameters. How an individual parameter correlates with the network output is almost incomprehensible to human interpretation. This latter limitation is extremely important in areas where machine learning is used in sensitive issues, for example policy decisions or medical diagnoses. We want to know \emph{why} the model is doing what it is doing.

It is largely hoped that quantum computers may be able to help with at least some of these problems. For example, the speedups promised by quantum machine learning algorithms (those based on high dimensional linear algebra) are claimed to be more interpretable also than classical counterparts, as the ability to run them many more times can give insights into the decisions being made by them. It is also hoped that quantum devices may be able to aid with `small data, big compute' problems\footnote{The small data part is perhaps more of a \emph{necessity}; in order to run a QML algorithm the data must be first loaded onto the quantum computer which may be tricky and time consuming for large datasets.}, which fits nicely into situations where large amounts of data are not accessible, for example in diagnosing patients with rare medical conditions, of which there may only be a handful of examples. As mentioned in the previous section, it is clear that the nature of a quantum advantage in machine learning is also subtle, and to date it is largely unknown how quantum computers may help the field. Regardless, it also clear is that the rewards are great for finding such a thing, which makes it a very exciting goal to strive for.

\section[\texorpdfstring{\color{black}}{} Thesis overview]{Thesis overview}
Before diving into background material, let us provide a brief summary of the contributions from the primary chapters of the Thesis. Each chapter provides one application and model and the chapters are ordered relative to an increasing complexity of the data presented to the application in question.

\begin{itemize}
    \item \chapref{chap:classifier} focuses on the use of the variational quantum algorithm (VQA) and the parametrised quantum circuit (PQC) for the supervised learning task of classification. Here, our aim is to study the means in which data can be \emph{encoded} into PQCs in a way to be \emph{robust} to some of the noise sources present in NISQ computers. We find that by focusing on the use of the quantum device for an application specific task, we can gain some noise robustness simply by careful construction of the classifier model. By robustness in this context, we mean the preservation of classification results before and after the noise channel is applied. We find that data encodings which preserve classification will always exist, and discuss the trade offs in finding suitable encodings in practice. We provide several theoretical results and extensive numerics to supplement this question. The work of this chapter was based on a collaboration with Ryan LaRose from Michigan State University, and resulted in the publication \href{https://link.aps.org/doi/10.1103/PhysRevA.102.032420}{\textit{Physical Review A 102, 032420 (2020)} - Robust data encodings for quantum classifiers.}
    
    \item \chapref{chap:born_machine} is concerned with a PQC for the purpose of generative modelling, which falls into the category of unsupervised learning. The specification of the PQC is referred to as a Born machine, since the statistics it generates originate directly from Born's rule of quantum mechanics. We study several aspects of the application of this model to the problem of generative modelling. Firstly, an argument about provable quantum advantage with a Born machine is presented. We then describe new training methods for the Born machine. Finally, we provide extensive numerics on three datasets. Here, we begin by demonstrating the effectiveness of the training methods. We next provide an example of a real world use case with a financial dataset; a Born machine as a \emph{market generator}, and compare against the restricted Boltzmann machine for this problem. We finally turn to a quantum dataset, and propose the use of the Born machine as a weak method of quantum compilation. The discussions of quantum advantage, and the training methods for the Born machine (plus related numerics) were the result of a collaboration between Daniel Mills, Vincent Danos and Elham Kashefi from the University of Edinburgh. This resulted in the publication \href{https://www.nature.com/articles/s41534-020-00288-9}{\textit{npj QI 6, 60} - The Born Supremacy: Quantum Advantage and Training of an Ising Born Machine.}. The part of this chapter containing the numerics for the Born machine relating to the financial dataset, and the comparison with the restricted Boltzmann machine were the result of a collaboration with Niraj Kumar and Elham Kashefi from the University of Edinburgh, and Max Henderson, Justin Chan and Marco Paini from Rigetti computing. This resulted in the publication \href{https://iopscience.iop.org/article/10.1088/2058-9565/abd3db/meta}{\textit{QST, 6(2)} - Quantum versus Classical Generative Modelling in Finance.}
    
    \item Finally, \chapref{chap:cloning} introduces our third application, the use of a PQC in a quantum foundations problem, resulting in a new variational algorithm for the approximate cloning of quantum states, $\VQC$. For this algorithm, we prove notions of faithfulness and derive gradients for the cost functions we propose. We also discuss the existence of barren plateaus in the algorithm. As a new research direction, we propose \emph{variational quantum cryptanalysis}; the merging of quantum cryptography with quantum machine learning, and demonstrate the applicability of $\VQC$ in this context. Concretely, we study quantum protocols whose security reduces to quantum cloning (specifically quantum key distribution, and quantum coin flipping), and show how $\VQC$ can be used to discover new attacks on these protocols which are directly \emph{implementable}, given only a specification of the available resources from a particular quantum device. For quantum coin flipping, we also provide new theoretical analyses of two example protocols, into which $\VQC$ can be inserted. This chapter is the result of a collaboration with Mina Doosti, Niraj Kumar and Elham Kashefi from the University of Edinburgh and resulted in the preprint \href{https://arxiv.org/abs/2012.11424}{\textit{ArXiv: 2012.11424} - Variational Quantum Cloning: Improving Practicality for Quantum Cryptanalysis.}
\end{itemize}

\chapter[Preliminaries I: Quantum information]{Preliminaries I: Quantum information}

\section[\texorpdfstring{\color{black}}{} Quantum computing]{Quantum computing} \label{sec:prelim_quantum_computing}

\begin{chapquote}{Prof. Scott Aaronson}
``Quantum computing is really ``easy" when you take the physics out of it.''
\end{chapquote}

In the year 2000, David DiVincenzo proposed seven ingredients, known as the `DiVincenzo Criteria'~\cite{divincenzo_physical_2000}, required  to construct a physical quantum computer. The first five are the following (the final two refer to quantum communication and are less relevant for our purposes):

\begin{tcolorbox}
\textbf{The DiVincenzo criteria:}
\begin{enumerate}
    \item A scalable physical system with well characterised qubits. (Quantum bits).
    
    \item The ability to initialise the state of the qubits to a simple fiducial state. (Quantum state preparation).
    
    \item A universal set of quantum gates (Quantum operations).
    
    \item A qubit specific measurement capability (Quantum measurement).
    
    \item Long relevant decoherence times (Quantum noise).

\end{enumerate}
\end{tcolorbox}

These criteria refer to the \emph{physical} implementation of each ingredient, and so are relevant for quantum physicists and engineers who directly work with the quantum hardware and qubits. The physical platforms in which these criteria can be realised have many forms, and qubits have been realised in many (competing) technologies. Ions, superconducting circuits or photons are among the most ubiquitous (at the time of writing) mediums in which qubits are realised. We remark a refinement and generalisation of DiVincenzo's criteria have been proposed by Ladd, Jelezko, Laflamme, Nakamura, Monroe and O'Brien (LJLNMO)~\cite{ladd_quantum_2010}. The LJLNMO criteria are only three: \emph{scalability}, \emph{universal logic} and \emph{correctability}, which allow for the possibility for quantum computation to be performed with more general building blocks than qubits (e.g. qu\emph{dits} or continuous variable systems) and allow alternative logic operations than quantum gates (e.g. adiabatic quantum evolution~\cite{farhi_quantum_2000} or measurement-based quantum computation~\cite{raussendorf_one-way_2001}), among other generalisations. In this Thesis, we abstract away the physical implementations or the LJLNMO generalisations. This abstraction is the mathematical model of quantum computation, and in the following sections, we describe the relevant ingredients of it, which match closely with the requirements in DiVincenzo's criteria.

\subsection[\texorpdfstring{\color{black}}{} Quantum states]{Quantum states} \label{ssec:prelim/qc/quantum_states}
A fundamental object in quantum mechanics is the \emph{quantum state}. This quantum state resides in a Hilbert space, $\mathcal{H}$, whose dimension is denoted $d$. A Hilbert space is a vector space equipped with an inner product, and is complete with respect to the norm defined by the inner product. A \emph{qubit} is a quantum state with dimension $d=2$. We can define a basis for the corresponding Hilbert space, whose elements are vectors, $\ket{0}, \ket{1} \in \mathcal{H}$:
\begin{equation} \label{eqn:comp_basis_states_as_vectors}
    \ket{0} = \left(\begin{array}{c}
         1  \\
         0 
    \end{array}\right) \qquad    
    \ket{1} = \left(\begin{array}{c}
         0  \\
         1 
    \end{array}\right)
\end{equation}
The notation $\ket{\psi}$ is known as `Dirac notation' and is called a `ket'. Since $\mathcal{H}$ is a vector space, we can define a `dual' space, $\hilb^{\perp}$, and states in the dual space are known as `bras', denoted $\bra{\psi}$. A vector in the dual space is obtained by taking the complex conjugate transpose of $\ket{\psi}$: $\bra{\psi} := \ket{\psi}^{\dagger}$. Since $\mathcal{H}$ is a vector space, linear combinations of these two vectors also reside in $\hilb$: $\ket{\psi} := \alpha\ket{0} + \beta \ket{1} \in \mathcal{H}$. However, in order for these states to be \emph{valid} quantum states, we impose the restriction that $\ket{\psi}$ must be a vector with norm $1$, which implies that $|\alpha|^2 + |\beta|^2 = 1$. States in the general form of $\ket{\psi}$ are referred to be in \emph{superposition}, since a measurement of this qubit will reveal one of the two possible states, $\ket{0}, \ket{1}$ with some \emph{probability}. We return to this point in \secref{ssec:prelim/qc/quantum_measurements}. This probability is defined by the \emph{amplitudes}, $\alpha, \beta$, of the state which can be complex numbers in general, $\alpha, \beta \in \mathbb{C}$. 

A vector representation of a general qubit in superposition can be written as:
\begin{equation} \label{eqn:single_qubit_state_as_vectors}
    \ket{\psi} =  \alpha\ket{0} + \beta \ket{1} =  \alpha\left(\begin{array}{c}
         1  \\
         0
    \end{array}\right) + \beta \left(\begin{array}{c}
         0  \\
         1 
    \end{array}\right) = \left(\begin{array}{c}
         \alpha  \\
         \beta
    \end{array}\right) 
\end{equation}
The qubit is a fundamental building block of quantum logic, and is named to draw parallels between the classical logic unit, the \emph{bit}. Unlike the qubit, a classical bit has only definite states it can reside in, i.e. a bit, $b$, can only be `on' ($b=1$) or `off' ($b=0$), with no intermediate possibilities. Since $\hilb$ is a Hilbert space, it is equipped with an inner product defined between two states, $\ket{\phi} =\left[
    \begin{array}{cc}
         \gamma & \delta 
    \end{array}\right]^T, \ket{\psi}= \left[
    \begin{array}{cc}
         \alpha & \beta 
    \end{array}\right]^T$ as:
\begin{equation}\label{eqn:inner_product}
    \braket{\phi}{\psi} 
    = \left[
    \begin{array}{cc}
         \gamma^*  & \delta^*
    \end{array}
    \right]\left[
    \begin{array}{c}
         \alpha  \\
         \beta 
    \end{array}\right] = \gamma^*\alpha + \delta^*\beta 
\end{equation}
The vectors, $\ket{0}, \ket{1}$ are of course not a unique choice for a qubit basis. Any spanning set of linearly independent vectors will suffice to build a basis, but those in \eqref{eqn:comp_basis_states_as_vectors} are usually called the `computational basis'. Two other important bases are $\{\ket{+}, \ket{-}\}$ and $\{\ket{+\irm}, \ket{-\irm}\}$, given by:
\begin{align}
     \ket{+} :=  \frac{1}{\sqrt{2}}\ket{0} +\ket{1}), \qquad 
    &\ket{-} :=  \frac{1}{\sqrt{2}}(\ket{0} - \ket{1})\\ 
    \ket{+\irm} :=  \frac{1}{\sqrt{2}}(\ket{0} + \irm\ket{1}), \qquad &\ket{-\irm} :=  \frac{1}{\sqrt{2}}(\ket{0} -\irm\ket{1})
\end{align}
These three sets of states, $\{\ket{\sfrac{0}{1}}\}$, $\{\ket{\sfrac{+}{-}}\}$ and $\{\ket{\sfrac{+\irm}{-\irm}}\}$ are the eigenstates of the \emph{Pauli} matrices, which we shall introduce shortly.


\subsubsection[\texorpdfstring{\color{black}}{} Mixed states]{Mixed states} \label{sssec:prelim/qc/quantum_states/mixed_states}
The formalism described above is actually not sufficient to capture the full generality of a possible quantum state. Specifically, the state presented in \eqref{eqn:single_qubit_state_as_vectors} is an example of a \emph{pure} quantum state - any correlations present in this state are `fundamentally quantum'\footnote{This will be an important distinction in~\chapref{chap:born_machine}.}. In general, we may have a quantum state, which also contains some \emph{classical} randomness, or uncertainty. For example, we could imagine instead of having a single (pure) quantum state, $\ket{\psi}$, we may have an \emph{ensemble} of $N$ (pure) states, $\{\ket{\psi_i}\}_{i=1}^N$. Furthermore, we may have some probability distribution, $\{p_i\}_{i=1}^N$ over the elements of this ensemble. From this, we can construct a \emph{mixed} quantum state:
\begin{equation} \label{eqn:mixed_quantum_state_definition}
    \rho := \sum\limits_{i=1}^N p_i \ketbra{\psi_i}{\psi_i}   
\end{equation}
which is the effective state we would generate if we chose to prepare one of the pure states, $\ketbra{\psi_i}{\psi_i}$\footnote{The notation $\ketbra{\psi}{\psi}$ is nothing more than the \emph{outer} product of the state, $\ket{\psi}$, with itself. Since $\ket{\psi}$ is a column vector, and $\bra{\psi}$ is a row vector, $\ketbra{\psi}{\psi}$ is a matrix.}, with probability $p_i$. The density matrix formalism allows us to model our \emph{uncertainty} about which (pure) state the system is actually in. Formally speaking, a density matrix is an \emph{operator} on the Hilbert space: $\rho: \hilb \rightarrow \hilb$ and we denote the space of density matrices\footnote{We later use the shorthand notation $\mathcal{S}_n$ to represent the space of $n$-qubit density matrices.} to be $\mathcal{S}(\hilb)$, which is a convex set. One important property of density matrices is that they have trace $1$; $\tr(\rho) = 1$, which ensures probability conservation.


\subsubsection[\texorpdfstring{\color{black}}{} More qubits]{More qubits} \label{sssec:prelim/qc/quantum_states/multiple_qubits}
One qubit, however, is not usually sufficient to do anything interesting - the evolution of a single two level quantum system can be easily simulated classically by multiplying $2\times 2$ matrices. The power of quantum computing comes into being when multiple quantum systems are added, and it turns out quantumly\footnote{We use this widely adopted phrase to mean anything done via quantum mechanical means.} (unlike with classical systems) a many body quantum system is worth more than the sum of its parts. 

The most common formalism used in quantum computing to describe multiple systems is the \emph{tensor product} model. Formally, a tensor product between two Hilbert spaces, $\hilb_1$ and $\hilb_2$ is denoted by $\hilb_{1,2} := \hilb_1 \otimes \hilb_2$, which is also a vector space. A basis of $\hilb_{1,2}$ is formed by taking the tensor product of the basis elements of the component Hilbert spaces.

However, not all states in $\hilb_{1,2}$ can be written in the form $\rho_1 \otimes \rho_2$ for some $\rho_1 \in S(\hilb_1), \rho_2 \in S(\hilb_2)$. Such states are by definition \emph{entangled}. More precisely:
\begin{defbox}
\begin{definition}[Entangled \& separable states]~ \\
    Any state $\rho_{1,2}$ in $\mathcal{S}(\hilb_{1,2})$ which can be written as: 
    \begin{equation}\label{eqn:entangled_state_defin}
        \rho_{1,2} = \sum\limits_i p_i \rho_1^i\otimes \rho_2^i
    \end{equation}
    for some $\{p_i\}, \sum_i p_i =1$ and for some states $\{\rho_1^i\}$ in $\mathcal{S}(\hilb_1)$ and $\{\rho_2^i\}$ in $\mathcal{S}(\hilb_2)$ is \emph{separable} and any state which is not separable is \emph{entangled}.
\end{definition}
\end{defbox}
A special case of separability occurs when there is just one $p_i$, in which the state is called a \emph{product} state.

Using the tensor product and density matrix formalism, we can also describe individual subsystems of a multi-qubit state. This is achieved using the \emph{partial trace} and \emph{reduced} density operators. Let us take $\rho_{1,2}$ from above. We can recover the subsystem $1$ by taking the partial trace over subsystem $2$ (\cite{nielsen_quantum_2010})
\begin{equation}\label{eqn:prelim/qc/quatum_states/reduced_density_mat}
    \rho_1 := \Tr_{2}\left(\rho_{1,2}\right)
\end{equation}
defined by:
\begin{equation}\label{eqn:prelim/qc/quatum_states/partial_trace}
    \Tr_{2} \left(\ketbra{\alpha_1}{\beta_1} \otimes \ketbra{\gamma_2}{\delta_2}\right) :=   \ketbra{\alpha_1}{\beta_1} \Tr \left(\ketbra{\gamma_2}{\delta_2}\right) = \ketbra{\alpha_1}{\beta_1} \braket{\gamma_2}{\delta_2} 
\end{equation}
where $\ket{\alpha_1}, \ket{\beta_1} \in \hilb_1$, $\ket{\gamma_1}, \ket{\delta_2} \in \hilb_2$. This action is referred to as `tracing out' the subsystem $2$.

\subsection[\texorpdfstring{\color{black}}{} Quantum operations]{Quantum operations} \label{ssec:prelim/qc/quantum_operations}
A general quantum operation is known as a \emph{channel}, which maps quantum states to quantum states in two different Hilbert spaces:
\begin{equation} \label{eqn:quantum_channel}
    \E : \mathcal{S}(\hilb_1) \rightarrow \mathcal{S}(\hilb_2), \qquad \E(\rho_1) = \rho_2
\end{equation}
There are multiple ways to think about the interpretation of such channels~\cite{nielsen_quantum_2010}. For example, as an interaction between the quantum system and some \emph{environment}. Alternatively, in a more mathematical sense using the \emph{operator-sum} formalism. Finally, we could build an interpretation from physically motivated principles or axioms we would expect quantum processes to obey. It turns out that these three viewpoints are equivalent, and they may each have their own utility in a particular scenario \cite{nielsen_quantum_2010}. For mathematical usefulness, we primarily use the operator-sum formalism for the remainder of this Thesis.

In this formalism, the channel $\E$, can be represented as:
\begin{equation} \label{eqn:operator_sum_rep}
    \rho' := \E(\rho) = \sum\limits_k E_k \rho E_k^\dagger
\end{equation}
where the action on $\rho$ is specified by $k$ \emph{operation elements} (or \emph{Kraus} operators~\cite{kraus_states_1983}) $\{E_k\}$ which can be specified as complex-valued matrices. In this form, the operator $\E$ can be proven to be \emph{completely positive} (CP) and we also require a completeness relation on the operators:
\begin{equation} \label{eqn:kraus_completeness_relation}
    \sum_k E_k^\dagger E_k \leq \mathds{1}
\end{equation}
In order to ensure the conservation of probabilities through $\E$. Furthermore, $\E$ becomes a completely positive \emph{trace preserving} (CPTP) map if we have an equality in \eqref{eqn:kraus_completeness_relation}. The relationship to the trace of the input and output quantum states can be seen as follows:
\begin{equation}
    \Tr(\rho') = \Tr\left(\sum\limits_k E_k \rho E_k^\dagger\right) = \Tr\left(\left[\sum\limits_k  E_k^\dagger E_k\right] \rho\right) \leq  \Tr\left(\mathds{1}\rho\right) = \Tr(\rho)
\end{equation}
Beginning with definitions of quantum channels in this way allows in the following sections to look at special cases of the above to tease out the parts relevant to us. An example of trace \emph{decrease} in the system $\rho'$ is where information is lost to an environment (see below) as a result of a measurement.

However, for the remainder of this Thesis, we are concerned only with CPTP maps, and \emph{unitary} operations, which make up a core element of quantum computation.
As mentioned above, we can also envisage a quantum operation via an interaction with an `environment' Hilbert space, denoted $\hilb_E$. However, in accordance with quantum mechanics, this interaction must be \emph{unitary}, and can be described by a unitary matrix, $U$.
A unitary matrix, is a complex, square matrix defined by the property,
\begin{equation} \label{eqn:unitarity_definition}
    U^{-1} = U^\dagger
\end{equation}
In order to implement a general channel, $\E$, we can imagine a quantum state in the environment Hilbert space, $\rho_E$, and then a unitary operation acting on both quantum states. The action of the channel on the target, $\rho$, is recovered by tracing out the environment subsystem:
\begin{equation} \label{eqn:quantum_operation_defn_with_environment}
    \E(\rho) = \tr_E\left(U \rho \otimes \rho_E U^\dagger\right)
\end{equation}
A special case of the above is when we have \emph{no} environment, and the unitary acts on the target system directly. If we choose $k=1$ and $E_1=U$ from \eqref{eqn:operator_sum_rep} we get:
\begin{equation}
    \rho' = U\rho U^\dagger
\end{equation}
In this case, the system, $\rho$, is \emph{closed}. A further simplification occurs if $\rho$ is a pure state, in which case we can represent the transition simply as:
\begin{equation}
  \ket{\psi'} = U\ket{\psi}
\end{equation}
A neat (classical) comparison with unitary evolution is the analogue of stochastic matrices acting on probability vectors. The fundamental difference is that in quantum mechanics, the probability vector is replaced with an \emph{amplitude} vector, whose elements may be complex.

Unitary evolution is therefore the key driving element in quantum mechanics, and indeed in quantum computation, where the state $\ket{\psi'}$ is the `output' of a quantum processor, on input $\ket{\psi}$. Given this, the next relevant question is how can we actually implement such unitaries on quantum devices to drive computation, especially given the apparent exponential complexity of the problem\footnote{For example, a unitary acting on an $n$ qubit quantum state (containing $2^n$ complex amplitudes) has dimension $2^n\times 2^n$, which is quite a big matrix.}. It turns out that this can be done in terms of simple single and two-qubit quantum operations on quantum processors, and we shall discuss this in the next section.

\subsection[\texorpdfstring{\color{black}}{} Quantum gates]{Quantum gates} \label{ssec:prelim/qc/quantum_gates}
In classical computation, the \emph{circuit model} is a useful computational model in which large operations are built by composing smaller ingredients (`gates') in a `circuit' acting on bit registers. In quantum computation, the analogue is, perhaps not surprisingly, the \emph{quantum} circuit model\footnote{Just as with classical computation, the circuit model is not the unique way to describe computation. Quantum computers could alternatively be driven by adiabatic evolution~\cite{farhi_quantum_2000}, measurement-based quantum evolution~\cite{raussendorf_one-way_2001} which are equivalent to the circuit model from a complexity point of view. In this Thesis, we only require the circuit model so we neglect further discussion of alternatives.}. Here, we require the ability to generate and implement the unitary transformations discussed in the previous section. It turns out that we can do so with the help of a \emph{universal} set of \emph{quantum} gates. Quantum gates are the logic operations which act on quantum information, analogous to \texttt{AND}, \texttt{OR} and \texttt{NOT} gates in the classical circuit model.
\begin{defbox}
\begin{definition}[Universal set of quantum gates~\cite{nielsen_quantum_2010}]\label{defn:prelim/qc/universal_gate_set}~ \\
A set of quantum gates, $\mathcal{G} = \{\mathsf{G}_i\}$, is said to be universal for quantum computation if any unitary operation may be approximated to an arbitrary accuracy using a quantum circuit involving only those gates.
\end{definition}
\end{defbox}
There are many candidates for universal gate sets, each of which has advantages and disadvantages. For practicality of implementation\footnote{We do not take into account the \emph{efficiency} of implementing an arbitrary unitary in terms of this universal set. Many unitaries may require exponentially many operations from the set to be implemented~\cite{nielsen_quantum_2010}. We therefore hope that at least some of the unitaries which \emph{can} be implemented using polynomially many gates (i.e. those that we can implement on a quantum computer) are useful in solving problems of interest.}, it is also sufficient to restrict to a \emph{discrete} set of gates and the price we pay for this is an error in the approximation of the unitary\footnote{In order to \emph{exactly} build an arbitrary unitary, the gate set is required to be infinite in size (e.g. to contain \emph{all} single qubit operations, of which there are infinitely many).}. 

We first list some useful single and two qubit quantum gates, and then comment on their universality. Firstly, we have the canonical \emph{Pauli} matrices\footnote{The specific representation of these unitary matrices is basis dependent. In this Thesis we assume that everything is relative to the computational (or `Pauli-$\ZG$') basis - i.e. those matrices which are diagonal have computational basis states as eigenvalues.}:
\begin{equation}\label{eqn:pauli_matrices}
 \XG = \left(\begin{array}{cc}
      0 & 1 \\
      1 & 0
 \end{array} \right),  \qquad
 \ZG = \left(\begin{array}{cc}
      1 & 0 \\
      0 & -1
 \end{array} \right),\qquad 
\YG = \left(\begin{array}{cc}
      0 & -\irm \\
      \irm    & 0
\end{array} \right)
 \end{equation}
Each of these gates induces the following transition on the computational basis states:
\begin{align}
    &\XG\ket{0} = \ket{1}, \YG\ket{0} = -\irm\ket{1},  \ZG\ket{0} = \ket{0} \\
    &\XG\ket{1} = \ket{0}, \YG\ket{1} = \irm\ket{0},  \ZG\ket{1} = -\ket{1} 
\end{align}
The `$\XG$' gate is the only one has some classical analogue, also being known as the `bit flip' gate (or $\mathsf{NOT}$ gate). It flips the computational basis state `0' to a `1' state and vice versa. Also to be noted is the effect of the Pauli-$\ZG$ gate (or $\mathsf{PHASE}$ gate), which only adds a phase (of $-1$) to the $\ket{1}$ state. The strangest action is that of the Pauli-Y gate, which flips the computational basis state but \emph{also} adds an imaginary (!) phase.

Of use for our purposes, are the following operations generated by these matrices, which are intuitively rotations around the corresponding axes of the Bloch sphere\footnote{The Bloch sphere is a convenient illustrative tool to represent single qubit states, which completely breaks down if we introduce multiple qubits. Nevertheless, we use it extensively in \chapref{chap:classifier}.}:
\begin{equation}\label{eqn:single_qubiht_rotations}
 \begin{split}
 &\RX(\theta) := e^{-\irm\frac{\theta}{2} \XG} = \cos\left(\frac{\theta}{2}\right)\mathds{1} - \irm\sin\left(\frac{\theta}{2}\right)\XG = \left(\begin{array}{cc}
      \cos\left(\frac{\theta}{2}\right) &  -\irm\sin\left(\frac{\theta}{2}\right)  \\
      -\irm\cos\left(\frac{\theta}{2}\right)  & \cos\left(\frac{\theta}{2}\right) 
 \end{array} \right),  \\
 &\RZ(\theta) := e^{-\irm\frac{\theta}{2} \ZG} = \cos\left(\frac{\theta}{2}\right)\mathds{1} - \irm\sin\left(\frac{\theta}{2}\right)\ZG = \left(\begin{array}{cc}
      e^{-\frac{\irm\theta}{2}} &  0  \\
      0  & e^{\frac{\irm\theta}{2}}
 \end{array} \right),  \\
 &\RY(\theta) := e^{-\irm\frac{\theta}{2} \YG} = \cos\left(\frac{\theta}{2}\right)\mathds{1} -\irm\sin\left(\frac{\theta}{2}\right)\YG = \left(\begin{array}{cc}
\cos\left(\frac{\theta}{2}\right) &  -\sin\left(\frac{\theta}{2}\right)  \\
      -\cos\left(\frac{\theta}{2}\right)  & \cos\left(\frac{\theta}{2}\right)
 \end{array} \right). 
\end{split}
 \end{equation}
The final distinct relevant gate is the Hadamard gate:
\begin{equation}
\HG = \frac{1}{\sqrt{2}}
\left(
\begin{array}{cc}
      1 & 1 \\
      1   & -1
\end{array} 
\right)
 \end{equation}
which translates between the Pauli-$\XG$ and Pauli-$\ZG$ basis (i.e. transforming eigenvalues from one basis to the other):
\begin{equation}
    \HG\ket{0} = \ket{+} = \frac{1}{\sqrt{2}}\left(\ket{0} + \ket{1}\right)\qquad 
    \HG\ket{1} = \ket{-} = \frac{1}{\sqrt{2}}\left(\ket{0} - \ket{1}\right)
 \end{equation}
Finally, we can list some useful two qubit gates. The two most common are the controlled-$\ZG$ ($\CZ$) and the controlled-$\XG$ ($\CX$) gates, defined in matrix representation as:
\begin{align} \label{eqn:cnot_matrix}
 \CX_{0, 1} := \left(\begin{array}{cccc}
    1  & 0 & 0 & 0 \\
    0  & 1 & 0 & 0 \\
    0  & 0 & 0 & 1 \\
    0  & 0 & 1 & 0 \\
 \end{array}\right) \qquad &= \qquad
 \Qcircuit @C=1em @R=2em {
\lstick{q_0} & \ctrl{1} & \qw \\
\lstick{q_1} & \targ & \qw
}\\  \label{eqn:cz_matrix}
  \CZ_{0, 1} := \left(\begin{array}{cccc}
    1  & 0 & 0 & 0 \\
    0  & 1 & 0 & 0 \\
    0  & 0 & 1 & 0 \\
    0  & 0 & 0 & -1 \\
 \end{array}\right) \qquad &= \qquad
 \Qcircuit @C=1em @R = 2em {
\lstick{q_0} & \ctrl{1} & \qw \\
\lstick{q_1} & \ctrl{-1} & \qw
}
\end{align}
In the above we also include the quantum circuit representation of these gates acting on quantum `registers' or `wires', $q_i$. The circuit representation of the single qubit gates above is simply given by their label (e.g. $\ZG, \HG, \RX(\theta)$) in a box acting on a particular qubit wire. As the name of these two qubit gates suggests, they are \emph{controlled}, meaning they require a `control' qubit and a `target' qubit. The reason for this nomenclature can be more easily seen if the gates are written in the following form:
\begin{equation} \label{eqn:controlled_u_gate}
    \CU_{0, 1} = \ketbra{0}{0}_0 \otimes \mathds{1}_1 +  \ketbra{1}{1}_0 \otimes \mathsf{U}_1 
\end{equation}
Where $\mathsf{U}$ is an arbitrary single\footnote{In fact, there is no need for this restriction.} qubit unitary. If we choose, $\mathsf{U} \equiv \ZG$, it can be checked that the above reduces to the $\CZ$ gate in \eqref{eqn:cz_matrix}. This representation is useful since it can be seen that if the `control' qubit (qubit $q_0$) is in the state $\ket{0}$, the identity operation will be applied to the target qubit (qubit $q_1$), and if it is in the state $\ket{1}$, the desired operation will be applied\footnote{Based on the `projectors' into these states, $\ketbra{0}{0}, \ketbra{1}{1}$. See \secref{ssec:prelim/qc/quantum_measurements} for further discussion about projectors.}. Hence, the gate is controlled on the `control' qubit being `1'. Similarly, $\CX$ flips the qubit $q_1$ if qubit $q_0$ is in the $\ket{1}$ state. For this reason, the $\CX$ gate is also commonly called the $\CNOT$ gate (controlled-NOT), to draw comparison with the classical analogue (the $\mathsf{NOT}$ gate). The subscripts in $\CU_{0, 1}$ are used to keep track of which qubit is the control, and which is the target, but we can usually drop these when the ordering is clear from context. Finally, let us introduce another very useful two qubit gate, the $\SWAP$ gate:
\begin{equation} \label{eqn:swap_gate_matrix_and_circuit}
  \SWAP := \left(\begin{array}{cccc}
    1  & 0 & 0 & 0 \\
    0  & 0 & 1 & 0 \\
    0  & 1 & 0 & 0 \\
    0  & 0 & 0 & 1 \\
 \end{array}\right) \qquad = \qquad
 \Qcircuit @C=1em @R = 2.8em {
\lstick{q_0} & \qswap & \qw \\
\lstick{q_1} & \qswap \qwx & \qw
}= \qquad
 \Qcircuit @C=1em @R = 2em {
\lstick{q_0} & \ctrl{1} & \targ & \ctrl{1} & \qw \\
\lstick{q_1} & \targ & \ctrl{-1} & \targ &  \qw
}
\end{equation}
This gate swaps the two states $\ket{01} \longleftrightarrow \ket{10}$ and therefore has the effect of swapping the quantum information between two registers, $q_0, q_1$\footnote{Since the other two possible states, $\ket{00}, \ket{11}$, are symmetric in $q_0$ and $q_1$.}. This gate is especially useful in near term quantum computers, where qubits (upon which we want to apply a joint operation) may not be directly connected to each other. Therefore, the $\SWAP$ gate serves to \emph{route} qubits around a quantum chip to bring them close enough to interact.

Now, we are somewhat in a position to discuss the universality of some of these gates. As mentioned above, it is sufficient to restrict our attention to a discrete set of universal gates. One of the most common such sets is:
\begin{equation}\label{eqn:discrete_set_universal_gates}
    \mathcal{G}_{\mathrm{discrete}} = \{\HG, \CNOT, \SG, \TG \}
\end{equation}
The $\TG$ and $\SG$ are just rotations around the Pauli-$\ZG$ axis, i.e. $\RZ(\theta)$ for particular choices of the angle $\theta$, \emph{up to a global phase}. We say two unitaries, $\UG, \VG$ are equivalent up to a global phase if we can write $\UG = \erm^{\irm \delta}\VG$, for some `global' phase, $\delta$. Such global phases are typically unimportant, since they do not manifest in a physical way (they cancel out via complex conjugation), and so we do not observe them. The choice of $\theta$ which makes the $\TG$ and $\SG$ is given by $\theta_{\TG} = \pi/8, \theta_{\SG} = \pi/4$. The above choice is not unique for a universal gate set, and there are others one could choose. 

One of the holy grails of quantum computation is the \emph{Solovay-Kitaev} theorem, which says that given an arbitrary single qubit gate, there exists a sequence of single qubit gates from the above set~\eqref{eqn:discrete_set_universal_gates} that can approximate the original unitary up to a precision $\epsilon$. Most importantly, this sequence has only polylogarithmic in $\epsilon^{-1}$, implying that the discreteness is not a problem in practice. This theorem is essential for building scalable quantum computers, but it is not relevant for the remainder of this Thesis so we conclude our discussion of universality with it. For further details on these topics, see the excellent overview given in~\cite{nielsen_quantum_2010} or the original\footnote{Solovay proved and announced the result via a mailing list in 1995, so is unpublished. Kitaev proved the result independently.} paper of~\cite{kitaev_quantum_1997}.

\subsection[\texorpdfstring{\color{black}}{} Quantum measurements]{Quantum measurements} \label{ssec:prelim/qc/quantum_measurements}
In the previous section, we discussed unitary evolution, a special case of a quantum operation in which there is only one operator, $E_1$, in \eqref{eqn:operator_sum_rep}. Let us discuss another extreme, which will cover the case of quantum \emph{measurement}. Measurements in quantum mechanics are crucial ingredients as they are the mechanism by which (classical) information can be extracted from a quantum system, and so are necessary in extracting \emph{answers} from quantum computations.

In contrast to unitary evolution, one can \emph{lose} information as a result of measuring a quantum system. This is because after a measurement, the quantum state of a system will irreversibly change and any information which was not extracted by the measurement will disappear, except in some special cases. 

A quantum measurement is defined by a collection of measurement operators, $\{M_m\}$, where $m$ denotes the outcome of the measurement. For example, if the measurement is binary, only two possible values for $m$ can be observed, $m\in \{0, 1\}$. The \emph{probabilities} of observing these outcomes is given by \emph{Born's rule}:
\begin{tcolorbox}
\textbf{Born's rule of quantum mechanics:}\\
An outcome, $m$, from a measurement on a quantum system, $\rho$, occurs with a probability given by:
    \begin{equation} \label{eqn:prelim/qc/born_measurement_rule}
        p_{\rho}(m) = \Tr(M_m\rho M_m^\dagger)
    \end{equation}
\end{tcolorbox}
Furthermore, if $\rho$ is a pure state, and if $M_m$ describes the special case of a \emph{projective} measurement (meaning the operators, $M_m$, obey $M_m M_{m'}= \delta_{m, m'}M_m$), we have:
\begin{align} \label{eqn:prelim/qc/born_measurement_rule_pure}
    p_{\rho}(m) = \Tr(M_m\rho M_m^\dagger) &= \Tr(M_m\ketbra{\psi}{\psi} M_m^\dagger) = \Tr(\bra{\psi} M_m^\dagger M_m\ket{\psi}) \\
    &= \braket{\psi}{\phi_m}\braket{\phi_m}{\psi} = |\braket{\phi_m}{\psi}|^2 
\end{align}

In this case $\{M_m = \ketbra{\phi_m}{\phi_m}\}$ are called \emph{projectors}\footnote{In general, a projector can be any operator, $P$, such that $P^2 = \mathds{1}$.}, which \emph{project} onto the states, $\ket{\phi_m}$. These states are the eigenvalues of an \emph{observable}, $\mathsf{O}$ that can be written as a spectral decomposition in terms of these eigenvectors, and their eigenvalues, (the outcomes $m$):
\begin{equation} \label{eqn:prelim/qc/observable_defined_by_projectors}
    \mathsf{O} = \sum_{m} m \ketbra{\phi_m}{\phi_m}
\end{equation}
The result of the quantum state after a measurement is:
\begin{equation}
    \ket{\psi} \rightarrow \ket{\psi}_m:= \frac{M_m\ket{\psi}}{\sqrt{\bra{\psi}M_m^\dagger M_m\ket{\psi}}}
\end{equation}
and so a measurement can be interpreted as taking the state $\ket{\psi}$ and replacing it with the state $\ket{\psi}_m$ with a probability given by \eqref{eqn:prelim/qc/born_measurement_rule_pure}. However, if we do not care about the state of a quantum system after a single measurement, there is a useful mathematical formalism which we can use. The \emph{positive operator value measure} formalism (POVM) describes more general measurements and here we define POVM elements as $E_m = M_m^{\dagger}M_m$. We can see from \eqref{eqn:prelim/qc/born_measurement_rule} that the outcome probabilities $p(m)$ can now be completely described in terms of the operators $E_m$\footnote{See \cite{nielsen_quantum_2010} for further discussion about the usefulness of the POVM formalism.}. A simple example of a measurement is the so-called `computational basis measurement', or a `measurement in the Pauli-$\ZG$ basis'. Here, the measurement operators, $M_m$, are given by the outer products of the computational basis states, $M_0 := \ketbra{0}{0}, M_1 := \ketbra{1}{1}$ (it is simple to check that this measurement is projective as well). The corresponding observable for this measurement is exactly the Pauli-$\ZG$ operator, hence the name:
\begin{equation} \label{eqn:prelim/qc/meas/pauli_z_observable}
    \ZG = +1 \ketbra{0}{0} -1 \ketbra{1}{1} 
    = +1\left(\begin{array}{cc}
        1 & 0  \\
        0 & 0
    \end{array}\right) -1 \left(\begin{array}{cc}
        0 & 0  \\
        0 & 1
    \end{array}\right) 
    = \left(\begin{array}{cc}
        1 & 0  \\
        0 & -1
    \end{array}\right) 
\end{equation}
The measurement of observables will be of crucial interest later in this Thesis, as it is one of the key ingredients in \emph{variational} quantum algorithms, which we introduce in \secref{sec:variational_quantum_algorithms}. In particular, we care about the expectation values, and variance of the observables which are defined in the usual way with respect to a specific state, $\ket{\psi}$:
\begin{align}
    \mathbb{E}[\mathsf{O}] &:= \langle \mathsf{O} \rangle_{\psi} := \bra{\psi} \mathsf{O} \ket{\psi} = \Tr\left(\ketbra{\psi}{\psi}\mathsf{O}\right)\label{eqn:observable_expectation_value}\\
    \text{Var}[\mathsf{O}] &:= \langle \mathsf{O}^2 \rangle_{\psi} -  \langle \mathsf{O} \rangle_{\psi}^2 \label{eqn:observable_variance}
\end{align}
As a simple example, take the expectation of the Pauli-$\ZG$ observable on the state $\ket{+} = 1/\sqrt{2}\left(\ket{0} + \ket{1}\right)$. Computing $\langle \ZG\rangle_{+}$ gives:
\begin{equation*}
  \langle \ZG\rangle_{+} = \bra{+}\ZG\ket{+} = \frac{1}{2}\left(\bra{0}\ZG\ket{0} + \bra{0}\ZG\ket{1}+ \bra{1}\ZG\ket{0} + \bra{1}\ZG\ket{1}\right) = 1/2\left(1-1\right) = 0
\end{equation*}
since $\ZG$ has no off-diagonal terms. Furthermore, since $\ZG^2= \mathds{1}$, we can also simply compute $\text{Var}[\mathsf{O}] = 1$.

\subsection[\texorpdfstring{\color{black}}{} Quantum noise]{Quantum noise} \label{ssec:prelim/qc/quantum_noise}
The final special case of a quantum operator (\eqref{eqn:quantum_operation_defn_with_environment}) that is of interest to us is \emph{quantum noise}. Noise in a quantum system is often detrimental to useful quantum computation as (in one form) it can cause decoherence (see Divencenzo's criteria). Roughly speaking this is the loss of information from the quantum state via interaction with its environment~\cite{schlosshauer_decoherence_2005}, i.e. the system $\rho_E$ in~\eqref{eqn:quantum_operation_defn_with_environment}. The preservation of information in a quantum system is crucial for the operation of quantum algorithms, and as such the fields of quantum \emph{error-correction} (QEC) ~\cite{brun_quantum_2019} and \emph{fault-tolerance} has evolved to find useful ways to correct the errors that occur in a physical quantum computer. Fundamental theoretical results such as the \emph{fault-tolerant threshold theorem} (\cite{ knill_resilient_1998, kitaev_fault-tolerant_2003, aharonov_fault-tolerant_2008}) are important to believe in the scalability of quantum computers.  This theorem roughly states that quantum errors can be corrected arbitrarily as long as the physical error rate is below a certain threshold, $p_{\text{th}}$, and implies that errors can be corrected faster than they accumulate.

In the long term, building error-corrected quantum computers is a universally accepted end goal, since we have algorithms with \emph{provable} quantum speedup\footnote{To name the few obvious candidates: Shor factoring (\cite{shor_algorithms_1994}), Grover search (\cite{grover_fast_1996}), and the HHL algorithm (\cite{harrow_quantum_2009}) among others~\cite{montanaro_quantum_2016}.}
which can be run on them. However, for now (and for the foreseeable future), the only devices we have available do not have the ability to perform QEC. Such devices in the current era have been dubbed \emph{noisy intermediate-scale quantum} (NISQ)~\cite{preskill_quantum_2018} computers and they have on the order of $10^1$-$10^3$ noisy qubits. These devices cannot implement QEC and fault-tolerant encoding of quantum information simply as a question of resources, the required amount is significantly higher than allowed by these meager numbers. The usual suspects for the resource hungry nature of QEC and fault tolerance, include but are not limited to, ancilla qubits for error detection, magic state factories~\cite{gidney_efficient_2019} and the requirement for many physical qubits per logical qubit. For example, a recent estimate for factoring $2048$ bit numbers using Shor's algorithm puts the required number of qubits at $\sim 20$ million (\cite{gidney_how_2019}), which is a factor of $200,000$ more than we have available now. Deriving and building applications for such devices is not only an important goal in its own right, but can also be very useful for benchmarking the hardware~\cite{mills_application-motivated_2021}. Since the central theme of this Thesis is to find such applications, 
we will not discuss further the expansive topic of quantum error correction, and instead focus our attention towards applications which can be run \emph{in the presence} of quantum noise and on small-scale quantum devices.

Finally, we note that quantum noise on a device may be either \emph{passive} or \emph{malicious} (or a mixture of both).
The former type of noise is that which occurs naturally in quantum devices - via interaction with outside sources, information is lost in a somewhat `random' fashion (the $E$ in $\rho_E$ really means an environment). In contrast, with malicious (or `adversarial') noise the computation is being \emph{deliberately} corrupted by a malicious adversary in order to gain information about the parties performing the computations. Here $E$ can refer to an `eavesdropper', Eve, who is actively manipulating the ancillary state, $\rho_E$\footnote{The fields of quantum verification~\cite{gheorghiu_verification_2019} and quantum cryptography~\cite{pirandola_advances_2020} have emerged to deal with such adversarial noise.}. When we use the term `noise' in this Thesis we mean the former, passive situation.

Let us begin by introducing some common simple noise channels which we use (primarily in \chapref{chap:classifier}). These models are generally not representative of the true noise on a given NISQ machine\footnote{Real devices may have noise which is correlated across qubits and may be time-dependent, meaning that a single quantum device may even have different performances across different days of the week! As such modelling, characterising and mitigating quantum errors is extremely challenging in practice and is an active area of research.} in isolation, but they are extremely useful as a starting point, and are frequently used in theoretical studies.

In the following for clarity, we use the notation $\rho$ to represent an $n$-qubit quantum state, and $\sigma$ for a single qubit state. Recall in the operator-sum formalism (\eqref{eqn:operator_sum_rep}) a quantum channel can be written as:
\begin{equation} \label{eqn:quantum_noise/operator_sum_rep}
    \rho \mapsto \E(\rho) := \sum_{k = 1}^{K} E_k \rho E_k^\dagger 
\end{equation}
Choosing the Kraus operators in~\eqref{eqn:quantum_noise/operator_sum_rep} gives us the noise channels of interest to us. The first specific instance of which is the Pauli channel.
\begin{defbox}
\begin{definition}[Pauli noise channel]\label{def:pauli_channel}~ \\
    The Pauli channel maps a single qubit state $\sigma$ to $\paul (\sigma)$ defined by
    \begin{equation} \label{eqn:pauli_channel}
        \paul (\sigma) := p_{\mathds{1}}  \sigma + p_X \XG \sigma \XG + p_Y \YG \sigma \YG + p_Z \ZG \sigma \ZG
    \end{equation}
    where $p_{\mathds{1}} + p_X + p_Y + p_Z = 1$, $\mathbf{p} := (p_X, p_Y, p_Z)$. 
\end{definition}
\end{defbox}



Two special cases of the Pauli channel are the bit-flip and phase-flip (dephasing) channel.

\begin{defbox}
\begin{definition}[Bit-flip noise channel]\label{def:bit_flip_channel}~ \\
    The bit-flip channel maps a single qubit state $\rho$ to $\flip (\rho)$ defined by
    \begin{equation} 
        \flip (\sigma) := (1 - p) \sigma + p \XG \sigma \XG
    \end{equation}
    where $0 \leq p \leq 1$. 
\end{definition}
\end{defbox}

While a bit-flip channel flips the computational basis state with probability $p$, the phase-flip channel introduces a relative phase with probability $p$.
\begin{defbox}
\begin{definition}[Dephasing noise channel]\label{def:dephasing_channel}~ \\
    The phase-flip (dephasing) channel maps a single qubit state $\sigma$ to $\deph (\sigma)$ defined by
    \begin{equation} \label{eqn:dephasing_channel}
        \deph (\sigma) := ( 1 - p) \sigma + p \ZG \sigma \ZG 
    \end{equation}
    where $0 \leq p \leq 1$. 
\end{definition}
\end{defbox}

The final special case of the Pauli channel is the depolarising channel which occurs when each Pauli is equiprobable $p_X = p_Y = p_Z = p$ and $p_{\mathds{1}} = 1-3p$.
This channel can be equivalently thought of as replacing the state $\rho$ by the maximally mixed state $\frac{\mathds{1}}{2}$\footnote{A maximally mixed state is one which is fully random, contains no useful information and so is mostly useless for computation.} with probability $p$.
\begin{defbox}
\begin{definition}[Depolarising noise channel]~ \\
\label{def:depolarizing_channel}
    The depolarising channel maps a single qubit state $\sigma$ to $\depo (\sigma)$ defined by
    \begin{equation} \label{eqn:depolarizing_channel}
        \depo (\sigma) := (1 - p) \sigma + p \frac{\mathds{1}}{2}
    \end{equation}
    where $0 \leq p \leq 1$. 
\end{definition}
\end{defbox}
The $d = 2^n$-dimensional generalisation of \defref{def:depolarizing_channel} is straightforward.
\begin{defbox}
\begin{definition}[Global depolarising noise channel]\label{def:global_depolarizing_channel}~ \\
    The global depolarising channel maps an $n$-qubit state $\rho$ to $\depon (\rho)$ defined by
    \begin{equation} \label{eqn:global_depolarizing_channel}
        \depon (\rho) := (1 - p) \rho + p \frac{\mathds{1}}{d}
    \end{equation}
    where $0 \leq p \leq 1$, $d = 2^n$, and $\mathds{1} := \mathds{1}_d$ is the $d$-dimensional identity operation.  
\end{definition}
\end{defbox}

Finally, we consider amplitude damping noise which models decay to the $\ket{0}$ state (for example, in a physical implementation this could be the decay from an excited state to the ground state via spontaneous emission of a photon).
\begin{defbox}
\begin{definition}[The amplitude damping channel]\label{def:amp_damp_channel}~ \\
    The amplitude damping channel maps a single qubit state $\sigma$ to $\damp (\sigma)$ defined by
    \begin{equation} \label{eqn:amp_damp_channel}
        \damp (\sigma) := \left[ \begin{matrix}
            \sigma_{00} + p \sigma_{11}     & \sqrt{1 - p} \sigma_{01} \\
            \sqrt{1 - p} \sigma_{10}      & (1 - p) \sigma_{11} \\
        \end{matrix} \right]
    \end{equation}
    where $0 \leq p \leq 1$. 
\end{definition}
\end{defbox}
One could also consider these channels interacting in parallel on an $n$ qubit state, via a decomposable quantum channel. For example, a two qubit (separable) quantum state, $\rho = \sigma_1 \otimes \sigma_2$, could be acted on by a bit flip channel individually on each qubit with different magnitudes; $\E(\rho) = \E^{\text{BF}}_{p_1}\otimes \E^{\text{BF}}_{p_2}\left(\sigma_1 \otimes \sigma_2\right) = \E^{\text{BF}}_{p_1}(\sigma_1) \otimes \E^{\text{BF}}_{p_2}(\sigma_2)$. As mentioned above, these noise channels are not completely representative of real quantum noise. Realistic quantum noise is generally time dependent, gate dependent and includes multi-qubit effects such as crosstalk~\cite{sarovar_detecting_2020}, which are correlated and non-local (meaning operating on qubits which may not be spatially close). Defining, detecting and mitigating this multitude of errors is one of the major challenges in building useful quantum processors.

\subsection[\texorpdfstring{\color{black}}{} Quantum hardware]{Quantum hardware} \label{ssec:born_machine/quantum_hardware}
Now that we have some idea about quantum noise and how it affects realistic computations, let us discuss some \emph{real} quantum hardware, on which all of the noise sources described in~\secref{ssec:prelim/qc/quantum_noise} are present. 

At the time of writing, there are many competing candidates for the physical manifestation of quantum computation which satisfy the Divencenzo criteria, or the LJLNMO criteria. The primary drivers are companies specialising in quantum computation (startups), or companies which have large quantum computing efforts with roadmaps to build large fault tolerant (universal) devices. For example, \href{https://quantumai.google/hardware}{Google}, \href{https://www.ibm.com/quantum-computing/systems}{IBM} and \href{https://www.rigetti.com/}{Rigetti} are some of the main developers of quantum computers built from \emph{superconducting} qubits\footnote{See~\cite{huang_superconducting_2020, kjaergaard_superconducting_2020} for an overview of recent advances in superconducting hardware.}. On the other hand, companies such as \href{https://www.xanadu.ai/hardware}{Xanadu}, \href{https://psiquantum.com/}{PsiQuantum} and \href{https://www.quix.nl/technology/}{QuiX} are developing quantum computers based on \emph{photons}\footnote{See \cite{arrazola_quantum_2021, bombin_interleaving_2021} for example.}. Others such as \href{https://ionq.com/technology}{IonQ}, \href{https://www.honeywell.com/us/en/company/quantum/quantum-computer}{Honeywell} and \href{https://universalquantum.com/}{Universal Quantum} are developing ion-trap quantum computers\footnote{See~\cite{lekitsch_blueprint_2017, wright_benchmarking_2019, pino_demonstration_2021}.}, while \href{https://pasqal.io/technology/}{Pasqal} focuses on neutral atoms. Of course, this list is absolutely non-exhaustive and not to mention countless smaller endeavours, plus the efforts of governmental and university labs and programmes around the world. Each of these different technologies have their advantages and disadvantages, and it is not clear which is the `most scalable'. Only time will tell.

Later in this thesis, we develop and study algorithms which are particularly suited to the near-term devices, like those held by the companies listed above. In order to determine the performance of our variational quantum algorithms and quantum machine learning models in reality, we must implement them on the actual hardware. For this Thesis, we primarily focus on the \computerfont{Aspen} series of Rigetti quantum chips. We refer to these as \emph{quantum processing units} (QPUs) and we call a simulator of such hardware a \emph{quantum virtual machine} (QVM)\footnote{Most hardware providers have their nomenclature for quantum simulators/hardware; QPU/QVM is that given to those of Rigetti.}. As mentioned above, these qubits are superconducting in nature, which is the same technology that demonstrated \emph{quantum computational supremacy} in 2019~\cite{arute_quantum_2019}\footnote{Followed shortly by a demonstration on a photonic platform~\cite{zhong_quantum_2020}.}. At the time of writing, the current iteration of the chip series is the $\aspennine$, but we use primarily slightly older versions in this thesis - the $\aspenfour$, $\aspenseven$ and $\aspeneight$ models. 

\begin{figure}[t]
    \centering
    \includegraphics[width=0.9\columnwidth, height=0.2\columnwidth]{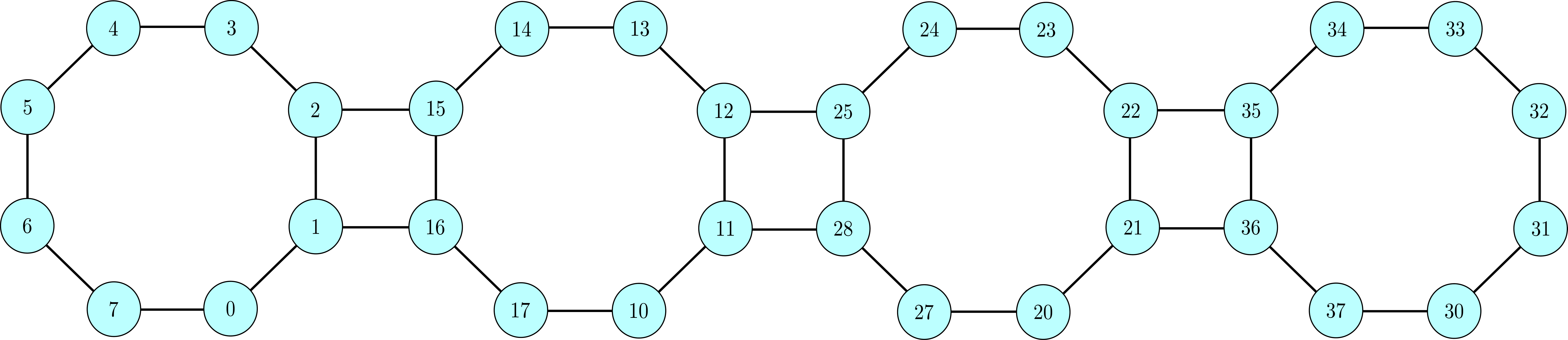}
    \caption[\color{black} Skeleton design of the $\aspenseven$/$\aspeneight/\aspennine$ $32$ qubit chip series of Rigetti.]{\textbf{Skeleton design of the $\aspenseven$/$\aspeneight/\aspennine$ $32$ qubit chip series of Rigetti.} Note that while $32$ qubits are shown here, not all qubits are actually usable due to defects (for example the $\aspenseven$ has only $28$ usable qubits, while the $\aspennine$ has $31$. Furthermore, not all connections shown above are directly accessible on the chip itself. The $\aspenfour$ chip has the same connectivity structure, but only has $16$ available qubits. We specify the exact connectivity we use in this Thesis when relevant.}
    \label{fig:aspen_7_chip}
\end{figure}

\begin{figure}
        \centering
        \includegraphics[width=0.7\columnwidth, height=0.5\textwidth]{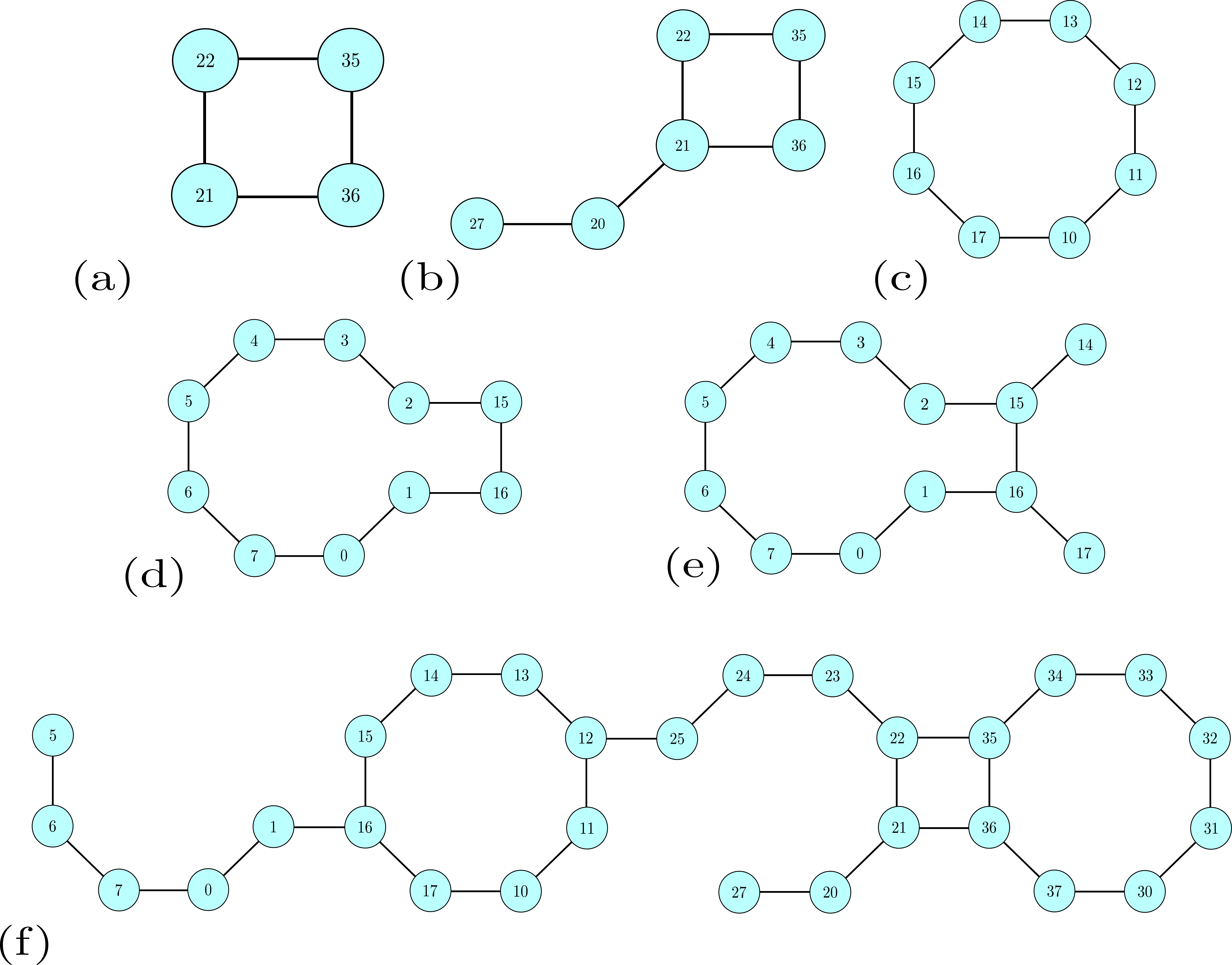}
    \caption[\color{black}Select sublattices from the $\aspenseven$ and $\aspeneight$ chips.]{\textbf{Select sublattices from the $\aspenseven$ and $\aspeneight$ chips, corresponding to different problem sizes.} Figure shows the (a) $\aspenseven$\computerfont{-4Q-C}, (b) $\aspenseven$\computerfont{-6Q-C}, (c) $\aspenseven$\computerfont{-8Q-C}, (d) 10 qubit $\aspeneight$, (d) 12 qubit $\aspeneight$ and (f) $\aspenseven$\computerfont{-28Q-A} sublattices.}
    \label{fig:aspen_sublattices}
\end{figure}

In~\figref{fig:aspen_7_chip}, we show the full ideal connectivity\footnote{This sparse connectivity is one of the disadvantages of superconducting technologies - in comparison to trapped ion qubits, which may have all-to-all native connectivity,~\cite{wright_benchmarking_2019}.} of the \computerfont{Aspen} quantum chip series, while in \figref{fig:aspen_sublattices}, some select sublattices of the chip are shown (we use these specific sublattices in \secref{sec:born_machine/finance}). As we alluded to above, on these devices, we have \emph{native} gatesets, which are the basic operations that non-native quantum programs must be compiled into\footnote{A native gateset is distinct from a universal gate set~\defref{defn:prelim/qc/universal_gate_set}. Even if one has decomposed a quantum algorithm into a universal gateset, these universal gates must be further decomposed into the native gates, or even further to pulse-level operations~\cite{abrams_implementation_2020, alexander_qiskit_2020}.}. The native single qubit gates are $\RZ(\theta), \RX(\pm\pi/2)$ and for an entangling two qubit gate, we can choose either $\CZ$ (\eqref{eqn:cz_matrix}) or the $\XY$ gate~\cite{abrams_implementation_2020}. For this Thesis, we use only the $\CZ$ gate, but one may prefer to use instead the $\XY$ gate depending on the application. We can also get the $\RY$ gate via the following decomposition, which is `almost' native:
\begin{equation} \label{eqn:rigetti_ry_gate_decomp}
    \RY(\theta) = \RX\left(\frac{\pi}{2}\right)\RZ(\theta)\RX\left(-\frac{\pi}{2}\right)
\end{equation}
Now, the state of the art (SOTA) in quantum hardware is constantly evolving and therefore this section will be quickly out of date. That said, let us highlight some specifications given on the Rigetti website (\url{https://www.rigetti.com/}) for the $\aspennine$ chip for illustration. Firstly, we have single and two qubit gate fidelities of $99.8\%$ and $99\%$ respectively. We also have the qubit lifetimes which are referred to as $T_1$ and $T_2$ times. The $T_1$ time is related to the probability of a qubit decaying from the $\ket{1}$ state to the $\ket{0}$ state as a function of time. For example with a simplified model, one may have $\Pr(\ket{1}) = \erm^{t/T_1}$, so the larger $T_1$ is, the slower the decay will be. This corresponds to the ability to create and maintain superposition states, $\propto \ket{0} + \ket{1}$ for as long as necessary. As such, the $T_1$ time is related to the amplitude damping channel in~\defref{def:amp_damp_channel}. In contrast, the $T_2$ time can be modelled as a dephasing quantum noise channel in~\defref{def:dephasing_channel} and related to the time taken for the state $\ket{+} \propto \ket{0}+ \ket{1}$ to acquire a relative phase. For the $\aspennine$ chip, the median estimated values for these quantities are: $T_1 = 27\mu s, T_2 = 19\mu s$. For comparison, the median time to operate a one qubit gate on this chip is $T_{1Q} = 48 ns$, while a two qubit gate requires about $T_{2Q} = 168 ns$.

One final point to note - in this Thesis we also run experiments on \emph{simulated} quantum hardware (via QVMs), in which case we can extract exact (noiseless) calculations, and have an arbitrary connectivity between qubits. For small scale problems, using both simulators and real hardware is an essential part of quantum algorithm research, particularly NISQ algorithms.

\subsection[\texorpdfstring{\color{black}}{} Distance measures]{Distance measures} \label{ssec:prelim/qc/distance_measures}
We extensively use measures of distance in this Thesis. Having concise methods to characterise closeness of objects is of fundamental importance, and it is central to applications. For example, when verifying correctness of the output of a quantum computation, we need some measure to tell us how well we are doing, and whether the computation is performing as we expect. Furthermore, as we revisit when discussing machine learning in \secref{sec:prelim_machine_learning}, distance measures usually provide a \emph{cost} function which we use to improve the output of a learning algorithm. In this latter example in particular, the strengths and weaknesses of certain distance measures determines the effectiveness of the learning procedure. We go to some detail here in order to impress the importance of this on the reader. Let us begin with classical concepts first, before moving to quantum measures.

Firstly, for completeness, the definition of a metric:
\begin{defbox}
\begin{definition}[Metrics]\label{defn:metric}~ \\
A metric, $\text{d}$, on a set, $\X$, is a function:
\begin{equation}
    \text{d}: \X \times \X \rightarrow [0, \infty)
\end{equation}
with the following properties:
\begin{itemize}
    \item `Faithfulness': $\text{d}(x, y) = 0 \iff x \equiv y$
    \item `Symmetry': $\text{d}(x, y) = \text{d}(y, x)$
    \item `Triangle Inequality': $\text{d}(x, y) \leq \text{d}(x, z) + \text{d}(z, y)$
\end{itemize}
For all $x, y, z \in \X$.
\end{definition}
\end{defbox}
There is usually no unique choice for a metric, and common choices for $\mathbb{R}^n$ are the $\ell_1$ and $\ell_2$ metrics\footnote{We may also refer to these interchangeably as `norms'; for example the $\ell_2$ `norm', $||\xbs||_2$ which measures the (Euclidean) length of a vector. Many distance measures can be induced from their corresponding norm.}:
\begin{equation}\label{eqn:distance_meas/l1_metric}
    \text{d}_{\ell_1}(\xbs, \ybs) := ||\xbs -\ybs||_1 := \sum_i |x_i - y_1|   \qquad \ell_1 \text{ metric}
\end{equation}
\begin{equation}\label{eqn:distance_meas/l2_metric}
    \text{d}_{\ell_2}(\xbs, \ybs)  := ||\xbs -\ybs||_2 := \sqrt{\sum_i (x_i - y_1)^2}   \qquad \ell_2 \text{ metric}
\end{equation}
The latter is also called the Euclidean distance, and these can be generalised to $\ell_p$ metrics in an obvious way. Here $\xbs := [x_1, \dots, x_n]^T, \ybs := [y_1, \dots, y_n]^T$ are $n$ dimensional real vectors. Throughout this Thesis, we use $\xbs$ to denote a vector, and $x$ to denote a scalar value.

\subsubsection[\texorpdfstring{\color{black}}{} Probability distance measures]{Probability distance measures} \label{ssec:prelim/qc/distance_measures/probability}

We are also interested in metrics over \emph{probability} spaces, describing similarity between probability distributions. For this Thesis, we assume the sample space under which the distribution is defined is finite (more generally countable)\footnote{This will always be the case in binary quantum computation, where the space is typically the space of binary strings, $\{0, 1\}^n$, which we label (technically incorrectly) as the Boolean hypercube.}. Let us define the space of distributions as $\mathcal{D}$, defined over a sample space, $\X$. Then, for a particular distribution, $D \in \mathcal{D}$, we can use $p(\xbs)$ to denote the probability density function, or probability mass function, of the outcome $\xbs \in \X$. The \emph{support} of $D$ is the subset of $\X$ for which $p(\xbs)>0$. Since we assume $\X$ is countable, we can define a vector according to $p$ for each outcome, so $D := [p(\xbs_1),  \dots, p(\xbs_n)]^T$\footnote{We use the notation $D$ and $p(x)$ interchangeably in this Thesis.}. When viewed in this manner, the $\ell_1$ metric above (\eqref{eqn:distance_meas/l1_metric}) also defines a metric over the probability simplex, and becomes the \emph{total variation} metric, also known as the statistical distance. We define it between two distributions, $p, q$, as:
\begin{equation}\label{eqn:total_variation_dist_defn}
    \text{d}_{\TV} = \TV(p, q) := \frac{1}{2} \sum_{\xbs} |p(\xbs) - q(\xbs)|
\end{equation}
If $p(\xbs)$ is the outcome distribution from a quantum circuit, $\xbs$ will be a bitstring of length $n$, and so the sum in \eqref{eqn:total_variation_dist_defn} will have exponentially ($2^n$) many terms. We revisit this point in detail later in the Thesis.

A second relevant distance measure is the \emph{Kullback-Leibler} ($\KL$) divergence (\cite{kullback_information_1951}), defined as:
\begin{equation}\label{eqn:kl_divergence_defn}
    \begin{split}
    \text{d}_{\KL}(p, q) := \KL(p, q) := \sum_{\xbs} p(\xbs)  \log\left(\frac{p(\xbs)}{q(\xbs)}\right) &= \sum_{\xbs} p(\xbs) \log p(\xbs)-  \sum_{\xbs} p(\xbs) \log q(\xbs) \\
    &=: \Hsf(p) - \XE(p, q)
    \end{split}
\end{equation}
$\Hsf(p)$ is the entropy of the distribution, $p(\xbs)$, and $\XE(p, q)$ is the cross entropy of $p$ and $q$, which can be seen as the expectation value of $\log q(\xbs)$ under $p$. Note that this is \emph{not} a metric, since it is not symmetric in its arguments\footnote{Switching the arguments in \eqref{eqn:kl_divergence_defn} results in the \emph{reverse} $\KL$ divergence.}. However, it has an important history in machine learning and is useful in information theory. Importantly, it provides an upper bound on the total variation distance via Pinkser's Inequality:
\begin{equation}\label{eqn:pinskers_inequality}
    \TV(p, q) \leq \sqrt{\frac{1}{2} \KL(p, q)}
\end{equation}

The next important measure we use in this Thesis is the \emph{maximum mean discrepancy} ($\MMD$). To introduce this metric, we require a quantity known as a \emph{feature map} and a \emph{kernel}. Kernel methods are important techniques in machine learning, and in order to do them justice we defer a more lengthy introduction to~\secref{ssec:prelim/machine_learning/kernel_methods} in~\chapref{chap:prelim_machine_learning}. For now, it is sufficient to think of a feature map as a non-linear `embedding' function applied to the sample space to map into a high dimensional Hilbert space, $\phi:\mathcal{X} \rightarrow \mathcal{H}$. Then from this, a kernel can simply be defined as the inner product between two feature vectors, in this Hilbert space, $\kappa(\xbs,\ybs)  := \langle\phi(\xbs),\phi(\ybs) \rangle_{\mathcal{H}}$.

Let us also define the `\emph{mean embedding}', $\mu_{p} \in \hilb$ which weights the feature map according to a distribution, $p$, as follows:
\begin{equation}\label{eqn:kernel_mean_embedding}
    \mu_{p}  := \mathbb{E}_{p}(\kappa(\xbs, \cdot)) =\mathbb{E}_{p}[\phi(\xbs)] = \sum\limits_{\xbs} \phi(\xbs) p(\xbs)
\end{equation}
We can use this mean embedding (which essentially takes an average with respect to $p$) then as the root of a distribution comparison measure, which becomes the maximum mean discrepancy\footnote{Note, this mean embedding was generalised to use quantum feature maps in~\cite{kubler_quantum_2019}.}. The $\MMD$ is defined via the \emph{difference} between two mean embeddings of two distributions~\cite{borgwardt_integrating_2006, gretton_kernel_2007} to be compared:
\begin{equation}\label{eqn:mmd_mean_embedding}
    \text{d}_{\MMD}(p, q)   \coloneqq ||\mu_{p} - \mu_{q}||_{\hilb}
\end{equation}
We use the notation $\text{d}_{\MMD}$ here since the object we refer to as the `$\MMD$' later in this Thesis is actually the square of the $\text{d}_{\MMD}$\footnote{$\text{d}_{\MMD}$ itself is the metric on the space of probability distributions.}:
\begin{equation}\label{eqn:distance_measurems/mmd_definition}
    \MMD^{\phi}(p, q) \coloneqq \text{d}_{\MMD}(p, q)^2 = \left|\left|\mathbb{E}_{p}[\phi(\xbs)] - \mathbb{E}_{q}[\phi(\xbs)]\right|\right|_{\hilb}^2
\end{equation}
The square in \eqref{eqn:distance_measurems/mmd_definition} can be expanded to reveal the more computationally useful form of the $\MMD$:
\begin{equation}\label{eqn:mmd_exact}
  \MMD^{\kappa}(p, q) = \underset{\substack{\xbs \sim p\\ \ybs \sim p}}{\mathbb{E}}(\kappa(\xbs,\ybs)) + \underset{\substack{\xbs \sim q \\\ybs \sim q }}{\mathbb{E}}(\kappa(\xbs,\ybs)) -\underset{\substack{\xbs \sim p\\ \ybs \sim q}}{2\mathbb{E}}(\kappa(\xbs,\ybs)) 
\end{equation}
In this form, the computation of the $\MMD$ exploits the so-called `\emph{kernel trick}' (where $\kappa$ is a kernel defined in \thmref{thm:kernel_definition}), which relieves us of the necessity in evaluating the feature maps themselves at every point, $\phi(\xbs)$, but only the inner products between pairs of points (i.e. computing the kernels), which is designed to be simpler. The same kernel trick is also applied in the support vector machine.

In order to estimate \eqref{eqn:mmd_exact}, we can draw $N$ independent samples from $p$, $\hat{\xbs} := \{\xbs^1,\dots, \xbs^N\} \sim p$, $M$ samples from $q$, $\hat{\ybs} = \{\ybs^1, \dots, \ybs^M\} \sim q$ and compute the following unbiased quantity (\cite{sriperumbudur_integral_2009}):
\begin{multline}\label{eqn:mmd_estimator}
      \widetilde{\MMD}^{\kappa}(p_N, q_M)   \coloneqq \widetilde{\text{d}}_{\MMD}(p_N, q_M)^2\\ =\frac{1}{N(N-1)}\sum\limits_{i \neq j}^N \kappa(\xbs^i, \xbs^j) + \frac{1}{M(M-1)}\sum\limits_{i \neq j}^M \kappa(\ybs^i, \ybs^j) -\frac{2}{MN}\sum\limits_{i ,j}^{M, N} \kappa(\xbs^i, \ybs^j)  
\end{multline}
A crucial property of this $\MMD$ estimator is the fast convergence rate in probability it enjoys to its true value (where we denote $p_N$ to be the empirical estimate of $p$ with $N$ samples, and likewise for $q$)\footnote{The notation $\leq \mathcal{O}_{p}$ indicates convergence in probability. A sequence of random variables, $\{X_n\}$ is said to converge in probability towards a random variable, $X$ if $\forall \epsilon >0, \lim\limits_{n\rightarrow \infty} \text{Pr}\left(|X_n - X| > \epsilon\right) = 0$.}:
\begin{equation}\label{eqn:mmd_samplecomplexity}
    |\widetilde{\text{d}}_{\MMD}(p_N, q_M)  - \text{d}_{\MMD}(p, q)| \leq \mathcal{O}_{p,q} \left(N^{-1/2} + M^{-1/2}\right)
\end{equation} 
This quadratic convergence rate is highly desirable\footnote{This convergence rate means that if we require the estimator to be $\epsilon$ close to the true value, we require $\mathcal{O}\left(1/\epsilon^2\right)$ samples from the distributions (assuming $N=M$ for simplicity).}, since it does not depend on the dimension of the space from which the samples are drawn. This is due to a very important property of the $\MMD$ (which we generalise later) - it is written solely in terms of \emph{expectation values} over the distributions. In contrast, both the $\TV$ (\eqref{eqn:total_variation_dist_defn}) and $\KL$ divergence (\eqref{eqn:kl_divergence_defn}) require the evaluation of the \emph{probabilities} themselves (since they cannot be written in terms of expectation values of \emph{both} $p, q$). The consequence is that if any outcome, $\xbs^*$, in (say $p$) has exponentially low probability (e.g. $p(\xbs^*) = \mathcal{O}(1/2^n)$), it will require exponentially many samples to estimate the specific probability of $\xbs^*$.

Let us now discuss the final relevant distribution measure relevant for this Thesis. To introduce this, we move onto the rich subject of \emph{optimal transport}. The `optimal transport problem' was formalised by Gaspard Monge in 1781 and asks how one can \emph{transport} one mass (the distribution, $p$) into another (the distribution, $q$), in an efficient manner. We do not attempt to provide a comprehensive overview of this expansive topic (see \cite{villani_optimal_2009, peyre_computational_2019} for a starting point), but only highlight the parts of it which are relevant for our purposes. 

In the discrete case, the optimal transport ($\OT$) distance is given by:
\begin{equation}  \label{eqn:optimal_transport}
    \OT^c(p, q)  \coloneqq \min\limits_{U \in \mathcal{U}(p, q)}\sum\limits_{(\xbs, \ybs) \in \mathcal{X}\times\mathcal{Y}} c(\xbs, \ybs) U(\xbs, \ybs)
\end{equation}
where $p, q$ are the marginal distributions of $U$ (called a `coupling'), i.e. $\mathcal{U}(p, q)$ is the space of joint distributions over $\mathcal{X}\times\mathcal{Y}$ such that:
\begin{equation}
\sum_{\xbs}U(\xbs, \ybs) = q(\ybs),\qquad \sum_{\ybs}U(\xbs, \ybs) = p(\xbs),
\end{equation}
in the discrete case. $c(\xbs, \ybs)$ is the `\textit{cost}' of transporting an individual `point', $\xbs$, to another point $\ybs$. If we take the optimal transport `cost', to be a metric on the sample space, $\mathcal{X}\times \mathcal{Y}$, i.e. $c(\xbs, \ybs) = \delta(\xbs, \ybs)$\footnote{Note, we confusingly switch between $d$ and $\delta$ here - both are metrics, but we use $\delta$ to be a metric on a vector space while $d$ becomes a metric on a probability space.} we get the `Wasserstein' metric, which turns out to be equivalent to  the Kantorovich metric due to the Kantorovich-Rubinstein theorem~\cite{dudley_real_2002}:
\begin{equation} \label{eqn:1_wasserstein_distance}
    \text{d}^\delta_{\W}(p, q)  := \min\limits_{U \in \mathcal{U}(p, q)}\sum\limits_{(\xbs, \ybs) \in \mathcal{X}\times \mathcal{Y}} \delta(\xbs, \ybs) U(\xbs, \ybs) 
\end{equation}
Unfortunately, unlike the $\MMD$ with its quadratic sample complexity, the optimal transport metric has exponential sample complexity, $\mathcal{O}(N^{-1/k})$~\cite{sriperumbudur_integral_2009}, where $k$ is the dimension of the underlying space, $\X$ (again assuming $k > 2$):
\begin{equation}\label{eqn:wasserstein_samplecomplexity}
    |\widetilde{\text{d}}^\delta_{\W}(p_N, q_M)  - \text{d}^\delta_{\W}(p, q)| \leq \mathcal{O}_{p,q} \left(N^{-1/k} + M^{-1/k}\right)
\end{equation} 

All of the probability measures we have introduced above actually can be categorised into two distinct families. The $\KL$ divergence is an example of an `$f$\emph{-divergence}', whereas the $\MMD$ and the Kantorovich metric ($\OT$) fall into the category of \emph{integral probability metrics} (IPMs). Interestingly, the $\TV$ distance is the only measure which is \emph{both} and $f$-divergence and an IPM.

$f$-divergences are a family parametrised by a function, $f$, as follows~\cite{ali_general_1966, csiszar_information-type_1967, sriperumbudur_integral_2009}:
\begin{defbox}
\begin{definition}[$f$-divergence]\label{defn:f_divergence}~ \\
    Let $f$ be a convex function on $(0, \infty)$ with $f(1) = 0$. The $f$-divergence $\text{d}_f$ between two distributions, $p, q$ defined on $\mathcal{X}$ is:
    \begin{equation} \label{eqn:f_divergence_defn}
        \text{d}_{f}(p || q) := \sum_{\xbs} f\left(\frac{p(\xbs)}{q(\xbs)}\right)p(\xbs)
    \end{equation}
\end{definition}
\end{defbox}
The choice of the function, $f$ can have a variety of effects on the properties of the divergence, and in order to recover the two special cases mentioned above, we can take $f(t) = t\log t$ for the $\KL$ divergence, and $f(t) = |t-1|$ to give the $\TV$ distance.

In contrast, IPMs are defined as follows~\cite{sriperumbudur_integral_2009}:
\begin{defbox}
\begin{definition}[Integral probability metrics]\label{defn:integral_probability_metrics}~ \\
    Let $\mathcal{F}$ be a class of real-valued bounded measurable functions on $\mathcal{Z}$. The IPM, $\text{d}_{\mathcal{F}}$, between two distributions, $p, q$ defined on $\mathcal{Z}$ is:
    \begin{equation} \label{eqn:ipm_defn}
        \text{d}_{\mathcal{F}}(p, q) := \sup_{\phi\in \mathcal{F}}\left| \sum\limits_{\xbs}  \phi(\xbs) p(\xbs) - \sum\limits_{\xbs} \phi(\xbs) q(\xbs)\right| = \sup_{\phi\in \mathcal{F}}\left|\underset{p}{\mathbb{E}}\left[ \phi\right] -
        \underset{q}{\mathbb{E}}\left[ \phi\right]\right|
    \end{equation}
\end{definition}
\end{defbox}
We use the definitions assuming $\mathcal{X}$ is discrete, but the generalisation to general measurable spaces, $\mathcal{M}$, is straightforward by replacing the sums with integrals, see~\cite{sriperumbudur_integral_2009}. For IPMs, their individual properties are given by the choice of the function family, $\mathcal{F}$. For example, we can recover the above Wasserstein, $\MMD$ and $\TV$ distances by taking the function families as follows:
\begin{align}
    \text{$\MMD$} \rightarrow \mathcal{F}_{\MMD}  &:= \{\phi \in
            \mathcal{H}: ||\phi||_{\mathcal{H}}\leq 1\}
            \rightarrow \text{d}_{\MMD}       
            \label{eqn:ipm_mmd_functions}\\
    \text{$\TV$} \rightarrow \mathcal{F}_{\TV}  &:=  \{\phi: ||\phi||_{\infty} \leq 1\} \label{eqn:ipm_tv_functions}\\
    \text{Wasserstein/$\OT$} \rightarrow  \mathcal{F}_{\W} &:=  \{\phi: ||\phi||_{L} \leq 1\} \label{eqn:ipm_wasserstein_functions}
\end{align}
$||\cdot||_{\mathcal{H}}, ||\cdot||_{\infty}, ||\cdot||_{L}$ in \eqref{eqn:ipm_wasserstein_functions} are the norm in the Hilbert space, $\mathcal{H}$, the infinity norm (or max/sup norm) and the Lipschitz semi-norm respectively. The latter two are defined by $||\phi||_{\infty} = \sup\{|\phi(\xbs)|: \xbs \in \mathcal{X}\}$ and $||\phi||_{L} := \sup\{|\phi(\xbs) - \phi(\ybs)|/\delta(\xbs,\ybs) : \xbs\neq \ybs \in \mathcal{X}\}$, where $\delta$ is a metric on $\mathcal{X}$.

\subsubsection[\texorpdfstring{\color{black}}{} Quantum distance measures]{Quantum distance measures} \label{sssec:prelim/qc/distance_measures/quantum}
Not surprisingly, we can generalise the definitions of distribution measures into the quantum realm, defining metrics which operate on quantum states in $\hilb$. The first one which can be considered is the \emph{trace distance}, which is a generalisation of $\TV$ to quantum states $\rho, \sigma$:
\begin{equation}\label{eqn:trace_distance_defn}
    \text{d}_{\tr}(\rho, \sigma) := \frac{1}{2}\tr|\rho - \sigma|
\end{equation}
Here, $|A| := \sqrt{A^\dagger A}$ is defined as the positive square root of a matrix $A$. The trace distance can be easily shown to reduce to the total variation distance if the two states, $\rho, \sigma$ commute~\cite{nielsen_quantum_2010}. \eqref{eqn:trace_distance_defn} can also be written in terms of the absolute values of the eigenvalues of the matrix $\rho-\sigma$. Just as the $\TV$ is a strong distance measure on distributions, $\text{d}_{\tr}$ is a strong distance measure on quantum states.

One can also generalise the $\ell_2$ norm to quantum states, where is becomes the \emph{Hilbert-Schmidt} distance:
\begin{equation}\label{eqn:hs_distance_defn}
    \text{d}_{\HS}(\rho, \sigma) := \tr\left[(\rho - \sigma)^2\right]
\end{equation}
We also have the following interesting relationship between $\text{d}_{\HS}$ and $\text{d}_{\tr}$~\cite{coles_strong_2019}:
\begin{equation}\label{eqn:hs_trace_relation}
    \frac{1}{2}\text{d}_{\HS} \leq \text{d}^2_{\tr} \leq r\text{d}_{\HS},~ r := \frac{\text{rank}(\rho)\text{rank}(\sigma)}{\text{rank}(\rho)+\text{rank}(\sigma)}
\end{equation}
The next important measure is the \emph{fidelity}~\cite{jozsa_fidelity_1994} between two quantum states, defined as\footnote{A version of the fidelity sometimes appears with a square root~\cite{nielsen_quantum_2010}, but we stick to the `squared' version in this Thesis.}:
\begin{equation}\label{eqn:fidelity_defn}
    F(\rho, \sigma) := \left[\tr\sqrt{\left(\sqrt{\rho} \sigma \sqrt{\rho}\right)}\right]^2
\end{equation}
The fidelity is symmetric, perhaps surprisingly, but it is not a metric on the space of quantum states. One can be defined from it however called the Bures angle\footnote{The Bures angle is not the only way one can define a metric using the fidelity.} (or quantum angle), defined by:
\begin{equation}\label{eqn:bures_angle_definition}
    \text{d}_{\BA} = \arccos\left(\sqrt{F(\rho, \sigma)}\right)
\end{equation}
This metric has a nice intuition as being the \emph{angle} between quantum states in $\hilb$. If $\rho, \sigma$ are pure states, the Bures angle corresponds to the Fubini-Study distance~\cite{zyczkowski_average_2005}. The fidelity can be simplified in some special cases. Firstly, let's consider if $\sigma = \ketbra{\psi}{\psi}$ is a pure state, then the fidelity becomes:
\begin{equation}\label{eqn:fidelity_defn_one_state_pure}
    F(\rho, \sigma) := \bra{\psi}\rho\ket{\psi}
\end{equation}
which is also the \emph{overlap} between $\rho$ and $\sigma$. Furthermore, if $\rho =\ketbra{\phi}{\phi}$ is also pure, we get:
\begin{equation}\label{eqn:fidelity_defn_both_states_pure}
    F(\rho, \sigma) := |\braket{\psi}{\phi}|^2
\end{equation}
As with the trace distance, the fidelity also generalises its classical counterpart. However the classical version is not relevant for this Thesis so we exclude it.

As a final remark, we note that in general (i.e. for two general mixed states) it is likely both the fidelity and the trace distance are exponentially hard to compute. This is due to the following relationship between the two:
\begin{equation}\label{eqn:fidelity_trace_dist_bound}
    1 - \sqrt{F(\sigma, \rho)} \leq  \text{d}_{\tr}(\rho, \sigma) \leq    \sqrt{1 - F(\sigma, \rho)}, 
\end{equation}
and the fact that the trace distance defines the complexity class \emph{quantum statistical zero knowledge} ($\mathsf{QZSK}$)\footnote{This class has a complete problem in deciding whether two quantum states are close or far in trace distance, which is an exponentially hard problem and not believed to be solvable by quantum computers in general (it is believed the containment $\BQP \subseteq \mathsf{QZSK}$ is strict). It generalises $\mathsf{SZK}$, which is the classical analogue - deciding whether two probability distributions are close or far in total variation distance.}. The difficulty in computing fidelity is related to the fact that the expression~\eqref{eqn:fidelity_defn} contains non-integer powers of $\sigma$ and $\rho$. Instead, to get estimates of the fidelity one can compute \emph{bounds} on it (see~\cite{cerezo_variational_2020-2}). These sub- and super-fidelity bounds are made of quantities which are integer powers of the states, For example, the \emph{purity} of a quantum state, $\Tr\left(\rho^2\right)$, or the overlap between two states, $\Tr\left(\rho\sigma\right)$ can be efficiently computed By `efficient' here we mean using polynomially many copies of the states, we can compute an estimate to a polynomial precision and exponential confidence. This can be done via the $\SWAP$ test in the next section.

We have given analogues of the classical $\TV$ distance, but not surprisingly we can generalise the other classical distance measures to the quantum case also. Above, we defined the classical entropy of a probability distribution (also called the Shannon entropy). The analogous version for quantum states is the von Neumann entropy, defined for a quantum state $\rho$ as:
\begin{equation} \label{eqn:von_neumann_entropy}
    S(\rho) := -\Tr\left[\rho \log \rho\right]
\end{equation}
Not surprisingly, this definition is used to define the \emph{quantum} relative entropy between two quantum states, $\rho, \sigma$, which is given by:
\begin{equation} \label{eqn:quantum_relative_entropy}
    S(\rho||\sigma) := -\Tr\left[\rho \log \sigma\right] - S(\rho) 
\end{equation}
By staring at \eqref{eqn:quantum_relative_entropy} for a few seconds, we can see it looks suspiciously like the classical relative entropy, or the $\KL$ divergence. Indeed, in the case where $\rho$ and $\sigma$ commute, it becomes exactly this, as with the $\TV$ and trace distance. Finally, we mention that one can also consider quantum generalisations of the Wasserstein distance between quantum states (see ~\cite{de_palma_quantum_2020}, among others), but again we do not directly use them in this Thesis so we leave further investigation to the interested reader.

\subsubsection[\texorpdfstring{\color{black}}{} The \texorpdfstring{$\SWAP$}{} test]{The \texorpdfstring{$\SWAP$}{} test} \label{ssec:prelim/qc/swap_test}
In the previous section, we described and discussed some features of the quantum fidelity~\eqref{eqn:fidelity_defn}, and in particular how it reduces to the overlap between two quantum states when one of the states is pure. Here, we discuss a very important algorithmic subroutine which is commonly used in quantum information, called the $\SWAP$ test, introduced by~\cite{buhrman_quantum_2001}. This is a method to extract overlaps (and hence fidelities) using an ancillary qubit, and a $\mathsf{C}\SWAP$ gate (see ~\eqref{eqn:controlled_u_gate}).

\begin{figure}[ht]
    \centering
    \includegraphics[width=0.6\columnwidth]{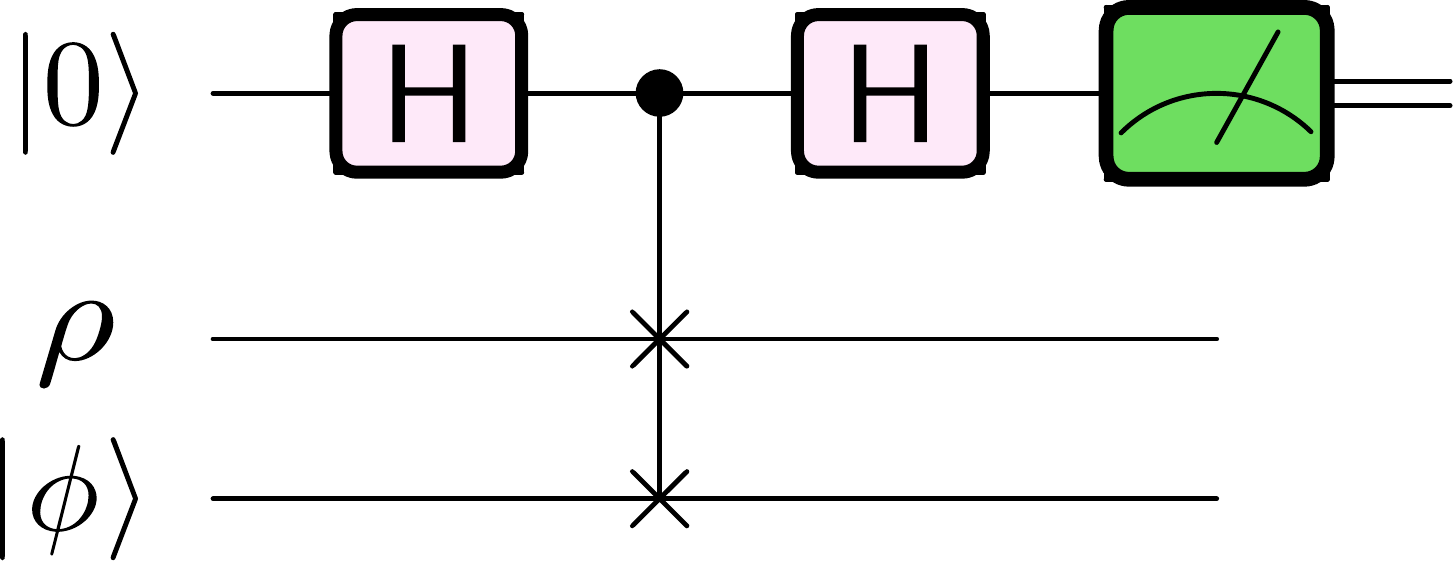}
    \caption[\color{black}The $\SWAP$ test.]{\textbf{The $\SWAP$ test.} The ancillary qubit is measured and the corresponding probability of getting outcome $\ket{0}$ gives the overlap between $\rho$ and $\ket{\psi}$, $\bra{\psi}\rho\ket{\psi}$.} \label{fig:prelims/qc/swap_test}
\end{figure}

We can directly relate the probability of the ancilla qubit, $\ket{\text{anc}}$ being in the state $\ket{0}$ after the transformation depicted in \figref{fig:prelims/qc/swap_test} to the overlap between two input states, $\rho$, $\ket{\psi}$, one of which may be a mixed state~\cite{buhrman_quantum_2001}:
\begin{equation} \label{eqn:swap_test_probability}
    \bra{\psi}\rho\ket{\psi} = 1 - 2\text{Pr}[\ket{\text{anc}} = \ket{0}] 
\end{equation}
A related subroutine to the $\SWAP$ test is called the `Hadamard' test, and can be used to extract real and imaginary parts of quantum expectation values. This test has many uses in quantum computation, but related to the topics of this Thesis, notably for extracting gradients with respect to quantum circuit parameters. Specifically, the Hadamard test can be used as an alternative method to the `parameter-shift' rule~\cite{banchi_measuring_2021}, which we present in \chapref{sec:prelim_quantum_machine_learning}.

\section[\texorpdfstring{\color{black}}{} Quantum cloning]{Quantum cloning} \label{ssec:prelim/qc/quantum_cloning}
A fundamental primitive in quantum theory is the famous impossibility result: the `no-cloning' theorem\footnote{The theorem was proven originally by~\cite{park_concept_1970} in 1970, and independently rediscovered by \cite{wootters_single_1982} and \cite{dieks_communication_1982} in 1982.}, stated as follows:
\begin{thmbox}
\begin{theorem}(The no-cloning theorem)\label{thm:no-cloning_theorem}~ \\
    For an arbitrary state, $\ket{\phi}$, in a Hilbert space $\hilb$, there exists no unitary, $U$, operating on $\hilb \otimes \hilb$ (e.g. the state $\ket{\phi}$ tensored with some `reference' state $\ket{R}\in \hilb$) with the following behaviour:
    \begin{equation}\label{eqn:no_cloning_theorem}
        U\left(\ket{\phi} \otimes \ket{R}\right) = \erm^{\irm \alpha(\phi, R)} \ket{\phi}\otimes \ket{\phi},
    \end{equation}
    where $\alpha$ is some global phase which may depend on the input states, $\ket{\phi}, \ket{R}$.
\end{theorem}
\end{thmbox}
A simple proof of the above theorem can be seen by observing the action of the unitary, $U$, on two different states, $\ket{\psi}, \ket{\phi}$. Assume without loss of generality that we set $\alpha = 0$. Then, take the inner product of $U$ acting on each of $\ket{\psi}$ and $\ket{\phi}$. On one hand, we have:
\begin{align}
    \left(\bra{\psi}\otimes \bra{R}\right) \left[U^{\dagger}U\right] \left(\ket{\phi}\otimes \ket{R}\right) &= \left(\bra{\psi}\otimes \bra{R}\right)\left(\ket{\phi}\otimes \ket{R}\right) \nonumber \qquad \text{since } U^{\dagger}U = \mathds{1} \\
    &= \braket{\psi}{\phi}\braket{R}{R} = \braket{\psi}{\phi} \nonumber
\end{align}
where we use the assumption that $\ket{R}$ is normalised, hence $\braket{R}{R} = 1$. On the other hand, we have:
\begin{align}
    \left(\bra{\psi}\otimes \bra{R}U^{\dagger}\right)\left(U\ket{\phi}\otimes \ket{R}\right) &= \bra{\psi}\otimes \bra{\psi}\ket{\phi}\otimes \ket{\phi} ~ \text{since} ~ U\ket{\sfrac{\psi}{\phi}} \otimes \ket{R} = \ket{\sfrac{\psi}{\phi}} \otimes \ket{\sfrac{\psi}{\phi}}   \nonumber  \\
     &= |\braket{\psi}{\phi}|^2 \nonumber
\end{align}
Putting the two together implies:
\begin{equation} \label{eqn:no_cloning_impossibility_theorem}
    \braket{\psi}{\phi} = |\braket{\psi}{\phi}|^2 
\end{equation}
The derived condition is only true if $\ket{\psi}$
and $\ket{\phi}$ are either identical (in which case $\braket{\psi}{\phi} = 1$), or they are orthogonal (so $\braket{\psi}{\phi} = 0$). Clearly, this does not cover all possibilities and we arrive at a contradiction, so quantum cloning is forbidden. While this seems like a strange consequence of quantum theory, it is actually not so different from the classical world. Indeed, there also exists a `no-cloning' theorem for classical probability distributions. While this is outside the scope of this Thesis, some intuition can be gained by imagining a single flip of a coin. From a single coin toss, we cannot know what the full probability distribution over possible coin outcomes can be, we only observe the outcome from the singular coin toss. The coin could be fair (i.e. both heads and tails occurring with equal probability) or biased (i.e. some skew towards one outcome or the other). The true answer is hidden from us unless we flip the coin multiple times. Analogously, since quantum theory is a generalisation of classical probability theory, we cannot know the contents of an arbitrary quantum state with only a single copy of it. Clearly, if we had the ability to clone quantum states perfectly, a single copy \emph{would} be sufficient to learn it completely (since we could make many clones, and measure in all possible bases to reconstruct the classical description of the state). This impossibility is manifested in the no-cloning theorem.

\subsection[\texorpdfstring{\color{black}}{} Beyond the no-cloning theorem]{Beyond the no-cloning theorem} \label{ssec:beyond_no_cloning}
Fortunately (for this Thesis at least), the no-cloning theorem is far from the end of the story. It turns out there some hidden assumptions in the theorem as presented above. Specifically, we require the `cloning machine' to be \emph{perfect}\footnote{The output of the cloning unitary must be \emph{exactly} the same state as the input: $\ket{\phi}\otimes \ket{\phi}$.} and \emph{deterministic}\footnote{The unitary must be successful with probability one.}. Both of these assumptions can be relaxed, and doing so has resulted in a rich field of study. 

The first work to address these assumptions was \cite{buzek_quantum_1996}, where the `perfectness' requirement was relaxed. Here the idea of \emph{approximate} cloning was born, meaning that the output clones do not have to be \emph{exactly} the same as the input state, but only somehow `\emph{close}'. This closeness is typically measured in terms of the fidelity \eqref{eqn:fidelity_defn} which we revisit in greater detail in \secref{sssec:approximate cloning_preliminaries}. 

The second assumption (determinism) was relaxed originally by \cite{duan_two_1997, duan_probabilistic_1998}. This resulted in \emph{probabilistic} quantum cloning, where the unitary is able to prepare perfect clones of the input state, but is only able to succeed with a certain probability. Probabilistic quantum cloning is less relevant for this Thesis so we neglect further discussion of it in order to focus on the approximate version.

\subsection[\texorpdfstring{\color{black}}{} Approximate cloning]{Approximate cloning} \label{sssec:approximate cloning}
\subsubsection[\texorpdfstring{\color{black}}{} Approximate cloning preliminaries \& notation]{Approximate cloning preliminaries \& notation} \label{sssec:approximate cloning_preliminaries}
Even once one has decided to focus on the approximate version of quantum cloning, further subdivisions are still possible. To keep track of these, let us begin with some notation.

The three most common properties of quantum cloning machines (QCMs) are:
\begin{tcbraster}[raster columns=3,raster equal height,nobeforeafter,raster column skip=1cm]
\begin{mybox}{red}{}
  \textbf{Universality.}
  \end{mybox}
  \begin{mybox}{blue}{}
  \textbf{Locality.}
  \end{mybox}
  \begin{mybox}{green}{}
  \textbf{Symmetry.}
  \end{mybox}
\end{tcbraster}

These properties manifest themselves in the \emph{comparison metric} which is used to compare the clones outputted from the QCM, relative to the ideal input states. Almost exclusively in the literature, the fidelity (\eqref{eqn:fidelity_defn}) is the comparison metric of choice, and we revisit this shortly.\\

\noindent  {\color{BrickRed}{\underline{\textbf{Universality}}}}

\vspace{2 mm}
This refers to the family of states which QCMs are designed for, $\mathcal{S}  \subseteq \mathcal{H}$ ($\mathcal{H}$ is the full Hilbert space), as this has a significant effect on their performance. Based on this, QCMs are subdivided into two categories, \emph{universal} (UQCM), and  \emph{state-dependent} (SDQCM) quantum cloning machines. In the former case \emph{all} states must be cloned equally well ($\mathcal{S} = \mathcal{H}$). In the latter, the cloning machine will be tailored to the family of states fed into it, so $\mathcal{S} \subset \mathcal{H}$.\\ 

\noindent {\color{RoyalPurple}{\underline{\textbf{Locality}}}}

\vspace{2 mm}
Locality refers to whether a QCM optimises a local comparison measure (i.e. check the quality of \emph{individual} output clones - a \emph{one-particle test criterion} (\cite{werner_optimal_1998, scarani_quantum_2005}) or a global one (i.e. check the quality of the global output state from the QCM - an \emph{all-particle test criterion}).

As in the example above, the state to be cloned, $\ket{\psi}_{\mathsf{A}}$\footnote{We suggestively use the notation `\textsf{A}', to make the connection to quantum cryptographic protocols later in this Thesis. In this context, it refers to the mythical `Alice' of Alice \& Bob - the canonical protagonists in cryptographic protocols. The first reference to Alice and Bob dates back to the seminal RSA paper \cite{rivest_method_1978}.}, will be inputted into the QCM with some `blank' state (say $\ket{0}$) to carry the clone, plus possibly some ancillary state $\ket{R}$ to aid with the cloning\footnote{This will behave as an `environment' as in \secref{ssec:prelim/qc/quantum_operations} in order to implement general quantum operations.}. For local cloners, the clones will be the reduced density matrices of the output states from the first two registers:
\begin{equation} \label{eqn:local_clones_definition}
    \begin{split}
        \rho^1 = \Tr_{2R}\left(U \ketbra{\psi}{\psi} \otimes  \ketbra{0}{0} \otimes  \ketbra{R}{R} U^\dagger\right) \\
        \rho^2 = \Tr_{1R}\left(U \ketbra{\psi}{\psi} \otimes \ketbra{0}{0} \otimes  \ketbra{R}{R} U^\dagger\right)
    \end{split}
\end{equation}
In contrast, for \emph{global} cloners, we care about the \emph{full} output state of the QCM, only neglecting the ancilla state, $\ket{R}$:\\
\begin{equation} \label{eqn:global_clones_definition}
    \begin{split}
        \rho^{1,2} = \Tr_{R}\left(U \ketbra{\psi}{\psi} \otimes  \ketbra{0}{0} \otimes  \ketbra{R}{R} U^\dagger\right)
    \end{split}
\end{equation}
\noindent  {\color{SeaGreen}{\underline{\textbf{Symmetry}}}}

\vspace{2 mm}
Symmetry in a QCM compares one clone relative to the other. Symmetric QCMs require each `clone' outputted from the QCM to be the \emph{same} relative to the comparison measure, which practically means that both clones will have the same local fidelity (see below). However, in some scenarios we may wish for \emph{asymmetric} output for the clones (meaning they have different fidelities). This property obviously only applies to local QCMs. 

Next, let us revisit the fidelity as the comparison metric of choice, in the context of these criteria. When dealing with local cloners, we define the \emph{local} fidelity, $F^j_{\Lbs} := F_{\textsf{L}}(\rho^j, \rho_{\text{A}}),  j \in\{1, 2\}$. This compares the ideal input state, $\rho_A := \ketbra{\psi}{\psi}_A$\footnote{For the purposes of this Thesis, the input state will always be a pure state. A notion of `cloning' can also be defined for mixed input states, where it is referred to as \emph{broadcasting}. The \emph{no-broadcasting} theorem (\cite{barnum_noncommuting_1996}) generalises the no-cloning theorem in this context.}, to the output clones, $\rho^j$.

In contrast, the \emph{global} fidelity  compares the entire output state of the QCM to a product state of input copies, $F_{\textsf{G}}(\rho^{1,2}, \rho_A \otimes \rho_A)$.
It may seem at first like the most obvious choice to study is the local fidelity, however the global fidelity is a relevant quantity for some cryptographic protocols. We revisit this explicitly later in the Thesis in \chapref{chap:cloning}.\\

\noindent \underline{\textbf{{From one to many}}}

\vspace{2 mm}
In the above, we only assumed a single state was given, and two clones were required. As a generalisation, we can instead receive $M > 1$ copies of the input state. In this case, the task is to produce $N > M$ output clones\footnote{A note on notation: we primarily use $m, n$ in this Thesis to denote the number of qubits, and $N, M$ to be typically a number of samples (as, for example, in~\secref{ssec:prelim/qc/distance_measures/probability}) or a vector dimension. In this section, and in~\chapref{chap:cloning}, we switch to $M, N$ indicating the number of qubits, for historical consistence.}. As such, the input state to the cloning unitary is (we assume the ancillary register, $\ket{R}$, may have an arbitrary dimension):
\begin{equation}
    \ket{\psi}_A^{\otimes M} \ket{0}^{\otimes (N-M)} \ket{R}
\end{equation}

This scenario is referred to as $M\rightarrow N$ cloning (\cite{gisin_optimal_1997}), and the standard scenario corresponds to $1\rightarrow 2$\footnote{It is interesting to note that in the limit $M\rightarrow \infty$, an optimal cloning machine becomes equivalent to a quantum state estimation machine~\cite{scarani_quantum_2005} for universal cloning.}. Now returning to the symmetry criterion concretely, a symmetric $M\rightarrow N$ cloning machine will require:
\begin{equation} \label{eqn:local_fidelity_symmetry_condition}
F^j_{\Lbs} = F^k_{\Lbs}, \qquad \forall j, k \in \{1, \dots N\}.
\end{equation}
Now that we have our notations in place, let us move onto some concrete examples in the next section.

\subsubsection[\texorpdfstring{\color{black}}{} State-dependent approximate cloning]{State-dependent approximate cloning} \label{ssec:prelim/qc/cloning/state_dependent_cloning}
The earliest result~\cite{buzek_quantum_1996} in approximate cloning was that a \emph{universal} symmetric cloning machine for qubits can be designed to achieve an optimal cloning fidelity of $ 5/6 \approx 0.8333$, which is notably higher than trivial copying strategies\footnote{Examples of trivial cloning machines are where one simply measures the state, and prepares two copies according to the measurement result, or so-called \emph{trivial amplification} where one simply prepares a random state as the second clone, and the first is allowed to pass through undisturbed. These trivial strategies can achieve an average fidelity of $75\%$ (\cite{scarani_quantum_2005}), so the reader should keep this number in mind, particularly in \chapref{chap:cloning}.} \cite{scarani_quantum_2005}. In other words, if $\ket{\psi}_{\mathsf{A}}$ is an arbitrary single qubit (pure) state on the Bloch sphere, there exists a UQCM which can achieve the local fidelity $F_{\Lbs, \opt}^{\text{U}, j} = 5/6, j \in \{1, 2\}$. This is not the end of the story however, and much higher fidelities can be achieved using a \emph{state-dependent} QCM. Roughly speaking, higher cloning fidelities correspond to extra degrees of freedom which are `known' to the cloner. The two most common examples are `phase-covariant' states and `fixed-overlap' states, which we describe next.\\\\

\noindent \textbf{Phase-Covariant Cloning}

\vspace{2mm}
\emph{Phase-covariant} (\cite{brus_phase-covariant_2000}) states are those states which are, roughly speaking, confined to a particular two dimensional plane in the Bloch sphere. The canonical choice is  to restrict to the $\XG$ - $\YG$ plane which supports states of the form:
\begin{equation} \label{eqn:x_y_plane_states}
    \ket{\psi_{xy}(\eta)} = \frac{1}{\sqrt{2}}\left(\ket{0} + e^{i\eta}\ket{1}\right)
\end{equation}
A SDQCM for these states can be constructed (\cite{brus_phase-covariant_2000}) with $F_{\Lbs,  \text{opt}}^{\text{PC}, j} \approx 0.85$ which notably is higher than the fidelity achievable with a UQCM.

These states are relevant since they are used in BB84 QKD protocols and also in universal blind quantum computation (UBQC)\footnote{Similar acronyms become confusing.} protocols (\cite{bennett_quantum_2014, broadbent_universal_2009}). Interestingly, the cloning of phase-covariant states can be accomplished in an \emph{economical} manner, meaning without needing an ancilla register, $\ket{R}$ (\cite{niu_two-qubit_1999}). However, as noted in \cite{scarani_quantum_2001, scarani_quantum_2005}, removing the ancilla has consequences for quantum cryptography and protocol attacks, which we revisit later in this Thesis. 

An alternative parameterisation of phase-covariant states are those in the $\XG$ - $\ZG$ plane:
\begin{equation} \label{eqn:x_z_plane_states}
    \ket{\psi_{xz}(\theta)} = \cos(\theta)\ket{0} + \sin(\theta)\ket{1}
\end{equation}
The optimal cloning fidelity here is the same as with the states (\eqref{eqn:x_y_plane_states}), since in both cases, one degree of freedom is revealed.

\begin{figure}[ht]
    \centering
\includegraphics[width=0.95\columnwidth]{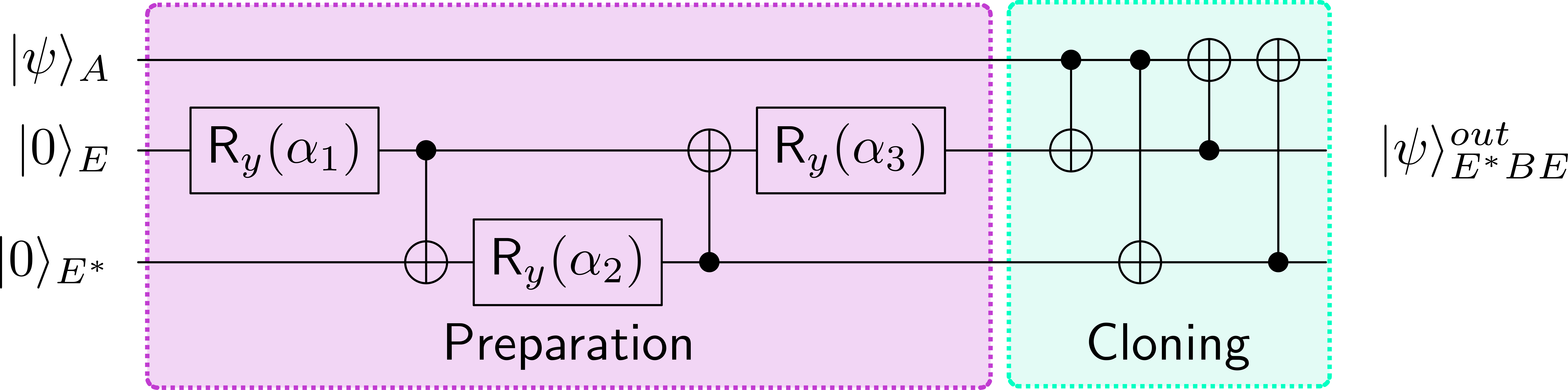}
\caption[\color{black}Ideal cloning circuit for universal and phase-covariant cloning.]{\textbf{Ideal cloning circuit for universal and phase-covariant cloning.} The \textsf{Preparation} circuit prepares the registers to receive the cloned states, while the \textsf{Cloning} circuit transfers information. Notice, the output registers which contain the two clones of $\ket{\psi}_A$ are the last and penultimate registers, while the ancilla, $R$ appears at the first register.} \label{fig:qubit_cloning_ideal_circ}
\end{figure}

To round off this discussion, let us present an explicit quantum circuit which implements the above cloning transformations. A unified circuit (\cite{buzek_quantum_1997, fan_quantum_2014, fan_quantum_2001}) for all above cases (universal, $\XG$-$\YG$ and $\XG$-$\ZG$ phase-covariant cloning) can be seen in~\figref{fig:qubit_cloning_ideal_circ}. The parameters of the circuit, $\boldsymbol{\alpha} = \{\alpha_1, \alpha_2, \alpha_3\}$, are chosen depending on the family of states to be cloned (\cite{buzek_quantum_1997, fan_quantum_2001, fan_quantum_2014}). 

Explicitly, we have suitable angles for universal cloning:
\begin{equation}\label{eqn:qubit_uqcm_angles}
    \alpha^U_1 = \alpha^U_3 = \frac{\pi}{8} , \qquad \alpha^U_2 = -\arcsin\sqrt{\left(\frac{1}{2}- \frac{\sqrt{2}}{3} \right)} 
\end{equation}
whereas for phase-covariant cloning of $\XG$-$\YG$ states, we require:
\begin{multline}\label{eqn:x_y_circuit_angles}
    \alpha^{xy}_1 = \alpha^{xy}_3 = \arcsin\sqrt{\left(\frac{1}{2}- \frac{1}{2\sqrt{3}}\right)} \approx 0.477, \\
    \alpha^{xy}_2 = -\arcsin\sqrt{\left(\frac{1}{2}- \frac{\sqrt{3}}{4}\right)}\approx -0.261
\end{multline}

\subsection[\texorpdfstring{\color{black}}{} Cloning of fixed overlap states]{Cloning of fixed overlap states}\label{ssec:non-ortho-states}

An alternative family of well known states are those which are not defined by their inhabitancy of a particular plane of the Bloch sphere, but instead by their \emph{overlap}. In this case, we usually care about only a finite set of states\footnote{As opposed to phase-covariant states, which have a continuous parameterisation.}. For our purposes, we usually assume we have only two\footnote{We can also consider this type of cloning with \emph{four} input states instead, which are usually the two given in~\eqref{eqn:state_dependent_cloning_states} plus their orthogonal counterparts. We use both examples in~\chapref{chap:cloning} of this Thesis.} states in this set, typically parametrised as follows:
\begin{equation} \label{eqn:state_dependent_cloning_states}
\begin{split}
    \ket{\psi_1} = \cos\phi\ket{0} +\sin\phi\ket{1}\\
    \ket{\psi_2} = \sin\phi\ket{0} +\cos\phi\ket{1} 
\end{split}
\end{equation}%
which have a fixed overlap, $s := \braket{\psi_1}{\psi_2} = \sin 2\phi$. We refer to this set of states as \emph{fixed-overlap} states\footnote{Confusingly, cloning states with this property is historically referred to as `state-dependent', but as we discussed above state-dependent cloning is a more general scenario where the cloner generically \emph{depends} on the input state properties.}. Interestingly fixed-overlap cloning was one of the original scenarios studied in the realm of approximate cloning \cite{brus_optimal_1998}, but was difficult to tackle analytically so was somewhat sidelined. Furthermore, fixed-overlap cloning has been used to demonstrate advantage related to quantum contextuality~\cite{lostaglio_contextual_2020}. 

Let us now turn to the optimal achievable cloning fidelities for these states. It was shown early on (\cite{brus_optimal_1998}) that the optimal \emph{local} cloning fidelity for $1\rightarrow 2$ cloning\footnote{This messy expression gives hints for the tricky analytical nature of this problem, even in the simplest scenario.} is given by:
\begin{multline} \label{eqn:local_optimal_non_ortho_fidelity_1to2}
    F^{\text{FO}, j}_{\Lbs, \opt}  = \frac{1}{2} + \frac{\sqrt{2}}{32 s}(1+s)\left(3−3s+\sqrt{1−2s+ 9s^2}\right)\\
     \times\sqrt{−1 + 2s + 3s^2 + (1−s)\sqrt{1−2s+ 9s^2}}, ~j \in \{1, 2\}
\end{multline}
It can be shown that the \emph{minimum} value for this expression is achieved when $s=\frac{1}{2}$ and gives $F^{\text{FO}, j}_{\Lbs, \opt} \approx 0.987$, which is much better than the symmetric phase-covariant cloner (recall $F_{\Lbs,  \text{opt}}^{\text{PC}, j} \approx 0.85$. 

To complement this, optimal \emph{global} fidelity of cloning the two states in \eqref{eqn:state_dependent_cloning_states} is given by:
\begin{equation}\label{eqn:optimal_global_non_ortho_state_fidelity}
    F^{{\text{FO}}}_{\Gbs, \opt}(M,N) = \frac{1}{2}\left( 1 + s^{M+N} + \sqrt{1-s^{2M}}\sqrt{1-s^{2N}} \right)
\end{equation}
At this point, it is necessary to draw attention to a certain fact which we will encounter in great detail later in this Thesis. Namely, it can be shown that the SDQCM which achieves this optimal \emph{global} fidelity, does \textbf{not} actually saturate the optimal \emph{local} fidelity (i.e. the individual clones do not have a fidelity given by \eqref{eqn:local_optimal_non_ortho_fidelity_1to2}). Instead, if we take the QCM which achieves an optimal value of~\eqref{eqn:optimal_global_non_ortho_state_fidelity}, and compute what the local fidelity of each clone would be, we get~\cite{brus_optimal_1998}:
\begin{multline}\label{eqn:state_dep_local_fidelity_from_global}
    F_{\Lbs, *}^{\text{FO}, j}(M, N) = \frac{1}{4}\left(
    \frac{1+s^M}{1+s^N}\left[1+s^2+2s^N\right] +\right.\\
     \left.\frac{1-s^M}{1-s^N}\left[1+s^2-2s^N\right] +
     2\frac{1-s^{2M}}{1-s^{2N}}\left[1-s^2\right]
    \right)    
\end{multline}
which (taking $M=1, N=2$) is actually a \emph{lower} bound for the optimal local fidelity, $F^{\text{FO}, j}_{\Lbs, \opt} $ in \eqref{eqn:local_optimal_non_ortho_fidelity_1to2}. As such, optimising either the local or global fidelities to find the optimal values will lead to different answers, in general\footnote{This is not always the case however, for universal and phase-covariant cloning, local and global optimisation are equivalent.}. We return to this issue in \chapref{chap:cloning}.

\chapter[Preliminaries II: Machine learning]{Preliminaries II: machine learning} \label{chap:prelim_machine_learning}

\section[\texorpdfstring{\color{black}}{} Machine learning]{Machine learning} \label{sec:prelim_machine_learning}

\begin{chapquote}{Nick Bostrom, TED, 2015}
 ``Machine intelligence is the last invention that humanity will ever need to make.''
\end{chapquote}

Having introduced most of the prerequisites in quantum computation, it is time to turn our attention to second pillar of this Thesis, the expansive field of machine learning (ML). There are many ways of describing what this field is, and what it aims to do, but roughly speaking, machine learning is an arm of artificial intelligence (AI) which aims to train machines to learn to solve some problem. Most commonly, they (computers) learn from data and solve problems without being explicitly programmed to do so. As such, the algorithms developed by machine learning practitioners are usually \emph{data-driven}, and in many cases, their success is dependent on the quality of data available to them. The methods and techniques used in ML algorithms draw heavily on the field of statistics and the type of machine learning we use here can also be referred to as \emph{statistical} learning. For an excellent and gentle introduction to this topic, see \cite{james_introduction_2013}.

Machine learning algorithms are canonically split into three main categories\footnote{Although, by modern standards, these categories are oversimplified. For example, there exists a plethora of intermediate categories, and merging of the ideas in each one. A major example is \emph{semi}-supervised learning.}:
\begin{tcbraster}[raster columns=3,raster equal height,nobeforeafter,raster column skip=1cm]
\begin{mybox}{red}{}
  \textbf{Supervised\\ learning.}
  \end{mybox}
  \begin{mybox}{blue}{}
  \textbf{Unsupervised\\ learning.}
  \end{mybox}
  \begin{mybox}{green}{}
  \textbf{Reinforcement learning.}
  \end{mybox}
\end{tcbraster}

Each of these subfields are large, and we do not attempt to provide a comprehensive overview here. In this immediate section, we are also only referring to \emph{classical} tasks, i.e. those where the underlying data is classical (vectors, matrices, etc.). We generalise this when discussing \emph{quantum} machine learning in~\chapref{sec:prelim_quantum_machine_learning}. 

The typical differences between these learning types is that supervised learning learns with \emph{labels} (we give a more precise definition in \secref{ssec:prelims/machine_learning/classification}), unsupervised learning learns \emph{without} labels, and reinforcement learning learns from \emph{experience} or via interaction. To avoid unnecessary tangents, we focus in this Thesis on the two particular examples which we use explicitly, one from supervised learning, and the other from unsupervised learning. We also take the viewpoint that (un)supervised learning is learning with(out) access to \emph{the correct answer} (or more technically, the `\emph{ground truth}') for given problem instances, rather than with(out) labels. The reason is that with this modification, the definitions also then encompass the contents of~\chapref{chap:cloning} of this Thesis.

However, for now, let us focus on the two following (classical) tasks:
\begin{enumerate}
    \item Classification (Supervised learning).
    \item Generative modelling (Unsupervised learning).
\end{enumerate}

Before diving into these applications, let us first introduce some terminology. Firstly, we have a \emph{machine learning model}. A model may be deterministic or probabilistic, and may be supervised or unsupervised in nature. In the supervised setting, a model is usually a proxy to represent a function, $f(\xbs)$, defined on some space $\xbs \in \mathcal{X}$. A generative model, on the other hand, is a proxy for a probability distribution, $p$. One of the most widespread classical machine learning models are \emph{neural networks}, which we discuss in \secref{ssec:prelim/machine_learning/neural_networks} which can be used to solve both of the problems above. At the time of writing, the closest\footnote{We use the term `closest' here lightly - there is still discussion in the community about what should be considered the quantum analogue of a neural network. For a (now outdated) discussion of this topic, see~\cite{schuld_quest_2014}.} counterpart in the quantum realm to a neural network is the \emph{parametrised quantum circuit} (PQC), which we introduce in \secref{sec:variational_quantum_algorithms}. These models usually belong to a \emph{model family}, $\mathcal{M}$; for example a family of functions, $\mathcal{M} := \mathcal{F} := \{f\}$ or probability distributions, $\mathcal{M} := \mathcal{D} := \{p\}$\footnote{This is actually slightly misleading since supervised learning problems \emph{also} are defined with a distribution over the input domain.}. It is also common to \emph{parametrise} the model family, $\mathcal{M} \rightarrow \mathcal{M}_{\paramtheta}$. A machine learning \emph{algorithm} then optimises these parameters in order to \emph{learn}, which corresponds to finding the optimal candidate in the model family to solve the problem of interest. As we use neural networks as our benchmark \emph{model}, we  view \emph{gradient descent} as our canonical machine learning \emph{algorithm}, since we use it almost exclusively in this Thesis.

The parameters, $\paramtheta$, define the model, but there is another type of parameter which one must also be concerned with. These are \emph{hyperparameters}, which we define as follows:
\begin{defbox}
\begin{definition}[Hyperparameters]\label{defn:hyperparams}~ \\
    In a machine learning algorithm, a \emph{hyperparameter} is a parameter used to control the learning procedure. It is distinct from a model parameter in that it is `outside' the model, and usually associated to the learning algorithm itself.
\end{definition}
\end{defbox}
Next, we have the concept of (labelled) data.
\begin{defbox}
\begin{definition}[A dataset]\label{defn:dataset}~ \\
    A dataset is a collection of datapoints, $\{x^i\}_i$, from some input domain, $\mathcal{X}$. Each element of a \emph{labelled} dataset will also have a corresponding \emph{label}, $y^i$, coming from some domain, $\Y$, for each datapoint, $x^i$. We may also have a probability distribution, $D \in \mathcal{D}$ defined over $\X$, so $x^i \sim D$.
\end{definition}
\end{defbox}
The above definition of data is intentionally flexible, as in general datapoints may be binary, real or complex-valued scalars ($x^i \in \{0, 1\}/\mathbb{R}/\mathbb{C}$), vectors ($x^i := \zbs^i = [z_1, \dots z_n]^T \in \mathbb{R}^n/\mathbb{C}^n$) or even quantum states in a Hilbert space, $x^i := \rho^i \in \hilb$. In this Thesis, data will take the form of each of the above. Generally speaking, the difference between supervised and unsupervised learning is the access to, or lack of, the labels, $y^i$.
\begin{defbox}
\begin{definition}[Train/test/validation data]\label{defn:train_test_val_data}~ \\
    Before a dataset is used in a ML algorithm, it is split into a \emph{training} set, a \emph{testing} set, and a \emph{validation} set. The training set is used to train the model, the test set acts as `unseen' data and tests the models generalisation capability and the validation set is used to tweak hyperparameters of the ML model.
\end{definition}
\end{defbox}

\section[\texorpdfstring{\color{black}}{} Neural networks]{Neural networks} \label{ssec:prelim/machine_learning/neural_networks}

\begin{figure}[!ht]
\centering
\includegraphics[width =0.9\columnwidth]{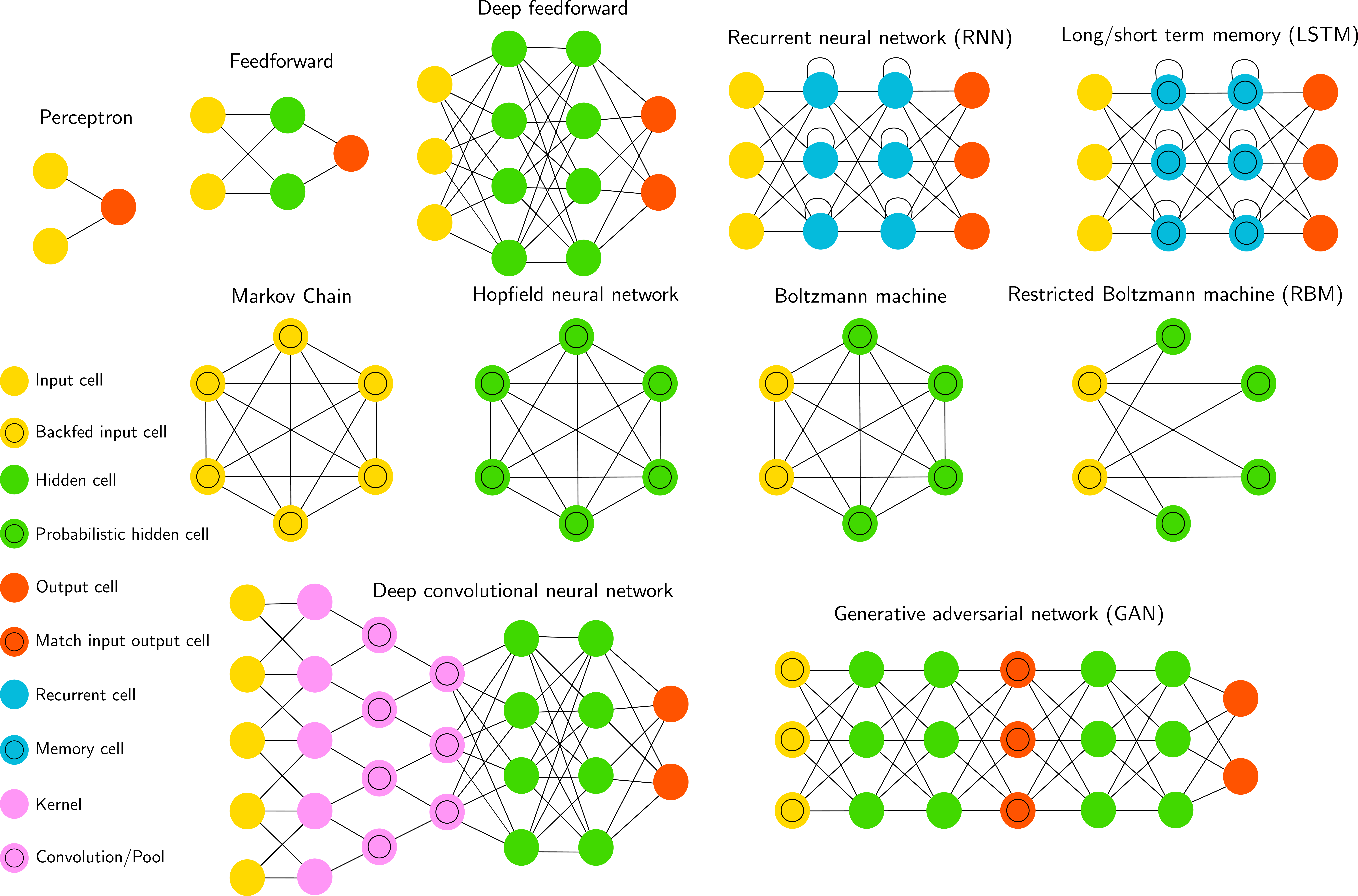}
\caption[\color{black} Examples of neural network architectures.]{\textbf{Examples of neural network architectures.} Image adapted from the `mostly complete chart of neural networks' from \url{https://www.asimovinstitute.org/neural-network-zoo/}.}
\label{fig:neural_network_examples}
\end{figure}

\begin{figure}[!ht]
\centering
\includegraphics[width =\textwidth]{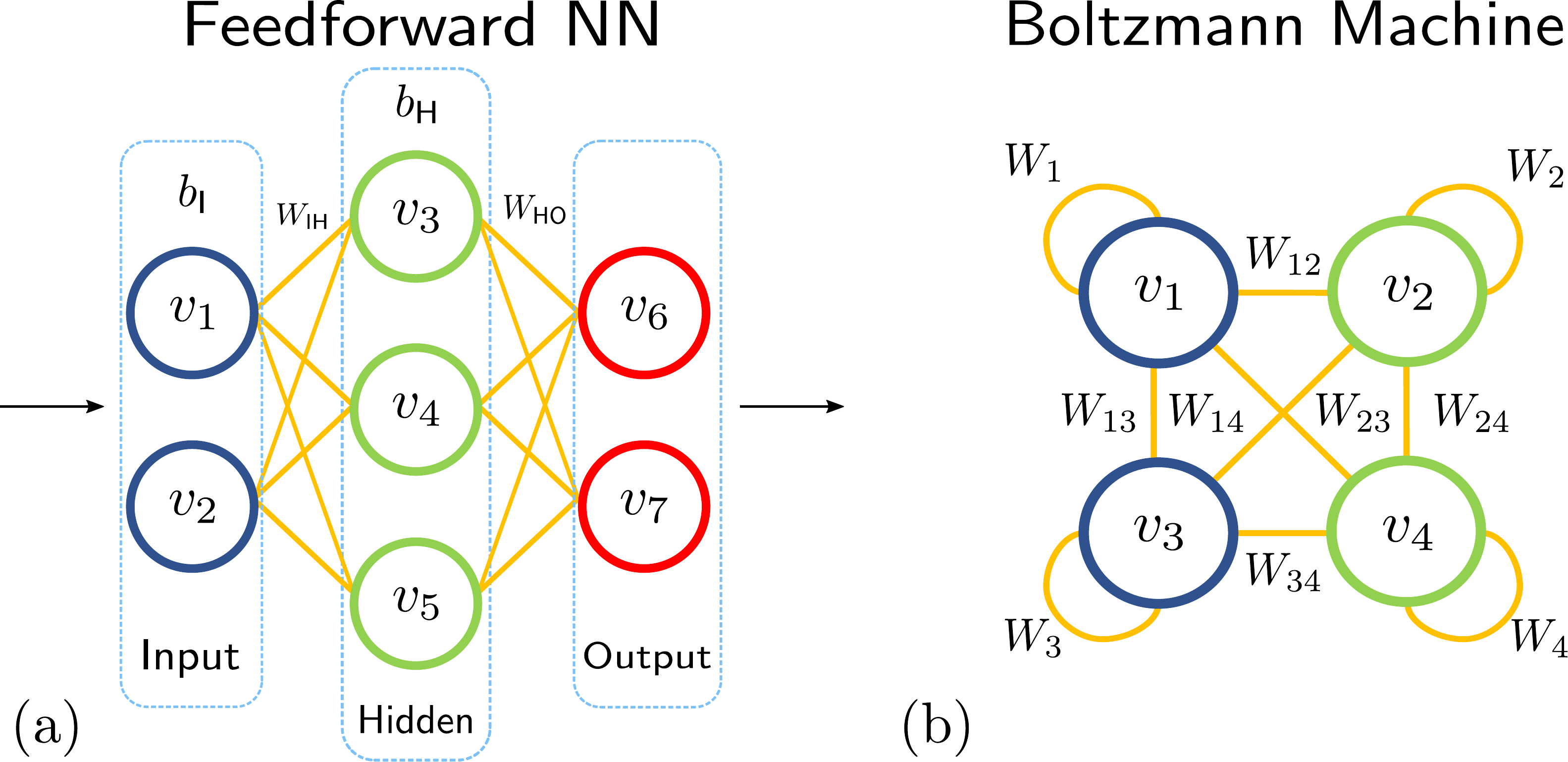}
\caption[\color{black} Two canonical neural networks.]{\textbf{Two canonical neural networks (NN).} (a) A feedforward NN with an input, hidden and output layer. (b) a fully connected Boltzmann machine with visible (blue) and hidden (green) units.}
\label{fig:feedforward_vs_boltzmann}
\end{figure}

One of the main workhorses of modern machine learning is the \emph{neural network} (NN). In its most basic form, a NN is a collection of nodes of \emph{neurons}\footnote{The inspirations for `artificial' neural networks came from neurons in the brain. This was formalised mathematically by McCulloch \& Pitts in 1943 in the development of an artificial neuron (\cite{mcculloch_logical_1943}).} which are connected in a graph structure. This graph structure is one of the defining features of NNs. If the graph has no loops (acyclic), it is referred to as a \emph{feedforward} neural network, while NNs with loops are often called \emph{recurrent}. There exists a plethora of examples of NN architectures, some examples are given in \figref{fig:neural_network_examples}. Neural networks are usually built by stacking \emph{layers} of simpler ingredients, and \emph{deep} neural networks (i.e.\@ those with several layers) are the primary cause of the deep learning revolution (\cite{schmidhuber_deep_2015, goodfellow_deep_2016}), and have seen great success in solving difficult real life tasks of practical interest. Each layer consists of trainable parameters which are optimised to solve the task in hand.
In particular, two we wish to highlight are the simple feedforward neural network, and the Boltzmann machine, shown in \figref{fig:feedforward_vs_boltzmann}. As observed, the key difference between the two is the graph structure; the Boltzmann machine contains loops whilst the feedforward network is acyclic. The parameters in both models consist of trainable \emph{weights}, $W$, and \emph{biases}, $b$ (or self-loops). We focus on these two examples, because they are canonical examples of classical models for the two primary (classical) ML tasks in which we are interested in this Thesis. Specifically, the feedforward network is used for classification tasks, and the Boltzmann machine is used for generative modelling of probability distributions. We explicitly use the Boltzmann machine in \chapref{chap:born_machine}.\\

\subsection[\texorpdfstring{\color{black}}{} Feedforward neural networks]{Feedforward neural networks} \label{ssec:prelim/machine_learning/neural_networks/ffnn}


Let us begin with the feedforward NN. This consists of an `input' layer, into which the data is inserted, a `hidden' layer which increases the representational power of the model, and a final output layer. These layers are connected by the weights, which form a matrix. For example, in \figref{fig:feedforward_vs_boltzmann}(a), $W_{13}$ is the weight connecting vertices $v_1$ in the input layer and $v_3$ in the hidden layer. Putting these weights together forms the matrix $W^{\mathsf{IH}}$. Each node in \figref{fig:feedforward_vs_boltzmann}(a) may also have its own bias, $b$. These weights and biases are the trainable parameters of the network, $\paramtheta = \{W^{\mathsf{IH}}, W^{\mathsf{HO}}, \boldsymbol{b}^{\mathsf{I}}, \boldsymbol{b}^{\mathsf{H}}\}$. Given an input data vector, $\xbs$, the output of this network is given by:
\begin{equation} \label{eqn:ffnn_layers}
\begin{split}
    f_{\paramtheta}(\xbs) := \ybs^{\text{out}}(\xbs) &:= \sigma(W^{\mathsf{HO}}\ybs^{1}(\xbs)  + \boldsymbol{b}^{\mathsf{H}})\\
    \ybs^{1}(\xbs) &:= \sigma(W^{\mathsf{IH}}\xbs + \boldsymbol{b}^{\mathsf{I}})\\
    \end{split}
\end{equation}
The output vector can be a vector of probabilities, for example (as we shall see in \secref{ssec:prelims/machine_learning/classification}) the probabilities that the input sample belongs to each of the desired classes. The function, $\sigma$, is crucially important in a neural network as it provides a non-linearity to the model\footnote{Without this non-linear function after each layer in deep neural networks, the model would be entirely linear, and a deep network would be equivalent to a single layer model. Implementing such a non-linearity to design \emph{quantum} neural networks is an active area of research, and is problematic due to the naturally linear nature of quantum mechanics.}, called an activation\footnote{Coming from the `activation' of a biological neuron when sufficient input signal is applied.} function. Common choices for $\sigma$ are the \textsf{sigmoid}, \textsf{tanh} or the \textsf{ReLU} (recitified linear unit) activation functions:
\begin{equation} \label{eqn:nn_activation_functions}
    \sigma_{\text{\textsf{sigmoid}}}(x) = \frac{1}{1+e^{-x}} \qquad 
      \sigma_{\text{\textsf{tanh}}}(x) = \tanh(x)\qquad
        \sigma_{\text{\textsf{ReLU}}}(x) = \max\{0, x\}
\end{equation}
These functions provide differentiable alternatives to the more obvious binary step function which could be used as an activation function.\\

\subsection[\texorpdfstring{\color{black}}{} The Boltzmann machine]{The Boltzmann machine} \label{ssec:prelim/machine_learning/neural_networks/boltzmann_machine}


The Boltzmann machine is an example of a physics inspired model known as an \emph{energy based} model. This means it is completely defined by an energy function, which in the case of the Boltzmann machine is given as an Ising model:
\begin{equation} \label{eqn:ising_model_classical_energy}
    E(\zbs) := -\left(\sum\limits_{i<j} W_{ij} z_iz_j + \sum\limits_{k}b_k z_k\right)
\end{equation}
The variables $z_i \in \{+1, -1\}$\footnote{This is inherited from the notation of atomic spins in the Ising model of ferromagnetism, named after Ernst Ising. It is sometimes convenient to associate these variables to be $0$ or $1$ instead. The weights between nodes $W_{ij}$ are also called `interaction' terms, and denoted $J_{ij}$, and the biases, $b_k$, are called local (external) magnetic field strengths and denoted $h_k$ for historic reasons.} are the values that the nodes, $v_i$ in \figref{fig:feedforward_vs_boltzmann}(b) can take. Each configuration of the spins will have its associated energy, and the model is defined by the distribution over these energy values. In other words, the probability of a given energy (a given configuration of the $n$ variables, $\zbs = [z_1, \dots, z_n]^T$) is given by the Boltzmann distribution:
\begin{equation} \label{eqn:boltzmann_distribution}
    p_{\paramtheta}(\zbs) := \frac{\erm^{E(\zbs)}}{\mathcal{Z}},
\end{equation}
where $\mathcal{Z} := \sum_{\zbs} \erm^{E(\zbs)}$ is the \emph{partition function}, a normalisation constant which is intractable to compute in general, due to the exponentially many terms in the sum over $2^n$ possible configurations of $\zbs$. We use the notation, $z_i$, to denote a specific \emph{realisation} of the value of node $i$ and we use $v_i$ to label the node itself. As observed in \figref{fig:feedforward_vs_boltzmann}(b), some of the nodes in the Boltzmann machine may be hidden also\footnote{Hidden nodes are one feature which generalise the Boltzmann machine from its simpler counterpart, the \emph{Hopfield} network, named after John Hopfield for his 1982 paper~\cite{hopfield_neural_1982}. The Hopfield network is also a deterministic model.} which are used to increase the representational power of the model, and as such the data is associated only to the remaining `visible' nodes. In \secref{ssec:prelims/machine_learning/generative_modelling}, we will see how the Boltzmann machine can be used for the second of our tasks: generative modelling. In this context, the parameters, $\paramtheta = \{W_{ij}, b_k\}$ of the Boltzmann machine will be used to fit the probability distribution of the model to a data distribution.

\section[\texorpdfstring{\color{black}}{} Machine learning tasks]{Machine learning tasks} \label{ssec:prelim/machine_learning/ml_tasks}

\subsection[\texorpdfstring{\color{black}}{} Classification as supervised learning]{Classification as supervised learning} \label{ssec:prelims/machine_learning/classification}
As mentioned above, we will use classification as the primary example of supervised learning. Before diving into the special case, let us define the general problem in this domain~\cite{schuld_supervised_2018}.

\begin{defbox}
\begin{definition}[Supervised learning problem]\label{defn:prelim/ml/supervised_learn_prob}~ \\
    Given an input domain, $\X$, and an output domain, $\Y$, a training data set:
    \begin{equation} \label{eqn:labeled_data_for_classifier}
        \mathcal{D} = \{(x^1, y^1), \dots, (x^M, y^M)\}
    \end{equation} 
    of training pairs, $(x^m, y^m) \in \X \times \Y$ with $m = 1, \dots, M$ of training inputs, $x^m$ and target outputs $y^m$ as well as a new unclassified input, $x^* \in \X$, guess or predict the corresponding output $y^*\in \Y$.
\end{definition}
\end{defbox}
Two of the most common examples of supervised learning problems are \emph{regression} and \emph{classification}. Roughly speaking, the difference is in the nature of the variables or labels, $y_i$. In regression problems, the labels are continuous (or quantitative), whereas in classification, the labels are discrete (meaning categorical or qualitative) (\cite{james_statistical_2013}). There are a plethora of methods to solve these problems, and in this thesis we use a quantum computer to do so (for the classification version)

The most basic form of classification is \emph{binary} classification, where the labels are only two possible values i.e. $y^i \in \{0, 1\}$, so each data point, $x^i$, is assigned a value of $0$ or $1$. The goal of the `classifier' is to learn this function, $f$, mapping the data to their corresponding labels. This function may come from a particular function family, $\mathcal{F}$, which defines our model class. As discussed at the start of this section, we can parametrise this function family, $f\rightarrow f_{\paramtheta}$. If we use a feedforward neural network for this task the parameters correspond to the weights and biases of the network. In \secref{ssec:prelim/machine_learning/learning_theory} we will introduce a slightly more formal discussion for completeness.

A very common baseline example (in the classical world) of such a classification task is the classification of digits, using the famous MNIST database, which consists of $60,000$ training images and $10,000$ test images of handwritten digits from $0$ to $9$. The goal is to learn the mapping from image to digit label, so to classify all images in the test set correctly.

\begin{figure}[!ht]
\centering
\includegraphics[width=\columnwidth]{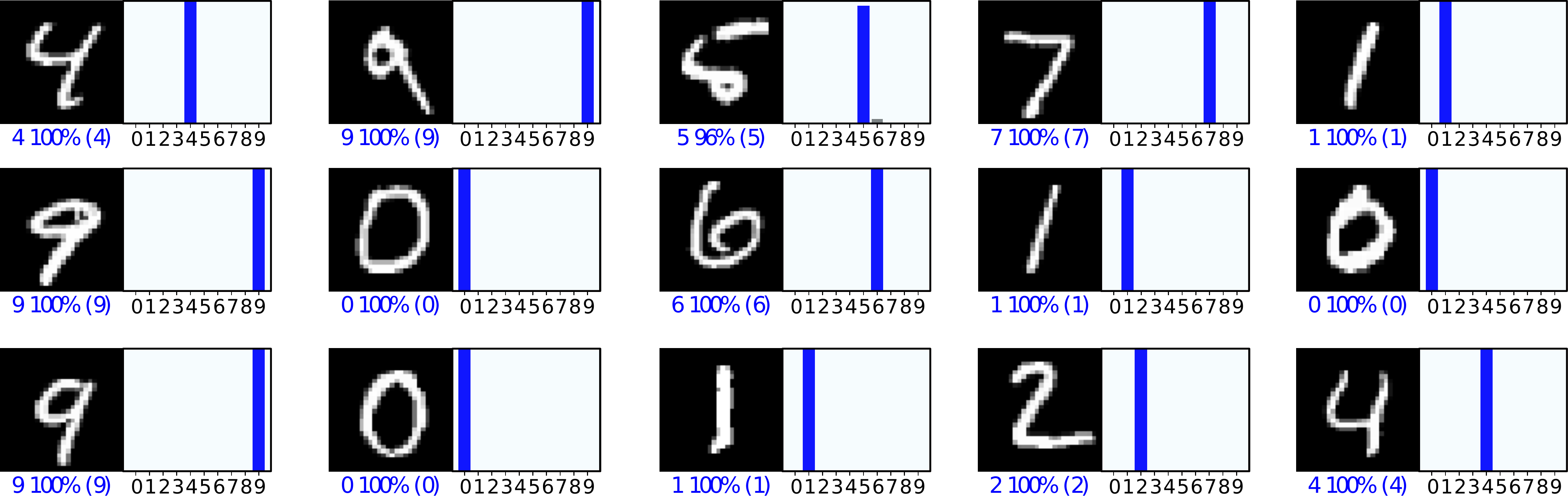}
\caption[\color{black} A feedforward neural network trained on the MNIST training dataset]{\textbf{A feedforward neural network trained on the MNIST training dataset}, classifying example images in the test dataset. The boxes to the right of each digit show the probability assigned to each possible label. We see all of these digits were correctly labelled, since the trained model assigns a high probability weight to the correct label. The model in question is a simple neural network consisting of an input layer, a hidden layer with $128$ nodes, and a final output layer with $10$ nodes, where each node in the output represents a digit, $\{0, \dots, 9\}$. This figure was produced using an example tutorial in Tensorflow~\cite{abadi_tensorflow_2016}.}
\label{fig:tensorflow_mnist_classification}
\end{figure}

Training machine learning models has become very accessible in recent years with the development of deep learning software, particularly in Python, such as TensorFlow~\cite{abadi_tensorflow_2016}, Pytorch~\cite{paszke_pytorch_2019} and Keras~\cite{chollet_keras_2015}. Using Tensorflow, we can very easily load the above MNIST dataset, and train a feedforward NN (as in \figref{fig:feedforward_vs_boltzmann}(a)) to classify images. \figref{fig:tensorflow_mnist_classification} shows the results of $15$ digits from the trained model. Under each image, we see what the model labelled the test digit as, and the corresponding probability of doing so. The particular trained model had $100\%$ confidence about each image, except the digit $5$ in the first row and third column, which has a $4\%$ chance of being the digit $6$, according to the model. From this example, one can imagine how neural networks have become so powerful, and widely used - modern architectures can have billions of parameters~\cite{trask_modeling_2015} and are applied on much more challenging tasks than digit recognition. As quantum computers scale in size, we hope they will be able to tackle similar problems.

\subsection[\texorpdfstring{\color{black}}{} Generative modelling as unsupervised learning]{Generative modelling as unsupervised learning} \label{ssec:prelims/machine_learning/generative_modelling}
\chapref{chap:born_machine} of this Thesis will be focused on using quantum computers for \emph{generative modelling}, which as mentioned is our primary example of unsupervised learning. 

Unlike the supervised learning models discussed in the previous section, generative models can be significantly more complex, since they have the ability to also \emph{generate} new data observations, sometimes called `synthetic data generation'. While the former deals with parametrised \emph{functions}, generative models use parametrised \emph{probability distributions}, $p_{\paramtheta}$, which again come from some \emph{distribution} model family, $\mathcal{D}$. The distribution, $p_{\paramtheta}$ may be continuous, in which case $p_{\paramtheta}$ is the probability \emph{density} function (PDF) over the sample space, $\X$. In this Thesis however, we exclusively use \emph{discrete} distributions over $\X$, so $p_{\paramtheta}$ is a probability \emph{mass} function (PMF). In particular, we will take $\X = \{0, 1\}^n$.

\begin{figure}[!ht]
\centering
\includegraphics[width=\textwidth]{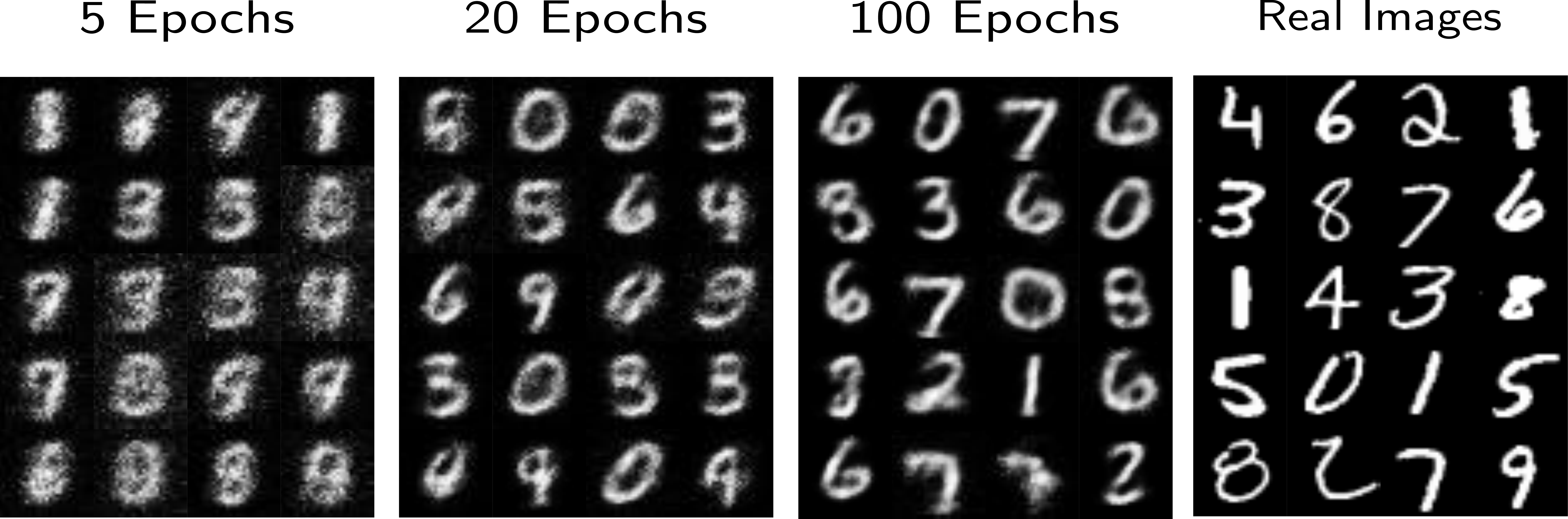}
\caption[\color{black} A restricted Boltzmann machine learning to generate MNIST digits.]{\textbf{A restricted Boltzmann machine learning to generate MNIST digits.} From left to right, we see improving performance (in terms of resolution), indicating that the model is learning to generate digits which look similar to those in the test set (the rightmost image). In the RBM, every (visible) spin can by used to represent each pixel being black or white (up or down).}
\label{fig:rbm_mnist_generated_images}
\end{figure}

In the generative modelling problem, we are required to learn some `data' distribution, which we denote $\pi$ over some space $\Y$. We may be given direct access to the probabilities, $\pi(\ybs), \ybs \in \Y$, but more commonly we will only have a dataset of $M$ vectors sampled according to $\pi$, $\{\ybs^i\}_{i=1}^M, \ybs \sim \pi$. We further assume $\pi$ is a discrete distribution over $\{0, 1\}^n$, so for efficiency we only have $M = \mathcal{O}\left(\poly(n)\right)$ samples. Given our model, $p_{\paramtheta}$, the goal is then to fit $p_{\paramtheta}$ to $\pi$, using the parameters, $\paramtheta$. If we were to use a Boltzmann machine for this task, we would fit the weights and biases in the energy function~\eqref{eqn:ising_model_classical_energy} and train the model distribution, $p_{\paramtheta}(\zbs)$. In~\figref{fig:rbm_mnist_generated_images}, we demonstrate generated results using a \emph{restricted} Boltzmann machine on the same MNIST dataset as in \figref{fig:tensorflow_mnist_classification}. In the next section, we provide some further detail on how to train such neural network models. This will be especially relevant for the contents of this Thesis, since the methods we use to train our \emph{quantum} models incorporate these exact same techniques (with suitable modifications).

\section[\texorpdfstring{\color{black}}{} Training a neural network]{Training a neural network} \label{ssec:prelim/machine_learning/training_NN}
There are a multitude of methods to train machine learning models, such as the neural networks we presented in the previous section. For particular models and algorithms, such as the support vector machine (SVM) the training procedure is a (relatively) simple convex optimisation problem, provided one can easily compute the underlying kernel matrix (e.g. the one defined in~\thmref{thm:kernel_definition} in \secref{ssec:prelim/qc/distance_measures/probability}). Unfortunately, for neural networks, the problem is not so simple. This is because the \emph{loss} or \emph{cost}\footnote{We use these terms interchangeably here; some may consider the cost function to be the loss function, but with the addition of \emph{regularisation}.} function landscape may become highly non-convex due the choice of our parameterisation, $\paramtheta$. In \secref{ssec:cost_functions} we will discuss desirable properties of cost functions in more detail, since they apply to both quantum and classical machine learning models. For now, let us simply assume that a \emph{cost function} is some function of our model parameters, $\Cbs(\paramtheta)$, and that it gives us a method to measure how well we are doing in the training process. The goal of training is to minimise\footnote{We can assume this WLOG since if one may want to \emph{maximise} $\Cbs(\paramtheta)$, we can convert to a minimisation problem by defining $\widetilde{\Cbs}_{\paramtheta} = - \Cbs_{\paramtheta}$ and minimising $\widetilde{\Cbs}_{\paramtheta}$.} $\Cbs(\paramtheta)$, i.e.\@ solve $\paramtheta^* = \argmin_{\paramtheta}\Cbs(\paramtheta)$. In many common examples, the cost can be defined as an expectation over the dataset which, in the limit $M \rightarrow \infty$, approaches the expectation over the full distribution from which each datapoint is drawn, $\xbs^i \sim D$:
\begin{equation} \label{eqn:per_sample_cost_function}
    \Cbs(\paramtheta) = \mathbb{E}_{\xbs \sim D} \Cbs_{\xbs}(\paramtheta) \approx \underbrace{\frac{1}{M}\sum_{i=1}^M \Cbs_i(\paramtheta)}_{R_{\text{emp}}}
\end{equation}
$\Cbs_i(\paramtheta)$ is the term evaluated using a single sample from the dataset, $\xbs_i$, and $\mathcal{R}_{\text{emp}}$ is known as the \emph{empirical risk}. Minimising this quantity is known as \emph{empirical risk minimisation}, and is commonly used as a proxy to minimise our true objective, $\Cbs$.

The cost function defines the cost landscape as a function of $\paramtheta$, and the training procedure is to find the absolute (global) minimum of $\Cbs(\paramtheta)$. However, as we mentioned, this problem is non-convex, meaning many sub-optimal \emph{local} minima exist and may also be found. The goal of a training algorithm is usually to efficiently traverse this landscape to find the global minimum.

Let us now focus on NNs specifically. Methods to train NNs can be roughly split into two categories - \emph{gradient-free} and \emph{gradient-based}. Gradient-free methods are also called zeroth-order methods, since they optimise $\Cbs$ using only information about $\Cbs$ itself. A non-exhaustive list of gradient-free optimisation methods include Bayesian optimisation (\cite{mockus_global_1989}), evolutionary methods (\cite{beyer_evolution_2002}) or simulated annealing (\cite{laarhoven_simulated_1987, kirkpatrick_optimization_1983}), see~\cite{rios_derivative-free_2013} for an overview. Gradient-free methods are useful in cases where perhaps the gradients of $\Cbs$ are not available, or difficult to compute.

In cases where we \emph{can} compute gradients, we can use alternatively \emph{gradient-based} optimisation. As the name suggests, these methods use first-order gradient methods to optimise $\Cbs$. One may also consider training algorithms which use \emph{second-order} gradient methods require evaluation of also the Jacobian or Hessian of $\Cbs$. For this Thesis, first-order gradient based methods will suffice.

The main algorithm we use is \emph{stochastic gradient-descent} (SGD):
\begin{defbox}
\begin{definition}[Stochastic gradient descent]\label{defn:sgd_definition}~ \\
    Given a single sample from a dataset, $\xbs^i$, stochastic gradient descent updates the parameters, $\paramtheta$, as:
    \begin{equation}\label{eqn:sgd_update_rule}
        \paramtheta^{(j+1)} \leftarrow \paramtheta^{(j)} - \Delta\Cbs(\eta, j, \paramtheta) =  \paramtheta^{(j)} - \eta \nabla_{\paramtheta} \Cbs_i(\paramtheta) 
    \end{equation}
    where $j$ defines an epoch, $\eta$ is a \emph{learning rate} and $\nabla_{\paramtheta}$ denotes the gradient vector with respect to each parameter, $\paramtheta_k$.
\end{definition}
\end{defbox}
The number of epochs is typically a predefined number, and one epoch constitutes one complete pass over the training data. The learning rate, $\eta$, controls the speed of training and both of these are \emph{hyperparameters} (see~\defref{defn:hyperparams}) of the learning algorithm. While SGD is very fast, it may suffer from high variance since only a single training sample is used to estimate the true gradient. Instead, one can consider \emph{mini-batch} gradient descent, where a certain number of samples are taken (given by a \emph{batch size}, $b$) from the training dataset to estimate $\nabla_{\paramtheta} \Cbs(\paramtheta)$ for each update. Taking the batch size to be the same as the training data set size ($b= 1$) recovers SGD as in \defref{defn:sgd_definition}, while taking $b=M$ recovers `regular' gradient descent, where the \emph{entire} dataset is used in each parameter update. Clearly, choosing the batch size also heavily affects the speed of training, and so it is another hyperparameter.

Notice that we have written the `update rule' for SGD in \eqref{eqn:sgd_update_rule} in two different ways, one including a term $\Delta\Cbs(\eta, j, \paramtheta)$, and then a specification of this expression which is that used specifically for SGD, i.e.\@ an update using a term $\eta \nabla_{\paramtheta} \Cbs_i(\paramtheta)$. The former expression is written in this form to allow more general update rules. For example, one may consider that the update depends on the epoch, $j$, i.e.\@ time dependence, or a more complicated expression in terms of the parameters, $\paramtheta$. 

In this Thesis, we primarily use an update rule known as Adam, which uses an \emph{adaptive} learning rate, $\eta$, and incorporates the notion of \emph{momentum} to SGD.
\begin{defbox}
    \begin{definition}[Adam update rule (\cite{kingma_adam_2015})]  \label{defn:adam_definition}~ \\
        The adaptive momentum estimation (Adam) update rule to the parameters is given by the following:
        \begin{align}\label{eqn:adam_update_rule}
            \paramtheta^{(j+1)} &\leftarrow \paramtheta^{(j)} - \Delta_{\text{Adam}}\Cbs(\eta, j, \paramtheta) =  \paramtheta^{(j)} - \frac{\eta_{\text{init}}}{\sqrt{\hat{\boldsymbol{v}}_j} + \epsilon} \hat{\boldsymbol{m}}_j, \\
            &\hat{\boldsymbol{m}}_j = \frac{\boldsymbol{m}_j}{1- \beta_1^j},\qquad  \hat{\boldsymbol{v}}_j = \frac{\boldsymbol{v}_j}{1- \beta_2^j}, \\
             \boldsymbol{m}_j &= \beta_1\boldsymbol{m}_{j-1} + (1-\beta_1) \nabla_{\paramtheta^{(j)}} \Cbs(\paramtheta),\\ 
                \boldsymbol{v}_j &= \beta_1\boldsymbol{v}_{j-1} + (1-\beta_2) [\nabla_{\paramtheta^{(j)}} \Cbs(\paramtheta)]^2
        \end{align}
        where $\beta_1, \beta_2, \epsilon, \eta_{\text{init}}$ are hyperparameters. $\beta_1^k, k\in \{1, 2\}$ means $\beta_k$ raised to the power $j$, which results in a exponential decay.
    \end{definition}
\end{defbox}
The hyperparameters, $\beta_1, \beta_2$\footnote{The original Adam paper chooses $\beta_1 = 0.9$, $\beta_2 = 0.999$, $\epsilon= 1\times 10^{-8}$, which are the values we use throughout this Thesis.}, are heavily weighted towards a value of $1$, which means that random fluctuations in the gradient, if they are sufficient small, will contribute less than the weighted sum of previous gradients. Therefore, the optimisation is likely to keep moving in the same direction at each epoch (hence the term `momentum'). The vector, $\boldsymbol{v}$, keeps track of the second moment of the gradient, and the terms $\hat{\boldsymbol{m}_j}, \hat{\boldsymbol{v}_j}$ are computed in order to correct the tendency of $\boldsymbol{m}_j, \boldsymbol{v}_j$ to be biased towards the zero vector, $\boldsymbol{0}$, since they are both initialised at $\boldsymbol{0}$. One could also choose alternative update rules such as Adagrad~\cite{duchi_adaptive_2011}, Adadelta~\cite{zeiler_adadelta_2012}, NAdam~\cite{dozat_incorporating_2016} or a plethora of others. In many cases, the update rule is designed to encourage large values at the start of training, but decrease towards the end so we do not overshoot the minimum we are after. As we shall see in \secref{sec:variational_quantum_algorithms}, for training modern quantum machine learning models, one may wish to include \emph{quantum awareness} into such optimisers. To this end, versions of update rules have been proposed by incorporating quantum measurement error~\cite{kubler_adaptive_2020}, or to partially overcome linear scaling of computing quantum gradients~\cite{cade_strategies_2020, gacon_simultaneous_2021} (see~\eqref{eqn:parameter_shift_rule} in \secref{ssec:vqa_cost_function_optimisation}). We will not use any of these more complicated methods in this Thesis, so we will not discuss them further here. However, we remark that all of these methods could be applied and tested on each of the applications we study in the subsequent chapters.\\

\subsection[\texorpdfstring{\color{black}}{} Computing gradients]{Computing gradients} \label{ssec:prelim/machine_learning/computing_gradients}

Finally, let us remark on the computation of the gradient terms, $\nabla_{\paramtheta} \Cbs(\paramtheta)$ for the two classical models we have introduced above. For the feedforward NN, we do this to illustrate a difference between the quantum models we discuss in \secref{sec:variational_quantum_algorithms} and for the Boltzmann machine, we explicitly use it later in this Thesis in \chapref{chap:born_machine}.

\subsubsection[\texorpdfstring{\color{black}}{} Feedforward neural networks]{Feedforward neural networks} \label{ssec:prelim/machine_learning/computing_gradients_ffnn}

For training feedforward neural networks, a key breakthrough was the discovery of the \emph{backpropagation} algorithm (\cite{rumelhart_learning_1987}), which is short for the backpropagation of errors in the neural network. Recall the structure of the simple example feedforward NN in \figref{fig:feedforward_vs_boltzmann}. We ultimately want to compute the gradient with respect to each parameter, the weights, $W_{\mathsf{HO}}, W_{\mathsf{IH}}$ plus the biases of each node, $b_{\mathsf{I}}, b_{\mathsf{H}}$. Fortunately, the output vector, $\ybs_{\text{out}}$ is a concatenation of simple functions applied to the input. So the gradient can be computed using the simple chain rule (take for example the gradient with respect to the input bias vector, $b_{\mathsf{I}}$ with a slight abuse of notation):
\begin{align}
 \frac{\partial C(W, b)}{\partial \boldsymbol{b}^{\mathsf{I}}}  &=
  \frac{\partial C(W, b)}{\partial \ybs^{\text{out}}} \times   \frac{\partial \ybs^{\text{out}}}{\partial \ybs^{\text{1}}} \times    \frac{\partial \ybs^{1}}{\partial \boldsymbol{b}^{\mathsf{I}}}  \\
    &= \frac{\partial C(W, b)}{\partial \ybs^{\text{out}}} \times   \frac{\partial \sigma(W^{\mathsf{HO}}\ybs^{1}  + \boldsymbol{b}^{\mathsf{H}})}{\partial \ybs^{\text{1}}} \times    \frac{\partial \sigma(W^{\mathsf{IH}}\xbs + \boldsymbol{b}^{\mathsf{I}})}{\partial \boldsymbol{b}^{\mathsf{I}}} \label{eqn:backprop_expanded}
\end{align}
If we use a sigmoidal activation function, $\ybs(\zbs) := \sigma_{\textsf{sigmoid}}(\zbs) =  1/\left(1+e^{-\zbs}\right)$, the derivative is simple: $\sfrac{\partial \sigma_{\textsf{sigmoid}}(\zbs)}{\partial \zbs} = \sigma_{\textsf{sigmoid}}(\zbs)(1- \sigma_{\textsf{sigmoid}}(\zbs))$. Furthermore, since each layer before the activation function has a simple affine form, it can also be trivially computed. We must also obviously assume that the cost function is a differentiable function with respect to $\ybs_{\text{out}}$, but one can choose the $\ell_2$ norm,~\eqref{eqn:distance_meas/l2_metric} as a simple example. 

One crucial feature of the above derivation is the computational efficiency of the calculation. By staring at \eqref{eqn:backprop_expanded} and beginning with the parameters at the \emph{end} or the network, one can see how each gradient evaluation reuses the same computational steps. For example, to compute the gradient with respect to \emph{every} weight and bias, we must begin by computing $\sfrac{\partial C(W, b)}{\partial \ybs_{\text{out}}}$. Similarly, the computation of every parameter gradient will require an evaluation of a gradient for the layer immediately after it. Storing these intermediate computations as one trains the network, rather than re-computing them for \emph{each} parameter separately, massively improves the complexity. This general principle underpins the field of \emph{differentiable programming}, and \emph{automatic differentiation}. Most deep learning libraries such as Tensorflow and PyTorch contain an \computerfont{autodiff} function which can perform this differentiation automatically for many functions. We highlight this here to make the connection to the quantum case. We shall see in \secref{ssec:vqa_cost_function_optimisation} how the `parameter-shift' rule, and the black box nature of quantum computers prohibits these chain rule type of gradient computations (at least in the near term). As such, for training quantum models we must pay the price of a linear scaling in the number of parameters which may be a major bottleneck as problem sizes scale. We will revisit this discussion in \secref{ssec:vqa_cost_function_optimisation}.

\subsubsection[\texorpdfstring{\color{black}}{} Boltzmann machine]{Boltzmann machine} \label{ssec:prelim/machine_learning/computing_gradients_boltz}

\begin{figure}[!ht]
    \centering
    \includegraphics[width=0.7\columnwidth]{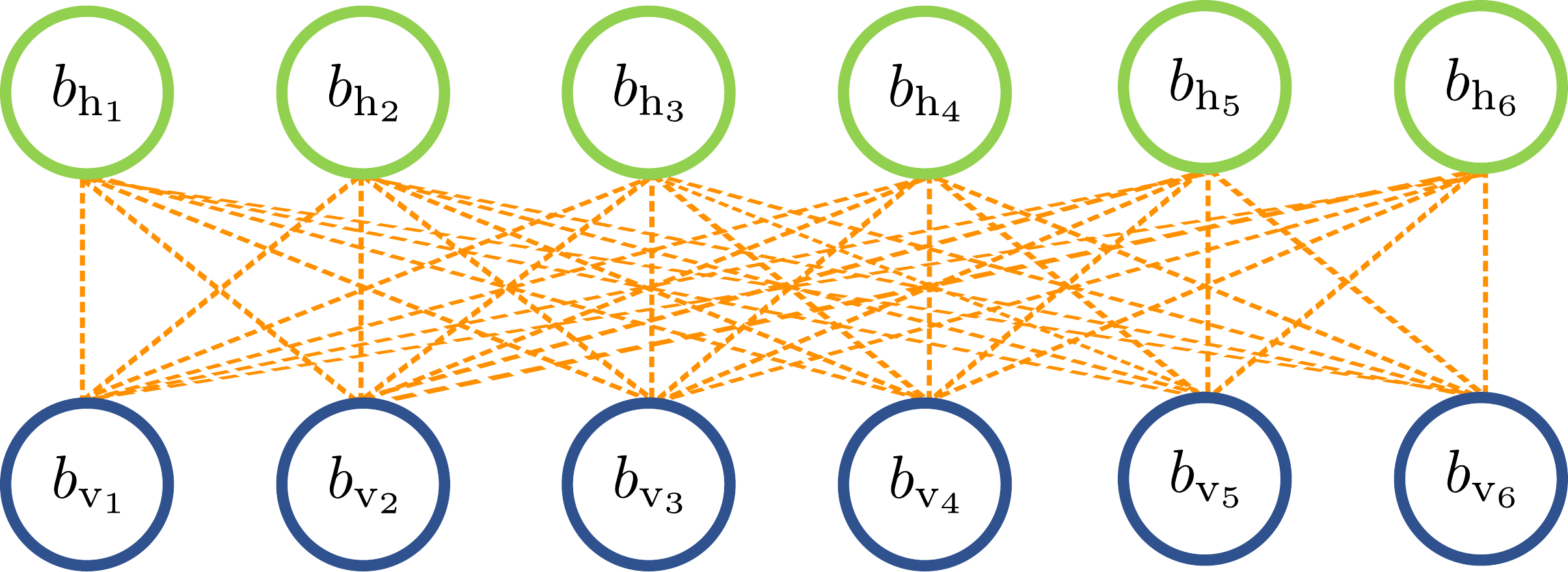}
    \caption[\color{black} An example $6$ visible node restricted Boltzmann machine]{\textbf{An example $6$ visible node restricted Boltzmann machine.} The RBM has hidden nodes and $12$ parameters overall. For this illustration, we assume the weights between nodes, $W_{ij}$, are not trainable. Biases for visible and hidden nodes, $b_{\mathrm{v}_i}, b_{\mathrm{h}_i}$ correspond to the self-loop weights in \figref{fig:feedforward_vs_boltzmann}(b). We shall explicitly use this RBM structure in \chapref{chap:born_machine}.}
    \label{fig:rbm_6_node_structure}
\end{figure}

Now, let's turn to the training of the Boltzmann machine. Recall that here we are trying to match the visible variables of the network (see~\figref{fig:feedforward_vs_boltzmann}) to a data distribution $\pi$, given some $M$ samples from $\pi, \{\ybs_i\}_{i=1}^M, \ybs_i \sim \pi(\ybs)$. As with training a feedforward neural network, we require a cost function, $\Cbs(\paramtheta)$, to indicate how well our model is training. Contrary to the previous example, now we must use cost functions which deal with probability distributions, rather than vectors or labels. Here is where the probability metrics we introduced in~\secref{ssec:prelim/qc/distance_measures/probability} will become useful. For training Boltzmann machines, we use the $\KL$ divergence~\eqref{eqn:kl_divergence_defn}, as this is one of the most common cost functions. Specifically, if we recall the definition of the $\KL$ divergence between the Boltzmann distribution, $p_{\paramtheta}(\vbs)$ and the data, $\pi$ in terms of the cross entropy:
\begin{equation}\label{eqn:prelim/ml/kl_divergence_defn}
    \KL(\pi, p_{\paramtheta}) = \sum_{\xbs} \pi(\xbs) \log \pi(\xbs)-  \sum_{\xbs} \pi(\xbs) \log p_{\paramtheta}(\xbs)
\end{equation}
Now the first term is the entropy of the data distribution, which does not depend on our model parameters, hence we ignore it in the optimisation problem. As such, our cost function then can be defined as:
\begin{equation}\label{eqn:boltzmann_machine_cost_function}
    \Cbs(\paramtheta) := -  \sum_{\xbs} \pi(\xbs) \log p_{\paramtheta}(\xbs) \approx -\frac{1}{M}\sum_{i=1}^M \log p_{\paramtheta}(\ybs_i) 
\end{equation}
which is the empirical negative log-likelihood. Minimising this expression, maximises the `likelihood' that the training data vectors, $\{\ybs_i\}$ are observed from our model distribution. Now, in this case, our trainable parameters are the weights and self-loops in the Boltzmann machine graph (\figref{fig:feedforward_vs_boltzmann}(b)). Due to the relatively simple form of the probability distribution generated by a Boltzmann machine~\eqref{eqn:ising_model_classical_energy}, \eqref{eqn:boltzmann_distribution}, we can compute the derivatives with respect to an edge of the graph as (the edge $ij$ connecting nodes $i$ and $j$):
\begin{equation} \label{eqn:boltzmann_machine_gradients}
    \frac{\partial \Cbs(\paramtheta)}{\partial W_{ij}} = \langle\vbs_i\vbs_j\rangle_{\pi} - \langle\vbs_i\vbs_j\rangle_{p_{\paramtheta}}, \qquad 
        \frac{\partial \Cbs(\paramtheta)}{\partial W_{k}} = \langle\vbs_k\rangle_{\pi} - \langle\vbs_k\rangle_{p_{\paramtheta}}
\end{equation}
which essentially matches the expectation value of each node $\langle\vbs_k\rangle$ and the correlations between pairs of nodes, $\langle\vbs_i\vbs_j\rangle$ between the model and data distributions. One might  expect that with these expressions, the training is straightforward. Unfortunately, it is not so simple. In order to \emph{evaluate} \eqref{eqn:boltzmann_machine_gradients}, we need to be able to efficiently sample from our model distribution, $p_{\paramtheta}$. A fully connected Boltzmann machine is not trivial to sample from, unfortunately. In order to bypass this problem, practitioners commonly restrict the connectivity to define \emph{restricted} Boltzmann machines (RBMs), which are defined with only a bipartite graph between visible and hidden nodes rather than a fully connected one. An example of an RBM is given in \figref{fig:rbm_6_node_structure}.

For an RBM, there exists an algorithm to generate samples from $p_{\paramtheta}$, which is called $k$-step contrastive divergence~\cite{hinton_training_2002, carreira-perpin_contrastive_2005, hinton_practical_2012}. A \emph{Gibbs sampler} is an algorithm used to prepare a sample on the visible units as the end result of a Markov chain Monte Carlo (MCMC)\footnote{A Markov chain MC algorithms are a family of probabilistic algorithms which enable sampling from a desired distribution which is constructed as the equilibrium distribution of a Markov chain. A Gibbs sampler is a special case which requires the conditional distributions of the target to be exactly samplable, since these are used to update the elements of the sample.} algorithm. Due to the bipartite nature of the RBM graph, the visible and hidden nodes are conditionally independent given each other. This means that all the nodes in each layer can be updated simultaneously, which significantly improves the time required to generate a sample. In contrast, a fully connected Boltzmann machine would require each node to be considered separately in the Gibbs sampling process. Running a Markov chain for an infinite number of steps would generate an exact sample from the desired distribution, however $k$-step contrastive divergence halts after $k$ steps of the chain, and produces and approximate sample. Hinton in his original work~\cite{hinton_training_2002} observed that taking simply $k=1$ was sufficient to generate good samples from the RBM. This $k=1$ contrastive divergence was the method used to generate samples from, and train, the RBM in~\figref{fig:rbm_mnist_generated_images}\footnote{This implementation uses code adapted from \url{https://github.com/meownoid/tensorflow-rbm}, which also adds a momentum term to the gradient update (see $\Delta \Cbs$ in ~\eqref{eqn:adam_update_rule}).}. We have discussed the above method to generate samples from the RBM for completeness, however we shall not actually use it directly in this Thesis. Instead, we use an alternative method based on path-integral Monte Carlo (PIMC)\footnote{In contrast to Markov chain Monte Carlo, the path-integral version computes a statistical estimate of a function using the path-integral formulation of quantum mechanics.}, which we revisit in \chapref{chap:born_machine}.

\section[\texorpdfstring{\color{black}}{} Kernel methods]{Kernel methods} \label{ssec:prelim/machine_learning/kernel_methods}

In~\secref{ssec:prelim/qc/distance_measures/probability} we briefly introduced feature maps and kernels, in terms of the mathematical machinery required to use them in probability distance measures. Let us now revisit kernel methods in their appropriate context and use in machine learning more generally. We then discuss the extension of kernel methods to the quantum world in~\secref{ssec:prelim/qml/q_kernel}.

Previously, we presented a feature maps and kernels as being simply functions, and their inner products respectively. In reality, the definition of kernels and their relationship to features maps is more of a subtle point which we address in the following section. We build to these discussions by working through an example in machine learning where the motivation to use kernels becomes apparent. The example we choose to illustrate this is the canonical \emph{support vector machine} (SVM), which was one of the most favourable classification algorithms before the advent of deep learning. Let us first introduce the basics of the SVM before discussing how kernels fit into the picture. The most basic version is the \emph{linear} SVM. Here, we have a labelled dataset $\{(\xbs^i, y^i), \xbs^i\in \mathcal{X}, y^i \in \{+1, -1\}\}_{i=1}^M$, and the linear SVM classifies points by fitting a maximum margin\footnote{The margin is the area between the fitted line, and the closest data points, which are the `support vectors'.} hyperplane (a straight line) in the sample space, $\mathcal{X}$. The line parameters are a weight vector and a bias, $\boldsymbol{w}, b$ with the hyperplane defined as $\boldsymbol{w}^T\xbs - b = 0$. This line must be constructed such that every point labelled $y_i=1$ has $\boldsymbol{w}^T\xbs - b \geq 1$ and those labelled $y_i=-1$ have $\boldsymbol{w}^T\xbs - b \leq -1$ where equality defines the parallel hyperplanes to the original line and describe the margin. Now, in order to actually find the maximum margin hyperplane in the SVM, one solves the following optimisation problem:
\begin{align}
    \boldsymbol{w}^* = &\min ||\boldsymbol{w}||^2 \label{eqn:linear_svm_problem}\\
    \text{subject to } &y^i(\boldsymbol{w}^T\boldsymbol{x}^i - b) \geq 1 \label{eqn:linear_svm_condition}
\end{align}
where the condition in~\eqref{eqn:linear_svm_condition} ensures that the predictions of the SVM (given by $\hat{y}^i = \text{sgn}\left(\boldsymbol{w}^T\boldsymbol{x}^i - b\right)$), match the true labels, $y^i$.

Of course, a linear SVM will only work\footnote{Although one can `soften' the hard margin requirement described above by introducing a parameter which controls the size of the margin.} if the data is linearly separable in the original space, $\mathcal{X}$, which will not be the case in many scenarios. To remedy this, the `\emph{non-linear}' SVM does not fit a hyperplane, but instead maps points $\xbs^i \in \mathcal{X}$ to the `feature' space, $\hilb$. We can then solve a similar optimisation problem to \eqref{eqn:linear_svm_problem} but replacing the original vectors with their feature mapped versions, $\phi(\xbs^i)$. As a result, the decision boundary generated in $\hilb$ will be still linear, but when mapped back into $\mathcal{X}$, will become non-linear, due to the non-linearity of $\phi$.

However, we have not yet mentioned how the kernels (the inner products of the feature vectors) fit into this picture. The answer lies in the \emph{dual} formulation of the optimisation problem~\eqref{eqn:linear_svm_problem}. In the original linear SVM, this dual contains terms involving only \emph{inner products} of the datapoints, $(\xbs^i)^T\xbs^j$. Therefore, when replacing $\xbs^i$ with their corresponding feature maps, $\phi(\xbs^i)$, and deriving the non-linear version of the dual problem, we get inner products of the feature vectors, $\phi(\xbs^i)^T\phi(\xbs^j)$ i.e. kernels\footnote{This is the logic behind `quantum enhance' SVMs, a `classical' kernel can be simply swapped out for a `quantum' kernel (computed on a quantum device). For suitable feature maps, we may hope to gain a quantum advantage.}. This is referred to as the `kernel trick', since by using the dual formulation of the non-linear SVM, we are never actually required to evaluate the actual feature maps themselves, only their overlaps.

The following theorem illustrates how kernels can be defined from feature maps, and the proof that the inner product produces a valid kernel can be found in several works (see e.g.~\cite{schuld_quantum_2019}), so we omit it here.
With this in mind, we realise that there is actually no need to restrict to kernels which result from the inner products of explicitly defined feature maps, which is a sufficient (see~\thmref{thm:kernel_definition}), but not necessary condition for a kernel definition. More generally, a kernel can be any symmetric function\footnote{Given a kernel in this form, a corresponding feature map, $\phi$ may be derivable from it, but it will not be unique in general.}, $\kappa:\mathcal{X}\times\mathcal{X} \rightarrow \mathbb{R}$, which is positive definite, meaning that the matrix it induces, called a \emph{Gram} matrix, is positive semi-definite\footnote{A symmetric matrix, $A$, is positive definite, if $\forall \xbs \in \mathbb{R}^d/\boldsymbol{0}, \xbs^T A \xbs > 0$. Positive \emph{semi-definiteness} also allows for  $\xbs^T A \xbs = 0$. More generally, if $A$ is Hermitian, positive definiteness can be defined by allowing $\xbs \in \mathbb{C}/\boldsymbol{0}$ and $\xbs^T \rightarrow \xbs^*$ (the complex conjugate transpose).}.

\begin{figure}[!ht]
    \centering
    \includegraphics[width=\columnwidth]{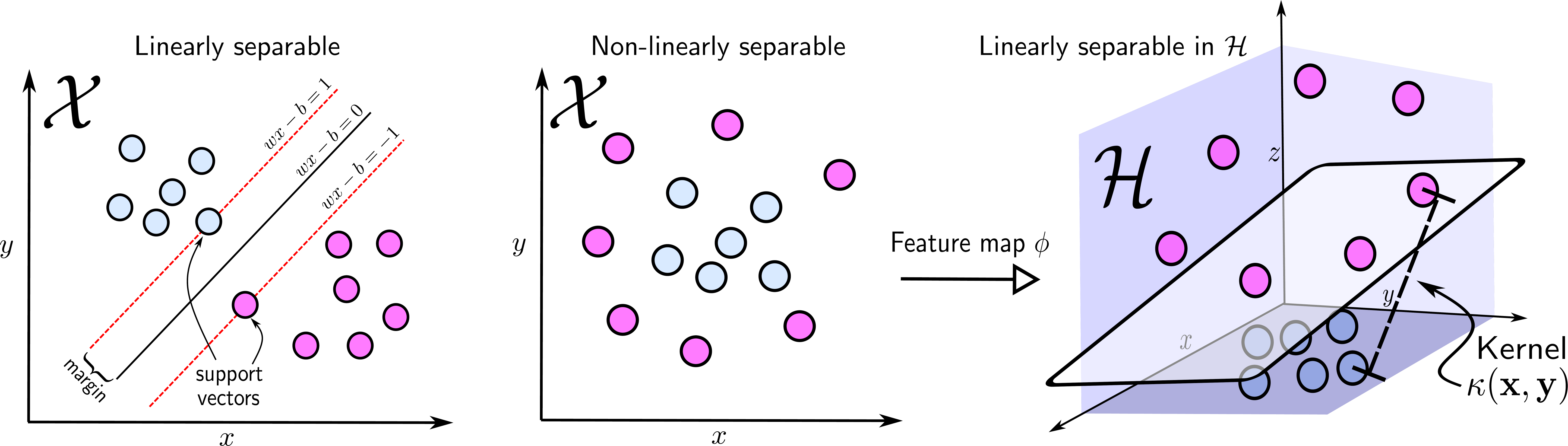}
    \caption[\color{black} The support vector machine and feature maps]{\textbf{The support vector machine and feature maps.} In the case where the data is linearly separable, datapoints are classified according to $\text{sgn}(wx-b)$. In the non-linearly separable case, a feature map is used to extend the dimension of the data vectors (in this particular example), $\xbs = (x, y) \rightarrow \phi(\xbs) = [x, y, x^2+y^2]$. Using the third dimension, the data becomes linearly separable again in the RKHS, $\hilb$. The kernel, $\kappa$, is related to the distances between these points in $\hilb$}
    \label{fig:svm}
\end{figure}
\newpage
For completeness, we also need to make sure that the inner products of feature maps do actually produce valid kernels (in the sense of being symmetric and positive definite. This fact is ensured by the following theorem, whose proof can be found in several works (see e.g.~\cite{schuld_quantum_2019}), so we omit it here:

\begin{thmbox}
\begin{theorem}(Feature map kernel)\label{thm:kernel_definition}~ \\
    Let $\phi: \mathcal{X} \rightarrow \mathcal{H}$ be a feature map. The inner product of two inputs, $\xbs, \ybs \in \mathcal{X}$, mapped to a feature space defines a kernel via:
    \begin{equation} \label{eqn:feature_map_kernel_defn}
        \kappa(\xbs,\ybs)  := \langle\phi(\xbs),\phi(\ybs) \rangle_{\mathcal{H}} 
    \end{equation}
\end{theorem}
\end{thmbox}
Note that the Hilbert feature space, $\hilb$, is usually called a \emph{reproducing kernel} Hilbert space (RKHS). The `\emph{reproducing}' part arises because the inner product of the kernel at a point in $\mathcal{X}$, with a function, $f \in \hilb$, in some sense \emph{evaluates} or reproduces the function at that point:  $\langle f, \kappa(\xbs, \cdot)\rangle  = f(\xbs)$ for $f \in \mathcal{H}, \xbs\in \mathcal{X}$. Now, let use return to the use of kernels in a context more relevant for the contents of this thesis, i.e. as introduced in distribution comparison measures such as the $\MMD$ in~\secref{ssec:prelim/qc/distance_measures/probability}. In this case also, we use the kernel trick to bypass difficult computations, and the feature maps allow us to map the sample points into a higher dimensional RHKS in which the distributions of interest may be more easily compared.

Let us conclude this section by mentioning some common examples of kernels. These include; the polynomial kernel (with degree $d$):
\begin{equation} \label{eqn:poly_kernel}
    \kappa_P(\xbs, \ybs) := (\xbs^T\ybs + c)^d,\qquad c > 0,
\end{equation}
The cosine kernel:
\begin{equation} \label{eqn:cos_kernel}
    \kappa_C(\xbs, \ybs) := \prod_{i=1^n}\cos(\xbs_i - \ybs_i),
\end{equation}
The Laplace kernel:
\begin{equation} \label{eqn:laplace_kernel}
    \kappa_L(\xbs, \ybs) :=  \erm^{-\frac{||\xbs - \ybs||_2}{\sigma}}
\end{equation}
Or the mixture of Gaussians kernel:
\begin{equation} \label{eqn:gaussian_kernel}
    \kappa_G(\xbs, \ybs) := \frac{1}{c}\sum_{i=1}^c \erm^{-\frac{||\xbs - \ybs||_2^2}{2\sigma_i}}
\end{equation}
where $\sigma_i$ are `bandwidth' (hyper)parameters. The choice of the kernel function/feature map (i.e. the choice of the RKHS) may allow different properties to be compared. For a comprehensive review of techniques in kernel embeddings, see~\cite{muandet_kernel_2017}. As mentioned above, we discuss the generalisation of kernels to the quantum scenario in~\secref{ssec:prelim/qml/q_kernel}.

\section[\texorpdfstring{\color{black}}{} Learning theory]{Learning theory} \label{ssec:prelim/machine_learning/learning_theory}

To conclude our preliminary material on (classical) machine learning, let us move from the completely practical in the training of neural networks, to the entirely theoretical realm at the other end of the spectrum and discuss \emph{learning theory}. Learning theory is the theoretical framework which underpins machine learning, and tells us what things (functions, distributions, quantum states, etc.\@) we can learn efficiently, and which things we cannot. It turns out that there is a deep relationship between learning theory and cryptography, since many cryptographic primitives are based on objects which cannot be efficiently learned. A canonical example is the \emph{learning with errors} problem, which is key to the field of \emph{post-quantum} cryptography\footnote{Post-quantum cryptography is the study of primitives which are secure even in the presence of quantum computers (hence in a `post-quantum' world). This was necessitated as a result of the `hackability' of RSA-based cryptosystems with Shor's algorithm. See e.g.~\cite{bernstein_post-quantum_2017} for a review of this field.}. While this problem is now considered to be in the cryptography domain, its origins actually were formed in the context of learning. 

While this topic may seem far removed from the contents of this Thesis, based on the topics introduced previously, we emphasise it is not. In fact, we make a small contribution to \emph{quantum} learning theory in \chapref{chap:born_machine} by discussing the learnability of distributions in a quantum setting, and also make a connection between quantum machine learning and quantum cryptography in \chapref{chap:cloning}.

For now, let us set the scene by introducing the most common and well studied learning scenario, that of \emph{function} learnability\footnote{Computational learning theory is a large field, and we do not attempt to even mention all possible extensions one may consider. For a (quite outdated now) overview of some topics in the field, see~\cite{angluin_computational_1992, kearns_introduction_1994}}. 
The primary question in this subfield are of the following form: given a function family, $\mathcal{F}$, can we efficiently learn a representation of any function, $f \in \mathcal{F}$? While one may consider learning $f$ \emph{exactly}, it is perhaps more common to instead consider learning \emph{approximately} and with \emph{high probability}. This learning model is called \emph{probably approximately correct} (PAC) learning. Furthermore, we usually refer to $\mathcal{F}$ as a \emph{concept class} instead of a function family, and the requirement of a learning algorithm is to output a candidate function, $h$ (a `hypothesis'), to a given `concept', $f \in \mathcal{F}$, which is not too different from $f$ on most possible inputs, $\xbs$, from some space, $\X$. For this section, we use the notation of ~\cite{arunachalam_guest_2017} and the more interested reader may look further into this paper and the references therein. Let us begin by defining what we mean by a learner:
\begin{defbox}
    \begin{definition}[PAC function learning]\label{defn:pac_function_learning}~ \\
        A learning algorithm, $\mathcal{A}$, is an $(\epsilon, \delta)$-PAC learning for a concept class, $\mathcal{F}$ if:\\
        \indent For every $f \in \mathcal{F}$, and every distribution, $D$, given access to a $\mathsf{PEX}(c, D)$ oracle, $\mathcal{A}$ outputs a hypothesis, $h$ such that:
        \begin{equation}
            \Pr\left(h(\xbs) \neq f(\xbs) \right) \leq \epsilon
        \end{equation}
        with probability at least $1-\delta$.
    \end{definition}
\end{defbox}
$\mathsf{PEX}(f, D)$ in the above is a `random example oracle', which when queried, outputs a \emph{labelled} example, $(\xbs, f(\xbs))$ where $\xbs$ is sampled from some distribution, $D:\{0, 1\}^n \rightarrow \{0, 1\}$. This model is suitable for the supervised learning problem we discussed in \secref{ssec:prelims/machine_learning/classification}, since the $\mathsf{PEX}$ oracle can be viewed as exactly providing one datapoint in a labelled dataset, see~\defref{defn:dataset}.

Besides a random example oracle, one may consider alternative query models to the function family, including \emph{membership} queries (where the learner provides $\xbs$ and the oracle returns $f(\xbs)$), or alternative learning models such as exact or agnostic learning~\cite{arunachalam_optimal_2017}, or learning under \emph{specific} distributions (rather than every distribution as in~\defref{defn:pac_function_learning}). 

One of the key quantities which one may care about in learning theory is the \emph{sample complexity} of learning the function. This is simply how many labelled examples does $\mathcal{A}$ require, in order to output a sufficiently good hypothesis. In the classical domain (and indeed in the quantum domain also), this quantity, $M$, is given by the so-called Vapnik Chervonenkis (VC) dimension, which is a combinatorial quantity depending on the concept class, defined as:

\begin{defbox}
    \begin{definition}(VC dimension~\cite{vapnik_uniform_2015}.)\label{defn:vc_dimension}~ \\
        Fix a concept class, $\mathcal{F}$, over $\{0, 1\}^n$. A set $\mathcal{S} = \{s_1, \dots, s_t\} \subseteq \{0, 1\}^n$ is said to be `\emph{shattered}' by a concept class, $\mathcal{F}$ if $\{[f(s_1) \cdots f(s_t)]: f \in \mathcal{F}\} = \{0, 1\}^t$. In other words, for every labelling, $\boldsymbol{\ell} \in \{0, 1\}^t$, there exists a $f\in \mathcal{F}$ such that $[f(s_1) \cdots f(s_t)] = \boldsymbol{\ell}$. The VC dimension of $\mathcal{F}$, $\mathsf{VCdim}(\mathcal{F})$, is the size of a largest $\mathcal{S}\subseteq \{0, 1\}^n$ that is shattered by $\mathcal{C}$.
    \end{definition}
\end{defbox}

The result of \cite{blumer_learnability_1989, hanneke_optimal_2016} shows that $M$ is exactly determined by $\mathsf{VCdim}(\mathcal{F}$):

\begin{thmbox}
    \begin{theorem}(\cite{blumer_learnability_1989,hanneke_optimal_2016,arunachalam_guest_2017})\label{thm:classical_pac_function_learnability}~ \\
        Let $\mathcal{F}$ be a concept class with $\mathsf{VCdim}(\mathcal{F})$. Then:
        \begin{equation}
            M = \Omega\left(\frac{\mathsf{VCdim}(\mathcal{F}) - 1}{\epsilon} + \frac{\log(1/\delta)}{\epsilon}\right)
        \end{equation}
        examples are necessary and sufficient for an $(\epsilon, \delta)$-PAC learner for $\mathcal{F}$.
    \end{theorem}
\end{thmbox}
With that, we conclude our discussion of function learning theory in the classical world - in the next section, \chapref{sec:prelim_quantum_machine_learning}, we will briefly introduce the quantum generalisations of the above definitions, and in \chapref{chap:born_machine}, we present the extension of these definitions into the \emph{distribution} (rather than function) learning setting.

\chapter[Preliminaries III: Quantum machine learning]{Preliminaries III: Quantum machine learning} \label{sec:prelim_quantum_machine_learning}

In the previous two sections, we introduced ideas from quantum computing, and machine learning independently. Now, let us merge some of them in \emph{quantum} machine learning (QML). The field of `QML' proper is only around 12 years old at the time of writing, and was kicked off by the seminal paper of Harrow, Hasidim and Lloyd (HHL)~\cite{harrow_quantum_2009}. This work introduced the HHL algorithm for solving systems of linear equations and performing matrix algebra. This algorithm promised (at the time) exponential speedups over classical algorithms for problems involving high dimensional matrix arithmetic, and a flurry of results followed for a variety of problems in supervised and unsupervised learning including least-squares fitting~\cite{wiebe_quantum_2012}, clustering~\cite{lloyd_quantum_2013}, principal component analysis~\cite{lloyd_quantum_2014}, support vector machines~\cite{rebentrost_quantum_2014}, dimensionality reduction~\cite{kerenidis_classification_2020} or Gaussian process regression~\cite{zhao_quantum-assisted_2019} to name a non-exhaustive list. However, will this paper may have sparked excitement in the research community, researchers have been considering for much longer how to quantize machine learning, and find advantages.

Roughly speaking, quantum machine learning algorithms can be divided into four categories, depending on the type of data on which they operate, and the nature of the algorithm. The notation, $\mathsf{ij}, \mathsf{i}, \mathsf{j} \in\{\mathsf{C}, \mathsf{Q}\}$ refers to the data type ($\mathsf{i}$) and the algorithm type ($\mathsf{j}$) either of which may be quantum, $\mathsf{Q}$, or classical, $\mathsf{C}$ in nature:
\begin{enumerate}
    \item $\mathsf{C}\mathsf{C}\rightarrow$ classical data, processed by classical algorithms (e.g. image classification/generation in \secref{sec:prelim_machine_learning} by neural networks).
    \item $\mathsf{C}\mathsf{Q} \rightarrow $ classical data encoded in quantum states or channels, processed by quantum algorithms (e.g. the HHL algorithm acting on data vectors),
    \item $\mathsf{Q}\mathsf{C} \rightarrow$ quantum data in a classical form, processed by classical algorithms (e.g. using neural networks to analyse measurement statistics from a quantum experiment),
    \item $\mathsf{Q}\mathsf{Q} \rightarrow$ quantum data, processed by quantum algorithms (e.g. feeding quantum states from quantum experiments \emph{directly} into quantum algorithms, for example quantum neural networks).  
\end{enumerate}
In this Thesis, we touch on all of these sub-fields.

While the HHL algorithm drew tremendous attention to the field, foundational ideas in QML have been around almost as long as quantum computing itself. For example, the typography of QML algorithms discussed above was introduced early on in $2006$ by~\cite{aimeur_machine_2006}. Bshouty and Jackson initiated quantum learning theory  (the quantum analogue of that presented in \secref{ssec:prelim/machine_learning/learning_theory}) in 1995 by generalising probably approximately correct (PAC) learning to the quantum setting~\cite{bshouty_learning_1995} (we will return to this topic later in this Thesis). Finally, proposals for `quantum' neural networks appeared as early as 1996~\cite{behrman_quantum_1996}, the same year as the canonical Grover search algorithm~\cite{grover_fast_1996}. Indeed, a well-accepted definition for a quantum version of a neural network is still elusive to date~\cite{schuld_quest_2014}, with the `non-linearity' (in \eqref{eqn:nn_activation_functions} for example) being difficult to emulate quantumly. As mentioned in~\secref{sec:prelim_machine_learning}, the most widely accepted model (at the current time) for a quantum neural network is based on the parametrised quantum circuit (PQC) which is the key component of this Thesis. 

Since the development of HHL, QML development has experienced what could be called two `waves' of progress. The first wave focused around `coherent' algorithm development, primarily using methodologies pioneered by HHL (such as the examples listed at the start of this chapter). The second wave was born with the introduction of the PQC and the \emph{variational} algorithm. The algorithms in the former case are widely believed not to be implementable on quantum hardware for many years to come, as they require large numbers of error corrected qubits (hence the nomenclature `coherent'), and deep circuits to guarantee the speedups they promise. In contrast, variational quantum algorithms\footnote{Variational algorithms are actually used for more general purposes than simply machine learning (meaning data-driven) applications. However, the core of both variational algorithms and heuristic quantum machine learning algorithms reduces to the optimisation of some parametrised object (typically a parametrised quantum circuit). Hence, we use both terms interchangeably in this Thesis to refer to the same concept.} (also dubbed `modern' quantum algorithms~\cite{biamonte_theory_2020}) in the second wave are heuristic in nature, and as such they do not usually have provable guarantees of runtime complexity. However, they are readily implementable on the quantum computers available now, many of which are accessible remotely through the `quantum cloud'~\cite{larose_overview_2019}. Early work in QML algorithms almost exclusively focused around quantum speedups in runtime. However, speed is not the only way to measure a `quantum advantage' over classical methods for machine learning tasks. Indeed, one may consider other avenues along which to search for a quantum advantage (but not limited to: expressibility, accuracy, interpretability, privacy and generalisability). The simultaneous rapid development of quantum computing hardware since $\sim 2010$ means that building and experimenting with real quantum models is accessible to anyone with an internet connection, and as such current research has partially shifted away from that of purely theoretical algorithm proofs to more experimental and exploratory in nature. For completeness, we mention some examples of quantum machine learning which do not quite fit into these categories, including speedups (however not exponential) in training Boltzmann machines~\cite{wiebe_quantum_2015}, or building quantum models based on quantum annealing hardware, including the \emph{quantum} Boltzmann machine~\cite{kieferova_tomography_2017, amin_quantum_2018, benedetti_quantum-assisted_2018, wilson_quantum-assisted_2019, wiebe_generative_2019}. One may also consider advances in quantum reinforcement learning~\cite{dong_quantum_2008, dunjko_advances_2017, dunjko_exponential_2018, lockwood_reinforcement_2020, jerbi_quantum_2021, jerbi_variational_2021, skolik_quantum_2021}

For coherent algorithms in the first wave, a crucial ingredient is required. This is the ability to access data `in superposition'. Specifically, these algorithms usually require one to be able to prepare states of the form:
\begin{equation} \label{eqn:amplitude_encoding}
    \ket{\xbs} = \sum\limits_{i=1}^{n} x_i \ket{i}
\end{equation}
Where $\xbs = [x_1, \dots, x_n]^T \in \mathbb{R}^n$, is a data vector\footnote{Assuming the vectors are normalised such that $||\xbs||_2 = 1$.}. This is the so-called \emph{amplitude} (\cite{schuld_supervised_2018, zhao_smooth_2021}) encoding and the procedure to prepare such a state is called a `data-loading' routine. We return to data encoding in \secref{sssec:vqa_input_and_ouput} and \chapref{chap:classifier}. 
Once amplitude encoded states are prepared, they are further processed by the quantum algorithm. However, when considering the true runtime of such algorithms, one must also factor in the cost of preparing such states. As stated by~\cite{aaronson_read_2015}, if a routine to prepare~\eqref{eqn:amplitude_encoding} required $n^c$ steps (for some constant $c$), then speedups are lost.

In principle, such data-loading issues can be sidestepped by the existence of a quantum random access memory\footnote{Similarly to a classical RAM, which returns the contents of a memory register when accessed.} (QRAM)~\cite{giovannetti_quantum_2008}, acting as follows on a superposition of `address' registers, $\ket{i}$:
\begin{equation}
    \frac{1}{\sqrt{N}}\sum\limits_{i=1}^{n} \ket{i} \xrightarrow[QRAM]{}  \frac{1}{\sqrt{N}}\sum\limits_{i=1}^{n} \ket{i}\ket{x_i}
\end{equation}
However, constructing such a quantum memory which can be accessed in superposition may be as hard as building a fully fault tolerant quantum computer. As we hinted at above, there are also many other practical issues with these `coherent' QML algorithms (such as HHL) that need to be carefully addressed in order determine whether speedups are actually feasible~\cite{aaronson_read_2015}. For an overview of technical details used in such linear system algorithms, see ~\cite{dervovic_quantum_2018} which also discusses constructions for QRAM objects. 

Finally, we would be remiss if this Thesis included a section introducing quantum machine learning without mentioning the `dequantisations'. In 2018,~\cite{tang_quantum-inspired_2018} proposed a \emph{fully classical} algorithm to kill the exponential speedup of one of these coherent algorithms - the `quantum recommendation system' algorithm of~\cite{kerenidis_quantum_2016}. Previously, this quantum algorithm was one of the most promising candidates for an end-to-end exponential speedup over its classical counterparts. The classical algorithm of Tang however demonstrated this was not the case, and the advantage of~\cite{kerenidis_quantum_2016} was reduced to only polynomial\footnote{However, we remark that this is not so much of a blow as it seems, the constant factors in the original algorithm of~\cite{tang_quantum-inspired_2018} indicated that the quantum algorithm could still be significantly faster. The most recent analysis by~\cite{shao_faster_2021} indicated that this speedup is only quadratic in reality for a special case of the problem.} We highlight this to demonstrate how subtle the nature of quantum advantage is in QML. For those interested in further reading of `dequantised' quantum algorithms, see~\cite{tang_quantum-inspired_2018-1, gilyen_quantum-inspired_2018, chia_quantum-inspired_2018, chia_sampling-based_2020, shao_faster_2021} and also~\cite{arrazola_quantum-inspired_2020} for a study of these algorithms in practice.

Before moving to the next section, we conclude by highlighting some reviews of quantum machine learning including~\cite{wittek_quantum_2014, schuld_introduction_2015, biamonte_quantum_2017, dunjko_machine_2018, benedetti_parameterized_2019} which discuss many of the topics we neglected here.

\section[\texorpdfstring{\color{black}}{} Variational quantum algorithms]{Variational quantum algorithms} \label{sec:variational_quantum_algorithms}
\begin{wrapfigure}{l}{0.5\textwidth}
%
\centering
\includegraphics[width=0.45\textwidth, height=0.4\textwidth]{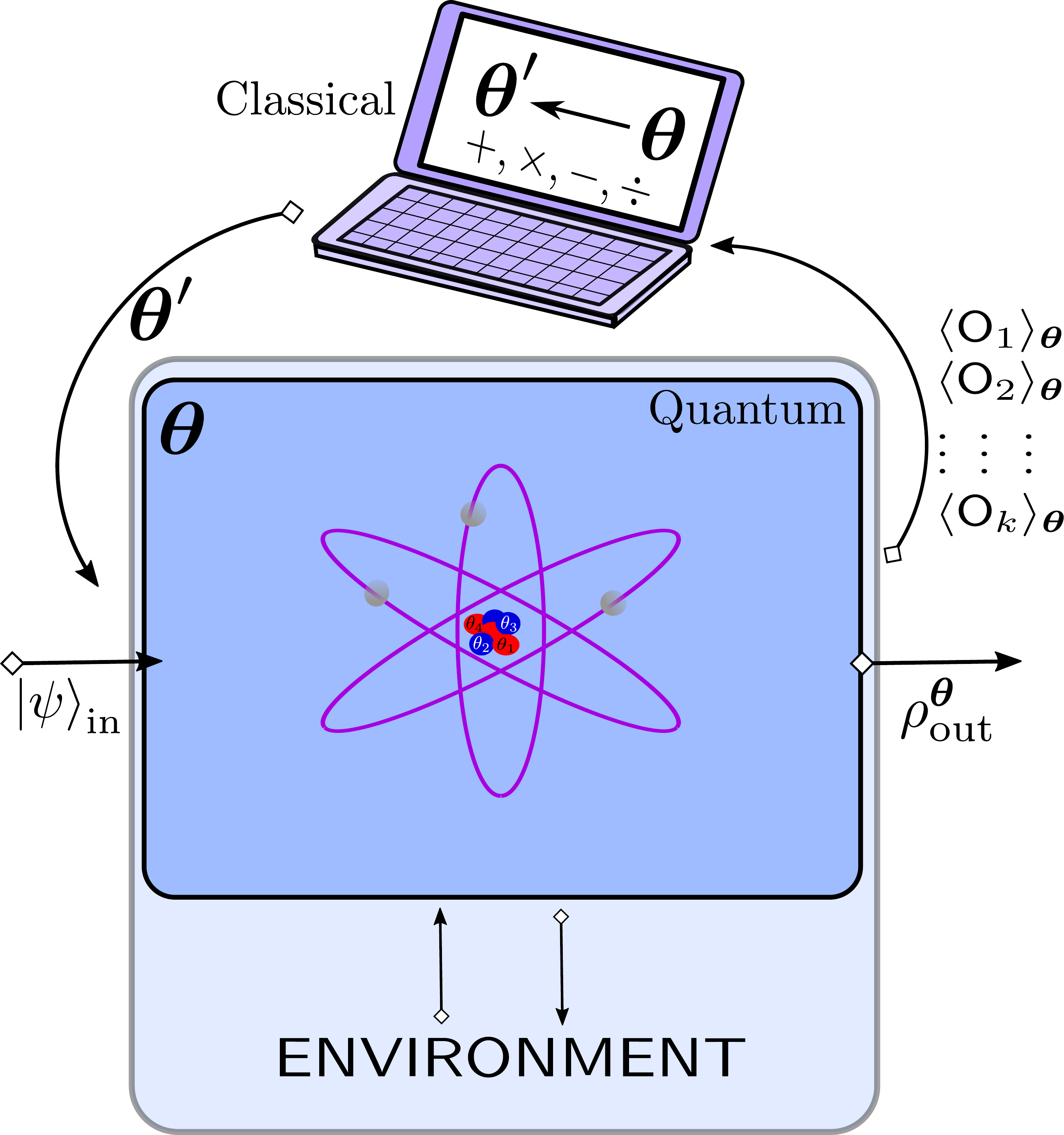}
\caption[\color{black}Cartoon illustration of a variational quantum algorithm.]{\textbf{Cartoon illustration of a variational quantum algorithm.}}
\label{fig:vqa_overview}
\end{wrapfigure}

This Thesis however is focused solely on algorithms in the second wave, the variational quantum algorithms ($\VQA$s). As mentioned above, $\VQA$s (also called `near-term' quantum machine learning algorithms) are a powerful model for quantum algorithm design on NISQ devices. 
\figref{fig:vqa_overview} gives an overview of some of the key ingredients involved. $\VQA$s are characterised by a quantum component, and a classical component, which work synergistically to solve the problem of interest. In order to maximise the use of quantum coherence time (which is limited by the lack of error correction in NISQ devices), the majority of the computation is carried out by the classical device, which performs the task of optimising the parameters of a quantum object (typically a parametrised quantum state) to solve the problem.
%
%
The quantum component may have some input, $\ket{\psi}_{\text{in}}$, and produces some output state, $\rho_{\text{out}}^{\paramtheta}$, which is parametrised by $\paramtheta$. In order to apply a general quantum transformation, the quantum component may interact with an environment (described by ancillary registers), and from the output state several observables may be measured, $\mathsf{O}_i$. These classical results are then passed into a classical computer which computes some cost function of these observables and finds parameter updates $\paramtheta \rightarrow \paramtheta'$ (via gradient descent or otherwise) to drive the computation. These approaches are clearly heuristic in nature, however some effort has been made to study them from a theoretical point of view~\cite{mcclean_theory_2016, biamonte_theory_2020} and the computational model they exhibit is actually also universal for quantum computation~\cite{biamonte_universal_2019} despite only using short depth quantum circuits.

At this point, if we replace the quantum device by a neural network\footnote{In fact, a more general scenario is where some function is applied to the output measurements from the quantum circuit before passing it to a classical optimiser. This function may be simple, or may be the output of another classical neural network. We do not consider this possibility in this Thesis, so we neglect it for simplicity.}, the connection to machine learning becomes clear. To introduce the relevant concepts in this section, we use and refer to the following reviews, \cite{cerezo_variational_2021, bharti_noisy_2021, endo_hybrid_2021}. Key concepts involved in variational quantum algorithms, are:

\begin{itemize}
    \item The cost function defining the problem.
    \item The $\Ansatz$ used by the quantum computer.
    \item The optimisation of the cost function.
    \item The input to the $\VQA$ (some form of \emph{data encoding} for classical data).
    \item The desired output - measurement readout, or quantum states.
\end{itemize}
In the subsequent sections, we introduce each of these to the degree of detail for which they are relevant in this Thesis.
\subsection[\texorpdfstring{\color{black}}{} Cost functions]{The cost function} \label{ssec:vqa_cost_functions}
Just as in classical machine learning, the problem to be solved can be reduced to optimising a cost function, which is some function of the parametrised state:
\begin{equation}\label{eqn:vqa_cost_function_general}
    \min_{\paramtheta} \Cbs(\paramtheta)
\end{equation}
The choice of cost function is crucial for the success of $\VQA$s, and not surprisingly, this fact originates from their machine learning origins. Concretely, four desirable qualities for $\VQA$ cost functions are (\cite{cerezo_variational_2021}):

\begin{tcbraster}[raster columns=4,raster equal height,nobeforeafter,raster column skip=0.2cm]
\begin{mybox}{red}{}
  \textbf{Faithfulness.}
  \end{mybox}
  \begin{mybox}{cyan}{}
  \textbf{\textbf{Efficiently\\ computable.} }
  \end{mybox}
  \begin{mybox}{green}{}
  \textbf{\textbf{Operational\\ meaning.}}
  \end{mybox}
    \begin{mybox}{violet}{}
  \textbf{ \textbf{Trainable.} }
  \end{mybox}
\end{tcbraster}

\begin{enumerate}
    \item \textbf{Faithfulness:}  The minimum point $\paramtheta^* := \argmin \Cbs(\paramtheta)$, is the solution to the problem of interest in the parameter space\footnote{The parameter space is the multidimensional space described by the parameters, $\paramtheta$.}.
    \item \textbf{Efficiently computable:} It should be possible to estimate $\Cbs$ to a reasonable precision with polynomial resources (number of qubits, gates in the circuit, etc.).
    \item \textbf{Operational meaning:} Smaller cost values correspond to higher quality solutions.
    \item \textbf{Trainable:} The cost should be optimisable in an efficient way (efficiently computable gradients, navigatable parameter space, etc.).
\end{enumerate}

For many variational algorithms, the problem can be encoded as a \emph{ground state} optimisation problem, meaning that the problem parameters are encoded into a Hamiltonian, $H$, and the solution is encoded in the ground state of this Hamiltonian (the state with the minimum energy). In this case the optimisation becomes:
\begin{equation}\label{eqn:vqa_cost_function_vqe}
    \min_{\paramtheta} \bra{\psi(\paramtheta)}H\ket{\psi(\paramtheta)} := E_{\paramtheta^*}
\end{equation}
The goal is to find the eigenstate, $\ket{\psi(\paramtheta^*)}$, with the lowest energy, $E_{\paramtheta^*}$. $\VQA$s in this form are known as variational quantum \emph{eigensolvers} ($\VQE$), which was one of the original $\VQA$ proposals by~\cite{peruzzo_variational_2014} that pioneered the field. $\VQE$ caused a great deal of excitement due to its applicability to quantum chemistry~\cite{mcardle_quantum_2020} problems and quantum simulation~\cite{cade_strategies_2020}. In these cases, the Hamiltonian, $H$, is the one defined by the physical system to be simulated, and is the source of the observables, $\mathsf{O}_i$, to be measured. For example, in finding the ground state of the Ising model Hamiltonian acting on $N$ spins:
\begin{equation} \label{eqn:ising_model_hamiltonian}
    H_{\text{ising}} :=  \sum\limits_{i < j}^N J_{ij} \ZG_i \otimes \ZG_j +  \sum\limits_{k=1}^N b_{k} \ZG_k,
\end{equation}
the observables to be measured are the correlations $\langle\ZG_i \otimes \ZG_j\rangle_{\psi}$ and the local expectations $\langle\ZG_k\rangle_{\psi}$\footnote{$\langle\mathsf{O}\rangle_{\psi}$ denotes the expectation value of the operator, $\mathsf{O}$, with respect to the state, $\ket{\psi}$: $\bra{\psi}\mathsf{O}\ket{\psi}$.}. Each of these observables is measured\footnote{Since all terms in this Hamiltonian commute, these statistics can all be collected using a single measurement setting, by measuring every spin (qubit) in the computational basis.}, and the results are collected (and weighted by the coefficients $J_{ij}, b_i$) to compute the energy of the system when in the state, $\ket{\psi}$.

Cost functions are highly problem dependent, those in the form of  \eqref{eqn:vqa_cost_function_vqe} are among the simplest, but in the case of machine learning problems particularly, the choice of cost function is complex and crucial to the success of the algorithm. As such, one can generalise cost functions into the following form~\cite{cerezo_variational_2021}:
\begin{equation} \label{eqn:general_vqa_cost_function}
    \Cbs(\paramtheta) = \sum_k f_k(\Tr[\mathsf{O}_k \rho^{\paramtheta}_{\text{out}} ])
\end{equation}
where $\{f_k\}_k$ are a set of functions applied to the output. In this Thesis, cost function choice will play an important role. \\

\noindent \textbf{Local versus Global Cost Functions}

\vspace{2mm}
Before moving on, let us highlight one important generic feature of $\VQA$ cost functions; their \emph{locality}. Namely, whether the observables involved in computing $\Cbs$ are either \emph{local} or \emph{global}. Roughly speaking, local observables are those consisting of a function of terms which each can be computed by only measuring a handful of qubits, whereas global cost functions require information about the \emph{entire} quantum state. For example, the cost function defined by $\VQE$ on the Ising model Hamiltonian, \eqref{eqn:ising_model_hamiltonian} is $2$-local since the terms in it require measuring no more than $2$ qubits at a time (the terms $\ZG_i \otimes \ZG_j$ only operate on two qubits, $i, j$, and $Z_k$ only acts on qubit $k$). In the case of $\VQE$ problems, the locality of the cost function directly relates to the locality of the Hamiltonian in question which is a well-studied topic~\cite{kempe_complexity_2006}, but it is a more general concept.

Specifically a $k$-local cost function observable can be written as (\cite{cerezo_cost_2021}):
\begin{equation} \label{eqn:local_vqa_cost_observable_generic}
    \mathsf{O} = c_0\mathds{1} + \sum_{i=1}^N c_i \widehat{\mathsf{O}}^k_i
\end{equation}
where each $\widehat{\mathsf{O}}_i^k$ are $k$-local operators, acting at most on $k \subset [n]$ qubits. In contrast, a global cost function has the form:
\begin{equation} \label{eqn:global_vqa_cost_observable_generic}
    \mathsf{O} = c_0\mathds{1} + \sum_{i=1}^N c_i \widehat{\mathsf{O}}_{1i} \otimes \widehat{\mathsf{O}}_{i2} \otimes \dots \otimes \widehat{\mathsf{O}}_{in}
\end{equation}
In both cases, $N$ is the number of terms given by the problem of interest and $c_0, c_i$ are coefficients which can assumed to be real WLOG (e.g. corresponding to $\{J_{ij}, b_k\}$ in \eqref{eqn:ising_model_hamiltonian}).

Cost function locality is also important regarding operational meaning. Typically, from this point of view, global cost functions are usually more favourable since they can be directly interpreted. In variational compilation~\cite{khatri_quantum-assisted_2019}, the cost function compares the similarity of two global unitaries, and is zero when the unitaries are identical. In contrast, local cost functions are perhaps less interpretable, but can usually be used as a bound to optimise a global cost. Furthermore, as we shall discuss in~\secref{ssec:vqa_cost_function_optimisation}, local unitaries are usually easier to optimise than their global counterparts (\cite{cerezo_cost_2021}). We return to this discussion in~\chapref{chap:cloning}.

\subsection[\texorpdfstring{\color{black}}{} Ans\"{a}tze]{Ans\"{a}tze} \label{ssec:vqa_ansatzse}

\vspace{2mm}
The second important ingredient in $\VQA$s is the $\Ansatze$, for the parametrised state, i.e. the operation that performs the mapping $\ket{\psi}_{\text{in}} \rightarrow \rho_{\text{out}}^{\paramtheta}$ in \figref{fig:vqa_overview}. In $\VQA$s, the $\Ansatz$\footnote{Historically, an $\Ansatz$ in physics is an `educated guess' for a problem solution. This has translated into quantum computing in preparing the `trial' state for a $\VQA$.} is usually a unitary transformation:
\begin{equation} \label{eqn:vqa_parametrised_state}
    \ket{\psi(\paramtheta)} = U(\paramtheta)\ket{\psi}_{\text{in}},\qquad~ \rho_{\text{out}}^{\paramtheta} := \ketbra{\psi(\paramtheta)}{\psi(\paramtheta)}
\end{equation}
This unitary, $U(\paramtheta)$, is usually referred to as a parametrised quantum circuit (PQC), although it need not be strictly a quantum circuit. For example, as we shall see momentarily, it may be implemented using a direct Hamiltonian evolution.
\begin{figure}[ht]
    \centering
        \includegraphics[width=\textwidth]{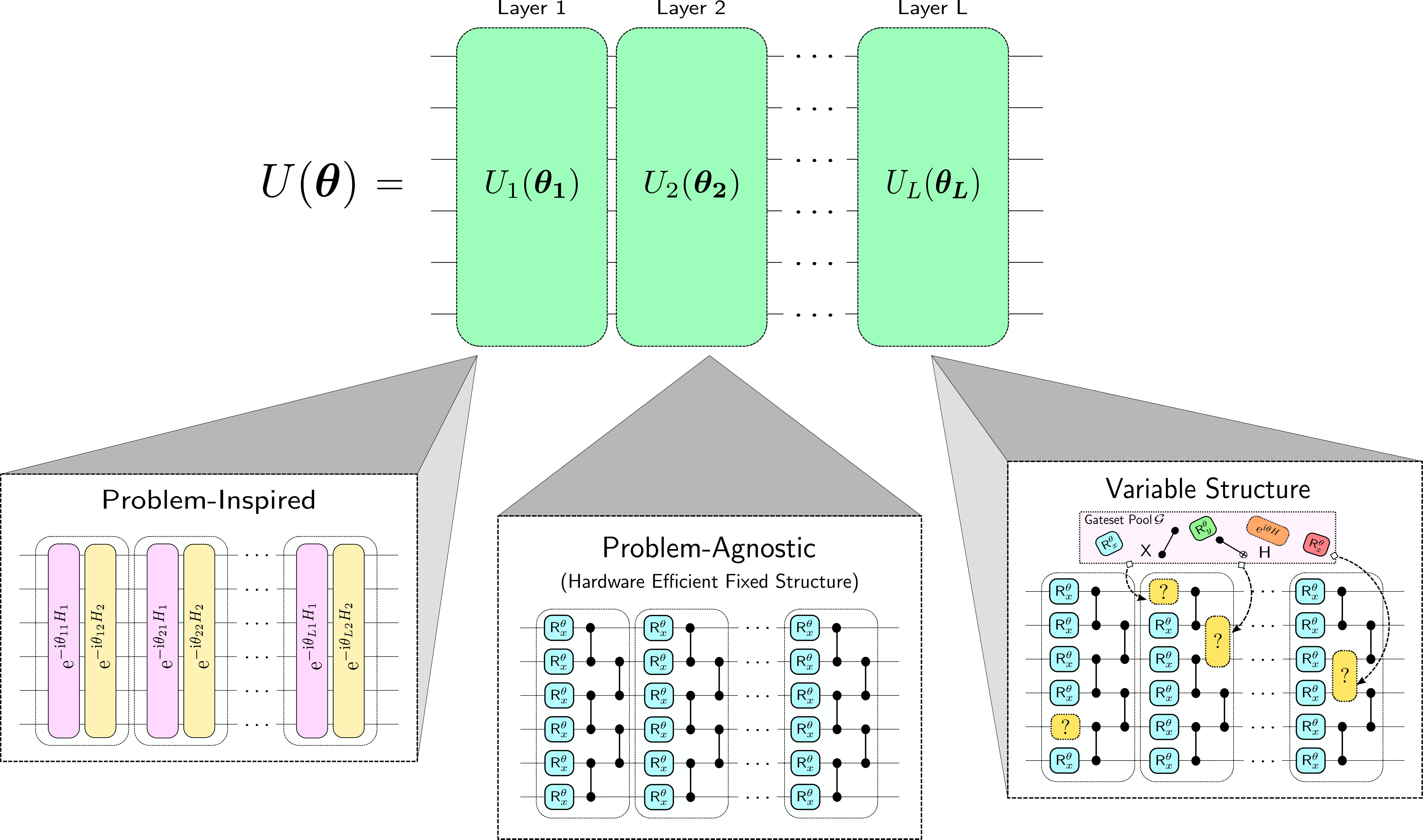}
            \caption[\color{black}Examples of $\VQA$ $\Ansatze$.]{\textbf{Examples of $\VQA$ $\Ansatze$}. The general unitary transformation $U(\paramtheta)$ is usually divided into $L$ layers $U(\paramtheta) = U_L(\paramtheta_L)\cdots U_2(\paramtheta_2) U_1(\paramtheta_1)$. Problem-inspired $\Ansatze$ typically encode problem information - an example shown with two non-commuting Hamiltonian terms, $H_1, H_2$, which may apply to the $\QAOA$ or as a Hamiltonian variational $\Ansatz$ for $\VQE$. Problem-agnostic $\Ansatze$ use native gatesets and connectivities (for example, shown is an nearest neighbour connected $\Ansatz$ using $\CZ$ as the entangling gate, with $\RX^\theta$ being the native single qubit parametrised gate). Finally, variable structure $\Ansatze$ may be problem-inspired, as in the ADAPT-$\VQE$ algorithm~\cite{grimsley_adaptive_2019}, or problem-agnostic with hardware native operations~\cite{cincio_machine_2021}. Note that the layer structure is not strictly necessary in the definition of an $\Ansatz$, and each layer need not have the same specific structure, but it provides a nice conceptual connection to neural network layers in machine learning.}
        \label{fig:vqa_ansatz_types}
\end{figure}
Two important families of $\Ansatze$ are \emph{problem-inspired} and \emph{problem-agnostic}~\cite{cerezo_variational_2021} (see~\figref{fig:vqa_ansatz_types}), where the former includes information from the problem specification in building the $\Ansatz$, whereas the latter is generic and applies to all problems. A canonical example of a problem inspired $\Ansatz$ is that used in the quantum approximate optimisation algorithm ($\QAOA$) \cite{farhi_quantum_2014}. The $\QAOA$ is a special case of the $\VQE$ algorithm, where the Hamiltonian, $H$, is restricted to be diagonal in the computational basis (an example is the Ising Hamiltonian, $H_{\text{ising}}$ in \eqref{eqn:ising_model_hamiltonian}. The $\QAOA$ is applied primarily to (classical) combinatorial optimisation algorithms, such as MAXCUT on a graph. In its original form, the $\QAOA$ applies $p$ layers of alternating `problem' and `driver' Hamiltonians, starting from a state which is \emph{not} diagonal in the computational basis, for example the state $\ket{\psi}_{\text{in}}^{\paramtheta} = \ket{+}^{\otimes n}$ on $n$ qubits. If one translates an optimisation problem of interest into finding the ground state of the Ising model Hamiltonian above, the corresponding $\Ansatze$ would then be:
\begin{equation} \label{eqn:qaoa_ansatz_basic}
    U(\paramtheta) = U(\boldsymbol{\gamma}, \boldsymbol{\beta})= \prod_{i=1}^p \erm^{-\irm\beta_i H_{\text{driver}} } \erm^{-\irm \gamma_i H_{\text{ising}}},
\end{equation}
where $H_{\text{driver}} =\bigotimes_{j=1}^n \XG_j$ is the driver Hamiltonian for which the initial state, $\ket{+}^{\otimes n}$ ,is an eigenstate. Each `layer' in the $\Ansatz$ has $2$ free parameters, so the optimisation requires finding $2p$ optimal parameters.

This $\Ansatz$ type has been generalised to the alternating \emph{operator} $\Ansatz$\footnote{Which also offers the acronym $\QAOA$, confusingly.}~\cite{hadfield_quantum_2019} which can encode different symmetries of the problem and may improve performance. A second type of $\Ansatz$ is a further generalisation of the $\QAOA$ $\Ansatz$, called the Hamiltonian variational $\Ansatz$, and is used when the problem Hamiltonian is not diagonal, i.e. in $\VQE$ problems. In this case, the Hamiltonian can be split into multiple non-commuting terms, $H_k$~\cite{cerezo_variational_2021},
\begin{equation} \label{eqn:hamiltonain_variational_ansatz}
    U(\paramtheta) = \prod_{l} \left(\prod_k \erm^{-\irm \theta_{l, k} H_{k}}\right),
\end{equation}
Again, the $\Ansatz$ is applied for a number of layers. Both of these $\Ansatze$ derive inspiration from the quantum adiabatic algorithm~\cite{farhi_quantum_2000}, and Trotterization methods for quantum simulation, which are used for simulating the evolution of a system. For the latter case in $\VQE$, the initial state is usually also problem inspired, for example the Hartree-Fock ground state, which is exactly solvable for a given system.

The second class of $\Ansatze$ are the problem-agnostic, which usually aim to reduce the circuit depth and number of gates required to implement them (in contrast to the alternating operator/Hamiltonian variational $\Ansatze$, which may require significant depth to implement). Furthermore, the operations in agnostic $\Ansatze$ are usually hardware-native. For example, to implement a hardware-native $\Ansatz$ on Rigetti~\cite{smith_practical_2017} one may use the $\CZ$ or $\XY$ gates for entanglement creation, whereas for IBM quantum computers, one would choose a directed $\CNOT$. Using native gates dispenses with the need to perform extensive \emph{compilation} and so keeps the circuits shallow. Furthermore, only allowing the connectivity between qubits which is directly accessible on the quantum device also is essential for high performing results on NISQ hardware\footnote{Compilation may significantly increase the circuit depth in a NISQ circuit implementation. If an algorithm requires a $\CNOT$, but only $\CZ$ is available natively, two $\HG$ gates will need to be added to change the basis. Secondly, performing qubit routing to apply a transformation between two qubits which are not directly connected is expensive - a single $\SWAP$ gate requires $3$ $\CNOT$ gates to implement.}. 

The most common problem-agnostic $\Ansatze$ are called \emph{hardware-efficient}. Two specific forms of hardware efficient $\Ansatze$ are used in this Thesis. The first are \emph{fixed-structure}, which contain a number of layers, where each layer has a fixed structure of entangling and single qubit rotations\footnote{For ease of implementation, when dealing with hardware efficient $\Ansatze$ in this Thesis, we assume that two qubit unitaries are un-parametrised (meaning implemented with a fixed parameter), and the trainable parameters, $\paramtheta$, are only contained in single qubit rotations. This is not a necessary assumption in general.}

The second class of hardware-efficient $\Ansatze$ are \emph{variable-structure} $\Ansatze$. In these circuits, not only are the continuous gate rotations optimised over, but also the gates in the circuit \emph{themselves}. A fixed-structure $\Ansatz$ may require several layers to solve a given problem, in contrast variable-structure $\Ansatze$ are capable of finding circuits with the \emph{absolute minimum} number of quantum gates. Formally, this corresponds to a more general optimisation problem than \eqref{eqn:vqa_cost_function_general}:
\begin{align} \label{eqn:variable_structure_ansatz_optimisation_problem}
    (\paramtheta^{*}, \boldsymbol{g}^{*}) = \argmin_{\paramtheta, \boldsymbol{g} \in \mathcal{G}} \Cbs(\paramtheta, \boldsymbol{g})
\end{align}
Here, $\mathcal{G}$ is a gateset \emph{pool}, $\boldsymbol{g}$ corresponds to a certain sequence of gates in the pool. Clearly, the problem now becomes a discrete optimisation problem as well as a continuous one (we still need to optimise over the gate parameters, $\paramtheta$).

The variable-structure approach was given proposed in~\cite{cincio_learning_2018}, and variations on this idea have been given in many forms~\cite{ostaszewski_structure_2021, rattew_domain-agnostic_2020, li_quantum_2020, pirhooshyaran_quantum_2021}. Most recently, these ideas have been given the broad classification of \emph{quantum architecture search} (QAS)~\cite{zhang_differentiable_2020} to draw parallels with neural architecture search~\cite{yao_taking_2019, liu_darts_2019}, (NAS)\footnote{NAS is the analogous task for finding optimal structures in neural network architectures, for example tuning the width and depth of the network.} in classical ML. We also mention that the variable-structure approach is not only restricted to be used with a hardware-efficient $\Ansatz$. Indeed, a variation has been used on the Hamiltonian variational $\Ansatz$ for $\VQE$ (\cite{grimsley_adaptive_2019, chivilikhin_mog-vqe_2020}) and for $\QAOA$~\cite{zhu_adaptive_2020}. One interpretation of QAS is its use in discovering novel primitives (we apply it to the primitive of quantum cloning in~\chapref{chap:cloning}) for quantum algorithms, protocols or experiments; QML for primitive discovery if you will. Interestingly, a parallel and related line of work has also been progressing using purely \emph{classical} machine learning in discovering novel quantum protocols and experiments~\cite{krenn_automated_2016, melnikov_active_2018, wallnofer_machine_2019}.

Finally, it is also crucial to mention that for \emph{machine learning} applications, good $\Ansatze$ design is still an open area of research. A primary reason for this is the data-driven nature of ML which makes definition of a `problem-inspired' $\Ansatz$ difficult. The choice is perhaps more natural when examining quantum machine learning for \emph{quantum data}, where one can use a physically inspired $\Ansatz$ which may encode certain problem symmetries. For example, a proposal for a quantum convolutional neural network~\cite{cong_quantum_2019} (QCNN) uses a translationally invariant structure which may be suitable for physical systems with this property, or defining Hamiltonian-based models~\cite{verdon_quantum_2019} as $\Ansatze$ which can learn to represent mixed quantum states.

\subsection[\texorpdfstring{\color{black}}{} Cost function optimisation]{Cost function optimisation} \label{ssec:vqa_cost_function_optimisation}
Once a suitable cost function has been chosen, the question then becomes how one can optimise it to solve the problem. In the above $\VQE$ example, this corresponds to finding the optimal parameter setting, $\paramtheta^*$, as a proxy for optimising over the set of possible quantum states to find the ground state of the Hamiltonian in question.

As with neural network training, optimisation routines for cost functions can be broadly grouped into two categories: gradient-free and gradient-based. Criteria number four of defining useful cost functions indicates they must be efficiently trainable (similarly again to the neural network training in~\secref{ssec:prelim/machine_learning/neural_networks}. This means (in the case of gradient-based optimisation) that they must have efficiently\footnote{Efficient here usually refers to with respect to the number of qubits, $N$, but also all other relevant parameters of the problem.} computable gradients.\\

\subsubsection[\texorpdfstring{\color{black}}{} Gradients]{Gradients} \label{sssec:vqa_gradients}

Fortunately, for many $\VQA$s, this is the case. Specifically, if the gates in the $\VQA$ $\Ansatz$, $U(\paramtheta)$, are of a particular form, the gradients of the cost function can be computed \emph{analytically}, using what is known as the \emph{parameter-shift} rule~\cite{mitarai_quantum_2018, schuld_evaluating_2019, bergholm_pennylane_2020, sweke_stochastic_2020, harrow_low-depth_2021} which we alluded to previously:
\begin{tcolorbox}
\textbf{The parameter-shift rule:}~ \\
    Given a cost function, $\Cbs$, which is evaluated using an $\Ansatz$ (with $L$ parameters) containing quantum operations of the form $U(\theta_i) = \erm^{-\irm (\theta_i/2) \Sigma}$, where $\Sigma^2= \mathds{1}$, the gradient of $\Cbs$ with respect to a particular parameter $\theta_i$ is given by:
    \begin{equation} \label{eqn:parameter_shift_rule}
        \frac{\partial \Cbs(\paramtheta)}{\partial \theta_i} := \partial_{\theta_i} \Cbs(\paramtheta) =  \frac{1}{2}\left[\Cbs(\paramtheta_i^+) - \Cbs(\paramtheta_i^-)\right],
    \end{equation}
    where $\paramtheta^{\pm}_i = (\theta_1, \theta_2, \dots, \theta_i\pm \frac{\pi}{2}, \dots, \theta_L)$
\end{tcolorbox}

Note also that the expression in \eqref{eqn:parameter_shift_rule} is \emph{exact}, and the shift constant $\pi/2$ is large, in contrast to the infinitesimally small shifts and approximate nature of methods such as finite-difference\footnote{See \cite{mari_estimating_2021} for a comparison between parameter-shift and finite-difference methods.}. However, since $\Cbs$ is estimated by running a circuit on a quantum device, we cannot extract the exact value $\Cbs$ without an infinite number of measurement shots. As such, we will only have access to estimates, $\widehat{\Cbs}$ and $\widehat{\partial \Cbs}$ of the cost and its gradients.

Interpreting \eqref{eqn:parameter_shift_rule}, we see that we can evaluate the gradient of a cost with respect to any particular parameter by simply computing the cost twice in a `parameter-shifted' fashion. As such, if we assume a complexity of $T$ for evaluating $\Cbs$, then we will occur an overhead of $2L\times T$ to evaluate the gradients for all $L$ parameters in the $\Ansatz$, and so if the cost is efficient to compute, so are its gradients. However, comparing this black-box type computation (since the parameters must be evaluated independently) to the efficiency of the backpropagation algorithm we discussed in \secref{ssec:prelim/machine_learning/training_NN}, where we can `reuse' computations from one layer to another, we see this linear scaling may be an expensive\footnote{If we treat the cost function in the quantum and classical cases as an oracle, we are only counting the number of evaluations which must be done for each parameter. A discussion of the quantum/classical computability of the cost is a separate question.} price to pay. Nonetheless, many quantum software libraries including TensorFlow Quantum~\cite{broughton_tensorflow_2020} by Google, or Pennylane~\cite{bergholm_pennylane_2020} by Xanadu offer these gradient computations incorporated to the standard deep learning libraries such as TensorFlow or PyTorch. One can circumvent this linear scaling by using a quantum \emph{simulator} to train quantum models, as direct access to the wavefunction allows the implementation of \computerfont{autodiff} methods. However, this is clearly not possible if training a (near term) model on quantum hardware, and at the time of writing, this linear scaling is the best we have.

A final point to note is the assumption that $\Sigma^2 = \mathds{1}$ in the above theorem. This is not in fact necessary, and the theorem has been extended in many aspects, including stochastic versions for more general operations~\cite{crooks_gradients_2019}
 higher order~\cite{mari_estimating_2021} and \emph{natural} gradients~\cite{stokes_quantum_2020, koczor_quantum_2020-1}, imaginary time evolution~\cite{mcardle_variational_2019}, and analytic methods~(\cite{ostaszewski_structure_2021, koczor_quantum_2020, cade_strategies_2020}) have been proposed. This is an active area of research, and this Thesis will only require the simplest forms of gradient descent using the parameter-shift rule in the form given above, so we neglect further discussion of other methods.\\

\newpage

\subsubsection[\texorpdfstring{\color{black}}{} Barren plateaus]{Barren plateaus} \label{sssec:vqa_barren_plateaus}


The problems mentioned relating to $\VQA$ optimisation are also common to machine learning, but the nature of $\VQA$s also introduces unique problems. These problems manifest in the cost function landscape and two phenomena are called \emph{barren plateaus} (BPs) and \emph{narrow gorges}. Barren plateaus are exponentially flat regions in the parameter space, in which gradient-based\footnote{This is not in fact limited to gradient methods, it has also been shown that gradient-free methods which rely on cost function \emph{differences} will also fail\cite{arrasmith_effect_2020}} optimisation will fail, since the gradient of the cost function is close to zero in all directions.
\begin{defbox}
    \begin{definition}[Barren plateau (from \cite{arrasmith_equivalence_2021})] \label{defn:barren_plateaus}~\\ 
        Consider a $\VQA$ cost function defined as in \eqref{eqn:general_vqa_cost_function} with $f_k(x) ;= a_k x, a_k \in \mathbb{R}$. The cost exhibits a barren plateau if $\forall \theta \in \paramtheta$, the variance of its partial derivative vanishes exponentially with $n$, the number of qubits:
        \begin{equation}
            \text{Var}_{\paramtheta}[\partial_{\theta_i} \Cbs(\paramtheta)] \leq F(n),\qquad \text{with}\qquad F(n) \in \mathcal{O}(b^{-n})
        \end{equation}
        for some $b>1$. The expectation values are taken with respect to the parameters, $\paramtheta$: $\text{Var}_{\paramtheta}[\partial_{\theta_i} \Cbs(\paramtheta)] = \langle [\partial_{\theta_i} \Cbs(\paramtheta)]^2\rangle_{\paramtheta}$\footnote{ For proofs it is usually assumed that the expectation is computed over unitaries that contain the parameters, and furthermore that these unitaries form exact (\cite{mcclean_barren_2018}) or approximate (\cite{holmes_connecting_2021}) quantum $t$-designs~\cite{dankert_exact_2009} in order to make the integration tractable.}.
    \end{definition}
\end{defbox}
This phenomenon was first observed by~\cite{mcclean_barren_2018} which found that such plateaus occurred in sufficiently deep \emph{random} (and hardware-efficient) PQCs. This was extended in \cite{cerezo_cost_2021} and found to occur with much shallower circuit depths and be highly dependent on the locality of the cost function. Entanglement, a key resource in quantum computation, was also proven to be detrimental to training PQCs as too highly entangled systems were shown to exhibit also a BP~\cite{marrero_entanglement_2021}.

Of interest to note, is the $\Ansatz$ dependent nature of barren plateaus - they are rigorously proven to exist primarily for hardware efficient $\Ansatze$, but have proven to be absent in quantum convolutional neural networks~\cite{pesah_absence_2020} or tree tensor network based $\Ansatze$~\cite{zhao_analyzing_2021, zhang_toward_2020}. Furthermore, using problem-inspired $\Ansatze$ such as the Hamiltonian variational $\Ansatz$ has been shown numerically to be useful for mitigating BPs~\cite{wiersema_exploring_2020}. Finally, the barren plateaus referenced above originate from the randomness in such circuit designs, and as such would exist even if $\VQA$s were implemented on fault-tolerant quantum devices. For near term NISQ hardware unfortunately, an alternate BP rears its head. Dubbed `noise-induced' barren plateaus~\cite{wang_noise-induced_2021}, these arise simply due to hardware noise on quantum devices, and may prove to be an alternative roadblock in the usefulness of $\VQA$s.

Hope, however, is not lost. While the discovery of BPs have proliferated, so too have strategies to mitigate them, ranging from clever initialisation strategies~\cite{grant_initialization_2019} and correlating parameters~\cite{volkoff_large_2021} to using entanglement as a resource~\cite{patti_entanglement_2020}.

Complementary to BPs, we also have the `narrow gorge' phenomenon. This states that the problem solution occurs in a deep gorge with a small width which shrinks exponentially quickly as the problem size increases. The narrow gorge has been less well studied to date, so we will not focus on it in this Thesis, but its appearance in a cost landscape was proven to be synonymous with a barren plateau - one never occurs without the other~\cite{arrasmith_equivalence_2021}. This observation led to the definition of quantum \emph{landscape} theory~\cite{arrasmith_equivalence_2021}, which aims to thoroughly investigate parameter landscapes. 

Given the above discussions, barren plateaus and landscape study is an active area of study and highly interesting. This is particularly true given their impact on discovering practical applications of quantum computers, which is the main goal of this Thesis. We return to a discussion of BPs later for our specific use case.

\subsection[\texorpdfstring{\color{black}}{} \texorpdfstring{$\VQA$}{} inputs and outputs]{\texorpdfstring{$\VQA$}{} inputs and outputs} \label{sssec:vqa_input_and_ouput}
The nature of the inputs and outputs of a $\VQA$ are highly problem dependent. We have touched on some of these in the above sections, but now let us highlight some key examples in more detail. When using $\VQA$s as machine learning applications on classical data, the data must be \emph{encoded} in quantum states to be processed by the $\Ansatz$. Similarly, in these cases the outputs must also be classical, which corresponds to performing some measurements on the quantum state. However, for QML/$\VQA$ applications on quantum data, quantum states are directly provided as training data $\{\rho_i\}_{i=1}^M$ to the algorithm. Depending on the nature of the problem, the output for quantum data may be still classical, but in some cases the $\VQA$ may output quantum states themselves as the solution\footnote{Of course, even in these cases some classical information is usually extracted for the purposes of training the PQC. There are also proposals for `quantum training', which is done more coherently than is typical in $\VQA$s~\cite{verdon_universal_2018, kerenidis_quantum_2020, liao_quantum_2021}}. We shall encounter an example later in this Thesis which exactly falls in this latter category. 

In order to motivate key concepts in this area, we proceed by discussing three examples, which will play central roles in the concepts used later in this Thesis. Let us begin with one of the simplest possible:
\begin{itemize}
    \item Data encoding for quantum classifiers.
\end{itemize}
A quantum classifier is one of the simplest applications one could consider for a $\VQA$/QNN. To be clear, here we refer to quantum classifiers for \emph{classical} data. As such, the problem statement is exactly the same as in~\secref{ssec:prelims/machine_learning/classification}. Quantum classifiers for quantum data are an extremely interesting area of study (see ~\cite{cong_quantum_2019} for example) but outside the scope of this Thesis. 

Given a classical dataset of $N$ dimensional (real) vectors, $\mathcal{D} = \{\xbs^m\}_{m=1}^M$, we must have some method to \emph{encode} or \emph{embed} the datapoints, $\xbs^m \in \mathbb{R}^N$ into quantum states for further processing in the quantum classifier. For the remainder of this Thesis, we primarily focus on data embeddings which encode only a data point per quantum state\footnote{As opposed to data encodings which consider also a superposition over datapoints in the dataset, which may be required for some QML algorithms. An example is the amplitude encoding given in~\eqref{eqn:amplitude_encoding}.}. Such an encoding process is achieved by a state preparation operation, $E$, which acts as follows:
\begin{equation} 
    E_{\phi} : \xbs \rightarrow \rho^\phi_{\xbs}
\end{equation}
In this Thesis, we assume an implementation of $E_{\phi}$ as a unitary, $S_{\phi}$, acting on a fixed initial (pure) state, so:
\begin{equation} 
    S_{\phi}(\xbs)\ket{0}^{\otimes n}= \ket{\phi(\xbs)}
\end{equation}
The function $\phi$ is a `feature map' which maps from the original data space, $\mathcal{X}$, into a `feature space' (similar in nature to the feature map discussed for kernel methods in~\secref{ssec:prelim/qc/distance_measures/probability}). The resulting state, $\ket{\phi(\xbs)}$ is therefore sometimes referred to as a \emph{quantum} feature vector. For $S_{\phi}({\xbs})$ to be useful as a data encoding, it should have several desirable properties. First, $S_{\phi}({\xbs})$ should have a number of gates which is at most polynomial in the number of qubits as well as the size of the dataset, dimension of the data to be encoded, and all other relevant input parameters. For machine learning applications, we want the family of state preparation unitaries to have enough free parameters such that there is a unique quantum state $\rhox$ for each feature vector $\xbs$ --- i.e., such that the encoding function $E_{\phi}$ is bijective. 
Additionally, for NISQ applications, sub-polynomial depth is even more desirable, and we want $S_{\phi}({\xbs})$ to be hardware-efficient.

We have already encountered such a state preparation routine earlier in this Thesis in the form of the amplitude encoding (\eqref{eqn:amplitude_encoding}). Data encoding methods are of key interest to the success of QML algorithms in general, see \cite{schuld_supervised_2018} for an overview of encoding strategies and techniques. For completeness, we list some example encoding methods (taken from \cite{schuld_supervised_2018, schuld_quantum_2019}). 
\begin{defbox}
    \begin{definition}[Basis encoding]\label{defn:basis_encoding}~\\ 
        Given a feature vector $\xbs = [x_1, ..., x_N]^T \in \{0, 1\}^N$, the basis encoding maps $\xbs \mapsto E_{\phi}^{\text{Basis}}$ as:
        \begin{equation}\label{eqn:basis_encoding_single_datapoint}
        \begin{split}
          &E_{\phi}^{\text{Basis}} : \xbs \in \{0, 1\}^N \rightarrow \ket{\phi(\xbs)} = \ket{\xbs}\\
            \implies &S^{\text{Basis}}_{\phi}(\xbs) \ket{0}^{\otimes n} = \XG^{x_0} \otimes \dots \otimes \XG^{x_N} \ket{0}^{\otimes n} 
        \end{split}
        \end{equation}
    \end{definition}
\end{defbox}
\begin{defbox}
    \begin{definition}[Amplitude encoding]\label{defn:amplitude_encoding}~\\ 
    Given a feature vector $\xbs = [x_1, ..., x_N]^T \in \mathbb{C}^N$, the basis encoding maps $\xbs \mapsto E_{\phi}^{\text{Amp}}$ as:
        \begin{equation}\label{eqn:amplitude_encoding_single_datapoint}
        \begin{split}
          &E_{\phi}^{\text{Amp}}: \xbs \in \mathbb{C}^n \rightarrow \ket{\phi(\xbs)} = \frac{1}{||\xbs||_2}\sum\limits_{i=1}^N x_i\ket{i}\\
        \end{split}
        \end{equation}
        Here, we use $n = \mathcal{O}(\log(N))$ qubits to encode $N$-dimensional datapoints.
    \end{definition}
\end{defbox}
\begin{defbox}
    \begin{definition}[Product (qubit) encoding]\label{defn:product_encoding}~\\ 
         Given a feature vector $\xbs = [x_1, ..., x_N]^T \in \mathbb{R}^N$, the basis encoding maps $\xbs \mapsto E_{\phi = (f_i, g_i)_{i=1}^n}^{\text{Prod}}$ given by
        \begin{equation}\label{eqn:product_encoding_single_datapoint}
          E_{\phi = (f_i, g_i)_{i=1}^n}^{\text{Prod}} : \xbs \in \mathbb{R}^N \rightarrow  \bigotimes_{i=1}^N \left[f_i(x_i)\ket{0} + g_i(x_i)\ket{1}\right], ~  |f_i|^2 + |g_i|^2 = 1, ~\forall i
        \end{equation}
    \end{definition}
\end{defbox}
The canonical example for a product encoding is choosing $f_i = \cos(x_i), g_i = \sin(x_i), ~\forall i$. In this case we have $S_{\phi}(\xbs) := \bigotimes_{i=1}^n\RY(2x_i)$, and the resulting state is:
\begin{equation}\label{eqn:angle_encoding_multiqubit}
    \begin{split}
    \ket{\phi(\xbs)} = \bigotimes_{i=1}^N \left[\cos(x_i)\ket{0} + \sin(x_i)\ket{1}\right]
    \end{split}
\end{equation}

Finally, let us end this discussion with a reflection on the dimension, $N$, of the datapoints. In order to actually \emph{use} the NISQ computers available today for classification dataset, we are limited to relatively small datasets. For example, if one tried to directly perform a classification task on the MNIST dataset (the example in \secref{ssec:prelims/machine_learning/classification}), each image consists of $28\times 28 = 784$ pixels, which means $N=784$. However, no (universal) quantum computer currently exists with $784$ qubits, so encoding using the product or basis encodings is not possible. In contrast, qubit efficient encodings such as the amplitude encoding would only require $10 > \log(784)$ qubits, but may require a linear (relative to the length of vector to be encoded) depth for the state preparation circuit, making it impractical to run on NISQ devices with short coherence times. An alternative name for state preparation circuits are `data loaders', and a family of such data loaders have been proposed with $q$ qubits, and depth $D$ with overall complexity $qD = \mathcal{O}(N\log N)$~\cite{johri_nearest_2021}, so it is likely either one must sacrifice circuit depth or qubit number to load data. Alternative strategies for fitting larger datasets on NISQ computers are, for example, using dimensionality reduction or feature extraction techniques such as principal component analysis (PCA)~\cite{grant_hierarchical_2018} to extract the most relevant features.

In summary, loading classical data into quantum states in NISQ devices is challenging in practice, but it is an active and interesting area of research, which we contribute to in \chapref{chap:classifier}.

Next, let us examine $\VQA$ \emph{outputs} taking now the following example:
\begin{itemize}
    \item Distribution extraction for quantum circuit Born Machines
\end{itemize}
A second key example of a $\VQA$ relevant to this Thesis (specifically in \chapref{chap:born_machine}) is the \emph{quantum circuit Born machine} (QCBM), which we define here for convenience\footnote{The original proposal~\cite{cheng_information_2018} for a Born machine was actually more general than in the definition of the QCBM. Instead,~\cite{cheng_information_2018} simply proposed a parametrised quantum \emph{state} as the model, which could be prepared by arbitrary means. The specification to the quantum circuit model came from the works of~\cite{liu_differentiable_2018, benedetti_generative_2019}.}. Just as with the quantum classifier above, the most basic example of a QCBM involves a unitary evolution by a PQC, $U(\paramtheta)$, on a reference state, again, usually taken to be $\ket{0}^{\otimes n}$. The QCBM is used for the task of \emph{distribution learning} (as introduced in \secref{ssec:prelims/machine_learning/generative_modelling}). Since quantum mechanics is naturally probabilistic, a measurement of the prepared state of the QCBM in the computational basis, produces a sample, $\zbs$ according to the distribution $p_{\paramtheta}$:
\begin{equation} \label{eqn:born_machine_pure_measurement_rule}
    \ket{0}^{\otimes n} \rightarrow U(\paramtheta)\ket{0}^{\otimes n} \xrightarrow{\text{Measure} ~ \ZG^{\otimes n}} \zbs \sim p_{\paramtheta}(\zbs) =  |\bra{\zbs}U(\paramtheta)\ket{0}^{\otimes n}|^2
\end{equation}
generated according to Born's rule of quantum mechanics (hence the name of the model). The sample $\zbs$ is a binary string of length $n$, and when discussing the QCBM, the primary object we are interested in is this distribution, $p_{\paramtheta}$.

This output distribution (and in general output probability distributions from quantum circuits) is intractable to compute, and so the QCBM is an example of an \emph{implicit} generative model~\cite{mohamed_learning_2017, diggle_monte_1984}. An implicit model is one which is easy to sample from, but whose corresponding probabilities are intractable to compute, or we do not have access to. A popular example of an implicit model is the generative adversarial network (GAN) which transforms a latent random variable via a deterministic function to a random variable distributed according to the distribution of choice. In the case of the QCBM, the ease of sampling is obvious; a measurement of all qubits produces a sample. However, the corresponding probabilities may be exponentially small, i.e. $p_{\paramtheta}(\zbs) = \mathcal{O}(1/2^n)$ for $n$ qubits. Since a straightforward method to estimate probabilities is to simply construct an empirical distribution by drawing $S$ samples from the QCBM (preparing the same circuit and measuring $S$ times):
\begin{equation} \label{eqn:empiricial_qcbm_distribution}
    \widehat{p_{\paramtheta}(\zbs)} := \frac{1}{S} \sum_{s=1}^S \delta(\zbs - \zbs^{(s)}),
\end{equation}
which counts the number of times a particular sample, $\zbs^{(s)}$, occurs in the $S$ samples, $\{\zbs^{(1)}, \zbs^{(2)}, \dots,\zbs^{(S)}\}$. If a particular sample, (say for example $\zbs^{(*)} = \underbrace{100010\dots 001}_{{n}}$) has a probability $p_{\paramtheta}(\zbs^{(*)}) = 1/2^n$, then by only running the QCBM circuit $\text{poly}(n)$ times (as we would to keep the runtime efficient), we would never observe the particular sample (unless we were incredibly lucky). This intuitive argument explains why probability estimation of quantum circuits is exponentially\footnote{Exponential here meaning in terms of sample complexity.} hard in the worst case\footnote{For example, if we are required to fully characterise a QCBM distribution which has a large number of these probabilities being exponentially small, we will not be able to observe a substantial probability mass. These type of distributions are exactly those used to demonstrate `quantum computational supremacy'~\cite{arute_quantum_2019}, which we will discuss later in the Thesis. Of course, this will not always be the case - quantum circuits can also produce distributions which are also very concentrated and may have many probabilities being zero.}.

In summary, when using $\VQA$s for generative modelling, the `outputs' we care about are the entire probability distribution of the quantum circuit, $p_{\paramtheta}(\zbs)$. Since we do not have access to $p_{\paramtheta}$ we must deal exclusively with samples, $\zbs$. As such, when dealing with these models, we must find powerful but efficient methods of doing so in order to train them. This will be one of the primary contributions of \chapref{chap:born_machine}, where we return to a discussion of Born machines in greater detail.

Finally, we note that a number of extensions to the simple Born machine model described above have been considered in the literature\footnote{Sometimes under different names, but the ideas could be adapted.}. In its standard form, a Born machine \emph{does not require} an input (in contrast to the quantum classifier). This is in contrast to comparable classical generative models such as GANs~\cite{goodfellow_generative_2014} which transform uniform input randomness into the desired distribution. A Born machine, in contrast, is naturally probabilistic, due to the inherent randomness in quantum mechanics. The first possible extension one could consider for the Born machine is to also include \emph{classical} input randomness, or by including latent variables as an input space. For example, one could consider a Born machine defined by a \emph{mixed} state~\cite{romero_variational_2021, verdon_quantum_2019, benedetti_variational_2021}:
\begin{equation} \label{eqn:mixed_state_born_machine}
        \rho^{\text{Born}}_{\paramtheta} = \sum_{\xbs} q(\xbs) U(\paramtheta)\ketbra{\psi_{\xbs}}{\psi_{\xbs}}U(\paramtheta)
\end{equation}
Where $q(\xbs)$ is a distribution over latent variables, $\xbs$, which are encoded into input states, $\ket{\psi_{\xbs}}$. This classical distribution $q(\xbs)$ could also be parametrised, and output from, say for example, another (classical) neural network~\cite{verdon_quantum_2019}. In this case, extracting samples from the machine requires the more general rule than \eqref{eqn:born_machine_pure_measurement_rule}:
\begin{equation} \label{eqn:born_machine_mixed_measurement_rule}
    p_{\paramtheta}(\zbs) = \Tr\left(\ketbra{\zbs}{\zbs}\rho^{\text{Born}}_{\paramtheta}\right)
\end{equation}

A second extension, is to allow a measurement in different bases. For example, there is no requirement for the output strings, $\zbs$, to be generated by computational basis measurements, $\ZG$. Measurement in alternate bases can be achieved by appending gates to $U(\paramtheta)$ to rotate the basis. For example, by adding $\HG$ to each qubit, and measuring $\ZG$, this is equivalent to measuring in the Pauli-$\XG$ basis. With this in mind, a neat method to extend the output space of the QCBM was proposed in~\cite{rudolph_generation_2020} by measuring in \emph{both} bases and concatenating the resulting samples together\footnote{In~\cite{rudolph_generation_2020}, the Born machine was actually used as a prior distribution for a GAN, and the resulting model was shown to be able to generate high-quality MNIST digits (see \chapref{sec:prelim_machine_learning}}). To summarise this idea, the QCBM is measured once in the Pauli-$\ZG$ basis to produce an $n$-bit sample $\zbs_1$. The same state is then re-prepared and measured in the Pauli-$\XG$ basis which gives another $n$-bit sample, $\zbs_2$. The final sample is then the concatenation of these two: an $2n$-bit sample $\zbs := \zbs_1||\zbs_2$, and the model is dubbed a \emph{basis-enhanced} Born Machine. Such a scheme has a limited expressibility since a basis-enhanced QCBM using $n$ qubits cannot generate all the possible distributions which are expressible using a $2n$ qubit (non basis-enhanced) QCBM, but it is a useful method to possibly enhance the capabilities of quantum generative models on NISQ hardware.

A final possible extension is to remove the `QC' from `QCBM'. It is plausible that advantages could be gained by studying alternative methods to prepare the final state, $\ket{\psi}$ outside of the circuit model. For example, via a Hamiltonian evolution, measurement based quantum evolution  or tensor network methods (see~\cite{gao_quantum_2018} for example).

Finally, let us conclude our discussion with purely \emph{quantum} input and output. We mentioned above that quantum data comes in the form of quantum states, $\{\rho_i\}_{i=1}^M$. In reality, what actually constitutes `quantum data' is slightly ambiguous. Let us illustrates this ambiguity with some examples of $\VQA$s which are designed specifically with quantum inputs in mind:
\begin{tcbraster}[raster columns=3,raster equal height,nobeforeafter,raster column skip=0.2cm]
\begin{mybox}{red}{}
{\begin{small}
  \textbf{Quantum\\ autoencoder}\\ \cite{romero_quantum_2017}.
\end{small}}
  \end{mybox}
  \begin{mybox}{cyan}{}
  {\begin{small}
  \textbf{Variational\\ quantum\\ thermaliser}\\ \cite{verdon_quantum_2019}.
    \end{small}}
  \end{mybox}
  \begin{mybox}{green}{}
  {\begin{small}
  \textbf{Quantum \\ convolutional\\ neural networks}\\ \cite{cong_quantum_2019}.
    \end{small}}
  \end{mybox}
    \begin{mybox}{violet}{}
    {\begin{small}
  { \textbf{QVECTOR}\\ \cite{johnson_qvector_2017}.}
     \end{small}}
  \end{mybox}
      \begin{mybox}{orange}{}
      {\begin{small}
  { \textbf{Variational \\quantum state\\ diagonalisation.}\\ \cite{larose_variational_2019}.}
     \end{small}}
  \end{mybox}
        \begin{mybox}{magenta}{}
        {\begin{small}
  \textbf{Variational \\compilation \&\\ unsampling.}\\ \cite{khatri_quantum-assisted_2019, carolan_variational_2020}
  \end{small}}
  \end{mybox}
\end{tcbraster}
We have presented a fairly broad selection of algorithms here; at one end of the spectrum, we have the quantum autoencoder and the quantum convolutional NN (QCNN). These can be considered specific quantum neural network models, whose definition of quantum data is exactly that given above, a set of quantum states, $\{\rho_i\}_{i=1}^M$. The goal of the autoencoder is to compress quantum information (as in its classical counterpart), and the QCNN is used to classify quantum states based on their properties. In the middle, we have variational quantum state diagonalisation and the variational quantum thermaliser which are the quantum analogues of principal component analysis, and a generative model (for mixed quantum states) respectively. 

Finally, at the other end of the spectrum we have QVECTOR, variational compilation (also~\cite{heya_variational_2018, jones_quantum_2018}) and unsampling~\cite{carolan_variational_2020}. These tend to fall more into the category of `variational algorithm' than QML models, and they have a slightly different feel to the others. The reason for this is the goal of these $\VQA$s is to learn \emph{subroutines} or \emph{primitives}. In other words, the `quantum data' in these cases is not strictly a set of quantum states, but instead we care about the actual \emph{unitary} in the PQC itself. For example, QVECTOR aims to learn unitaries which can perform quantum error correction, and protect quantum memories, whereas compilation outputs a compiled version of an input unitary. Finally, quantum unsampling (\cite{carolan_variational_2020}) aims to learn the underlying unitary\footnote{Unitary learning in particular is an interesting problem in QML - quantum versions of the `no-free lunch' theorem~\cite{wolpert_no_1997} have been proven in this context~\cite{poland_no_2020, sharma_reformulation_2020}.} which \emph{produced} certain quantum states. Now, hopefully it is clear that quantum data can have a variety of meanings, depending on the context. In~\chapref{chap:cloning}, we present an example of a $\VQA$ which operates on quantum data; in this case it is the unitary itself we are interested in so this application is similar to that of variational compilation.
To round off this section, let us describe one method which is useful in dealing with quantum data in a $\VQA$. Since the algorithms of this nature require a classical feedback loop, even if the desired input-output relationship is purely quantum, we must still extract classical information from the output states from the $\VQA$, if for no other purpose than to train the model. In \secref{ssec:prelim/qc/swap_test}, we introduced the $\SWAP$ test, which could be used to estimate overlaps between quantum datapoints. Alternatively, one could simply extract a \emph{full description} of the quantum state and use this as the classical signal. This process is called \emph{quantum tomography}, and in general is inefficient. Regardless, it is extremely important. Let us illustrate tomography with a single qubit example. 

Any qubit state, $\rho$, can be decomposed into a linear combination of Pauli matrices:
\begin{equation}\label{eqn:qubit_state_as_paulis}
    \rho_{\boldsymbol{r}} = \frac{1}{2}\left(\mathds{1} + r_x\XG + r_y\YG + r_z\ZG \right)
\end{equation}
Now, we can reconstruct $\rho$, by estimating the Bloch vector, $\boldsymbol{r}:=[r_x, r_y, r_z]^T$. This can be done by estimating the expectation value of each of the Pauli observables, $\Tr(A\rho), \mathsf{A} \in \{\XG, \YG, \ZG\}$ and conglomerating the results. Of course, due to statistical errors, the final state computed using an estimate, $\hat{\boldsymbol{r}}$ of $\boldsymbol{r}$, may not be a valid quantum state, so methods such as direct inversion tomography, or maximum likelihood tomography are used to project onto the `best' quantum state. (See~\cite{schmied_quantum_2016} for a discussion of single qubit tomography methods, and \cite{dariano_quantum_2003, blume-kohout_optimal_2010, gross_quantum_2010} and references therein for a discussion of the more general scenario). For the purposes of this Thesis, it will suffice to use an implementation of quantum tomography from the \computerfont{forest-benchmarking} library~\cite{gulshen_forest_2019}. We use it particularly in \chapref{chap:cloning} where we adopt the technique of direct linear inversion.

\section[\texorpdfstring{\color{black}}{} Quantum kernel methods]{Quantum kernel methods} \label{ssec:prelim/qml/q_kernel}
In~\secref{ssec:prelim/qc/distance_measures/probability}, we introduced \emph{kernel methods} in the context of generative modelling. However, we only briefly alluded to what is, arguably, their main use - in support vector machines. Here, the feature maps (as in~\thmref{thm:kernel_definition}) are typically used to project data vectors into a higher dimensional space in which they are linearly separable via a hyperplane, and so can be easily classified. Kernel methods have also become popular in the quantum domain, beginning with the works of~\cite{schuld_quantum_2019, havlicek_supervised_2019} which noticed the connection between feature maps and quantum states, since both live in Hilbert spaces, $\hilb$. In particular,~\cite{schuld_quantum_2019} proved the relevant formalism and~\cite{havlicek_supervised_2019} conjectured that a quantum advantage could be gained if one used feature maps which correspond to classically intractable kernels\footnote{We return to a related discussion on classical intractablility and the relationship to quantum advantage in~\chapref{chap:born_machine}.}, meaning those which could be efficiently evaluated by a quantum computer, but not by any classical one.

Just as in the classical case, one may define a quantum RKHS, as follows:
\begin{defbox}
    \begin{definition}[Quantum RKHS]\label{defn:quantum_rkhs}~ \\
    Let $\phi:\mathcal{X}\rightarrow \hilb$ be a feature map over an input set $\mathcal{X}$, giving rise to a \emph{real} kernel: $\kappa(\xbs, \ybs) = |\braket{\phi(\xbs)}{\phi(\ybs)}|^2$. The corresponding RKHS is therefore:
    \begin{equation}
        \mathcal{H}_\kappa  \coloneqq \{f:\mathcal{X} \rightarrow \mathbb{R}|\ f(\xbs) = |\braket{w}{\phi(\xbs)}|^2, ~\forall \xbs \in \mathcal{X}, w\in \mathcal{H}\} 
    \end{equation}
    \end{definition}
\end{defbox}

For example, one may consider the state $\ket{\phi(\xbs)}$ being prepared by a quantum circuit, in the case of~\cite{havlicek_supervised_2019}, one inspired by circuits in the family of \emph{instantaneous quantum polynomial time} ($\IQP$,~\cite{shepherd_temporally_2009}), which we return to in \chapref{chap:born_machine}.
\begin{align}
    \phi: \xbs \in \mathcal{X}^n \rightarrow &\ket{\phi(\xbs)} \label{quantumfeaturemap}\\
    \ket{\phi(\xbs)}  &:= \mathcal{U}_{\phi(\xbs)}\ket{0}^{\otimes n} = U_{\phi(\xbs)}H^{\otimes n}U_{\phi(\xbs)}H^{\otimes n}\ket{0}^{\otimes n} \label{kernelcircuit}\\
    U_{\phi(\xbs)}  &:=  \exp\left(\irm\sum\limits_{S\subseteq [n]}\phi_S(\xbs)\prod\limits_{i \in S}\ZG_i\right) \label{kernelcircuitunitary}
\end{align}
so the resulting kernel is:
\begin{equation}\label{eqn:quantum_kernel}
    \kappa_Q(\xbs, \ybs)  := |\braket{\phi(\xbs)}{\phi(\ybs)}|^2 
\end{equation}
The unitary, $U_{\phi(\xbs)}$, is the key ingredient in the class $\IQP$ since it is diagonal in the computational basis, and so when decomposed into single and two qubit unitaries, can be implemented in a temporally unstructured manner (`instantaneous'). One may also imagine simply extracting a kernel via a \emph{complex} quantum amplitude (rather than as a \emph{probability}, as in \defref{defn:quantum_rkhs}, which is real valued). We mention this here to foreshadow a result in~\secref{ssec:born_machine/stein_discrepancy} in which an adaption of a proof technique will allow for complex valued kernels. 

To round off this discussion, let us mention some important follow up works in the space of quantum kernels, which indicate their promise as a source of quantum advantage. Firstly,~\cite{havlicek_supervised_2019, schuld_supervised_2021} demonstrated how quantum kernels were actually equivalent to QNNs in the sense that the space occupied by quantum models\footnote{A quantum model is defined as one which outputs a function as $f_{\paramtheta}(\xbs) = \Tr\left(\rho(\xbs)E\right)$ where $\rho$ is a quantum encoding of the data vector, $\xbs$ and $E$ is a suitable (POVM) measurement in the notation of~\secref{ssec:prelim/qc/quantum_measurements}.} (e.g. the quantum classifiers discussed in~\secref{sssec:vqa_input_and_ouput} and which form the focus of~\chapref{chap:classifier}) is equivalent to the quantum RKHS from~\defref{defn:quantum_rkhs}. This is perhaps surprising, since on the surface both methods appear quite different, but these differences appear more in the practical aspects of dealing with both methods, for example in trainability. Secondly,~\cite{hubregtsen_training_2021} proposed parameterising and training the quantum feature maps in~\eqref{quantumfeaturemap} to improve classification (similar to the idea of~\cite{lloyd_quantum_2020} and that which we give in~\chapref{chap:classifier} for quantum classifiers). Finally~\cite{huang_power_2021} (among others) noticed that since quantum models embedded in a form similar to~\eqref{eqn:quantum_kernel} are simply quadratic functions in the data, $\xbs$, one could devise a classical algorithm which could completely reproduce the predictions of the model, given sufficient data\footnote{The specific statement in~\cite{huang_power_2021} relates to the average prediction error of a quantum model relative to a classical one, which is not too dissimilar to the \emph{risk}, $\mathcal{R}$, mentioned in~\secref{ssec:prelim/machine_learning/training_NN}.}. This can be seen by writing the functions be generated by a quantum model (with amplitude encoded data,~\eqref{eqn:amplitude_encoding}), as the result of the measurement of an observable as:
\begin{equation}\label{eqn:quantum_model_quadratic_fit}
    f(\xbs) = \Tr\left(\rho(\xbs)E\right) = \left(\sum\limits_{k=1}^n\xbs^{k*}\bra{k}\right)U^{\dagger}\mathsf{O}U\left(\sum\limits_{l=1}^n\xbs^{l}\ket{l}\right) = \sum\limits_{k=1}^n\sum\limits_{l=1}^n B_{kl}\xbs^{k*}\xbs^{l}
\end{equation}
which is simply a quadratic function in the data point with $n^2$ coefficients and can be fit classically using $n^2/\epsilon$ data examples (as shown in~\cite{huang_power_2021}) 

This indicates that even though a kernel or quantum model may be intractable to compute exactly, it may not actually outperform a classical model on the task of interest (a true quantum advantage in machine learning). However, one may expect that classical intractability is at least a prerequisite for quantum advantage, since if one has the ability to efficiently simulate the model, solving the learning task is also classically possible. We make a similar observation in the context of generative modelling in~\chapref{chap:born_machine}.

In light of this,~\cite{huang_power_2021} proposes a \emph{projected} quantum kernel which is a function of the reduced density matrices of two encoded feature maps. With such a kernel, and a measure known as the \emph{geometric difference},~\cite{huang_power_2021} was able to also demonstrate a sizeable outperformance over a comparable classical machine learning algorithm.

\section[\texorpdfstring{\color{black}}{} Quantum learning theory]{Quantum learning theory} \label{sec:prelim/qml/q_learning_theory}

To round off this chapter, let us circle back to the discussion of classical learning theory presented in \secref{ssec:prelim/machine_learning/learning_theory}. The field of \emph{quantum} learning theory is almost as old as quantum computation itself, having been initiated by~\cite{bshouty_learning_1995}. In this work, the \emph{quantum} random example oracle ($\mathsf{QPEX}$) was introduced (recall~\defref{defn:pac_function_learning}), which instead of outputting a single labelled example when queried, produces a \emph{superposition} of all possible labelled examples:

\begin{defbox}
\begin{definition}[Quantum random access oracle~\cite{bshouty_learning_1995}]\label{defn:qpex_oracle}~ \\
     A quantum random access oracle, $\mathsf{QPEX}$, when queried outputs a quantum state of the following form:
     \begin{equation}\label{eqn:quantum_random_example_oracle}
         \mathsf{QPEX} \rightarrow \ket{\psi} := \sum\limits_{\xbs}\sqrt{D(\xbs)}\ket{\xbs, f(\xbs)}
      \end{equation}
\end{definition}
\end{defbox}
The work of~\cite{bshouty_learning_1995} demonstrated how, when given \emph{quantum} access to such examples, certain learning problems (for example the problem disjunctive normal form (DNF) could be solved with exponentially fewer queries than classically possible. There is an important caveat however, in that this is achievable relative to a \emph{specific} distribution (the uniform distribution) and in general it is known that such exponential reductions are not possible. The result of \cite{servedio_equivalences_2004} demonstrates how if there exists a quantum learning algorithm for a given problem, then there also exists a \emph{classical} algorithm for the same problem, which requires at most polynomially more samples than its quantum counterpart\footnote{The result of~\cite{arunachalam_optimal_2017} subsequently proved optimal sample complexity bounds for quantum learning.}. So, we have another \emph{no free lunch} for quantum learning theory. This is perhaps not so surprising - we also only have exponential speedups with a handful of quantum algorithms, so why should there be any difference with learning theory problems?

To complement~\cite{servedio_equivalences_2004}, the work of~\cite{arunachalam_optimal_2017} proved the following theorem, the quantum analogue~\thmref{thm:classical_pac_function_learnability}:

\begin{thmbox}
    \begin{theorem}(\cite{arunachalam_optimal_2017, arunachalam_guest_2017})    \label{thm:quantum_pac_function_learnability}~ \\
    Let $\mathcal{F}$ be a concept class with $\mathsf{VCdim}(\mathcal{F})$. Then, for every $\delta \in \left(0, \frac{1}{2}\right)$ and $\epsilon \in \left(0, \frac{1}{20}\right)$:
    \begin{equation}
        M = \Omega\left(\frac{\mathsf{VCdim}(\mathcal{F}) - 1}{\epsilon} + \frac{\log(1/\delta)}{\epsilon}\right)
    \end{equation}
    examples are necessary and sufficient for an $(\epsilon, \delta)$-PAC learner for $\mathcal{F}$.
    \end{theorem}
\end{thmbox}
This theorem illustrates that quantum provides no sample complexity advantage when we require learning to be with respect to \emph{every} distribution. In the last few years, there have been a flurry of results in quantum learning theory, for example, ~\cite{atici_improved_2005, kothari_optimal_2014, arunachalam_quantum_2019, arunachalam_two_2020, arunachalam_quantum_2020, arunachalam_quantum_2020} and perhaps most excitingly, the theoretical demonstration of an exponential speedup using support vector machines~\cite{liu_rigorous_2021}. The latter result is somewhat contrived and uses an engineered kernel in the spirit of~\secref{ssec:prelim/qml/q_kernel} based on the discrete logarithm. As such it is not as `natural' as one might like. However it is a very promising start. In \chapref{chap:born_machine}, we will discuss a work which proved a similar result in the \emph{distribution learning} framework. For an excellent overview of topics in quantum learning theory, we encourage the reader to see~\cite{arunachalam_guest_2017}.

\chapter{Robust data encodings for quantum classifiers} \label{chap:classifier}

\begin{chapquote}{Grad-CAM (\cite{selvaraju_grad-cam_2020})}
\textbf{The Authors}: ``. . . when today’s intelligent systems fail, they often fail spectacularly disgracefully
without warning or explanation, leaving a user staring at an
incoherent output, wondering why the system did what it did.''
\end{chapquote}

\section[\texorpdfstring{\color{black}}{} Introduction]{Introduction} \label{sec:classifiers/introduction}
In~\secref{ssec:prelims/machine_learning/classification} we introduced the problem of classification, one of the most common problems for which classical neural networks are deployed. In~\secref{sssec:vqa_input_and_ouput}, we then described a notion of a \emph{quantum} classifier, one based on a PQC which required data vectors, $\xbs$, to be encoded into quantum states after which a label is extracted via a single qubit measurement. However, everything in this previous discussion assumed access to a \emph{perfect} quantum computer - one in which data uploading and processing is achieved perfectly, with no errors in the process. Of course, this assumption is not valid in the NISQ regime. 

In~\secref{ssec:prelim/qc/quantum_noise}, we introduced some example noise models which are present on NISQ devices, and also discussed the overheads required to implement `coherent' or `fault-tolerant' quantum algorithms, which put them out of the reach of current devices. We also touched on the overhead required to \emph{detect} and \emph{correct} errors, which are vital subroutines. Given this, the main question we ask in this chapter is the following. Given an \emph{application specific} mindset, is it possible to gain any \emph{natural} noise tolerance (or robustness) \emph{for free}.
The application we have in mind is the variational quantum classifier, and we shall see in this chapter how even posing this question raises interesting ideas.

As a spoiler, the answer we give to this question is: \textbf{yes} (\emph{but} we may need to be slightly pathological in our model designs).

Before we begin, let us outline the structure of the chapter. We begin by presenting a more formal definition of the variational quantum classifier we consider in~\secref{ssec:classifiers/quantum_classifiers}. We elaborate on the definitions and examples of data encodings from~\secref{sssec:vqa_input_and_ouput}. After this, we present analytic results and proofs for robustness in~\secref{sec:noise_robustness}. We begin by showing that different encodings lead to different classes of learnable decision boundaries, then characterise the set of robust points for example quantum channels (those presented in~\secref{ssec:prelim/qc/quantum_noise}). We state and prove robustness results, and discuss the existence of robust encodings. Finally, we prove a lower bound on the number of robust points in terms of fidelities between noisy and ideal states. Lastly, we include several numerical results in \secref{sec:classifier/numerical_results} that reinforce and extend our findings. Finally, we conclude this chapter with some musings in \secref{sec:classifier/conclusions}.


\section[\texorpdfstring{\color{black}}{} Quantum classifiers]{Quantum classifiers} \label{ssec:classifiers/quantum_classifiers}
Let us begin by defining specifically what we mean by a \emph{quantum} classifier. We also in this chapter deal only with \emph{binary} classifiers, where the label takes only two possible values,  ${y}^i \in \{0, 1\}$ for a corresponding datapoint (or feature vector), $\xbs^i \in \mathcal{X}$\footnote{In practice, we typically have $\mathcal{X} = \mathbb{R}^N$, but other sets --- e.g., $\mathcal{X} = \mathbb{Z}^N$ or $\mathcal{X} \in \mathbb{Z}_2^N$ --- are possible, so we write $\mathcal{X}$ for generality.}. Let $M$ denote the number of such datapoints in a dataset.

\begin{figure}[ht]
    \centering
    \includegraphics[width=0.6\columnwidth]{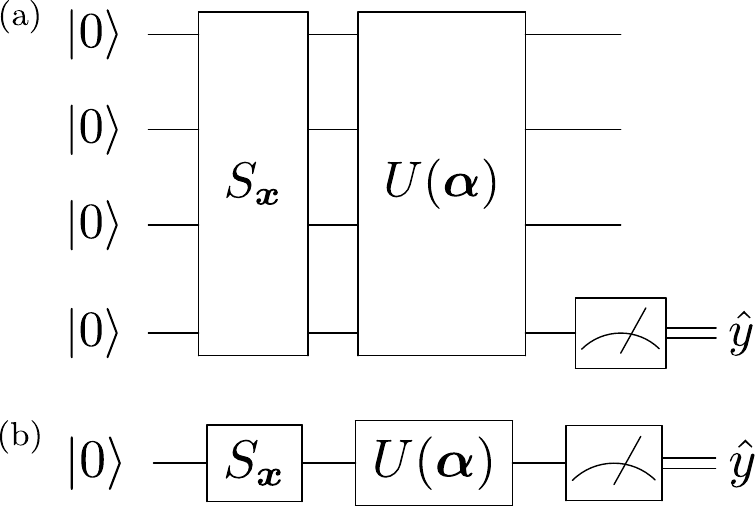}
    \caption[\color{black} A binary quantum classifier.]{\textbf{A common architecture for a binary quantum classifier that we study in this work.} The general circuit structure is shown in (a) and the structure for a single qubit is highlighted in (b). In both, a feature vector $\xbs$ is encoded into a quantum state $\rhox$ via a ``state preparation'' unitary $S_{\xbs}$. The encoded state $\rhox$ then evolves to $U \rhox U^\dagger =: \rhotildex$ where $U(\vec{\alpha})$ is a unitary $\Ansatz$ with trainable parameters $\vec{\alpha}$. A single qubit of the evolved state $\rhotildex$ is measured to yield a predicted label $\yhat$ for the vector $\xbs$.}
    \label{fig:classifier}
\end{figure}

Given this data, the goal (as discussed in \secref{ssec:prelims/machine_learning/classification}) is to output a rule $f: \mathcal{X} \rightarrow \{0, 1\}$ which accurately classifies the data and can predict the labels of unseen data.

For quantum classification, outputting information (predictions) can be done in a relatively straightforward manner. As several authors have noted~\cite{farhi_classification_2018, schuld_circuit-centric_2020, schuld_supervised_2018, grant_hierarchical_2018, perez-salinas_data_2020}, it is natural to use the measurement outcome of a single qubit, which produces a binary outcome, as a class prediction\footnote{An alternative strategy could be to use the parity ($\pm 1$)of outcome measurements from all qubits, as proposed by~\cite{abbas_power_2021}. However, this is a \emph{global} property and therefore may suffer from barren plateau problems as introduced in~\secref{ssec:vqa_cost_function_optimisation}}. We adopt this strategy in this chapter. 

Informally, we define a (binary) quantum classifier as a procedure for encoding data into a quantum circuit, processing it through trainable PQC, and outputting a (binary) predicted label. Given a feature vector $\xbs \in \mathcal{X}$, a concise description of such a classifier can be written as:
\begin{tcbraster}[raster columns=3,raster equal height,nobeforeafter,raster column skip=0.5cm]
\begin{mybox}{red}{}
  \textbf{Encoding}
  \begin{equation}\label{eqn:encoding}
      \xbs \mapsto \rhox
  \end{equation}
  \end{mybox}
  \begin{mybox}{blue}{}
  \textbf{Processing}
  \begin{equation} \label{eqn:processing}
    \rhox \mapsto \rhotildex
  \end{equation}
  \end{mybox}
  \begin{mybox}{green}{}
  \textbf{Prediction}
    \begin{equation}\label{eqn:prediction}
        \rhotildex \mapsto \yhat [\rhotildex]
     \end{equation}
  \end{mybox}
\end{tcbraster}

Several remarks are in order.
First, a given data point $\xbs$ in the training set is encoded in a quantum state $\rho_{\xbs} \in \mathcal{S}_n$ via a state preparation unitary $S_{\xbs}$ (introduced in \secref{sssec:vqa_input_and_ouput} and we elaborate in~\secref{sec:data_encodings}). We remark that each (unique) $\xbs$ in the training set leads to a (unique) $S_{\xbs}$, so the state preparation unitary can be considered a parameterized family of unitary $\Ansatze$.  

For the processing step~\eqref{eqn:processing}, several $\Ansatz$ architectures have been proposed in the literature, including quantum convolutional neural networks (mentioned in~\secref{sssec:vqa_input_and_ouput}) \cite{cong_quantum_2019, henderson_quanvolutional_2020}, strongly entangling $\Ansatze$ \cite{schuld_circuit-centric_2020}, and more \cite{stoudenmire_supervised_2016, grant_hierarchical_2018}. In this chapter, we allow for a general unitary evolution $U(\vec{\alpha})$ such that:
\begin{equation} \label{eqn:unitary_evolution}
     \rhotildex =  U(\vec{\alpha}) \rhox U^\dagger(\vec{\alpha})
\end{equation}
We remark that some classifier $\Ansatze$ involve intermediate measurements and conditional processing (notably \cite{cong_quantum_2019}) and so do not immediately fit into~\eqref{eqn:unitary_evolution}. Our techniques for showing robustness could be naturally extended to such architectures, however, and so we consider~\eqref{eqn:unitary_evolution} as a simple yet general model. We primarily focus on data encodings and their properties in this chapter, and not on other $\VQA$ features such as the choice of cost function. For this reason we often suppress the trainable parameters $\vec{\alpha}$ and write $U$ for $U(\vec{\alpha})$. 

Finally, the remaining step is to extract information from the state $\rhotildex$ to obtain a predicted label. As mentioned, a natural method for doing this is to measure a single qubit which yields a binary outcome $0$ or $1$ taken as the predicted label $\yhat$. Since measurements are probabilistic, we measure $N_m$ times and take a ``majority vote.'' That is, if $0$ is measured $N_0$ times and $N_0 \geq N_m / 2$, we take $0$ as the class prediction, else $1$. Generalising the finite statistics\footnote{We discuss details arising from finite statistics in \secref{ssec:finite_sampling_error}. The results we prove can be easily modified to incorporate finite statistics as shown in this section, but they are simpler to state in terms of probabilities.}, this condition can be expressed analytically as
\begin{equation} \label{eqn:decision_rule}
    \yhat [\rhotildex] = \begin{cases}
        0 \qquad \text{if   }  \Tr [ \Pi_0^c \rhotildex] \geq \frac{1}{2} \\
        1 \qquad \text{otherwise} 
    \end{cases}, \qquad     \Pi_0^c := \ketbra{0}{0}_c \equiv \ketbra{0}{0}_c \otimes \mathds{1}_{\bar{c}}
\end{equation}
%
is the projector onto the ground state of the classification qubit, labelled $c$, and the remaining qubits are labelled $\bar{c}$. For brevity we often omit these labels when it is clear from context.
Throughout this chapter, we use $\yhat$ for predicted labels and $y$ for true labels, and we refer to~\eqref{eqn:decision_rule} as the \emph{decision rule} for the classifier. \eqref{eqn:decision_rule} is not the only choice for such a decision rule. In particular, one could choose a different ``weight'' $\lambda$ such that $\yhat = 0$ if $\Tr [ \Pi_0 \rhotildex] \geq \lambda$ as in~\cite{perez-salinas_data_2020}, add a bias to the classifier as in~\cite{schuld_circuit-centric_2020}, or measure the classification qubit in a different basis. The latter shall be an example of a technique to gain robustness to certain noise models later. Let us also remark that this decision boundary is `harsh', in the sense that it does not discriminate, in principle, between points which have only $\Tr [ \Pi_0^c \rhotildex] = 1 / 2 + \epsilon$ versus those which have $\Tr [ \Pi_0^c \rhotildex] = 1 - \epsilon$. As such, one could also introduce a sequence of soft decision rules by defining the model based on the distance of the single qubit probability\footnote{In practice, this is easily achieved by defining the decision rule based on the positivity, or negativity of the expectation value of $\ZG$ on the classification qubit, $\langle \ZG_c\rangle_{\rhotildex}$. We deal with the probabilities rather than expectation values for simplicity, but the results could be extended in a straightforward manner.} to $1/2$. Such decision rules also allow the classifier to express a `confidence' in its choice for a given label. Our techniques for showing robustness (\secref{subsec:robustness_results}) could be easily adapted for such alternate decision boundaries.

The preceding discussion is summarised with the following formalisation:
\begin{defbox}
\begin{definition}[(Variational) quantum classifier]\label{def:binary_quantum_classifier}~ \\
    A (binary) variational quantum classifier consists of three functions: 
    \begin{enumerate}
    \item[(i)] An encoding function: 
    \begin{equation}\label{eqn:encoding_formal}
         E: \mathcal{X} \rightarrow \mathcal{S}_n, \qquad
         E(\xbs) = \rhox
    \end{equation}
    
    \item[(ii)] A function which evolves the state (possibly including ancillary qubits and/or measurement):
    \begin{equation}\label{eqn:evolution_formal} 
         \mathcal{U}: \mathbb{C}^{2^n \times 2^n} \rightarrow \mathbb{C}^{2^m \times 2^m}, \qquad
         \mathcal{U}(\rhox) = \rhotildex
    \end{equation}
    
    \item[(iii)] A decision rule:
    \begin{equation} \label{eqn:decision_rule_formal}
         \hat{y}: \mathbb{C}^{2^m \times 2^m} \rightarrow \{0, 1\} 
    \end{equation}
    \end{enumerate}
\end{definition}
\end{defbox}
Note that we have not included training data or a cost function in this definition, so a quantum classifier can be considered a hypothesis family. We can distinguish between the hypothesis family and the trained model --- in which optimisation has been performed to minimise a cost function over a specified training data set --- by referring to the latter as the \emph{realised} quantum classifier if it is not clear from context. Let us now concentrate our focus to the primary object of interest for the remainder of this chapter, step (i) in~\defref{def:binary_quantum_classifier}, the data encoding strategy.


\subsection[\texorpdfstring{\color{black}}{} Data encodings]{Data encodings} \label{sec:data_encodings}

\begin{figure}
    \centering
    \includegraphics[width=0.9\columnwidth]{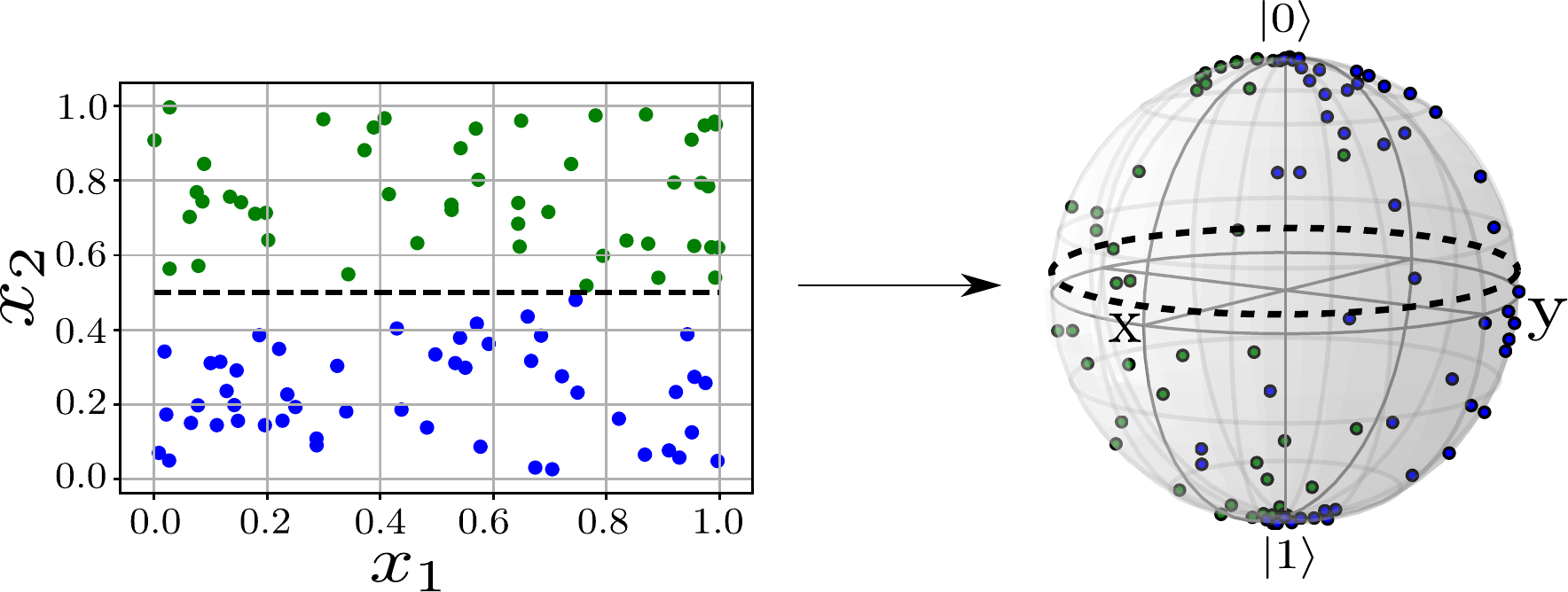}
    \caption[\color{black} A visual representation of classical data encoding for a single qubit.]{\textbf{A visual representation of data encoding~\eqref{eqn:encoding_formal} for a single qubit.} On the left is shown a set of randomly generated points $\{\xbs^i, y^i\}_{i = 1}^{M}$ normalised to lie within the unit square, $(\xbs^i = [x^i_{1}, x^i_{2}]^T)$, separated by a true decision boundary shown by the dashed black line. A data encoding maps each $\xbs^i \in \mathbb{R}^2$ to a point on the Bloch sphere $\rho_{\xbs^i} \in \mathbb{C}^2$. The dashed black line on the Bloch sphere shows the initial decision boundary of the quantum classifier. During the training phase, unitary parameters are adjusted to rotate the dashed black line to correctly classify as many training points as possible. Different data encodings lead to different learnable decision boundaries and different robustness properties.}
    \label{fig:data_encoding_schmatic_single_qubit}
\end{figure}



As discussed in~\secref{sssec:vqa_input_and_ouput}, a data encoding can be thought of as ``loading'' a data point $\xbs \in \mathcal{X}$ from memory into a quantum state so that it can be processed by a classifier. To present the work of this chapter, we primarily focus on the example of a single qubit classifier. This is highly illustrative and allows us to visualise some nice features of the model. We will also focus on variations of the `qubit' encoding in~\defref{defn:product_encoding}, i.e. data encodings whose state preparation unitary is of the form:
\begin{equation} \label{eqn:product_encoding_unitary_general}
    S_{\phi}(\xbs) = \bigotimes_{i=1}^n S_{\phi_i}(\xbs).
\end{equation}
However, we remark that some of the results presented towards the end of the chapter do in fact generalise to multi-qubit classifiers.

As a first step, let us generalise the product encoding of \eqref{eqn:angle_encoding_multiqubit} to reduce the number of qubits required by a factor of two. We do so by exploiting the relative phase degree of freedom in a single qubit and define a `dense angle'\footnote{The qubit encoding in \eqref{eqn:angle_encoding_multiqubit} is sometimes referred to as an `angle' encoding. Here we make it `dense' by utilising both degrees of freedom in a single qubit.}:
\begin{defbox}
\begin{definition}\label{def:dense_angle_encoding}[Dense angle encoding]~ \\
    Given a feature vector $\xbs = [x_1, ..., x_N]^T \in \mathbb{R}^N$, the dense angle encoding maps $\xbs \mapsto E^{\DAE}(\xbs)$ as:
    \begin{equation} \label{eqn:dense_angle_encoding_general}
        \ket{\phi(\xbs)} := \bigotimes_{i=1}^{\ceil*{N / 2}} \cos (\pi x_{2i -1})\ket{0} + e^{2 \pi i x_{2i}} \sin (\pi x_{2i - 1})\ket{1}
    \end{equation}
\end{definition}
\end{defbox}
%
For some of our analytic and numerical results, we highlight the dense angle encoding for two-dimensional data $\xbs \in \mathbb{R}^2$ with a single qubit given by
\begin{equation} \label{eqn:dae_single_qubit}
     \ket{\phi(\xbs)}_{\DAE}  := \cos (\pi x_{1}) \ket{0} + e^{2  \pi i x_{2}} \sin (\pi x_{1})\ket{1} 
\end{equation}
which has density matrix
\begin{equation*} 
    \rhox =
    \left[ \begin{matrix}
        \cos^2 \pi x_1  & e^{ - 2 \pi i x_2} \cos \pi x_1 \sin \pi x_1 \\
        \erm^{  2 \pi \irm x_2} \cos \pi x_1\sin \pi x_1& \sin^2 \pi x_1 \\
    \end{matrix} \right] 
\end{equation*}

Although the angle encoding in \eqref{eqn:angle_encoding_multiqubit} and dense angle encoding in \eqref{eqn:dense_angle_encoding_general} use sinuosoids and exponentials, there is nothing special about these functions (other than, perhaps, they appear in common parameterisations of qubits and unitary matrices~\cite{nielsen_quantum_2010}). We can easily abstract these to a general class of qubit encodings which use arbitrary functions\footnote{In fact, we already did this when introducing data encodings in \secref{sssec:vqa_input_and_ouput}. The main difference here is a definition with approximately half the required number of qubits.}.
\begin{defbox}
\begin{definition}\label{defn:general_qubit_encoding}[General qubit encoding]~ \\
    Given a feature vector $\xbs = [x_1, ..., x_N]^T \in \mathbb{R}^N$, the general qubit encoding maps $\xbs \mapsto E^{\GQE}(\xbs)$ as:
    \begin{equation} \label{eqn:qubit_encoding_general}
        \ket{\phi(\xbs)}_{\GQE} := \bigotimes_{i=1}^{\ceil*{N / 2}} f_i(x_{2i - 1}, x_{2i}) \ket{0} + g_i(x_{2i - 1}, x_{2i}) \ket{1}
    \end{equation}
    where $f, g : \mathbb{R} \times \mathbb{R} \rightarrow \mathbb{C}$ are such that $|f_i|^2 + |g_i|^2 = 1 \ \forall i$. 
\end{definition}
\end{defbox}
We remark that a similar type of generalisation was used in~\cite{perez-salinas_data_2020} with a single qubit classifier that allowed for repeated application of an arbitrary state preparation unitary. However, a key difference in our definition is that care needs to be taken to avoid information loss in the encoding - a single qubit only has two degrees of freedom, while a general $SU(2)$ unitary has three. Therefore, while it is technically possible to use three sequential rotation gates as an encoding unitary, each with a single feature encoded in the rotation angle, this is perhaps not advisable. While this strategy would reduce the number of qubits required by a factor of $3$ over the product encoding of \eqref{eqn:product_encoding_unitary_general}), information in the feature vectors will be lost since the input state is usually fixed at $\ket{0}^{\otimes n}$.

We can also similarly generalise the amplitude encoding to allow for parameterisations of features (amplitudes).
\begin{defbox}
\begin{definition}[Generalised amplitude encoding]\label{def:generalised_amplitude_encoding}~ \\
    For $\xbs \in \mathbb{C}^N$, the generalised amplitude encoding maps  $\xbs \mapsto E_{\phi = \{f_i\}_{i=1}^{N}}^{\GAE}$ as:
    \begin{equation}\label{eqn:amplitude_encoding_general}
    \begin{split}
      &E_{\phi = \{f_i\}_{i=1}^N}^{\GAE} : \xbs \in \mathbb{C}^N \rightarrow  \sum\limits_{i=1}^{N} f_i(\xbs)\ket{i},\qquad \sum_i |f_i|^2 = 1, ~\forall i
    \end{split}
    \end{equation}
\end{definition}
\end{defbox}
The functions $f_i$ could only act on the $i^{\text{th}}$ feature, e.g. $f_i(\xbs) = \sin x_i$, or could be more complicated functions of several (or all) features. Note, we are not commenting on the efficiency of preparing such a state here.

Thus far, we have formally defined a data encoding~\eqref{eqn:qubit_encoding_general} and its role in a quantum classifier (\defref{def:binary_quantum_classifier}), and we have given several examples. While we have discussed different properties of state preparation circuits which implement data encodings (depth, overhead, etc.), we have not yet discussed the two main properties of data encodings we consider in this chapter: \textit{learnability} and \textit{robustness}.
By \textit{learnability}, we mean the expressive power~\cite{raghu_expressive_2017} of a given hypothesis family in its ability to find a set of parameters such that the realised quantum classifier can (optimally) separate the data classes. In other words, learnability measures the extent to which the quantum classifier can represent certain functions. We show in~\secref{ssec:classes_learnable_decision_boundaries} that different data encodings lead to different classes of learnable decision boundaries. For \textit{robustness}, we show that different data encodings lead to different sets of robust points (to be defined) in~\secref{ssec:characterize_robust_points} ---~\secref{subsec:existence_of_robust_encodings}. As a teaser for the statement made in the introduction of this chapter, we find that encodings always \emph{exist} which satisfy our robustness definitions, but these may come at the expense of the learnability of our model, and hence are pathological.

\newpage

\subsection[\texorpdfstring{\color{black}}{} Robust data encodings]{Robust data encodings} \label{subsec:robustness_definition}

\begin{figure*}[t!]
    \centering
    \includegraphics[width=1\textwidth, height =0.2\textwidth]{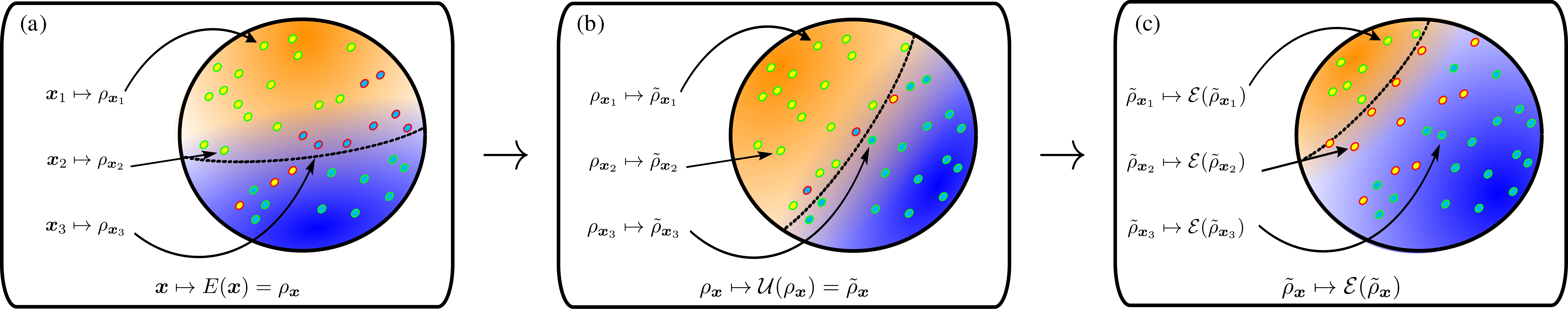}
    \caption[\color{black} Cartoon illustration of robust points for a single qubit classifier]{\textbf{Cartoon illustration of robust points for a single qubit classifier.} In panel (a), input training data points $\xbs_i$ with classes $y_i \in \{ \text{yellow}, \text{blue} \}$ are mapped into quantum states $\rhoxi$ according to some encoding function $E$. The dashed line through the Bloch sphere indicates the initial decision boundary. Points with a green outline are classified correctly, while points with a red outline are misclassified. In (b), data points are processed by the PQC with optimal unitary parameters (after minimising a cost function to find such parameters). For clarity, we keep data points fixed and adjust the location of the decision boundary, which is now rotated to correctly classify more points (fewer points with red outlines). In (c), a noise process $\E$ occurs which shifts the location of the final processed points (or location of decision boundary), causing some points to be misclassified. The set of points which maintain the same classification in (b) and (c) are the robust points. Example points $\xbs_1$ and $\xbs_2$ are correctly classified in (a) and (b) then misclassified in (c) due to the noise. Example point $\xbs_3$ is incorrectly classified in (a), correctly classified in (b) after propagating through the PQC, and remains correctly classified in (c).}
    \label{fig:single_qubit_binary_classifier_with_noise}
\end{figure*}

In this section, we define what we mean for a classifier to be `robust' against noise. In the analytical and numerical results that we present in the following sections, we focus on the simple noise models derived in \secref{ssec:prelim/qc/quantum_noise}. 

Let us begin by introducing our definitions for robust points and robust data encodings of quantum classifiers. Informally, the intuition is as follows: the quantum classifier with decision rule~\eqref{eqn:decision_rule} requires only a ``coarse-grained'' measurement to extract a predicted label. For example, with a single qubit classifier, all points in the ``top'' hemisphere of the Bloch sphere are predicted to have label $0$, while all points in the ``bottom'' hemisphere are predicted to have label $1$. The effect of noise is to shift and contract\footnote{For example, `shifts' would result from coherent noise sources, whereas `contractions' typically arise as a result of stochastic noise.} points within the Bloch sphere, but certain points can get shifted such that they get assigned the same labels they would \emph{without} noise.
This is the idea of robustness, represented schematically in \figref{fig:single_qubit_binary_classifier_with_noise}. For classification purposes, we only require that the point remain ``in the same hemisphere'' in order to get the same predicted label.

Formally, we define a robust point as follows.
\begin{defbox}
    \begin{definition}[Robust point]\label{def:robust_point}~ \\
        Let $\mathcal{E}$ be a quantum channel, and consider a (binary) quantum classifier with decision rule $\yhat$ as defined in~\eqref{eqn:decision_rule}. We say that the state $\rhox \in \mathcal{S}_n$ encoding a data point $\xbs \in \mathcal{X}$ is a \textit{robust point} of the quantum classifier if and only if
        \begin{equation} \label{eqn:robust_point_definition}
            \yhat [ \mathcal{E} ( \rhotildex )] = \yhat[\rhotildex] 
        \end{equation}
        where $\rhotildex$ is the processed state via~\eqref{eqn:evolution_formal}. %
    \end{definition}
\end{defbox}

As mentioned, for the purpose of classification, \eqref{eqn:robust_point_definition} is a well-motivated and reasonable definition of robustness. We remark that \eqref{eqn:robust_point_definition} is expressed in terms of probability; in practice, additional measurements may be required to reliably determine robustness. We discuss this point further in \secref{ssec:finite_sampling_error}.

Further, we note that \eqref{eqn:robust_point_definition} assumes that noise occurs only after the evolution $\rhox \mapsto \rhotildex$. While this may be a useful theoretical assumption, in practice noise happens throughout a quantum circuit. We can therefore consider robustness for an \textit{ideal data encoding} as in \defref{def:robust_point}, or for a \textit{noisy data encoding} in which some noise process $\mathcal{E}_1$ occurs after encoding and another noise process $\mathcal{E}_2$ occurs after evolution:
\begin{equation} \label{eqn:robust_point_noisy_encoding}
    \yhat [\mathcal{E}_2 (\mathcal{U} (\mathcal{E}_1(\rhox)))] =
    \yhat [\rhotildex]
\end{equation}
For our results, we primarily consider \eqref{eqn:robust_point_definition}, although we show robustness for \eqref{eqn:robust_point_noisy_encoding} in some cases. 

Robust points (\eqref{eqn:robust_point_definition}) are related but not equivalent to (density operator) fixed points of a quantum channel, and can be considered an application-specific generalisation of fixed points. In~\secref{ssec:characterize_robust_points}, we characterise the set of robust points for example channels, and in~\secref{subsec:existence_of_robust_encodings} we use this connection to prove the existence of robust data encodings. 

For classification, we are concerned with not just one data point, but rather a set of points (e.g., the set $\mathcal{X}$ or training set \eqref{eqn:labeled_data_for_classifier}). We therefore define the set of robust points, or robust set, in the following natural way.
\begin{defbox}
    \begin{definition}[Robust set]\label{def:robust_set}~ \\
        Consider a (binary) quantum classifier with encoding $E : \mathcal{X} \rightarrow \mathcal{S}_n$ and decision rule $\yhat$ as defined in~\eqref{eqn:decision_rule}. Let $\mathcal{E}$ be a quantum channel.
        The set of robust points, or simply \textit{robust set}, is
        \begin{equation} \label{eqn:robust_point_set_definition}
            \mathcal{R} (\mathcal{E}, E, \yhat) := \left\{ \xbs \in \mathcal{X}: \yhat [ \mathcal{E} ( \rhotildex )] = \yhat[\rhotildex] \right\}
        \end{equation}
        where $\rhotildex$ is the processed state via \eqref{eqn:evolution_formal} and $\rho_{\xbs} = E(\xbs)$.
    \end{definition}
\end{defbox}
%
While the robust set generally depends on the encoding $E$, there are cases in which $\mathcal{R}$ is independent of $E$.
In this scenario, we say all encodings are robust to this channel.
Otherwise, the size of the robust set (i.e., number of robust points) can vary based on the encoding, and we distinguish between two cases. If the robust set is the set of all possible points,
we say that the encoding is \textit{completely robust} to the given noise channel. 
\begin{defbox}
    \begin{definition}[Completely robust data encoding]\label{def:complete_noise_robustness}~ \\
        Consider a (binary) quantum classifier with encoding $E$ and decision rule $\yhat$ as defined in~\eqref{eqn:decision_rule}. Let $\xbs \in \mathcal{X}$ and let $\mathcal{E}$ be a quantum channel. We say that $E$ is a \textit{completely robust data encoding} for the quantum classifier if and only if
        \begin{equation} \label{eqn:completely_robust_encoding_condition}
            \mathcal{R}(\mathcal{E}, E, \yhat) = \mathcal{X}
        \end{equation}
    \end{definition}
\end{defbox}
We note that in practice (e.g. for numerical results), complete robustness is determined relative to the training set \eqref{eqn:labeled_data_for_classifier}. That is, we empirically observe that $E$ is a completely robust data encoding if and only if
\begin{equation}
    \mathcal{R}(\mathcal{E}, E, \yhat) = \{\xbs_i\}_{i = 1}^{M}
\end{equation}

Complete robustness can be a strong condition, so we also consider a partially robust data encoding, defined as follows.
\begin{defbox}
    \begin{definition}[Partially robust data encoding]\label{def:partial_noise_robustness}~ \\
        Consider a (binary) quantum classifier with encoding $E$ and decision rule $\yhat$ as defined in~\eqref{eqn:decision_rule}. Let $\xbs \in \mathcal{X}$ and let $\mathcal{E}$ be a quantum channel. We say that $E$ is a \textit{partially robust data encoding} for the quantum classifier if and only if
        \begin{equation} \label{eqn:partially_robust_encoding_condition}
            \mathcal{R}(\mathcal{E}, E, \hat{y}) \subsetneq \mathcal{X}
        \end{equation}
    \end{definition}
\end{defbox}
Similar to complete robustness, partial robustness is determined in practice relative to the training set. For $0 \le \delta \le 1$, we say that $E$ is a $\delta$-robust data encoding if and only if
\begin{equation}\label{eqn:delta_robust_encoding_condition}
    \left| \mathcal{R}(\mathcal{E}, E, \hat{y})  \right| = \delta M 
\end{equation}
where $|\cdot|$ denotes cardinality so that $ | \mathcal{R}(\mathcal{E}, E, \hat{y}) | \in [M]$.

\section[\texorpdfstring{\color{black}}{} Analytic results]{Analytic results} \label{sec:noise_robustness}

Using the definitions from \secref{ssec:prelim/qc/quantum_noise}, we now state and prove results about data encodings\footnote{In actual fact, our results generally apply to the \emph{entire} model as a whole, not just the data encoding.}. First, we show that different encodings lead to different classes of decision boundaries in \secref{ssec:classes_learnable_decision_boundaries}. Next, we characterise the set of robust points for example quantum channels in~\secref{ssec:characterize_robust_points}. In \secref{subsec:robustness_results}, we prove several robustness results for different quantum channels, and in \secref{subsec:existence_of_robust_encodings} we discuss the existence of robust encodings as well as an observed tradeoff between learnability and robustness.
Finally, in \secref{ssec:fidelity_bounds}, we prove an upper bound on the number of robust points in terms of fidelities between noisy and noiseless states.

Before diving into these results, let us first derive some useful identities for a single qubit classifier. Let $\rho$ be a single qubit state with matrix elements $\rho_{ij}$, i.e.
\begin{equation*}
    \rho := \left( \begin{matrix}
        \rho_{00} & \rho_{01} \\
        \rho_{10} & \rho_{11} \\
    \end{matrix} \right)
\end{equation*}
Then we have:
\begin{equation*}
    \XG \rho \XG = \left( \begin{matrix}
        \rho_{11} & \rho_{10} \\
        \rho_{01} & \rho_{00} \\
    \end{matrix} \right)   \qquad
    \YG \rho \YG = \left( \begin{matrix}
        \rho_{11} & - \rho_{10} \\
        - \rho_{01} & \rho_{00} \\
    \end{matrix} \right)   \qquad
    \ZG \rho \ZG = \left( \begin{matrix}
        \rho_{00} & - \rho_{01} \\
        - \rho_{10} & \rho_{11} \\
    \end{matrix} \right)   
\end{equation*}
Next, defining the projectors $\Pi_0 := \ketbra{0}{0}$ and $\Pi_1 := \ketbra{1}{1}$, one can show
\begin{equation*}
    \Tr [ \Pi_0 \XG \rho \XG] = \Tr [ \Pi_1 \rho],  \qquad
    \Tr [ \Pi_0 Y \rho \YG] = \Tr [ \Pi_1 \rho] \qquad
    \Tr [ \Pi_0 \ZG \rho \ZG] = \Tr [ \Pi_0 \rho] 
\end{equation*}
For any Hermitian matrix $A = [A_{ij}]$ and any unitary matrix $U = [U_{ij}]$, we have
\begin{equation} \label{eqn:useful_00mtx_elt}
    \Tr [ \Pi_0 U A U^\dagger] =
    |U_{00}|^2 A_{00} + 2 \Re [ U_{00}^* U_{01} A_{10}] + |U_{01}|^2 A_{11} .
\end{equation}
Similarly, one can show that
\begin{equation} \label{eqn:useful_11mtx_elt}
    \Tr[ \Pi_{1} U A U^\dagger ] =
     |U_{10}|^{2} A_{00} + 2 \Re [ U_{11}^* U_{10} A_{01}] + |U_{11}|^2 A_{11} .
\end{equation}
If we further assume the single qubit unitary, $U(\boldsymbol \alpha)$, has the decomposition: $\RZ(2\alpha_1)\RY(2\alpha_2)\RZ(2\alpha_3)$ (up to a global phase) \cite{nielsen_quantum_2010}, we get:
\begin{equation}\label{eqn:single_qubit_unitary_decompostion}
    U(\boldsymbol\alpha) = \left( \begin{matrix}
        \erm^{\irm ( - \alpha_1  - \alpha_3)} \cos\alpha_2&
        -\erm^{\irm ( - \alpha_1+ \alpha_3 )} \sin \alpha_2\\
        \erm^{\irm ( \alpha_1  - \alpha_3 )} \sin \alpha_2 & 
        \erm^{\irm ( \alpha_1  + \alpha_3)} \cos \alpha_2 \\
    \end{matrix} \right)
\end{equation}
Therefore, we get the various terms in \eqref{eqn:useful_00mtx_elt}, \eqref{eqn:useful_11mtx_elt} t be:
\begin{align*}
    |U_{00}|^2 = \cos^2(\alpha_2) ,\qquad
    |U_{01}|^2 &= |U_{10}|^2 = \sin^2(\alpha_2), \qquad
    |U_{11}|^2 = \cos^2(\alpha_2)\\
     U_{00}^*U_{01} &= -\erm^{\irm 2\alpha_3}\cos(\alpha_2)\sin(\alpha_2) = -\frac{1}{2}\erm^{2\irm \alpha_2}\sin(2\alpha_2) \\
     U_{11}^*U_{10} &= \erm^{-\irm 2\alpha_3}\cos(\alpha_2)\sin(\alpha_2) = \frac{1}{2}\erm^{-2\irm \alpha_3}\sin(2\alpha_2)
\end{align*}
So the conditions become:
\begin{align*} 
  \Tr [ \Pi_0 U A U^\dagger] 
  &=|U_{00}|^2 A_{00} + 2 \Re [ U_{00}^* U_{01} A_{10}] + |U_{01}|^2 A_{11}  \\
  &= \cos^2\left(\alpha_2\right)A_{00} + \sin^2\left(\alpha_2\right)A_{11} - \Re[e^{2i\alpha_3}\sin\left(\alpha_2\right)A_{10}] 
\end{align*}
\begin{align*}
        \Tr[ \Pi_{1} U A U^\dagger ] 
        &=|U_{10}|^{2} A_{00} + 2 \Re [ U_{11}^* U_{10} A_{01}] + |U_{11}|^2 A_{11} \\
      &=\sin^2\left(\alpha_2\right)A_{00} +\cos^2\left(\alpha_2\right)A_{11}
      +\Re[\erm^{-2\irm\alpha_3}\sin\left(2\alpha_2\right)A_{01}] 
\end{align*}

\subsection[\texorpdfstring{\color{black}}{} Classes of learnable decision boundaries]{Classes of learnable decision boundaries} \label{ssec:classes_learnable_decision_boundaries}

In \secref{sec:variational_quantum_algorithms} and \secref{sec:data_encodings} we introduced several data encodings and discussed differences in the state preparation circuits which realise them. Here, we show that different encodings lead to different sets of decision boundaries for the quantum classifier, thereby demonstrating that the success of the quantum classifier in \defref{def:binary_quantum_classifier} depends \emph{crucially} on the data encoding \eqref{eqn:qubit_encoding_general}\footnote{This fundamental importance of the data encoding in quantum classifiers has also been observed and reinforced in previous and subsequent work to ours \cite{lloyd_quantum_2020, gil_vidal_input_2020, schuld_effect_2021}.}. 

The decision boundary according to the decision rule~\eqref{eqn:decision_rule} is implicitly defined by
\begin{equation} \label{eqn:decision_boundary_defining_rule}
     \Tr [ \Pi_0 \rhotildex] = \frac{1}{2}
\end{equation}
Consider a single qubit encoding~\eqref{eqn:qubit_encoding_general} so that:
\begin{equation*}
    \rhox = \left( \begin{matrix}
        f(x_1, x_2)^2 & f(x_1, x_2) g(x_1, x_2)^* \\
        f(x_1, x_2) g(x_1, x_2) & |g(x_1, x_2)|^2 
    \end{matrix} \right)
\end{equation*}
where we assumed without loss of generality that $f$ is real valued.
Let the unitary $U$ be such that $\rhotildex = U \rhox U^\dagger$ and has matrix elements $U_{ij}$. Then, one can write the decision boundary~\eqref{eqn:decision_boundary_defining_rule} as (see~\eqref{eqn:useful_00mtx_elt})
\begin{equation}\label{eqn:projector_expanded}
  |U_{00}|^2 f^2 + 2 \Re [ U_{00}^* U_{01} f g] + |U_{01}|^2 |g|^2 = \frac{1}{2}
\end{equation}
where we have let $f := f(x_1, x_2)$ and $g := g(x_1, x_2)$ for brevity. \eqref{eqn:projector_expanded} implicitly defines the decision boundary in terms of the data encoding $f$ and $g$. The unitary matrix elements $U_{ij}$ then define the hypothesis family.

\eqref{eqn:projector_expanded} can be solved numerically for different encodings, and we do so in \secref{ssec:classes_decision_boundaries_numerical} (\figref{fig:random_decision_boundaries_encodings}) to visualise decision boundaries for single qubit classifiers. At present, we can proceed further analytically with a few inconsequential assumptions to simplify the equations.

For the amplitude encoding, we have $f(x_1, x_2) = x_1$ and $g(x_1, x_2) = x_2$. Suppose for simplicity that matrix elements $U_{00} =: a$ and $U_{01} =: b$ are real. Then, \eqref{eqn:projector_expanded} can be written
\begin{equation} \label{eqn:learnable_decision_boundary_amplitude}
    (a x_1 + b x_2)^2 = \frac{1}{2} 
\end{equation}
which defines a line $x_2 = x_2(x_1)$ with slope $\sfrac{-a}{b}$ and intercept $\sfrac{1}{\sqrt{2} b}$. Thus, a single qubit classifier in \defref{def:binary_quantum_classifier} which uses the amplitude encoding \eqref{eqn:amplitude_encoding} can learn decision boundaries that are straight lines. 

Now consider the dense angle encoding \eqref{eqn:dense_angle_encoding_general} on a single qubit, for which $f(x_1, x_2) = \cos(\pi x_1)$ and $g(x_1, x_2) = e^{2 \pi i x_2} \sin (\pi x_1)$. Supposing again that matrix elements $U_{00} \equiv a$ and $U_{01} \equiv b$ are real, we can write~\eqref{eqn:projector_expanded} as
\begin{equation}
    a^2 \cos^2 \pi x_1 + 2 a b \cos \pi x_1 \sin \pi x_1 \cos 2 \pi x_2 + b^2 \sin^2 \pi x_1 = \frac{1}{2}
\end{equation}
This can be rearranged to

\begin{equation} \label{eqn:learable_decision_boundary_dae}
    \cos 2 \pi x_2 = \frac{1 - 2 a^2  + (2a^2 -  2b^2) \sin^2 \pi x_1}{a b \sin 2 \pi x_1} 
\end{equation}
which defines a class of sinusoidal functions $x_2 = x_2(x_1)$ (see \figref{fig:random_decision_boundaries_encodings} in~\secref{ssec:classes_decision_boundaries_numerical}).

The different decision boundaries defined by~\eqref{eqn:learnable_decision_boundary_amplitude} and~\eqref{eqn:learable_decision_boundary_dae} emphasise the effect that encoding has on learnability. A realised classifier may have poor performance due to its encoding, and switching the encoding may lead to better results. We note that a similar phenomenon occurs in classical machine learning --- a standard example being that a dot product kernel cannot separate data on a spiral, but a Gaussian kernel can. 
It may not be clear \textit{a priori} what encoding to use (similarly in classical machine learning with kernels), but different properties of the data may lead to educated guesses. We note that \cite{lloyd_quantum_2020} consider training over hyperparameters to find good encodings, and we propose a similar idea in \secref{ssec:encoding_learn_alg} to find good \emph{robust} encodings.

\subsection[\texorpdfstring{\color{black}}{} Characterisation of robust points]{Characterisation of robust points} \label{ssec:characterize_robust_points}

Given the model of a classifier, and the definitions of our noise channels we can begin to discuss our robustness results against these channels. Intuitively, for a given noise channel we wish to characterise the set of robust points (\defref{def:robust_set}). If we then know what conditions our classifier will be deployed in (meaning the noise model), we can then build the classifier to encode datapoints into this robust set. As a result, these points will be protected. As a first step, we look instead at the \emph{fixed points} of a quantum noise channel $\mathcal{E}$, defined as the states $\rho \in \mathcal{S}_n$ which obey the following:
\begin{equation} \label{eqn:fixed_point_definition}
    \mathcal{E}(\rho) = \rho
\end{equation}
To proceed, we will first use some toy examples to build intuition and demonstrate the relationship between robust points and fixed points. In \secref{subsec:existence_of_robust_encodings}, we formalise this intuition. We remark that the characterisations similar to the ones in this Section may be of independent interest from a purely theoretical perspective, as robust points can be considered a type of generalised fixed point, or symmetry, of quantum channels. 

The pure states which are fixed points of the dephasing channel~\eqref{eqn:dephasing_channel} are $\Pi_0 := |0\>\<0|$ and $\Pi_1 := |1\>\<1|$, and
\begin{equation} \label{eqn:fixed_points_of_dephasing_channel}
    \rho = a \Pi_0 + b \Pi_1 
\end{equation}
with $a + b = 1$ is the general mixed-state density operator fixed point. In contrast, let us now consider the robust points of the same dephasing channel, which satisfy
\begin{equation} \label{eqn:robust_condition_for_dephasing}
    \yhat[ \deph (\rho) ] = \yhat [\rho]
\end{equation}
instead of~\eqref{eqn:fixed_point_definition}. Certainly the state in~\eqref{eqn:fixed_points_of_dephasing_channel} will satisfy~\eqref{eqn:robust_condition_for_dephasing} --- i.e., any fixed point is a robust point --- but the set of robust points may be strictly larger. To completely characterise the robust set, we seek the set of $\rho \in \mathcal{S}_2$ such that
\begin{equation} \label{eqn:robust_condition_for_dephasing_inequality1}
    \Tr [ \Pi_0 \rho ] \geq 1/2 \implies \Tr[ \Pi_0 \deph (\rho)] \geq 1 / 2
\end{equation}
and
\begin{equation} \label{eqn:robust_condition_for_dephasing_inequality2}
    \Tr [ \Pi_0 \rho ] < 1/2 \implies \Tr[ \Pi_0 \deph (\rho)] < 1 / 2 .
\end{equation}
Using the simple properties of the trace and Pauli matrices from the beginning of this section, we can write
\begin{equation*}
    \Tr[\Pi_0 \deph(\rho)] = (1 - p) \Tr[\Pi_0 \rho] + p \Tr[\Pi_0 \ZG \rho \ZG] = \Tr[ \Pi_0 \rho ]. 
\end{equation*}
Thus~\eqref{eqn:robust_condition_for_dephasing_inequality1} and~\eqref{eqn:robust_condition_for_dephasing_inequality2} are satisfied for all density operators $\rho \in \mathcal{S}_{2}$.
That is, every data point $\xbs \in \mathcal{X}$ is a robust point of the dephasing channel (independent of the encoding) for the quantum classifier in~\defref{def:binary_quantum_classifier}. 

Consider now an amplitude damping channel~\eqref{eqn:amp_damp_channel} with $p = 1$, for which the only fixed point is the pure state $\Pi_0$. By expanding the decision rule~\eqref{eqn:decision_boundary_defining_rule} for a state under the amplitude damping channel as:
\begin{equation*}
    \Tr [ \Pi_0 \damp (\rho) ] = (1 - p) \Tr [ \Pi_0 \rho ] + p ,
\end{equation*}
we see that a robust point $\sigma$ must satisfy $\Tr[\Pi_0 \sigma] = 1$. That is, the only robust point is $\Pi_0$, and in this case the set of robust points is identical to the set of fixed points.

As expected from \eqref{eqn:robust_point_definition} and \eqref{eqn:fixed_point_definition}, these examples confirm that
\begin{equation} \label{eqn:fixed_points_are_subset_of_robust_points}
    \mathcal{F}(\mathcal{E}) \subseteq \mathcal{R}(E, \mathcal{E}, \hat{y}) 
\end{equation}
where $\mathcal{F}(\mathcal{E})$ denotes the set of fixed points of $\mathcal{E}$. In~\secref{subsec:existence_of_robust_encodings}, we use this connection to generalise the above discussion and prove the existence of robust data encodings.

\newpage
\subsection[\texorpdfstring{\color{black}}{} Robustness results]{Robustness results} \label{subsec:robustness_results}

Regarding analytic results for \emph{robustness}, we first demonstrate results for single qubit classifiers with several channels in~\secref{subsec:single-qubit-classifier_pauli} and~\secref{subsec:single-qubit-classifier_amplitude}. Then, we state and prove a robustness result for multi-qubit classifiers with (global) depolarising noise in \secref{subsec:multi-qubit-classifier}.

\subsubsection[\texorpdfstring{\color{black}}{} Single qubit classifier - Pauli noise]{Single qubit classifier - Pauli noise}
\label{subsec:single-qubit-classifier_pauli}

First, we consider when robustness can be achieved for a Pauli channel.
\begin{thmbox}
\begin{theorem} \label{thm:robustness_pauli_noise_xy}
    Let $\paul$ be a Pauli channel~\eqref{eqn:pauli_channel} and consider a quantum classifier on data from the set $\mathcal{X}$. Then, for any encoding $E: \mathcal{X} \rightarrow \mathcal{S}_2$, we have complete robustness
    \begin{equation*}
        \mathcal{R} (\paul, E, \yhat) = \mathcal{X}
    \end{equation*}
    if $p_X + p_Y \leq1/2$. (Recall that $\vec{p} := [p_{\mathds{1}}, p_X, p_Y, p_Z]$).
\end{theorem}
\end{thmbox}

\begin{proof}
    The predicted label in the noisy case is identical to \eqref{eqn:decision_rule} with $\rhotildex$ replaced by $\paul ( \rhotildex )$. That is,
    \begin{equation*} 
        \yhat [ \paul (\rhotildex) ] = \begin{cases}
            0 \qquad \text{if   } \Tr [ \Pi_0 \paul( \rhotildex )] \geq 1 / 2 \\
            1 \qquad \text{otherwise} 
        \end{cases} .
    \end{equation*}
    By definition~\eqref{eqn:pauli_channel}, we have
    \begin{equation}
    \label{eqn:pauli_proof1}
            \Tr [ \Pi_0  \paul( \rhotildex )] = p_{\mathds{1}} \Tr[ \Pi_0 \rhotildex] + p_X \Tr [ \Pi_0 \XG \rhotildex \XG]  + p_Y \Tr [ \Pi_0 \YG \rhotildex \YG] + p_Z \Tr [ \Pi_0 \ZG \rhotildex \ZG]
    \end{equation}
    Using the straightforward expressions given at the start of this section, we may write~\eqref{eqn:pauli_proof1} as:
    \begin{equation*}
        \Tr [ \Pi_0  \paul( \rhotildex )] = (p_{\mathds{1}} + p_Z) \Tr [ \Pi_0 \rhotildex ] + (p_X + p_Y) \Tr[ \Pi_1 \rhotildex ]
    \end{equation*}
    By resolution of the identity, $1 = \Tr [ \rhotildex ] = \Tr[\Pi_0 \rhotildex] + \Tr[ \Pi_1 \rhotildex ]$, we come to the simplified expression:
    \begin{equation*}
        \Tr [ \Pi_0  \paul( \rhotildex )] = \left[ 1 - 2 \nu \right] \Tr [ \Pi_0 \rhotildex ] + \nu
    \end{equation*}
    where $\nu := p_X + p_Y$. 
    
    Suppose the noiseless classification is $\yhat = 0$ so that $\Tr[ \Pi_0 \rhotildex ] \geq 1 / 2$. Since $\nu \leq1/2$, we have
    \begin{equation*}
        \Tr [ \Pi_0  \paul( \rhotildex )] \geq \left[ 1 - 2 \nu \right] \frac{1}{2} + \nu  = \frac{1}{2}
    \end{equation*}
    Hence, classification of data points with label $\yhat = 0$ is robust for any encoding.
    Suppose the noiseless classification is $\yhat = 1$ so that $\Tr[ \Pi_0 \rhotildex ] < 1 / 2$. Since $\nu \leq1/2$, we have
    \begin{equation*}
        \Tr [ \Pi_0  \paul( \rhotildex )] < \left[ 1 - 2 \nu \right] \frac{1}{2} + \nu = \frac{1}{2}.
    \end{equation*}
    Hence, classification of data points with label $\yhat = 1$ is also robust for any encoding. 
\end{proof}

Returning to the condition, $p_X + p_Y \leq 1/2$, one can imagine a NISQ computer in which either $p_X$ or $p_Y$ were large enough such that this condition is not satisfied. In this regard, we note two things. First, if this condition is \emph{not} satisfied, then not every encoding strategy will be robust to the Pauli channel in this model. In particular, the set of robust points will now be \textit{dependent} on the encoding strategy. This is similar to the behaviour of the amplitude damping channel which we illustrate in \secref{sec:classifier/numerical_results}. 

Second, the requirement $p_X + p_Y \leq 1/2$ appears because the decision rule uses a measurement in the computational basis. We note that if we measure in the Hadamard basis (see~\thmref{corr:pauli_x_robustness}) then we get a modified robustness condition, namely $p_Y + p_Z \leq 1/2$. A similar conclusion can be drawn from measurements in the Pauli-$\YG$ basis. These results suggest that device-specific encoding strategies and decision rules may be important for achieving robustness in practice on NISQ computers.

\begin{figure}[!ht]
\centering
    \includegraphics[width = 0.65\columnwidth, height=0.4\columnwidth]{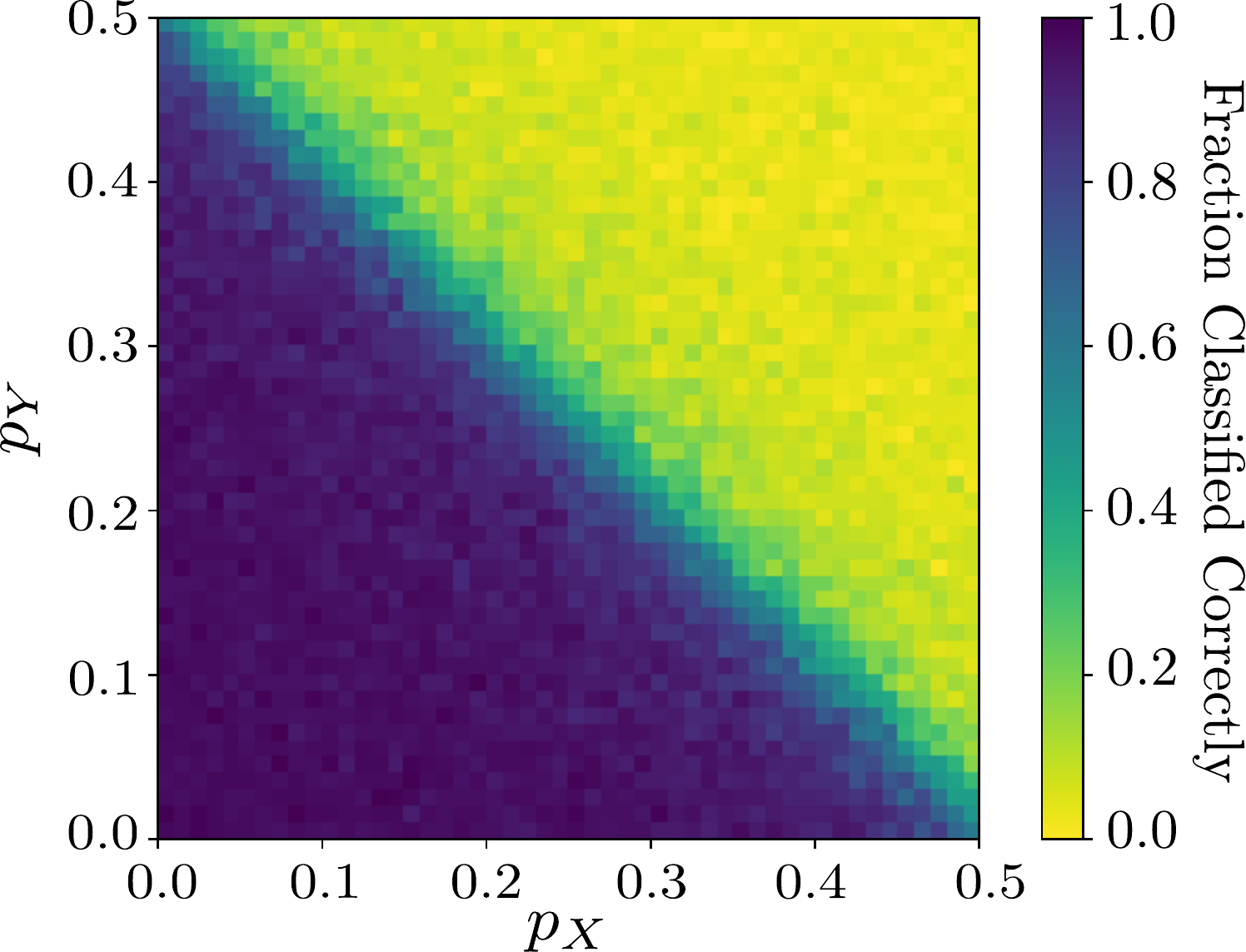}
        \caption[\color{black} Misclassification percentage as a result of Pauli noise.]{
        \textbf{Misclassification percentage as a result of Pauli noise with strengths $\{p_X, p_Y, p_Z = 0\}$.} As expected, classification is robust for $p_X + p_Y < 1/2$ based on Theorem \ref{thm:robustness_pauli_noise_xy}, and a sharp transition occurs when this constraint is violated to give maximal misclassification. By classified `correctly' in this context, we mean the fraction of points which are classified \emph{the same} with and without noise.  
        }
\label{fig:pauli_noise_results}
\end{figure}

To illustrate the results of \thmref{thm:robustness_pauli_noise_xy}, we focus on the dense angle encoding (\defref{def:dense_angle_encoding}), and a dataset for which the $\DAE$ can achieve nearly $100\%$ accuracy  (specifically the ``vertical'' dataset - see \secref{sec:classifier/numerical_results}). We then compute the percentage which would be misclassified as a function of and Pauli noise parameters, $p_X$, $p_Y$. The results are seen in \figref{fig:pauli_noise_results}. We note here, that for values of $p_X + p_Y > 1/2$, one has two strategies to achieve robustness. The first is to adjust the measurement basis as per the discussion of the previous paragraph and requires changing the model itself. Alternatively, one can apply an extra step of post processing and relabel every output `$\yhat = 0$' to `$\yhat = 1$', and vice versa.

Three robustness results for different channels can be shown as corollaries of~\thmref{thm:robustness_pauli_noise_xy} as follows. 
\begin{enumerate}
    \item By setting $p_X = p_Y = 0$ in~\thmref{thm:robustness_pauli_noise_xy}, it follows that $\mathcal{R} (\deph, E, \yhat) = \mathcal{X}$. That is, all encodings are unconditionally robust to dephasing errors:
    \begin{corrbox}
    \begin{corollary} \label{thm:robustness_single_qubit_dephasing_noise}
        Let $\deph$ be a dephasing channel~\eqref{eqn:dephasing_channel}, and consider a quantum classifier on data from the set $\mathcal{X}$. Then, for any encoding $E: \mathcal{X} \rightarrow \mathcal{S}_2$,
        \begin{equation*}
            \mathcal{R} (\deph, E, \yhat) = \mathcal{X}
        \end{equation*}
    \end{corollary}
    \end{corrbox}
    \item By setting $p_Z = p_Y = 0$ and using a modified decision rule which measures in the Hadamard basis, denoted $\hat{z}$, it follows that $\mathcal{R} (\flip, E, \hat{z}) = \mathcal{X}$. That is, all encodings are unconditionally robust to bit-flip errors (using a modified decision rule). We first have the more general theorem, from which the above statement follows immediately ($\Pi_+ := \ketbra{+}{+}$):
    \begin{thmbox}
    \begin{theorem} \label{corr:pauli_x_robustness}
        Consider a quantum classifier on data from the set $\mathcal{X}$ with modified decision rule:
        \begin{equation} \label{eqn:classification_scheme_measure_hadamard_basis}
            \hat{z} [ \rhotildex ] = \begin{cases}
                0 \qquad \text{if   } \Tr [ \Pi_+ \rhotildex ] \geq 1 / 2 \\
                1 \qquad \text{otherwise} 
            \end{cases}
        \end{equation}
        Then, for any $E: \mathcal{X} \rightarrow \mathcal{S}_2$, with a Pauli channel $\paul$ such that $p_Y + p_Z \leq 1/2$, we have:
        \begin{equation*}
            \mathcal{R} (\paul, E, \hat{z}) = \mathcal{X}
        \end{equation*}
    \end{theorem}
    \end{thmbox}
    
    We can now specialise~\thmref{corr:pauli_x_robustness} to the case $p_Z = p_Y = 0$:
    \begin{corrbox}
    \begin{corollary} \label{corr:bit_flip_robustness}
        Consider a quantum classifier on data from the set $\mathcal{X}$ with modified decision rule $\hat{z}$ defined in~\eqref{eqn:classification_scheme_measure_hadamard_basis}. Then, for any encoding $E: \mathcal{X} \rightarrow \mathcal{S}_2$,
        \begin{equation*}
            \mathcal{R} (\flip, E, \hat{z}) = \mathcal{X} .
        \end{equation*}
    \end{corollary}
    \end{corrbox}
    \item By setting $p_X = p_Y = p_Z = p / 4$ and  $p_{\mathds{1}} := (1 - 3p/4)$, we then have that $\mathcal{R} (\depo, E, \yhat) = \mathcal{X}$. That is, all encodings are unconditionally robust to depolarising noise:
    \begin{corrbox}
    \begin{corollary} \label{corr:robustness_to_depolarizing_on_unitary_ansatz_single_qubit}
        Let $\depo$ be a depolarising channel~\eqref{eqn:depolarizing_channel}, and consider a quantum classifier on data from the set $\mathcal{X}$. Then, for any encoding $E: \mathcal{X} \rightarrow \mathcal{S}_2$:
        \begin{equation*}
            \mathcal{R} (\depo, E, \yhat) = \mathcal{X} 
        \end{equation*}
    \end{corollary}
    \end{corrbox}
\end{enumerate}
Interestingly, we remark that fact (3) holds with measurements in \emph{any} basis, not just the computational basis. In~\secref{subsec:multi-qubit-classifier}, we generalise this result to multi-qubit classifiers and as well as to noisy data encodings~\eqref{eqn:robust_point_noisy_encoding}.

\subsubsection[\texorpdfstring{\color{black}}{} Single qubit classifier - amplitude damping]{Single qubit classifier - amplitude damping}
\label{subsec:single-qubit-classifier_amplitude}

Let us now move to amplitude damping noise, for which the robust set $\mathcal{R}$ depends on the encoding $E$. From the channel definition~\eqref{eqn:amp_damp_channel}, it is straightforward to see that:
\begin{equation} \label{eqn:amp_damp_simple_expansion}
    \Tr [ \Pi_0 \damp ( \rhotildex )] = \Tr [ \Pi_0 \rhotildex ] + p \Tr [ \Pi_1 \rhotildex ]
\end{equation}
Suppose first that the noiseless prediction is $\yhat = 0$ so that $\Tr [ \Pi_0 \rhotildex ] \geq 1/2$. Then, certainly $\Tr [ \Pi_0 \damp ( \rhotildex )] \geq 1/2$ because $p \geq 0$ and $\Tr [ \Pi_1 \rhotildex ] \geq 0$. Thus, the noisy prediction is always identical to the noiseless prediction when the noiseless prediction is $\yhat = 0$. This can be understood intuitively because an amplitude damping channel only increases the probability of the ground state. Now, suppose that the noiseless prediction is $\yhat = 1$. From~\eqref{eqn:amp_damp_simple_expansion}, we require:
\begin{equation*}
    \Tr [ \Pi_0 \damp ( \rhotildex )] = \Tr [ \Pi_0 \rhotildex ] + p \Tr [ \Pi_1 \rhotildex ] < 1/2
\end{equation*}

to achieve robustness. Using the resolution of the identity, $\Tr[ \Pi_1 \rhotildex] = 1 - \Tr [ \Pi_0 \rhotildex]$, we arrive at the condition:

\begin{equation} \label{eqn:amp_damp_robustness_condition}
    \Tr [ \Pi_1 \rhotildex ] > \frac{1}{2(1 - p)}
\end{equation}
Let $\rhox$ be given by the general qubit encoding~\eqref{eqn:qubit_encoding_general} so that~\eqref{eqn:amp_damp_robustness_condition} can be written (see~\eqref{eqn:useful_11mtx_elt})
\begin{equation*}
    |U_{10}|^2 f^2 + 2 \Re [ U_{11}^* U_{10} f g^*] + |U_{11}|^2 |g|^2 > \frac{1}{2(1 - p)}
\end{equation*}

where $U_{ij}$ denote the optimal unitary matrix elements. Formally, we then have:
\begin{thmbox}
\begin{theorem} \label{thm:robustness_amp_damp}
    Consider a quantum classifier on data from the set $\mathcal{X}$, and let $\damp$ denote the amplitude damping channel~\eqref{eqn:amp_damp_channel}. Then, for any qubit encoding $E$ defined in~\eqref{eqn:qubit_encoding_general} which satisfies
    \begin{equation}\label{eqn:amp_damp_robustness_condition_expanded_unitary}
        |U_{10}|^2 f^2 + 2 \Re [ U_{11}^* U_{10} f g^*] + |U_{11}|^2 |g|^2 > \frac{1}{2(1 - p)} ,
    \end{equation}
    we have
    \begin{equation*}
        \mathcal{R} (\damp, E, \yhat) = \mathcal{X} .
    \end{equation*}
    If $E$ is not completely robust, the set of points $\xbs$ such that~\eqref{eqn:amp_damp_robustness_condition_expanded_unitary} holds define the partially robust set.
\end{theorem}
\end{thmbox}
%
We note that~\eqref{eqn:amp_damp_robustness_condition_expanded_unitary} depends on the optimal unitary $U$ as well as the encoding $E$. This is expected as the final state $\rhotildex$ has been processed by the PQC. In practice, since we do not know the optimal unitary parameters \textit{a priori}, it remains a question of how large the (partially) robust set will be for a given an encoding. To address this point, we discuss in~\secref{ssec:encoding_learn_alg} how training over hyperparameters in the encoding function can help find the robust region even after application of the unknown optimal unitary. Additionally, in the next Section we discuss whether we can find an encoding which satisfies~\eqref{eqn:amp_damp_robustness_condition_expanded_unitary}, or more generally whether a robust encoding exists for a given channel. 

Given the robustness condition~\eqref{eqn:amp_damp_robustness_condition_expanded_unitary} for the amplitude damping channel, it is natural to ask whether such an encoding exists. In~\secref{subsec:existence_of_robust_encodings}, we show the answer is yes by demonstrating there always exists a robust encoding for any trace preserving quantum operation. This encoding may be trivial, which leads to the idea of a tradeoff between learnability and robustness.

\subsubsection[\texorpdfstring{\color{black}}{} Multi-qubit classifier]{Multi-qubit classifier}
\label{subsec:multi-qubit-classifier}

We now consider global depolarising noise on a multi-qubit classifier. It turns out that any encoding is completely robust to this channel applied at any point throughout the circuit. To clearly state the theorem, we introduce the following notation. First, let 
\begin{equation} \label{eqn:short_notation_depo_channel}
    \E_{p_i} (\rho) = p_i \rho + (1 - p_i) \frac{\mathds{1}_d}{d} 
\end{equation}
be shorthand for a global depolarising channel with probability $p_i$. (Note $p_i$ and $1 - p_i$ are intentionally reversed compared to~\defref{def:global_depolarizing_channel} to simplify the proof.) Then, let
\begin{equation} \label{eqn:depo_robustness_notation_of_unitary_noise_application}
    \tilde{\rho}_{\xbs}^{(m)} \equiv \left[ \prod_{j = 1}^{J} U_i \circ \E_{p_i} \right] \circ \rhox
\end{equation}
denote the state of the encoded point $\rhox$ after $J$ applications of a global depolarising channel and unitary channel. For instance, $J = 1$ corresponds to 
\begin{equation*}
    U_1 \circ \E_{p_1} \circ \rhox \equiv U_1( \E_{p_1} ( \rhox ) )
\end{equation*}
and $m = 2$ corresponds to
\begin{equation*}
    U_2 \circ \E_{p_2} \circ U_1 \circ \E_{p_1} \circ \rhox \equiv U_2 ( \E_{p_2} ( U_1( \E_{p_1} ( \rhox ) ) ) ) 
\end{equation*}
We remark that $U_i$ can denote any unitary in the circuit.

With this notation, we state the theorem as follows.

\begin{thmbox}
\begin{theorem} \label{thm:robustness_to_depolarizing_on_unitary_ansatz_multiple_qubits}
    Consider a quantum classifier on data from the set $\mathcal{X}$ with decision rule $\hat{y}$ defined in \eqref{eqn:classification_scheme_measure_hadamard_basis}. Then, for any encoding $E: \mathcal{X} \rightarrow \mathcal{S}_n$,
    \begin{equation*}
        \mathcal{R} \left(\depon , E, \hat{y} \right) = \mathcal{X}
    \end{equation*}
    where $\depon$ denotes the composition of global depolarising noise acting at any point in the circuit --- i.e. such that the final state of the classifier is given by~\eqref{eqn:depo_robustness_notation_of_unitary_noise_application}. 
\end{theorem}
\end{thmbox}
To prove~\thmref{thm:robustness_to_depolarizing_on_unitary_ansatz_multiple_qubits}, we use the following lemma.
\begin{lembox}
\begin{lemma} \label{lem:lemma_for_global_depo_proof}
    The state in~\eqref{eqn:depo_robustness_notation_of_unitary_noise_application} can be written as (adapted from~\cite{sharma_noise_2020})
    \begin{equation} \label{eqn:lemma_for_global_depo_proof}
        \tilde{\rho}_{\xbs}^{(J)} = 
        \prod_{j = 1}^{J} p_i U_J \cdots U_1 \rhox U_1^\dagger \cdots U_J^\dagger + \left( 1 - \prod_{j = 1}^{J} p_j \right) \frac{\mathds{1}_d}{d}
    \end{equation}
    where $d = 2^n$ is the dimension of the Hilbert space. 
\end{lemma}
\end{lembox}
\begin{proof}
    Using the definition of the global depolarising channel~\eqref{eqn:short_notation_depo_channel}, it is straightforward to evaluate (with $\mathds{1} \equiv \mathds{1}_d$):
    \begin{equation*}
        \tilde{\rho}_{\xbs}^{(1)} = U_1 \circ \E_{p_1} \circ \rhox = p_1 U_1 \rhox U_1^\dagger + (1 - p_1)\frac{\mathds{1}}{d}
    \end{equation*}
    Thus~\eqref{eqn:lemma_for_global_depo_proof} is true for $J = 1$. Assume~\eqref{eqn:lemma_for_global_depo_proof} holds for $J = k$. Then, for $k + 1$ we have
    \begin{align*}
        \tilde{\rho}_{\xbs}^{(k + 1)} &= U_{k + 1} \circ \E_{p_{k + 1}} \circ \tilde{\rho}_{\xbs}^{(k)} 
        = p_{k + 1} U_{k + 1} \tilde{\rho}_{\xbs}^{(k)} U_{k + 1}^\dagger + (1 - p_{k + 1})\frac{\mathds{1}}{d}
    \end{align*}
    The last line can be simplified to arrive at
    \begin{equation}
        \tilde{\rho}_{\xbs}^{(k + 1)} = 
        \prod_{i = 1}^{k + 1} p_i U_{k + 1} \cdots U_1 \rhox U_1^\dagger \cdots U_{k + 1}^\dagger \\ 
        + \left( 1 - \prod_{i = 1}^{k + 1} p_i \right) \frac{\mathds{1}}{d} 
        \end{equation}
    which completes the proof.
\end{proof}

We can now prove~\thmref{thm:robustness_to_depolarizing_on_unitary_ansatz_multiple_qubits} as follows. Let $l$ denote the total number of alternating unitary gates with depolarising noise in the classifier circuit so that \eqref{eqn:lemma_for_global_depo_proof} can be written
\begin{equation} \label{eqn:final_state_global_depo_circuit}
    \tilde{\rho}_{\xbs}^{(l)} = \bar{p} \rhotildex + (1 - \bar{p}) \frac{\mathds{1}}{d}.
\end{equation}
Here, we have defined $\bar{p} := \prod_{i = 1}^{l} p_i$ and noted that $U_l \cdots U_1 \rhox U_1^\dagger \cdots U_l^\dagger = \rhotildex$ is the final state of the noiseless circuit before measuring. \eqref{eqn:final_state_global_depo_circuit} is thus the final state of the noisy circuit before measuring. Then:
\begin{equation} \label{eqn:n_qubit_depolarising_noise_expression}
    \Tr [ \Pi_0 \tilde{\rho}_{\xbs}^{(l)} ] = \bar{p} \Tr[ \Pi_0 \rhotildex ] + \frac{(1 - \bar{p})}{2}
\end{equation}
using $\Tr[ \Pi_0 \mathds{I}_d ] = 2^{d - 1}$. 
To prove robustness, suppose that $\yhat [ \rhotildex ] = 0$ so that $\Tr[ \Pi_0 \rhotildex ] \geq 1/2$. Then,
\begin{equation*}
    \Tr [ \Pi_0 \tilde{\rho}_{\xbs}^{(l)} ] \geq \frac{\bar{p}}{2} + \frac{(1 - \bar{p})}{2} = \frac{1}{2}
\end{equation*}
so that $\yhat [ \tilde{\rho}_{\xbs}^{(l)} ] = 0$. Similarly for the case $\yhat[ \rhotildex ] = 1$, which completes the proof of~\thmref{thm:robustness_to_depolarizing_on_unitary_ansatz_multiple_qubits}.

Thus, any encoding strategy exhibits complete robustness to global depolarising noise. We remark again that our definition of robustness (\defref{def:robust_point}) is in terms of probabilities, meaning that more measurements for sampling may be required to reliably evaluate robustness. With this remark, we note an interesting connection to explain a phenomenon observed in recent literature: the authors of \cite{grant_hierarchical_2018}, found that classification accuracy decreased under the presence of depolarising noise.
~\thmref{thm:robustness_to_depolarizing_on_unitary_ansatz_multiple_qubits} implies this was exclusively due to the finite shot noise used to obtain the predicted label. We discuss errors due to finite sampling in more detail in \secref{ssec:finite_sampling_error}.

While global depolarising noise admits a clean robustness result for an arbitrary $d$-dimensional circuit, general channels can lead to complicated equations which are best handled numerically. We include several numerical results in~\secref{sec:classifier/numerical_results}, and we discuss avenues for proving more analytical results with certain classes of channels in future work in~\secref{sec:classifier/conclusions}. To close the present discussion, we highlight the special case of multi-qubit classifiers with ``factorisable noise'', for which it is straightforward to apply previous results proved in this section.

In particular, suppose that $\E : \mathcal{S}_n \rightarrow \mathcal{S}_n$ is a noise channel which factorises into single qubit channels, e.g.
\begin{equation} \label{eqn:factorizable_noise_channel_into_single_qubits_definition}
    \E = \E_1 \otimes \cdots \otimes \E_n
\end{equation}
where $\E_i : \mathcal{S}_2 \rightarrow \mathcal{S}_2$ for $i \in [n]$. Without loss of generality, let the classification qubit be the $n^{\text{th}}$ qubit. Then, if the processed state of the classification qubit is robust to the channel $\E_n$, the encoded state will be robust to the entire channel $\E$ in~\eqref{eqn:factorizable_noise_channel_into_single_qubits_definition}. Specifically, we have the following (proof in~\appref{app_ssec:noise_robustness}):
\begin{thmbox}
\begin{theorem}\label{thm:factorizable_noise_before_meas}
If $\E$ is any noise channel which factorises into a single qubit channel, and a multiqubit channel as follows:
\begin{equation*}
    \E(\rho) = \E_{\Bar{c}}(\rhotildex^{\Bar{c}}) \otimes \E_c(\rhotildex^c)
\end{equation*}
where WLOG $\E_c$ acts only on the classification qubit $(\rhotildex^c = \Tr_{\Bar{c}}(\rhotildex))$ after encoding and unitary evolution, and $\E_{\Bar{c}}$ acts on all other qubits arbitrarily, $(\rhotildex^{\Bar{c}} = \Tr_c(\rhotildex))$. Further assume the state meets the robust classification requirements for the single qubit error channel $\E_c$. Then the classifier will be robust to $\E$.
\end{theorem}
\end{thmbox}
The above theorem is a simple consequence of causality in the circuit, only errors which have to happen before the measurement can corrupt the outcome. As such, outside of single qubit errors, we only need to consider errors before the measurement which specifically involve the classification qubit. This result also holds for general $n - 1$ qubit channels which act on every qubit except the classification qubit. Although this is relatively straightforward, the result could be used as a building block to better understand more intricate robustness properties of quantum classifiers.


\subsubsection[\texorpdfstring{\color{black}}{} Robustness to measurement noise]{Robustness to measurement noise} \label{app_ssec:meas_noise}

Just as the case of quantum compilation~\cite{sharma_noise_2020}, we can deal with measurement noise in the classifier:

\begin{defbox}
\begin{definition}[Measurement noise]\label{defn:measurement_noise}~ \\
\emph{Measurement noise} is defined as a modification of the standard POVM basis elements, $\{\Pi_0 = \ketbra{0}{0}, \Pi_1 = \ketbra{1}{1}\}$ by the channel $\mathcal{E}_{\vec{p}}^{\textnormal{meas}}$ with assignment matrix $\vec{p}$ for a single noiseless qubit:
\begin{align}
    \Pi_0 &= \ketbra{0}{0} \overset{\mathcal{E}_{\vec{p}}^{\textnormal{meas}}}{\rightarrow} \tilde{\Pi}_0 = p_{00}\ketbra{0}{0} + p_{01}\ketbra{1}{1}\nonumber\\
    \Pi_1 &= \ketbra{1}{1} \overset{\mathcal{E}_{\vec{p}}^{\textnormal{meas}}}{\rightarrow} \tilde{\Pi}_1 = p_{10}\ketbra{0}{0} + p_{11}\ketbra{1}{1} \label{eqn:noisy_povm_elements}\\
    \vec{p}& :=  \left(\begin{array}{cc}
         p_{00} & p_{01}  \\
         p_{10} & p_{11} 
    \end{array}\right)\nonumber
\end{align}
where $p_{00} + p_{10} = 1, p_{10} +p_{11} = 1$, and hence $p_{kl}$ is the probability of getting the $k$ outcome given the input $l$. Furthermore, we assume that $p_{kk} > p_{kl}$ for $k \neq l$.
\end{definition}
\end{defbox}

The definition for the general case of $n$ qubit measurements can be found in \cite{sharma_noise_2020}, but we shall not need it here, since we only require measuring a single qubit to determine the decision function. More general classifiers which measure multiple qubits (e.g. and then take a majority vote for the classification) could also be considered, but these are outside the scope of this work. Now, we can show the following result in a similar fashion to the above proofs (explicit proof in~\appref{app_ssec:noise_robustness}):

\begin{thmbox}
\begin{theorem} \label{thm:measurement_noise_robustness}
Let $\mathcal{E}_{\vec{p}}^{\textnormal{meas}}$ define measurement noise acting on the classification qubit and consider a quantum classifier on data from the set $\mathcal{X}$. 
Then, for any encoding $E: \mathcal{X} \rightarrow \mathcal{S}_2$, we have complete robustness
    \begin{equation*}
        \mathcal{R} (\mathcal{E}_{\vec{p}}^{\textnormal{meas}}, E, \yhat) = \mathcal{X}
    \end{equation*}
     if the measurement assignment probabilities satisfy $p_{00} > p_{01}, p_{11} > p_{10}$.
\end{theorem}
\end{thmbox}

Just as above, we can replace the ideal state, $\rhotildexi$ with a noisy state, $\mathcal{E}(\rhotildexi)$, where the operator accounts for other forms of noise, not including measurement noise. We can see this allows us to take a model which is robust without measurement noise, and `upgrade' it to one which is. However, we may be able to find looser restrictions by considering different types of noise together, rather than in this modular fashion.

\begin{figure}
\centering
\includegraphics[width =0.5\columnwidth, height=0.4\columnwidth]{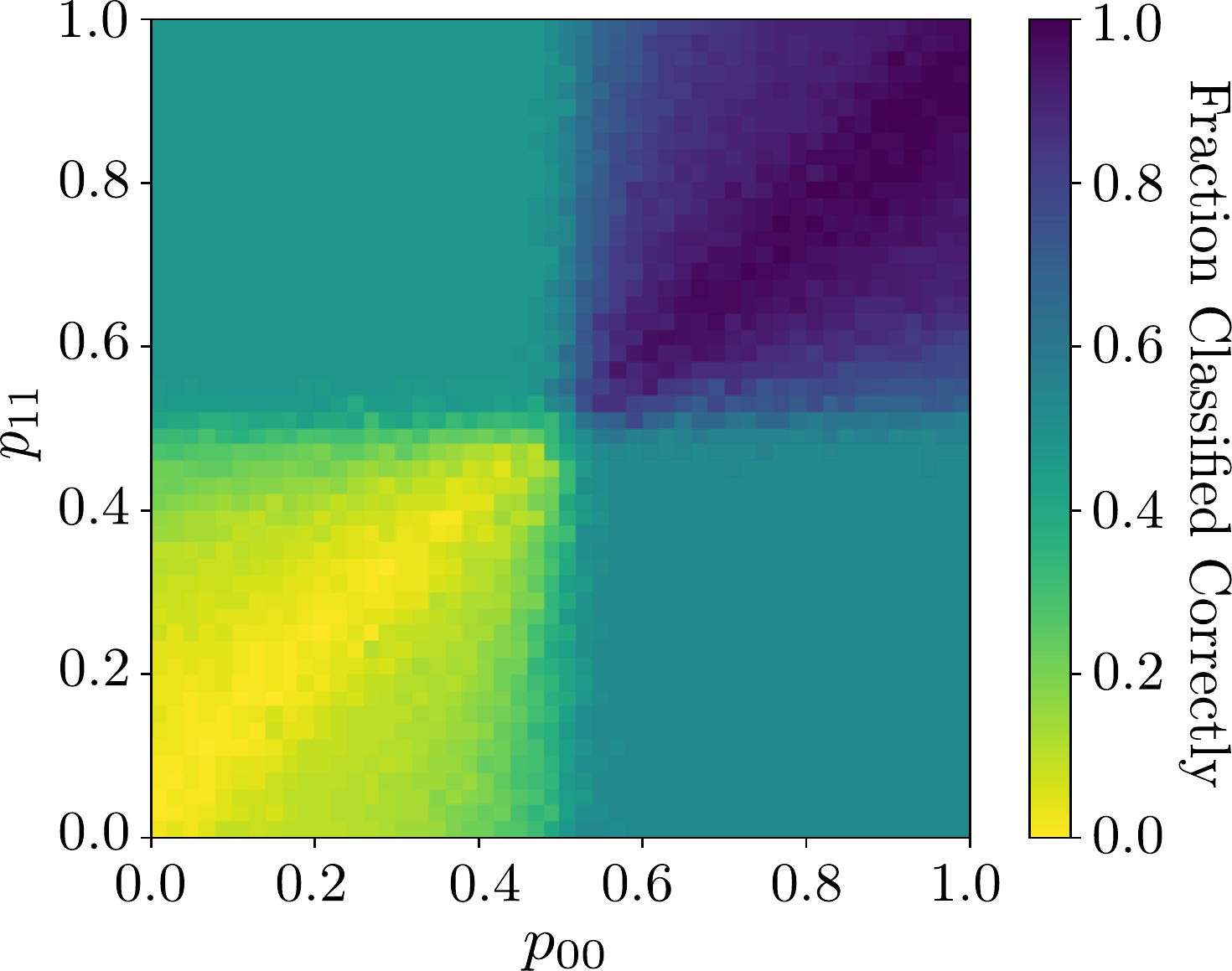}
\caption[\color{black} Misclassification percentage as a result of measurement noise.]{\textbf{Misclassification percentage as a result of measurement noise as a function of probabilities $\{p_{00}, p_{11}\}$.} If either $p_{00}$ or $p_{11}$ is less than $1/2$, then half the correctly classified points will be misclassified, with the probability increasing as expected, with the number of misclassified points increasing as the off diagonal terms, $p_{01}, p_{10} \rightarrow 1$, as expected from~\thmref{thm:measurement_noise_robustness}. By classified `correctly' in this context, we mean the fraction of points which are classified \emph{the same} with and without noise. 
}
\label{fig:measurement_noise_results}
\end{figure}

To illustrate the results of~\thmref{thm:measurement_noise_robustness} we focus on the dense angle encoding, which can achieve nearly $100\%$ accuracy on the ``vertical'' dataset. We then compute the percentage which would be misclassified as a function the assignment probabilities in the noisy projectors in~\eqref{eqn:noisy_povm_elements}. The results are seen in~\figref{fig:measurement_noise_results}.

Of course, all we have simply done here is include the uncertainty from measurement in the definition of the robust point. If one were to do the same analysis for the original definition of a robust point (\defref{def:robust_point}), we would have a robustness result only for those points which are sufficiently far outside the $\epsilon$ region (say for example those points, $\xbs$, for which $\Tr [ \Pi_0^c \rhotildex ] \geq 1/2 + \epsilon$)

\subsection[\texorpdfstring{\color{black}}{} Modifications for finite sampling]{Modifications for finite sampling} \label{ssec:finite_sampling_error}
In practical applications, we of course do not have access to exact probabilities but rather samples from the underlying distribution. Specifically, we sample a finite number $N$ times from the state and get an estimate of the probabilities in terms of recorded frequencies. In this section, we discuss implications of this finite sampling error and show how to modify results from the previous sections to account for finite sampling.

The first modification is to the decision rule~\eqref{eqn:decision_rule}. Here, we replace the true probability $p_{\xbs} := \Tr [ \Pi_0^c \rhotildex ]$ with an unbiased estimator $\bar{p}_{\xbs}(N)$ obtained from $N$ samples. The modified decision rule is then
\begin{equation} \label{eqn:modified-decision-rule-finite-sampling}
    \yhat_N [\rhotildex] = \begin{cases}
        0 \qquad \text{if   } \bar{p}_{\xbs} (N) \geq 1 / 2 \\
        1 \qquad \text{otherwise} 
    \end{cases} .
\end{equation}
Suppose we find that $\bar{p}_{\xbs} (N) \geq 1/2$. Then, by Hoeffding's inequality
\begin{equation}\label{eqn:hoeffding_sampling_error}
    \Pr\left(|\bar{p}_{\xbs}(N) - \pzero| \leq \epsilon\right) \geq 1 - \gamma 
\end{equation}
we have:

\begin{align}
  &|\pzero - \bar{p}_{\xbs}(N)| \leq \epsilon \\
 \implies &-\epsilon \leq \pzero - \frac{1}{2}  \leq \epsilon \\
 \implies &\frac{1}{2} - \epsilon \leq \pzero \leq \frac{1}{2} + \epsilon
\end{align}
Therefore (taking the left hand inequality above), for any $\epsilon > 0$, we know that $p_{\xbs} \geq 1/2 - \epsilon$ with probability at least $1-\gamma$ (we refer to $1 - \gamma$ as the confidence) if we take:
\begin{equation} \label{eqn:N-epsilon-gamma}
    N \geq N_c (\epsilon, \gamma) := \frac{1}{(2 \epsilon^2)}\log \frac{1}{\gamma} 
\end{equation}
samples from $p_{\xbs}$. This follows since each measurement of the observable gives a Bernoulli random variable with expectation $\Tr [ \Pi_0^c \rhotildex]$.

We note that it is in principle \#P-hard to exactly determine the predicted class label since points could be within any $\epsilon > 0$ from the decision boundary. In practice we do not expect this to be a limiting issue aside from pathological examples, but it is important to note. One could introduce a penalty term into the cost function such that the distance $\epsilon$ between the boundary and the closest point is maximised (as in support vector machines) to mitigate this effect.

Now, to begin the discussion of robustness in this context, we first note that clearly, finite sampling effects can be treated as a form of noise. If we consider $N\rightarrow \infty$ in \eqref{eqn:N-epsilon-gamma}, we recover the prediction of the ideal channel, $\yhat_{N\rightarrow \infty} [\rhotildex] = \yhat[\rhotildex]$. Then, if $N$ being finite (and fixed) is treated as a noise channel so, $\yhat_N [\rhotildex] = \yhat[\mathcal{E}^{\textnormal{samp}}(\rhotildex)]$, all points $\xbs$ for which $\yhat[\rhotildex] \geq 1/2 + \epsilon$ (assuming the label is $0$) will be robust against $\mathcal{E}_N^{\textnormal{samp}}$.

Now, unfortunately those $\xbs \in \X$ for which $1/2 \leq\yhat[\rhotildex]\leq 1/2 + \epsilon$ may or may not be in the robust set for the `finite sampling channel', $\mathcal{R} (\mathcal{E}_N^{\textnormal{samp}}, E, \yhat)$.

Let $\mathcal{E}_{N_1}^{\textnormal{samp}}$ be the noise channel induced by a certain number of samples, $N_1$, such that a particular classifier does not have the entirety of the dataset in the robust set, $\mathcal{R} (\mathcal{E}_{N_1}^{\textnormal{samp}}, E, \yhat) \neq \mathcal{X}$. In order to make a classifier for which \emph{all} points in the dataset are in the robust set, we simply need to increase the number of samples to some sufficiently large value, $N_2 > N_1$, and we shall recover complete robustness, i.e. we have $\mathcal{R} (\mathcal{E}_{N_2}^{\textnormal{samp}}, E, \yhat) = \mathcal{X}$. However, this comes at a caveat that $N_2$ may not be a polynomial function of the problem parameters. We leave the investigation of the value $N_2$ should be for a particular problem (as it will in general be highly problem- and dataset-dependent) to future work.

Now, in the above discussions we have taken care of the cases `ideal classifier $\rightarrow$ noisy classifier' and `ideal classifier $\rightarrow$ finite sampling classifier'. The final case to deal with is `finite sampling classifier $\rightarrow$ finite sampling \emph{noisy} classifier'. Coupled with the discussion of the number of shots required for finite-sampling robustness, this final step will allow us to uplift the robustness argument for any other noise channel into the finite sampling regime, i.e. the case, `ideal classifier $\rightarrow$ finite sampling noisy classifier'. To do so, we present a modified definition of a robust point to account for errors due to finite sampling, specifically one which incorporates the $\epsilon, \delta$ parameters.

\begin{defbox}
    \begin{definition}[$(\epsilon, \gamma)$-Robust point]~ \\
    \label{def:epsilon_robust_point}
        Let $0 < \epsilon < 1 / 2$, $0 < \gamma < 1$, and $\mathcal{E}$ be a quantum channel. We say that the state $\rhox \in \mathcal{S}_n$ encoding a data point $\xbs \in \mathcal{X}$ is an $(\epsilon, \gamma)$-robust point of the quantum classifier if and only if
        \begin{equation} \label{eqn:epsilon-delta-robust-point-definition}
            \yhat_N[ \mathcal{E}( \rhotildex ) ] = \yhat_N[ \rhotildex ]
        \end{equation}
        with confidence $1 - \gamma$. Explicitly, \eqref{eqn:epsilon-delta-robust-point-definition} gives two conditions:
        \begin{align}
            \Tr [ \Pi_0^c \rhotildex ] \geq 1/2 - \epsilon \implies \Tr [ \Pi_0^c \mathcal{E} ( \rhotildex ) ] \geq 1/2 - \epsilon\\
             \Tr [ \Pi_0^c \rhotildex ] < 1/2 + \epsilon \implies \Tr [ \Pi_0^c \mathcal{E} ( \rhotildex ) ] < 1/2 + \epsilon
        \end{align}
        with confidence $1 - \gamma$. This is achieved using $N \geq N_c(\epsilon, \gamma) := 1 / (2 \epsilon^2)\log 1 / \gamma $ measurements in the decision rule~\eqref{eqn:modified-decision-rule-finite-sampling}.
    \end{definition}
\end{defbox}

Procedurally, we can think of this definition as follows. We are provided with an $0 < \epsilon < 1/2$ which corresponds to how close to the boundary we want to be able to distinguish robust points (the farthest we can be from the boundary is $1/2$), and a $0 < \gamma < 1$ which corresponds to how certain we want to be about points within this region. Every point $\sigma$ which satisfies
\begin{equation} \label{eqn:epsilon-region}
    \left| \Tr [ \sigma \Pi_0^c] - 1 / 2 \right| < \epsilon
\end{equation}
is classified correctly with confidence $1 - \gamma$ using at least $N_c(\epsilon, \gamma)$ measurements. Here, $\sigma$ could be either $\rhotildex$ or $\mathcal{E}(\rhotildex)$ as in~\eqref{eqn:epsilon-delta-robust-point-definition}. 

For a specific example of how finite sampling considerations may be incorporated into the previous results in a natural way, consider Corollary (3) of~\thmref{thm:robustness_pauli_noise_xy} which states that a single qubit classifier is unconditionally robust to depolarising noise. The modified statement of this theorem which accounts for finite sampling is as follows.
\begin{thmbox}
\begin{theorem} \label{epsilon-gamma-robustness-for-single-qubit-depolarizing-noise}
    Let $\depo$ be a single qubit depolarising channel and consider a quantum classifier on the set $\mathcal{X}$. For any encoding $E : \mathcal{X} \rightarrow \mathcal{S}_2$, let $\xbs \in \mathcal{X}$ and $\epsilon, \gamma$ be such that
    \begin{equation}
        \bar{p}_{\xbs} (N (\epsilon, \gamma)) \geq \frac{1}{2} .
    \end{equation}
    Then, $\rhox = E(\xbs)$ is an $(\epsilon, \gamma)$-robust point.
\end{theorem}
\end{thmbox}
\begin{proof}
    Given $\xbs \in \mathcal{X}$ and $\epsilon, \gamma$ such that
    \begin{equation}
        \bar{p}_{\xbs} (N (\epsilon, \gamma)) \geq 1 / 2 ,
    \end{equation}
    we have $\yhat_N [ \rhotildex ] = 0$ with confidence $1 - \gamma$. This means that
    \begin{equation}
        \Tr [ \Pi_0^c \rhotildex ] \geq 1/2 - \epsilon
    \end{equation}
    with confidence $1 - \gamma$. It follows that
    \begin{align*}
        \Tr [ \Pi_0^c \depo ( \rhotildex )] &= p / 2 + (1 - p) \Tr [ \Pi_0^c \rhotildex ] \\
        &\geq p / 2 + (1 - p) (1 / 2 - \epsilon) \\
        &= 1/2 - \epsilon + p\epsilon\\ 
        &\geq \frac{1}{2} - \epsilon 
    \end{align*}
    with confidence $1 - \gamma$. The analogous result can be shown when $\bar{p}_{\xbs} (N (\epsilon, \gamma)) < 1 / 2$ which completes the proof.
\end{proof}

The other preceding results could similarly be uplifted to incorporate finite sampling in a similar manner.

\subsection[\texorpdfstring{\color{black}}{} Existence of robust encodings]{Existence of robust encodings} \label{subsec:existence_of_robust_encodings}

In~\secref{ssec:characterize_robust_points}, we considered example channels and characterised their robust points and fixed points. We found that the set of fixed points $\mathcal{F}(\mathcal{E})$ is always a subset of the robust set $\mathcal{R}(\mathcal{E}, E, \yhat)$ in~\eqref{eqn:fixed_points_are_subset_of_robust_points}. Here, we use this connection to show that there always exists a robust encoding for a trace-preserving channel $\mathcal{E}$ (regardless of optimal unitary parameters which may appear in the robustness condition, e.g.~\eqref{eqn:amp_damp_robustness_condition_expanded_unitary}).
\begin{thmbox}
    \begin{theorem}[Existence of fixed points~\cite{schauder_fixpunktsatz_1930, nielsen_quantum_2010}]\label{thm:schauder_fixed_point}~ \\
    Any trace-preserving quantum operation has at least one density operator fixed point~\eqref{eqn:fixed_point_definition}.
    \end{theorem}
\end{thmbox}
Using this and the observation that $\mathcal{F}(\mathcal{E}) \subseteq \mathcal{R}(\mathcal{E}, E, \yhat)$, we have the following existence theorem for robust encodings.
\begin{thmbox}
\begin{theorem}
\label{thm:robust_encodings_existence}
    Given a data point $\xbs \in \mathcal{X}$, a trace-preserving quantum channel $\mathcal{E}$, and decision rule $\yhat$ defined in~\eqref{eqn:decision_rule}, there exists an encoding $E$ such that
    \begin{equation} \label{eqn:existenstence_robust_encoding}
        \yhat [ \mathcal{E} ( E(\xbs) ) ] = \yhat[ E(\xbs) ] .
    \end{equation}
\end{theorem}
\end{thmbox}
We note that the optimal unitary of the PQC affects the ``location'' of the robust set, but not the existence.
We emphasise that~\thmref{thm:robust_encodings_existence} is with respect to a single data point $\xbs \in \mathcal{X}$. As mentioned in~\secref{subsec:robustness_definition}, it is more relevant for applications to consider the training set~\eqref{eqn:labeled_data_for_classifier} or entire set $\mathcal{X}$. 

Appropriately, one can ask whether a completely robust encoding (\defref{def:complete_noise_robustness}) exists for a given channel $\E$. This answer also turns out to be yes, but in a potentially trivial way.

In particular, suppose that there is a unique fixed point $\sigma$ of the channel $\E$, e.g. depolarising noise or amplitude damping noise with $p = 1$. Then, consider the encoding
\begin{equation*}
    \E (\xbs) = \sigma 
\end{equation*}
for all $\xbs \in \mathcal{X}$. From a robustness perspective, this has the desirable property of complete robustness. From a machine learning perspective, however, this has very few desirable properties: all training data is mapped to the same point so that it is impossible to successfully train a classifier\footnote{In principle, one can achieve an encoding which is completely robust and able to correctly classify all data if there are at least two orthogonal fixed points in $\mathcal{F}(\E)$. For example, if $\E$ is the bit-flip channel, the encoding $\xbs_i \mapsto \ket{0} + (-1)^{y_i} \ket{1}$ is both completely robust and completely learnable (the optimal unitary is a Hadamard gate), but assumes the true labels $y_i$ are known.}.

The previous example, while extreme, serves to illustrate the tradeoff between learnability (expressive power) and robustness. By expressive power, we mean the ability of the classifier (hypothesis family) to predict correct labels without regard to noise. By robustness, we mean the property of retaining the same label (without regard to correctness) in the presence of noise. These two properties can be schematically connected as below:
\begin{equation*}
    y[\xbs] \ \ \xleftrightarrow{\text{Learnability}} \ \ \yhat [ \rhotildex ] \ \
    \xleftrightarrow{\text{Robustness}} \ \ \yhat [ \E( \rhotildex ) ] 
\end{equation*}

The tradeoff we observe is that it is possible to maintain robustness by means of a certain encodings, but these encodings generally reduce the expressive power of the classifier. Succinctly, more robustness leads to less expressive power, and \textit{vice versa}. We discuss this point more in~\secref{ssec:encoding_learn_alg}. 

\subsection[\texorpdfstring{\color{black}}{} Lower bounds on partial robustness]{Lower bounds on partial robustness} \label{ssec:fidelity_bounds}

In this section, we consider a slightly modified binary quantum classifier which embeds the cost function in the circuit and computes the cost by measuring expectation values. In contrast to the classifier in~\defref{def:binary_quantum_classifier}, the output of this circuit is thus the cost $C$ instead of an individual predicted label $\yhat$. Correspondingly, the input to the circuit is all data points in the training set~\eqref{eqn:labeled_data_for_classifier} (using a ``mixed state encoding'' discussed below) instead of a single data point $\xbs$. Such a classifier was recently introduced
by~\cite{cao_cost_2019} and presents an interesting framework to analyse in the context of noise. In the remainder of this section, we prove a lower bound on the size of the robust set in terms of fidelities between noisy and noiseless states.

Before precisely stating this theorem, we formally define the mixed state encoding and cost function of this modified classifier.
\begin{defbox}
    \begin{definition}[Mixed state encoding]\label{def:mixed_state_encoding}~\\ 
        Let $\dataset$ be a dataset and $E$ be an encoding. For each feature vector $\xbs_i$, let
        \begin{equation} \label{eqn:pure_state_of_mixed_state_encoding_for_embedded_cost_classifier}
        \sigmaxi := E(\xbs_i) \otimes \ketbra{y_i}{y_i} = \rhoxi \otimes \ketbra{y_i}{y_i} .
        \end{equation}
        The mixed state encoding is then defined by
        \begin{equation} \label{eqn:mixed-state-encoding-def}
            \sigma := \frac{1}{M} \sum_{i = 1}^{M} \sigmaxi .
        \end{equation}
    \end{definition}
\end{defbox}
We note that~\eqref{eqn:mixed-state-encoding-def} may be realised by preparing one of the pure states~\eqref{eqn:pure_state_of_mixed_state_encoding_for_embedded_cost_classifier} with equal probability. We use the indicator cost function given by
\begin{equation} \label{eqn:indicator_cost_over_dataset}
    \Cbs := \frac{1}{M} \sum_{i = 1}^{M}  \mathds{1} (\hat{y}_i(\rhotildexi) \neq y_i) .
\end{equation}
Here, the indicator $\mathds{1}$ evaluates to the truth value of its argument --- i.e., $\mathds{1}(\yhat_i \neq  y_i) = 0$ if $y_i = \yhat_i$, else $1$. This is clearly a special case of the general VQA cost in \eqref{eqn:general_vqa_cost_function} where $f$ is the indicator, and the observables to be measured are the $\Pi_0$ projector on the classification qubit.

We now state and prove the following theorem which provides a lower bound on the size of the robust set.
\begin{thmbox}
\begin{theorem} \label{thm:partial_upper_bound}
    Consider a quantum classifier using the mixed state encoding~\eqref{eqn:mixed-state-encoding-def} and indicator cost function~\eqref{eqn:indicator_cost_over_dataset}. Assuming that a noise channel $\E$ acts only on the encoded feature vectors $\rhoxi$ (and not the encoded labels $|y_i\rangle \langle y_i |$), then
    \begin{equation} \label{eqn:partially_robust_set_bound}
         |\mathcal{R}(\mathcal{E}, E, \hat{y})| \geq M\left[1 -  \sqrt{1 - F( \E(\tilde{\sigma}), \tilde{\sigma})}\right] 
    \end{equation}
    where $F$ is the fidelity between quantum states, \eqref{eqn:fidelity_defn}.
\end{theorem}
\end{thmbox}

This theorem is useful since, if one has an expression for the output fidelity of a quantum circuit after a noise channel is applied, one can directly determine a bound on how the classification results will be affected. In order to prove it, we relate the difference between the noisy cost and the noiseless cost to the size of the robust set.

We can also show the following tighter bound based on the average trace distance between the individual encoded states (i.e., not using the mixed state encoding) for the original cost function, $\Cbs$:

\begin{equation} \label{eqn:fidelity_bound_average}
    \Delta_\E \Cbs \leq \frac{2}{M} \sum_{i=1}^M \sqrt{1-F(\E(\rhotildexi), \rhotildexi)} .
\end{equation}

\begin{proof}
\begin{align}
    \Delta_\E \Cbs := \left| \Cbs_\E - \Cbs \right| &= \left|  \Tr [ D ( \E (\tilde{\sigma} ) - \tilde{\sigma}) ] \right| \nonumber\\   
    &\leq \frac{1}{M}\sum\limits_{i=1}^M| \Tr(D\left[ \E(\rhotildexi)\otimes\ketbra{y_i}{y_i} - \rhotildexi\otimes\ketbra{y_i}{y_i}\right])| \nonumber\\
    &\leq \frac{1}{M}\sum\limits_{i=1}^M||D||_{\infty} ||\left[\E(\rhotildexi)- \rhotildexi\right]\otimes\ketbra{y_i}{y_i}||_1 \nonumber\\
    &\leq \frac{2}{M}\sum\limits_{i=1}^M \sqrt{1-F(\E(\rhotildexi), \rhotildexi)} \label{eqn:fidelity_bound_average_appendix}
\end{align}
\end{proof}

In~\secref{ssec:fidelity_analysis_exp}, we use these inequalities to bound the size of the robust set for several different encodings on an example implementation.


\section[\texorpdfstring{\color{black}}{} Numerical results]{Numerical results} \label{sec:classifier/numerical_results}

In this section, we present numerical evidence to reinforce the theoretical results proved in~\secref{sec:noise_robustness} and build on the discussions. In~\secref{ssec:classes_decision_boundaries_numerical}, we show classes of learnable decision boundaries for example encodings, building on the previous discussion in~\secref{ssec:classes_learnable_decision_boundaries}. We then plot the robust sets for partially robust encodings in~\secref{subsec:robust_sets_for_partially_robust_encodings} to visualise the differences that arise from different encodings. We also generalise some encodings defined~\secref{sec:data_encodings} to include hyperparameters and study the effects. This leads us to attempt to train over these hyperparameters, and we present an ``encoding learning algorithm'' in~\secref{ssec:encoding_learn_alg} to perform this task. Finally, in~\secref{ssec:fidelity_analysis_exp} we compute lower bounds on the size of robust sets based on \secref{ssec:fidelity_bounds}. We note that we include code to reproduce all results in this section at~\cite{coyle_noiserobustclassifier_2020}.
For all numerical results in the following sections related to the single qubit classifier, we use three simple datasets; the first is the ``moons'' dataset from \computerfont{scikit-learn}, \cite{pedregosa_scikit-learn_2011}, and two we denote ``vertical'' and ``diagonal''. These datasets can be seen in~\figref{fig:single_qubit_datasets}.

\subsection[\texorpdfstring{\color{black}}{} Decision boundaries and implementations]{Decision boundaries and implementations} \label{ssec:classes_decision_boundaries_numerical}

\begin{figure}[!ht]
    \includegraphics[width=\columnwidth]{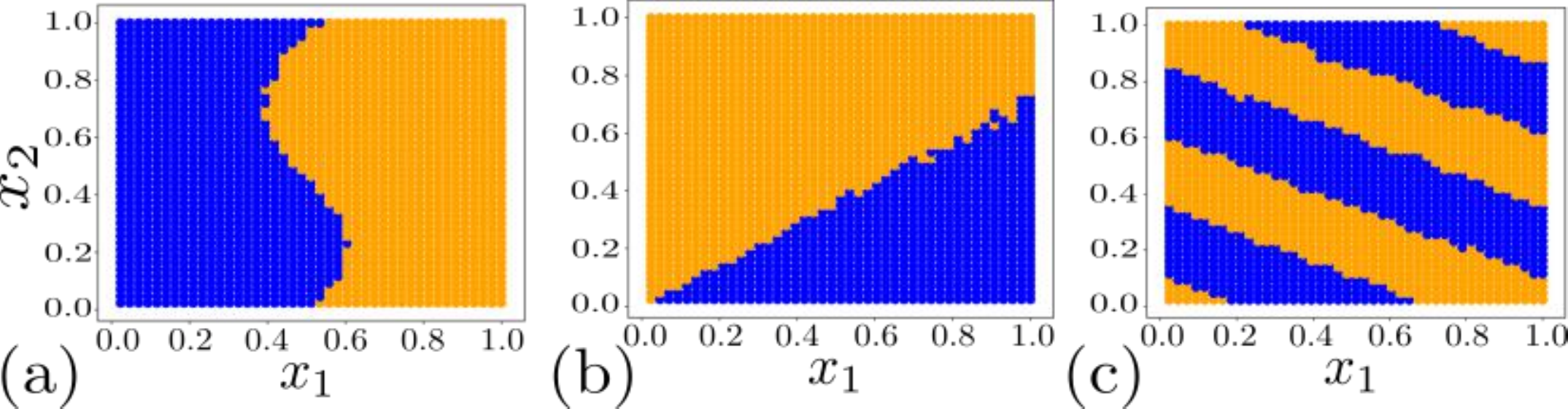}
    \caption[\color{black} Examples of learnable decision boundaries for a single qubit classifier.]{\textbf{Examples of learnable decision boundaries for a single qubit classifier.} We show the (a) dense angle encoding, (b) amplitude encoding, and (c) superdense angle encoding where $\theta = \pi$ and $\phi = 2 \pi$.
    Colours denote class labels. The PQC used here consisted of an arbitrary single qubit rotation (see~\figref{fig:specific_classifier_circuits}) with random parameters.}
    \label{fig:random_decision_boundaries_encodings}
\end{figure}

In~\secref{sec:data_encodings}, we defined an encoding~\eqref{eqn:encoding_formal} and gave several examples. In~\secref{ssec:classes_learnable_decision_boundaries}, we showed that a classifier with the amplitude encoding~\eqref{eqn:amplitude_encoding_single_datapoint} can learn decision boundaries that are straight lines, while the same classifier with the dense angle encoding~\eqref{eqn:dae_single_qubit} can learn sinusoidal decision boundaries. We show this in~\figref{fig:random_decision_boundaries_encodings}, and we build on this discussion in the remainder of this section.

\figref{fig:random_decision_boundaries_encodings}(c) shows a ``striped'' decision boundary learned by a ``superdense'' angle encoding, defined below. The superdense encoding introduces a linear combination of features into the qubit (angle) encoding~\eqref{eqn:angle_encoding_multiqubit} and is another special case of the general qubit encoding~\eqref{eqn:qubit_encoding_general}.
\begin{defbox}
\begin{definition}[Superdense angle encoding ($\SDAE$)]\label{def:sdae}~ \\
    Let $\xbs = [x_1, ..., x_N]^T \in \mathbb{R}^N$ be a feature vector and $\boldsymbol{\theta}, \boldsymbol{\phi} \in \mathbb{R}^N$ be parameters. Then, the superdense angle encoding maps $\xbs \mapsto E(\xbs)$ given by
    \begin{equation} \label{eqn:sdae_general}
        \ket{\xbs} = \bigotimes_{i=1}^{\ceil*{N / 2}}\cos (\theta_i x_{2i-1} + \phi_i x_{2i}) \ket{0} + \cos (\theta_i x_{2i-1} + \phi_i x_{2i}) \ket{1} .
    \end{equation}
\end{definition}
\end{defbox}
For a single qubit, the $\SDAE$ is
\begin{equation*}
    \ket{\xbs} := \cos \left(\theta x_1+ \phi x_2\right) \ket{0} + \sin \left(\theta x_1+ \phi x_2\right)\ket{1} .
\end{equation*}
We observe that $\phi = 0$ recovers the qubit (angle) encoding~\eqref{eqn:angle_encoding_multiqubit} considered by \cite{stoudenmire_supervised_2016, schuld_supervised_2018, cao_cost_2019} and~\eqref{eqn:sdae_general} encodes two features per qubit.

We note that~\defref{def:sdae} includes hyperparameters $\boldsymbol{\theta}$ and $\boldsymbol{\phi}$. The reason for this will become clear in~\secref{ssec:encoding_learn_alg} when we consider optimizing over encoding hyperparameters to increase robustness. As previously mentioned, a similar idea was investigated in~\cite{lloyd_quantum_2020} for the purpose of (in our notation) learnability.

\begin{figure}[!ht]
\includegraphics[width = \columnwidth]{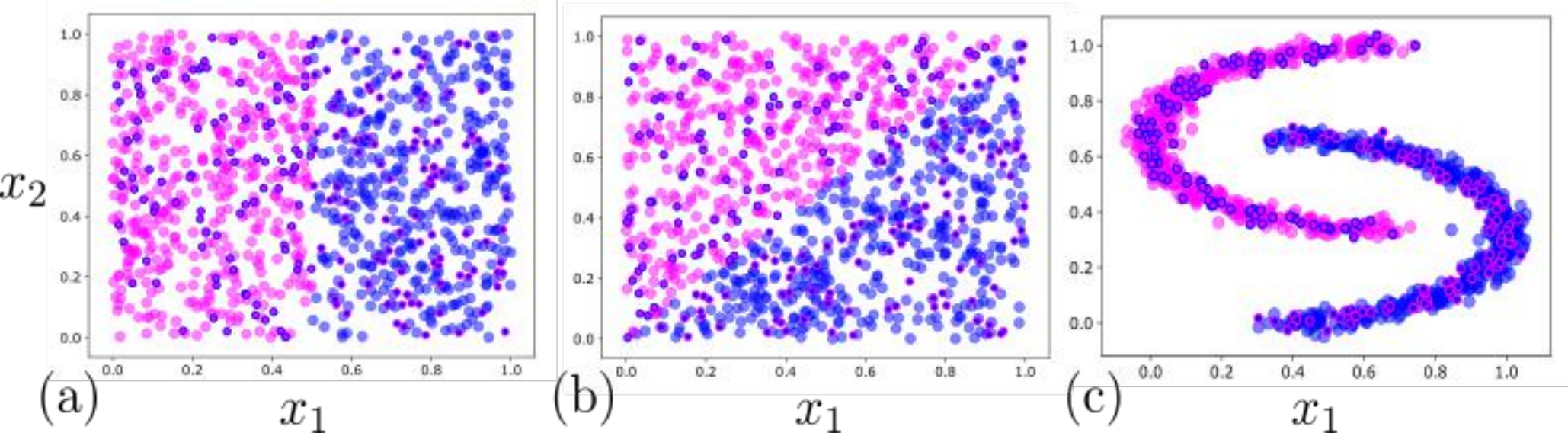}
\caption[\color{black} Three single qubit (two dimensional) datasets.]{\textbf{Three single qubit (two dimensional) datasets which we use.} (a) ``vertical'', (b) ``diagonal'' and (c) the ``moons''  dataset from scikit-learn \cite{pedregosa_scikit-learn_2011} rotated by $90^{\circ}$ with a noise level of $0.05$. 20\% of each set is test data, indicated by the points circled with the opposite colour. We choose the latter two due to the fact that the ``moons'' and ``vertical'' datasets can be well classified by the dense angle encoding, while the ``diagonal'' dataset can be well classified by the amplitude encoding, which can be seen by studying the decision boundaries generated in \figref{fig:random_decision_boundaries_encodings}.}
\label{fig:single_qubit_datasets}
\end{figure}

\figref{fig:single_qubit_datasets}, we illustrate the three single qubit datasets we employ in this chapter to demonstrate our results, namely the ``vertical'', ``diagonal'' and ``moons''. The former two are linearly separable, whereas the ``moons'' dataset is nonlinear.

As a final example to explore the importance of encodings, we consider an example implementation on a standard dataset using different encodings. The dataset we consider is the Iris flower dataset~\cite{fisher_use_1936} in which each flower is described by four features ($\xbs \in \mathbb{R}^4$). The original dataset includes three classes (species of flower) but we only consider two for binary classification. A quantum classifier using the qubit angle encoding~\eqref{eqn:angle_encoding_multiqubit} and a tree tensor network (TTN) $\Ansatz$ was considered in~\cite{grant_hierarchical_2018}. Using this encoding and PQC, the authors were able to successfully classify all points in the dataset. 

Since the angle encoding maps one feature into one qubit, a total of four qubits was used for the example in~\cite{grant_hierarchical_2018}. Here, we consider encodings which map two features into one qubit and thus require only two qubits. Descriptions of the encodings, PQC $\Ansatze$, and overall classification accuracy are shown in Table~\ref{table:two_qubit_iris_class_accuracy}. 

\begin{table}[ht!]
    \centering
     \begin{tabular}{|| c | c | c | c | c ||} 
     \hline
     \textbf{Encoding} & \textbf{PQC $\Ansatz$} & $\boldsymbol{N_P}$ & $\boldsymbol{n} $ & \textbf{Accuracy} \\ [0.5ex] 
     \hline\hline
     Angle & TTN & 7 & 4 & 100\% \\ 
     \hline
     Dense Angle & $U(4)$ & 12 & 2 & 100\% \\
     \hline
     Amplitude  & $U(4)$ & 12 & 2 & 100\% \\
     \hline
     Superdense Angle & $U(4)$ & 12 & 2 & 77.6\% \\
      \hline
    \end{tabular}
    \caption[\color{black} Classification accuracy achieved on the Iris dataset using different encodings and PQCs in the quantum classifier.]{\textbf{Classification accuracy achieved on the Iris dataset using different encodings and PQCs in the quantum classifier.} The top row is from~\cite{grant_hierarchical_2018} and the remaining rows are from this work. The heading $N_p$ indicates number of parameters in the PQC and $n$ is the number of qubits in the classifier. The accuracy is the overall performance using a train-test ratio of $80\%$ on classes $0$ and $2$. (See~\cite{coyle_noiserobustclassifier_2020} for full implementation details).}
    \label{table:two_qubit_iris_class_accuracy}
\end{table} 

\begin{figure}[!ht]
\centering
    \includegraphics[width = 0.9\textwidth, height = 0.3\textwidth]{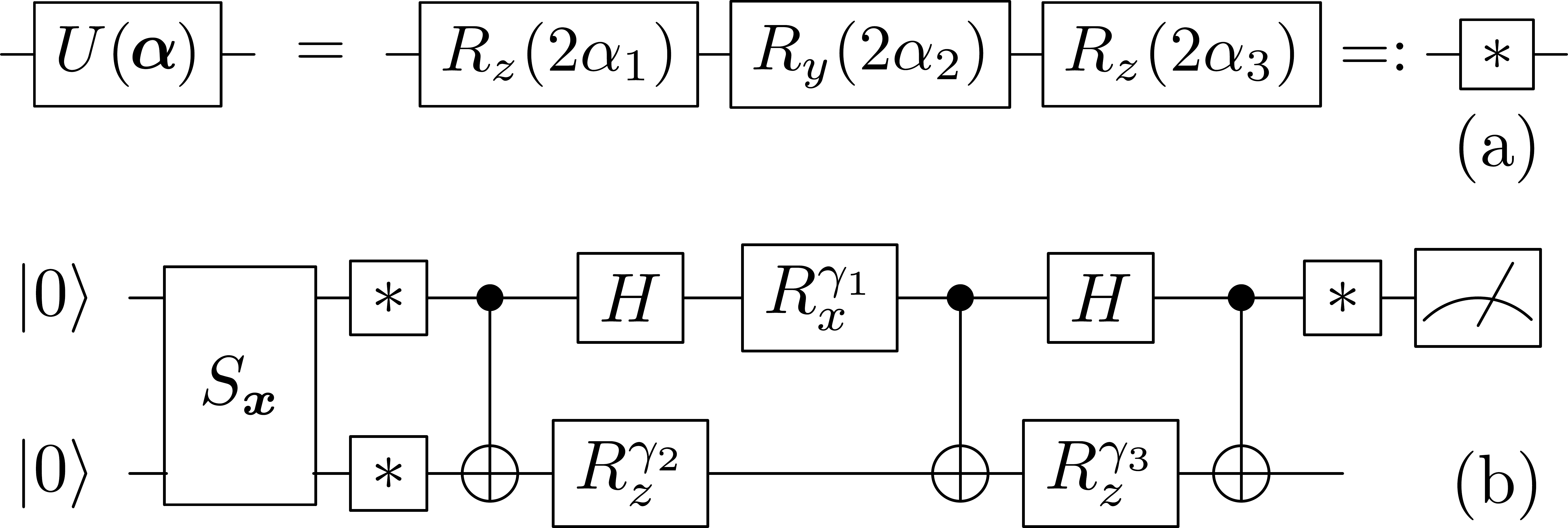}
    \caption[\color{black} Circuit diagrams for the PQC $\Ansatze$ used for classification.]{\textbf{Circuit diagrams for the PQC $\Ansatze$ we use to obtain numerical results.} (a) Ansatz for the single qubit classifier. Note that any element of $U(2)$ can be represented by this $\Ansatz$~\cite{nielsen_quantum_2010}. (b) Ansatz for the two qubit classifier, with $12$ parameters, $\{\alpha_k\}_{k=1}^{12}$. The first $6$ parameters, $\alpha_1,\dots,\alpha_6$, are contained in the first two single qubit unitaries, and $\alpha_{10}, \dots, \alpha_{12}$ are the parameters of the final single qubit gate. The parameters of the intermediate rotations are defined by $\{\gamma_1, \gamma_2, \gamma_3\} := \{2\alpha_7+\pi, 2\alpha_8, 2\alpha_9\}$. This decomposition can realise any two qubit unitary~\cite{vidal_universal_2004} up to global phase with the addition of a single qubit rotation on the bottom qubit at the end of the circuit. We omit this rotation since we only measure the first qubit for classification. As such, we reduce the number of trainable parameters ($\vec{\alpha}$) from $15$ to $12$.}
\label{fig:specific_classifier_circuits}
\end{figure}

\figref{fig:specific_classifier_circuits} illustrates the specific decompositions for the single and two qubit classifiers we utilise for the numerical results in this section. For the matrix representation of the circuit shown in \figref{fig:specific_classifier_circuits}(a), see~\eqref{eqn:single_qubit_unitary_decompostion}.

As can be observed, we are able to achieve 100\% accuracy using the amplitude and dense angle encoding. For the $\SDAE$, the accuracy drops. 
Because the $\SDAE$ performs worse than other encodings, this implementation again highlights the importance of encoding on learnability. Additionally, the fact that we can use two qubits instead of four highlights the importance of encodings from a resource perspective. Specifically, NISQ applications with fewer qubits are less error prone due to fewer two-qubit gates, less crosstalk between qubits, and reduced readout errors. The reduction in the number of qubits here due to data encoding parallels, e.g., the reduction in the number of qubits in quantum chemistry applications due to qubit tapering~\cite{bravyi_tapering_2017}. For QML, such a reduction is not always beneficial as the encoding may require a significantly higher depth. For this implementation, however, the dense angle encoding has the same depth as the angle encoding, so the reduction in number of qubits is meaningful.

\subsection[\texorpdfstring{\color{black}}{} Robust sets for partially robust encodings]{Robust sets for partially robust encodings} \label{subsec:robust_sets_for_partially_robust_encodings}

In~\secref{subsec:robustness_results}, we proved conditions under which an encoding is robust to a given error channel. Typically in practice, encodings may not completely satisfy such robustness criteria, but will exhibit partial robustness --- i.e., some number of training points will be robust, but not all. In this section, we characterise such robust sets for different partially robust encodings. We emphasise two points that (i) the number of robust points is different for different encodings, and (ii) the ``location'' of robust points is different for different encodings.

To illustrate the first point, we consider amplitude damping noise --- which has robustness condition~\eqref{eqn:amp_damp_robustness_condition_expanded_unitary} --- for two different encodings: the dense angle encoding and the amplitude encoding. For each, we use a dataset which consists of 500 points in the unit square separated by a vertical decision boundary at $x_1 = 0.5$.

\begin{figure}[!ht]
{\includegraphics[width = 0.9\columnwidth]{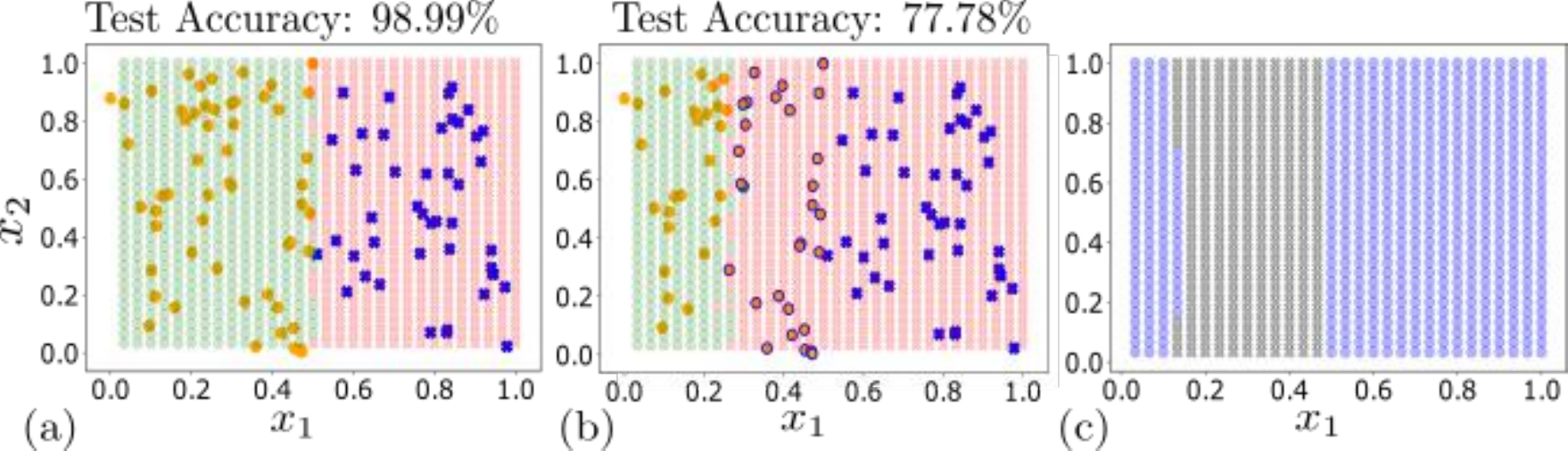}}
\caption[\color{black} Partial robustness for the dense angle encoding.]{\textbf{Partial robustness for the dense angle encoding.} The dataset consists of 500 points in the unit square separated by a vertical decision boundary, and we use a train-test split of $80\%$. Red crosses/green circles indicate decision boundary location. Orange circles/blue crosses indicate points labelled as each class. Misclassified points have opposite colour border.  Panel (a) shows the classifier test accuracy after optimizing the unitary without noise. Panel (b) shows the reduced accuracy after amplitude damping noise of strength $p = 0.4$ is added. The robust set is at the far left and far right of the unit square, explicitly shown in Panel (c). Here, blue circles indicate the robust set and black crosses indicate its complement.
} 
\label{fig:dae_encoding_linear_boundary}
\end{figure}

The results for the dense angle encoding are shown in~\figref{fig:dae_encoding_linear_boundary}. Without noise, the classifier is able to reach an accuracy of $\sim 99\%$ on the training set. When the amplitude damping channel with strength $p = 0.2$ is added, the test accuracy reduces to $\sim 78\%$. This encoding is thus partially robust, and the set of robust points is shown explicitly in~\figref{fig:dae_encoding_linear_boundary}(c).

\begin{figure}
{\includegraphics[width = 0.9\columnwidth]{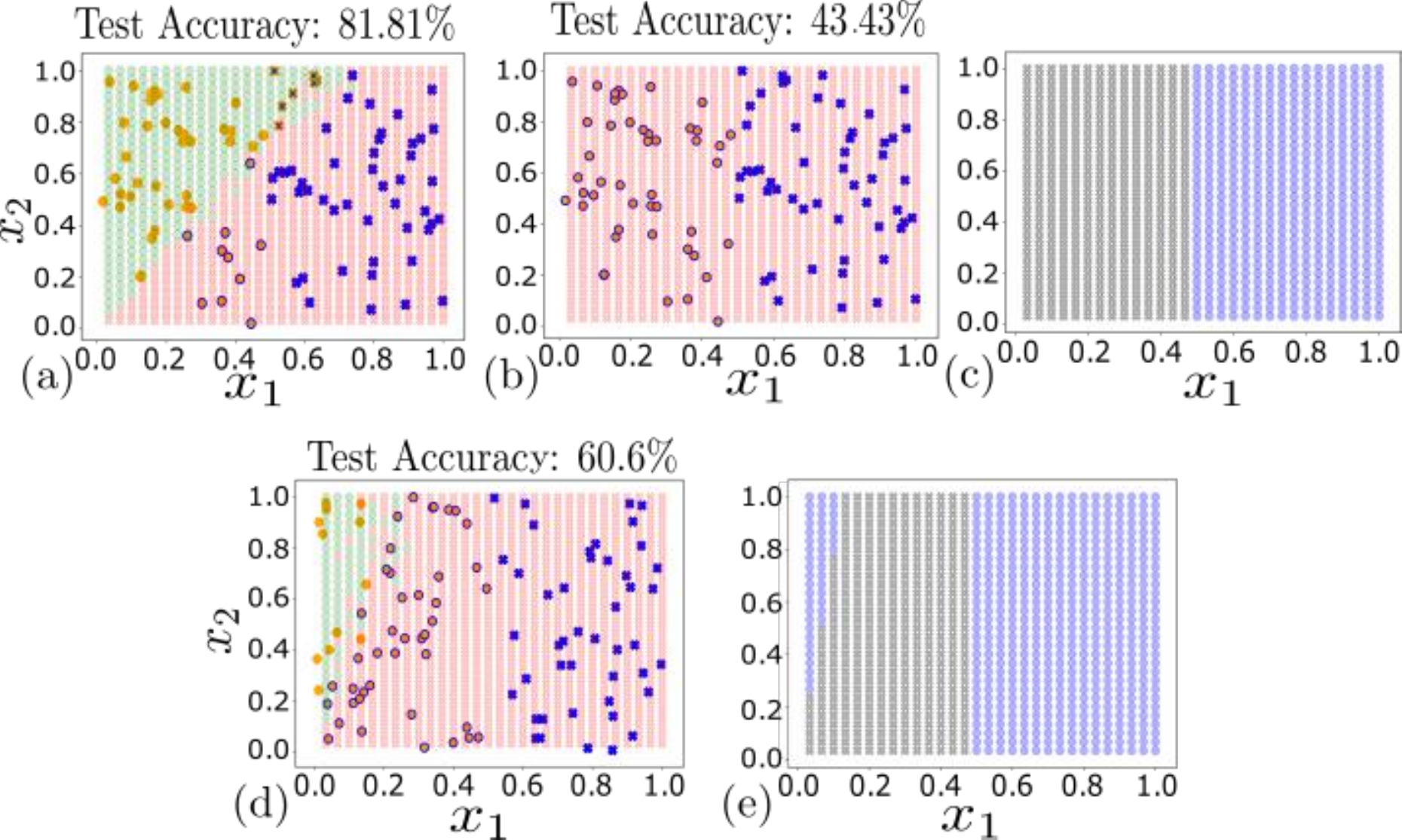}}
\caption[\color{black} Partial robustness for the amplitude encoding.]{\textbf{Partial robustness for the amplitude encoding.} The dataset consists of 500 points in the unit square separated by a vertical decision boundary, and we use a train-test split of $80\%$. Panel (a) shows classifier test accuracy after optimizing the unitary without noise. Panel (b) shows the reduced accuracy after adding amplitude damping noise with strength $p = 0.4$. The robust set is shown explicitly in Panel (c) where a blue circle indicates a robust point and a black cross indicates a misclassified point.
Panel (d) is the same as (b) but with decreased strength $p = 0.2$ of the amplitude damping channel. Test accuracy reduces from $81\%$ to $60\%$ in this case. Panel (e) shows the robust set for (d).
}
\label{fig:wf_encoding_linear_boundary}
\end{figure}

The results for the amplitude encoding are shown in~\figref{fig:wf_encoding_linear_boundary}. Here, the classifier is only able to reach $\sim 82\%$ test accuracy without noise. When the same amplitude damping channel with strength $p = 0.4$ is added, the test accuracy drops to $\sim 43\%$. We consider also the effect of amplitude damping noise with strength $p = 0.2$ in~\figref{fig:wf_encoding_linear_boundary}, for which the classifier achieves test accuracy $\sim 61\%$. The robust set for both channels is also shown in~\figref{fig:wf_encoding_linear_boundary}.

\subsection[\texorpdfstring{\color{black}}{} An encoding learning algorithm]{An encoding learning algorithm}\label{ssec:encoding_learn_alg}

\begin{figure}[!ht]
\centering
    \includegraphics[width=0.9\columnwidth, height=0.6\columnwidth]{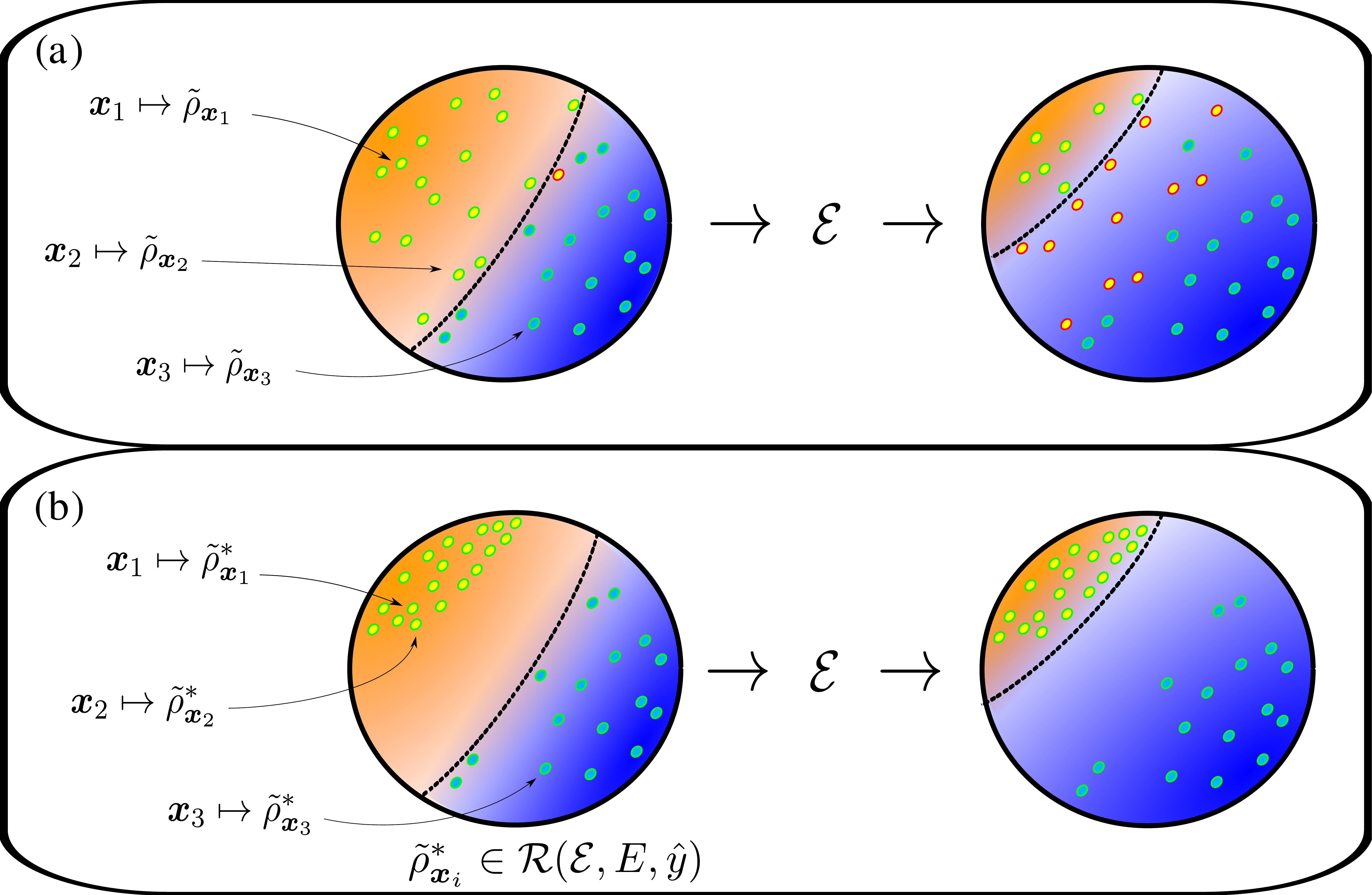}
    \caption[\color{black} Cartoon illustration of the encoding learning algorithm.]{\textbf{Cartoon illustration of the encoding learning algorithm with a single qubit classifier.} In (a), a preset encoding with no knowledge of the noise misclassifies a large number of points. In (b), the encoding learning algorithm detects misclassifications and tries to adjust points to achieve more robustness, attempting to encode into the robust set for the channel.}
    \label{fig:single_qubit_binary_classifier_encoding_learning}
\end{figure}

Given the emphasis that we have placed on the importance of finding suitable encodings, but the apparent difficulty in doing so, we introduce an ``encoding learning algorithm'' to try and search for good encodings. The goal is crudely illustrated in~\figref{fig:single_qubit_binary_classifier_encoding_learning}. As mentioned above,~\cite{lloyd_quantum_2020} trains over hyperparameters using the re-uploading structure of~\cite{perez-salinas_data_2020} to increase learnability. Here, the encoding learning algorithm adapts to noise to increase robustness. We note the distinction that in our implementations we train the unitary in a noiseless environment and do not alter its parameters during the encoding learning algorithm. By training the encoding we change $E$ and hence also the model family. This is actually desirable in this case since, as we have demonstrated, $E$ directly impacts the robustness of the model family. This could be incorporated alongside training the unitary itself to deal with coherent noise, for example as seen in \cite{omalley_scalable_2016, sharma_noise_2020}.

The encoding learning algorithm is similar to the Rotoselect\footnote{We implement a similar idea in~\chapref{chap:cloning} based on quantum structure learning to search for good quantum cloning circuits.} algorithm which is used to learn circuit structure~\cite{ostaszewski_structure_2021}. For each function pair $(f_j, g_j)$ from a discrete set of parametrised functions $\{f_k(\paramtheta_k), g_k(\paramtheta_k)\}_{k = 1}^{K}$  we train the unitary $U(\vec{\alpha})$ to minimise the cost while keeping the encoding (hyper)parameters $\paramtheta_j$ fixed. Next, we add a noise channel $\E$ which causes some points to be misclassified. Now, we optimise the encoding parameters $\paramtheta_j$ in the noisy environment. For this optimisation, the same cost function is used, and the goal is to further decrease the cost (and hence increase the set of robust points) by varying the encoding hyperparameters. Algorithm 1 contains pseudocode for the procedure used.

\begin{small}
\begin{algorithm}[ht]
\SetAlgoLined
\SetKwInOut{Input}{Input}\SetKwInOut{Output}{output}
\Input{Noise parameters, $\vec{p}$, parametrised quantum circuit, $U(\boldsymbol\alpha)$, $M$ Labelled data examples $(\xbs_i, y_i)_{i=1}^M$, encoding set $\{f_k, g_k\}_{k=1}^K$, cost function C.}
\KwResult{Optimised encoding for noise and dataset.}
 Initialise encoding, $(f^*, g^*) \leftarrow \{f_k, g_k\}_{k=1}^K$ and parameters, $\{\paramtheta_j\} \leftarrow (0, 2\pi ]_j ~\forall j$ heuristically or at random. Initialise $C^*= M$\; 
 \For{j=1\dots K}{ \do
  Select subset of data: $(\xbs_i, y_i)_{i=1}^D \leftarrow(\xbs_i, y_i)_{i=1}^M$\;
  Encode each sample using encoding choice, $(f_j, g_j)$: prepare $\{\rhotildexi^{\boldsymbol\alpha, \paramtheta_j}\}_{i=1}^D$\;
  $\boldsymbol\alpha^* \leftarrow  \argmin_{\boldsymbol\alpha} C_D(\boldsymbol\alpha, \paramtheta_j)$\;
    Add noise with parameters $\vec{p}$: $\{\rhotildex^{\boldsymbol\alpha^*, \paramtheta_j}\}_{i=1}^D \leftarrow \{\E_{\vec{p}}(\rhotildex^{\boldsymbol\alpha^*, \paramtheta_j^*})\}_{i=1}^D$\;
    $\{\paramtheta^*_j\}  \leftarrow \argmin_{\paramtheta_j} C_D(\boldsymbol\alpha^*, \paramtheta_j)$\; 
    $C^*_j \leftarrow  C_D(\boldsymbol\alpha^*, \paramtheta_j^*)$\;
  \If{$C^*_j \leq C^*$}{
  $C^* \leftarrow C^*_j$\;
  $(f^*, g^*) \leftarrow f_j(\paramtheta_j^*), g_j(\paramtheta_j^*$) \;
  }
  }
 \Output{$C^*, \vec{\alpha}^*, f^*, g^*$}
 \caption{Quantum encoding learning algorithm (QELA)}
\end{algorithm}
\end{small}

\begin{figure}[ht!]
    \includegraphics[width = \columnwidth, height=0.5\columnwidth]{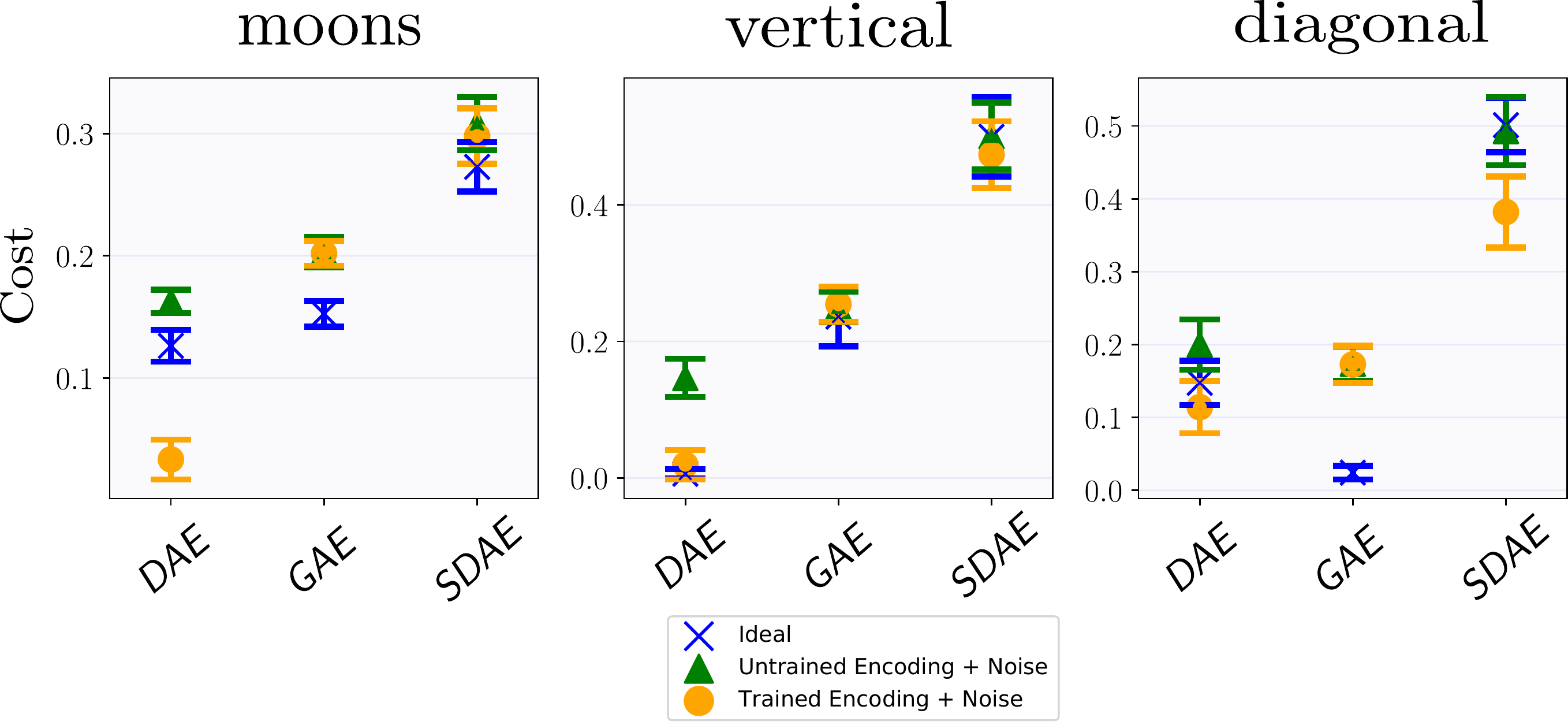}
        \caption[\color{black} Minimum cost achieved from applying the encoding learning algorithm to three example datasets.]{
            \textbf{Minimum cost achieved (vertical axis) from applying the encoding learning algorithm to three example datasets (each plot) using three different encodings (horizontal axis).} The blue crosses [\crule[blue]{0.2cm}{0.2cm}] show the minimum cost achieved when training over only unitary parameters without any noise present (ideal).  The green triangles [\crule[ForestGreen]{0.2cm}{0.2cm}] show the same case with the addition of amplitude damping noise of strength $p = 0.3$. The orange circles [\crule[orange]{0.2cm}{0.2cm}] show minimum cost after applying the encoding learning algorithm. The dense angle encoding ($\DAE$) is seen to perform well on all datasets and is capable of adapting well to noise, even outperforming the ideal case without noise and fixed encoding. The superdense angle encoding ($\SDAE$) does not perform well on any shown dataset since the generated decision boundary is highly nonlinear and cannot correctly classify more than half the dataset. The generalised amplitude encoding ($\GAE$) performs well on the diagonal boundary, since it generates a suitable decision boundary. Each datapoint shows the mean and standard deviation of the costs after ten independent runs of the learning procedure.
        }
\label{fig:encoding_learning_algorithm}
\end{figure}

We test the algorithm on linearly separable and non-linearly separable datasets in~\figref{fig:encoding_learning_algorithm}. In particular, we use three different encodings on three datasets. The encodings used are the dense angle encoding, superdense angle encoding, and a specific instance of the generalised amplitude encoding from~\defref{def:generalised_amplitude_encoding} given by
\begin{equation*} 
    \ket{\xbs}:= \frac{\sqrt{1+\theta x_2^2}}{||\xbs||_2} x_1 \ket{0} + \frac{\sqrt{1-\theta x_1^2}}{||\xbs||_2} x_2 \ket{1} 
\end{equation*}
Here, $\theta$ is a free parameter which is trained over.

Using these encodings and the datasets given at the beginning of this section, we study performance for the noiseless case, noisy case, and the effect of the encoding learning algorithm. We observe that the algorithm is not only capable of recovering the noiseless classification accuracy achieved, but is actually able to outperform it in some cases, as can be seen in~\figref{fig:encoding_learning_algorithm}.

\begin{figure}[ht!]
\includegraphics[width = 0.95\columnwidth, height=0.29\columnwidth]{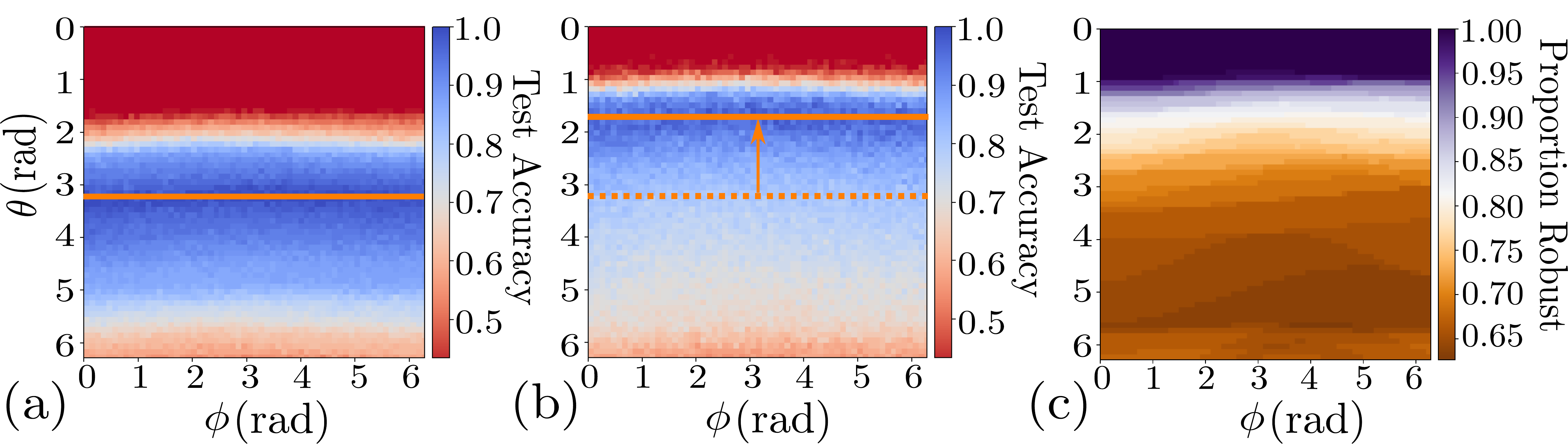}
\caption[\color{black} Learnability vs.\ robustness on the ``vertical'' dataset using the parametrised dense angle encoding.]{\textbf{Learnability vs.\ robustness on the ``vertical'' dataset using the parametrised dense angle encoding.} The horizontal and vertical axes show the encoding hyperparameters $\phi$ and $\theta$, respectively. Panels (a) and (b) show the classifier accuracy while Panel (c) shows the proportion of robust points. Panel (a) shows accuracy without noise as a function of encoding parameters.
Panel (b) shows accuracy with the addition of an amplitude damping channel of strength $p = 0.3$. Panel (c) shows $\delta$-robustness for different parameter values. As expected, the robust set is largest when all points are encoded into the zero state, i.e. $\theta=0$. This leads to all points labelled $1$ being misclassified, with a resulting accuracy of approximately $50\%$. The orange [\crule[orange]{0.2cm}{0.2cm}] filled line indicates optimal $\theta$ parameters in each panel. From (a) to (b), the $\theta$ parameters corresponding to highest accuracy are shifted towards the robust points (i.e., towards $\theta = 0$) in (c). The optimal parameters in each case are given in Table~\ref{table:vertical_boundary_encodin_params_plot}}
\label{fig:dae_encoding_learnability_versus_robustness}
\end{figure}

Finally, we consider the discussion in~\secref{subsec:existence_of_robust_encodings} the tradeoff between learnability (expressive power) and robustness. We make this quantitative  and explicit in~\figref{fig:dae_encoding_learnability_versus_robustness} by plotting accuracy (percent learned correctly) and robustness against hyperparameters $\theta$ and $\phi$ in a generalised dense angle encoding: 
\begin{equation}
    \ket{\xbs} = \cos (\theta x_{1}) \ket{0} + e^{i \phi x_{2}} \sin (\theta x_{1})\ket{1}
\end{equation}
More specifically, in~\figref{fig:dae_encoding_learnability_versus_robustness}, we illustrate how the noise affects the hyperparameters, $\theta^*$ and $\phi^*$, which give maximal classification \emph{accuracy} in both the noiseless and noisy environments, and also those which give maximal \emph{robustness} (in the sense of \defref{def:partial_noise_robustness}).~\figref{fig:dae_encoding_learnability_versus_robustness}(a), shows the percentage misclassified in the noiseless environment, where red indicates the lowest accuracy on the test set, and blue indicates the highest accuracy. We then repeat this in~\figref{fig:dae_encoding_learnability_versus_robustness}(b) and~\figref{fig:dae_encoding_learnability_versus_robustness}(c) to find the parameters which maximise accuracy in the presence of noise, and the maximise robustness. As expected, for the amplitude damping channel, the best parameters (with noise) are closer to the fixed point of the channel (i.e. $\theta^* \rightarrow 0$ implies encoding in the  $\ket{0}$ state), thereby demonstrating the tradeoff between learnability and robustness. 
%

Table~\ref{table:vertical_boundary_encodin_params_plot} provides the best parameters found in the procedure,  corresponding to \figref{fig:dae_encoding_learnability_versus_robustness}. Each set of parameters (each row, measured in radians) performs optimally in one of three areas. The first is the noiseless environment, in which a $\theta\approx 2.9$ parameter performs optimally. The second is the amplitude-damped environment, in which $\theta\approx 1.6$ achieves the best accuracy, and finally, $\theta=0$ is the most robust point to encode in, for the whole dataset. For each of these parameter sets, we also test them in the other scenarios, for example, the best parameters found in the noisy environment ($[\theta, \phi] = [1.6, 3.9]$) have a higher $\delta$-robustness ($81\%$) than those in the noiseless environment ($70\%$), since these parameters force points to be encoded closer to the $\ket{0}$ state, i.e., the fixed point of the channel in question.  
\begin{table}[ht]
\centering
 \begin{tabular}{||c | c | c | c||} 
 \hline
 \textbf{Parameters} &  \textbf{Accuracy} & \textbf{Accuracy}  & \textbf{$\delta$-Robustness}\\
  &   \textbf{w/o noise} & \textbf{w/ noise} & \\ [0.5ex] 
 \hline\hline
 $[\theta, \phi] = [2.9, 2.9]$ & $100\%$ & $84\%$ & $70\%$ \\ 
 \hline
 $[\theta, \phi] = [1.6, 3.9]$ & $49\%$ & $100\%$ & $81\%$ \\ 
 \hline
 $[\theta, \phi] = [0, 0]$ & $43\%$ & $43\%$ & $100\%$\\ 
 \hline
\end{tabular}
\caption[\color{black} Optimal parameters for dense angle encoding optimising accuracy and robustness in noisy and noiseless environments.]{\textbf{Optimal parameters $[\theta, \phi]$ for dense angle encoding (with parameters in $U(\vec{\alpha})$ trained in noiseless environment)} in (a) noiseless environment, (b) noisy environment (i.e. amplitude damping channel is added) and (c) for maximal robustness. Optimal parameters in noisy environment are closer to fixed point of amplitude damping channel ($\ket{0}$, i.e. $\theta \equiv 0$) and give a higher value of $\delta$-robustness. }\label{table:vertical_boundary_encodin_params_plot}
\end{table}
\subsection[\texorpdfstring{\color{black}}{} Fidelity bounds on partial robustness]{Fidelity bounds on partial robustness} \label{ssec:fidelity_analysis_exp}

\begin{figure*}[!t]
\includegraphics[width = \columnwidth, height = 0.4\textwidth]{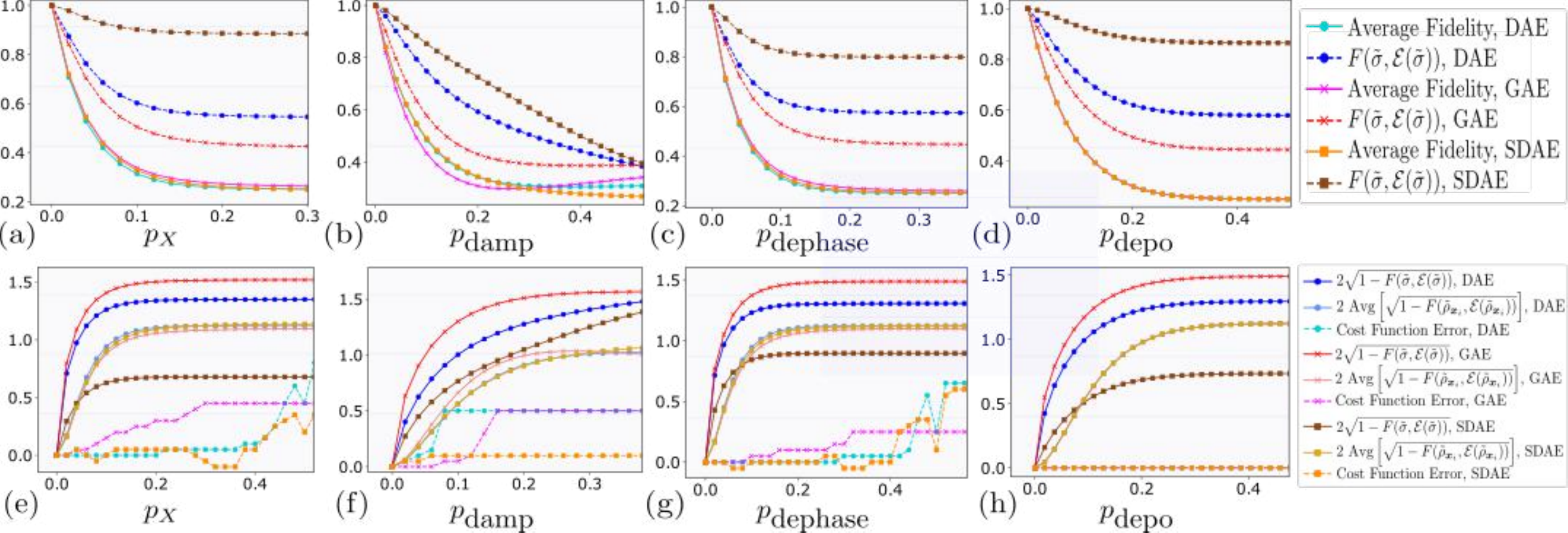}
\caption[\color{black} Fidelity bounds on robustness]{\textbf{Fidelity bounds on robustness}. Top row: Fidelity between noisy and noiseless states for different encodings on the Iris dataset. Comparing average fidelity over every encoded state in dataset, versus fidelity of noisy and noiseless encoded mixed states. From left to right, the noise models are bit-flip noise, amplitude damping noise, dephasing noise, and global depolarising noise, with strengths varied across the horizontal axis. Bottom row: Upper bounds on partial robustness in terms of fidelity using the bounds~\eqref{eqn:fidelity_bound_mixed_state} and~\eqref{eqn:fidelity_bound_average}. For each plot (in both rows), three different data encodings are considered --- curves corresponds to three product qubit encodings: dense angle encoding, amplitude encoding, and superdense angle.}
\label{fig:fidelity_analysis_iris}
\end{figure*}

As a final numerical implementation, we compute the bounds on partial robustness proved in~\secref{ssec:fidelity_bounds} for several different encodings and error channels. The implementation we consider is the previously-discussed Iris dataset classification problem using two qubits. The results are shown in~\figref{fig:fidelity_analysis_iris}. In this Figure, each plot corresponds to a different error channel with strength varied across the horizontal axis. Each curve in the top row corresponds to the fidelity of noisy and noiseless states using different encodings. Each curve in the bottom row shows the lower bounds on partial robustness proved in~\secref{ssec:fidelity_bounds}. 


As can be seen in the bottom row of \figref{fig:fidelity_analysis_iris}, upper bounds on partial robustness are different for different encodings, particularly at small noise values. (Recall that a trivial upper bound on the size of partial robustness is one so that curves at large channel strengths above one are mostly uninformative.) %

For such low values of noise, they give us some information about the maximum cost function deviation we can expect.  Based on the average fidelity over the datasets, in Figs.~(\ref{fig:fidelity_analysis_iris}(a) - \ref{fig:fidelity_analysis_iris}(d)) all three encodings behave qualitatively the same. However, the cost function error for the three encodings is significantly different, especially for bit flip and dephasing errors, Figures~(\ref{fig:fidelity_analysis_iris}(f) - \ref{fig:fidelity_analysis_iris}(g)). As expected, a depolarising channel causes no misclassification, as seen in \ref{fig:fidelity_analysis_iris}(h), despite the decrease in fidelity of the states. Recall that the superdense angle encoding was not able to achieve perfect classification accuracy on the Iris dataset, so under amplitude damping noise, e.g. the cost function error can only decrease by about $25\%$ ($\sim 77\% \rightarrow 50 \%$).
We can also observe that the dense angle encoding is less susceptible to bit flip and phase errors than the amplitude encoding in~\figref{fig:fidelity_analysis_iris}(f) and~\figref{fig:fidelity_analysis_iris}(g).

\section[\texorpdfstring{\color{black}}{} Discussion and conclusions]{Discussion and conclusions} \label{sec:classifier/conclusions}

As our first contribution for a NISQ application in machine learning, in this chapter we focused on the binary quantum classifier, a special case of a variational quantum algorithm  used in supervised learning. The basic ingredients of the model are common to those in recent literature, and we began by examining different data encoding strategies for classical data in the model. Our novel contributions in this area was an investigation of the model in the presence of quantum noise models, in particular in the search for \emph{robust} data encoding strategies. To reiterate the statement from the Introduction of this chapter, we found that it is indeed possible to design quantum classifiers which are somewhat robust to noise, but perhaps at the expense of the \emph{learnability} of the model (pathological design). Specifically, we provided some intuition for how to best design such models and backed these intuitions up by rigorous proofs and supplementary numerical results, at least for special cases and simple noise models. 

One of the major open questions raised by this work is to what extent will such methods scale to larger problem sizes. Certainly in theory, something must be possible (due to the existence result we proved), but a \emph{necessary} condition for robustness is elusive. Perhaps ingredients and theory from quantum error correction will prove useful - we know at least by bringing the entire theory of QEC we will achieve \emph{full} robustness, but what are the minimum elements we require to at least gain \emph{some} (but still remain NISQy). For example, one could study encodings into decoherence-free subspaces \cite{lidar_decoherence_1998, lidar_review_2014}. An alternative question which may be of independent is to extend the notion of a robust point to a ``generalised fixed point'' of a channel, i.e.\@ points which satisfy $f(\E(\rho)) = f(\rho)$ for some function $f$. (When $f = \yhat$, the ``generalised fixed point'' is a robust point, but other arbitrary functions $f$ could be considered.) 
In a more empirical research direction, perhaps using machine learning itself to \emph{find} robust encodings will be fruitful (as initiated with our $\QELA$), or even just extending this work to larger datasets (e.g., the MNIST dataset~\cite{lecun_gradient-based_1998}) and more sophisticated model design (for example by incorporating ``data re-uploading''~\cite{perez-salinas_data_2020} - more about this in~\secref{ssec:classifier/subsequent}). 

In conclusion, this chapter evidences that application-specific robustness (for (variational) quantum algorithms in general) is a promising and interesting direction of study.

\subsection[\texorpdfstring{\color{black}}{} Subsequent work]{Subsequent work} \label{ssec:classifier/subsequent}
In order to properly insert our work from this chapter in its proper context in the literature, we believe it is important to not only discuss the work which came before, but also subsequent work which came after. To this end, we briefly summarise some we find interesting, which may be merged with our work, and the other future research directions given the previous section. First, and perhaps most importantly,~\cite{schuld_effect_2021, gil_vidal_input_2020} highlighted explicitly the effect of the data encoding on the function families representable by quantum models such as ours. Interestingly, the authors found that by using data-reuploading~\cite{perez-salinas_data_2020} the model can be expressed as a Fourier series, whose frequencies and coefficients are given by the data uploading and parameterised unitaries respectively. This reinforces our findings regards the importance of clever data encodings for the purpose of robustness. Secondly, on the theoretical side, there has been a flurry of work studying such quantum models regarding their \emph{generalisation ability}. For example, metrics such as the effective dimension~\cite{abbas_power_2021}, the capacity~\cite{wright_capacity_2020}, the pseudo-dimension~\cite{caro_pseudo-dimension_2020} among others~\cite{bu_effects_2021, bu_statistical_2021, bu_rademacher_2021, gyurik_structural_2021, huang_power_2021} have proven useful in deriving bounds on the generalisation of such models, which are key in further understanding them. It would be particularly interesting to study the relationship between these concepts, and the ideas in this chapter. For example, one may examine the effect on these generalisation bounds when one is restricted to a specific set of robust points for a given noise channel.

Secondly, there have also been several works following ours considering the link between noise and defences against adversarial attacks in a quantum setting. For example,~\cite{liu_vulnerability_2019} demonstrated how, in high dimensions, it is easy for an adversary to perturb a quantum state such that it is misclassified. The work of~\cite{du_quantum_2021} showed how the simple quantum noise models studied in this chapter could lead to `natural' versions of differential privacy. Finally,~\cite{guan_robustness_2021} and \cite{weber_optimal_2021} derive robustness bounds to complement our own based on the distance measures discussed in~\secref{sssec:prelim/qc/distance_measures/quantum}. The latter also demonstrated a link between robust classification and quantum hypothesis testing, revealing an interesting avenue with which to explore in future work. 

Finally, one may consider the more general influence of noise on NISQ algorithms. As we touched on in~\secref{ssec:vqa_cost_function_optimisation}, noise may exacerbate the effect of barren plateaus (\cite{wang_noise-induced_2021}), or enable the efficient classical simulation of such quantum models~\cite{franca_limitations_2020}. Alternatively, quantum error mitigation~\cite{endo_hybrid_2021} may also be considered as a complementary technique to robustness, which is a rapidly expanding area of study.

\chapter{Generative modelling with quantum circuit Born machines} \label{chap:born_machine}

\section[\texorpdfstring{\color{black}}{} Introduction]{Introduction} \label{sec:born_machines/introduction}

\begin{chapquote}{Ian Goodfellow, paraphrased from Richard Feynman}
 ``What an AI cannot create, it cannot understand.''
\end{chapquote}

In the previous chapter, we discussed an application of $\VQA$s in machine learning using a quantum classifier. These models are relatively simple, in the sense that supervised learning (classically) is quite well explored both from a theoretical and from an implementation point of view. In contrast, \emph{unsupervised} learning is significantly more challenging and this chapter is dedicated to the generative modelling unsupervised learning problem (discussed in \secref{ssec:prelims/machine_learning/generative_modelling}), and the model for this task which we introduced in \secref{sssec:vqa_input_and_ouput}, called the \emph{quantum circuit Born machine} (QCBM).

We begin by recalling most `vanilla' form of the quantum circuit Born machine (QCBM) from~\secref{sssec:vqa_input_and_ouput}. Specifically, this involved the preparation of a parametrised state, $\ket{\psi(\paramtheta)}$, via a quantum circuit, $U(\paramtheta)$, acting on a fixed initial state $\ket{0}^{\otimes n}$. The generative model which this produces is the distribution given by measuring all qubits in the computational basis:
\begin{equation} \label{eqn:born_machine/born_machine_pure_measurement_rule}
    p_{\paramtheta}(\xbs) := |\bra{\xbs}U(\paramtheta)\ket{0}^{\otimes n}|^2
\end{equation}
As a reminder, the task of generative modelling is to learn a representation of some data distribution, $q(\ybs)$, given a set of samples, $\{\ybs_j\}_{j=1}^M$ from $q$. In practice, this means fitting the parameters of the generative model (the QCBM), $\paramtheta$, such that $p_{\paramtheta}(\xbs) \approx q(\xbs)~ \forall \xbs$. This is usually achieved via the minimisation of some `distance', $\text{d}$, between $p_{\paramtheta}$ and $q$ (so the problem is `solved' when $\text{d}(p_{\paramtheta}, q) \rightarrow 0$, recall the requirement of faithfulness from~\secref{ssec:vqa_cost_functions}).

Given this problem statement, the goal of this chapter is threefold. We investigate the QCBM in some detail and in doing so we provide the following contributions:
\begin{itemize}
    \item
    A discussion of quantum advantage using the QCBM. In simple terms, the ability to solve a generative modelling problem better than any classical model could. Slightly more precisely, given a distribution, $q$, output $p_{\paramtheta}$ which is `closer' - with respect to some distance measure - than any classical generative model could achieve.
    \item 
    Improved differentiable training methods for the QCBM. Informally, better methods to `fit' the model parameters $\paramtheta$ so that $p_{\paramtheta}$ is closer to $q$.
    \item 
    A numerical comparison study of the QCBM versus a comparable classical model (specifically the restricted Boltzmann machine (RBM) discussed in \secref{ssec:prelims/machine_learning/generative_modelling}) in a close-to-real-world use case in finance.
\end{itemize}

Let us begin our adventure with a discussion of quantum advantage in generative modelling, using a QCBM.

\section[\texorpdfstring{\color{black}}{} Quantum advantage in generative modelling]{Quantum advantage in generative modelling} \label{ssec:born_machine/quantum_advantage}

Before diving into quantum advantage may arise in generative modelling, let us first motivate why such a thing might be feasible in the first place. 

The reason is because the a core ingredient in generative models is \emph{sampling}. It turns out, due to the probabilistic nature of quantum mechanics, sampling from a quantum distribution is a very straightforward thing to do (as we discussed in \secref{sssec:vqa_input_and_ouput}). This observation prompted the development of quantum computational \emph{supremacy} (QCS)\footnote{Also called quantum superiority, or more commonly quantum advantage to avoid any controversies.}, which refers to a specific task: the generation of samples from a quantum state, which \emph{could not} be generated by \emph{any} classical algorithm, in a reasonable time. The reason for defining such an abstract problem, is primarily to illustrate the power of NISQ computers. NISQ computers cannot perform complex algorithms due to noise limitations (see \secref{ssec:prelim/qc/quantum_noise}), but what they \emph{can} do is prepare a quantum state via a short-depth quantum circuit, $U$, and measure all qubits in the computational basis. This then defines a natural sampling problem. The question to demonstrate QCS is then twofold:
\begin{itemize}
    \item Ensure $U$ is simple enough to implement physically on NISQ devices.
    \item Prove rigorously that the sampling problem \emph{cannot} be solved classically.
\end{itemize}
To achieve the former, a number of quantum circuit families have been proposed, including instantaneous quantum polynomial time ($\IQP$) circuits, $\BosonSampling$, and random circuit sampling ($\RCS$). All of these proposals define \emph{sub-universal} complexity classes in that they cannot implement an arbitrary quantum computation in $\BQP$ (see \secref{ssec:prelim/qc/quantum_operations}), but yet they are still complex enough to be difficult to simulate classically (or so we believe). Excluding a simple preparation and measurement operation, $\IQP$ contains only gates which are diagonal in the computational (Pauli-$\ZG$) basis, $\BosonSampling$ involves interacting photonic modes in an interferometer without feedforward, and $\RCS$ involves implementing random layers of hardware native locally connected gates.

The next ingredient is to prove such sampling tasks cannot be achieved classically\footnote{By `achieved' here, we mean cannot by simulated in \emph{polynomial time} - if classical simulators are allowed to take exponential time then we know then would be able to simulate quantum computers by just keeping track of the exponentially large wavefunction.}. The most common route to do so, is to assume it can be simulated classically, and then derive an unlikely consequence in complexity theory, i.e. the collapse of the \emph{polynomial hierarchy} (\cite{stockmeyer_polynomial-time_1976})\footnote{An introduction to (quantum) complexity theory is not necessary for reading this theses, and so for a comprehensive overview of this topic see \cite{nielsen_quantum_2010} as well as \href{https://complexityzoo.net/Complexity_Zoo}{the complexity zoo} for a list of complexity classes.}. Such a collapse occurring is a generalisation of $\Pee$ being equal to $\NP$, which is widely believed to be not the case.

\subsection[\texorpdfstring{\color{black}}{} The Ising Born machine]{The Ising Born machine} \label{ssec:generative_modelling/born_machine/ising_born_machine}
In the work of a previous dissertation~\cite{coyle_project_2018}, we introduced a specification of a QCBM based on a generalisation of $\IQP$ circuits which we dubbed the quantum circuit \emph{Ising} Born machine ($\QCIBM$). This model was a generalisation because we added the possibility of measurement in alternative bases (standard $\IQP$ circuits only measures in the Hadamard basis). Specifically, the circuits considered for the $\QCIBM$ had the following structure (we reintroduce it here from~\cite{coyle_project_2018} since we use this same model in our numerical results later in this thesis):
\begin{align} \label{circuit:ising_born_machine}
    \Qcircuit @C=0.6em @R=0.8em {
    \lstick{\ket{0}}    & \gate{\HG}  & \multigate{3}{U_z(\boldsymbol\alpha)} & \gate{U^1_f(\Gamma_1, \Delta_1, \Sigma_1)}  & \meter &\cw & \rstick{x_1} \\
    \lstick{\ket{0}}    & \gate{\HG}  & \ghost{U_z(\boldsymbol\alpha)}        & \gate{U^2_f(\Gamma_2, \Delta_2, \Sigma_2)}  & \meter &\cw & \rstick{x_2} \\
    \cdots              &           &                                       & \cdots                                    & \cdots &    &  \\
    \lstick{\ket{0}}    & \gate{\HG}  & \ghost{U_z(\boldsymbol\alpha)}        & \gate{U^n_f(\Gamma_n, \Delta_n, \Sigma_n)}  & \meter &\cw & \rstick{x_n} 
    }
\end{align}
where: $x_i \in \{0, 1\}$; the unitaries are defined by~\eqref{eqn:diagonalunitary} and~\eqref{eqn:finalmeasurementgate}; $S_j$ indicates the subset of qubits on which each operator, $j$, is applied; and $\boldsymbol\alpha := \{\alpha_j\}$ are the trainable parameters.
\begin{equation}\label{eqn:diagonalunitary}
  U_z(\boldsymbol\alpha) \coloneqq  \prod_j U_z \left( \alpha_j, S_j \right) = \prod_j\exp \left(  \mathrm{i} \alpha_j \bigotimes_{k \in S_j} Z_k \right)
 \end{equation}
  \begin{equation}\label{eqn:finalmeasurementgate}
    U_f \left( \mathbf{\Gamma}, \mathbf{\Delta}, \mathbf{\Sigma} \right) \coloneqq \exp\left( \mathrm{i}\sum\limits_{k=1}^n \Gamma_k \XG_k + \Delta_k \YG_k +\Sigma_k \ZG_k\right)
\end{equation}
The model was dubbed `Ising' since restricting to the case $|S_j| \leq 2$ (since only single and two-qubit gates are required for universal quantum computation) the term in the exponential of~\eqref{eqn:diagonalunitary} becomes exactly the Ising Hamiltonian discussed in~\eqref{eqn:ising_model_hamiltonian}.

Furthermore, we also showed in~\cite{coyle_project_2018} how the shallowest depth ($p=1$) version of the $\QAOA$ algorithm (see~\eqref{eqn:qaoa_ansatz_basic} in~\secref{ssec:vqa_ansatzse}) could also be recovered by a specific setting of the final measurement angles in~\eqref{eqn:finalmeasurementgate}.

By inheriting the `hardness' results of standalone $\IQP$ and $p=1$ $\QAOA$ circuits~ \cite{bremner_classical_2011, fujii_commuting_2017, farhi_quantum_2016}, we could show that the $\QCIBM$ remained in the same complexity class as the the standalone versions during training of the generative model, and then also sampling from a $\QCIBM$ could not be simulated classically (up to multiplicative error).

However, while hardness-of-simulating the generative model is at least an important step on the road to true quantum advantage\footnote{It surely is at least a \emph{necessary} condition; if it was possible to classically simulate the generative quantum model at each point during training - one could just use the classical simulation instead of having to actually build a quantum computer (which would be significantly more expensive, from a practical sense).}, it is not the end of the story. One would also like to make a case for the \emph{hardness of learning}, i.e. that the $\QCIBM$ (or indeed any quantum generative model) could learn a distribution which could not be \emph{learned} classically. Such a `quantum learning advantage' (especially in a near-term NISQy regime) would be concrete proof of the usefulness of quantum machine learning.

\subsection[\texorpdfstring{\color{black}}{} The supremacy of quantum learning]{The supremacy of quantum learning} \label{ssec:generative_modelling/born_machine/qls}
The contribution of this section is to formalise this question, using concepts from learning theory, and provide an initial attempt at showing such an advantage, by focusing on the `hard-to-simulate' quantum supremacy circuits discussed above. We call this \emph{quantum learning supremacy} (QLS) to parallel the corresponding QCS which deals exclusively with the underlying sampling problem. Informally a generative quantum machine learning algorithm is said to have demonstrated QLS, if it is possible for it to efficiently learn a representation of a distribution for which there does not exist a classical learning algorithm achieving the same end. More specifically, the quantum device has the ability to produce samples according to a distribution that is close in total variation distance (\eqref{eqn:total_variation_dist_defn})\footnote{The reason for this choice will become apparent shortly.} to some target distribution, using a polynomial number of samples from the target.

To begin, we must understand the inputs and outputs to learning algorithm, for which we have a certain amount of freedom. The inputs are samples, either classical vectors, or quantum states encoding a superposition of such bitstring states, i.e. \textit{qsamples} \cite{schuld_supervised_2018} (a special case of the amplitude encoding from \defref{defn:amplitude_encoding}):
\begin{defbox}
    \begin{definition}[Qsample]\label{defn:qsample_definition}~ \\
    Given a discrete probability distribution, $[p(\xbs^1),p(\xbs^2), \dots, p(\xbs^{2^n})]^T,~ \xbs^i \in \{0, 1\}^n$, a qsample encoding of this distribution maps $p(\xbs) \rightarrow E^{\text{qsamp}}$ given by:
    \begin{equation} \label{eqn:qsample_encoding_defn}
        E^{\text{qsamp}} : \{p(\xbs^i)\}_{i} \rightarrow \ket{p(\xbs)} := \sum\limits_{i=1}^{2^n} \sqrt{p(\xbs^i)}\ket{\xbs^i}
    \end{equation}
    \end{definition}
\end{defbox}
The state in \eqref{eqn:qsample_encoding_defn} is clearly a generalisation of the classical distribution, since we can measure all qubits in the computational basis, and we will recover the sample $\xbs^i$ with the correct probability, $\sqrt{p(\xbs^i)}^2 = p(\xbs^i)$. Giving the learning algorithm to classical samples, $\xbs_i$ or a quantum superposition over all possible samples (i.e. a qsample, \eqref{eqn:qsample_encoding_defn}) can lead to advantages in supervised learning theory~\cite{servedio_equivalences_2004}, so it is not unreasonable to expect an analogue is possible for generative modelling. 

While the sample complexity of quantum versus classical learning has been studied in the supervised learning sense (see the discussions in~\secref{ssec:prelim/machine_learning/learning_theory} and \secref{sec:prelim/qml/q_learning_theory}), the same cannot be said about unsupervised learning. Fortunately, we do have a framework in the classical setting for studying distribution learning problems, so-called \emph{distribution} learning theory, introduced by \cite{kearns_learnability_1994}. Here we bring this forward to the quantum setting.

Unlike the supervised setting, there are more than one type of `learner' (or more than two if one also counts the quantum generalisation). We can define distribution learners which much learn solely from samples from the distribution (learning with \emph{Generators}), and those which can learn from the probabilities themselves (learning with \emph{Evaluators}). For simplicity, we assume the distributions classes in question (as in \cite{kearns_learnability_1994}), are discrete over binary vectors of length $n$, denoted $\mathcal{D}_n$. Let us now define Generators and Evaluators:

\begin{defbox}
    \begin{definition}[Generator~\cite{kearns_learnability_1994}]\label{defn:generator_supp}~ \\
    A class of distributions, $\mathcal{D}_n$, has efficient Generators, $\GEN_{D}$, if for every distribution $D \in \mathcal{D}_n$, $\GEN_{D}$ produces samples in $\{0, 1\}^n$ according to the exact distribution $D$, using polynomial resources.
    The generator may take a string of uniformly random bits, of size polynomial in $n$, $r(n)$, as input.
    \end{definition}
\end{defbox}

Notice this definition allows, for example, for the Generator to be either a classical circuit, or a quantum circuit, with polynomially many gates. Further, in the definition of a classical Generator~\cite{kearns_learnability_1994} a string of uniformly random bits is taken as input, and then transformed into the randomness of $D$. However, a quantum Generator would be able to produce its own randomness and so no such input is necessary. In this case the algorithm could ignore the input string $r(n)$. 
\begin{defbox}
    \begin{definition}[Evaluator~\cite{kearns_learnability_1994}]\label{defn:evaluator_supp}~ \\
         A class of distributions, $\mathcal{D}_n$ has efficient Evaluators, $\EVAL_{D}$, if for every distribution $D \in \mathcal{D}_n$, $\EVAL_{D}$ produces the weight of an input $\xbs$ in $\{0, 1\}^n$ under the exact distribution $D$, i.e. the probability of $\xbs$ according to $D$. The Evaluator is \emph{efficient} if it uses polynomial resources.
    \end{definition}
\end{defbox}
The distinction between $\EVAL$ and $\GEN$ is important and interesting since the output probabilities of $\IQP$ circuits are $\#\Pee$-Hard to compute and also hard to sample from by classical means, as discussed above~\cite{bremner_classical_2011}, yet the distributions they produce can be sampled from efficiently by a quantum computer. This draws parallels to examples in \cite{kearns_learnability_1994} where certain classes of distributions are shown not to be learnable efficiently with an Evaluator, but they \emph{are} learnable with a Generator.

For our purposes, the following definitions of learnable will be used. In contrast to \cite{kearns_learnability_1994}, which was concerned with defining a `good' generator to be one which achieves closeness relative to the $\KL$ divergence (\eqref{eqn:kl_divergence_defn}), we wish to expand this to general cost functions, $d$. The reason for this will become apparent later on, when dive into quantum sampling hardness results (high typically strive for closeness in $\TV$) in more detail, and when we introduce our new training methods for the $\QCIBM$ .

\begin{defbox}
    \begin{definition}[$\left( \text{d} , \epsilon \right)$-Generator]~ \\
        For a cost function, $\text{d}$, let $D \in \mathcal{D}_n$. Let $\GEN_{D'}$ be a Generator for a distribution $D'$. We say $\GEN$ is a $\left( \text{d} , \epsilon \right)$-Generator for $D$ if $\text{d}(D, D') \leq \epsilon$.
    \end{definition}
\end{defbox}

\noindent A similar notion of an $\epsilon$-good Evaluator could be defined.

\begin{defbox}
    \begin{definition}[$\left(\text{d} , \epsilon, \C \right)$-Learnable]\label{def:weak_learnable_supp}~ \\
    For a metric $\text{d}$, $\epsilon > 0$, and complexity class $\C$\footnote{The introduction of a complexity class here should not be considered rigorous, but is used to allow the algorithm to be either quantum or classical in nature.}, a class of distributions $\mathcal{D}_n$ is called $\left( \text{d} , \epsilon, \C \right)$-learnable (with a Generator) if there exists an algorithm $\mathcal{A} \in \C$, called a learning algorithm for $\mathcal{D}_n$, which given $0 < \delta < 1$ as input, and given access to $\GEN_{D}$ for any distribution $D \in \mathcal{D}_n$, outputs $\GEN_{D'}$, a $\left( \text{d} , \epsilon \right)$-Generator for $D$, with high probability:
    \begin{equation}
        \Pr \left[ \text{d} \left( D, D' \right) \leq \epsilon \right] \geq 1 - \delta
    \end{equation}
    $\mathcal{A}$ should run in time $\mathsf{poly}(1/\epsilon, 1/\delta, n)$.
    \end{definition}
\end{defbox}

\noindent Finally, we define what it would mean for a quantum algorithm to be superior to any classical algorithm for the problem of distribution learning, before moving onto our initial attempt to achieve such a thing:

\begin{defbox}
    \begin{definition}[Quantum learning supremacy (QLS)]\label{defn:quantum_learning_supremacy}~ \\
        An algorithm $\mathcal{A} \in \BQP$ is said to have demonstrated the supremacy of quantum learning over classical learning if there exists a class of distributions $\mathcal{D}_n$ for which there exists $d ,\epsilon$ such that $\mathcal{D}_n$ is $\left( d ,\epsilon, \BQP \right)$-Learnable, but $\mathcal{D}_n$ is not $\left( d ,\epsilon, \BPP \right)$-Learnable.
    \end{definition}
\end{defbox}

\subsection[\texorpdfstring{\color{black}}{} A learning advantage via quantum computational supremacy]{Quantum learning supremacy via quantum computational supremacy} \label{ssec:generative_modelling/born_machine/qls_via_qcs}
Let us now move onto an initial attempt to achieve QLS. Since this question of distribution learning advantage is motivated by the classical hardness of sampling from the distributions generated by a quantum circuit, a natural distribution class, $\mathcal{D}_n$ to learn would be exactly these `quantum supremacy distributions'. We focus on $\IQP$ circuits as a concrete example, although a similar logic could in principle be applied to alternative QCS candidates such as $\BosonSampling$ or $\RCS$. For QCS, there are a number of further subtleties which must be address including: worst versus average case hardness of classical simulation, and the notions of the `error' of such simulations, which relate to the probability metrics introduced in \secref{ssec:prelim/qc/distance_measures/probability}. For our present discussion, the latter is more relevant, although for discussions of the former and a comparison between the different QCS proposals, see~\cite{harrow_quantum_2017, bouland_complexity_2019}.

The most common error models used in proofs of QCS are \emph{multiplicative} and \emph{additive} (or total variation) error:

\begin{defbox}
    \begin{definition}[Multiplicative error]\label{defn:defnmulterror}~ \\
        A circuit family is weakly simulatable within multiplicative (relative) error, if there exists a classical probabilistic algorithm, $Q$, which produces samples, $\xbs$, according to the distribution, $q(\xbs)$,  in time which is polynomial in the input size, which differs from the ideal quantum distribution, $p(\xbs)$, by a multiplicative constant, $c > 1$:
        \begin{equation}
            \frac{1}{c}p(\xbs) \leq q(\xbs) \leq c p(\xbs) \qquad \forall \xbs \label{eqn:multerror}
        \end{equation}
    \end{definition}
\end{defbox}

It would be desirable to have a quantum sampler which could achieve the bound of \eqref{eqn:multerror}, but this is not believed to be experimentally achievable, i.e. it is not believed that a \textit{physical} quantum device could achieve such a multiplicative error bound on its probabilities, relative to its ideal functionality (i.e. replacing $q$ in \eqref{eqn:multerror} by the output distribution of a noisy quantum device). That is why much effort has been put into trying to find systems for which QCS could be provably demonstrated according to the total variation distance error condition, \eqref{eqn:tv_simulation_error}, which is easier to achieve on near term quantum devices. 

\begin{defbox}
    \begin{definition}[Total variation ($\TV$) error]\label{defn:tv_simulation_error}~ \\
        A circuit family is weakly simulable within variation distance  error, $\epsilon$, if there exists a classical probabilistic algorithm, $Q$, which produces samples, $\xbs$, according to the distribution, $q(\xbs)$, in polynomial time, such that it differs from the ideal quantum distribution, $p(\xbs)$ in total variation distance, $\epsilon$:
        \begin{equation}\label{eqn:tv_simulation_error} 
          \TV(p, q)  \coloneqq \frac{1}{2}\sum\limits_{\xbs}|p(\xbs)- q(\xbs)| \leq \epsilon 
        \end{equation}
    \end{definition}
\end{defbox}

It has been proven that weakly simulating $\IQP$ circuits to a fixed $\TV$ error is classically hard. Specifically, if a classical algorithm could sample from a distribution, $q$, for which $\TV(p, q) \leq 1/384$, where $p$ is the output distribution from an $\IQP$ circuit, then the polynomial hierarchy would collapse to the third level (see \cite{bremner_average-case_2016} for a formal statement of the relevant theorem), assuming the hardness of computing the Ising partition function.

Now, where does QLS fit into this? Well, let us revisit the relationship between these definitions, and those for distribution learning in \secref{ssec:generative_modelling/born_machine/qls}. Firstly, note that there is a direct and obvious connection to the `strength' of classical simulators of quantum circuits.

The two primary types of classical simulation are \emph{strong} and \emph{weak}\footnote{We already used this terminology in \defref{defn:defnmulterror} and \defref{defn:tv_simulation_error}.} simulators:

\begin{defbox}
    \begin{definition}[Strong and weak classical simulation~\cite{bremner_classical_2011, fujii_commuting_2017}]\label{defn:strong_weak_sim}~\\ 
        A uniformly generated quantum circuit, $C$, from a family of  circuits, with input size $n$, is \emph{weakly simulatable} if, given a classical description of the circuit, a classical algorithm can produce samples, $\xbs$, from the output distribution, $p(\xbs)$, in $\poly(n)$ time. 
        
        On the other hand, a \emph{strong simulator} of the family would be able to compute the output probabilities, $p(\xbs)$, and also all the marginal distributions over any arbitrary subset of the outputs. Both of these apply with some notion of error, $\epsilon$.
    \end{definition}
\end{defbox}
As mentioned in \cite{bremner_classical_2011}, strong simulation is a harder task than weak simulation, and it is this weak simulatability which we want to rule out as being classically hard, since it better captures the process of sampling. The suitable notion of error, $\epsilon$, for strong simulation would be the precision to which the probabilities can be computed. Recalling the definitions in \secref{ssec:generative_modelling/born_machine/qls}, it is clear that an Evaluator for a quantum circuit would be a strong simulator of it, and a Generator would be a weak simulator. 

Next, let us take the learning metric in \defref{defn:quantum_learning_supremacy}, to be the total variation distance, $\text{d} = \TV$, and let $\mathcal{D}_n$ be the family of $\IQP$ distributions. Imagine, we were able to concoct a learning algorithm, $\mathcal{A}$ (for example the $\QCIBM$), to output a generator for a distribution, $D_{\QCIBM}$, ($\GEN_{D_{\QCIBM}}$) given $\poly$ many samples (classical or quantum) from an $\IQP$ distribution, $D_{\IQP}$. Furthermore, assume we have $\TV(D_{\QCIBM}, D_{\IQP})\leq \gamma$. 

Now, assuming $D_{\QCIBM}$ distributions are classically \emph{easy} to emulate\footnote{In other words, proof by contradiction.} within a fixed $\TV$ error, $\delta$, meaning there exists a classical probabilistic algorithm, $\mathcal{C}$, could output a distribution, $D_{\mathcal{C}}$ such that:
\begin{equation}
    \TV(D_{\mathcal{C}}, D_{\QCIBM}) \leq \delta
\end{equation}
Then, by a simple triangle inequality argument, we have:

\begin{align}
    &\TV(D_{\IQP}, D_{\mathcal{C}}) = \frac{1}{2}\sum_{\xbs}|D_{\IQP}(\xbs)-D_{\mathcal{C}}(\xbs)| \\
    &= \frac{1}{2}\sum_{\xbs}|D_{\IQP}(\xbs) - D_{\QCIBM}(\xbs) + D_{\QCIBM}(\xbs)-D_{\mathcal{C}}(\xbs)|\\
    &\leq \frac{1}{2}\sum_{\xbs}|D_{\IQP}(\xbs) -D_{\QCIBM}(\xbs)| +\frac{1}{2}\sum_{\xbs}|D_{\QCIBM}(\xbs)-D_{\mathcal{C}}(\xbs)|\\
    &\leq \delta+\gamma =: \epsilon = \frac{1}{384}
\end{align}
Therefore, if we are able to make $\delta + \gamma$ smaller than $1/384$, we arrive at a contradiction to the result of \cite{bremner_average-case_2016}, and as such, $D_{\QCIBM}$ must also be difficult to simulate within a small fixed $\TV$ distance. As a result, something classically infeasible has been achieved\footnote{A very related idea was proposed by~\cite{rocchetto_learning_2018}, where the authors studied the learnability of classical hard distributions using (classical) variational autoencoders.}.

This above discussion covers the `classical hardness' of the learning task, but we have yet to discuss the feasibility of the learning problem itself, i.e. outputting a generator distribution achieving $\gamma$ closeness in $\TV$ to an $\IQP$ distribution. 

Unfortunately, it is here the argument appears to fail. A key ingredient in proofs of the hardness of classically simulating quantum supremacy distributions is the property of \emph{anti-concentration}, which formally means that given a random instance from a `hard' circuit family, for example an instance of an $\IQP$ circuit, then it does not become too unlikely that the probability of any fixed outcome, $\xbs$, by measuring the state $U\ket{0}^{\otimes n}$, is much smaller than it would be if $\xbs$ was drawn from the uniform distribution. In other words, anti-concentrated distributions are `exponentially flat':
\begin{equation}
    \exists \alpha, \gamma > 0, \text{s.t.} \forall \xbs \in \{0, 1\}^n, \qquad \underset{U \sim \mu}{\text{Pr}}\left(p_U(\xbs) := |\bra{\xbs}U\ket{0}^n|^2 \geq \frac{\alpha}{2^n}\right) \geq \gamma, 
\end{equation}
where the unitary is drawn according to a measure $\mu$ on the unitary group. 

Now comes the rub; the work of \cite{hangleiter_sample_2019} proved that \emph{verifying} such anti-concentrating distributions, for example $\IQP$\footnote{It was also proven for $\BosonSampling$ and $\RCS$. For $\IQP$ this anti-concentration behaviour has been explicitly proven, for $\BosonSampling$ on the other hand it is only a (believeable) conjecture.} would require exponentially many samples from the distribution to be verified. Verification in this sense means building a `testing' algorithm which can distinguish whether the distribution to be verified is far or close to the ideal distribution in $\TV$. 

As a consequence of this no-go result, one can conjecture that also building an efficient Generator for an anti-concentrating distribution is also not feasible. This is because one could imagine a verification algorithm to be a subroutine in any learning algorithm attempting to demonstrate QLS. The key property of QCS circuit families being hard to simulate, is also what makes them hard to learn. However, the above discussion assumed efficient \emph{classical} sample access to the QCS distribution - it may be possible to bypass the classical no-go theorem given \emph{qsample} access. 

To conclude, we note that at the time of writing the question of a \emph{near-term} quantum advantage in generative modelling is still an open question. However, since the completion of the works in this chapter, some interesting progress was made towards this goal. We return to this discussion in the conclusion.

\section[\texorpdfstring{\color{black}}{} Training a quantum circuit Born machine]{Training a quantum circuit Born machine} \label{ssec:born_machine/training}

Let us now introduce our second contribution in the subject of quantum generative modelling. As mentioned in \secref{sec:born_machines/introduction} this will be the introduction of new gradient based training methods for the quantum circuit Born machine. We will use primarily the $\QCIBM$ as a specific instance, but remark that everything presented in this section applies generally to QCBMs. We already introduced much of the relevant details for this section in \secref{ssec:prelim/qc/distance_measures/probability} which described various measures on the space of probability distributions. 

Recall in \secref{ssec:prelim/qc/distance_measures/probability} we introduced two families of distribution measures, specifically $f$-divergences (\eqref{defn:f_divergence}), and integral probability metrics (IPMs) (\eqref{defn:integral_probability_metrics}). In order to train generative models, these probability measures will serve as our `cost functions', $\Cbs$, as in \secref{ssec:vqa_cost_functions}. Since quantum circuit Born machines are just a special case of variational algorithms applied to generative modelling, we similarly wish for the cost functions we use to have the same well defined properties as with any $\VQA$ (discussed in \secref{ssec:vqa_cost_functions}), namely faithfulness, efficient computability and operational meaning. We go into significant detail with these properties in this sections since they turn out to be \emph{extremely} important for training generative models. 

Let us begin by discussing $f$-divergences, any why they may not satisfy the `efficient computability' criterion. Let us take the $\KL$ divergence as a special case. Rewriting ~\eqref{eqn:kl_divergence_defn} in terms of the model distribution (specifically the $\QCIBM$ output distribution), $p_{\paramtheta}$, and let us use the notation, $\pi$ to denote the `data' distribution we are trying to fit:
\begin{equation}\label{eqn:born_machines/kl_divergence_defn}
    \Cbs_{\KL}(\paramtheta) := \KL(\pi, p_{\paramtheta}) := \sum_{\xbs} \pi(\xbs) \log \pi(\xbs) -  \sum_{\xbs} \pi(\xbs) \log p_{\paramtheta}(\xbs)
\end{equation}
Since the first term (the entropy of $\pi$) is constant with respect to $p_{\paramtheta}$, when solving the optimisation, $\argmin_{\paramtheta} \Cbs_{\KL}(\paramtheta)$, it can be ignored. Therefore, we now solve instead $\argmax_{\paramtheta} \XE(\pi, p_{\paramtheta})$. However, recall in \secref{ssec:born_machine/quantum_advantage}, we discussed how computing the outcome probabilities of quantum circuits was $\#\Pee$-hard, hence we may require exponential resources to get a good estimate of $p_{\paramtheta}(\xbs)$ for all $\xbs \in \{0, 1\}^n$. Since the expression in~\eqref{eqn:born_machines/kl_divergence_defn} cannot be written as a sampling expression over $p_{\paramtheta}$\footnote{By this we mean that the $\KL$ divergence cannot be written in terms of quantities like $\underset{\xbs\sim p_{\paramtheta}}{\mathbb{E}}(\cdot)$, which can be estimated by simply drawing samples from $p_{\paramtheta}$. Terms of this form can usually be estimated more efficiently since we only need to sample the high-probability values, $\xbs$, to get a good estimate of the expectation value - we do not necessarily need to know the probability at \emph{every} $\xbs$ (of which there are exponentially many)}, we do not expect to be able to efficiently compute the term $\XE(\pi, p_{\paramtheta})$, and hence we cannot compute $\KL(\pi, p_{\paramtheta})$. This was realised specifically for the case of QCBMs by \cite{liu_differentiable_2018}, where the corresponding gradients of the $L$ parameters in a QCBM were computed and found to be\footnote{An important point to note here, we have a slightly different functional form for the gradients than those presented in \cite{liu_differentiable_2018}. This is because we will be using these gradients to train the $\QCIBM$ in~\eqref{circuit:ising_born_machine}, and the gates in this circuit are of the form $\erm^{\irm \theta\Sigma}$, rather than $\erm^{-\irm (\theta/2) \Sigma}$. These are essentially equivalent at any rate, and only introduce an alternative factor in front of the gradient term to change its magnitude. We will use this formulation of the gradient for all the cost functions described here.}:
\begin{equation}\label{eqn:qcbm_kldiv_gradients}
    \frac{\partial \KL}{\partial \theta_k} = -\sum\limits_{\xbs} \pi(\xbs) \frac{p_{\paramtheta^-_k} - p_{\paramtheta^+_k}}{p_{\paramtheta}}
\end{equation}
where we use the notation $\paramtheta^{\pm}_k := (\theta_1, \theta_2, \dots, \theta_k\pm \sfrac{\pi}{2}, \dots, \theta_L)$ to denote the parameter shift rule in~\eqref{eqn:parameter_shift_rule}. Hence, if we cannot compute the probabilities $p_{\paramtheta}(\xbs)$ efficiently, we will not be able to compute neither $f$-divergences, nor their gradients. A solution to this was realised by ~\cite{liu_differentiable_2018} (which we alluded to in \secref{ssec:prelim/qc/distance_measures/probability}), which was to instead use the $\MMD$~\eqref{eqn:mmd_exact} to efficiently train QCBMs in a differentiable manner. Since the $\MMD$ can be written exclusively as expectation values over the data distribution and the QCBM distribution it can be evaluated efficiently (specifically with the quadratic sample complexity of ~\eqref{eqn:mmd_samplecomplexity}). The same can be said for its gradients, which were computed by ~\cite{liu_differentiable_2018} to be: 
\begin{equation} \label{eqn:mmd_gradient_exact}
    \frac{\partial \Cbs_{\MMD}(p_{\paramtheta}, \pi)}{\partial \paramtheta_k} =\underset{\substack{\boldsymbol{a} \sim p_{\paramtheta_k}^-\\ \xbs \sim p_{\paramtheta}}}{2\mathbb{E}}(\kappa(\boldsymbol{a}, \xbs))- \underset{\substack{\boldsymbol{b} \sim p^+_{\paramtheta_k}\\ \xbs \sim p_{\boldsymbol\paramtheta}}}{2\mathbb{E}}(\kappa(\boldsymbol{b}, \xbs)) 
    - \underset{\substack{\boldsymbol{a} \sim p^-_{\paramtheta_k}\\ \ybs \sim \pi}}{2\mathbb{E}}(\kappa(\boldsymbol{a}, \ybs))  +\underset{\substack{\boldsymbol{b} \sim p^+_{\paramtheta_k}\\ \ybs \sim \pi}}{2\mathbb{E}}(\kappa(\boldsymbol{b}, \ybs))
\end{equation}
where we define $\Cbs_{\MMD}:=  \MMD^{\kappa}$ from~\eqref{eqn:mmd_exact}.

Both of the cost functions we have discussed so far are `faithful', meaning in this case that $\Cbs(p_{\paramtheta}, \pi) = 0 \iff p_{\paramtheta} = \pi$ (by definition, since they are a divergence, and a metric between probability distributions respectively\footnote{A subtle point for the diligent reader: it may be expected that the faithfulness of the $\MMD$ must have something to do with the kernel function and feature map - indeed this is true, the $\MMD$ is only faithful if the corresponding kernel is `universal'. We return to a discussion of kernel universality later in this Thesis.}). We have established how the $\MMD$ is an efficient method to train QCBMs (and generative models generally), whereas the $\KL$ divergence may not be. The final quality we have not touched upon is the \emph{operational meaning} of these cost functions, which we take in this sense to be how `powerful' they are as training mechanisms. 

By this, we mean their performance and properties relative to $\TV$\footnote{We take this viewpoint specifically in light of our discussion of QLS in \secref{ssec:born_machine/quantum_advantage}, where total variation distance is the relevant measure.}. Firstly, it is clear that the $\KL$ divergence is stronger than $\TV$ since the upper bound it provides via Pinsker's inequality~\eqref{eqn:pinskers_inequality}. Hence, if we were able to efficiently train a QCBM with the $\KL$ divergence to a particular value, we can directly upper bound the $\TV$ from this value - closeness in $\KL$ implies closeness in $\TV$. 

The same cannot be said unfortunately about the $\MMD$. Specifically, the $\MMD$ provides only a \emph{lower} bound on $\TV$ from~\cite{sriperumbudur_integral_2009}:
\begin{equation} \label{eqn:tvd_mmd_lower_bound}
    \TV(p_{\paramtheta}, \pi) \geq \frac{\sqrt{\MMD(p_{\paramtheta}, \pi)}}{\sqrt{k}}
\end{equation}
if we have that $k := \sup_{\xbs \in \mathcal{X}^n} \kappa(\xbs, \xbs) < \infty$. For example, taking the Gaussian kernel~\eqref{eqn:gaussian_kernel} (or even the quantum kernel~\eqref{eqn:quantum_kernel}), we have:
\begin{equation}
\kappa_G(\xbs, \xbs) =  \erm^{-\frac{1}{2}|\xbs - \xbs|^2} = 1
\end{equation}
hence $k = 1$ and the lower bound is immediate. With this in mind (we will circle back to this operational meaning afterwards), let us move onto the new cost functions we propose for training a QCBM. These are the \emph{Stein} discrepancy, and the \emph{Sinkhorn} divergence.

\subsection[\texorpdfstring{\color{black}}{} Training with the Stein discrepancy]{Training with the Stein discrepancy} \label{ssec:born_machine/stein_discrepancy}

Firstly, let's look at the Stein discrepancy ($\SD$). The $\SD$ has become popular for goodness-of-fit tests~\cite{liu_kernelized_2016}, i.e. testing whether samples come from a particular distribution or not\footnote{In contrast, the $\MMD$ is typically used for kernel two-sample tests~\cite{gretton_kernel_2007}.}. This discrepancy is based on Stein's method~\cite{stein_bound_1972}, which is a way to bound distance metrics between probabilities including, for example, IPMs.

The original formulation of the discrepancy was done in the the continuous case. However, in the case of a QCBM, we require a \emph{discrete} version of the discrepancy (since the output are samples, $\xbs \in \{0, 1\}^n$). 

To set the scene, let us begin with the continuous version. A key ingredient is Stein's identity (in the case where the sample space is one-dimensional, $x\in \mathcal{X} \subseteq \mathbb{R}$) given by:
\begin{equation}\label{eqn:stein_identity}
    \mathbb{E}_\pi\left[\mathcal{A}_\pi\phi(x)\right] = \mathbb{E}_\pi \left[s_\pi(x)\phi(x)+ \nabla_x\phi(x)\right]  = 0 
\end{equation}
where $s_\pi(x) = \nabla_x\log(\pi(x))$ is the \emph{Stein score} function of the distribution $\pi$, $\nabla_x f(x)$ is the gradient of a function, $f$ and $\mathcal{A}_\pi$ is a so-called \emph{Stein operator} of $\pi$. The functions, $\phi$, which obey~\eqref{eqn:stein_identity}, are said to be in the \emph{Stein class}\footnote{These functions, $\phi$ are essentially those which allow the identity to be proven via integration by parts, i.e. those which are smooth and have appropriate boundary conditions.} of the distribution $\pi$. From Stein's identity, one can  define a discrepancy  between the two distributions, $p_{\paramtheta}, \pi$, by the following optimisation problem, \cite{yang_goodness--fit_2018}:
\begin{equation} \label{eqn:stein_discrepancy}
    \text{d}_{\SD}(p_{\paramtheta}|| \pi)  \coloneqq \sup_{\phi \in \mathcal{F}}\left(\mathbb{E}_{p_{\paramtheta}}[\mathcal{A}_\pi\phi] - \mathbb{E}_{p_{\paramtheta}}[\mathcal{A}_{p_{\paramtheta}}\phi]\right)^2 
\end{equation}
If $p_{\paramtheta} \equiv \pi$, then $\text{d}_{\SD}(p_{\paramtheta}|| \pi) = 0$ by~\eqref{eqn:stein_identity}. Exactly as with the $\MMD$, the power of the discrepancy in~\eqref{eqn:stein_discrepancy}, will depend on the choice of the function space, $\mathcal{F}$. By choosing it to be a RKHS, a kernelised form which is computable in closed form can be obtained. Also, this form of~\eqref{eqn:stein_discrepancy} is very reminiscent of that of the integral probability metrics,~\eqref{defn:integral_probability_metrics} (and indeed the $\SD$ can be written in such a form where the function family in the IPM is distribution-dependent).

From~\eqref{eqn:stein_discrepancy}, the problem arises. Due to the gradient term, $\nabla_x$, in~\eqref{eqn:stein_identity}, the above expressions are only defined for smooth probability densities, $p_{\paramtheta}, \pi$, which are supported on \textit{continuous} domains (e.g. $\mathbb{R}$). Therefore, in order to make the $\SD$ compatible with training a QCBM, we must perform a discretisation procedure.

Fortunately, this has been addressed by~\cite{yang_goodness--fit_2018}, which adapted the kernelised $\SD$ to the discrete domain. This was achieved by introducing a discrete gradient `shift' operator. Just as above, let us assume we have $n$-dimensional sample vectors, $\xbs \in \mathcal{X}^n \subseteq \mathbb{R}^n$ (where $\mathcal{X}$ is a discrete set). First of all, we shall need some definitions~\cite{yang_goodness--fit_2018}:

\begin{defbox}
\begin{definition}[Cyclic permutation]\label{defn:cyclic_permutation}~ \\
    For a set $\mathcal{X}$ of finite cardinality, a cyclic permutation $\neg:\mathcal{X} \rightarrow \mathcal{X}$ is a bijective function such that for some ordering $x^{[1]},x^{[2]}, \dots, x^{[|\mathcal{X}|]}$ of the elements in $\mathcal{X}$, $\neg = \neg_i: x^{[i]} \mapsto x^{[(i+1)\mod|\mathcal{X}|]}, \forall i = 1,2,\dots, |\mathcal{X}|$
\end{definition}
\end{defbox}
\begin{defbox}
\begin{definition}[Partial difference operator and difference score function]\label{defn:partial_diff_op}~ \\
Given a cyclic permutation  $\neg$ on $\mathcal{X}$, for any vector, $\xbs = [x_1, \dots, x_n]^T \in \mathcal{X}^n$. For any function $f:\mathcal{X}^n \rightarrow \mathbb{R}$, denote the (partial) difference operator as:
\begin{equation}
    \Delta_{x_i}f(\xbs) := f(\xbs) - f(\neg_i\xbs) \qquad \forall i=1, \dots, d \label{discreteshiftoperator}
\end{equation}
with $\Delta f(\xbs) = (\Delta_{x_1}f(\xbs), \dots \Delta_{x_n}f(\xbs))^T$. Define the (difference) score function for a positive probability mass function, $p(\xbs) > 0 ~\forall \xbs$ as:

\begin{equation} \label{eqn:discrete_stein_score_function_defn}
    \mathbf{s}_p(\xbs) \coloneqq \frac{\Delta p(\xbs)}{p(\xbs)}, \qquad
    (\mathbf{s}_p(\xbs))_i = \frac{\Delta_{x_i} p(\xbs)}{p(\xbs)} = 1 -  \frac{p(\neg_i\xbs)}{p(\xbs)}
\end{equation}
\end{definition}
\end{defbox}

Furthermore, \cite{yang_goodness--fit_2018} defines the inverse permutation by $\revneg = \revneg_i: x^{[i]} \mapsto x^{[(i-1)\mod|\mathcal{X}|]}$, and the \textit{inverse} shift operator by:
\begin{equation}\label{eqn:inverse_discrete_shift_operator}
    \Delta^*_{x_i}f(\xbs)  \coloneqq f(\xbs) - f(\revneg_i\xbs) ~\forall i=1, \dots, n
\end{equation}
For our purposes, this generalisation is actually not necessary. Since the sample space for a single qubit is binary, $\mathcal{X} = \{0, 1\}$, the forward and reverse permutations are identical, so $\Delta = \Delta^*$. With this change to the gradient operator, \cite{yang_goodness--fit_2018} defines discrete versions of the Stein identity, and the kernelised Stein discrepancy which we can now state.
\begin{thmbox}
\begin{theorem}
[Difference Stein's identity for complex valued functions, adapted from Theorem 2 in \cite{yang_goodness--fit_2018}]\label{thm:complex_discrete_stein_identity}~ \\
    For any function\footnote{We have implicitly generalised the input space to be a vector space as in~\cite{yang_goodness--fit_2018}, but we have also taken the possibility that the feature maps may be complex valued. This allows us the freedom to use \emph{quantum} kernels for $\kappa$ in place of purely classical ones. This is also possible due to the nature of the discretisation which results in \emph{all} functions $\Phi:\mathcal{X}^n \rightarrow \mathbb{C}^m$ to be in the Stein class of the operator $\mathcal{A}_p$. The proof of this is a simple adaptation of the corresponding proof in \cite{yang_goodness--fit_2018} for real valued functions, and we omit it here.} $\Phi:\mathcal{X}^n \rightarrow \mathbb{C}^m$, and a probability mass function $p$ on $\mathcal{X}^n$, the discrete Stein identity is given by:
    \begin{equation} \label{eqn:vector_discrete_steinidentity}
            \underset{\xbs\sim p}{\mathbb{E}}\left[\mathcal{A}_p\Phi(\xbs)\right] =     \underset{\xbs\sim p}{\mathbb{E}}\left[\mathbf{s}_p(\xbs)\Phi(\xbs)^T - \Delta\Phi(\xbs)\right] = \mathbf{0} 
    \end{equation}
    where $\Delta\Phi(\xbs)$ is an $n\times m$ matrix: $(\Delta\Phi)_{ij} = \Delta_{x_i}\Phi_j(\xbs)$, i.e. shifting the $i^{th}$ element of the $j^{th}$ function value.
\end{theorem}
\end{thmbox}
\newpage
We can then reproduce the following Theorem:
\begin{thmbox}
\begin{theorem}[Theorem 7 in \cite{yang_goodness--fit_2018}]\label{thm:yangkerneldiscretestein}~ \\
The discrete kernelised $\SD$ between two distributions, $p, q$ is given by:
\begin{equation}\label{eqn:stein_discrepancy_cost}
    \Cbs_{\SD}(p, q)  := \mathbb{E}_{\xbs, \ybs\sim p}\left[\kappa_q(\xbs, \ybs)\right] 
\end{equation}
where $\kappa_q$ is the `Stein kernel':
\begin{multline} \label{eqn:stein_kernel}
    \kappa_q(\xbs, \ybs)   \coloneqq s_q(\xbs)^T\kappa(\xbs, \ybs)s_q(\ybs) -s_q(\xbs)^T\Delta_{\ybs}^*\kappa(\xbs, \ybs) \\
    - \Delta_{\xbs}^*\kappa(\xbs, \ybs)^Ts_q(\ybs) + \Tr(\Delta_{\xbs, \ybs}^*\kappa(\xbs, \ybs))
\end{multline}
\end{theorem}
\end{thmbox}

Let us pause briefly here to return to the discussion above regarding valid kernels. As mentioned above, for the $\MMD$ to be faithful, its kernel must be \emph{characteristic} or universal~\cite{fukumizu_kernel_2007, sriperumbudur_injective_2008}. The Gaussian kernel~\eqref{eqn:gaussian_kernel} is indeed one which is characteristic \cite{fukumizu_kernel_2007}, and some effort has been made to find conditions under which a kernel is characteristic \cite{sriperumbudur_universality_2011}. If the sample space is discrete, it turns out the condition for determining universality is simpler. In this case we only require the Gram matrix for the kernel, $K_{ij} = \kappa(\xbs^i, \ybs^j)$ to be positive definite\footnote{Recall, a matrix, $K$, is positive definite $\iff$ $\forall \xbs \in \mathbb{R}^d/\mathbf{0},~\xbs^TK\xbs > 0$.}~\cite{sriperumbudur_integral_2009}. If the Gram matrix for a kernel is positive definite we refer to the kernel itself as being \emph{strictly} positive definite (confusingly). Just as with the $\MMD$, it turns out this requirement of kernel-positive-definiteness is again exactly what makes the $\SD$ a valid discrepancy measure~\cite{yang_goodness--fit_2018}.

Next, we can compute the gradients of~\eqref{eqn:stein_discrepancy_cost} as with those of the $\MMD$ (replacing $p\rightarrow p_{\paramtheta}, q \rightarrow \pi$ for the QCBM and data distributions):
\begin{multline}
  \frac{\partial \Cbs_{\SD}}{\partial \theta_k} = \sum\limits_{\xbs, \ybs} \frac{\partial p_{\paramtheta}(\xbs)}{\partial \theta_k} \kappa_\pi(\xbs, \ybs) p_{\paramtheta}(\ybs) +  \sum\limits_{\xbs, \ybs} p_{\paramtheta}(\xbs)\kappa_\pi(\xbs, \ybs)   \frac{\partial p_{\paramtheta}(\ybs)}{\partial \theta_k}\\
    = \sum\limits_{\xbs, \ybs}p^-_{\theta_k}(\xbs)  \kappa_\pi(\xbs, \ybs) p_{\paramtheta}(\ybs) -  \sum\limits_{\xbs, \ybs}p^+_{\theta_k}(\xbs)  \kappa_\pi(\xbs, \ybs) p_{\paramtheta}(\ybs)\\ +\sum\limits_{\xbs, \ybs} p_{\paramtheta}(\xbs)\kappa_\pi(\xbs, \ybs)   p^-_{\theta_k}(\ybs) 
      - \sum\limits_{\xbs, \ybs} p_{\paramtheta}(\xbs)\kappa_\pi(\xbs, \ybs)p^+_{\theta_k}(\ybs)
\end{multline}
\begin{equation} \label{eqn:stein_gradient_exact}
    \frac{\partial \Cbs_{\SD}}{\partial \theta_k} 
    = \underset{\substack{\xbs \sim p^-_{\theta_k} \\ \ybs\sim p_{\paramtheta}}}{\mathbb{E}}[\kappa_\pi(\xbs, \ybs)] - \underset{\substack{\xbs \sim p^+_{\theta_k} \\ \ybs\sim p_{\paramtheta}}}{\mathbb{E}}[\kappa_\pi(\xbs, \ybs)] +\underset{\substack{\xbs \sim p_{\paramtheta}\\ \ybs\sim p^-_{\theta_k}}}{\mathbb{E}}[\kappa_\pi(\xbs, \ybs)] - \underset{\substack{\xbs \sim p_{\paramtheta} \\ \ybs\sim p^+_{\theta_k}}}{\mathbb{E}}[\kappa_\pi(\xbs, \ybs)] 
\end{equation}
Using again the parameter-shift rule~\eqref{eqn:parameter_shift_rule} and that the Stein kernel, $\kappa_{\pi}$, of~\eqref{eqn:stein_kernel} does not depend on the parameter, ${\theta_k}$.

Clearly, the primary difference between the $\MMD$ and the $\SD$ is the form of the kernel which must be computed in each case. To compute the Stein kernel, we must evaluate the base kernel $\kappa$ for each pair of samples. Let us assume this costs time $T(n)$ (a polynomial function of the number of qubits, $n$).

However, adding to the computational burden, we also must compute the following `shifted' versions of the kernel for each pair of samples, $(\xbs, \ybs)$, in both parameters:
\begin{equation}
    \Delta_{\xbs}\kappa(\xbs, \ybs) \qquad \Delta_{\ybs}\kappa(\xbs, \ybs) \qquad  \Tr\left[\Delta_{\xbs, \ybs}\kappa(\xbs, \ybs)\right]
\end{equation}
which are required in~\eqref{eqn:stein_kernel}. To compute one of these terms (for example, $\Delta_{\xbs}\kappa(\xbs, \ybs)$), we have:
\begin{equation} \label{eqn:stein_x_shifted_terms}
    \Delta_{\xbs}\kappa(\xbs, \ybs) = [\kappa(\xbs, \ybs), \dots, \kappa(\xbs, \ybs)]^T- [\kappa(\neg_1\xbs, \ybs), \dots, \kappa(\neg_n\xbs, \ybs)]^T 
\end{equation}
so we must evaluate $\kappa(\neg_i\xbs, \ybs)$ for $i = \{1,\dots, n\}$. Therefore, computing the shifted kernel operator in a single parameter takes $\mathcal{O}( T(n)\times (n+1))$. The same holds for the kernel gradient with respect to the second argument, $\Delta_{\ybs}\kappa(\xbs, \ybs)$.
For $\Delta_{\xbs, \ybs}\kappa(\xbs, \ybs)$, the process is slightly more involved because:

\begin{align}
    \Tr\left[\Delta_{\xbs, \ybs}\kappa(\xbs, \ybs)\right] &= \Tr\left(\Delta_{\xbs}[ \Delta_{\ybs}\kappa(\xbs, \ybs)] \right)= \Tr\left(\Delta_{\xbs}[\kappa(\xbs, \ybs) - \kappa(\xbs, \neg\ybs)]\right)   \\
    &= n\kappa(\xbs, \ybs) - \sum\limits_{i=1}^n\kappa(\xbs, \neg_i\ybs) - \sum\limits_{i=1}^n\kappa(\neg_i\xbs, \ybs) + \sum\limits_{i=1}^n\kappa(\neg_i\xbs, \neg_i\ybs)
\end{align}
Each individual term in the respective sums requires the same complexity, i.e. $\mathcal{O}(T(n))$ so the term $\Tr\left[\Delta_{\xbs, \ybs}\kappa(\xbs, \ybs)\right]$ overall requires $\mathcal{O}( T(n)\times (3n+1))$. 

However, we have not yet discussed the computation of the score function, $s_{\pi}$. If we are given oracle access to the probabilities, $\pi(\ybs)$, then there is no issue and $\SD$ will be computable. Unfortunately, in many practical applications this will not be the case. To deal with this scenario, in \secref{ssec:identity_computing_stein_score} and \secref{ssec:spectral_computing_stein_score}, we give two methods to approximate the score function, given samples from the dataset,  $\pi$. Notice that even with this hurdle (difficulty in compute the score), the $\SD$ is still more suitable than the $\KL$ divergence to train these models, since the latter requires computing the \textit{circuit} probabilities, $p_{\paramtheta}(\xbs)$, which is in general intractable, and so could not be done for \textit{any} dataset. In contrast, the Stein discrepancy only requires the \emph{data} probabilities, which may make it amenable for generative modelling using some datasets.

\subsection[\texorpdfstring{\color{black}}{} Computing the Stein score function]{Computing the Stein score function} 
\label{supp_matt:steinscoremethod}

Let us now discuss the final ingredient for the Stein discrepancy: the computability of the score function~\eqref{eqn:discrete_stein_score_function_defn}. For every sample, $\xbs\sim p_{\paramtheta}$, that we receive from the Born machine we require the score function of that \emph{same outcome} being outputted from the data distribution, $\xbs \sim \pi$. This involves computing $\pi(\xbs)$, and also $\Delta_{\xbs}\pi(\xbs)$, i.e. $\pi(\neg_i\xbs), \forall i \in\{1, \dots, n\}$.

This section will discuss our approach to approximate the score function of the data, $s_{\pi}$, from samples, $\{\ybs^i: \ybs \sim \pi(\ybs)\}$, based on the methods of ~\cite{li_gradient_2018, shi_spectral_2018}. We call these methods the `Identity', and `Spectral' methods for convenience. Of course, the most obvious approach to computing the score, using $M$ samples alone, would be to simply accumulate the empirical distribution which is observed by the samples, $\widehat{p_{\paramtheta}(\ybs)} := \frac{1}{M} \sum_{m=1}^S \delta(\ybs - \ybs^{(m)})$ and compute the score from this distribution. However, this immediately has a severe drawback. Since the score for a given outcome, $s_{\pi}(\xbs^m)$, requires \textit{also} computing the probabilities of all shifted samples, $\pi(\neg_i\xbs^m) ~ \forall i$, if we have not seen any of the outcomes $\neg_i\xbs^m$ in the observed data, we will not have values for these outcomes in the empirical distribution, and hence we cannot compute the score. This would be a major issue as the number of qubits grows, since we will have exponentially many outcomes, many of which we will not see with $\poly(n)$ samples.

\subsubsection[\texorpdfstring{\color{black}}{} Identity approximation of Stein score]{Identity approximation of Stein score} \label{ssec:identity_computing_stein_score}

As a first attempt, we adopt the method of \cite{li_gradient_2018}. This involves noticing that the score function appears explicitly in Stein identity, and inverting Stein's identity gives a procedure to approximate the score (hence we dub this approach the `Identity' method).

Of course, we shall need to use the discrete version of Stein's identity in our case, and re-derive the result of \cite{li_gradient_2018} but there are no major difficulties in doing so. 

Let us define the `score matrix', $G$, using $M$ samples drawn from $\pi$:
\begin{equation}
G^{\pi}  \coloneqq \left(\begin{array}{cccc}
    \mathbf{s}^1_\pi(\xbs^1)  & \mathbf{s}^1_\pi(\xbs^2)  &\dots &\mathbf{s}^1_\pi(\xbs^M)   \\
     \mathbf{s}^2_\pi(\xbs^1)  & \mathbf{s}^2_\pi(\xbs^2)  &\dots &\mathbf{s}^2_\pi(\xbs^M)     \\
         \vdots & \vdots &\ddots &\vdots   \\
    \mathbf{s}^n_\pi(\xbs^1)  & \mathbf{s}^n_\pi(\xbs^2)  &\dots &\mathbf{s}^n_\pi(\xbs^M)    \\
\end{array}\right)\label{steinscorematrix}
\qquad G^{\pi}_{i,j} := \mathbf{s}^i_\pi(\xbs^j) = \frac{\Delta_{x_i^j}\pi(\xbs^j)}{\pi(\xbs^j)}
\end{equation}
Each column is the term which corresponds to the score function for the distribution, $\pi$, and that given sample.

Now, to compute an approximation of $G^{\pi}$: $\widetilde{G}^{\pi} \approx G^{\pi}$ we can invert the discrete version of Stein's Identity,~\eqref{eqn:vector_discrete_stein_identity_method} (as in the continuous case of ~\cite{li_gradient_2018}):
\begin{equation} \label{eqn:vector_discrete_stein_identity_method}
    \underset{\xbs\sim \pi}{\mathbb{E}}[\mathbf{s}_{\pi}(\xbs)\Phi(\xbs)^T - \Delta \mathbf{f}(\xbs)] = \mathbf{0} 
\end{equation}
where $\mathbf{f}$ is a complex vector valued `test' function.

Rearranging~\eqref{eqn:vector_discrete_stein_identity_method} in terms of the score, and following ~\cite{li_gradient_2018}:
\begin{equation}
    \underset{\xbs\sim \pi}{\mathbb{E}}[\mathbf{s}_\pi\phi(\xbs)^T] = \underset{\xbs\sim \pi}{\mathbb{E}} [\Delta \mathbf{f}(\xbs)] \\
    \implies \sum\limits_{\xbs}\pi(\xbs)\mathbf{s}_\pi(\xbs)\mathbf{f}(\xbs)^T = \sum\limits_{\xbs}\pi(\xbs)\Delta \mathbf{f}(\xbs)
\end{equation}
Taking a Monte Carlo estimate on both sides with $M$ samples, we have:
\begin{equation}
    \implies \sum\limits_{i=1}^{M}\mathbf{s}_\pi(\xbs^i)\mathbf{f}(\xbs^i)^T \approx \frac{1}{M}\sum\limits_{i=1}^{M}\Delta_{\xbs^i} \mathbf{f}(\xbs^i)
\end{equation}
Next, defining:

\begin{align}
  F &\coloneqq [\mathbf{f}(\xbs^1), \mathbf{f}(\xbs^2), \dots, \mathbf{f}(\xbs^M)]^T,\quad
  \widetilde{G}^\pi \coloneqq  [\mathbf{s}^\pi(\xbs^1), \mathbf{s}^\pi(\xbs^2), \dots, \mathbf{s}^\pi(\xbs^M)]^T,\\
    \overline{\Delta_{\xbs}\mathbf{f}} &= \frac{1}{M}\sum_{i=1}^M \Delta_{\xbs^i}\mathbf{f}(\xbs^i), \qquad \Delta_{\xbs^i}\mathbf{f}(\xbs^i) \coloneqq   [\Delta_{\xbs^i}f_1(\xbs^i),\dots,\Delta_{\xbs^i}f_l(\xbs^i)]^T
\end{align}
Now the optimal value for the approximate Stein matrix, $\widetilde{G}^\pi$ will be the solution to the following ridge regression problem, and adding a regularisation term, with parameter, $\eta$, to avoid the matrix being non-singular:
\begin{equation}
    \widetilde{G}^{\pi} = \argmin_{\widetilde{G}^{\pi} \in \mathbb{R}^{M\times n}}\left|\left|\overline{\Delta_{\xbs}\mathbf{f}} - \frac{1}{M}F\widetilde{G}^{\pi}\right|\right|_F^2 + \frac{\eta}{M^2}\left|\left|\widetilde{G}^{\pi}\right|\right|_F^2 \label{redregprob_supp}
\end{equation}
Where $||\cdot||_F$ is the Frobenius norm: $||A||_F = \sqrt{\tr\left(A^TA\right)}$.
The analytic solution of this ridge regression problem is well known and can be found by differentiating the above~\eqref{redregprob_supp} with respect to $\widetilde{G}^{\pi}$ and setting to zero:

\begin{align}
    \widetilde{G}^{\pi} &=  M(K+\eta\mathds{1})^{-1}F^T\overline{\Delta_{\xbs}\mathbf{f}}\\
    \widetilde{G}^{\pi} &=  M(K+\eta\mathds{1})^{-1}\langle\Delta, K\rangle \label{eqn:approx_score_li_derivation}
\end{align}
The approach of~\cite{li_gradient_2018} involves implicitly setting the test function to be a feature map in a RKHS, $\mathbf{f} = \Phi$. In this case, we get $K = F^TF$, and also $\langle \Delta, K\rangle_{ab} = \frac{1}{M}\sum_{i=1}^M \Delta_{x^i_b}\kappa(\xbs^a, \xbs^i)$. Unfortunately, there is no motivation given in~\cite{li_gradient_2018} for which choice of feature map should be used to compute~\eqref{eqn:approx_score_li_derivation}. A sensible choice might be the exponentiated Hamming kernel suggested as a suitable kernel to use in (binary) discrete spaces by \cite{yang_goodness--fit_2018}, where $\text{d}_{\mathsf{HN}}(\xbs, \ybs)$ is the \emph{normalised} Hamming distance (or $\ell_1$ distance) between binary vectors, $\xbs, \ybs$:
\begin{equation}\label{eqn:hamming_kernel}
    \kappa_H(\xbs, \ybs)  := \exp\left(-\text{d}_{\mathsf{HN}}(\xbs, \ybs)\right), \qquad
    \text{d}_{\mathsf{HN}}(\xbs, \ybs)  :=\frac{1}{n} \sum\limits_{i=1}^n|x_i - y_i|
\end{equation}
Any of these kernels could be used, since the only requirement on the above method is that the feature map obeys the discrete Stein identity, which we have seen is the case for \textit{any} complex vector valued function.

\subsubsection[\texorpdfstring{\color{black}}{} Spectral approximation of Stein score]{Spectral approximation of Stein score} \label{ssec:spectral_computing_stein_score}

While the method used to approximate the score function method which was shown in \secref{ssec:identity_computing_stein_score} is straightforward, it does not give a method of computing the score accurately at sample points from the $\QCIBM$ which have \textit{not} been seen in the data distribution, $\pi$ (which becomes exponentially more likely as $n$ grows). If this were to occur during training, a possible solution~\cite{li_gradient_2018} is simply to add that sample to the sample set, and recompute the score function by the Identity method. However, this is expensive, so more streamlined approaches would be desirable. Worse still, this tactic would potentially introduce bias to the data, since there is no guarantee that the given sample from the Born machine, does not have zero probability in the true data, and hence would \emph{never} occur. 

The approach to resolve this is that of \cite{shi_spectral_2018} (which we dub the `Spectral' method). This uses the Nystr{\"o}m method as a subroutine to approximate the score, which is a technique to approximately solve integral equations~\cite{nystrom_uber_1930}. The Nystr{\"o}m method works by finding eigenfunctions of a given kernel with respect to the target probability mass function, $\pi$. As in the case of thew Identity method, the Spectral method was defined when $\pi$ is a continuous probability measure, so we must again perform a discretisation procedure.

We summarise the parts of~\cite{shi_spectral_2018} which are necessary in the discretisation. For the most part the derivation follows cleanly from~\cite{shi_spectral_2018}, and from \secref{ssec:identity_computing_stein_score}. Firstly, the eigenfunctions in question are given by the following summation equation:
\begin{equation}  \label{eqn:nystrom_summation_equation}
    \sum_{\ybs}\kappa(\xbs, \ybs)\psi_j(\ybs)\pi(\ybs) = \mu \psi_j(\xbs)
\end{equation}
where $\{\psi_j\}_{j= 1}^N \in \ell^2(\mathcal{X}, \pi)$, and $\ell^2(\mathcal{X}, \pi)$ is the space of all square-summable\footnote{Sequences, $\{X_n | X_n\in \mathcal{X}\}_n$, in $\ell^2(\mathcal{X}, \pi)$ satisfy $\sum\limits_{n=1}^{\infty}|X_n|^2\pi(X_n) < \infty$} sequences with respect to $\pi$, over the discrete sample space, $\mathcal{X}$. If the kernel is a quantum one, as in~\eqref{eqn:quantum_kernel}, the feature space has a basis, $\{\psi_j = \braket{\boldsymbol{s}_j}{\psi}\}_{j= 1}^N \in \ell^2(\mathcal{X}, \pi)$, where $\ket{s_j}$ are for example computational basis states. We also have the constraint that these functions are orthonormal under the discrete $\pi$:
\begin{equation}
    \sum_{\xbs}\psi_i(\xbs)\psi_j(\xbs) \pi(\xbs) = \delta_{ij} \label{nystromorthomal}
\end{equation}
Approximating~\eqref{eqn:nystrom_summation_equation} by a Monte Carlo estimate drawn with $M$ samples, and finding the eigenvalues and eigenvectors of the covariance kernel matrix, $K_{ij} = \kappa(\xbs^i, \ybs^j)$, in terms of the approximate ones given by the Monte Carlo estimate, exactly as in ~\cite{shi_spectral_2018}, we get:
\begin{equation} \label{eqn:nystrom_eigenfunctions}
    \psi_j(\xbs) \approx \widetilde{\psi}_j(\xbs) = \frac{\sqrt{M}}{\lambda_i}\sum\limits_{m = 1}^M u_j(\xbs^m)\kappa(\xbs, \xbs^m) 
\end{equation}
$\{u_j\}_{j = 1, \dots J}$ are the $J^{th}$ largest eigenvalues of the kernel matrix, $K$, with eigenvalues, $\lambda_j$. The true eigenfunctions are related to these `sampled' versions by: $\psi_j(\xbs^m) \approx \sqrt{M} u_{jm} \forall m \in\{1, \dots, M\}, \mu_j \approx \lambda_j/M$.

Assuming that the discrete score functions are square-summable with respect to $\pi$, i.e. $s^i(\xbs) \in \ell^2(\mathcal{X}, \pi)$, we can expand the score in terms of the eigenfunctions of the $\ell^2(\mathcal{X}, \pi)$:
\begin{equation}
    s^i(\xbs) = \sum\limits^{N}_{j=1}\beta_{ij}\psi_j(\xbs) \label{scoreexpansionineigenbasis}
\end{equation}
Since the eigenfunctions, $\psi_j$ are complex valued, they automatically obey the discrete Stein's identity \thmref{thm:complex_discrete_stein_identity} and we get the same result as ~\cite{shi_spectral_2018}:
\begin{equation}
    \beta_{ij} = -\mathbb{E}_{\pi}\Delta_{x_i}\psi_j(\xbs)
\end{equation}
Proceeding, we apply the discrete shift operator, $\Delta_{x_i}$, to both sides of~\eqref{eqn:nystrom_summation_equation} to give an approximation for the term, $\widetilde{\Delta}_{x_i}\psi(\xbs)  \approx \Delta_{x_i}\psi(\xbs)$:
\begin{equation} \label{eqn:approx_eigenfunction_shifts}
    \widetilde{\Delta}_{x_i}\psi(\xbs) = \frac{1}{\mu_j M}\sum\limits_{m=1}^M\Delta_{x_i}\kappa(\xbs, \xbs^m) 
\end{equation}
It can also be shown in this case that $\widetilde{\Delta}_{x_i}\psi(\xbs) \approx \Delta_{x_i}\widetilde{\psi}(\xbs)$, by comparing~\eqref{eqn:approx_eigenfunction_shifts} with~\eqref{eqn:nystrom_eigenfunctions}, and hence we arrive at the estimator for the score function:
\begin{equation} \label{eqn:spectral_estimation_functions}
    \widetilde{s}^i(\xbs) = \sum\limits_{j=1}^J \widetilde{\beta}_{ij}\widetilde{\psi}_j(\xbs) \widetilde{\beta}_{ij} = -\frac{1}{M}\Delta_{x_i}\widetilde{\Psi}_j(\xbs^m) 
\end{equation}
If the sample space is the space of binary strings of length $n$, the number of eigenfunctions, $N$ will be exponentially large, $N = 2^n$, and so the sum in~\eqref{scoreexpansionineigenbasis} is truncated to only include the $J^{th}$ largest eigenvalues and corresponding eigenvectors.

\subsection[\texorpdfstring{\color{black}}{} Training with the Sinkhorn divergence]{Training with the Sinkhorn divergence} \label{ssec:born_machine/sinkhorn_divergence}

We have left the best until last. While the above Stein discrepancy was an improvement in terms of computability on the $\KL$ divergence, it still left a lot to be desired - particularly in the cumbersome and expensive methods required to estimate the Stein kernel and score function. It also lacked a theoretical understanding of its operational meaning relative to $\TV$\footnote{Although based on numerical experiments, it seems to provide a desirable upper bound on $\TV$}. The next cost function (the `Sinkhorn divergence' ($\SH$)) we introduce circumvents these issues:
\begin{itemize}
    \item It is provably efficient to compute (depending on a hyperparameter),
    \item It provides an upper bound on $\TV$ (again, depending on the same hyperparameter).
\end{itemize}
Before introducing the Sinkhorn divergence, we must revisit optimal transport ($\OT$) from \secref{ssec:prelim/qc/distance_measures/probability}. Recall, the definition of $\OT$ (for comparing the QCBM and data distributions):
\begin{align}  \label{eqn:born_machine/optimal_transport}
\OT^c(p, q)  &:= \min\limits_{U \in \mathcal{U}(p_{\paramtheta}, \pi)}\sum\limits_{(\xbs, \ybs) \in \mathcal{X}\times\mathcal{Y}} c(\xbs, \ybs) U(\xbs, \ybs)\\
\sum_{\xbs}U(\xbs, \ybs) &= \pi(\ybs), \qquad \sum_{\ybs}U(\xbs, \ybs) = p_{\paramtheta}(\xbs)
\end{align}
If we used this directly to train a QCBM, we would incur an exponential sampling cost in the number of qubits (recall the sample complexity of computing~\eqref{eqn:born_machine/optimal_transport} scales as $\mathcal{O}(N^{-1/n})$, given $N$ samples from $p_{\paramtheta}/\pi$). However, it does provide the operational meaning we are seeking, via the following inequality~\cite{gibbs_choosing_2002}:
\begin{equation}
    \text{d}_{\text{min}} \TV(p_{\paramtheta}, \pi) \leq \OT^{\delta}(p_{\paramtheta}, \pi) \leq \text{diam}(\mathcal{X})\TV(p_{\paramtheta}, \pi) \label{tv_wasserstein_inequality}
\end{equation}
where $\text{diam}(\mathcal{X}^n) = \max\{\delta(\xbs, \ybs), \xbs, \ybs \in \mathcal{X}^n\}$, $  \text{d}_{\text{min}} = \min_{x\neq y}\delta(\xbs, \ybs)$, and $\delta(\xbs, \ybs)$ is metric on the space, $\mathcal{X}^n$. We can achieve this bound by choosing $c = \delta$ in~\eqref{eqn:born_machine/optimal_transport} since in this case optimal transport reduces to the Wasserstein metric~\eqref{eqn:1_wasserstein_distance}, $\OT^{\delta} = \text{d}_{\mathsf{W}}$. 

If, for instance, we were to choose $\delta(\xbs, \ybs)$ to be the $\ell_1$ metric between the binary vectors of length $n$ (a.k.a. the  Hamming distance), then we get that $ \text{d}_{\text{min}} = 1, \text{diam}(\mathcal{X}) = n$, and so:
\begin{equation} \label{eqn:tvd_optimal_transport_bound}
     \TV(p_{\paramtheta}, \pi) \leq \OT^{\ell_1}(p_{\paramtheta}, \pi) \leq n \TV(p_{\paramtheta}, \pi)
\end{equation}
To overcome the exponential sampling cost, we can introduce forms of \emph{regularisation}. Regularisation is typically introduced in ML to prevent overfitting. In this case, regularisation will effectively smooth the problem, which enables the solution to be found more quickly.

A particular choice for this regularisation was proposed in~\cite{cuturi_sinkhorn_2013}, where an \emph{entropy} term, in the form of the $\KL$ divergence, is introduced to the optimal transport optimisation problem~\eqref{eqn:born_machine/optimal_transport} as follows:
\begin{align} \label{eqn:regularised_optimal_transport}
    \OT^c_\epsilon(p_{\paramtheta}, \pi) & :=  \min\limits_{U \in \mathcal{U}(p_{\paramtheta}, \pi)}\left(\sum\limits_{(\xbs, \ybs) \in \mathcal{X}\times\mathcal{Y}} c(\xbs, \ybs)U(\xbs, \ybs) + \epsilon \KL(U|p \otimes q)\right) \\
    \KL(U|p_{\paramtheta} \otimes \pi) & =  \sum\limits_{(\xbs, \ybs) \in \mathcal{X}\times\mathcal{Y}} U(\xbs, \ybs) \log\left(\frac{U(\xbs, \ybs)}{(p_{\paramtheta} \otimes \pi )(\xbs, \ybs)}\right)
\end{align}
The effect of the entropy term is controlled by the hyperparameter $\epsilon > 0$. As $\epsilon \rightarrow 0$, we obviously recover the original, unregularised, problem, and if $\epsilon$ is large, the coupling will be penalised the further it is from a product distribution on $\mathcal{X} \times \mathcal{Y}$ (i.e. $U$ in the form $p_{\paramtheta}\otimes \pi$). See~\cite{peyre_computational_2019} for an overview of methods to compute optimal transport metrics. 

Using~\eqref{eqn:regularised_optimal_transport} as a cost function by itself however is perhaps not desirable since, by introducing $\epsilon$, we lose the faithfulness property: $\OT^c_\epsilon(p, p) \neq 0$ in general. However, this can be remedied by introducing two symmetric terms and defining:
\begin{equation} \label{eqn:sinkhorn_divergence}
    \Cbs_{\SH}^\epsilon(p_{\paramtheta}, \pi)  \coloneqq \OT^c_\epsilon(p_{\paramtheta}, \pi) - \frac{1}{2} \OT^c_\epsilon(p_{\paramtheta}, p_{\paramtheta}) -\frac{1}{2}\OT^c_\epsilon(\pi, \pi)
\end{equation}
This quantity is the Sinkhorn divergence ($\SH$), and we can simply check that it is indeed faithful.

Now, similarly to the Stein discrepancy and $\MMD$, we can derive gradients of the Sinkhorn divergence, with respect to the given parameter, ${\paramtheta}_k$. According to \cite{feydy_interpolating_2019}, each term in~\eqref{eqn:sinkhorn_divergence} can be written as:
\begin{equation}
    \OT^c_\epsilon(p_{\paramtheta}, \pi) = \langle p_{\paramtheta}, f \rangle +  \langle \pi, g \rangle  \\
     =\sum\limits_{\xbs} p_{\paramtheta}(\xbs)f(\xbs) + \pi(\xbs)g(\xbs)
\end{equation}
$f$ and $g$ are the so-called optimal Sinkhorn potentials, arising from a primal-dual\footnote{The original $\OT$ problem was a linear programming (LP) problem and is called the `primal' problem. The current formulation is the `dual' version of the primal which (by the weak duality theorem~\cite{bot_introduction_2009}) at least provides an upper bound to the true solution of the LP.} formulation of optimal transport. These are computed using the Sinkhorn algorithm\footnote{This gives the divergence its name~\cite{sinkhorn_relationship_1964}.}. These vectors are initialised at $f^0(\xbs) = 0 = g^0(\xbs)$, and iterated in tandem according to~\eqref{eqn:sinkhorn_dual_vectors_1} and~\eqref{eqn:sinkhorn_dual_vectors_2} for a fixed number of `Sinkhorn iterations' until convergence. The number of iterations required  depends on $\epsilon$, and the specifics of the problem. Typically, smaller values of epsilon will require more iterations, since this is bringing the problem closer to unregularised optimal transport, which is more challenging to compute. For further discussions on regularised optimal transport and its dual formulation, see~\cite{feydy_interpolating_2019, peyre_computational_2019}. Now, following~\cite{feydy_interpolating_2019}, the $\SH$ can be written on a discrete space as:
\begin{equation} \label{eqn:sinkhorn_dual_formulation}
    \Cbs_{\SH}^\epsilon(p_{\paramtheta}, \pi) = \sum\limits_{\xbs} \left[p_{\paramtheta}(\xbs)\left(f(\xbs) - s(\xbs)\right)
    + \pi(\xbs)\left(g(\xbs) - t(\xbs)\right)\right] 
\end{equation}
$s, t$ are the `\textit{autocorrelation}' dual potentials, arising from the terms $\OT^c_\epsilon(p_{\paramtheta}, p_{\paramtheta})$, $\OT^c_\epsilon(\pi, \pi)$ in~\eqref{eqn:sinkhorn_divergence}.

Again, following~\cite{feydy_interpolating_2019}, we can see how by discretising the situation based on $N, M$ samples from $p_{\paramtheta}, \pi$ respectively: $ \widetilde{\xbs} := \{\xbs^1, \dots, \xbs^N\}\sim p_{\paramtheta}(\xbs)$,  $\widetilde{\ybs} :=\{\ybs^1, \dots, \ybs^M\} \sim \pi(\ybs)$. With this, the optimal dual vectors, $f, g$ are given by:
\begin{align}
    f^{l+1}(\xbs^i) &= -\epsilon \text{LSE}_{k=1}^M\left(\log\left(\pi(\mathbf{\ybs}^k) + \frac{1}{\epsilon}g^{l}(\ybs^k) - \frac{1}{\epsilon} C_{ik}(\xbs^i, \ybs^k)\right)\right)\label{eqn:sinkhorn_dual_vectors_1}\\
    g^{l+1}(\ybs^j) &= -\epsilon \text{LSE}_{k=1}^N\left(\log\left(p_{\paramtheta}(\xbs^k) + \frac{1}{\epsilon}f^{l}(\xbs^k) - \frac{1}{\epsilon} C_{kj}(\xbs^k, \ybs^j)\right)\right) \label{eqn:sinkhorn_dual_vectors_2}
\end{align}
$C(\xbs, \ybs)$ is the so-called optimal transport \textit{cost matrix} derived from the cost function applied to all samples, $C_{ij}(\xbs^i, \ybs^j) = c(\xbs^i, \ybs^j)$ and 
\begin{equation}
    \text{LSE}_{k=1}^N(\boldsymbol{v}_k) := \log\sum\limits_{k=1}^N\exp(\boldsymbol{v}_k)
\end{equation}

is a log-sum-exp\footnote{The log-sum-exp function has a gradient which is given by the softmax function, which is also used as a neural network activation function~\eqref{eqn:nn_activation_functions}.} reduction for a vector $\boldsymbol{v}$, used to give a smooth approximation to the true dual potentials.

The autocorrelation potential, $s$, is given by:
\begin{equation} \label{eqn:sinkhorn_autocorrelation_term}
    s(\xbs^i) = -\epsilon \text{LSE}_{k=1}^N\left(\log\left(p_{\paramtheta}(\mathbf{\xbs}^k) + \frac{1}{\epsilon}s(\xbs^k) - \frac{1}{\epsilon} C(\xbs^i, \xbs^k)\right)\right)
\end{equation}
$t(\ybs^i)$ can be derived similarly by replacing $p_{\paramtheta} \rightarrow \pi$ in~\eqref{eqn:sinkhorn_autocorrelation_term} above. However, the autocorrelation dual can be found using a well-conditioned fixed point update \cite{feydy_interpolating_2019}, and convergence to the optimal potentials can be observed with much fewer Sinkhorn iterations:
\begin{equation}
    s(\xbs^i) \leftarrow \frac{1}{2}\left[s(\xbs^i)-\epsilon \text{LSE}_{k=1}^N\left(\log\left(p_{\paramtheta}(\xbs^k) + \frac{1}{\epsilon}s(\xbs^k) - \frac{1}{\epsilon} C(\xbs^i, \xbs^k)\right)\right)\right] \label{autocorrelationsinkhorntermsupdate_supp}
\end{equation}

Finally, let us derive the gradient of $\Cbs^\epsilon_{\SH}$. The derivative with respect to a single probability of the \textit{observed} samples, $p_{\paramtheta}(\xbs^i)$, is given by \cite{feydy_interpolating_2019}:
\begin{equation}
    \frac{\partial \Cbs_{\SH}^\epsilon(p_{\paramtheta}, \pi)}{\partial p_{\paramtheta}(\xbs^i)} = f(\xbs^i) - s(\xbs^i)
\end{equation}
However, this only applies to the samples which have been \emph{used} to compute $f, s$ in the first place. If one encounters a sample from $p_{{\paramtheta}_k^{\pm}}$ (which we shall in the gradient), $\xbs^s \sim p_{{\paramtheta}_k^{\pm}}$, which one has not seen in the vectors sampled from $p_{\paramtheta}$ (i.e. $\xbs^s \notin \widetilde{\xbs}$, one has no value for the corresponding vectors at this point: $f(\xbs^s), s(\xbs^s)$. Fortunately, as shown in \cite{feydy_interpolating_2019}, the gradient does extend smoothly to this point (and all points in the sample space) and in general is given by:
\begin{equation}
    \frac{\partial \Cbs_{\SH}^\epsilon(p_{\paramtheta}, \pi)}{\partial p_{\paramtheta}(\xbs)} = \varphi(\xbs) 
\end{equation}
\begin{multline}
    \varphi(\xbs) = -\epsilon \text{LSE}_{k=1}^M\left(\log\left(\pi(\mathbf{\ybs}^k) + \frac{1}{\epsilon}g^{0}(\ybs^k) - \frac{1}{\epsilon} C(\xbs, \ybs^k)\right)\right)\\
    + \epsilon \text{LSE}_{k=1}^N\left(\log\left(p_{\paramtheta}(\xbs^k) + \frac{1}{\epsilon}s^{0}(\xbs^k) - \frac{1}{\epsilon} C(\xbs, \xbs^k)\right)\right)
\end{multline}
where $g^{(0)}, s^{(0)}$, are the optimal vectors which solve the original optimal transport problem,~\eqref{eqn:sinkhorn_dual_vectors_2} and~\eqref{eqn:sinkhorn_autocorrelation_term} at convergence, given the samples, $\widetilde{\xbs}, \widetilde{\ybs}$ from $p_{\paramtheta}, \pi$ respectively.
Given this, the gradient of the Sinkhorn divergence with respect to the parameters, ${\paramtheta}_k$ is given by:
\begin{align}
    \frac{\partial \Cbs_{\SH}^\epsilon}{\partial {\paramtheta}_k} &= \sum\limits_{\xbs}\frac{\partial \Cbs_{\SH}^\epsilon(p_{\paramtheta}, \pi)}{\partial p_{\paramtheta}(\xbs)}\frac{\partial p_{\paramtheta}(\xbs)}{\partial {\paramtheta}_k} \\
    & = \sum\limits_{\xbs}\varphi(\xbs)\left(p_{{\paramtheta}^-_k}(\xbs) - p_{{\paramtheta}^+_k}(\xbs)\right)  = \underset{\substack{\xbs \sim p_{{\paramtheta}_k^-} }}{\mathbb{E}}[\varphi(\xbs)] 
    -\underset{\substack{\xbs \sim p_{{\paramtheta}_k^+}  }}{\mathbb{E}}[\varphi(\xbs)] \label{sinkhorngradient_supp}
\end{align}
Where again we have used the parameter shift rule~\eqref{eqn:parameter_shift_rule}. Therefore, one can compute the gradient by drawing samples from the distributions, $\widetilde{\xbs} \sim p_{{\paramtheta}^\pm}$, and computing the vector $\varphi(\xbs)$ , for each sample, $\xbs \in \widetilde{\xbs}$, using the vectors, $g^{(0)}, s^{(0)}$ already computed during the evaluation of $\SH$ at each epoch.

\subsubsection[\texorpdfstring{\color{black}}{} Sinkhorn divergence sample complexity]{Sinkhorn divergence sample complexity} \label{ssec:born_machine/sinkhorn_divergence_samp_complex}

The sample complexity of the Sinkhorn divergence is of great interest to us as we claim that the $\TV$ and the $\KL$ are not suitable to be directly used as cost functions. This is due to the difficulty of computing the outcome probabilities of quantum circuits efficiently. We now motivate why the $\MMD$ is a weak cost function, and why the Sinkhorn divergence should be used as an alternative. This will depend critically on the regularisation parameter $\epsilon$, which allows a smooth interpolation between the $\OT$ metric and the $\MMD$. 

Firstly, we will require the Sinkhorn divergence to be \emph{faithful}. Fortunately, by the results of~\cite{feydy_interpolating_2019}, we can be assured that no matter which value of $\epsilon$ is chosen, the $\SH$ has this property:
\begin{align}
    \Cbs_{\SH}^\epsilon(p, q)  = 0 \iff p \equiv q \label{sinkhornzeroiff}\\
    \Cbs_{\SH}^\epsilon(p, q) \geq \Cbs_{\SH}^\epsilon(p, p) = 0 \label{sinkhorngreaterzero_supp}
\end{align}

It also metrizes convergence in law, effectively meaning it can be estimated using $M$ samples, and will converge to its true value in the limit of large samples:
\begin{equation}
    \Cbs_{\SH}^\epsilon(\widetilde{p}_M, p) \rightarrow 0 \iff \widetilde{p}_M \rightharpoonup p \label{sinkhornconvergenceinlaw_supp}
 \end{equation}
Secondly, we also know that the Sinkhorn divergence inherits the sample complexities of both the $\MMD$ and unregularised $\OT$, since it becomes both of these metrics in the extreme limits of epsilon~\cite{ramdas_wasserstein_2017}:
\begin{align}
\underline{\epsilon \rightarrow 0}:& \qquad\Cbs_{\SH}^0(p, q)\rightarrow \OT_0^c(p, q)\\
\underline{\epsilon \rightarrow \infty}:& \qquad \Cbs_{\SH}^\epsilon(p, q) \rightarrow \MMD^{\kappa}(p, q),  \qquad \kappa(\xbs, \ybs) = -\delta(\xbs, \ybs)
\end{align}
Therefore, we expect that the sample complexity becomes large $(\mathcal{O}(N^{-1/n}))$ as $\epsilon\rightarrow 0$ and small as $\epsilon \rightarrow \infty$ $(\mathcal{O}(N^{-1/2}))$, with some complicated function of $\epsilon$ in between. Fortunately again, this has also been addressed, this time by~\cite{genevay_sample_2019}. First, however, we will need to introduce the concept of \emph{Lipschitz continuity}:
\begin{defbox}
    \begin{definition}[Lipschitz continuity]~ \\
    \label{defn:lipschitz_continuity}
    Given two metric spaces, $(\mathcal{A}, \text{d}_{\mathcal{A}}), (\mathcal{B}, \text{d}_{\mathcal{B}})$ where $\text{d}_{\mathcal{C}}$ denotes a metric on the space $\mathcal{C} \in \{\mathcal{A}, \mathcal{B}\}$, a function, $f:\mathcal{A}\rightarrow\mathcal{B}$, is called \emph{Lipschitz continuous} if there exists an $L \in \mathbb{R}$ (called the Lipschitz constant) such that $\forall \xbs, \ybs \in \mathcal{A}$:
    \begin{equation}
        \text{d}_{\mathcal{B}}(f(\xbs),f(\ybs)) \leq L d_{\mathcal{A}}(\xbs, \ybs) \label{lipschitzcontinuous_supp}
    \end{equation}
    \end{definition}
\end{defbox}

Now, we can address the sample complexity results from~\cite{genevay_sample_2019}: Specifically, we have the following two results\footnote{Note the original theorems technically apply to the regularised $\OT$ cost, rather than $\Cbs_{\SH}$, but the addition of the symmetric terms in~\eqref{eqn:sinkhorn_divergence} will not affect the asymptotic sample complexities since they add only constant overheads.} The first is the mean difference between the true Sinkhorn divergence, $\Cbs_{\SH}^\epsilon(p, q)$ and its estimator derived from the empirical distributions, $\Cbs_{\SH}^\epsilon(\widetilde{p}_M, \widetilde{q}_M)$ is given by:
\begin{thmbox}
    \begin{theorem}(Theorem 3 from~\cite{genevay_sample_2019})\label{thm:sinkhornexpectationsamplecomplexity_supp}~ \\
        Consider the Sinkhorn divergence between two distributions, $p$, and $q$ on two bounded subsets, $\mathcal{X}, \mathcal{Y}$ of $\mathbb{R}^n$, with a $C^{\infty}$, $L$-Lipshitz cost $c$. One has:
        \begin{equation}
            \mathbb{E}|\Cbs_{\SH}^\epsilon(p,q) - \Cbs_{\SH}^\epsilon(\widetilde{p}_M, \widetilde{q}_M)| = \mathcal{O}\left(\frac{\erm^{\frac{\kappa}{\epsilon}}}{\sqrt{M}}\left(1+\frac{1}{\epsilon^{\lfloor n/2\rfloor}}\right)\right)\label{sinkhornexpectationsamplecomplexity_supp}
        \end{equation}
        where $\kappa := 2L|\mathcal{X}|+||c||_{\infty}$ and constants only depend on $|\mathcal{X}|, |\mathcal{Y}|, c$ and $||c^{l}||_{\infty}$ for $l = 0, \dots, \lfloor n/2\rfloor$.
    \end{theorem}
\end{thmbox}
The second is the following concentration result:
\begin{corrbox}
\begin{corollary}
(Corollary 1 from~\cite{genevay_sample_2019})\label{thm:sinkhornconcentrationsamplecomplexity_supp}~ \\
With probability at least $1-\gamma$, we have that:
\begin{equation}
    \left|\Cbs_{\SH}^\epsilon(p,q) - \Cbs_{\SH}^\epsilon(\widetilde{p}_M, \widetilde{q}_M)\right| \leq 12B\frac{\lambda K}{\sqrt{M}} + 2C\sqrt{\frac{2\log\frac{1}{\gamma}}{M}}\label{sinkhornconcentrationsamplecomplexity_supp}
\end{equation}
where $\kappa := 2L|\mathcal{X}|+||c||_{\infty}$, $C  := \kappa + \epsilon \erm^{\frac{\kappa}{\epsilon}}$, $B\leq 1+\erm^{\left(2\frac{L|\mathcal{X}|+||c||_{\infty}}{\epsilon}\right)}, \lambda = \mathcal{O}(1+ \frac{1}{\epsilon^{\lfloor n/2\rfloor}})) \text{ and } K  \coloneqq \max_{\xbs \in \mathcal{X}}\kappa_S(\xbs, \xbs)$.
\end{corollary}
\end{corrbox}
$\kappa_S$ is the Matern or the Sobolev kernel, associated to the Sobolev space, $\mathcal{H}^s(\mathbb{R}^n)$, which is a RKHS for $s > n/2$, but we will not go into further detail here.

The more exact expression for~\eqref{sinkhornexpectationsamplecomplexity_supp} is given by:
\begin{equation}\label{samplecomplexityexpanded_supp}
     \mathbb{E}|\Cbs_{\SH}^\epsilon - \widetilde{\Cbs}_{\SH}^\epsilon| \leq 12\frac{B\lambda\sqrt{K}}{\sqrt{M}}
     = \mathcal{O}\left(\frac{1}{\sqrt{M}}\left(1+\erm^{\left(2\frac{L|\mathcal{X}|+||\delta||_{\infty}}{\epsilon}\right)}\right)\left(1+\frac{1}{\epsilon^{\lfloor n/2\rfloor}}\right)\right)
\end{equation}
where we use $\Cbs_{\SH}^\epsilon$, rather than $\OT_\epsilon^c$ as in~\cite{genevay_learning_2018}, through the use of the triangle inequality.

Now, for our particular case, we would like to choose the Hamming distance as a metric on the Hamming hypercube (for a fixed dimension, $n$). However, due to the smoothness requirement of the above theorems, $c \in C^\infty$, this would not hold in the discrete case we are dealing with. However, we can take a broader view to simply use the $\ell_1$ distance, and embed the Hamming hypercube in a larger space. This is possible since, as we mentioned above, the Hamming distance, $\text{d}_{\text{H}}$ is exactly the $\ell_1$ metric restricted to binary vectors.

In this scenario, formally, we are dealing with the general hypercube in $\mathbb{R}^n$, but where the probability masses are strictly concentrated on the vertices of the hypercube. Now, we can compute directly some of the constants in the above,~\thmref{thm:sinkhornexpectationsamplecomplexity_supp} and \corrref{thm:sinkhornconcentrationsamplecomplexity_supp}. Taking $\mathcal{X}$ to be the unit hypercube in $\mathbb{R}^n$, and $c = \delta=\ell_1$, which is Lipschitz continuous, we can compute the following:
\begin{equation}
    |\mathcal{X}| = \sup_{\xbs, \ybs \in \mathcal{X}}||\xbs - \ybs||_1= n,\qquad
    ||\delta(\xbs, \ybs)||_\infty = \sup\{|\delta(\xbs, \ybs)| : (\xbs, \ybs) \in \mathcal{X}\times \mathcal{Y}\} = n\label{sinkhorn_sample_complexity_calcs}
\end{equation}
A rough upper bound for the Lipschitz constant, $L$, can be obtained as follows.
If we take $\text{d}_{\mathcal{A}}$ to be the sum metric on the product space, $\mathcal{X}\times\mathcal{Y}$, and $f$ to be the $\ell_1$ distance, we get:
\begin{equation}
    |\delta(\xbs^1, \ybs^1) - \delta(\xbs^2, \ybs^2)| \leq L \left[\text{d}_{\mathcal{X}}(\xbs^1, \xbs^2)+ \text{d}_{\mathcal{Y}}(\ybs^1, \ybs^2)\right] \label{costfunctionlipschitzcont_supp}
\end{equation}
For two points, $(\xbs^1, \ybs^1), (\xbs^2, \ybs^2) \in \mathcal{X}\times \mathcal{Y}$, and the cost $c:\mathcal{X}\times\mathcal{Y}\rightarrow \mathbb{R}, c = ||\cdot||_1$.
Now, 
\begin{equation}
    \frac{\left|\sum_i |\xbs_i^1 - \ybs_i^1| - \sum_i |\xbs_i^2 - \ybs_i^2| \right|}{\left[\sum_i |\xbs_i^1 - \xbs_i^2| + \sum_i |\ybs_i^1 - \ybs_i^2| \right]} \leq L \label{lipshitzderivation2_supp}
\end{equation}
We want to find an upper bound for the left hand side of~\eqref{lipshitzderivation2_supp}, assuming that $\xbs^1 \neq \xbs^2$ and $\ybs^1 \neq \ybs^2$. In this case, both numerator and denominator are zero, so any non-zero $L$ will satisfy~\eqref{costfunctionlipschitzcont_supp}. Now, applying the trick of embedding the Hamming hypercube in the general hypercube, we can assume $\xbs^{1,2}_i, \ybs^{1,2}_i \in \{0, 1\}~\forall i$. To derive such a bound, we can bound both the numerator and the denominator of the LHS of~\eqref{lipshitzderivation2_supp} independently. We find the denominator is as small as possible, when only one element of $\xbs^{1,2}$ or $\ybs^{1,2}$ is equal to one, and all the are equal to zero.  The numerator is as large as possible when one of $\xbs^{1,2}$ or $\ybs^{1,2}$ is the all-one vector. In this case, the LHS is upper bounded by $n$ (for a fixed $n$), so we can choose $L = n$, which is a constant for a fixed $n$.

Plugging these quantities into~\eqref{samplecomplexityexpanded_supp} we arrive at:
\begin{equation}
     \mathbb{E}|\Cbs_{\SH}^\epsilon - \widetilde{\Cbs}_{\SH}^\epsilon| = \mathcal{O}\left(\frac{1}{\sqrt{M}}\left(1+\erm^{\left(2\frac{n^2+n}{\epsilon}\right)}\right)\left(1+\frac{1}{\epsilon^{\lfloor n/2\rfloor}}\right)\right)
\end{equation}
The constants in $\mathcal{O}\left(1+\frac{1}{\epsilon^{\lfloor n/2\rfloor}}\right)$, depend on $|\mathcal{X}|, |\mathcal{Y}|, n, \text{ and } ||\delta^{(k)}||_\infty$ which are at most linear in $n$. 
Clearly, due to the asymptotic behaviour of the Sinkhorn divergence, we would like to choose $\epsilon$ sufficiently large in order to remove as much dependence on the dimension, $n$, as possible. This is because, in our case, the dimension of the space is equivalent to the number of qubits, and hence to derive a favourable sample complexity, we would hope for the dependence on $n$ to be polynomial in the number of qubits. Ignoring the constant terms, we can see that by choosing $\epsilon = \mathcal{O}(n^{2})$, we get:
\begin{equation}
     \mathbb{E}|\Cbs_{\SH}^{\mathcal{O}(n^{2})} - \widetilde{\Cbs}_{\SH}^{ \mathcal{O}(n^{2})}| =~ \mathcal{O}\left(\frac{1}{\sqrt{M}}\right) \label{sinkhorn_expectation_sample_chosen_MAIN}
\end{equation}
Similarly the concentration bound from~\corrref{thm:sinkhornconcentrationsamplecomplexity_supp} is: 
\begin{align}
    &|\Cbs_{\SH}^\epsilon - \widetilde{\Cbs}_{\SH}^\epsilon| \leq 12B\frac{\lambda K}{\sqrt{M}} + 2C\sqrt{\frac{2\log\frac{1}{\gamma}}{M}}\nonumber \\
    &\leq \frac{12K}{\sqrt{M}}\left(1+\erm^{\left(2\frac{L|\mathcal{X}|+||\delta||_{\infty}}{\epsilon}\right)}\right)\mathcal{O}\left(1+\frac{1}{\epsilon^{\lfloor n/2\rfloor}}\right)+ 2\kappa\sqrt{\frac{2\log\frac{1}{\gamma}}{M}} + 2\epsilon \erm^{\frac{\kappa}{\epsilon}}\sqrt{\frac{2\log\frac{1}{\gamma}}{M}}\nonumber \\
    &= \frac{1}{\sqrt{M}}\left[12\left(1+\erm^{\left(\frac{\mathcal{O}(n^{2})}{\epsilon}\right)}\right)\mathcal{O}\left(1+\frac{1}{\epsilon^{\lfloor n/2\rfloor}}\right)+\mathcal{O}(n^{2})\sqrt{2\log\frac{1}{\gamma}} + 2\epsilon \erm^{\frac{\mathcal{O}(n^{2})}{\epsilon}}\sqrt{2\log\frac{1}{\gamma}}\right]\nonumber \\
    &= \mathcal{O}\left(\frac{1}{\sqrt{M}}\left[\left(1+\erm^{\left(\frac{n^{2}}{\epsilon}\right)}\right)\left(1+\frac{1}{\epsilon^{\lfloor n/2\rfloor}}\right)\right.+ \left.n^{2}\sqrt{2\log\frac{1}{\gamma}}  + \epsilon \erm^{\frac{n^{2}}{\epsilon}}\sqrt{2\log\frac{1}{\gamma}}\right]\right) \label{sinkhornbornsamplecomplexity_supp}
\end{align}
since $\kappa = 2n^2+n = \mathcal{O}(n^{2})$. Now, choosing the same scaling for $\epsilon$ as in~\eqref{sinkhorn_expectation_sample_chosen_MAIN} we get that with probability $1-\gamma$:
\begin{align}
    |\Cbs_{\SH}^{\mathcal{O}(n^{2})} - \widetilde{\Cbs}_{\SH}^{\mathcal{O}(n^{2})}| 
    &= \mathcal{O}\left(\frac{1}{\sqrt{M}}\left[\left(1+\frac{1}{n^{\lfloor n/2\rfloor}}\right)\right.\right.+ \left.\left.n\sqrt{2\log\frac{1}{\gamma}}\right]\right) \label{sinkhornbornsamplecomplexitychoosed_supp}\\
    &=\mathcal{O}\left(\frac{n}{\sqrt{M}}\log(1/\gamma)^{1/2}\right) \label{sinkhornborn_samplecomplexity_choosed_MAIN}
\end{align}

It is likely in practice however, that a much smaller value of $\epsilon$ could be chosen, without blowing up the sample complexity. This is evidenced by numerical results in~\cite{genevay_learning_2018, feydy_interpolating_2019, genevay_sample_2019}.

Up to this point, we have discussed the favourable parts of the sample complexity of the Sinkhorn divergence relating to \emph{computability}. Let us now revisit it from the point of view of quantum advantage. Recall that from~\eqref{eqn:tvd_optimal_transport_bound}, the unregularised optimal transport provides an upper bound on $\TV$ (meaning it is `\emph{strong}', as evidenced by its sample complexity), while on the other hand we have from~\eqref{eqn:tvd_mmd_lower_bound}, the $\MMD$ provides only a \emph{lower} bound (and hence is `weak'). Given the interpolation between these two metrics, there must be a crossover at some value of $\epsilon$ for which the Sinkhorn divergence also provides an upper bound. As with to our judicious choice of $\epsilon$ to enable a favourable sample complexity, we can perform a similar exercise to search for an upper bound to $\TV$.

Again, we can adapt results from~\cite{genevay_sample_2019} for this purpose. Specifically, we have the following bound on regularised and unregularised optimal transport (Theorem 1 in~\cite{genevay_sample_2019}):
\begin{equation}
    0 \leq \OT^c_\epsilon(p, q) -\OT^c_0(p, q)  \leq 2\epsilon\log\left(\frac{\erm^2LD}{n\epsilon}\right)
    \sim_{\epsilon \rightarrow 0} 2\epsilon\log\left(1/\epsilon\right)
\end{equation}
where the size of the sample space is bounded by $D, |\mathcal{X}| <D$, as measured by the metric, and $L$ is the Lipschitz constant. As above, we can choose $D = n$, $L = n$ (for a Born machine) to arrive at:
\begin{equation}
    0 \leq \OT^{\ell_1}_\epsilon(p_{\paramtheta}, \pi) - \OT^{\ell_1}_0(p_{\paramtheta}, \pi)\leq 2\epsilon\log\left(\frac{\erm^2n}{\epsilon}\right)
\end{equation}
The $\log$ term will be positive as long as $\epsilon \leq n\erm^2$, in which case regularised $\OT$ will give an upper bound for the Wasserstein metric, and hence the $\TV$ through~\eqref{tv_wasserstein_inequality} so finally have:
\begin{equation}
     \TV(p_{\boldsymbol\theta}, \pi) \leq  \OT^{\ell_1}_0(p_{\boldsymbol\theta},\pi) \leq  \OT^{\ell_1}_{\epsilon \leq n \erm^2}\label{tv_wasserstein_sinkhorn_inequality}
\end{equation}
Unfortunately, comparing this with~\eqref{sinkhorn_expectation_sample_chosen_MAIN} and~\eqref{sinkhornborn_samplecomplexity_choosed_MAIN}, we can see that with this scaling of $\epsilon$, the sample complexity would pick up an exponential dependence on the dimension, $n$, so it would not be efficiently computable. As such, we cannot hope to use the Sinkhorn divergence as an efficient means to both bound $\TV$ (and therefore provide a verification that the Born machine has learned a `classically hard' distribution, as in~\secref{ssec:born_machine/quantum_advantage}), and to also be provable efficiently computable.

\section[\texorpdfstring{\color{black}}{} Numerical Results]{Numerical results} \label{ssec:born_machine/numerics}

We conclude this discussion with a series of numerical results in training QCBMs. This section will proceed in two parts. 

Firstly, we provide a comparison between the training methods described in \secref{ssec:born_machine/training} using a small scale example and a toy dataset. To summarise the results of this comparison, we surprisingly find that both our new cost functions, the Stein discrepancy and the Sinkhorn divergence are able to outperform the $\MMD$ when training a QCBM, \emph{even at very small scales}. This provides an indication that as we scale the size of the quantum computer on which the QCBM is implemented, we should expect better and better results using our methods (which is reinforced by the theory of the previous section).

In the second part of this section, we change gears slightly, and compare a QCBM to a completely \emph{classical} model, the restricted Boltzmann machine, this time on a real-world dataset.
For this latter comparison, we find that a QCBM is able to outperform the RBM at a certain scale (when both models are compared in a fair way) and as such this provides some numerical evidence for the theoretical arguments given in \secref{ssec:born_machine/quantum_advantage} which argues for the supremacy of quantum generative modelling. Again, this is surprising, as there is nothing a-priori to suggest why a quantum model should perform better on this dataset, when it has no apparent `quantum' features. Of course, these results should only be taken as a preliminary study - to be validated as part of a much larger numerical benchmark between quantum and classical models. Finally, we implement an example of `quantum' data in generative modelling, and use a $\QCIBM$ to learn a distribution outputted by a quantum circuit.

\subsection[\texorpdfstring{\color{black}}{} The data]{The data} \label{ssec:born_machine/numerics/data}
Next, we introduce the datasets we use for generative modelling. As mentioned above, the first is a toy dataset, and the second is one which originates in a real-world use case, specifically in finance. The former has the advantage in that we have complete knowledge of the underlying probabilities, which is especially relevant since we use them to compute the $\TV$ distance to comparing training methods. This is also true for the quantum dataset, for which we can exactly simulate the `data' target circuit. For the finance dataset, we unfortunately do not have access to the underlying probabilities, but only a fixed number of samples.

\subsubsection[\texorpdfstring{\color{black}}{} A toy dataset]{A toy dataset} \label{sssec:born_machine/data/toy_dataset}
The toy distribution is the one given by \eqref{eqn:born_machine_toy_dataset}, which is used in both \cite{amin_quantum_2018, verdon_quantum_2017} to train versions of the quantum Boltzmann machine:
\begin{equation}
    \pi(\ybs) :=\frac{1}{T}\sum\limits_{k=1}^T 
    p^{n - d_H(\boldsymbol{s}^k, \ybs)}
    (1-p)^{d_H(\boldsymbol{s}^k, \ybs)}
\label{eqn:born_machine_toy_dataset}
\end{equation}
To generate this data, $T$ binary strings of length $n$, written $s_k$ and called `modes', are chosen randomly. A sample $\ybs \sim \pi(\ybs)$ is then produced with a probability which depends on its Hamming distance $\text{d}_{\mathsf{H}}(\boldsymbol{s}^k, \ybs) = \sum_i|\boldsymbol{s}^k_i - \ybs_i|$ to each mode.

\subsubsection[\texorpdfstring{\color{black}}{} A financial dataset]{A financial dataset} \label{sssec:born_machine/data/finance_data}
\begin{figure}[t]
    \centering
    \includegraphics[width=\columnwidth, height=0.3\columnwidth]{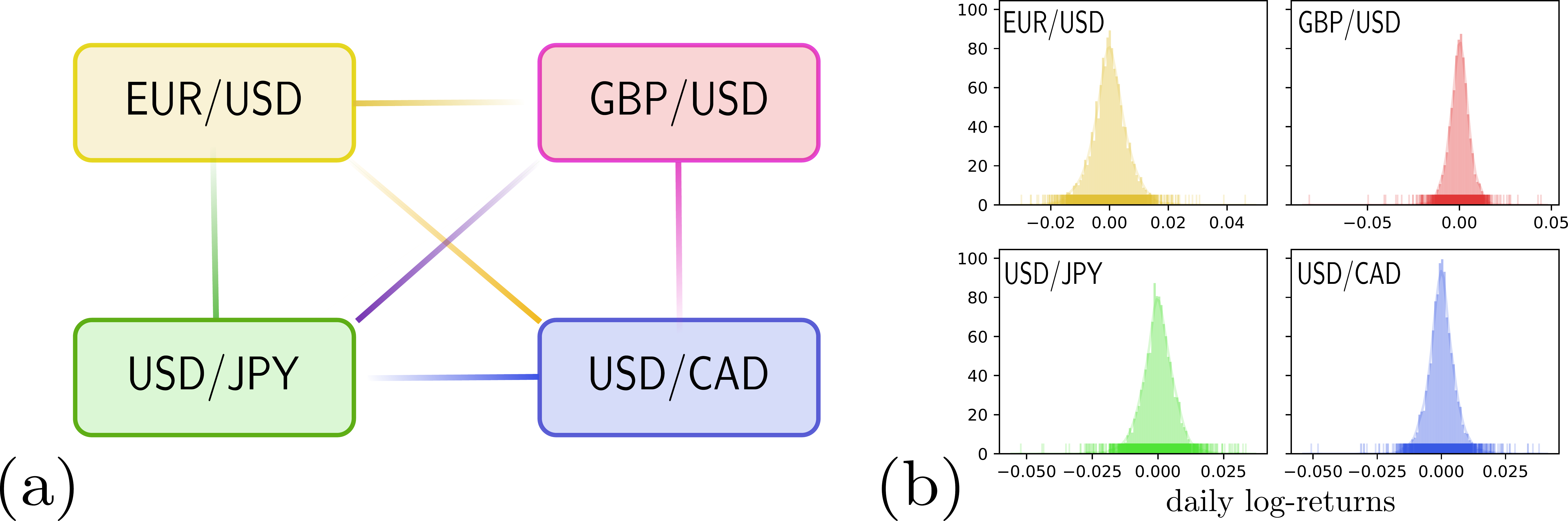}
    \caption[\color{black} Data generated from FX spot prices of the currency pairs.]{\textbf{Data generated from FX spot prices of the above currency pairs.} The generative model aims to learn correlations between each pair based on a $16$ bit binary representation. (a) The selection  of currency pairs we use, and (b) the marginal distributions of the log-returns of each pair over a $20$ year period. We aim to learn the joint distribution of subsets of the pairs.}
    \label{fig:currency_pairs}
\end{figure}

Our second dataset is one considered in~\cite{kondratyev_market_2019} and has a financial origin. This contains $5070$ samples of daily log-returns of $4$ currency pairs between $1999-2019$ (see \figref{fig:currency_pairs}). We use this dataset as a means to compare a quantum model (our Born machine) with a comparable classical model (the restricted Boltzmann machine (RBM), see~\secref{ssec:prelim/machine_learning/neural_networks}).

However, in order to fit on the binary architecture of the Born and Boltzmann machines, the spot prices of each currency pair are converted to $16$ bit binary values, resulting in samples of $64$ bits long. This discretisation provides a convenient method for fitting various problem sizes onto models with different numbers of qubits or visible nodes for the Born machine or RBM respectively. In particular, we can tune both the number of currency pairs ($i$), and the precision of each pair ($j$) so the problem size is described by a tuple $(i, j)$. For example, as we revisit in~\secref{sssec:born_machine/numerics/finance/numerics}, a $12$ qubit Born machine can be tasked to learn the distribution of $4$ currency pairs at $3$ bits of precision, $3$ pairs with $4$ bits or $2$ pairs at $6$ bits of precision.

\subsubsection[\texorpdfstring{\color{black}}{} A quantum dataset]{A quantum dataset} \label{sssec:born_machine/data/quantum_data}
As a final example, we examine a dataset generated by a \emph{quantum} process, in particular, we generate data using one instance of a Born machine (an $\IQP$-$\QCIBM$ specifically, and use \emph{another} Born machine (a $\QAOA$-$\QCIBM$) to learn this distribution. If one could make such a process theoretically rigorous at scale (accounting for technical details), it would satisfy our definition of QLS~\defref{defn:quantum_learning_supremacy}, since an $\IQP$-$\QCIBM$ is an example of a distribution family, $\mathcal{D}_n$, which is not $(\TV, \frac{1}{384}, \BPP)$-Learnable. 

Furthermore, one may also view this task as a form of \emph{weak circuit compilation}.
The major objective in the field of quantum compilation (or gate synthesis as it is sometimes referred to) is to \emph{compile} a given target unitary, $U$, into a gate sequence, $V := V_1V_2 \dots V_k$. For example, while we know from \secref{ssec:prelim/qc/quantum_operations}, that arbitrary unitaries can be built using a universal set of gates, a question remains\footnote{The Solovay-Kitaev is more of an \emph{existence} result; it only tells us that \emph{some} decomposition exists. It does not tell us how to find it.} in how do we actually do this \emph{efficiently}, relative to one parameter or another we care about. For example, one may wish to minimise the number of non-Clifford (e.g. $\TG$), or two qubit gates in a given quantum circuit. This task particularly important for building and running algorithms on quantum computers, especially near term ones. As we have seen above, particular hardware platforms (for example the Rigetti \computerfont{Aspen} QPU) may only be able to implement certain native gates (see~\secref{ssec:born_machine/quantum_hardware}) and usually\footnote{This is certainly true for superconducting platforms, others may be more flexible. For example, ion-trap based processors boast all-to-all connectivity.} have a limited qubit connectivity, how can we implement unitaries in an efficient and noise tolerant manner. For further reading on the topic of compilation, see~\cite{fowler_constructing_2011, booth_jr_quantum_2012, campbell_unified_2017, heyfron_efficient_2018, venturelli_compiling_2018, nam_automated_2018, haner_software_2018, amy_t-count_2019} for a non-exhaustive list of relevant resources.

We have already alluded to an interesting approach to the compilation problem using using variational quantum algorithm techniques~\cite{jones_quantum_2018, khatri_quantum-assisted_2019, heya_variational_2018}. These propose approximating the target unitary by assuming that $V$ is a PQC built from the native gates of a particular quantum device, which can be trained. Using a Born machine for this task is a similar methodology, but without introducing any extra quantum resources to perform the compilation\footnote{For example, the approach of ~\cite{khatri_quantum-assisted_2019} requires that the unitary to be compiled, $U$, is actually applied on the quantum device.}. Specifically, we train a Born machine to reproduce the output distribution given when a problem unitary is applied on an input state. Of course, this is not true compilation in the sense that the underlying unitaries (the problem unitary and that of the Born machine) may not be equivalent (since for this to be the case the output distribution of the two cases would have to be the same for all possible input states), which is why we refer to it as \emph{weak}.

As a particular example for this problem, we revert back to the $\QCIBM$ $\Ansatz$ and compile a $\QAOA$-$\QCIBM$ to the output distribution of an $\IQP$-$\QCIBM$:
\begin{multline} \label{eqn:ibm_compilation_qaoa_to_IQP}
    \IBM\left(\left\{J^{\QAOA}_{ij}, b^{\QAOA}_k\right\}, \left\{\Gamma_k = \frac{\pi}{4}\right\}, 0,  0\right) \\
    \overset{Compile}{\rightarrow}   \IBM\left(\left\{J^{\IQP}_{ij}, b^{\IQP}_k\right\}, \left\{\Gamma_k = \frac{\pi}{2\sqrt{2}}\right\}, 0, \left\{\Sigma_k = \frac{\pi}{2\sqrt{2}}\right\}\right)
\end{multline}

In other words trying to find the optimal parameters, $\boldsymbol{\alpha}^{\QAOA}$ which fit the distribution of the $\IQP$-$\QCIBM$ with parameters, $\boldsymbol{\alpha}^{\IQP}$. The `quantum' dataset, $\{\ybs^j\}_{j=1}^M$, is generated by running the following circuit $M$ times:
\begin{align} \label{circuit:qaoa_born_machine}
    \Qcircuit @C=0.6em @R=0.8em {
    \lstick{\ket{0}}    & \gate{H}  & \multigate{3}{U_z(\boldsymbol\alpha^{\IQP})} & \gate{U^1_f\left(\frac{\pi}{2\sqrt{2}}, 0, \frac{\pi}{2\sqrt{2}}\right)}  & \meter &\cw & \rstick{y_1} \\
    \lstick{\ket{0}}    & \gate{H}  & \ghost{U_z(\boldsymbol\alpha^{\IQP})}        & \gate{U^2_f\left(\frac{\pi}{2\sqrt{2}}, 0, \frac{\pi}{2\sqrt{2}}\right)}  & \meter &\cw & \rstick{y_2} \\
    \cdots              &           &                                       & \cdots                                    & \cdots &    &  \\
    \lstick{\ket{0}}    & \gate{H}  & \ghost{U_z(\boldsymbol\alpha^{\IQP})}        & \gate{U^n_f\left(\frac{\pi}{2\sqrt{2}}, 0, \frac{\pi}{2\sqrt{2}}\right)}  & \meter &\cw & \rstick{y_n} 
    }
\end{align}
Of course, one may take the view that this problem is not \emph{strictly} one with quantum data, but classical data which originated in a quantum process. As an extension, one may consider the training of a QCBM, where the learning signal is derived \emph{directly} from the quantum state produced by \eqref{circuit:qaoa_born_machine} (before measurement). In this case, the dataset would consist of qsamples (\defref{defn:qsample_definition})\footnote{Obviously, with the modification that this is not a quantum encoding of a classical distribution, but a purely quantum sample.}. Indeed, this is the driving principle underlying models such as quantum generative adversarial networks (QGANs)~\cite{lloyd_quantum_2018}. As an example for how one may use qsamples for training a Born machine\footnote{Since clearly the cost functions we propose in \secref{ssec:born_machine/training} will no longer be directly suitable.}, the learning signal may be derived by estimating the overlaps between the quantum data and the Born machine states using, for example, the $\SWAP$ test (see~\secref{ssec:prelim/qc/swap_test}). Alternatively, one might choose to use the quantum generalisations of the distance measures we introduced in \secref{sssec:prelim/qc/distance_measures/quantum}.

\subsection[\texorpdfstring{\color{black}}{} Comparison between training methods]{Comparison between training methods} \label{ssec:born_machine/numerics/training_methods}

Let us begin by implementing the training of the $\QCIBM$ using the new cost functions introduced in \secref{ssec:born_machine/training}, and comparing to the best previous known differentiable training method, using the $\MMD$. We aim for two properties that our cost functions should exhibit in order to claim they have outperformed the $\MMD$:
\begin{itemize}
    \item \textit{Speed of Convergence:} Both of the cost functions, $\SD$ and $\SH$, should achieve equal or lower total variation distance ($\TV$) than the $\MMD$ in a \emph{shorter} time period (even accounting for various learning rates).
    \item \textit{Accuracy:} Since the cost functions we employ are in some sense `stronger' than $\MMD$, we would like for them to achieve a \emph{smaller} $\TV$ than is possible with the $\MMD$ in an equal or quicker time.
\end{itemize}

$\TV$ was chosen as an objective benchmark for several reasons. Firstly, it is typically the notion of distance which is required by quantum supremacy experiments where one wants to prove hardness of classical simulation (as we mentioned in \secref{ssec:born_machine/quantum_advantage}). Secondly, we use it in the definitions of quantum learning supremacy. Finally, it is one of the strongest notions of convergence in probability (as discussed in \secref{ssec:born_machine/training}) one can ask for, so it follows that a training procedure which can more effectively minimise $\TV$, in an efficient way, should be better for generative modelling.

Since we are using an artificial dataset to compare our methods, and since we are directly simulating the $\QCIBM$ (so we have access to the outcome probabilities), we can directly compute the $\TV$ during training of the model. Note that this would not be possible in general as the number of qubits scales. When we examine the financial dataset \secref{sec:born_machine/finance}, we cannot use $\TV$ as an objective measure, since we do \emph{not} have the exact data probabilities. Therefore we shall need to devise an alternative third party metric to gauge the relative performance of the model.

For all the numerical results shown in this section, we use the $\QCIBM$ structure shown in \eqref{circuit:ising_born_machine} with trainable `Ising' parameters (the parameters, $\boldsymbol{\alpha}$), and fixed `$\QAOA$'-type measurement angles in \eqref{eqn:finalmeasurementgate}. We choose this $\Ansatz$\footnote{This can be considered a hardware efficient fixed-structure $\Ansatz$ as introduced in \secref{ssec:vqa_ansatzse}. Alternatively, one may consider it as a `quantum-advantage-inspired' $\Ansatz$, since at the minimum to gain a quantum advantage in machine learning, one would desire to use a PQC structure which is not classically simulatable.} to connect to the discussions of quantum computational/learning supremacy in \secref{ssec:born_machine/quantum_advantage}.

\begin{figure*}
    \centering
    \includegraphics[width=\columnwidth, height=0.7\columnwidth]{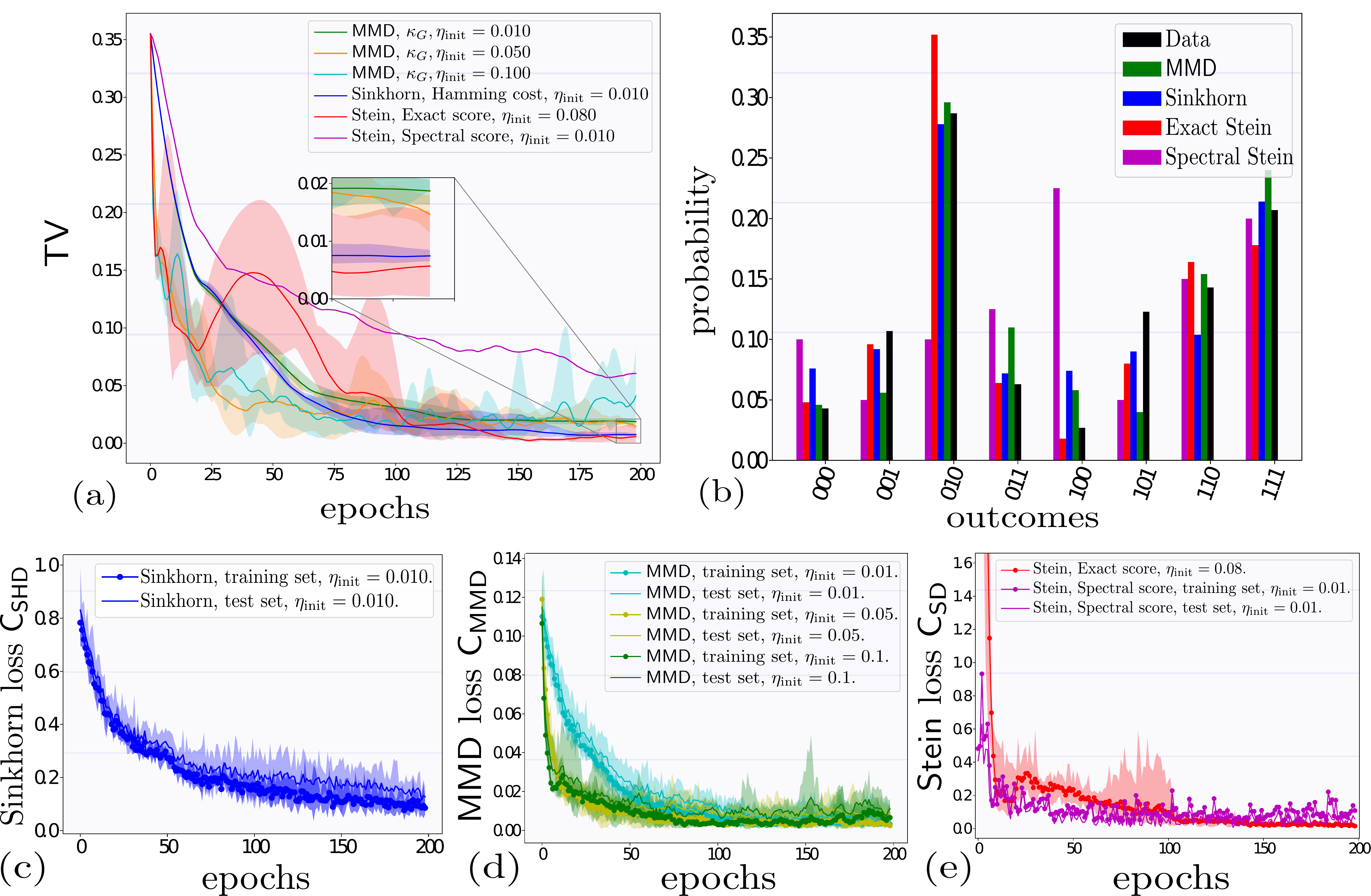}
\caption[\color{black} Comparison of Born machine training methods for $3$ qubits, $\MMD$ Sinkhorn and Stein training with exact and spectral score functions.]{\textbf{$\MMD$ [\crule[cyan]{0.2cm}{0.2cm}, \crule[yellow]{0.2cm}{0.2cm}, \crule[ForestGreen]{0.2cm}{0.2cm}] vs. Sinkhorn [\crule[blue]{0.2cm}{0.2cm}] and Stein training with exact score function [\crule[red]{0.2cm}{0.2cm}] and spectral score method [\crule[magenta]{0.2cm}{0.2cm}] for 3 qubits.} We use a with fully connected topology, Rigetti \computerfont{3q-qvm},
\protect\threeqqvm \ \  , trained on the data, \eqref{eqn:born_machine_toy_dataset}. 500 data points are used for training, with 400 used as a training set, and 100 used as a test set. (a) $\TV$ difference between training methods, with regularisation parameter $\epsilon = 0.1$ for $\SH$, and 3 eigenvectors for Spectral Stein method. Both Sinkhorn divergence and Stein discrepancy are able to achieve a lower $\TV$ than the $\MMD$. Inset shows region of outperformance on the order of $\sim 0.01$ in $\TV$. We observe that the Spectral score method was not able to minimise $\TV$ as well as the exact Stein discrepancy, potentially indicating the need for better approximation methods. (b) Final learned probabilities of each training method.  (c) $\Cbs^{0.08}_{\SH}$ on train and test set. (d) $\Cbs_{\MMD}$ with three different initial learning rates. (e) $\Cbs_{\SD}$ using the exact and spectral score methods. Training on test set observed as thin lines without markers for all methods, excluding the exact score method, since this uses exact probabilities. }
    \label{fig:MMDvSinkvStein3}
\end{figure*}

\begin{figure}
    \centering
    \includegraphics[width=\columnwidth, height=0.5\columnwidth]{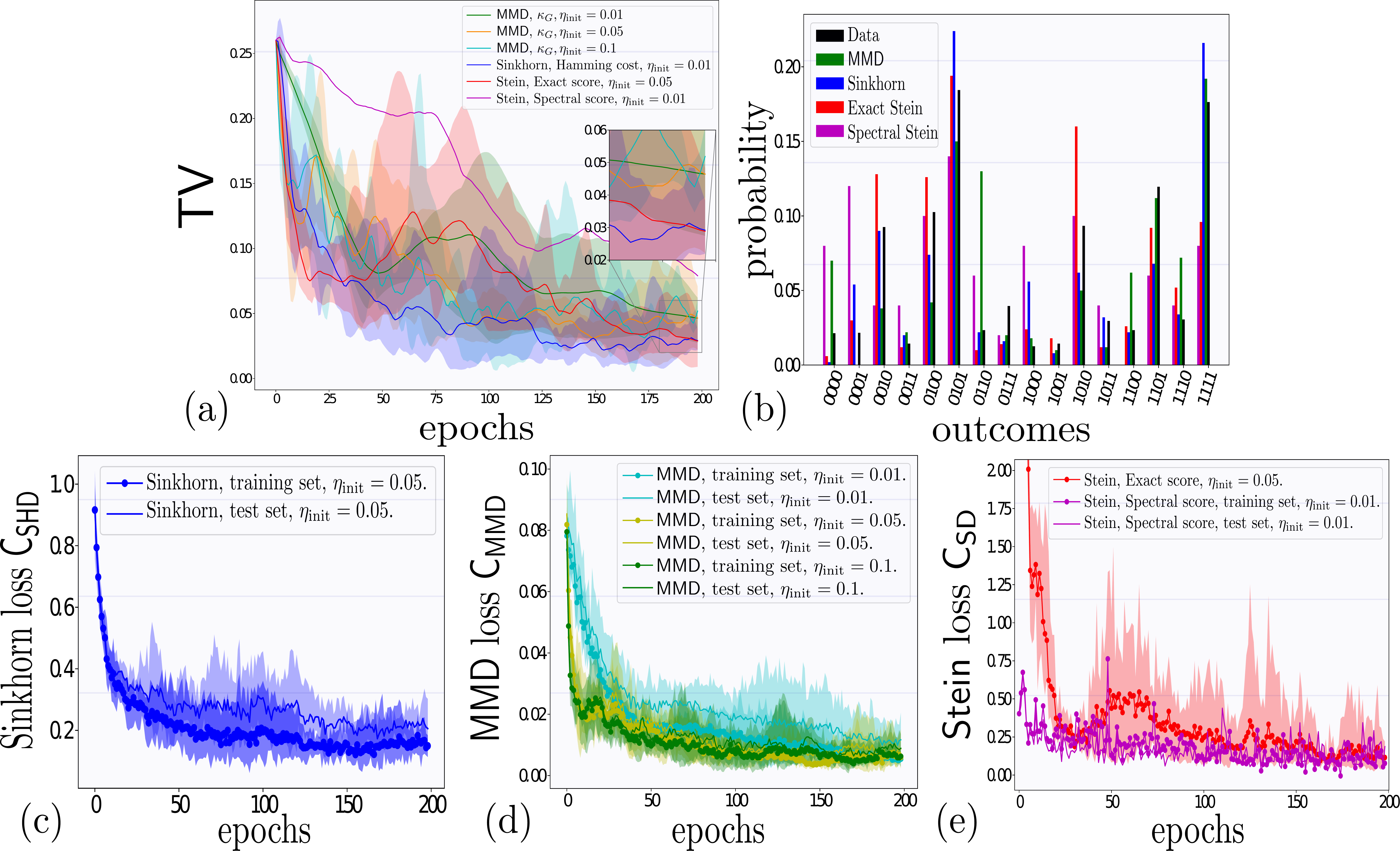}
    \caption[\color{black} Comparison of Born machine training methods for $4$ qubits, $\MMD$ Sinkhorn and Stein training with exact and spectral score functions.]{\textbf{$\MMD$ [\crule[cyan]{0.2cm}{0.2cm}, \crule[yellow]{0.2cm}{0.2cm}, \crule[ForestGreen]{0.2cm}{0.2cm}] vs. Sinkhorn [\crule[blue]{0.2cm}{0.2cm}] and Stein training with spectral score method [\crule[magenta]{0.2cm}{0.2cm}] using 6 eigenvectors for 4 qubits.} We use qubit topology in fully connected graph for four qubits, i.e. Rigetti {\fontfamily{cmtt}\selectfont 4q-qvm},  \protect\fourqqvm \ \ \ . (a) $\TV$ difference between training methods. Both Sinkhorn divergence and Stein discrepancy can achieve lower $\TV$ values than the $\MMD$. (b) Final learned probabilities of target data, \eqref{eqn:born_machine_toy_dataset} [\crule[black]{0.2cm}{0.2cm}]. (c) $\Cbs^{1}_{\SH}$ with regularisation parameter $\epsilon = 1$. Trained using Hamming optimal transport cost function. (d) $\Cbs_{\MMD}$ with three different initial learning rates. (e) $\Cbs_{\SD}$ with the exact and spectral scores using $\eta_{\text{init}} = 0.05$ for Adam.}
    \label{fig:MMDvSinkvStein4}
\end{figure}

In \figref{fig:MMDvSinkvStein3}, \figref{fig:MMDvSinkvStein4}, we illustrate the superior performance of our alternative training methods, as measured by the total variation distance\footnote{In both figures, for all methods we use $500$ samples, with batches of $250$, except for the Stein discrepancy with the spectral method. For this method, for $3$ qubits we use $40$ samples and a batch size of $20$, while for $4$ qubits in \figref{fig:MMDvSinkvStein4} we use $50$ samples and batches of $25$. In all cases we have a $80\%/20\%$ train/test split.}. For these comparisons, we use a basic three and four qubit simulator provided by the Rigetti forest platform~\cite{smith_practical_2017}, the \computerfont{3q-qvm} and \computerfont{4q-qvm} respectively. In all cases, we use the analytic gradients derived in \secref{ssec:born_machine/training}, and the Adam optimiser (\defref{defn:adam_definition}) with varying initial learning rates, $\eta_{\text{init}}$. For the $\MMD$, we use the mixture of Gaussians kernel (\eqref{eqn:gaussian_kernel}) with the same bandwidth parameters as in \cite{liu_differentiable_2018}. In all the plots (where error bars are present), we show the mean, maximum and minimum values over 5 independent training runs on the same dataset. We choose these error bars rather than the variance to highlight the worst and best case performance of all models.

In particular, we wish to highlight the noticeable out-performance observed for the Sinkhorn divergence\footnote{To compute the Sinkhorn divergence and its gradients, we adopted the library of~\cite{jeanfeydy_mmd_2018}.} and the Stein discrepancy relative to training with the $\MMD$ (using a Gaussian kernel), as measured by $\TV$ in \figref{fig:MMDvSinkvStein3}. Furthermore we observed that the gap (highlighted in the inset in \figref{fig:MMDvSinkvStein3}(a)) which separates the Sinkhorn divergence and Stein discrepancy (red and blue lines) from the $\MMD$ (green, yellow and cyan lines) grows as the number of qubits grows. Unfortunately, the Spectral method to approximate the Stein score does not outperform the $\MMD$, despite training successfully. The discrepancy between the true and approximate versions of the Stein score is likely due to the low number of samples used to approximate the score, with the number of samples limited by the computational inefficiency. We leave tuning the hyperparameters of the model in order to get better performance to future work. 

\begin{figure}
    \centering
    \includegraphics[width=0.95\columnwidth, height=0.48\columnwidth]{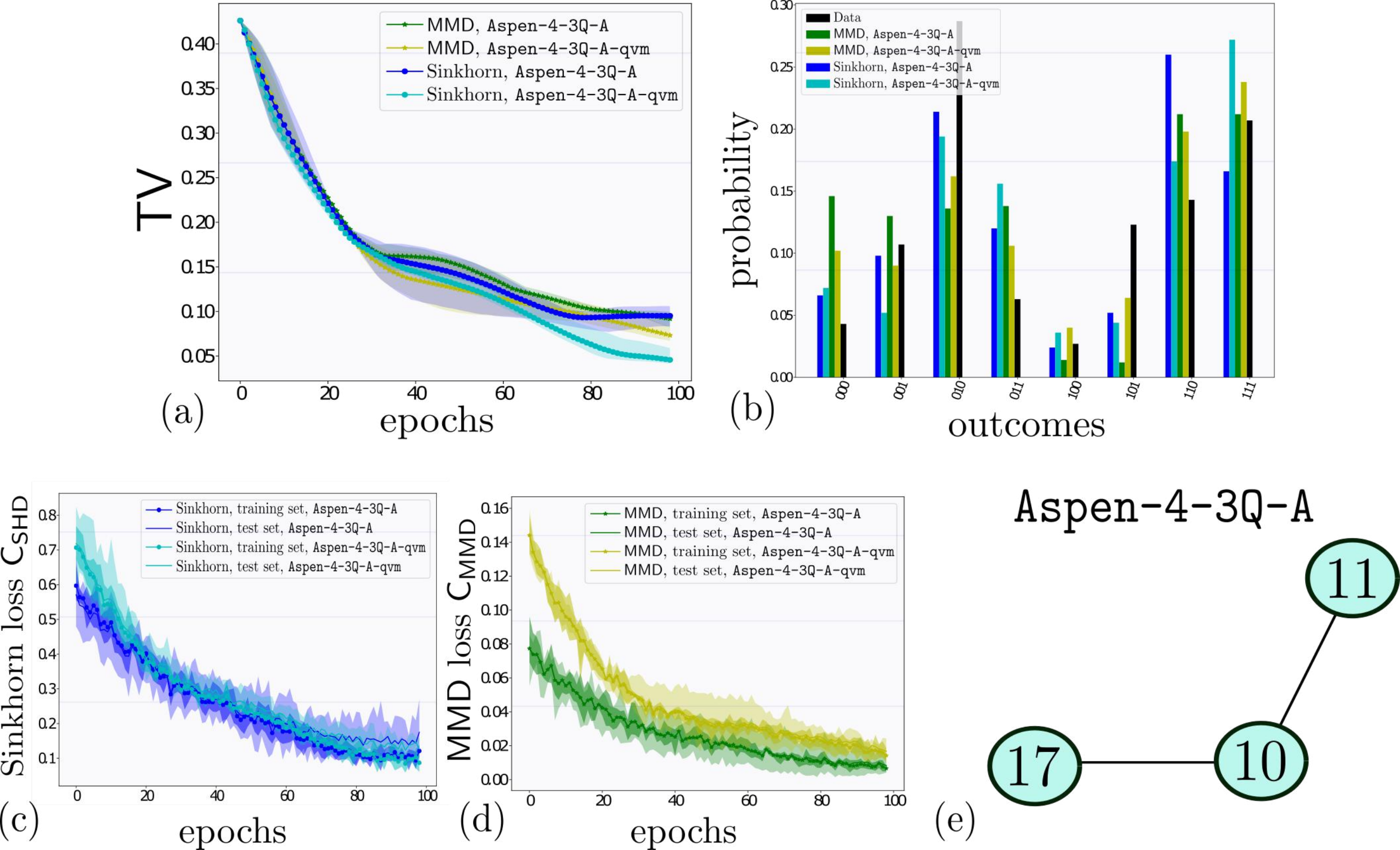}
    \caption[\color{black} $\MMD$ vs. Sinkhorn for $3$ qubits on QVM versus QPU.]{\textbf{$\MMD$ [\crule[ForestGreen]{0.2cm}{0.2cm}, \crule[yellow]{0.2cm}{0.2cm}] vs. Sinkhorn [\crule[blue]{0.2cm}{0.2cm}, \crule[cyan]{0.2cm}{0.2cm}] for 3 qubits comparing performance on the QPU (\computerfont{Aspen-4-3Q-A-qvm}) vs. simulated behaviour on QVM (\computerfont{Aspen-4-3Q-A-qvm})}. Target data given in [\crule[black]{0.2cm}{0.2cm}]. (a) $\TV$ Difference between $\MMD$ [\crule[ForestGreen]{0.2cm}{0.2cm}, \crule[yellow]{0.2cm}{0.2cm}], and Sinkhorn [\crule[blue]{0.2cm}{0.2cm}, \crule[cyan]{0.2cm}{0.2cm}] with regularisation parameter $\epsilon = 0.1$ on QVM vs QPU. (b) Final learned probabilities of target data [\crule[black]{0.2cm}{0.2cm}] using $\MMD$ [\crule[ForestGreen]{0.2cm}{0.2cm}, \crule[yellow]{0.2cm}{0.2cm}] LR $\eta_{\mathsf{\text{init}}} = 0.2$ and Sinkhorn [\crule[blue]{0.2cm}{0.2cm}, \crule[cyan]{0.2cm}{0.2cm}] with $\epsilon = 0.1, \eta_{\mathsf{\text{init}}} = 0.08$. (c) $\Cbs^{0.2}_{\SH}$ on the QVM  [\crule[cyan]{0.2cm}{0.2cm}] vs. QPU  [\crule[blue]{0.2cm}{0.2cm}]. (e) $\Cbs_{\MMD}$ on QVM  [\crule[yellow]{0.2cm}{0.2cm}] vs. QPU  [\crule[ForestGreen]{0.2cm}{0.2cm}]. (f) Qubit `line' topology in Rigetti {\fontfamily{cmtt}\selectfont Aspen-4-3Q-A} chip, using qubits, $(10, 11, 17)$. }
    \label{fig:MMDvSinkvStein3_real}
\end{figure}

\begin{figure}
    \centering
    \includegraphics[width=\columnwidth, height=0.25\columnwidth]{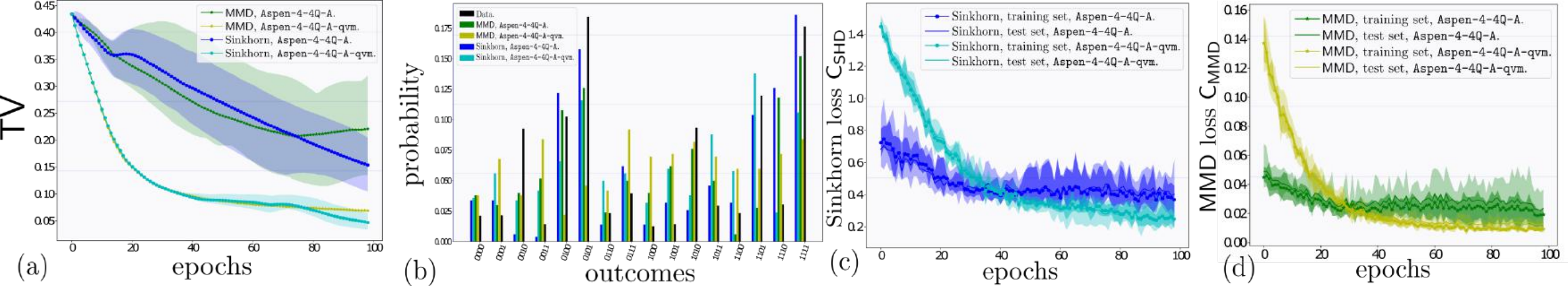}
\caption[\color{black} $\MMD$ vs. Sinkhorn
for 4 qubits on QVM versus QPU.]{\textbf{$\MMD$ [\crule[ForestGreen]{0.2cm}{0.2cm}, \crule[yellow]{0.2cm}{0.2cm}] vs. Sinkhorn [\crule[blue]{0.2cm}{0.2cm}, \crule[cyan]{0.2cm}{0.2cm}] for 4 qubits comparing performance on the QPU ({\fontfamily{cmtt}\selectfont Aspen-4-4Q-A}) vs. simulated behaviour on QVM ({\fontfamily{cmtt}\selectfont Aspen-4-4Q-A-qvm}).} Target data given in [\crule[black]{0.2cm}{0.2cm}]. (a) $\TV$ Difference between training methods with regularisation parameter $\epsilon = 0.08$, (b) Final learned probabilities. The probabilities given by the coloured bars are those achieved after training the model with either the $\MMD$ or $\SH$ on the simulator or the physical Rigetti chip, on an average run. The probabilities of the model are generated by simulating the entire wavefunction after training. (c) $\Cbs^{0.08}_{\SH}$ on QVM  [\crule[cyan]{0.2cm}{0.2cm}] vs. QPU  [\crule[blue]{0.2cm}{0.2cm}]. (d) $\Cbs_{\MMD}$ on QVM  [\crule[yellow]{0.2cm}{0.2cm}] vs. QPU  [\crule[ForestGreen]{0.2cm}{0.2cm}]. In both latter cases, trained model performance on 100 test samples is seen as the thin lines without markers. Again it can be seen that the Sinkhorn divergence outperforms the $\MMD$ both simulated and on chip, with the deviation apparent towards the end of training. Similar behaviour observed after 100 epochs, but not shown due to limited QPU time.}
    \label{fig:MMDvSink4_real}
\end{figure}

This behaviour is shown to persist on the QPU, as demonstrated in \figref{fig:MMDvSinkvStein3_real} and \figref{fig:MMDvSink4_real}\footnote{In both of these figures, we use $\eta_{\text{init}} = 0.01$ for Adam, a train size of $500$ and a test size of $250$ datapoints.} for three and four qubits. In both cases, we show: training of the model with both the $\MMD$ and $\SH$ relative to $\TV$, the learned probabilities of both methods on, and off, the QPU and the behaviour of the cost functions associated to both methods. This reinforces our theoretical argument that the Sinkhorn divergence is able to better minimise $\TV$ to achieve superior results. For these experiments, we used two sublattices of the $\aspenfour$ devices, specifically the \computerfont{Aspen-4-3Q-A} and \computerfont{Aspen-4-4Q-A}, and their respective \computerfont{qvm} versions. The available qubits and connectivity for the former sublattice is shown in \figref{fig:MMDvSinkvStein3_real}(e). For the $\QCIBM$ ansatz used on the hardware, we only use a $\CZ$ connectivity which matches this restricted topology. This is one reason why we observe high quality results even on the quantum chip. In \secref{sec:born_machine/finance} and \chapref{chap:cloning}, we take this hardware-restricted methodology much further.

Interestingly, in \figref{fig:MMDvSinkvStein3_real}(d), we note that the hardware actually trains to lower values of $\MMD$ than the simulator, despite the reverse situation relative to total variation seen in \figref{fig:MMDvSinkvStein3_real}(a) (i.e. the simulator outperforms the hardware). This potentially indicates that the training with the $\MMD$ on noisy hardware could lead to overconfident, and incorrect results. However, we do note that the Born machine model does appear to demonstrate a form of `\emph{optimal parameter resilience} (OPR)\footnote{This concept of OPR was introduced by ~\cite{sharma_noise_2020} which evokes a particular type of noise resilience of VQAs.}', since training directly on the QPU also provides circuit parameters which also perform well when simulated (comparing the plots of $\TV$ in \figref{fig:MMDvSinkvStein3_real}(a), \figref{fig:MMDvSink4_real}(a)).

Given the performance noted above, we would recommend the Sinkhorn divergence as the primary candidate for future training of these models, due to its simplicity and competitive performance. One should also note that our goal here is not to exactly fit the data, only to compare the training methods. Specifically, we only use a shallow fixed circuit structure for training (i.e. a $\QAOA$-based $\QCIBM$ circuit) which we do not alter. For the financial dataset, we experiment further with a variable number of parameters.

Now that we have warmed up with small examples using three and four qubits, let us ramp up the scale in the next section, where we run experiments with up to $28$ qubits.

\subsection[\texorpdfstring{\color{black}}{} Quantum versus classical generative modelling in finance]{Quantum versus classical generative modelling in finance} \label{sec:born_machine/finance}
The goal of this section is twofold. Firstly, as mentioned previously we experiment with larger quantum Born machines for the task of generative modelling. Secondly, we target a much more real-world use case, focusing on the financial data distribution given in \secref{sssec:born_machine/data/finance_data}. Finally, we provide another dimension to our `quantum learning supremacy' arguments presented in \secref{ssec:born_machine/quantum_advantage}. We do this by producing a rigorous numerical comparison between the QCBM as a quantum model, versus the restricted Boltzmann machine (RBM) as a classical model in learning the same dataset.

\subsubsection[\texorpdfstring{\color{black}}{} The \texorpdfstring{$\Ansatz$}{}]{The \texorpdfstring{$\Ansatz$}{}} \label{sssec:born_machine/finance/numerics/ansatz}
\begin{figure}[t]
    \centering
    \includegraphics[width=0.9\columnwidth, height=0.3\columnwidth]{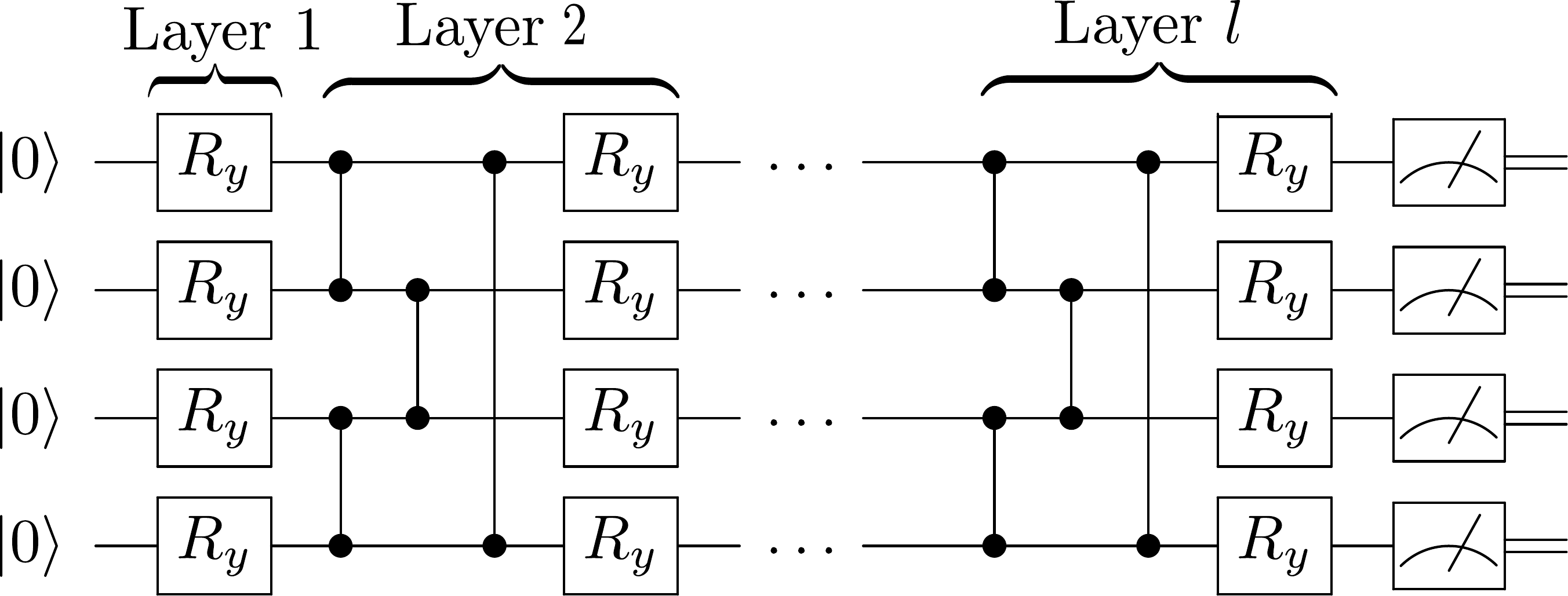}
    \caption[\color{black} Hardware efficient circuit for the \computerfont{Aspen}\texttrademark-7-4Q-C device.]{\textbf{Hardware efficient circuit for the \computerfont{Aspen}\texttrademark-7-4Q-C.} Here we show the model with $l$ layers using the native entanglement structure native to the chip.}
    \label{fig:aspen_7_4q_circuit_ansatz}
\end{figure}

\begin{figure}[t]
    \centering
    \includegraphics[width=0.9\columnwidth, height=0.7\columnwidth]{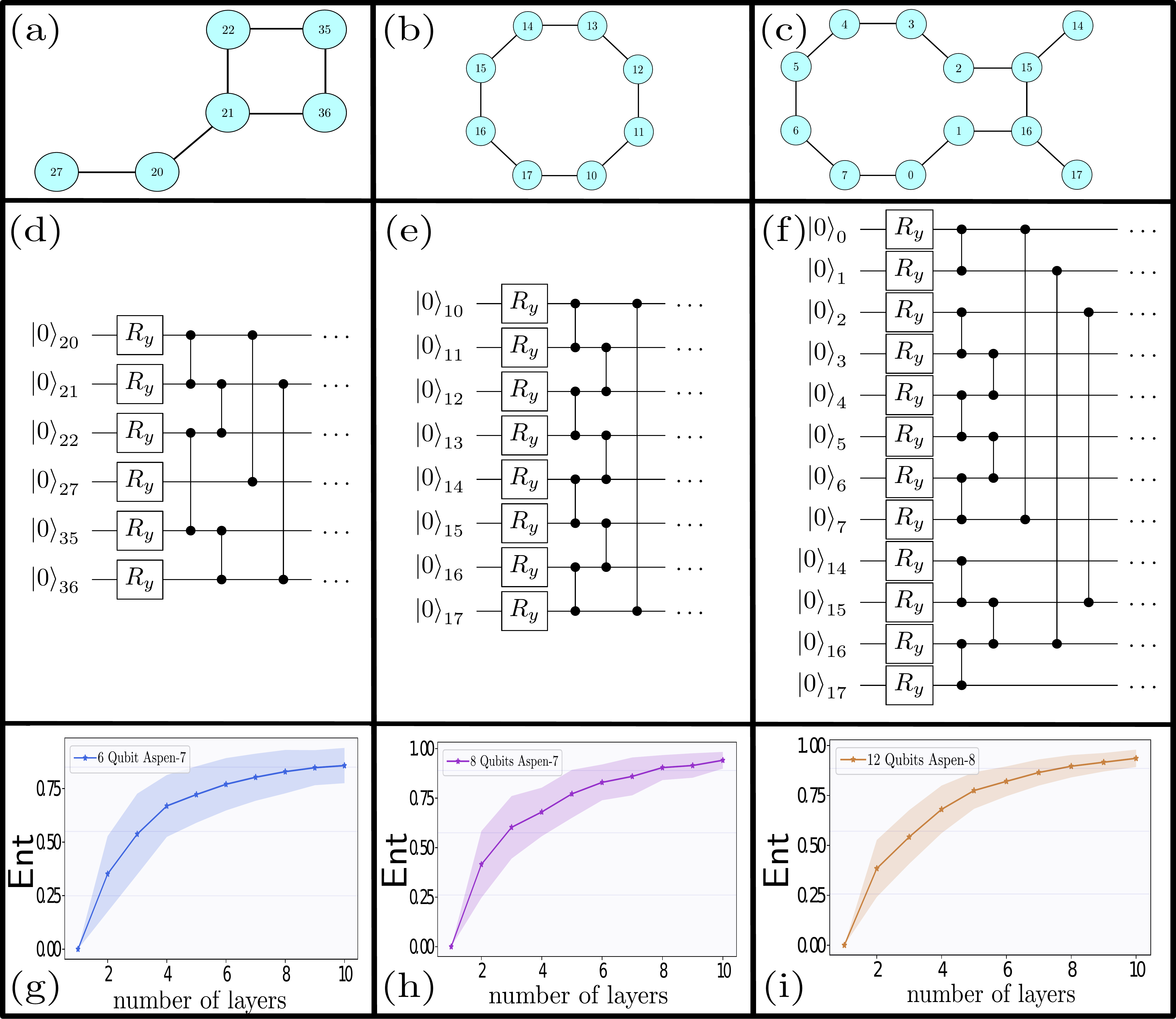}
    \caption[\color{black} Hardware efficient circuits for $6$, $8$, $12$ qubit Born machine $\Ansatz$.]{\textbf{Hardware efficient circuits for $6$, $8$, $12$ qubit Born machine $\Ansatz$.} (a)-(c) show $\aspenseven\text{\computerfont{-6Q-C}}$ ,  $\aspenseven\text{\computerfont{-8Q-C}}$  from the $\aspenseven$  chip and a $12$ qubit sublattice from the $\aspeneight$  chip which we consider. (d) - (f) illustrate the the entanglement structure in a single layer, which tightly matches the chip topology. (g) - (i)  show the average entangling capability, $\mathsf{Ent}$ in \eqref{eqn:average_ent_capability_definition} as a function of the number of layers in the circuit for each of the entangling structures shown in (d)-(f). Error bars show mean and standard deviation over $100$ random parameter instances, $\{\paramtheta^i\}_{i=1}^{100}, \paramtheta^i_j\sim U(0, 2\pi)$, in the single qubit rotations. $U$ is the uniform distribution over the interval $[0, 2\pi]$.}
    \label{fig:aspen_7_6q_8q_12q_circuit_ansatze}
\end{figure}

We move away from the $\QCIBM$ here, and instead focus on a more flexible $\Ansatz$, which is closer to the hardware-efficient fixed structure (HEFS) $\Ansatze$ discussed in \secref{ssec:vqa_ansatzse}. Specifically, we fit the entangling structure of the QCBM to the sublattices of the Rigetti $\aspenseven$ and $\aspeneight$ as shown in \figref{fig:aspen_sublattices}. For example, \figref{fig:aspen_7_4q_circuit_ansatz} shows a four qubit HEFS suited to the $\aspenseven\text{\computerfont{-4Q-C}}$ in \figref{fig:aspen_sublattices}(a). We also show one layer of the matching $\Ansatze$ for three other sublattices in \figref{fig:aspen_7_6q_8q_12q_circuit_ansatze}.

For each circuit in \figref{fig:aspen_7_6q_8q_12q_circuit_ansatze}, we compute a quantity, $\mathsf{Ent}$ which is the average Meyer-Wallach entanglement capacity~\cite{meyer_global_2002}.

This is a measure of entanglement in quantum states proposed as a method of comparing different circuit $\Ansatze$ by \cite{sim_expressibility_2019}. This measure has been used in a similar context by \cite{hubregtsen_evaluation_2021} in order to draw connections between $\Ansatz$ structure and classification accuracy. We first define an entanglement measure $Q$, for a given input state $\ket{\psi}$, as:
\begin{align}\label{eqn:meyer_wallach_entanglement_measure}
    Q(\ket{\psi}) &:= \frac{4}{n}\sum_{j=1}^n \text{d}_{\text{para}}(\mathsf{del}_j(0)\ket{\psi}, \mathsf{del}_j(1)\ket{\psi})\\
    \text{d}_{\text{para}}(\ket{\boldsymbol{u}}, \ket{\boldsymbol{v}}) &= \frac{1}{2} \sum\limits_{i, j}|\boldsymbol{u}_i\boldsymbol{v}_j - \boldsymbol{u}_j\boldsymbol{v}_i|^2\\
    \ket{\boldsymbol{u}} &:= \sum_i \boldsymbol{u}_i\ket{i}, \ket{\boldsymbol{v}} := \sum_j \boldsymbol{v}_j\ket{j}
\end{align}
where $\text{d}_{\text{para}}$ is a particular distance between two quantum states given by: $\ket{\boldsymbol{u}}$ and $\ket{\boldsymbol{v}}$. This distance can be understood as the square of the area of the parallelogram created by vectors $\ket{\boldsymbol{u}}$ and $\ket{\boldsymbol{v}}$. The notation $\mathsf{del}_j(b)$ is a linear map (a sort of `deletion' operator) which acts on computational basis states as follows:
\begin{equation} \label{eqn:iota_definition}
    \mathsf{del}_j(b)\ket{b_1\dots b_n} := \delta_{bb_j}\ket{b_1\dots \hat{b}_j\dots b_n}
\end{equation}
where $\hat{\cdot}$ indicates the absence of the $j^{th}$ qubit. For example, $ \mathsf{del}_2(0)\ket{1001} = \ket{101}$
However, to evaluate $Q$ for a quantum state, we instead use the equivalent formulation derived by \cite{brennen_observable_2003}, which involves computing the purities of each subsystem of the state $\ket{\psi}$:
\begin{equation}\label{eqn:meyer_wallach_entanglement_alternative}
    Q(\ket{\psi}) = 2\left(1-\frac{1}{n}\sum_{k=1}^{n}\Tr[\rho_k^2]\right)
\end{equation}
where $\rho_k := \Tr_{\Bar{k}}\left(\ketbra{\psi}{\psi}\right)$ is the partial trace over all $n$ subsystems of $\ket{\psi}$ \emph{except} $k$. This reformulation of $Q$ gives more efficient computation and operational meaning since the purity of a quantum state is efficiently computable. Given $Q$, we define \cite{sim_expressibility_2019} $\mathsf{Ent}$ as the average value of $Q$ over a set, $\mathcal{S}$ of $M$ randomly chosen parameter instances,  $S := \{\paramtheta^i\}_{i=1}^M$:
\begin{equation} \label{eqn:average_ent_capability_definition}
    \mathsf{Ent} := \frac{1}{|S|} \sum_{i} Q(\ket{\psi_{\paramtheta^i}})
\end{equation}
Quantities such as $\mathsf{Ent}$ are extremely valuable for measuring the performance of heuristic algorithms like $\VQA$s. Since we do not, in general, have theoretical arguments about performance, or the relative quality of one $\Ansatz$ over another, it is invaluable to have useful and interpretable metrics which one can compute in practice. With these in hand, we can begin the process of studying which features (e.g. entangling capability, expressibility, contextuality, etc.) of PQCs and $\VQA$s lead to advantages over classical methods.

\subsubsection[\texorpdfstring{\color{black}}{} Training \& evaluation]{Training \& evaluation} \label{sssec:born_machine/numerics/training}
Now, as mentioned at the start of this section, one of our goals here is to evaluate the performance of the QCBM relative to a comparable classical model - the restricted Boltzmann machine. For a fair comparison, we must compare both on a level ground. In order to do so, we fix the number of parameters in each model and then test which one performs better in practice. This methodology has been performed previously in two other works relating to a QCBM, in both cases the RBM was the classical model of choice to compare to. The first, \cite{alcazar_classical_2020}, compared an RBM to a QCBM using data from a portfolio optimisation problem, while the second, \cite{kondratyev_non-differentiable_2021}, targeted the same dataset as we do here (the currency pairs from \secref{fig:currency_pairs}). Both of these works found that the QCBM had the capacity to outperform the RBM. We supplement these findings, in particular, by extending the methodology of~\cite{kondratyev_non-differentiable_2021}. We include extensive numerical experiments both simulated, \emph{and} on quantum hardware. We also implement \emph{differentiable} training methods for the QCBM and an alternative method for sample generation from the RBM (recall the discussion from \secref{ssec:prelim/machine_learning/training_NN}) based on path-integral Monte Carlo.

To enforce the same number of parameters in the model, we first choose  QCBM HEFS $\Ansatze$ as in \figref{fig:aspen_7_4q_circuit_ansatz}, \figref{fig:aspen_7_6q_8q_12q_circuit_ansatze} which has $n\times l$ parameters for $l$ layers. Given this choice, we then build a corresponding RBM which has $n_{\mathrm{v}} := |\mathcal{N}_{\mathrm{v}}| = n$ visible nodes and $n_{\mathrm{h}} := |\mathcal{N}_{\mathrm{h}}| = nl-n = n\times (l-1)$ hidden nodes. Since adding trainable two qubit parameters to the $\Ansatz$ would increase our compilation overhead, we opt to fix also the RBM weights to have random (but \emph{constant}) values. Therefore, only the local biases of each node are trained. An example of such a fixed-weight RBM with $6$ visible nodes can be found in \figref{fig:rbm_6_node_structure}. We revisit weight training in \figref{fig:weight_training}.

Firstly, before detailing the training methods for both models, let us ask the question: how should we evaluate their performance? As we discussed above, computing a $\TV$ distance would be ideal since it is a strong benchmark of relative performance. However, we clearly cannot do so here, since we do not have access to the exact data probabilities, $\pi(\ybs)$. As an alternate strategy, we employ an \emph{adversarial discriminator}, adapted from ideas in generative adversarial networks\footnote{See~\cite{zeng_learning_2019, situ_quantum_2018, romero_variational_2021, anand_experimental_2020} related work on adversarial training methods for quantum generative models}. This discriminator is essentially a classifier who is given the responsibility of deciding whether a sample, $\zbs$, comes from the actual data distribution, $\zbs \sim\pi$, and hence is a `real' sample (labelled, say $1$), or from the model, $\zbs \sim p_{\paramtheta}$ and hence is `fake' (labelled $0$). The principle here is the same as that discussed in \chapref{chap:classifier}, with only the model being different (or indeed, the classifier we choose could be exactly the one presented in \chapref{chap:classifier}). We measure performance relative to the classification accuracy of this discriminator (or the error it makes in classifying an example) For an untrained generative model, the classifier should be able to easily discriminate between the real and synthetic samples, but as training progresses, the discriminator should increasingly make mistakes. The generative model is fully trained when the best strategy of the classifier is simply to guess randomly when presented a sample and so will have an error of $50\%$.

For simplicity, we use a common classical model called a \emph{random forest}~\cite{ho_random_1995}, an implementation of which is readily available in the  \computerfont{scikit-learn}~ \cite{pedregosa_scikit-learn_2011} python package. A random forest is an \emph{ensemble} classification method, as it takes a majority vote for a classification output over multiple \emph{decision tree} classifiers. We chose this since it was able to provide a separation between our models\footnote{If, on the contrary, a random forest was \emph{not} able to separate the performance of both models, we would be in an inconclusive situation - either there is truly no separation, or our classification model is not powerful enough to isolate suitable features which label the model data as `fake'.} so we do not need to look at `stronger' classifiers.

As for training methods, we primarily use the Sinkhorn cost function, $\Cbs^\epsilon_{\SH}$, \eqref{eqn:sinkhorn_divergence} plus its analytic gradients, ~\eqref{sinkhorngradient_supp} for training the QCBM\footnote{For completeness, we also considered training the QCBM using the $\MMD$, using a genetic algorithm as in~\cite{kondratyev_non-differentiable_2021} and finally also with respect to the adversarial discriminator. Of course, training with respect to the discriminator would not provide an unbiased comparison. The $\MMD$ is less favourable due to the arguments presented in \secref{ssec:born_machine/training} and we found the genetic algorithm to be significantly slower than differentiable training.}. For all the results presented int the following section, we use the Adam optimiser~\eqref{eqn:adam_update_rule} with an initial learning rate of $\eta_{\text{init}} = 0.05$ and a value for $\epsilon=0.5$, the latter of which we determined using a basic hyperparameter search using small problem instances. Also to note, is that in all simulated versions of the QCBM, we use the \computerfont{noisy-qvm} version of a quantum device, which incorporates a simple noise model including readout errors and standard $T_1$ and $T_2$ times. Details can be found in the source code of PyQuil~\cite{smith_practical_2017}. 

For the RBM, we use stochastic gradient descent (\defref{defn:sgd_definition}) and the gradient update rule given in \secref{ssec:prelim/machine_learning/training_NN}. As we discussed in the latter section, the most common method to generate samples using from the RBM is via contrastive divergence. Here, we opt for an alternative method to sample from the Gibbs distribution based the path-integral formulation of quantum mechanics (path-integral Monte Carlo (PIMC), as discussed in \secref{ssec:prelim/machine_learning/training_NN}). The PIMC method is the core ingredient in QxSQA (introduced in~\cite{padilha_qxsqa_2019}), which is a GPGPU-Accelerated simulated quantum annealer. In~\cite{padilha_qxsqa_2019}, QxSQA was shown to provide sufficiently good approximations to the required Boltzmann distribution to be considered near exact, for a large range of model sizes, even orders of magnitude larger than those considered here. To generate RBM samples, we use the following hyperparameters whose definitions can be found in~\cite{padilha_qxsqa_2019}: $\Gamma_0=3, \Gamma_\tau = 1\times 10^{-20}, N_{\text{sweeps}} = 1, PT = 0.1, N_{\text{anneals}} = 250, T_{\text{eff}} = 1$.

\subsubsection[\texorpdfstring{\color{black}}{} Financial dataset results]{Financial dataset results} \label{sssec:born_machine/numerics/finance/numerics}

\begin{figure}[t]
    \centering
    \includegraphics[width=0.95\columnwidth, height=0.7\columnwidth]{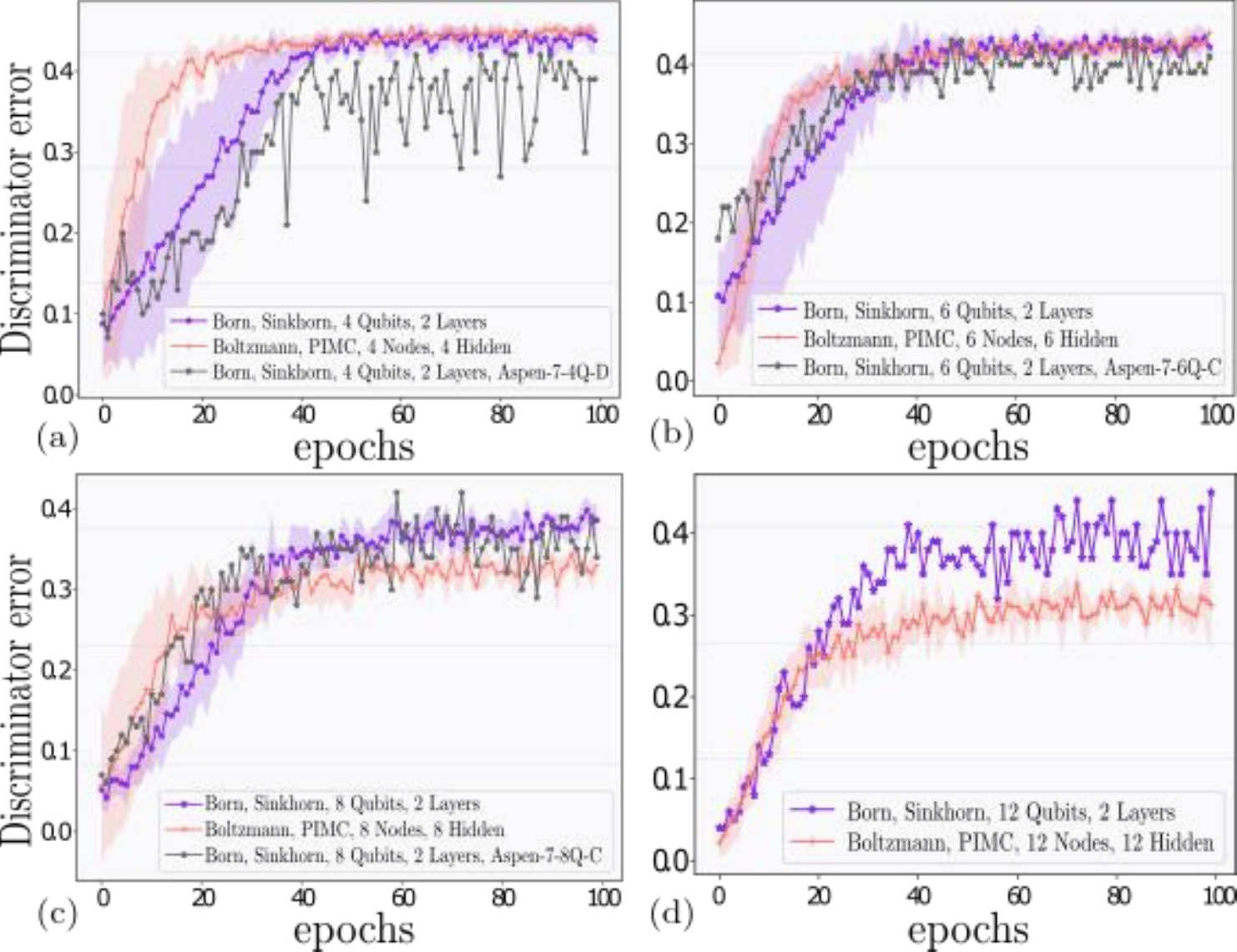}
    \caption[\color{black} 2 currency pairs at 2, 3, 4 bits and 6 bits of precision.]{\textbf{2 currency pairs (specifically \textsf{EUR/USD} and \textsf{GBP/USD}).} We compare at (a) 2 bits, (b) 3 bits, (c) 4 bits and (d) 6 bits of precision. Correspondingly, we use a QCBM [\crule[RoyalPurple]{0.2cm}{0.2cm}] of $4, 6, 8$ and $12$ qubits using the $\Ansatze$ described above, and an RBM [\crule[Salmon]{0.2cm}{0.2cm}] with the same numbers of visible nodes. The hidden units are scaled in each case to match 2 layers of the QCBM. Results when the QCBM is run on sublattices of the $\aspenseven$ QPU are shown in grey, whereas the simulated version is given by the purple line, with simple noise model.}
    \label{fig:2_currency_pairs_born_v_boltz}
\end{figure}

\begin{figure}[t]
    \centering
    \includegraphics[width=\columnwidth, height=0.4\columnwidth]{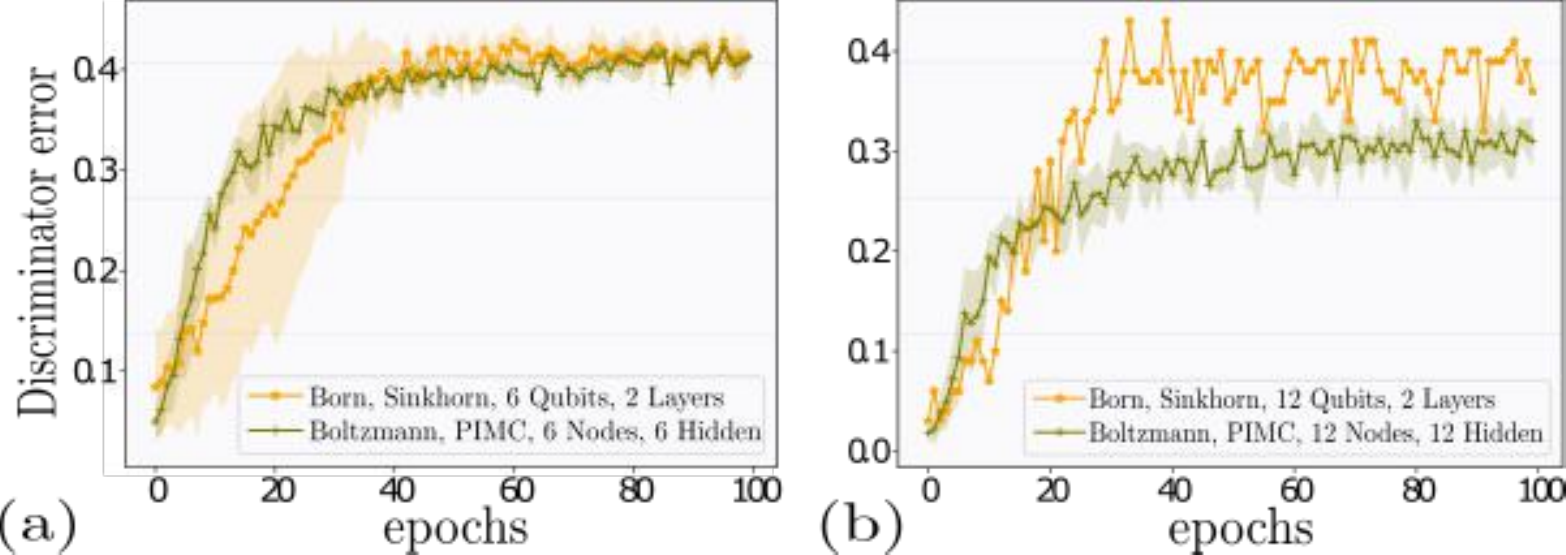}
    \caption[\color{black} 3 currency pairs at 2 bits and 4 bits of precision.]{\textbf{3 currency pairs}. We compare at (a) 2 bits and (b) 4 bits of precision, using a QCBM [\crule[YellowOrange]{0.2cm}{0.2cm}] of $6$ and $12$ qubits and an RBM [\crule[YellowGreen]{0.2cm}{0.2cm}] with the same numbers of visible nodes.}
    \label{fig:3_currency_pairs_born_v_boltz}
\end{figure}

\begin{figure}[t]
    \centering
    \includegraphics[width=\columnwidth, height=0.4\columnwidth]{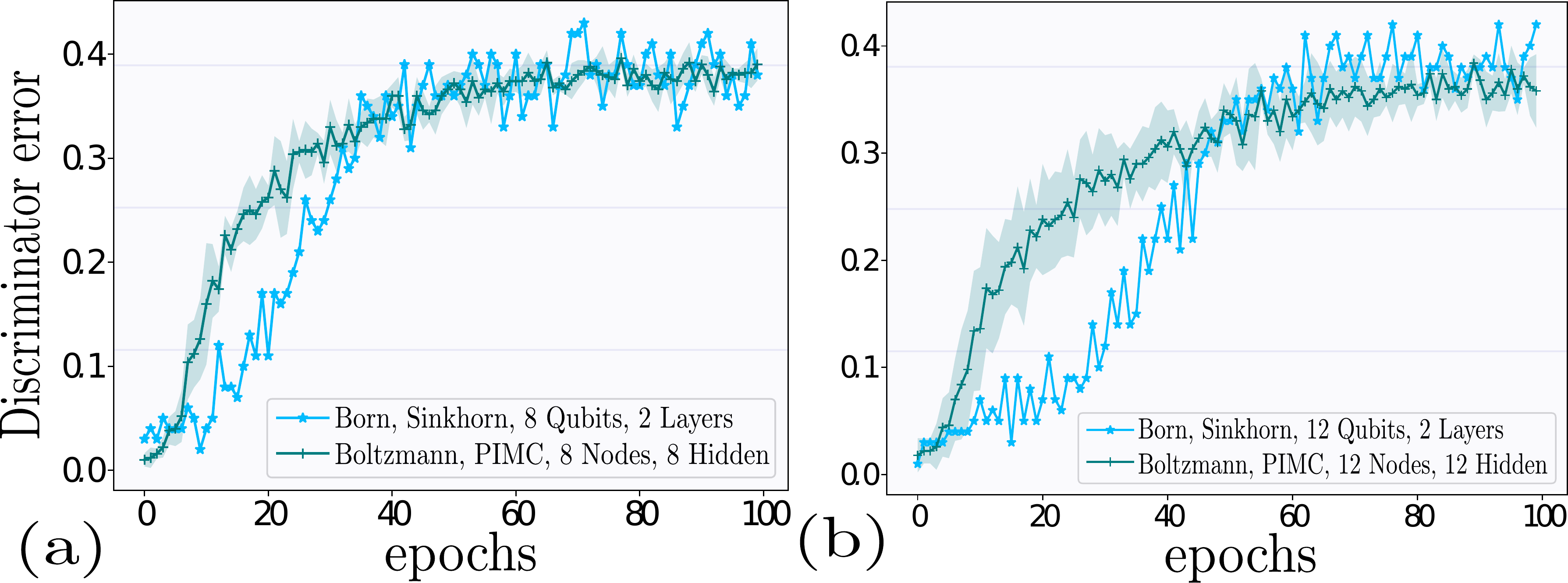}
    \caption[\color{black} 4 currency pairs at 2 bits and 3 bits of precision.]{ \textbf{All 4 currency pairs.} We compare at (a) 2 bits, (b) 3 bits of precision, using a QCBM [\crule[SkyBlue]{0.2cm}{0.2cm}] of $8$ and $12$ qubits and RBM [\crule[JungleGreen]{0.2cm}{0.2cm}] with the same numbers of visible nodes. We notice the RBM performs similarly here to the QCBM, with a possible advantage for the QCBM observed. The lack of a more pronounced gain in this case is likely due to the smaller bit precision used here.}
    \label{fig:4_currency_pairs_born_v_boltz}
\end{figure}

\begin{figure}[t]
    \centering
    \includegraphics[width=0.9\columnwidth, height=0.4\columnwidth]{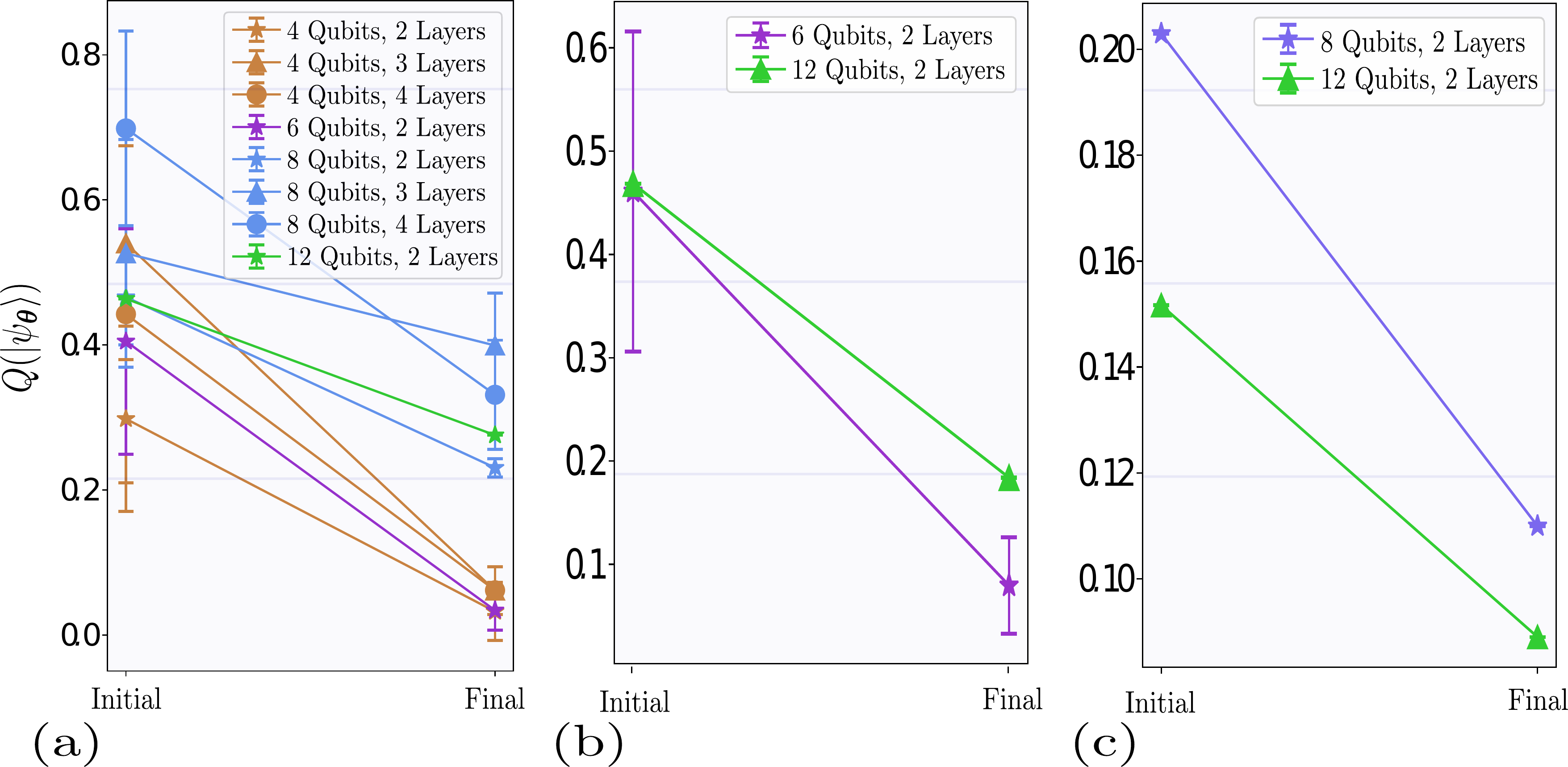}
    \caption[\color{black} Meyer-Wallach entangling capability for a random choice of parameters and the trained parameters in the same circuit.]{\textbf{Meyer-Wallach entangling capability \eqref{eqn:meyer_wallach_entanglement_alternative} for a random choice of parameters (Initial) and the trained parameters (Final) in the same circuit.} The above results are generated by simulating the corresponding circuits. Error bars represent mean and standard deviation over 5 independent training runs, where they are shown. The circuit $\ansatze$ used are those above in \figref{fig:aspen_7_6q_8q_12q_circuit_ansatze} closely matching the corresponding chip topology. In each panel we see the entanglement in the final states trained on (a) $2$ currency pairs at $2, 3, 4, 6$ bits of precision, (b) $3$ currency pairs at $2$ and $4$ bits of precision and (c) $4$ currency pairs at $2$ and $3$ bits of precision.}
    \label{fig:entangling_capability_in_training}
\end{figure}

Finally, let us present the results of the comparison between the RBM and the QCBM. 
In summary, we find the Born machine has the capacity to outperform the RBM as the precision of the currency pairs increases. In \figref{fig:2_currency_pairs_born_v_boltz}, we use data from $2$ currency pairs, at $2, 3, 4$ and $6$ bits of precision. We notice the Born machine outperforms the RBM around $4$ bits (measured by a higher discriminator error), and still performs relatively well when run on the QPU. Similar behaviour is observed for $3$ currency pairs in \figref{fig:3_currency_pairs_born_v_boltz}, which uses a precision of $2$ and $4$ bits, and with $4$ pairs in \figref{fig:4_currency_pairs_born_v_boltz} for a precision of $2$ and $3$ bits. In \figref{fig:entangling_capability_in_training} we plot the entangling capability (defined by \eqref{eqn:meyer_wallach_entanglement_alternative}) of the states generated by initial and final circuits learned via training. Curiously, we notice that in the problem instances in which the Born machine outperforms the Boltzmann machine (those with a higher level of precision), the trained circuits produce states which are more entangled (as measured by the Meyer-Wallach entanglement capacity, \eqref{eqn:average_ent_capability_definition}) than those that do not, despite the data being completely classical in nature. This is especially prominent for $2$ currency pairs in \figref{fig:entangling_capability_in_training}(a), in which the training drives the entanglement capability at $2$ and $3$ bits of precision close to zero (even for increased numbers of layers), but it is significantly higher for $4$ and $6$ bits of precision, when the Born machine outperforms the RBM, as seen in \figref{fig:2_currency_pairs_born_v_boltz}. Similar behaviour is seen for $3$ currency pairs, but not as evident for $4$ pairs. The latter effect is possibly correlated to the similar performance of both models for $4$ currency pairs up to $3$ bits of precision. The observed behaviour of the entangling capability of the QCBM states is one direction in exploring \emph{why} the model may demonstrate an advantage relative to the RBM and provides possibly the most important question raised by this work for future investigation. Seeking explanations for certain advantages could be beneficial in designing future QML algorithms which can actively exploit such features.

In all the figures presented here, where shown (we typically repeated runs only for small problem instances due to the overheads required to simulate the QCBM), errors bars indicate mean and standard deviations over $5$ independent training runs and deviations are due to the stochastic nature of the training procedure. In all figures, where shown, grey lines indicate training runs on the \computerfont{Aspen}\texttrademark\ QPU.

\begin{figure}[t]
    \centering
    \includegraphics[width=\columnwidth, height=0.4\columnwidth]{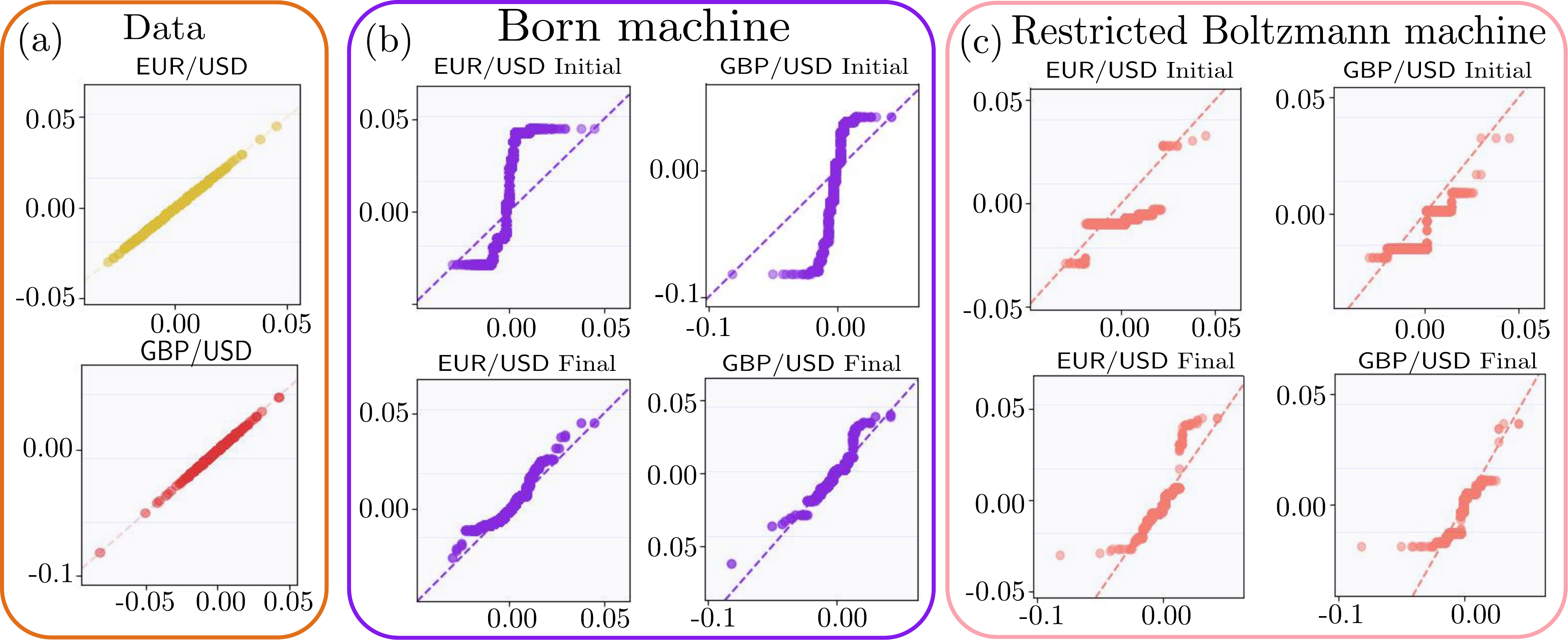}
    \caption[\color{black} QQ plots of the marginal distributions of $2$ currency pairs at $6$ bits of precision.]{\textbf{QQ Plots corresponding to \figref{fig:2_currency_pairs_born_v_boltz}(d) of the marginal distributions of $2$ currency pairs ($\mathsf{EUR/USD}$ and $\mathsf{GBP/USD}$) at $6$ bits of precision.} The QCBM distributions [\crule[RoyalPurple]{0.2cm}{0.2cm}] and those generated by the RBM [\crule[Salmon]{0.2cm}{0.2cm}]. (a) shows the QQ plot for the marginal distribution of each currency pair with respect to itself as a benchmark. (b) Born machine initial (top panels) and final (bottom panels) marginal distributions for both pairs and similarly in (c) for the RBM. While not able to completely mimic the data due to the low number of parameters, the Born machine clearly produces a better fit.}
    \label{fig:QQ_plots_2_pairs}
\end{figure}

As a further measure of visually comparing performance of the trained models, we show QQ (quantile-quantile) plots of the marginal output distributions from one of the outperforming QCBM cases ($2$ pairs at $6$ bits of precision) in \figref{fig:QQ_plots_2_pairs}. The QCBM clearly produces a better data fit than the RBM when trained (bottom panels), where a perfect fit would be a straight line, shown in \figref{fig:QQ_plots_2_pairs}(a) where the data is plotted against itself. In both cases, we use $5070$ samples from the QCBM and RBM, to match the size of the dataset.

In~\figref{fig:differing_layers} and~\figref{fig:differing_hidden_nodes} we investigate changing the parameter count in each model for fixed problem sizes. For both models, we find that increasing the model size past those shown in the main text did not have a significant impact on the expressiveness, indicating that saturation has occurred for these parameter numbers. We notice that increasing the number of hidden nodes (layers) for the RBM (QCBM) slightly decreases (increases) convergence speed, but it does not affect the final converged value appreciably.

\begin{figure}[t]
    \centering
    \includegraphics[width=0.9\columnwidth, height=0.6\columnwidth]{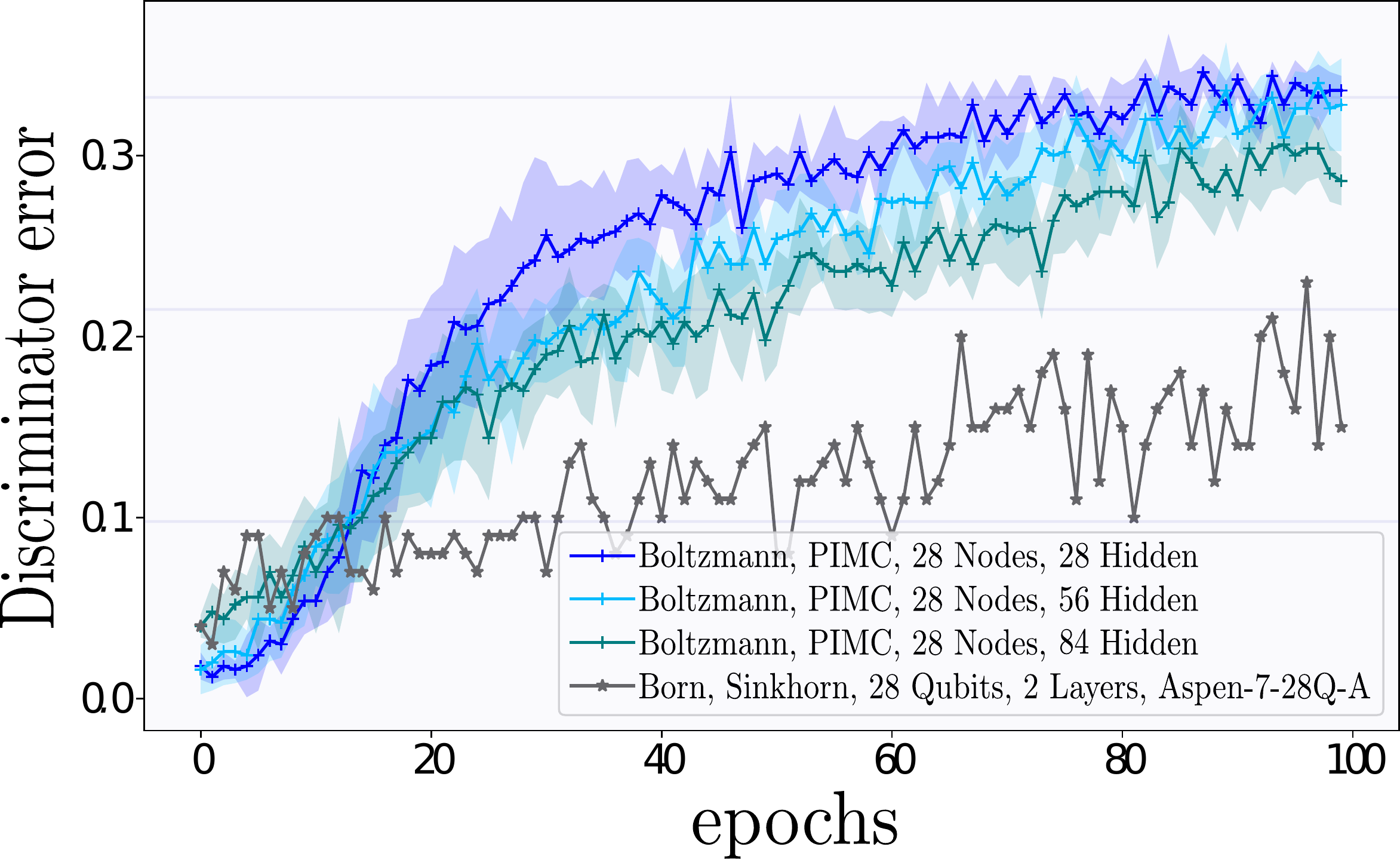}
    \caption[\color{black} $4$ currency pairs at $2$ bits of precision with a $28$ qubit QCBM.]{\textbf{Random forest discriminator error during training for a problem size of $4$ currency pairs at $7$ bits of precision.} We use $28$ visible nodes in the Boltzmann machine, [\crule[Blue]{0.2cm}{0.2cm}, \crule[SkyBlue]{0.2cm}{0.2cm}, \crule[JungleGreen]{0.2cm}{0.2cm}] and $28$ qubits in the Born machine [\crule[Gray]{0.2cm}{0.2cm}]. The $28$ qubit Born machine is run exclusively on the $\aspenseven$\computerfont{-28Q-A} chip using 2 layers of the hardware efficient $\Ansatz$ similar to those shown in \figref{fig:aspen_7_6q_8q_12q_circuit_ansatze}.}
    \label{fig:28_qubit_born_boltzmann}
\end{figure}

We are also able to somewhat successfully train the largest instance of a Born Machine
to date in the literature, namely one consisting of $28$ qubits on the Rigetti $\aspenseven$ chip (whose topology is shown in \figref{fig:aspen_sublattices}(e), and we find it performs surprisingly well. \figref{fig:28_qubit_born_boltzmann} shows the result of this, comparing a $28$ qubit QCBM with a $28$ visible node RBM. While the performance of the Born machine is significantly less than that of its counterpart, it is clear that the model is learning (despite hardware errors), up to a discriminator error of $20\%$. We emphasise that this result does not contradict those shown above since we are not able to simulate the QCBM at this scale in a reasonable amount of time. We would not necessarily expect the Born machine to match the performance of the RBM \emph{on hardware} at this scale for a number of reasons, the most likely cause for diminishing performance is quantum errors in the hardware. However we cannot rule out other factors, such as the $\Ansatz$ choice. We leave thorough investigation of improving hardware performance to future work, perhaps by including error mitigation~\cite{hamilton_error-mitigated_2020} to reduce errors, thorough error modelling and parametric compilation and active qubit reset~\cite{smith_practical_2017, karalekas_quantum-classical_2020} to improve running time and other performance metrics.

\subsubsection[\texorpdfstring{\color{black}}{} Alternative model structures]{Alternative model structures} \label{app_b:alternative_model_structure}

For completeness, we showcase here the effect of using alternative model structures for the QCBM and the RBM.\\

\noindent \textbf{Differing numbers of Born machine layers} \\

\begin{figure}[ht!]
    \centering
    \includegraphics[width=0.9\columnwidth, height=0.4\columnwidth]{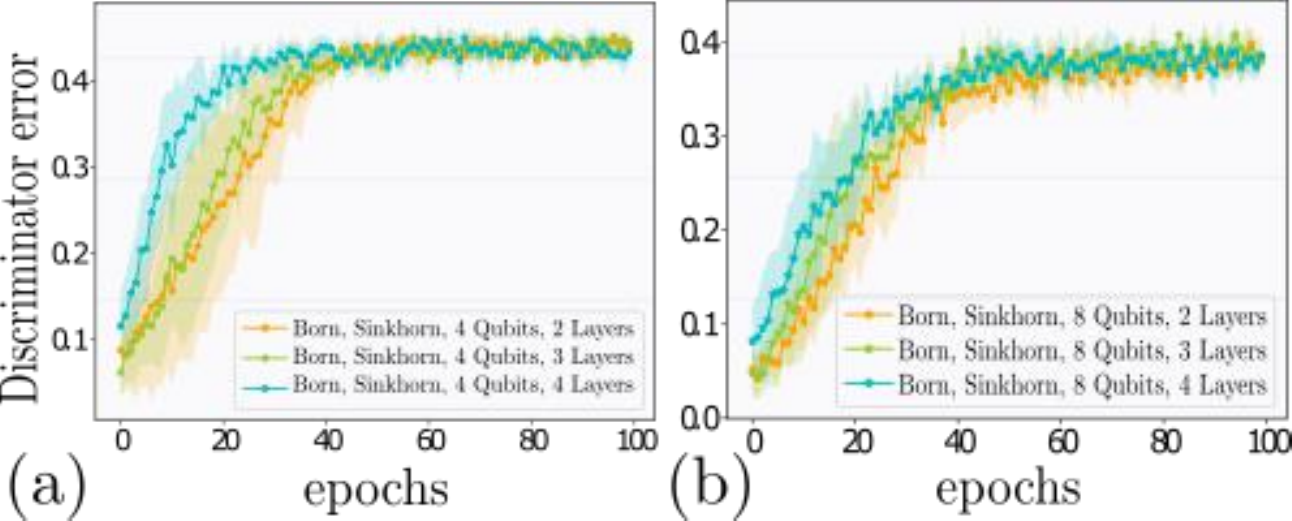}
    \caption[\color{black} Differing numbers of layers in the QCBM.]{\textbf{$2$ [\crule[YellowOrange]{0.2cm}{0.2cm}], $3$ [\crule[YellowGreen]{0.2cm}{0.2cm}] and $4$ [\crule[SkyBlue]{0.2cm}{0.2cm}] layers of the hardware-efficient $\Ansatz$ for (a) $4$ and (b) $8$ qubits.} Models are trained on $2$ currency pairs at $2$ and $4$ bits of precision respectively. No major advantage observed for using an increasing number of layers, except perhaps in convergence speed, suggesting that $2$ layers is sufficient for these problem instances.}
    \label{fig:differing_layers}
\end{figure} 

Firstly, in \figref{fig:differing_layers}, we show the effect of alternating the number of layers of the hardware efficient $\Ansatze$, shown in \figref{fig:aspen_7_6q_8q_12q_circuit_ansatze} for $4$ and $8$ qubits. In particular, we notice that increasing the number of layers does not have a significant impact, at least at these scales, except perhaps in convergence speed of the training. It is likely however, that at larger scales, increased parameter numbers would be required to improve performance.\\

\noindent \textbf{Differing numbers of Boltzmann hidden nodes} 

\begin{figure}[ht!]
    \centering
    \includegraphics[width=\columnwidth, height=0.3\columnwidth]{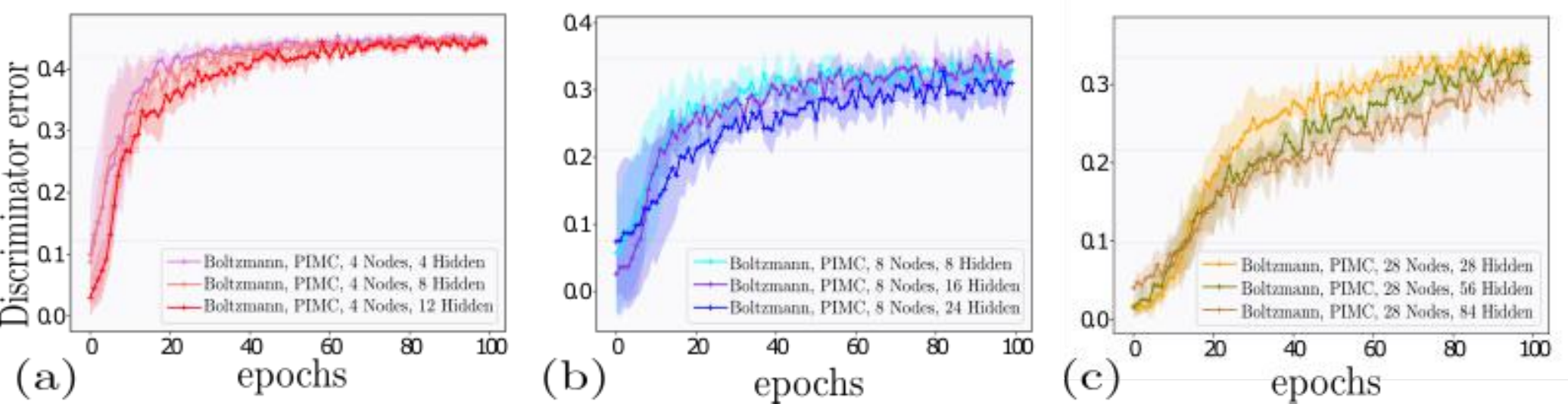}
    \caption[\color{black} Differing numbers of hidden nodes in the RBM.]{\textbf{Differing number of hidden nodes for RBMs.} We compare (a) $4$ (b) $8$ and (c) $28$ visible nodes. Enlarging the hidden space for the RBM again did not impact significantly for these problem sizes and in particular would not give a performance boost to outperform the QCBM.}
    \label{fig:differing_hidden_nodes}
\end{figure}

\vspace{2mm}
We also demonstrate the effect of changing the number of hidden nodes in the Boltzmann machine in \figref{fig:differing_hidden_nodes}, where we have $4, 8$ and $28$ visible nodes. Again, we observe that an increasing number of hidden nodes (and by extension, number of parameters) does not substantially affect the performance of the model, in fact it can hinder it, at least when training only biases of the Boltzmann machine. In particular, it does not substantially alter the final accuracy achieved by the model. We noticed similar behaviour when also training the weights of the Boltzmann machine. \\ 

\noindent \textbf{Weight training of Boltzmann machine}

\begin{figure}[ht!]
    \centering
    \includegraphics[width=\columnwidth, height=0.3\columnwidth]{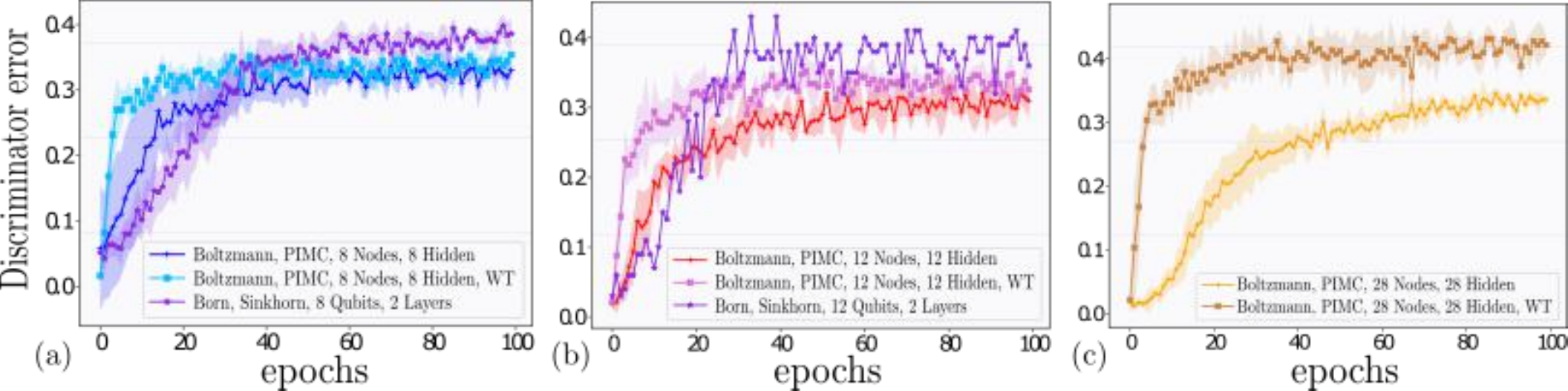}
    \caption[\color{black} Weight training on the Boltzmann machine along with the node biases.]{\textbf{Weight training (WT) on the Boltzmann machine along with the node biases.} We compare (a) $8$, (b) $12$ and (c) $28$ visible node RBMs along with the corresponding Born machine. The latter uses $4$ currency pairs, while the others use $2$, as in the text above. }
    \label{fig:weight_training}
\end{figure}

\vspace{2mm}
Finally, we compare the effect of weight training of the Boltzmann machine to training the bias terms alone or in other words, the training of the edge connections in the network, as opposed to just the parameters of the nodes. For the problem instances where the Boltzmann machine was able to converge to the best discriminator accuracy (i.e. in the small problem instances), we find training the weights has the effect of increasing convergence speed, and also increased accuracy where training the biases only was insufficient to achieve high discriminator error. Interestingly, we note that the Born machine still outperforms the $8$ and $12$ visible node RBMs, even when the weights are also trained, and this does not seem to majorly affect the performance. However, training the weights does make a large difference for $28$ nodes, as seen in \figref{fig:weight_training}(c), so again further investigation is needed in future work of this phenomenon.

\subsection[\texorpdfstring{\color{black}}{} Weak quantum compilation with a QCBM]{Weak quantum compilation with a QCBM} \label{ssec:born_machine/numerics/compilation}
\begin{figure}
    \centering
    \includegraphics[width=\columnwidth, height=0.5\columnwidth]{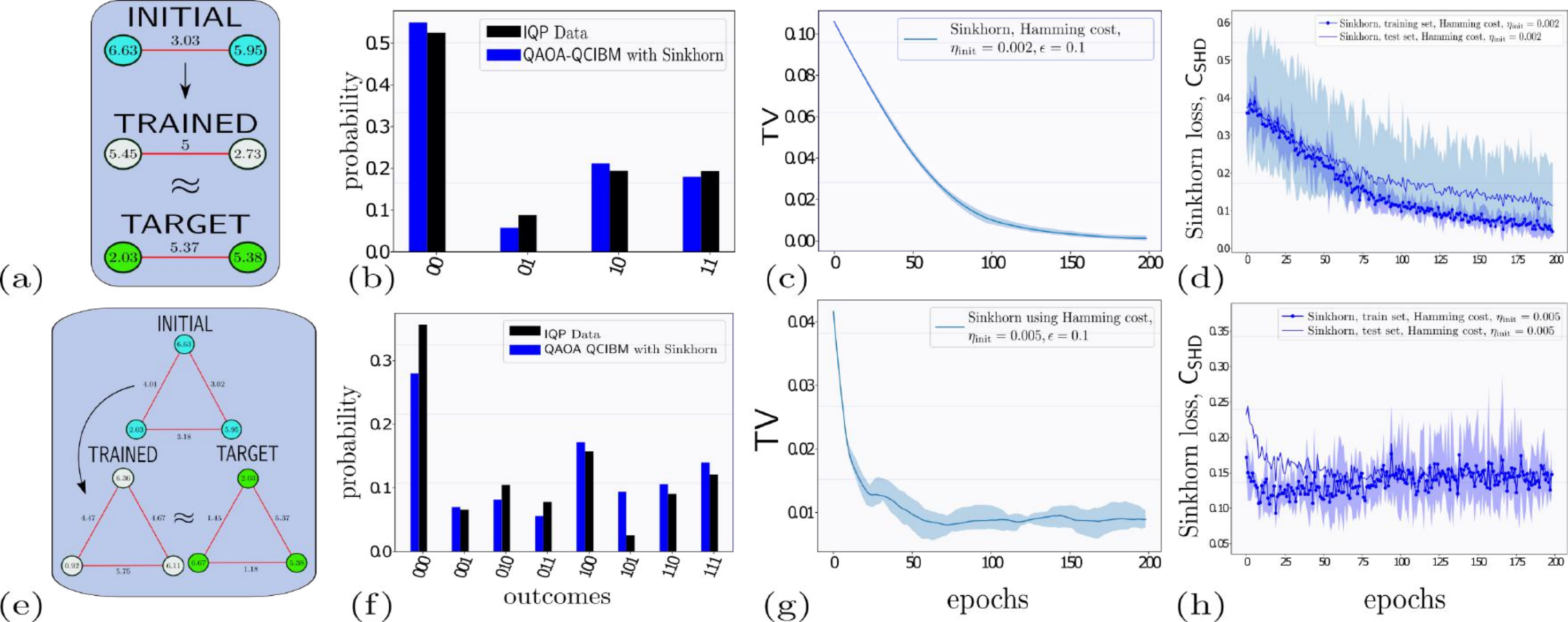}
\caption[\color{black} Compilation of a $p = 1\ \QAOA$-$\QCIBM$ circuit to a $\IQP$ circuit with two and three qubits.]{\textbf{Compilation of a $p = 1\ \QAOA$-$\QCIBM$ circuit to a $\IQP$ circuit with two and three qubits using $\Cbs_{\SH}^\epsilon$ with $\epsilon = 0.1$.} (a, e) Initial [\crule[cyan]{0.2cm}{0.2cm}] and trained [\crule[Lavender]{0.2cm}{0.2cm}] $\QAOA$ circuit parameters for two and three qubits. Target $\IQP$ circuit parameters [\crule[green]{0.2cm}{0.2cm}]. Parameter values scaled by a factor of 10 for readability. (b, f) Final learned probabilities of $\QAOA$-$\QCIBM$ [\crule[blue]{0.2cm}{0.2cm}] circuit versus `data' probabilities ($\IQP$-$\QCIBM$) [\crule[black]{0.2cm}{0.2cm}]. (c, g) Total variation distance and (d, h) Sinkhorn divergence on train and test set, using a Hamming optimal transport cost.}
    \label{fig:autocompilation_twothree_qubits}
\end{figure}

Finally, let us return to the `quantum' dataset from~\secref{sssec:born_machine/data/quantum_data}, which originated from our definition of `weak' quantum circuit compilation. In~\figref{fig:autocompilation_twothree_qubits} we demonstrate this for two and three qubits. Here the error bars represent mean, maximum and minimum values achieved over 5 independent training runs on the same data set. In both cases, the $\IBM$ circuit is able to mimic the target distribution well, even though actual parameter values, and circuit families are different. We also use $500$ data samples from the target circuit with a $400/100$ train/test split.

\section[\texorpdfstring{\color{black}}{} Discussion and conclusion]{Discussion and conclusion} \label{sec:born_machine/conclusions}

At first glance, the contents of this chapter may all seem to be incremental advances in almost disconnected topics. We discussed quantum advantage, new training methods for a quantum circuit Born machine, and extensive numerical experiments on both simulated and on NISQ hardware. Let us take a moment now and tie all of these together with a grander goal in mind: the search for a \textbf{useful} quantum machine learning algorithm on \textbf{near-term} quantum hardware with \textbf{provable} advantage over classical methods.

For \textbf{provable}, we laid the foundations for the study of quantum learning supremacy. Rooted in learning theory, we gave a formalism in which one may claim a true quantum advantage in the problem of generative modelling. We then discussed how one may (or may not) achieve such a proof with quantum computational supremacy distributions, and gave a numerical implementation, on a small scale, in learning such an example distribution (weak compilation).

We then provided new training methods for a particular instance of a generative model (the Born machine), which is also our second example of a variational quantum algorithm. This is relevant for \textbf{near-term}, since NISQ computers are notoriously temperamental, and so we must use all available resources to extract optimal performance from them. This includes, but is not limited to, any methods to better benchmark or improve convergence in training of quantum machine learning models.

Finally, regards \textbf{useful}, we studied a dataset of \emph{practical} relevance, and which is suitable for the generative modelling problem. Since finance is predicted to be a major application area for quantum computers in the long term, any method to solve problems (even incrementally) in this area is valuable. Returning again to (numerically) provable quantum advantage, we benchmarked our generative model relative to a comparable classical counterpart, the restricted Boltzmann machine, and found promising results.

In conclusion, while we do not claim to have found such a `killer application', we hope the contents of this chapter may help along the path in the search for such a thing.

\subsection[\texorpdfstring{\color{black}}{} Subsequent work]{Subsequent work} \label{ssec:born_machine/subsequent}
To round off this chapter, let us briefly highlight some (not all) related work which appeared after the completion of the work in this chapter, in particular those relevant to the discussion in the previous section. Firstly, relating to quantum learning supremacy, the work of~\cite{sweke_quantum_2021} demonstrated an instance of this, working from, and extending, the definitions provided in \secref{ssec:born_machine/quantum_advantage}. In particular, (similar to~\cite{liu_rigorous_2021}), the authors showed how to embed the discrete $\log$ problem into a generative modelling scenario. Doing so, allowed them to describe a distribution family, $\mathcal{D}_n$, which could be learned efficiently by quantum means, but not by any classical algorithm. This result does indeed satisfy our definition of QLS (\defref{defn:quantum_learning_supremacy}), and does so even on a classical data input, but importantly this advantage is not near-term in nature, and the problem it solves is somewhat contrived. Nevertheless, it is an exciting result. 

On a related note,~\cite{gao_enhancing_2021} used foundational ideas from quantum mechanics to demonstrate a provable separation between certain families of Bayesian networks, and their quantum generalisations. These networks are used in generative modelling and probabilistic inference (a very related problem to generative modelling).

Relevant to our heuristics for the QCBM, the work of~\cite{benedetti_variational_2021} used the method of the Stein discrepancy for probabilistic inference with a QCBM. Secondly, as we mentioned above,~\cite{rudolph_generation_2020} defined the basis-enhanced QCBM, which was able to effectively seed a convolutional generative adversial network (when used as a prior distribution) and generate high-resolution MNIST digits (see~\figref{fig:rbm_mnist_generated_images}). Finally,~\cite{kieferova_quantum_2021} proposed alternative training methods for (pre-measurement) Born machines based on different metrics, similarly to the methodology we propose in \secref{ssec:born_machine/training} (albeit with the goal of alleviating barren plateaus in QCBM training landscapes). Clearly, probabilistic quantum machine learning, and in particular quantum generative modelling continues to be an exciting and evolving area of study.

\chapter{Practical quantum cryptanalysis by variational quantum cloning} \label{chap:cloning}

\section[\texorpdfstring{\texorpdfstring{\color{black}}{}}{} Introduction]{Introduction} \label{sec:cloning/introduction}

\begin{chapquote}{Star Wars: Episode II – Attack of the Clones}
\textbf{Obi-Wan}: ``Your clones are very impressive. You must be very proud.''
\end{chapquote}

In the previous chapters, we introduced two quantum machine learning models, and discussed their implementations on NISQ computers. Both of these applications were \emph{machine learning} problems, i.e. data driven. In this chapter, we present our third, and final, application for NISQ computers, but the problem we address is not strictly a machine learning one, rather it is a problem in quantum \emph{information} and \emph{cryptography}. While, at first, this may seem quite far removed from the previous applications, in fact, we stress that it \emph{is not}. We shall see in this chapter how many of the concepts are transferable (with suitable adaptation), and the method we propose will simply boil down to the same variational quantum algorithm framework introduced in \secref{sec:variational_quantum_algorithms}. We also mention how this application may fall into the scope of either supervised or unsupervised learning\footnote{Recalling our categorisation from \secref{sec:prelim_machine_learning} of (un)supervised learning as (not) having the answer for the given problem beforehand.}, depending on what information the algorithm has access to.

Let us begin. In \secref{ssec:prelim/qc/quantum_cloning}, we went to great pains to introduce the background behind the no-cloning theorem (\thmref{thm:no-cloning_theorem}) and the subtleties surrounding it. We introduced approximate cloning unitaries, $U$, which could produce two (or more) copies of an input quantum state (from some family of states), which were approximately the same as the input. We also introduced a circuit which implements $U$ in the case of phase-covariant and universal cloning in~\figref{fig:qubit_cloning_ideal_circ}. However, in these discussions we neglected to mention how one could actually \emph{find} $U$ (or a circuit decomposition for $U$), given a particular family of quantum states, a very related problem to quantum compilation. The approach of the literature to date has been a theoretical one - specify the problem, assume the most general transformation, and solve for the parameters of the unitary given various symmetries of the problem to maximise the cloning fidelity. This needs to be done individually for each generalisation and every set of states one might want to consider. As we mentioned in~\secref{ssec:non-ortho-states}, the theoretical analyses may be tricky, as evidenced by the analytical forms for the fidelities in cloning fixed-overlap states. Worse still, if we are able to find such a unitary, there is no guarantee that it will perform well on near term quantum hardware to actually \emph{implement} the cloning transformation.

At this point, it is appropriate to take a slight detour and digress into the field of quantum cryptography, and illustrate intuitively how one may connect it to quantum cloning (we will be more concrete in~\secref{sec:cloning/variational_cryptanalysis}). Early proposals for quantum cryptographic protocols appeared as early as the 1980s with conceptual ideas including quantum money~\cite{wiesner_conjugate_1983} and quantum key distribution~\cite{bennett_quantum_2014}. Since then, new protocols are being developed at a staggering rate, exploiting quantum phenomena such as entanglement and non-locality of quantum correlations for security proofs~\cite{gottesman_quantum_2001, broadbent_universal_2009, aaronson_quantum_2009, vaccaro_quantum_2007}.
On the experimental side, rapid progress is being made as well, with the first violation of a loophole free Bell inequalities (\cite{bell_einstein_1964}) demonstrated in 2015 and the satellite Micius implementing quantum protocols over long distances, including quantum key distribution (QKD)~\cite{yin_satellite-based_2017, yao_taking_2019, ren_ground--satellite_2017}. For an overview of recent advances in quantum cryptography, see the topical review~\cite{pirandola_advances_2020}.

For each of these new protocols, they must be proven secure against malicious behaviour in order to be useful. A standard method to prove the security of a given quantum communication protocol (e.g. QKD\footnote{We use QKD as one of our examples in this chapter, but we note it is not the optimal use of the algorithm presented here. Instead, we use it as a baseline application and as a benchmark since it is well studied.}) implemented between two parties (say Alice ($A$) and Bob ($B$)) is to assume the existence of an adversary (usually an eavesdropper, Eve\footnote{We will use the notation $E$ later in the chapter to denote any states Eve keeps for herself for whatever purpose, and $E^*$ to denote an ancillary system she uses to aid her malevolent behaviour, but later discards.} ($E$)) with unlimited power (or some computationally bounded amount of resources) and to prove that the adversary cannot gain any (or a limited amount of) secret information about the protocol. In the case of \emph{quantum} protocols, these security analyses, in many cases, reduce to the ability (or lack thereof) of Eve to clone quantum states used in the protocol. Therefore, bounds on approximate quantum cloning have a direct implication for information leakage in a protocol. In some cases, the optimal \emph{attack} Eve may perform is actually a cloning-based one - she makes two approximate clones of a state sent by Alice, forwards one onto Bob and keeps the other for herself to cheat. There is nothing preventing Eve doing this - it is the job of the protocol designer to account for it, and set security parameters in such a way that if, for example, Eve \emph{disturbs} the state beyond allowable regimes, the protocol is aborted.

In the following sections, we describe our contribution to this area.
Specifically, we introduce an algorithm for finding suitable cloning unitaries, $U$. We dub it `variational quantum cloning' ($\VQC$), and we build the various ingredients for it. These include: differentiable cost functions for gradient-descent based optimisation, theoretical guarantees on these cost functions including notions of faithfulness and barren plateaus (\secref{sec:cloning/varqlone}), and variable-structure $\Ansatze$ with quantum architecture search (\secref{sec:cloning/numerics}). Importantly, using these techniques the unitaries found by $\VQC$ are \emph{short-depth} and hardware efficient, meaning they are suitable for implementation on NISQ devices. We explicitly, and more rigorously, make the connection between quantum cloning and cryptography in~\secref{sec:cloning/variational_cryptanalysis} by focusing on two distinct families of quantum protocols, quantum key distribution and quantum coin flipping. For these protocols, we show how $\VQC$ can improve the performance of attacks on them, and discover novel attacks, facilitated by our QML algorithm. However, we stress that while our focus in this chapter is towards cryptographic applications, the underlying primitive is that of quantum cloning. Therefore, the approach may have broader applications wherever the primitive of cloning arises, for example in quantum foundations, or perhaps in the demonstration or benchmarking of quantum hardware. Finally, in~\secref{sec:cloning/conclusions} we conclude and discuss future work.

\section[\texorpdfstring{\color{black}}{} Variational quantum cloning: cost functions and gradients]{Variational quantum cloning: cost functions and gradients} \label{sec:cloning/varqlone}

Let us now dive into some details and specifics of $\VQC$. It is here where we will most closely align with the terminology used in \secref{sec:variational_quantum_algorithms}. To reiterate, our motivation is to find short-depth circuits to clone a given family of states, and also use this toolkit to investigate state families where the optimal figure of merit is unknown. We note that a basic version of what the algorithm we propose in this chapter was put forward by~\cite{jasek_experimental_2019}, but we significantly extend their methods with state of the art approaches, provide theoretical arguments and make the explicit connection to cryptography. We comment on the differences where relevant through the chapter.

\begin{figure*}
    \includegraphics[width=0.95\columnwidth,height=0.35\textwidth]{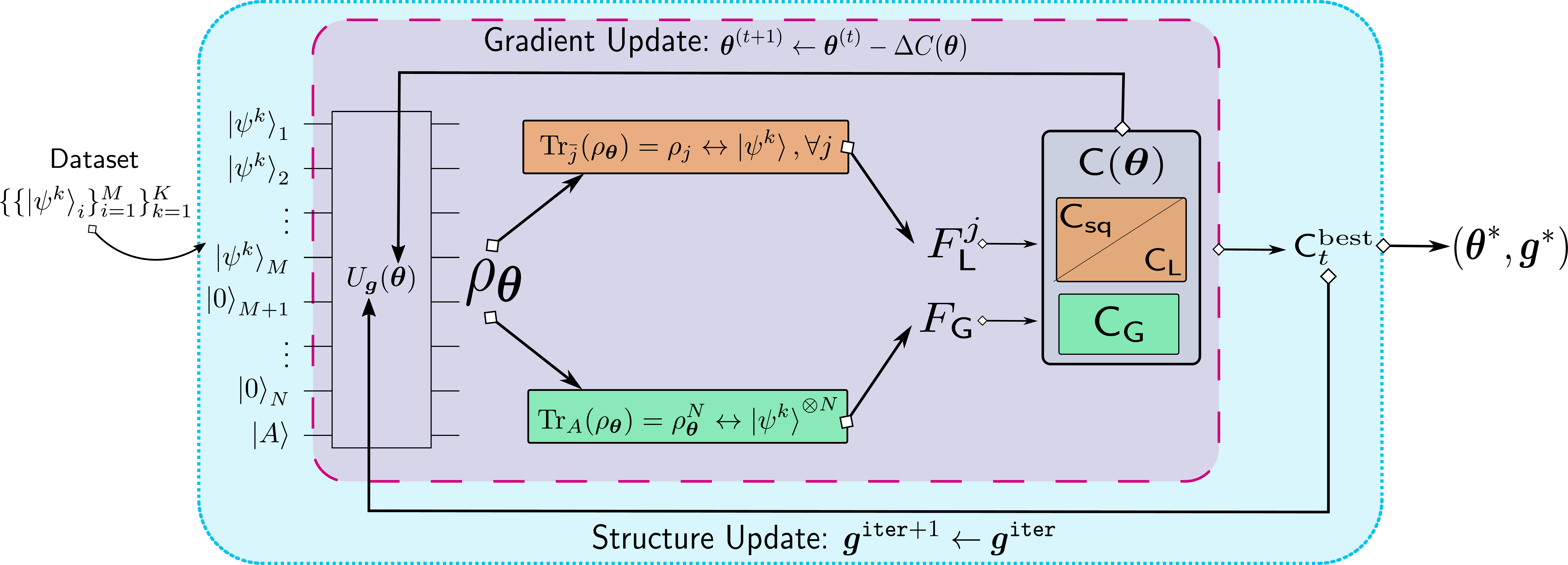}
    \caption[\color{black} Illustration of $\VQC$ for $M \rightarrow N$ cloning.]{\textbf{Illustration of $\VQC$ for $M \rightarrow N$ cloning.} A dataset of $K$ states is chosen from $S$, with $M$ copies of each. These are fed with $N-M$ blank states, and possibly another ancilla, $\ket{\phi}_A$ into the variable structure $\Ansatz$, $U_{\boldsymbol{g}}(\paramtheta)$ (more details about this in \secref{sec:cloning/numerics}). Depending on the problem, either the global, or local fidelities of the output state, $\rho_{\paramtheta}$, is compared to the input states, $\ket{\psi^k}$, and the corresponding local or global cost function, $\Cbs(\paramtheta)$ is computed, along with its gradient. We have two optimisation loops, one over the continuous parameters, $\paramtheta$, by gradient descent, and the second over the circuit structure, $\boldsymbol{g}$. Gradient descent over $\paramtheta$ in each structure update step outputs, upon convergence, the `minimum' cost function value, $\Cbs^{\text{best}}_t$, for the chosen cost function, $t \in \{\Lbs, \sq, \Gbs\}$.}
    \label{fig:vqc_overview}
\end{figure*}

\subsection[\texorpdfstring{\color{black}}{} Cost functions]{Cost functions} \label{ssec:cost_functions}
By this point in the thesis, we have hopefully sufficiently impressed the importance of cost functions in QML applications upon the reader. $\VQC$ is no different. Here, we define and study three cost functions which are suitable for approximate cloning tasks. We begin by stating the functions, and then discussing the various ingredients and their relative advantages.

The first, we call the `\emph{local cost}' for $M\rightarrow N$ cloning\footnote{Recall our footnote in~\secref{ssec:prelim/qc/quantum_cloning} about the change of notation for qubit number in this chapter.}, given by:
\begin{align} \label{eqn:local_cost_full}
    \Cbs_{\Lbs}^{M\rightarrow N}(\paramtheta) &:= \mathop{\mathbb{E}}_{\substack{\ket{\psi} \in \mathcal{S}}} \left[\Cbs^{\psi}_{\Lbs}(\paramtheta) \right], \qquad \Cbs^{\psi}_{\Lbs}(\paramtheta)  := \tr( \mathsf{O}^{\psi}_{\Lbs} \rho_{\paramtheta}) \\
    \mathsf{O}^{\psi}_{\Lbs}  &:= \mathds{1} - \frac{1}{N}\sum\limits_{j=1}^N\ketbra{\psi}{\psi}_j \otimes \mathds{1}_{\Bar{j}} \label{eqn:local_cost_observable}
\end{align}

where $\ket{\psi} \in \mathcal{S}$ is the family of states to be cloned. The subscripts $j\ (\Bar{j})$ indicate operators acting on subsystem $j$ (everything except subsystem $j$) respectively.

The second, we call the `\emph{squared local cost}' or just `squared cost' for brevity:
\begin{equation} \label{eqn:squared_local_cost_mton}
    \Cbs_{\sq}^{M\rightarrow N}(\paramtheta) := \mathop{\mathbb{E}}_{\substack{\ket{\psi} \in \mathcal{S}}}\left[ \sum\limits_{i=1}^N (1-F^i_{\Lbs}(\paramtheta))^2 + \sum\limits_{i<j}^N (F^i_{\Lbs}(\paramtheta)-F^j_{\Lbs}(\paramtheta))^2\right] 
\end{equation}

The notation $F^j_{\Lbs}(\boldsymbol{\theta}) := F_{\Lbs}(\ketbra{\psi}{\psi}, \rho^j_{\boldsymbol{\theta}})$ indicates the fidelity of Alice's input state, relative to the reduced state of output qubit $j$, given by $\rho^j_{\boldsymbol{\theta}} = \tr_{\Bar{j}}\left(\rho_{\paramtheta}\right)$. As in~\eqref{eqn:vqa_parametrised_state}, $\rho_{\paramtheta}$ is the state outputted by the PQC acting on some fixed input state (see~\figref{fig:vqc_overview}). These first two cost functions are related only in that they are both functions of \emph{local} observables, i.e. the local fidelities.

The third and final cost is fundamentally different to the other two, in that it instead uses global observables, and as such, we refer to it as the `\emph{global cost}':
\begin{equation} \label{eqn:global_cost_full}
    \Cbs_{\Gbs}^{M\rightarrow N}(\paramtheta) := \mathop{\mathbb{E}}_{\substack{\ket{\psi} \in \mathcal{S}}} \left[\tr(\mathsf{O}^{\psi}_{\Gbs}\rho_{\paramtheta})\right],~
     \mathsf{O}^{\psi}_{\Gbs} := \mathds{1} - \ketbra{\psi}{\psi}^{\otimes N}
\end{equation}
For compactness, we will drop the superscript $M \rightarrow N$ when the meaning is clear from context. 

Now, let us motivate our choices for the above cost functions. For \eqref{eqn:squared_local_cost_mton}, if we restrict to the special case of $1\rightarrow 2$ cloning (i.e. we have only two output parties, $j\in \{B, E\}$), and remove the expectation value over states, we recover the cost function used in \cite{jasek_experimental_2019}. A useful feature of this cost is that symmetry is explicitly enforced by the difference term $(F_i(\paramtheta) - F_j(\paramtheta))^2$. We highlight this importance later in \secref{sec:cloning/numerics}.

In contrast, the local and global cost functions are inspired by other variational algorithm literature~\cite{larose_variational_2019,  cerezo_variational_2020, cerezo_cost_2021, khatri_quantum-assisted_2019, sharma_noise_2020} where their properties have been extensively studied, particularly in relation to `barren plateaus' (see \defref{defn:barren_plateaus}). Since our primary cost functions (\eqref{eqn:local_cost_full}, \eqref{eqn:squared_local_cost_mton}) are \emph{local} in nature, they are less susceptible to barren plateaus than those which are global. We confirm this in our case by explicitly proving that \eqref{eqn:local_cost_full} is efficiently trainable with a $\mathcal{O}(\log N)$ depth hardware efficient $\Ansatze$~\cite{cerezo_cost_2021} in \secref{sssec:pc_cloning_fixed_hardware_efficient_ansatz}. 

In contrast to many previous applications, by the nature of quantum cloning, $\VQC$ allows the local cost functions to have immediate operational meaning (see~\secref{ssec:vqa_cost_functions}), illustrated through the following example (using the local cost, \eqref{eqn:local_cost_full}) for $1\rightarrow 2$ cloning:
\begin{align*}
    \Cbs^{\psi}_{\Lbs}(\paramtheta) &= \tr\left[\left(\mathds{1} - \frac{1}{2}\sum\limits_{j=1}^2\ketbra{\psi}{\psi}_j \otimes \mathds{1}_{\Bar{j}}\right) \rho_{\paramtheta}\right]\\
     \implies  \Cbs_{\Lbs}(\paramtheta)  &= 1 - \frac{1}{2}\mathbb{E}\left[F_{\Lbs}\left(\ketbra{\psi}{\psi}, \rho_{\paramtheta}^1\right) + F_{\Lbs}\left(\ketbra{\psi}{\psi}, \rho_{\paramtheta}^2\right)\right]
\end{align*}
where $\mathbb{E}[F_{\Lbs}]$ is the average fidelity (\cite{scarani_quantum_2005}) over the possible input states. The final expression of $\Cbs_{\Lbs}(\paramtheta)$ in the above equation follows from the expression of fidelity when one of the states is pure. Similarly, the global cost function relates to the global fidelity of the output state with respect to the input state(s).

To estimate the cost functions in practice, we can prepare a `dataset' of $K$ samples (drawn uniformly at random from $\mathcal{S}$), $\{\ket{\psi^k}\}^{K}_{k=1}$, and compute a Monte-Carlo estimate of the expectation values:
\begin{equation*}
     \Cbs_{\Lbs}(\paramtheta)  \approx 1 - \frac{1}{2K}\sum_{k=1}^{K}\left[F_{\Lbs}\left(\ketbra{\psi^k}{\psi^k}, \rho_{\paramtheta}^1\right) + F_{\Lbs}\left(\ketbra{\psi^k}{\psi^k}, \rho_{\paramtheta}^2\right)\right]
\end{equation*}
However, there is one caveat which we have not explicitly addressed as yet. Due to the hard limits on approximate quantum cloning, the above costs cannot have a minimum at zero\footnote{This is in contrast to the cost functions we used for the Born machine, for which we could achieve a minimum at zero by definition, or for other $\VQA$ applications such as variational quantum compilation~\cite{khatri_quantum-assisted_2019} or variational quantum state diagonalisation~\cite{larose_variational_2019} where again defining a zero-minimum cost function makes sense.}, but instead at some finite value (say $\Cbs^{\opt}_{\Lbs}$ for the local cost). It is for this reason why we view $\VarQlone$ as being either both a supervised, and unsupervised algorithm, based on our definitions of these in~\secref{sec:prelim_machine_learning}. If we have prior knowledge of the optimal cloning fidelity which can be achieved (the `correct' answer), this can be incorporated to achieve zero-minimum cost functions and we can train with `supervision'. If not, we can still use $\VarQlone$ in an unsupervised manner take the lowest value found to be the approximation of the cost minimum. The former is likely to be more valuable to extract the most hardware-friendly cloning circuit possible, whereas the latter viewpoint can be used to traverse uncharted water and discover cloning circuits for unknown families.

\subsection[\texorpdfstring{\color{black}}{} Cost function gradients]{Cost function gradients} \label{ssec:cost_function_gradients}
As with the applications in the previous sections, we will again opt for a gradient descent based approach to train the $\VQC$ PQC\footnote{This is one of the primary differences between our method and that of~\cite{jasek_experimental_2019}, which opts for gradient-free optimisation using Nelder-Mead (\cite{nelder_simplex_1965})}. We will again assume the same form for the $\Ansatz$ unitary, $U(\paramtheta)$; that it is composed of unitaries of the form: $\exp(i\theta_l \Sigma)$, where $\Sigma$ is a generator with two distinct eigenvalues. Given this, we again use the parameter-shift rule (\eqref{eqn:parameter_shift_rule}) to derive analytic gradients. We do this explicitly for the squared local cost,~\eqref{eqn:squared_local_cost_mton}, and for brevity neglect the gradient derivations for the others, since they are simpler and follow analogously. For this cost function we have the gradient with respect to a parameter, $\theta_l$ as:
\begin{multline}  \label{eqn:analytic_squared_grad_mton}
    \frac{\partial \Cbs_{\sq}(\boldsymbol{\theta})}{\partial \theta_l}  = 
    2{\mathbb{E}}\left[\sum\limits_{i<j} (F^i_{\Lbs}- F^j_{\Lbs}) \left(F^{i, l+ \frac{\pi}{2} }_{\Lbs} - F^{i,  l- \frac{\pi}{2} }_{\Lbs} -F^{j,  l+ \frac{\pi}{2}}_{\Lbs} + F^{j, l-\frac{\pi}{2}}_{\Lbs} \right) \right.\\
    \left. - \sum\limits_{i} (1- F^i_{\Lbs})(F^{i, l+\frac{\pi}{2}}_{\Lbs}
    - F^{i, l- \frac{\pi}{2}}_{\Lbs})  \right]
\end{multline}
where $F^{j, l\pm\pi/2}_{\Lbs}(\boldsymbol{\theta})$ denotes the fidelity of a particular state, $\ket{\psi}$, with $j^{th}$ reduced state of the $\VQC$ circuit which has the $l^{th}$ parameter shifted by $\pm \pi/2$. We suppress the $\paramtheta$ dependence in the above. We give the proof of this in~\appref{app_ssec:varqlone_gradients}.

Using the same method, we can derive the gradient of the local cost, \eqref{eqn:local_cost_full} with $N$ output clones as:
\begin{equation}\label{eqn:gradient_local_cost_full}
    \frac{\partial \Cbs_{\Lbs}(\paramtheta)}{\partial \theta_l} = \mathbb{E}\left(\sum_{i=1}^N \left[F_{\Lbs}^{i, l-\pi/2} -  F_{\Lbs}^{i, l+\pi/2} \right]\right)
\end{equation}
Finally, similar techniques result in the analytical expression of the gradient of the global cost function:
\begin{equation}\label{eqn:gradient_global_cost_full}
    \frac{\partial \Cbs_{\Gbs}(\paramtheta)}{\partial \theta_l} = \mathbb{E}\left(F_{\Gbs}^{l-\pi/2} - F_{\Gbs}^{l+\pi/2}\right)    
\end{equation}
where $F_{\Gbs}:= F(\ketbra{\psi}{\psi}^{\otimes N}, \rho_{\paramtheta})$ is the global fidelity between the parametrised output state and an $n$-fold tensor product of input states to be cloned.

%
\subsection[\texorpdfstring{\color{black}}{} Asymmetric cloning]{Asymmetric cloning} \label{ssec:asymmetric_cloning}
The local cost functions above are defined in a way to enforce \emph{symmetry} in the output clones. However, from the purposes of eavesdropping, this may not be the optimal attack for Eve to implement (or just generally one may not wish to produce symmetric clones). In particular, she may wish to impose less disturbance on Bob's state so she reduces the chance of getting detected. This subtlety was first addressed in \cite{fuchs_information_1996, fuchs_optimal_1997}. 

For example, if she wishes to leave Bob with a fidelity parametrised by a specific value, $p$ (given by $F^{p, B}_{\Lbs} = 1 - p^2/2$), one may define an asymmetric version of the $1\rightarrow 2$ squared local cost function in \eqref{eqn:squared_local_cost_mton}:
\begin{equation} \label{eqn:asymmetric_cost_function_maintext}
      \Cbs_{\Lbs, \text{asym}}(\paramtheta) := \mathbb{E}\left[F^{p, B}_{\Lbs} - F^{B}_{\Lbs}(\paramtheta)\right]^2 + \mathbb{E}\left[F^{p, E}_{\Lbs} - F^{E}_{\Lbs}(\paramtheta)\right]^2 \\
\end{equation}
Given a particular value for Bobs fidelity, Eves fidelity must obviously be similarly constrained. The corresponding value for Eve's fidelity ($F^{p, E}_{\Lbs}$) in this case can be derived from the `\emph{no-cloning inequality}'~\cite{scarani_quantum_2005}:
\begin{equation} \label{eqn:no_cloning_inequality}
 \sqrt{(1 - F^{p, B}_{\Lbs})(1 - F^{q, E}_{\Lbs})} \geqslant \frac{1}{2} - (1 - F^{p, B}_{\Lbs}) - (1 -F^{q, E}_{\Lbs})   
\end{equation}
where the output clones of Bob and Eve are denoted by $F^{p, B}_{\Lbs}$ and $F^{q, E}_{\Lbs}$ for the desired parameterisations $p$ and $q$.

It can be easily verified that the following fidelities saturate the above inequality:
\begin{equation}     \label{eqn:asymmetric_clones_appendix}
    F^{p, B}_{\Lbs} = 1 - \frac{p^2}{2}, \qquad F^{q, E}_{\Lbs} = 1 - \frac{q^2}{2}, \qquad p, q \in [0,1] , 
\end{equation}
where $p, q$ satisfy $p^2 + q^2 + pq = 1$. This implies that Eve is free to choose a desired fidelity for either clone, by varying the parameter, $p$. For example, suppose Eve wish to send a clone to Bob with a particular fidelity $F^{p, B}_{\Lbs} = 1 - p^2/2$, then from \eqref{eqn:no_cloning_inequality} her clone would have a corresponding fidelity: 
\begin{equation} \label{eqn:asym_fids_eve_fixed_q}
    F^{p, E}_{\Lbs} = 1 - \frac{1}{4}(2 - p^2 - p\sqrt{4 - 3p^2})
\end{equation}
which gives the term in \eqref{eqn:asymmetric_cost_function_maintext} as a function of $p$.

The asymmetric cost function can also be naturally generalised to the case of $M \rightarrow N$ cloning. We note that $\Cbs_{\Lbs, \text{asym}}(\paramtheta)$ can also be used for symmetric cloning by enforcing $p = \frac{1}{\sqrt{3}}$ in \eqref{eqn:no_cloning_inequality}. However, it comes with an obvious disadvantage in the requirement to have knowledge of the optimal clone fidelity values a-priori (hence, learning with this cost function would necessarily be \emph{supervised}). In contrast, our local cost functions (\eqref{eqn:local_cost_full}, \eqref{eqn:squared_local_cost_mton}) do not have this requirement, and thus are more suitable to be applied in general cloning scenarios, since they can work in an unsupervised manner also.


\subsection[\texorpdfstring{\color{black}}{} Cost function guarantees]{Cost function guarantees} \label{ssec:cost_function_guarantees}
As discussed above in \secref{ssec:vqa_cost_functions} and in the previous chapters, we would also like our variational algorithm to have certain theoretical guarantees. The primary one we focus on here is \emph{faithfulness}.

Since our cost functions, when defined in an unsupervised manner do not have a minimum at zero, it is slightly more tricky to prove faithfulness guarantees about them, than in other variational algorithms.We can still do so, but we must provide slightly relaxed definitions of faithfulness. Specifically, we consider notions of \emph{strong} and \emph{weak} faithfulness, relative to the error in our solution. In the following, we denote $\rho_{\opt}^{\psi, j}$ ($\rho^{\psi, j}_{\paramtheta}$) to be the optimal ($\VQC$ learned) reduced state for qubit $j$, for a particular input state, $\ket{\psi} \in \mathcal{S}$. If the superscript $j$ is not present, we mean the global state of all clones. 
\begin{defbox}
\begin{definition}[Strong faithfulness]\label{defn:strong_cloning_faithfulness}~ \\
    A cloning cost function, $\Cbs$, is strongly faithful if:
    \begin{equation}\label{eqn:strongly_faithful_cost_defn}
        \Cbs(\paramtheta) = \Cbs^{\opt} \implies \rho_{\paramtheta}^\psi = \rho_{\opt}^{\psi}, \qquad \forall \ket{\psi} \in \mathcal{S}
    \end{equation}
    where $\Cbs^{\opt}$ is the minimum value achievable for the cost, $\Cbs$, according to quantum mechanics.
\end{definition}
\end{defbox}

\begin{defbox}
\begin{definition}[$\epsilon$-weak local faithfulness]\label{defn:weak_local_cloning_faithfulness}~ \\
    A local cloning cost function, $\Cbs_{\Lbs}$, is $\epsilon$-weakly faithful if:
    \begin{equation}\label{eqn:weakly_faithful_local_cost_defn}
        |\Cbs_{\Lbs}(\paramtheta) - \Cbs_{\Lbs}^{\opt}| \leq \epsilon \implies \text{d}(\rho_{\paramtheta}^{\psi, j}, \rho_{\opt}^{\psi, j}) \leq f(\epsilon), \qquad  \forall \ket{\psi} \in \mathcal{S}, \forall j 
    \end{equation}
\end{definition}
\end{defbox}
where $\text{d}(\cdot, \cdot)$ is a chosen metric in the Hilbert space between the two states and $f$ is some function of $\epsilon$ (and perhaps other problem dimensions).
\begin{defbox}
\begin{definition}[$\epsilon$-weak global faithfulness]\label{defn:weak_global_cloning_faithfulness}~ \\
    A global cloning cost function, $\Cbs_{\Gbs}$, is $\epsilon$-weakly faithful if:
    \begin{equation}\label{eqn:weakly_faithful_global_cost_defn}
        |\Cbs_{\Gbs}(\paramtheta) - \Cbs_{\Gbs}^{\opt}| \leq \epsilon \implies \text{d}(\rho_{\paramtheta}^{\psi}, \rho_{\opt}^{\psi}) \leq f(\epsilon), \qquad \forall \ket{\psi} \in \mathcal{S}
    \end{equation}
\end{definition}
\end{defbox}
One could also define local and global versions of strong faithfulness, but this is less interesting so we do not focus on it here.

In the next few sections, we prove that each of our cost functions satisfy these definitions of faithfulness, taking the metric to be the Bures angle, $ \text{d}_{\text{BA}}$, as defined in \eqref{eqn:bures_angle_definition}~\cite{fubini_sulle_1904, study_kurzeste_1905, nielsen_quantum_2010}, and also the trace distance, $\text{d}_{\tr}$ as in~\eqref{eqn:trace_distance_defn}. Let us begin with the squared cost function, where we go into some details for the proofs, but we neglect this detail for the proofs of the other cost functions, as they more or less follow straightforwardly.


\subsubsection[\texorpdfstring{\color{black}}{} Squared cost function]{Squared cost function} \label{sssec:cloning/squared_cost_function_faithfulness}
We begin by writing the squared cost function as:
\begin{equation} \label{eqn:squared_local_cost_mton_supp_redo}
        \Cbs_{\sq}^{M\rightarrow N}(\paramtheta) = \frac{1}{\mathcal{N}}\int_{\mathcal{S}}\left[ \sum\limits_{j=1}^N (1-F_i(\paramtheta))^2 + \sum\limits_{i<j}^N (F_i(\paramtheta)-F_j(\paramtheta))^2\right]d \psi
\end{equation}
where the expectation of a fidelity $F_i$ over the states in distribution $\mathcal{S}$ is defined as $\mathbb{E}[F_i] = \frac{1}{\mathcal{N}}\int_{\mathcal{S}}F_i\cdot d\psi$, with the normalisation condition being $\mathcal{N} = \int_{\mathcal{S}}d\psi$. For qubit states, if the normalisation is over the entire Bloch sphere in $SU(2)$, then $\mathcal{N} = 4\pi$. For notation simplicity, we herein denote the $\Cbs_{\sq}^{M\rightarrow N}(\paramtheta)$ as $\Cbs_{\sq}(\paramtheta)$.\\

\noindent \textbf{1. Strong faithfulness}:

\vspace{2 mm}
\begin{thmbox}
\begin{theorem}\label{thm:squared_local_cost_squared_strong_faithful_local_appendix}
    The squared local cost function is locally strongly faithful:
    \begin{equation}\label{eqn:squared_cost_function_locally_faithful_appendix}
        \Cbs_{\sq}(\paramtheta) = \Cbs_{\sq}^{\opt} \implies \rho_{\paramtheta}^{\psi, j} = \rho_{\opt}^{\psi, j} \qquad \forall \ket{\psi} \in \mathcal{S}, \forall j \in [N]
    \end{equation}
\end{theorem}
\end{thmbox}
\begin{proof}
The cost function $\Cbs_{\text{sq}}(\paramtheta)$ achieves a minimum at the joint maximum of $\mathbb{E} [F_i(\paramtheta)]$ for all $i \in [N]$. In symmetric $M \rightarrow N$ cloning, the expectation value of all the $N$ output fidelities peak at $F_i = F_{\opt}$ for all input states $\ket{\psi}$. This corresponds to a unique optimal joint state $\rho_{\opt}^{\psi, j} = U_{\opt}\ket{\psi}^{\otimes M}\bra{0}^{\otimes N-M}\bra{\psi}^{\otimes M}\bra{0}^{\otimes N-M}U_{\opt}^{\dagger}$ for each $\ket{\psi} \in S$, where $U_{\opt}$ is the optimal unitary. Since the joint optimal state and the corresponding fidelities are unique for all $\ket{\psi} \in \mathcal{S}$, we conclude that the cost function achieves a minimum under precisely the unique condition i.e. $\mathbb{E}[F_j(\paramtheta)] = F_{\opt}$ for all $j \in [N]$. This condition implies that, 
\begin{equation}\label{eqn:strong_faithful_condition}
    \rho^{\psi,j}_{\paramtheta} = \rho^{\psi,j}_{\opt}, \qquad \forall \ket{\psi} \in S, \forall j \in [N]
\end{equation}
We note that since $F_{\opt}$ is the same for all the reduced states $j \in [N]$, this implies that the optimal reduced states are all the same for a given $\ket\psi \in \mathcal{S}$. Therefore, perfectly optimising the cost function results in the ideal clones being outputted.  
\end{proof}

\noindent \textbf{2. Weak faithfulness:}

\vspace{2 mm}
Before proving weak faithfulness for the cost function, we first need the following lemma:

\begin{lembox}
\begin{lemma} \label{lemma:trace_bound_squared_cost}
Suppose the cost function is $\epsilon$-close to the optimal cost in symmetric cloning
%
\begin{equation}
    \Cbs_{\sq}(\paramtheta) - \Cbs^{\opt}_{\sq} \leq \epsilon
    \label{eqn:cost_to_epsilon_squared_appendix}
\end{equation}
then we have,
\begin{equation}\label{eqn:tracecloseness_squared_local_appendix}
        \tr\left[(\rho_{\opt}^{\psi, j} - \rho^{\psi, j}_{\paramtheta})\ket{\psi}\bra{\psi}\right] \leq \frac{\mathcal{N}\epsilon}{2(1 - F_{\opt})}, \qquad \forall \ket{\psi} \in \mathcal{S}, \forall j \in [N]
\end{equation}
\end{lemma}
\end{lembox}
\begin{proof}
In $M \rightarrow N$ symmetric cloning, the optimal cost function value is achieved when each output clone achieves the fidelity $F_{\opt}$.  Thus, using  \eqref{eqn:squared_local_cost_mton_supp_redo}, the optimal cost function value is $\Cbs^{\opt}_{\sq} = N (1 - F_{\opt})^2$

The optimal cost function corresponds to all output clones having the same fidelity. This is explicitly enforced by taking the limit $\epsilon \rightarrow 0$, in which case the difference terms of \eqref{eqn:squared_local_cost_mton_supp_redo} vanish. Thus, the cost function explicitly enforces the symmetry property. Let us assume $\epsilon \rightarrow 0$, and consider the quantity $ \Cbs_{\sq}(\paramtheta) - \Cbs^{\opt}_{\sq}$:
\begin{equation}     \label{eqn:costtotrace_squared_appendix}
    \begin{split}
        \Cbs_{\sq}(\paramtheta) - \Cbs^{\opt}_{\sq} &=
        \frac{1}{\mathcal{N}}\int_{\mathcal{S}}\left[ \sum\limits_{i}^N (1-F_i(\paramtheta))^2 + \sum\limits_{i<j}^N  (F_i(\paramtheta)-F_j(\paramtheta))^2\right]d \psi  - N\cdot (1 - F_{\opt})^2\\
        &\underset{\epsilon \rightarrow 0}{\approx} \frac{1}{\mathcal{N}}\int_{\mathcal{S}}\left[ \sum\limits_{j}^N (1-F_j(\paramtheta))^2  - N\cdot (1 - F_{\opt})^2 \right]d \psi  \\
    &\approx \frac{1}{\mathcal{N}}\int_{\mathcal{S}}\left[ \sum\limits_{j}^N (F_{\opt} - F_j(\paramtheta))(2 - F_{\opt} - F_j(\paramtheta)) \right]d \psi  \\
     &\geqslant \frac{2(1 - F_{\opt})}{\mathcal{N}}\int_{\mathcal{S}}\left[ \sum\limits_{j}^N (F_{\opt} - F_j(\paramtheta)) \right]d \psi  \\
        &= \frac{2(1 - F_{\opt})}{\mathcal{N}}\left[ \sum\limits_{j}^N \int_{\mathcal{S}}\text{Tr}[(\rho_{\opt}^{\psi, j} - \rho^{\psi, j}_{\paramtheta})\ket{\psi}\bra{\psi}]d\psi \right]\\
    \end{split}
\end{equation}
The second line follows since $F_{\opt}$ is the same for each input state, $\ket{\psi}$. Utilizing the inequality in \eqref{eqn:cost_to_epsilon_squared_appendix} and \eqref{eqn:costtotrace_squared_appendix}, we obtain,
\begin{equation}
\begin{split}
      &  \sum\limits_{j}^N \int_{\mathcal{S}}\text{Tr}\left[(\rho_{\opt}^{\psi, j} - \rho^{\psi, j}_{\paramtheta})\ket{\psi}\bra{\psi}\right]d\psi  \leq \frac{\mathcal{N}\epsilon}{2(1 - F_{\opt})} \\
      & \implies \text{Tr}\left[(\rho_{\opt}^{\psi, j} - \rho^{\psi, j}_{\paramtheta})\ket{\psi}\bra{\psi}\right] \leq \frac{\mathcal{N}\epsilon}{2(1 - F_{\opt})}, \hspace{3mm} \forall \ket{\psi} \in \mathcal{S}, \forall j \in [N]
\label{eqn:tracecloseness_squared_local}
\end{split}
\end{equation}
\end{proof}

Using the above lemma, we can prove the following two theorems for the squared local cost function:
\begin{thmbox}
\begin{theorem}\label{thm:squared_local_cost_squared_FS_weak_faithful_appendix}
The squared cost function as defined \eqref{eqn:squared_local_cost_mton_supp_redo}, is $\epsilon$-weakly faithful with respect to the Bures angle distance measure $\text{d}_{\BA}$.
In other words, if we have:
\begin{equation}\label{eqn:squared_cost_to_epsilon_appendix}
    \Cbs_{\sq}(\paramtheta) - \Cbs^{\opt}_{\sq} \leq \epsilon
\end{equation}
where $\Cbs^{\opt}_{\sq} := (F_i(\paramtheta)-F_j(\paramtheta))^2 = N(1-F_{\opt})^2$ is the optimal cost, then the following fact holds:

\begin{equation}     \label{eqn:fubini_study_bound_squared_appendix}
    \text{d}_{\BA}(\rho^{\psi, j}_{\paramtheta}, \rho_{\opt}^{\psi, j}) \leq \frac{\mathcal{N}\epsilon}{2(1 - F_{\opt})\sin(F_{\opt})} := f_1(\epsilon),   \qquad \forall \ket{\psi} \in \mathcal{S}, \forall j \in [N]
\end{equation}
\end{theorem}
\end{thmbox}
The proof of this is lengthy so we defer it to~\appref{app_ssec:proof_faithfulness}.

To conclude this discussion, let us compute the constants in the special case of phase-covariant states in \eqref{eqn:x_y_plane_states}. For this example, we have $\mathcal{N}=4\pi$ and so:
\begin{equation} \label{eqn:fubini_study_bound_phase_cov_qubits_explicit}
     \text{d}_{\BA}(\rho^{\psi,j}_{\paramtheta}, \rho^{\psi,j}_{\text{opt}}) \leq 56 \cdot \epsilon
\end{equation}
Now, we can also prove a similar result relative to the trace distance~\eqref{eqn:trace_distance_defn}. Contrary to the previous result, the following only holds true when the input states are qubits\footnote{This is because we explicitly use the properties of a single qubit unitary. In contrast, for the Bures angle proof of \thmref{thm:squared_local_cost_squared_FS_weak_faithful_appendix}, we did not need to make any such assumption, and so would also hold for higher dimensional quantum states, i.e. qudits.}.
\begin{thmbox}
\begin{theorem} \label{thm:squared_local_cost_squared_trace_weak_faithful_appendix}
The squared cost function, \eqref{eqn:squared_local_cost_mton_supp_redo}, is $\epsilon$-weakly faithful with respect to the trace distance $\text{d}_{\tr}$.
\begin{equation}     \label{eqn:trace_distance_bound_squared_appendix}
        \text{d}_{\tr}(\rho_{\opt}^{\psi, j},  \rho^{\psi, j}_{\paramtheta})  \leq g_1(\epsilon), \qquad \forall j \in [N]
\end{equation}
where:
\begin{equation} \label{eqn:trace_distance_squared_cost_bound_function}
    g_1(\epsilon) \approx \frac{1}{2}\sqrt{4F_{\opt}(1 - F_{\opt}) + \epsilon\frac{\mathcal{N}(1 - 2F_{\opt})}{2(1 - F_{\opt})}}
\end{equation}
\end{theorem}
\end{thmbox}

Again, the proof of this is lengthy so we defer it to~\appref{app_ssec:proof_faithfulness_trace}. An immediate consequence of \eqref{eqn:trace_distance_squared_cost_bound_function} is that there is a non-vanishing gap of $\sqrt{F_{\opt}(1 - F_{\opt})}$ between the states. This is due to the fact that the two output states are only $\epsilon$-close in distance when projected on a specific state $\ket{\psi}$. However, the trace distance is a projection independent measure and captures the maximum over all the projectors.


\subsubsection[\texorpdfstring{\color{black}}{} Local cost function]{Local cost function} \label{sssec:local_cost_function_appendix}
Next, we prove analogous results for the local cost function, defined for $M \rightarrow N$ cloning to include the distribution $\mathcal{S}$ over the input states is,
\begin{equation} \label{eqn:local_cost_full_suppforproof}  
    \Cbs_{\Lbs}(\paramtheta) := \mathbb{E}\left[1 - \frac{1}{N}\left(\sum\limits_{j=1}^{N} F_j({\paramtheta})\right)\right] = 1 - \frac{1}{N\mathcal{N}}\int_{\mathcal{S}} \sum\limits_{j=1}^{N} F_j({\paramtheta}) d\psi
\end{equation}
where $\mathcal{N} = \int_{\mathcal{S}}d\psi$ is the normalisation condition. 
As above, we can show this cost function also exhibits strong faithfulness:\\

\noindent \textbf{1. Strong faithfulness:}

\vspace{2 mm}
\begin{thmbox}
\begin{theorem} \label{thm:squared_local_cost_local_FS_strong_faithful_appendix}
    The squared local cost function is locally strongly faithful:
    \begin{equation}\label{eqn:strongly_local_faithful_local_cost_defn_appendix}
        \Cbs_{\Lbs}(\paramtheta) = \Cbs_{\Lbs}^{\opt} \implies \rho_{\paramtheta}^{\psi, j} = \rho_{\opt}^{\psi, j} \qquad \forall \ket{\psi} \in \mathcal{S}, \forall j \in [N]
    \end{equation}
\end{theorem}
\end{thmbox}

\begin{proof}
Similar the faithfulness arguments of the squared cost function, one can immediately see that the cost function $\Cbs_{\Lbs}(\paramtheta)$ achieves a unique minimum at the joint maximum of $\mathbb{E}[F_j(\paramtheta)]$ for all $j \in [N]$. Thus, the minimum of $\Cbs_{\Lbs}(\paramtheta)$ corresponds to the unique optimal joint state with its unique local reduced states $\rho_{\opt}^{\psi, j}$ for each $j \in [N]$ for each input state $\ket{\psi} \in \mathcal{S}$. 
Thus the cost function achieves a minimum under precisely the unique condition i.e. the output state is equal to the optimal clone state. \\ 
\end{proof}

\noindent \textbf{2. Weak faithfulness:} 

\vspace{2 mm}
Now, we can also prove analogous versions of weak faithfulness. Many of the steps in the proof follow similarly to the squared cost derivations above, so we omit them for brevity where possible. As above, we first have the following lemma:
\begin{lembox}
\begin{lemma} \label{lemma:trace_bound_local_cost}
Suppose the cost function is $\epsilon$-close to the optimal cost in symmetric cloning
%
\begin{equation}
    \Cbs_{\Lbs}(\paramtheta) - \Cbs^{\opt}_{\Lbs} \leq \epsilon
    \label{eqn:costtoepsilon_local_appendix}
\end{equation}
where we assume $ \lim_{\epsilon \rightarrow 0} |\mathbb{E}[F_i(\paramtheta)] - \mathbb{E}[F_j(\paramtheta)]| \rightarrow 0, \forall i, j$, and therefore $\Cbs_{\opt} := 1-F_{\opt}$. Then,
\begin{equation}\label{eqn:tracecloseness_local_cost_appendix}
        \tr[(\rho_{\opt}^{\psi, j} - \rho^{\psi, j}_{\paramtheta})\ket{\psi}\bra{\psi}] \leq \mathcal{N}\epsilon, \qquad \forall \ket{\psi} \in \mathcal{S}, \forall j \in [N]
\end{equation}
\end{lemma}
\end{lembox}
The proof of \lemref{lemma:trace_bound_local_cost} follows identically to \lemref{lemma:trace_bound_squared_cost}, but with the exception that we can write $\Cbs_{\Lbs}(\paramtheta) - \Cbs^{\opt}_{\Lbs} = \mathbb{E}\left(F_{\opt} - F(\paramtheta)\right)$ in the symmetric case, assuming $F_i(\paramtheta) \approx F_j(\paramtheta)$,  $\forall i \neq j \in [N]$.

Now, we can prove the following theorem:
\begin{thmbox}
\begin{theorem} \label{thm:local_cost_FS_weak_faithful_appendix}
The local cost function, \eqref{eqn:local_cost_full_suppforproof}, is $\epsilon$-weakly faithful with respect to $\text{d}_{\BA}$. If we have
\begin{equation}  \label{eqn:cost_to_epsilon_local_appendix}
    \Cbs_{\Lbs}(\paramtheta) - \Cbs_{\Lbs}^{\opt} \leq \epsilon
\end{equation}
where $\Cbs_{\Lbs}^{\opt} := 1 - F_{\opt}$ then the following fact holds:
\begin{equation}     \label{eqn:fubini_study_bound_local_appendix}
    \text{d}_{\BA}(\rho^{\psi,j}_{\paramtheta}, \rho^{\psi, j}_{\opt}) \leq \frac{\mathcal{N}\epsilon}{\sin(F_{\opt})} =: f_2(\epsilon), \hspace{3mm} \forall \ket{\psi} \in \mathcal{S}, \forall j \in [N]
\end{equation}
\end{theorem}
\end{thmbox}

\begin{proof}
 We rewrite the \eqref{eqn:tracecloseness_local_cost_appendix} in terms of the Bures angle:
\begin{align} 
    F(\rho_{\opt}^{\psi, j}, \ket{\psi}) - F(\rho^{\psi,j}_{\paramtheta}, \ket{\psi}) &\leq \mathcal{N}\epsilon \nonumber \\
    \implies  \cos^2(\text{d}_{\BA}(\rho_{\opt}^{\psi, j}, \ket{\psi})) -  \cos^2(\text{d}_{\BA}(\rho^{\psi,j}_{\paramtheta}, \ket{\psi})) &\leq \mathcal{N}\epsilon   \label{eqn:fidelityFS_local_appendix}
\end{align}
Following the derivation in \secref{sssec:cloning/squared_cost_function_faithfulness}, we obtain:
\begin{equation}
    \text{d}_{\BA}(\rho^{\psi,j}_{\paramtheta}, \rho_{\opt}^{\psi, j}) \leq \frac{\mathcal{N}\epsilon}{\sin(F_{\opt})}, \qquad \forall \ket{\psi} \in \mathcal{S}, \forall j \in [N]
    \label{eqn:FSbound-standard-local}
\end{equation}
\end{proof}

Finally, we have \thmref{thm:local_cost_trace_weak_faithful_appendix} relating to the trace distance. The proof follows identically to \thmref{thm:squared_local_cost_squared_trace_weak_faithful_appendix}
so we simply state the result:
\begin{thmbox}
\begin{theorem} \label{thm:local_cost_trace_weak_faithful_appendix}
The local cost function, \eqref{eqn:local_cost_full_suppforproof}, is $\epsilon$-weakly faithful with respect to $\text{d}_{\tr}$ on qubits.
\begin{equation}     \label{eqn:trace_distance_bound_local_appendix}
  \text{d}_{\tr}(\rho_{\opt}^{\psi, j},  \rho^{\psi, j}_{\paramtheta})  \leq \frac{1}{2}\sqrt{4F_{\opt}(1 - F_{\opt}) + \mathcal{N}\epsilon(1 - 2F_{\opt})} =: g_2(\epsilon), \qquad \forall j \in [N]
\end{equation}

\end{theorem}
\end{thmbox}

\subsubsection[\texorpdfstring{\color{black}}{} Global cost function]{Global cost function} \label{app_sssec:global_faithfulness}

Finally, we show in the next theorems that the global cost function exhibits similar notions of faithfulness:
\begin{thmbox}
\begin{theorem} \label{thm:global_cost_strong_faithful_appendix}
    The global cost function is globally strongly faithful, i.e.:
    \begin{equation}\label{eqn:strongly_faithful_cost_defn_appendix}
        \Cbs_{\Gbs}(\paramtheta) = \Cbs^{\opt}_{\Gbs} \implies \rho_{\paramtheta}^\psi = \rho_{\opt}^{\psi} \qquad \forall \ket{\psi} \in \mathcal{S}
    \end{equation}
\end{theorem}
\end{thmbox}
\begin{proof}
 The global cost function $\Cbs_{\Gbs}(\paramtheta)$ achieves the minimum value $\Cbs^{\opt}_{\Gbs}$ at a unique point corresponding to $\mathbb{E}[F_{\Gbs}(\paramtheta)] = F_{\Gbs}^{\opt}$, where $F_{\Gbs}^{\opt}$ corresponds to the fidelity term for $\Cbs^{\opt}_{\Gbs}$. This corresponds to the unique global clone state $\rho^{\psi}_{\opt}$. Thus the cost function, achieves a unique minimum under precisely the unique condition i.e. the output global state is equal to the optimal clone state for all inputs in the distribution.
\end{proof}
Now, as usual, moving the global faithfulness, we have
\begin{lembox}
\begin{lemma} \label{lemma:global-state-closeness}
Suppose the global cost is $\epsilon$-close to optimality in symmetric cloning
%
\begin{equation}
    C_{\Gbs}(\paramtheta) - C^{\opt}_{\Gbs} \leq \epsilon
    \label{eqn:costtoepsilon_global_trace}
\end{equation}
where $\Cbs^{\opt}_{\Gbs} := 1 - F_{\Gbs}^{\opt}$. Then,
\begin{equation}\label{eqn:tracecloseness}
        \tr\left[(\rho^{\psi}_{\opt} - \rho^{\psi}_{\paramtheta})\ketbra{\psi}{\psi}^{\otimes N}\right] \leq \mathcal{N}\epsilon, \qquad \forall \ket{\psi} \in \mathcal{S}
\end{equation}
\end{lemma}
\end{lembox}
\begin{proof}
The proof of \lemref{lemma:global-state-closeness} follows identically to \lemref{lemma:trace_bound_local_cost} but with the exception that $\Cbs_{\Gbs}(\paramtheta) - \Cbs^{\opt}_{\Gbs} = \mathbb{E}[ F_{\Gbs}^{\opt} - F_{\Gbs}(\paramtheta)]$.
\end{proof}

\newpage
Finally, we  also have weak faithfulness:
\begin{thmbox}
\begin{theorem} \label{thm:trace_FS_bound_global}
Suppose the cost function is $\epsilon$-close to the optimal cost in symmetric cloning
%
\begin{equation}
    \Cbs_{\Gbs}(\paramtheta) - \Cbs^{\opt}_{\Gbs} \leq \epsilon
    \label{eqn:costtoepsilon_global_FS}
\end{equation}
where $\Cbs^{\opt}_{\Gbs} := 1 - F_{\Gbs}^{\opt}$. Then we have, $\forall \ket{\psi} \in \mathcal{S}$:
\begin{align}     \label{eqn:global_fubini_study_bound_squared_appendix}
    \text{d}_{\BA}(\rho^{\psi}_{\paramtheta}, \rho^{\psi}_{\opt}) &\leq \frac{\mathcal{N}\epsilon}{\sin(F_{\Gbs}^{\opt})} =: f_4(\epsilon), \\
\label{eqn:globaltracecloseness}
        \text{d}_{\tr}(\rho^{\psi}_{\opt}, \rho^{\psi}_{\paramtheta}) &\leq  \frac{1}{2}\sqrt{4F^{\opt}_{\Gbs}(1 - F^{\opt}_{\Gbs}) + \mathcal{N}\epsilon(1 - 2F^{\opt}_{\Gbs})}  =: g_4(\epsilon)
\end{align}
\end{theorem}
\end{thmbox}
\begin{proof}
The proof of \thmref{thm:trace_FS_bound_global} follows along the same lines as that of \thmref{thm:local_cost_FS_weak_faithful_appendix} and \thmref{thm:local_cost_trace_weak_faithful_appendix}.
\end{proof}

\subsubsection[\texorpdfstring{\color{black}}{} Global versus local faithfulness]{Global versus local faithfulness} \label{sssec:global_v_local_faithfulness}

This section explores the relationship between local and global cost function optimisation for different cloners (universal, phase-covariant, etc.). In particular, we address the question of whether optimizing a cloner with a local or a global cost function also achieves an optimal solution relative to the other cost (operational meaning). If the answer is affirmative, we can use whichever cost exhibits the most desirable qualities and be confident they will achieve the same results. If not, we must be more careful as the choice may not lead to the optimal behaviour we desire and will be application dependent.

We note that this relationship only manifests in \emph{symmetric} cloning, since there is no possibility to enforce asymmetry in the global cost function. As we shall see in~\eqref{eqn:asymmetric_cost_full_appendix}, the only way to enforce asymmetry is by constructing a cost function which optimises with respect to the local asymmetric optimal fidelites.  


The tradeoff between local and global faithfulness turns out to be subtle when dealing with cloning problems, and is in contrast to similar studies in analogous variational algorithm literature. Let us begin this discussion with the following theorem (proof given in~\appref{app_ssec:proof_local_gloabl_relationship}):
\begin{thmbox}
\begin{theorem}\label{thm:relationship-local-global-appendix}
For the general case of $M \rightarrow N$ cloning, the global cost function $\Cbs_{\Gbs}(\paramtheta)$ and the local cost function $\Cbs_{\Lbs}(\paramtheta)$ satisfy the inequality,
\begin{equation}
 \Cbs_{\Lbs}(\paramtheta) \leq \Cbs_{\Gbs}(\paramtheta) \leq N\cdot\Cbs_{\Lbs}(\paramtheta)  
\end{equation}
\end{theorem}
\end{thmbox}

A similar inequality was proven in the work of, for example, \cite{bravo-prieto_variational_2019}.
We however note that the inequality proven in \thmref{thm:relationship-local-global-appendix} (unlike in ~\cite{bravo-prieto_variational_2019}) does not allow us make statements about the similarity of individual clones from the closeness of the global cost function and vice versa.  This can be seen as follows:
\begin{equation}
\begin{split}
    \Cbs_{\Gbs}(\paramtheta) - \Cbs_{\Gbs}^{\opt} \leq \epsilon &\implies  \Cbs_{\Lbs}(\paramtheta) - \Cbs_{\Lbs}^{\opt} \leq \epsilon - (\Cbs_{\Gbs}(\paramtheta) - \Cbs_{\Lbs}(\paramtheta)) + (\Cbs_{\Lbs}^{\opt} - \Cbs_{\Gbs}^{\opt})\\ 
    &\implies \Cbs_{\Lbs}(\paramtheta) - \Cbs_{\Lbs}^{\opt} \leq \epsilon + (\Cbs_{\Lbs}^{\opt} - \Cbs_{\Gbs}^{\opt}) \\ 
    & \qquad \nRightarrow \Cbs_{\Lbs}(\paramtheta) - \Cbs_{\Lbs}^{\opt} \leq \epsilon
\end{split}    
\end{equation}
Here we have used the result of \thmref{thm:relationship-local-global-appendix} that $\Cbs_{\Gbs}(\paramtheta) \geq \Cbs_{\Lbs}(\paramtheta)$ and we note that $\Cbs_{\Lbs}^{\opt} - \Cbs_{\Gbs}^{\opt} \neq 0$ for all the $M \rightarrow N$ cloning. In particular, for $1 \rightarrow 2$ cloning, $\Cbs_{\Lbs}^{\opt} = 5/6$, while $\Cbs_{\Gbs}^{\opt} = 2/3$. This is due to the non-vanishing property of these cost functions, even at the theoretical optimum, and highlights the subtlety of the case in hand.

While we are unable to leverage generic inequalities for our purpose based on the cost functions, we can make statements in \emph{specific} cases. In other words, by restricting the cloning problem to a specific input set of states, we can guarantee that optimizing \emph{globally} will be sufficient to also optimise \emph{local} figures of merit. 
 
In particular, in the following we establish this strong and weak faithfulness guarantees for the \emph{special cases} of universal and phase-covariant cloning by analysing problem-specific features.
\begin{thmbox}
\begin{theorem}[\thmref{thm:local_clones_from_global_cost_universal_maintext_universal} in main text] \label{thm:local_clones_from_global_cost_universal_appendix_universal}
The \emph{global} cost function is \emph{locally} strongly faithful for a universal symmetric cloner, i.e,:
\begin{equation}\label{eqn:strongly_faithful_global_cost_local_clone_strong_universal_appendix}
    \Cbs_{\Gbs}(\paramtheta) = \Cbs^{\opt}_{\Gbs} \iff \rho_{\paramtheta}^{\psi, j} = \rho_{\opt}^{\psi, j} \qquad \forall \ket{\psi} \in \mathcal{H}, \forall j \in \{1,\dots,N\}
\end{equation}
\end{theorem}
\end{thmbox}
\begin{proof}
In the symmetric universal case, $\Cbs^{\opt}_{\Lbs}$ has a unique minimum when, each local fidelity saturates:
\begin{align} \label{eqn:univ-localfid}
    F_{\Lbs}^{\opt} = \frac{M(N+2) + N - M}{N(M+2)}
\end{align}
achieved by local reduced states, $\{\rho_{\opt}^{\psi, j}\}_{j=1}^N$. Now, it has been shown that the optimal global fidelity $F_{\Gbs}$ that can be reached~\cite{buzek_universal_1998, scarani_quantum_2005} is,
\begin{align} \label{eqn:univ-globalfid}
    F_{\Gbs}^{\opt} = \frac{N!(M+1)!}{M!(N+1)!}
\end{align}
which also is the corresponding unique minimum value for $\Cbs^{\opt}_{\Gbs}$, achieved by some global state $\rho_\opt^\psi$.

Finally, it was proven in~\cite{werner_optimal_1998, keyl_optimal_1999} that the cloner which achieves one of these bounds is unique and also saturates the other, and therefore must also achieve the unique minimum of both global and local cost functions, $\Cbs^{\opt}_{\Gbs}$ and $\Cbs^{\opt}_{\Lbs}$. Hence, the local states which optimise $\Cbs^{\opt}_{\Lbs}$  must be the reduced density matrices of the global state which optimises $\Cbs^{\opt}_{\Gbs}$ and so:
\begin{equation}    \label{eqn:local-global-state-closeness}
    \rho_{\opt}^{\psi, j} := \tr_{\Bar{j}}(\rho_{\opt}^{\psi}), \ \ \forall j
\end{equation}
Thus for a universal cloner, the cost function with respect to both local and global fidelities will converge to the same minimum.
\end{proof}

Now, before proving an analogous statement in the case of phase-covariant cloning, we first need the following lemma (we return to the notation of $B, E$ and $E^*$ for clarity):
\begin{lembox}
\begin{lemma}\label{lemma:phase-covariant-clone-uniqueness}
For any $1\rightarrow 2$ phase-covariant cloning machine which takes states $\ket{0}_{B}\otimes\ket{\psi}_{E}$ and an ancillary qubit $\ket{A}_{E^*}$ as input, where $\ket{\psi} := \frac{1}{\sqrt{2}}(\ket{0} + e^{i\theta}\ket{1})$, and outputs a 3-qubit state $\ket{\Psi_{BEE^*}}$ in the following form:
\begin{multline} \label{eqn:phasecov-tripartite-output}
    \ket{\Psi_{BEE^*}} = \frac{1}{2}\left[\ket{0,0} + e^{i\phi}(\sin\eta\ket{0,1} + \cos\eta\ket{1,0}))\ket{0}_{E^*}\right. \\
    \left.+ e^{i\phi}\ket{1,1} + (\cos\eta\ket{0,1} + \sin\eta\ket{1,0})\ket{1}_{E^*}\right]
\end{multline}
 the global and local fidelities are simultaneously maximised at $\eta = \frac{\pi}{4}$ where $0\leq \eta \leq \frac{\pi}{2}$ is the `shrinking factor'. 
\end{lemma}
\end{lembox}

\begin{proof}
To prove this, we follow the formalism adopted by~\cite{cerf_cloning_2002}. This uses the fact that a symmetric phase-covariant cloner induces a mapping of the following form~\cite{scarani_quantum_2005}:
\begin{align} \label{eqn:phasecov-map}
    \begin{split}
        & \ket{0}\ket{0}\ket{0} \rightarrow \ket{0}\ket{0}\ket{0} \\
        & \ket{1}\ket{0}\ket{0} \rightarrow (\sin\eta\ket{0}\ket{1} + \cos\eta\ket{1}\ket{0})\ket{0} \\
        & \ket{0}\ket{1}\ket{1} \rightarrow (\cos\eta\ket{0}\ket{1} + \sin\eta\ket{1}\ket{0})\ket{1} \\
        & \ket{1}\ket{1}\ket{1} \rightarrow \ket{1}\ket{1}\ket{1}
    \end{split}
\end{align}
Next, we calculate the global state by tracing out the ancillary state to get $\rho_{\Gbs}^{\opt}$:
\begin{align} \label{eqn:phasecov-global-density}
    \rho_{\Gbs}^{\opt} = \tr_{E^*}(\ket{\Psi_{BEE^*}}\bra{\Psi_{BEE^*}}) = \ket{\Phi_1}\bra{\Phi_1} + \ket{\Phi_2}\bra{\Phi_2}
\end{align}
where:
\begin{equation}
    \begin{split}
        \ket{\Phi_1} &:= \frac{1}{2}\left[\ket{0,0} + e^{i\phi}(\sin\eta\ket{0,1} + \cos\eta\ket{1,0})\right]\\ \ket{\Phi_2} &:= \frac{1}{2}\left[e^{i\phi}\ket{1,1} + (\cos\eta\ket{0,1} + \sin\eta\ket{1,0})\right]
    \end{split}
\end{equation} 
Therefore, the global fidelity can be found as:
\begin{align} \label{eqn:phasecov_global_fid_appendix}
    F_{\Gbs}^{\opt} = \tr(\ket{\psi}\bra{\psi}^{\otimes 2}\rho_{\Gbs}^{\opt}) = |\braket{\psi^{\otimes 2}}{\Phi_1}|^2 +|\braket{\psi^{\otimes 2}}{\Phi_2}|^2 = \frac{1}{8}(1 + \sin\eta + \cos\eta)^2
\end{align}
Now, optimising $F_{\Gbs}^{\opt}$ with respect $\eta$, we see that $F_{\Gbs}^{\opt}$ has only one extremum between $[0,\frac{\pi}{2}]$ specifically at $\eta=\frac{\pi}{4}$. We can also see that the optimal local fidelity is also achieved for the same $\eta$ and is equal to:
\begin{align} \label{eqn:phasecov_local_fid_appendix}
    F_{\Lbs}^{\opt} = \frac{1}{2}\left(1 + \frac{\sqrt{2}}{2}\right)
\end{align}
which is the upper bound for local fidelity of the phase-covariant cloner.
\end{proof}

With \lemref{lemma:phase-covariant-clone-uniqueness} established, we can prove:
\begin{thmbox}
\begin{theorem} [\thmref{thm:local_clones_from_global_cost_universal_maintext_phase_covariant} in main text] \label{thm:local_clones_from_global_cost_phase-covariant_appendix}
The global cost function is locally strongly faithful for phase-covariant symmetric cloning, i.e.:
\begin{equation}\label{eqn:strongly_faithful_global_cost_local_clone_strong_phase_covariant_appendix}
    \Cbs_{\Gbs}(\paramtheta) = \Cbs^{\opt}_{\Gbs} \iff \rho_{\paramtheta}^{\psi, j} = \rho_{\opt}^{\psi, j} \qquad \forall \ket{\psi} \in \mathcal{S}, \forall j \in \{B, E\}
\end{equation}
where $\mathcal{S}$ is the distribution corresponding to phase-covariant cloning.  
\end{theorem}
\end{thmbox}
\begin{proof}

Now, we have in \lemref{lemma:phase-covariant-clone-uniqueness} that the global and local fidelities of a phase-covariant cloner are both achieved with a cloning transformation of the form in~\eqref{eqn:phasecov-map}. Applying this transformation unitary to $\ket{\psi}\ket{\Phi^+}_{BE}$ (where $\ket{\Phi^+}_{BE}$ is a Bell state) leads to Cerf's formalism for cloning. Furthermore, we can observe that due to the symmetry of the problem, this transformation is unique (up to global phases) and so any optimal cloner must achieve it.

Furthermore, one can check that the ideal circuit in \figref{fig:qubit_cloning_ideal_circ} does indeed produce an output in the form of \eqref{eqn:phasecov-tripartite-output} once the preparation angles have been set for phase-covariant cloning. By a similar argument to the above, we can see that a variational cloning machine which achieves an optimal cost value, i.e. $\Cbs_{\Gbs}(\paramtheta) = \Cbs^{\opt}_{\Gbs}$ must also saturate the optimal cloning fidelities. Furthermore, by the uniqueness of the above transformation (\eqref{eqn:phasecov-map}) we also have that the local states of $\VQC$ are the same as the optimal transformation, which completes the proof.

\end{proof}

\subsubsection[\texorpdfstring{\color{black}}{} Asymmetric faithfulness]{Asymmetric faithfulness} \label{ssec:asymmetric_faithfulness}

Finally, let us prove the analogous results for the asymmetric cost function, \eqref{eqn:asymmetric_cost_function_maintext}, which can also be written as:
\begin{equation} \label{eqn:asymmetric_cost_full_appendix}
      \Cbs_{\Lbs, \text{asym}}(\paramtheta)
      = \frac{1}{\mathcal{N}}\int_{\mathcal{S}}\left((F_{\Lbs}^{p, B} - F_{\Lbs}^{B}(\paramtheta))^2 + (F_{\Lbs}^{p, E} - F_{\Lbs}^{E}(\paramtheta))^2\right)d\psi
\end{equation}

\noindent \textbf{1. Strong faithfulness:} 

\vspace{2mm}
\begin{thmbox}
\begin{theorem} \label{eqn:asymmetric_local_cost_squared_FS_strong_faithful}
    The asymmetric $1 \rightarrow 2$ local cost function is  strongly faithful:
    \begin{equation}\label{eqn:asymmetric_strong_local_faithfulness}
         \Cbs_{\Lbs, \mathrm{asym}}(\paramtheta) = \Cbs^{ \opt}_{\Lbs, \mathrm{asym}}(\paramtheta)  \implies \rho_{\paramtheta}^{\psi, i} = \rho_{\opt}^{\psi, i} \qquad \forall \ket{\psi} \in \mathcal{S}, \forall i \in \{B, E\}
    \end{equation}
\end{theorem}
\end{thmbox}

\begin{proof}
The cost function $\Cbs_{\Lbs, \mathrm{asym}}(\paramtheta)$ achieves the minimum value of zero, uniquely when $F_{\Lbs}^{B}(\paramtheta) = F_{\Lbs}^{p, B}$ and $F_{\Lbs}^{E}(\paramtheta) = F_{\Lbs}^{p, E}$ for all input states $\ket{\psi} \in \mathcal{S}$. This corresponds to the unique reduced states $\rho^{\psi,B}_{\opt}$ and $\rho^{\psi, E}_{\opt}$ for Bob and Eve. Thus the cost function, achieves a unique minimum of zero precisely when the output reduced state for Bob and Eve is equal to the optimal clones for all inputs in $\mathcal{S}$. \\
\end{proof}

\newpage
\noindent \textbf{2. Weak faithfulness:}

\vspace{2mm}
Returning again to $\epsilon$-weak faithfulness, we get similar results as in the symmetric case above (explicit proof given in~\appref{app_ssec:asymetric_faithful}):
\begin{thmbox}
\begin{theorem} \label{thm:asymmetric_cost_FS_weak_faithful_appendix} 
The asymmetric cost function, \eqref{eqn:asymmetric_cost_full_appendix}, is $\epsilon$-weakly faithful with respect to $\text{d}_{\BA}$
\begin{equation}  \label{eqn:asymmetric_cost_function_epsilon_guarantee}
 \Cbs_{\Lbs, \mathrm{asym}}(\paramtheta) - \Cbs^{\opt}_{\Lbs, \mathrm{asym}} \leq \epsilon   
\end{equation}
where $\Cbs^{\opt}_{\Lbs, \mathrm{asym}} = 0$. then the following fact holds for Bob and Eve's reduced states:
\begin{equation} \label{eqn:asymmetric_cost_function_FS_bound_appendix}
     \text{d}_{\BA}(\rho^{\psi,B}_{\paramtheta}, \rho^{\psi,B}_{\opt}) \leq \frac{\sqrt{\mathcal{N}\epsilon}}{\sin(1 - p^2/2)}, \hspace{3mm} \text{d}_{\BA}(\rho^{\psi,E}_{\paramtheta}, \rho^{\psi,E}_{\opt}) \leq \frac{\sqrt{\mathcal{N} \epsilon}}{\sin(1 - q^2/2)}
\end{equation}
Furthermore, we also have the following trace distance bounds:
\begin{align}    \label{eqn:asymmmetric_trace_distance_bound_Bob_Eve}
  \text{d}_{\tr}(\rho^{\psi, B},  \rho^{\psi, B}_{\paramtheta})  
        &\leq \frac{1}{2}\sqrt{p^2(2 - p^2) - \sqrt{\mathcal{N}\epsilon}(1 - p^2)}, \\
        \text{d}_{\tr}(\rho^{\psi, E},  \rho^{\psi, E}_{\paramtheta})  
        &\leq \frac{1}{2}\sqrt{q^2(2 - q^2) - \sqrt{\mathcal{N}\epsilon}(1 - q^2)} 
\end{align}

\end{theorem}
\end{thmbox}

\subsection[\texorpdfstring{\color{black}}{} Sample complexity of the algorithm]{Sample complexity of the algorithm} \label{app_ssec:sample_complexity_appendix}

\begin{figure}
    \centering
        \includegraphics[width=0.9\columnwidth, height=0.3\columnwidth]{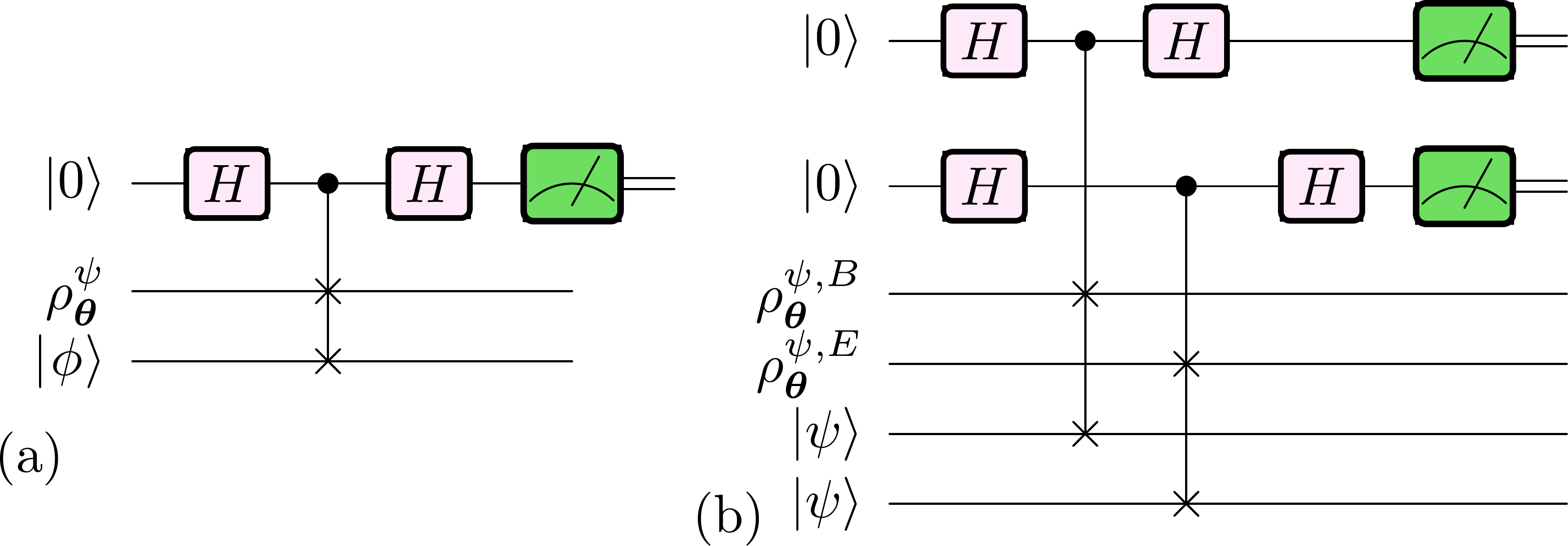}
    \caption[\color{black} The $\SWAP$ test circuit illustrated for $1\rightarrow 2$ cloning.]{\textbf{The $\SWAP$ test circuit illustrated for $1\rightarrow 2$ cloning.} In (a) for example, we compare the global state $\rho^{\psi}_{\paramtheta}$, with the state $\ket{\phi}$, where $\ket{\phi} := \ket{\psi}\otimes \ket{\psi}$ is the product state of two copies of $\psi$. (b) Local $\SWAP$ test with the reduced state of Bob and Eve separately. One ancilla is required for each fidelity to be computed. The generalisation for $N$ input states in $M \rightarrow N$ cloning is straightforward.}
    \label{circuit:swaptest_global_and_local}
\end{figure}

As discussed above (\secref{ssec:vqa_cost_functions}) we require the cost functions to be efficiently computable in general, which is why we went to such pains particularly in~\chapref{chap:born_machine}. However, luckily for $\VQC$, the situation is more straightforward. This is because there is a simple routine to estimate the fidelities in the cost functions, given by the $\SWAP$ test~\eqref{eqn:swap_gate_matrix_and_circuit} and for $\VQC$ more specifically in~\figref{circuit:swaptest_global_and_local}. This works since at least one of the states in the $\SWAP$ test is pure in $\VarQlone$. Given a particular cost function, $\Cbs^{\psi}(\paramtheta)$, for an input state, $\ket{\psi}$, to be cloned, let $L$ be the number of copies of $\ket{\psi}$ to estimate $\Cbs^{\psi}(\paramtheta)$. Then, let $K$ be the number of states from $\mathcal{S}$ required to estimate the expectation value over $\ket{\psi}$, $\Cbs(\paramtheta) = \mathop{\mathbb{E}}_{\substack{\ket{\psi} \in \mathcal{S}}}[\Cbs^{\psi}(\paramtheta)]$.

The number of states $L \times K$ is given by the following theorem.

\begin{thmbox}
\begin{theorem}\label{thm:sample-complexity-appendix}[Sample Complexity of $\VQC$]
The number of samples $L \times K$ required to estimate the cost function $\Cbs(\paramtheta)$ up to $\epsilon'$-additive error with a success probability $\delta$ is:
\begin{equation}\label{eqn:number-of-samples-appendix}
    L \times K =  \mathcal{O}\left(\frac{1}{\epsilon'^2}\log \frac{2}{\delta}\right)
\end{equation}
where $K$ is the number of distinct states $\ket{\psi}$ sampled uniformly at random from the distribution $\mathcal{S}$, and $L$ is the number of copies of each input state. 
\end{theorem}
\end{thmbox}
We give the proof of this in~\appref{app_ssec:vqc_sample_complexity} for the global cost function, which is a straightforward application of H\"{o}effding's inequality~\cite{hoeffding_probability_1963}. The proof and theorem could also be adapted straightforwardly to the other cost functions.

Finally, we note that the $\SWAP$ test, in practice, is somewhat challenging to implement on NISQ devices, predominately due to the overhead of compiling the $3$-local controlled $\SWAP$ into the native gateset of a particular quantum hardware (In our numerical results in~\secref{sec:cloning/numerics}, we will circumvent the $\SWAP$ test by directly extracting the quantum states via tomography to compute the required fidelities.) Furthermore, in this case, we have a strict need for the copy of the input state $\ket{\psi}$ to be kept coherent while implementing the $\SWAP$ test, due to the equivalence between fidelity and overlap if one state is pure. This is due to the fact that for \emph{mixed} quantum states, there is no known efficient method to compute the fidelity~\cite{watrous_quantum_2002} and one must resort to using bounds on it\footnote{Perhaps one may also use bounds discovered variationally~\cite{chen_alternative_2002, mendonca_alternative_2008, puchala_bound_2009, cerezo_variational_2020}.}. In light of this, one could use the shorter depth circuits to compute the overlap found using a variational approach similar to that implemented here~\cite{cincio_learning_2018}. This is the interesting consequence that if we were to try and perform \emph{broadcasting}~\cite{barnum_noncommuting_1996} instead of cloning, we would again need to use an alternative strategy to estimate the fidelities.

\section[\texorpdfstring{\color{black}}{} Variational quantum cryptanalysis]{Variational quantum cryptanalysis} \label{sec:cloning/variational_cryptanalysis}

Now that we have introduced $\VQC$, we can begin to discuss one of the main additional applications for it: \emph{quantum cryptanalysis}. We do this for two specific types of quantum cryptographic protocols, each of which uses one family of state-dependent cloning states introduced in \secref{ssec:prelim/qc/quantum_cloning}. We begin in the following sections by introducing the particular protocols we study here and providing theoretical analyses about them. We then demonstrate how $\VQC$ can be used to \emph{implement} the attacks we propose. Using $\VQC$ as a quantum machine learning toolkit in this way to supplement quantum cryptography, we refer to as a first step into what we dub \emph{variational quantum cryptanalysis}. 

\begin{figure}[ht]
    \centering
\includegraphics[width=\columnwidth]{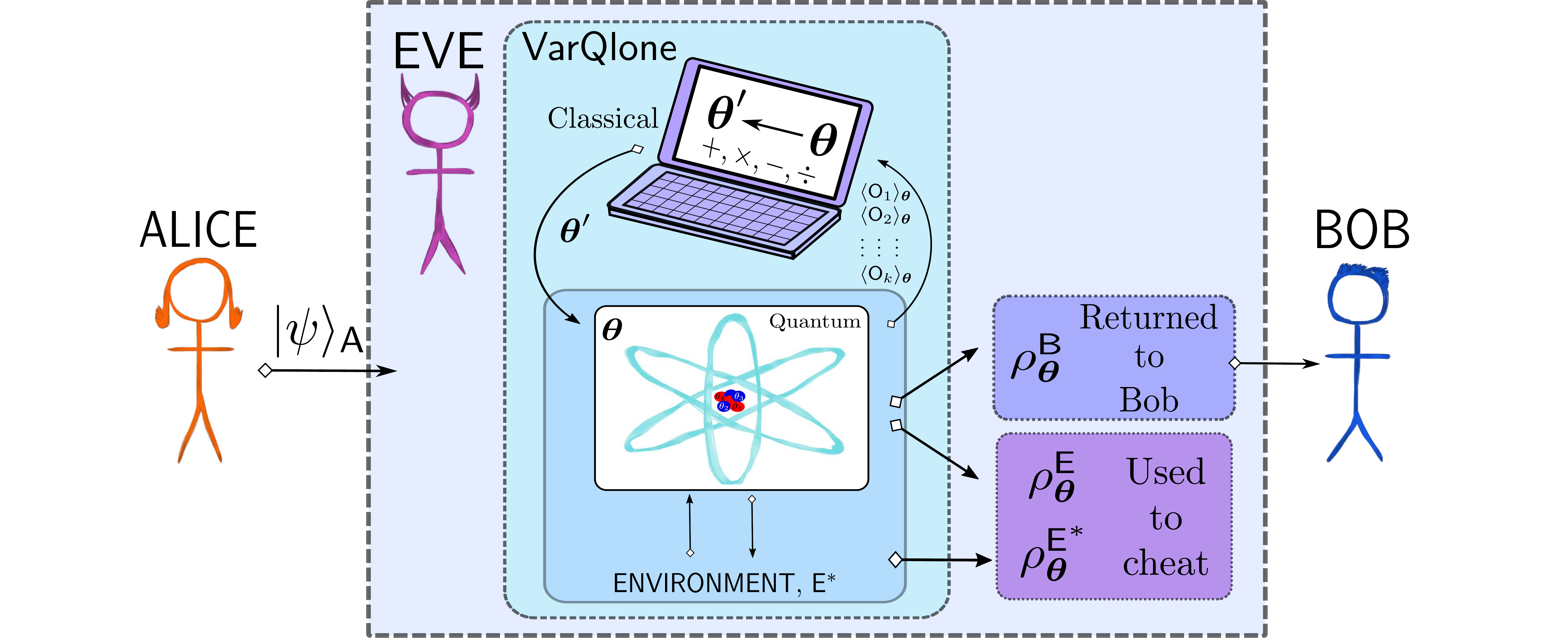}
    \caption[\color{black} Cartoon overview of $\VQC$ in a cryptographic attack.]{
    \textbf{Cartoon overview of $\VQC$ in a cryptographic attack.} Here an adversary Eve, $E$, implements a $1\rightarrow 2$ cloning attack on states used in a quantum protocol (for example QKD) between Alice and Bob. Eve intercepts the states sent by Alice $\ket{\psi}_A$ and may interact with an ancillary `environment', $E^*$. This interaction is trained (an optimal parameter setting $\paramtheta$ is found) by Eve to optimally produce clones, $\rho^B_{\paramtheta}, \rho^E_{\paramtheta}$. In order to attack the protocol, Eve will return $\rho^{B}_{\paramtheta}$ to Bob and use the rest (her clone, $\rho^{E}_{\paramtheta}$ plus the remaining environment state, $\rho^{E^*}_{\paramtheta}$) to cheat. The training procedure consists of using a classical computer to optimise the quantum parameters, via a cost function. The cost is a function of $k$ observables, $\mathsf{O}_k$, measured from the output states, which are designed to extract fidelities of the states to compare against the ideal state.
    }
    \label{fig:variational_algorithm_vqc}
\end{figure}

We already discussed at the beginning of this chapter the intimate relationship between quantum cloning and cryptography, and now \figref{fig:variational_algorithm_vqc} demonstrates at a high level how $\VQC$ may be inserted into an attack (on a QKD protocol specifically). The reader should keep this image in mind as we progress through the following sections, as it is useful for providing intuitions.

\subsection[\texorpdfstring{\color{black}}{} Quantum key distribution and cloning attacks]{Quantum key distribution and cloning attacks} \label{sec:qkd_and_cloning_attacks}

Let us begin with quantum key distribution protocols, and specifically the Bennett-Brassard 84 (BB84)~\cite{bennett_quantum_2014} protocol. In this protocol, one party (say Alice), sends single qubit states in two orthogonal bases (for instance, the eigenstates of the Pauli $\XG$ and Pauli $\YG$ matrices, $\ket{\pm}$ and $\ket{\pm \irm}$) to a second party, Bob, via a quantum channel that is susceptible to an eavesdropping adversary, Eve. Eve's goal is to extract the secret information sent between Alice and Bob, encoded in the states. 

Firstly, let us clarify the types of attacks one may consider on such protocols. The simplest attack by Eve is a so-called `incoherent' or individual attack, where Eve only interacts with the quantum states one at a time, and in the same fashion, and does so before the reconciliation phase of the protocol.  In such attacks, the security condition states that a secret key can no longer be extracted if Eve has as much information about the states as Bob, then a secret key can no longer be extracted. As such, the protocol requires a critical `error rate', $D_{\text{crit}}$, above which Alice and Bob will detect malfeasance, and abort the protocol.

For incoherent attacks, this error rate is achieved if the fidelity of the states received by Bob and stored by Eve is the same compared to the original state sent by Alice. Therefore, for BB84, we have the critical error rate to be $D^{\text{incoh}}_\text{crit} = 1 - F_{\Lbs,  \opt}^{\text{PC}, \mathrm{E}}= 14.6 \%$.

However, this criteria does not allow for comparison between a cloning machine using the ancilla, and one without\footnote{Recall in \secref{ssec:prelim/qc/quantum_cloning} we mentioned how phase-covariant cloning could be optimally achieved in both cases.}. This is important since, as discussed above and in~\cite{scarani_quantum_2005}, a phase-covariant cloning machine with ancilla provides the optimal attack on both Alice and Bob. In contrast, one \emph{without} an ancilla retains no information with which Eve can use to attack Bob's side of the protocol.

In order to have a better comparison we analyse the \emph{key rate}, given by the following expression~\cite{scarani_quantum_2005}:
\begin{equation} \label{eqn:qkd-key-rate}
    R = I(A:B) - \min\{\chi(A:E_Q), \chi(B:E_Q)\}
\end{equation}
Here, we have the term $\chi$, which is the so-called \emph{Holevo quantity}, written in terms of the von Neumann entropy (see~\eqref{eqn:von_neumann_entropy}):
\begin{equation} \label{eqn:qkd-holevo-quantity}
    \chi(Q:E) := S(\rho_E) - \frac{1}{2}S(\rho^0_E) - \frac{1}{2}S(\rho^1_E)
\end{equation}
In \eqref{eqn:qkd-key-rate}, we also have $I(A:B)$, which is the \emph{mutual information} (MI) between Alice and Bob and the index $Q$ denotes that Eve may employ general quantum strategies.  In \eqref{eqn:qkd-holevo-quantity}, $\rho_E$ denotes the mixed state of Eve over all of the combinations of Alice's choice of input, and $\rho^0_E$ and $\rho^1_E$ denotes the states of Eve for the random variables that encode $0$ and $1$ in the protocol respectively. 

To compute the critical error rate from this expression, $D_{\text{crit}}$, it is enough to calculate the Holevo quantity for Eve, set $R = 0$, and $I(A:B) = 1 - H(D_{\text{crit}})$ and to solve the resulting equation for $D_{\text{crit}}$. In \secref{sec:cloning/numerics}, we return to this calculation for the cloning transformations we learn using $\VQC$ and we shall see how the algorithm is able to approximately saturate the optimal bound for the best possible individual attack.

\section[\texorpdfstring{\color{black}}{} Quantum coin flipping and cloning attacks]{Quantum coin flipping and cloning attacks} \label{app_sec:coin_flipping_cloning_attacks}
Next, we move in this section to the primitive of \emph{quantum coin flipping}~\cite{mayers_unconditionally_1999, aharonov_quantum_2000} and the use of states which have a fixed overlap. Protocols of this nature are a nice case study for our purposes since they provide a testbed for cloning fixed-overlap states, and in many cases explicit security analyses are missing. In particular in this work, to the best of our knowledge, we provide the first purely cloning based attack on the protocols we analyse. We go into some detail in the following in order to properly introduce our explicit attacks and the corresponding analyses of their success.

The two protocols we consider are that of Mayers \textit{et al.}~\cite{mayers_unconditionally_1999}) and that of Aharonov \textit{et al.}~\cite{aharonov_quantum_2000}. Before introducing these protocols, we first describe more generally the goals of coin flipping.

\subsection[\texorpdfstring{\color{black}}{} Quantum coin flipping]{Quantum coin flipping}\label{ssec:quantum_coin_flipping}
In (quantum) coin flipping protocols, we have two parties, Alice and Bob (no Eve this time) who share a mutual distrust, but who wish to agree the outcome of a random `coin flip'.
The most relevant quantity in such protocols is the \emph{bias} of the resulting coin. A `biased coin' has one outcome more likely than the other, for example with the following probabilities:
\begin{equation}\label{eqn:coin_flipping_epsilon_biased}
    \Pr(y = 0) = \frac{1}{2} + \epsilon, \qquad
    \Pr(y = 1) = \frac{1}{2} -  \epsilon
\end{equation}
where $y$ is a bit outputted by the coin. We can associate $y=0$ to heads ($\mathsf{H}$) and $y=1$ to tails ($\mathsf{T}$). A fair coin would correspond to $\epsilon=0$, while the above coin is $\epsilon$-biased towards $\mathsf{H}$.

It has been shown that it is impossible\footnote{Meaning it is not possible to define a coin flipping protocol such that \emph{neither} party can enforce any bias.} in an information theoretic manner, to achieve a secure coin flipping protocol with $\epsilon=0$ in both the classical and quantum setting~\cite{blum_coin_1983, lo_why_1998, mayers_unconditionally_1999}. Furthermore, there are two notions of coin flipping studied in the literature: \emph{weak} coin flipping, (where it is a-priori known that both parties prefer opposite outcomes), and \emph{strong} coin flipping, (where neither party knows the desired outcome of the other party). In the quantum setting, the lowest possible bias achievable by any strong coin flipping
protocol is limited by $\sim 0.207$~\cite{alexei_kitaev_quantum_2003}. Although several different protocols have been suggested for $\epsilon$-biased strong coin flipping~\cite{mayers_unconditionally_1999, aharonov_quantum_2000, bennett_quantum_2014, berlin_fair_2009}, the states used in them share a common structure.

\subsubsection[\texorpdfstring{\color{black}}{} Quantum states for strong coin flipping]{Quantum states for strong coin flipping} \label{ssec:quantum_states_coin_flipping}
Multiple qubit coin flipping protocols utilise the following set of states:
\begin{equation}\label{eq:coinflip-st}
    \ket{\phi_{x, a}} = 
    \begin{cases}
    \ket{\phi_{x,0}} = \cos\phi\ket{0} + (-1)^x\sin\phi\ket{1} \\
    \ket{\phi_{x,1}} =  \sin\phi\ket{0} + (-1)^{x \oplus 1}\cos\phi\ket{1}
  \end{cases} 
\end{equation}
where $x \in \{0,1\}$. This set of states are all conveniently related through a reparametrisation of the angle $\phi$~\cite{brus_approximate_2008}, which makes them easier to deal with mathematically. 

On top of using the same family of states, many coin flipping protocols also have a common structure in the protocol itself. To begin, Alice encodes some random classical bits into a subset of the above states and Bob may do the same. Each party will then exchange classical or quantum information (or both) as part of the protocol. Attacks (attempts to bias the coin) by either party usually reduce to how much one party can learn about the classical bits of the other.

We explicitly treat two cases:
\begin{enumerate}
    \item The protocol of Mayers \textit{et al.}~\cite{mayers_unconditionally_1999} in which the states, $\{\ket{\phi_{0,0}}, \ket{\phi_{1,0}}\}$ are used (which have a fixed overlap $s = \cos\left(2\phi\right)$). We denote this protocol $\mathcal{P}_1$.
    
    \item The protocol of Aharonov \textit{et al.}~\cite{aharonov_quantum_2000}, which uses the full set, i.e. $\{\ket{\phi_{x, a}}\}$. We denote this protocol $\mathcal{P}_2$.
\end{enumerate} 

In all strong coin flipping protocols, the security or fairness of the final shared bit lies on the impossibility of perfect discrimination of the underlying non-orthogonal quantum states. In general, the protocol can be analysed with either Alice or Bob being dishonest. Here we focus, for illustration, on a dishonest Bob who tries to bias the bit by cloning the non-orthogonal states sent by Alice. 

For all of the below, the biases are computed assuming access to the \emph{ideal} cloning machine (i.e. the one which clones the input states with the optimal, analytic fidelities). In~\secref{sec:cloning/numerics}, we compare these ideal biases with those achievable using the quantum cloning machines learned by $\VQC$.

\subsection[\texorpdfstring{\color{black}}{} 2-state coin flipping protocol (\texorpdfstring{$\mathcal{P}_1$}{})]{2-state coin flipping protocol (\texorpdfstring{$\mathcal{P}_1$}{})} \label{app_ssec:mayers_protocol_and_attack}

The first quantum coin flipping protocol we consider is one of the earliest proposed by~\cite{mayers_unconditionally_1999}. The protocol was originally described with $k$ `rounds', but for simplicity we primarily analyse the version with a single round, $k=1$. This is sufficient to describe a cloning based attack, which we discuss in the next section.

In this protocol, Alice sends the states\footnote{Since the value of the overlap is the only relevant quantity, the slightly different parameterisation of these states to those of \eqref{eq:coinflip-st} does not make a difference for our purposes. However, we note that explicit cloning unitary would be different in both cases.} $\ket{\phi_0} := \ket{\phi_{0,0}}$ and $\ket{\phi_1} := \ket{\phi_{1, 0}}$ such that the angle between them is $\phi := \frac{\pi}{18} \implies s := \cos(\frac{\pi}{9})$. Let us first describe the general protocol.

\subsubsection[\texorpdfstring{\color{black}}{} \texorpdfstring{$\mathcal{P}_1$}{} with \texorpdfstring{$k$}{} rounds]{\texorpdfstring{$\mathcal{P}_1$}{} with \texorpdfstring{$k$}{} rounds}
\label{app_sssec:mayers_cloning_attack_general_krounds}

To begin, Alice and Bob choose $k$ random bits, $\{a_1, \dots, a_k\}$ and $\{b_1, \dots, b_k\}$ respectively. The final bit is then given by the \computerfont{XOR} of input bits over all $k$ rounds i.e.,
\begin{equation}
y = \bigoplus_j a_j  \oplus \bigoplus_j b_j   
\end{equation}
In each round $j = 1,\dots, k$ of the protocol, and for every step $i=1, \dots, n$ within each round, Alice uniformly picks a random bit $c_{i,j}$ and sends the state\footnote{We use the notation $\overline{x}$ to denote the conjugate of bit $x$, i.e. $\overline{0} = 1, \overline{1} = 0$.} $\ket{\phi_c^{i,j}} := \ket{\phi_{c_{i,j}}} \otimes\ket{\phi_{\overline{c_{i,j}}})}$ to Bob. Likewise, Bob uniformly picks a random bit $d_{i,j}$ and sends the state $\ket{\phi_d^{i,j}} := \ket{\phi_{d_{i,j}}} \otimes\ket{\phi_{\overline{d_{i,j}}}}$ to Alice. Hence, each party sends multiple copies of either $\ket{\phi_0}\otimes\ket{\phi_1}$ or $\ket{\phi_1}\otimes\ket{\phi_0}$\footnote{Note that if $c_{i,j}$ and $d_{i,j}$ are chosen independently of $a_j$ and $b_j$, no information about the primary bits has been transferred.}.

In the next step, for each $j$ and $i$, Alice announces the value $a_j\oplus c_{i,j}$. If it turns out that $a_j\oplus c_{i,j} = 0$, Bob returns the second state of the pair $(i,j)$ back to Alice, and sends the first state otherwise. Similarly Bob announces $b_j\oplus d_{i,j}$ and Alice returns one of the states back to Bob accordingly. Now, we come to why it is sufficient to only consider a single round in the protocol from the point of view of a cloning attack.  This is because a dishonest Bob can bias the protocol if he learns about Alice's bit $a_j$, which he can do by guessing $c_{i,j}$ with a probability better than $1/2$. With this knowledge, Bob only needs to announce a single false $b_j \oplus d_{i,j}$ in order to cheat, and so this strategy can be deferred to the final round~\cite{mayers_unconditionally_1999}. Hence a single round of the protocol is sufficient for analysis, and we herein drop the $j$ index. 

In the last phase of the protocol, after $a$ and $b$ are announced by both sides (so $y$ can be computed by both sides), Alice measures the remaining states with the projectors, $(E_{b}, E^{\perp}_{b})$ and the returned states by Bob with $(E_{\overline{a}}, E^{\perp}_{\overline{a}})$ (\eqref{eqn:mayers_povm_appendix}). She aborts the protocol if she gets the measurement result corresponding to $\perp$, and declares Bob as being dishonest. In this sense, the use of quantum states in this protocol is purely for the purpose of cheat-detection.
\begin{equation}\label{eqn:mayers_povm_appendix}
 E_l = \ket{\phi_l}\bra{\phi_l}^{\otimes n}, \qquad
     E_l^\perp = \mathds{1} - \ket{\phi_l}\bra{\phi_l}^{\otimes n}, \qquad l \in\{0, 1\}
\end{equation}

\subsubsection[\texorpdfstring{\color{black}}{} A cloning attack on \texorpdfstring{$\mathcal{P}_1$}{}]{A cloning attack on \texorpdfstring{$\mathcal{P}_1$}{}}
\label{app_sssec:mayers_cloning_attack}

Now we are in a position to discuss an explicit attack that can be implemented by Bob on $\mathcal{P}_1$. WLOG, we assume that Bob wishes to bias the bit towards $y = 0$. For clarity, we give the attack for when Alice only sends one copy of the state ($n=1$), but we discuss the general case in the next section:\\

\noindent\rule[0.5ex]{\linewidth}{1pt}
 \vspace{1em}
\noindent \textbf{Attack 1:} Cloning attack on $\mathcal{P}_1$ with $k=1$.\\
\noindent\rule[0.5ex]{\linewidth}{1pt}
 \vspace{0.5em}
\noindent\textit{Inputs.} Random bit for Alice ($a \leftarrow_R \{0, 1\}$) and Bob ($b\leftarrow_R  \{0, 1\}$). Bob receives a state $\ket{\phi_c^i}$.
\sbline
\textit{Goal.} A biased bit towards $0$, i.e. $p(y=0) > 1/2$.
\sbline
\textit{The attack:}
\begin{enumerate}
\item for $i = 1, \dots, n$: 
  \begin{enumerate}
    \item \textbf{Step 1:} Alice announces $a \oplus c_i$. If $a \oplus c_i = 0$, Bob sends the second qubit of $\ket{\phi_c^i}$ to Alice, otherwise he sends the first qubit. 
    \item \textbf{Step 2:} Bob runs a $1\rightarrow 2$ state-dependent (fixed-overlap) cloner on the qubit he has to return to Alice, producing two approximate clones. He sends her one clone and keeps the other.
    \item \textbf{Step 3:} Bob runs an optimal state discrimination on the remaining qubit and any other output of the cloner, and finds $c_1$ with a maximum success probability $P^{\opt}_{\mathrm{disc}, \mathcal{P}_1}$. He then guesses a bit $a'$ such that $P_{\mathrm{succ}, \mathcal{P}_1}(a' = a) := P^{\opt}_{\mathrm{disc}, \mathcal{P}_1}$.
     \item \textbf{Step 4:} If $a'\oplus b = 0$ he continues the protocol honestly and announces $b \oplus d_1$, otherwise he announces $a' \oplus d_1$. The remaining qubit on Alice's side is $\ket{\phi^i_{a}}$.
  \end{enumerate}
\end{enumerate}
\noindent\rule[0.5ex]{\linewidth}{1pt}

We return to this attack in \figref{fig:mayers_1to2_cloning_fidelities_variational_main_text} where we give a cartoon illustration, plus the corresponding $\VQC$ numerics. Now, we can evaluate the success probability of Bob's attack with the following theorem:
\begin{thmbox}
\begin{theorem}\label{thm:mayers_attack_bias_probability_appendix}[Ideal cloning attack bias on $\mathcal{P}_1$]~ \\
Bob can achieve a bias of $\epsilon \approx 0.27$ using an ideal state-dependent cloning attack on the protocol, $\mathcal{P}_1$ with a single copy of Alice's state.
\end{theorem}
\end{thmbox}

\begin{proof}

As mentioned in the previous section, the final measurements performed by Alice on her remaining $n$ states, plus the $n$ states returned to her by Bob allow her to detect his nefarious behaviour. If he performed a cloning attack, the $\perp$ outcomes would be occasionally detected by Alice. We must compute both the probability that he is able to guess the value of Alice's bit $a$ (by guessing the value of the bit $c_1$), and the probability that he is detected by Alice. This would provide us with Bob's final success probability in cheating, and hence the bias probability.

At the start of the attack, Bob has a product state of either $\ket{\phi_0}\otimes\ket{\phi_1}$ or $\ket{\phi_1}\otimes\ket{\phi_0}$ (but he does not know which). In Step 2, depending on Alice's announced bit, Bob proceeds to clone one of the qubits, sends one copy to Alice and keeps the other to himself. We again assume the announced bit of Alice is $0$. In this case, at this point in the attack, he has one of the following pairs: $\ket{\phi_0}\bra{\phi_0}\otimes\rho^1_c$ or $\ket{\phi_1}\bra{\phi_1}\otimes\rho^0_c$, where $\rho^1_c$ and $\rho^0_c$ are the approximate clones for $\ket{\phi_1}$ and $\ket{\phi_0}$ respectively. 

Bob must now discriminate between the following density matrices:
\begin{equation}\label{eqn:mayers_bob_pairs_discriminate_appendix}
  \rho_1 := \ket{\phi_0}\bra{\phi_0}\otimes\ket{\phi_1}\bra{\phi_1} \qquad \text{ and }  \qquad
  \rho_2 := \ket{\phi_1}\bra{\phi_1}\otimes\rho^0_c
\end{equation}
Alternatively, if Alice announced $a\oplus c_i = 1$, he would have:
\begin{equation}\label{eqn:mayers_bob_pairs_discriminate_alternate_appendix}
    \rho_1 := \ket{\phi_1}\bra{\phi_1}\otimes\ket{\phi_0}\bra{\phi_0} \qquad \text{ and }  \qquad  
    \rho_2 := \ket{\phi_0}\bra{\phi_0}\otimes\rho^1_c
\end{equation}
In either case, we have that the minimum discrimination error for two density matrices is given by the Holevo-Helstrom bound~\cite{holevo_statistical_1973, helstrom_quantum_1969} as follows\footnote{This also is because the we assume a symmetric cloning machine for both $\ket{\phi_0}$ and $\ket{\phi_1}$. If this is not the case, the guessing probability is instead the average of the discrimination probabilities over both states.}:
\begin{equation}\label{eqn:helstrom}
    P^{\opt}_{\mathrm{disc}} = \frac{1}{2} + \frac{1}{4}\norm{\rho_1 - \rho_2}_{\Tr} = \frac{1}{2} + \frac{1}{2}\text{d}_{\Tr}(\rho_1, \rho_2)
\end{equation}
The ideal symmetric cloning machine for these states will have an output of the form:
\begin{equation}\label{eqn:reduced_local_state_mayers_attack_appendix}
      \rho_c = \alpha\ket{\phi_0}\bra{\phi_0} + \beta \ket{\phi_1}\bra{\phi_1}
     + \gamma (\ket{\phi_0}\bra{\phi_1} + \ket{\phi_1}\bra{\phi_0})  
\end{equation} 
Where $\alpha, \beta$ and $\gamma$ are functions of the overlap $s = \braket{\phi_0}{\phi_1} = \cos{\frac{\pi}{9}}$. Now, using \eqref{eqn:mayers_bob_pairs_discriminate_appendix}, $\rho_2$ can be written as follows:
\begin{multline}\label{eqn:reduced_local_state_rho2_mayers_attack_appendix}
      \rho_2 = \alpha\ket{\phi_1}\bra{\phi_1}\otimes\ket{\phi_0}\bra{\phi_0} + \beta  \ket{\phi_1}\bra{\phi_1}\otimes\ket{\phi_1}\bra{\phi_1}
     \\
     + \gamma ( \ket{\phi_1}\bra{\phi_1}\otimes\ket{\phi_0}\bra{\phi_1} +  \ket{\phi_1}\bra{\phi_1}\otimes\ket{\phi_1}\bra{\phi_0})  
\end{multline} 
Finally by plugging in the values of the coefficients in \eqref{eqn:reduced_local_state_mayers_attack_appendix} for the optimal local cloning machine~\cite{brus_optimal_1998} and finding the eigenvalues of $\sigma := (\rho_1 - \rho_2)$, we can calculate the corresponding value for \eqref{eqn:helstrom}, and recover the following minimum error probability:
\begin{equation}\label{eqn:mayers-cloning-min-error}
P_{\mathrm{fail}, \mathcal{P}_1} = P^{\mathrm{er}}_{\mathrm{disc}, \mathcal{P}_1} = 1 - P^{\opt}_{\mathrm{disc}, \mathcal{P}_1} \approx 0.214
\end{equation}

\noindent This means that Bob can successfully guess $c_1$ with ${P}^{1}_{\text{succ}, \mathcal{P}_1} = 78.5\%$ probability.

Now let us examine the probability of a cheating Bob being detected by Alice. We note that whenever Bob guesses the bit, $a$, successfully, the measurements $(E_{b}, E^{\perp}_{b})$ will be passed with probability $1$, hence we use $(E_{\overline{a}}, E^{\perp}_{\overline{a}})$ where the states sent by Bob will be measured. Using \eqref{eqn:local_optimal_non_ortho_fidelity_1to2} with the value of overlap $s = \cos\left(\pi/9\right)$, the optimal fidelity is $F_{\mathsf{L}} \approx 0.997$ and so the probability of Bob getting caught is at most $1\%$. Putting this together with Bob's guessing probability for $a$ gives his overall success probability of $77.5\%$.

In the end, this implies that Bob is able to successfully create a bias of $\epsilon \approx 0.775 - 0.5 = 0.275$.

\end{proof}

We also have the following corollary, for a general number of states ($n$) exchanged, which shows the protocol can be completely broken and Bob can enforce an arbitrary bias:
\begin{corrbox}
\begin{corollary}
    The probability of Bob successfully guessing $a$ over all $n$ copies has the property:
    \begin{equation} \label{eqn:bob_guess_prob_n_rounds_mayers_appendix}
        \lim\limits_{n\rightarrow \infty}{P}^{n}_{\mathrm{succ}, \mathcal{P}_1} = 1
    \end{equation}
\end{corollary}
\end{corrbox}

\begin{proof}
If Bob repeats Attack $1$ over all $n$ copies, he will guess $n$ different bits $\{a'_{i}\}_{i=1}^n$. He can then take a majority vote and announce $b$ such that $a^* \oplus b = 0$, where we denote $a^*$ as the bit he guesses in at least $\frac{n}{2} + 1$ of the rounds.

If $n$ is even, he may have guessed $a'$ to be $0$ and $1$ an equal number of times. In this case, the attack becomes indecisive and Bob is forced to guess at random. Hence we separate the success probability for even and odd $n$ as follows: 
\begin{equation}\label{eqn:mayers_attack_guess_probability_appendix}
    P^n_{\text{succ}, \mathcal{P}_1} = 
    \begin{cases}
    \sum\limits_{k=\frac{n+1}{2}}^{n} \binom{n}{k}(1 - P_{\mathrm{fail}})^k P_{\mathrm{fail}}^{n-k} &n  \text{ odd,} \\
     \sum\limits_{k=\frac{n}{2} + 1}^{n} \binom{n}{k}(1 - P_{\mathrm{fail}})^k P_{\mathrm{fail}}^{n-k} + \frac{1}{2}\binom{n}{{n}/{2}}(1 - P_{\mathrm{fail}})^{\frac{n}{2}} P_{\mathrm{fail}}^{\frac{n}{2}}  & n \text{ even} \\
  \end{cases} 
\end{equation}
By substituting the value of $P_{\mathrm{fail}}$ one can see that the function is uniformly increasing with $n$ so $\lim\limits_{n\rightarrow \infty}{P}^{n}_{\text{succ}, \mathcal{P}_1} = 1$.
\end{proof}

Although as Bob's success probability in guessing correctly increases with $n$, the probability of his cheating strategy getting detected by Alice will also increase. We also note that this strategy is independent of $k$, the number of different bits used during the protocol.

\subsection[\texorpdfstring{\color{black}}{} 4-state coin flipping protocol (\texorpdfstring{$\mathcal{P}_2$}{})]{4-state coin flipping protocol (\texorpdfstring{$\mathcal{P}_2$}{})} \label{ssec:aharonov_protocol_attack}

Another class of coin flipping protocols are those which require all the four states in~\eqref{eq:coinflip-st}. One such protocol was proposed by Aharonov \textit{et al.}~\cite{aharonov_quantum_2000}, where $\phi$ is set as $\frac{\pi}{8}$:
\begin{equation}\label{eqn:aharonov_coinflip_states}
    \ket{\phi_{x, a}} = 
    \begin{cases}
    \ket{{\frac{\pi}{8}}_{x,0}} = \cos\left( \frac{\pi}{8} \right)\ket{0} + (-1)^x\sin\left( \frac{\pi}{8} \right)\ket{1} \\
    \ket{{\frac{\pi}{8}}_{x,1}} =  \sin\left( \frac{\pi}{8} \right)\ket{0} + (-1)^{x \oplus 1}\cos\left( \frac{\pi}{8} \right)\ket{1}
  \end{cases} 
\end{equation}

In protocols of this form, Alice encodes her bit in `basis information' of the family of states. For instance, we can take $\{\ket{\phi_{0,0}}, \ket{\phi_{1, 0}}\}$ to encode the bit $a = 0$ and $\{\ket{\phi_{0, 1}}, \ket{\phi_{1,1}}\}$ to encode $a = 1$. The goal again is to produce a final `coin flip' $y = a\oplus b$, while ensuring that no party has biased the bit, $y$. A similar protocol has also been proposed using BB84 states~\cite{bennett_quantum_2014} where $\ket{\phi_{0,0}} := \ket{0}, \ket{\phi_{0,1}} := \ket{1}, \ket{\phi_{1,0}} := \ket{+}$ and $\ket{\phi_{1,1}} := \ket{-}$. In this case, the states (also some protocol steps) are different but the angle between them is the same as those in $\mathcal{P}_2$. A fault-tolerant version of $\mathcal{P}_2$ has also been proposed in~\cite{berlin_fair_2009}, which uses a generalised angle as in~\eqref{eq:coinflip-st}.

The protocol proceeds as follows. First Alice sends one of the states, $\ket{\phi_{x, a}}$ to Bob. Later, one of two things will happen. Either, Alice sends the bits ($x, a$) to Bob, who measures the qubit in the suitable basis to check if Alice was honest, \emph{or} Bob is asked to return the qubit $\ket{\phi_{x, a}}$ to Alice, who measures it and verifies if it is correct. Now, example cheating strategies for Alice involve incorrect preparation of $\ket{\phi_{x, a}}$ and giving Bob the wrong information about $(x, a)$, or for Bob in trying to determine the bits $(x, a)$ from $\ket{\phi_{x, a}}$ before Alice has revealed them classically. We again focus only on Bob's strategies here to use cloning arguments. We note that the information theoretic achievable bias of $\epsilon = 0.42$ proven in~\cite{aharonov_quantum_2000} applies only to Alice's strategy since she has greater control of the protocol (she prepares the original state). In general, a cloning based attack strategy by Bob will be able to achieve a lower bias, as we show. As above, Bob randomly selects his own bit $b$ and sends it to Alice. He then builds a QCM to clone all 4 states in \eqref{eqn:aharonov_coinflip_states}.

We next sketch the two cloning attacks on Bob's side of $\mathcal{P}_2$. Again, as with the protocol, $\mathcal{P}_1$, Bob can cheat using as much information as he gains about $a$ and again, once Bob has performed the cloning, his strategy boils down to the problem of state discrimination. In both attacks, Bob will use a (variational) state-dependent cloning machine.

\subsubsection[\texorpdfstring{\color{black}}{} Cloning attacks on \texorpdfstring{$\mathcal{P}_2$}{}]{Cloning attacks on \texorpdfstring{$\mathcal{P}_2$}{}}
\label{app_sssec:aharonov_cloning_attack}

In the first attack model (which we denote I - see \figref{fig:aharonov_1to2_cloning_fidelities_variational_plus_attack_models}(a) in \secref{ssec:results_state_dep_cloning}) Bob measures \emph{all} the qubits outputted from the cloner to try and guess $(x, a)$. As such, it is the \emph{global} fidelity that will be the relevant quantity. This strategy is useful with the first possible challenge in the protocol, where Bob is not required to send anything back to Alice. We discuss in \appref{app_sssec:aharonov_cloning_attack_I} how the use of cloning in this type of attack can also reduce resources for Bob from a general POVM to projective measurements in the state discrimination, which may be of independent interest. The main attack here boils down to Bob measuring the global output state from his QCM using the projectors, $\{\ketbra{v}{v}, \ketbra{v^\perp}{v^\perp}\}$, and from this measurement, guessing $a$. These projectors are constructed explicitly relative to the input states using the Neumark theorem~\cite{bae_quantum_2015}.

The second attack model (which we denote II, see \figref{fig:aharonov_1to2_cloning_fidelities_variational_plus_attack_models}(a) in \secref{ssec:results_state_dep_cloning}) is instead a \emph{local} attack and as such will depend on the optimal local fidelity. It may also be more relevant in the scenario where Bob is required to return a quantum state to Alice. We note that Bob could also apply a global attack in this scenario but we do not consider this possibility here in order to give two contrasting examples. In the below, we compute a bias assuming he does not return a state for Alice for simplicity and so the bias will be equivalent to his discrimination probability. The analysis could be tweaked to take a detection probability for Alice into account also. In this scenario, Bob again applies the QCM, but now he only uses one of the clones to perform state discrimination (given by the \textsf{Discriminator} in \figref{fig:aharonov_1to2_cloning_fidelities_variational_plus_attack_models}(a)).

\subsubsection[\texorpdfstring{\color{black}}{} Attack I on \texorpdfstring{$\mathcal{P}_2$}{}]{Attack I on \texorpdfstring{$\mathcal{P}_2$}{}}
\label{app_sssec:aharonov_cloning_attack_I}

For attack I, which is a $4$ state \emph{global} attack on $\mathcal{P}_2$, we have:
\begin{thmbox}
\begin{theorem}\label{thm:aharonov_attack_I_bias_probability_appendix}[Ideal cloning attack (I) bias on $\mathcal{P}_2$]~ \\
Using a cloning attack on the protocol, $\mathcal{P}_2$, (in attack model I) Bob can achieve a bias:
\begin{equation}
    \epsilon^{\mathrm{I}}_{\mathcal{P}_2, \mathrm{ideal}} \approx 0.35
\end{equation}
\end{theorem}

\end{thmbox}
We note first that this attack model (i.e. using cloning) can be considered a constructive way of implementing the optimal discrimination strategy of the states Alice is to send. In order to bias the bit, Bob needs to discriminate between the four pure states in \eqref{eq:coinflip-st} or equivalently between the ensembles encoding $a=\{0,1\}$, where the optimal discrimination is done via a set of POVM measurements.

However, by implementing a cloning based attack, we can simplify the discrimination.
This is because the symmetric state-dependent cloner (which is a unitary) has the interesting feature that for either case ($a=0$ or $a=1$), the cloner's output is a pure state in the $2$-qubit Hilbert space. As such, the states (after going through the QCM) can be optimally discriminated via a set of projective measurements $\{P_v, P_{v^{\perp}}\}$, rather than general POVMs (as would be the case if the QCM was not used). So, using $\VQC$ to obtain optimal cloning strategies also is a means to potentially reduce resources for quantum state discrimination also. Now, let us prove \thmref{thm:aharonov_attack_I_bias_probability_appendix}:

\begin{proof}
The attack involves the global output state of the cloning machine. For this attack we can use the fixed overlap $1 \rightarrow 2$ cloner with the global fidelity given by \eqref{eqn:optimal_global_non_ortho_state_fidelity}:
\begin{equation}\label{eqn:optimal_fidelity_g_1_2}
    F^{{\mathrm{FO}, \text{opt}}}_{\mathsf{G}}(1,2) = \frac{1}{2}\left( 1 + s^{3} + \sqrt{1-s^{2}}\sqrt{1-s^{4}} \right) \approx 0.983
\end{equation}
where $s=\sin(2\phi) = \cos(\frac{\pi}{4})$ for $\mathcal{P}_2$. Alternatively we can use the 4-state cloner which clones the two states with a fixed overlap plus their orthogonal set. For both of these cloners we are interested in the global state of the cloner which we denote as $\ket{\psi_{x, a}^{1\rightarrow 2}}$ for an input state $\ket{\phi_{x, a}}$.

In order for Bob to guess $a$ he must discriminate between $\ket{\phi_{0, 0}}$ (encoding $a=0$) and $\ket{\phi_{1, 1}}$ (encoding $a=1$) or alternatively the pair $\{\ket{\phi_{0, 1}}, \ket{\phi_{1, 0}}\}$. This is since the pairs $\{\ket{\phi_{0, 0}}, \ket{\phi_{0, 1}}\}$ are orthogonal and $\{\ket{\phi_{0, 0}}, \ket{\phi_{1, 0}}\}$ both encode $a=0$, so the only choice is to discriminate between $\ket{\phi_{0, 0}}$ and $\ket{\phi_{1, 1}}$. Due to the symmetry and without an ancilla, the cloner preserves the overlap between each pairs i.e. $\braket{\psi_{0, 0}^{1\rightarrow 2}}{\psi_{1, 1}^{1\rightarrow 2}} = \braket{\phi_{0,0}}{\phi_{1,1}} = s$ (we also have $\braket{\psi_{0, 1}^{1\rightarrow 2}}{\psi_{1, 0}^{1\rightarrow 2}} = s$).

\noindent Now we select the projective measurements $P_v=\ket{v}\bra{v}$ and $P_{v^{\perp}}=\ket{v^{\perp}}\bra{v^{\perp}}$ such that $\braket{v}{v^{\perp}} = 0$. One can show that the discrimination probability is optimal when $\ket{v}$ and $\ket{v^{\perp}}$ are symmetric with respect to the target states (illustrated in \figref{fig:aharonov_1to2_cloning_fidelities_variational_plus_attack_models}(a)) according to the Neumark theorem. From the figure, we have that $\braket{v}{v^{\perp}} = 0$ so $2\theta + 2\phi = \frac{\pi}{2} \Rightarrow \theta = \frac{\pi}{4} - \phi$. Finally, writing the cloner's states for $\{\ket{\psi_{0, 0}^{1\rightarrow 2}}, \ket{\psi_{1,1}^{1\rightarrow 2}}\}$ in the basis $\{\ket{v}, \ket{v^{\perp}}\}$ gives:
\begin{equation}\label{eqn:coin_flip_output_cloner_states_in_v}
    \begin{split}
    & \ket{\psi_{0, 0}^{1\rightarrow 2}} = \cos\left(
    \frac{\pi}{4} - \phi\right)\ket{v} + \sin\left(
    \frac{\pi}{4} - \phi\right)\ket{v^{\perp}},\\
    & \ket{\psi_{1, 1}^{1\rightarrow 2}} = \cos\left(\frac{\pi}{4} - \phi\right)\ket{v} - \sin\left(\frac{\pi}{4} - \phi\right)\ket{v^{\perp}}
    \end{split}
\end{equation}
where it can be checked that $\braket{\psi_{0, 0}^{1\rightarrow 2}}{\psi_{1, 1}^{1\rightarrow 2}} = \cos\left(\frac{\pi}{2} - 2\phi\right) = \sin\left(2\phi\right) = s$. Hence $\ket{v}$ and $\ket{v^{\perp}}$ can be explicitly derived. Note that these bases are also symmetric with respect to the other pair i.e $\{\ket{\psi_{0, 1}^{1\rightarrow 2}}, \ket{\psi_{1,0}^{1\rightarrow 2}}\}$. Finally, the success probability of this measurement is then given by:
\begin{equation}\label{eqn:coin_flip_global_disc_probability_appendix}
P^{\opt, \mathrm{I}}_{\mathrm{disc}, \mathcal{P}_2} = \frac{1}{2} + \frac{1}{2}\braket{\psi_{0, 0}^{1\rightarrow 2}}{\psi_{1, 1}^{1\rightarrow 2}} =   \frac{1}{2} + \frac{1}{2}\sin{2\phi} = 0.853
\end{equation}
which is the maximum cheating probability for Bob. From this, we derive the bias as:
\begin{equation}
  \epsilon^{\mathrm{I}}_{\mathcal{P}_2, \mathrm{ideal}} =  P^{\opt, \mathrm{I}}_{\mathrm{disc}, \mathcal{P}_2} -   \frac{1}{2} = 0.353
\end{equation}
which completes the proof.
\end{proof}

\subsubsection[\texorpdfstring{\color{black}}{} Attack II on \texorpdfstring{$\mathcal{P}_2$}{}]{Attack II on \texorpdfstring{$\mathcal{P}_2$}{}}
\label{app_sssec:aharonov_cloning_attack_II}

Finally, we consider a second attack model (attack II) on the protocol, $\mathcal{P}_2$, which is in the form of a `local' attack. Here, we further consider two scenarios:
\begin{enumerate}
    \item A cloning machine which is able to clone \emph{all} $4$ states: $\ket{\phi_{0, 0}}, \ket{\phi_{1, 1}}$ \emph{and} $\ket{\phi_{0, 1}}, \ket{\phi_{1, 0}}$,
    \item A cloning machine tailored to only the two states, $\ket{\phi_{0, 0}}$ and $\ket{\phi_{1, 1}}$ (which Bob needs to discriminate between).
\end{enumerate}

We focus on the former scenario, since it connects more cleanly with the $\VQC$ clone fidelities, but scenario 2 facilitates a more optimal attack (in the ideal situation).\\

\noindent \textbf{\underline{Scenario 1:}}

\vspace{2mm}
In this case, we can compute an exact discrimination probability, but it will result in a less optimal attack.
\begin{thmbox}
\begin{theorem}\label{thm:aharonov_4state_attack_II_bias_probability_appendix}[Ideal cloning attack (II) bias on $\mathcal{P}_2$ in scenario $1$.]~ \\
Using a cloning attack on the protocol, $\mathcal{P}_2$, (in attack model II with $4$ states) Bob can achieve a bias:
\begin{equation}\label{eqn:attack_4_state_aharonov_success_probability_exact_appendix}
    \epsilon^{\mathrm{II}}_{\mathcal{P}_2, \mathrm{ideal}} = 0.25
\end{equation}
\end{theorem}
\end{thmbox}
As a sketch proof, we explicitly evaluate the output states from an ideal cloner, and then compute the success probability via the trace distance. The full proof for this theorem is given in~\appref{app_ssec:proof_aharonov_attack_II}. \\

\noindent \textbf{\underline{Scenario 2:}}

\vspace{2mm}
Here, we give a bound on the success probabilities of Bob in terms of the local fidelities of the QCM where the cloning machine is only tailored to clone two fixed-overlap states. Here we rely on the fact that Bob can discriminate between the two ensembles of states (for $a=0$, $a=1$) with equal probabilities. 
\begin{thmbox}
\begin{theorem}\label{thm:aharonov_2state_attack_II_bias_probability_appendix}
The optimal discrimination probability for a cloning attack on the protocol, $P_2$, (in attack model II, with $2$ states) is:
\begin{equation} \label{eqn:attack_2_state_aharonov_success_probability_bound_appendix}
     0.619 \leq P^{\opt, \mathrm{II}}_{\mathrm{disc}, \mathcal{P}_2} \leq 0.823
\end{equation}
\end{theorem}
\end{thmbox}

\begin{proof}

For each of the input states, $\ket{\phi_{i,j}}$, in \eqref{eqn:aharonov_coinflip_states}, we denote  $\rho_{ij}^c$ to be a clone outputted from the QCM. Due to symmetry, we only need to consider one of the two output clones. We can now write the effective states for each encoding ($a=0, a=1$) as:
\begin{equation}\label{eqn:ensembles}
\rho_{(a=0)} := \frac{1}{2}(\rho_{00}^c + \rho_{10}^c), \qquad \qquad \rho_{(a=1)} := \frac{1}{2}(\rho_{01}^c + \rho_{11}^c)
\end{equation}
Dealing with these two states is sufficient since it can be shown that discriminating between these two density matrices, is equivalent to discriminating between the entire set of $4$ states in \eqref{eq:coinflip-st}.

Again, we use the discrimination probability from the Holevo-Helstrom bound:
\begin{equation}\label{eqn:opt-helstrom-clone}
 P^{\opt,  \mathrm{II}}_{\mathrm{disc}, \mathcal{P}_2} := P^{\opt}_{\mathrm{disc}}(\rho_{(a=0)},\rho_{(a=1)}) := \frac{1}{2} + \frac{1}{2}\text{d}_{\Tr}(\rho_{(a=0)},\rho_{(a=1)})
\end{equation}
Now, we have:
\begin{equation*}
\begin{split}
  \text{d}_{\Tr}(\rho_{(a=0)},\rho_{(a=1)}) & = \frac{1}{2}\left|\left|\rho_{(a=0)} - \rho_{(a=1)}\right|\right|_{\Tr} = \frac{1}{2}\left|\left|\frac{1}{2}(\rho_{00}^c - \rho_{11}^c) + \frac{1}{2}(\rho_{10}^c - \rho_{01}^c)\right|\right|_{\Tr} \\
  & \leq \frac{1}{4}\left|\left|(\rho_{00}^c - \rho_{11}^c)\right|\right|_{\Tr} + \left|\left|(\rho_{10}^c - \rho_{01}^c)\right|\right|_{\Tr} \\
  & \leq \frac{1}{2} \left[\text{d}_{\Tr}(\rho_{00}^c, \rho_{11}^c) + \text{d}_{\Tr}(\rho_{01}^c,\rho_{10}^c)\right]
\end{split}
\end{equation*}
\begin{equation}\label{eqn:opt-trace-distance}
\begin{split}
  \implies    P^{\opt}_{\mathrm{disc}}(\rho_{(a=0)},\rho_{(a=1)}) &\leq  \frac{1}{2} (P^{\opt}_{\mathrm{disc}}(\rho_{00}^c,\rho_{11}^c) + P^{\opt}_{\mathrm{disc}}(\rho_{01}^c,\rho_{10}^c))  \\
  &= P^{\opt}_{\mathrm{disc}}(\rho_{00}^c,\rho_{11}^c) 
\end{split}
\end{equation}
The last equality follows since for both ensembles, $\{\ket{\phi_{0,0}},\ket{\phi_{1,1}}\}$ and $\{\ket{\phi_{0,1}},\ket{\phi_{1,0}}\}$, their output clones have equal discrimination probability:
\begin{equation}
    P^{\opt}_{\mathrm{disc}}(\rho_{00}^c,\rho_{11}^c)  = P^{\opt}_{\mathrm{disc}}(\rho_{01}^c,\rho_{10}^c) 
\end{equation}
This is because the QCM is symmetric, and depends only on the overlap of the states (we have in both cases $\braket{\phi_{00}}{\phi_{11}} = \braket{\phi_{01}}{\phi_{10}} = \sin(2\phi)$).

Furthermore, since the cloning machine can only lower the discrimination probability between two states, we have:
\begin{equation*}
  P^{\opt}_{\mathrm{disc}}(\rho_{00}^c,\rho_{11}^c) \leq P^{\opt}_{\mathrm{disc}}(\rho_{00}^c,\ket{\phi_{1,1}}\bra{\phi_{1,1}}) =: \overline{P^{\opt}_{\mathrm{disc}}}
\end{equation*}
Now, using the relationship between fidelity and the trace distance (\eqref{eqn:fidelity_trace_dist_bound}), we have the bounds:
\begin{equation}\label{eqn:opt_disc_probability_helstrom}
\frac{1}{2}+\frac{1}{2}\left(1 - \sqrt{\bra{\phi_{1,1}}\rho^c_{00}\ket{\phi_{1,1}}}\right) \leq\overline{P^{\opt}_{\mathrm{disc}}} \leq\frac{1}{2} + \frac{1}{2}\sqrt{1 - \bra{\phi_{1,1}}\rho^c_{00}\ket{\phi_{1,1}}}
\end{equation}
By plugging in the observed density matrix for the output clone, we can find this discrimination probability.
As in the previous section, the output density matrix from the QCM for an output clone can be written as \eqref{eqn:reduced_local_state_mayers_attack_appendix}:
\begin{equation}
  \rho^c_{00} = \alpha\ket{\phi_{0,0}}\bra{\phi_{0,0}} + \beta \ket{\phi_{1,1}}\bra{\phi_{1,1}} +
\gamma (\ket{\phi_{0,0}}\bra{\phi_{1,1}}+ \ket{\phi_{1,1}}\bra{\phi_{0,0}}) 
\end{equation}
This state has a local fidelity, $F_{\Lbs} = \bra{\phi_{0,0}}\rho^c_{00}\ket{\phi_{0,0}} = \alpha + s^2\beta + s\gamma$. On the other hand, we have $F(\rho^c_{00}, \ketbra{\phi_{1, 1}}{\phi_{1, 1}}) = s^2\alpha + \beta + s\gamma$.

Combining these two, we then have:
\begin{equation}
F(\rho^c_{00}, \ketbra{\phi_{1, 1}}{\phi_{1, 1}}) = F_{\Lbs} + (s^2 - 1)(\alpha - \beta)
\end{equation}
Plugging in $F_{\Lbs}$ from \eqref{eqn:local_optimal_non_ortho_fidelity_1to2}, and $\alpha - \beta = \sqrt{\frac{1-s^2}{1-s^4}}$ (for an optimal state-dependent cloner), we get:
\begin{small}
    \begin{equation} \label{eqn:attack_2_state_aharonov_success_probability_bound_not_filled_appendix}
        \frac{1}{2} + \frac{1}{2}\left[1 - \sqrt{F_{\Lbs} + (s^2 - 1) \sqrt{\frac{1-s^2}{1-s^4}}}\right] \leq P^{\opt, \text{II}}_{\mathrm{disc}, \mathcal{P}_2} \leq \frac{1}{2} + \frac{1}{2}\sqrt{1 - F_{\Lbs} - (s^2 - 1) \sqrt{\frac{1-s^2}{1-s^4}}}
    \end{equation}
\end{small}
To complete the proof, we use $F_{\Lbs} \approx 0.989$ and $s = 1/\sqrt{2}$ which gives the numerical discrimination probabilities above.

\end{proof}

\newpage

\section[\texorpdfstring{\color{black}}{} \texorpdfstring{$\VarQlone$}{} numerics]{\texorpdfstring{$\VarQlone$ numerics}{}} \label{sec:cloning/numerics}
In preceding sections, we introduced the $\VQC$ algorithm and its relationship to cryptography at a high level. We also introduced (theoretical) cryptographic attacks on several protocols. Here, we combine these two aspects and present the corresponding numerical results.

Before diving back into the cryptographic applications, let us revisit some specifics of the $\VarQlone$ algorithm, in particular the $\Ansatze$ we use.

The first two options are \emph{fixed-structure} $\Ansatze$. We include one which is `problem-inspired' (based on the theoretical derivation) and one which is `problem-agnostic' (a hardware-efficient fixed-structure (HEFS)) using the nomenclature of~\secref{ssec:vqa_ansatzse}. We then generalise to the primary one we opt for, a \emph{variable-structure} $\Ansatz$. Recall, in the latter case, both the continuous parameters \emph{and} the gates in the $\Ansatz$ are optimised over (see~\secref{ssec:vqa_ansatzse}).

\subsection[\texorpdfstring{\color{black}}{} Fixed-structure \texorpdfstring{$\Ansatze$}{}]{Fixed-structure \texorpdfstring{$\Ansatze$}{}} \label{app_ssec:fixed_structure_ansatz}

To demonstrate the fixed-structure $\Ansatze$, we use $1\rightarrow 2$ phase-covariant cloning as the primary example (recall this is the relevant cloning family to attack the BB84 protocol), which requires cloning the following states:
\begin{equation} \label{eqn:x_y_plane_states_supp_numerics}
    \ket{\psi_{xy}(\eta)} = \frac{1}{\sqrt{2}}\left(\ket{0} + e^{i\eta}\ket{1}\right)
\end{equation}

\subsubsection[\texorpdfstring{\color{black}}{} Phase-covariant cloning with a fixed ideal \texorpdfstring{$\Ansatz$}{}]{Phase-covariant cloning with a fixed ideal \texorpdfstring{$\Ansatz$}{}} \label{app_ssec:pc_cloning_fixed_ansatz}

\begin{figure}
    \centering
        \includegraphics[width=\columnwidth,height=0.4\textwidth]{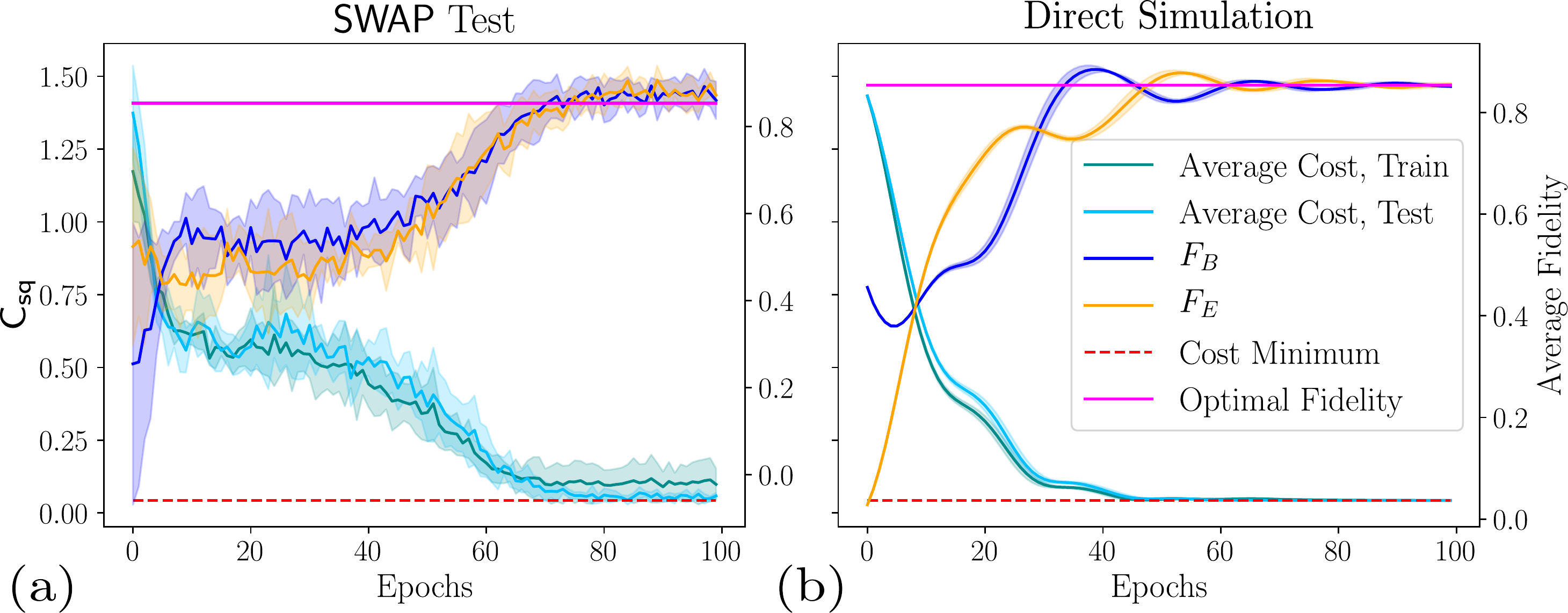}
    \caption[\color{black} Learning the parameters of the ideal cloning circuit.]{\textbf{Learning the parameters of the fixed circuit in \figref{fig:qubit_cloning_ideal_circ}.} We use $30$ random samples with an $80/20\%$ train/test split. To train, we use the analytic gradient, \eqref{eqn:analytic_squared_grad_mton}, and the Adam optimiser with a batch size of $10$ states and an initial learning rate of $0.05$. In all cases, the error bars show mean and standard deviation over $5$ independent training runs. Figure shows the results when the fidelity is computed using (a) the $\SWAP$ test (with $50$ measurement shots) and (b) using direct density matrix simulation. In both cases, we plot the average (squared) cost (\eqref{eqn:squared_local_cost_mton}) on the train and test set, and also the average fidelities of the output states of Bob, $F_B$ [\crule[Blue]{0.2cm}{0.2cm}], and Eve, $F_E$ [\crule[YellowOrange]{0.2cm}{0.2cm}], corresponding to this cost function value. Also plotted are the theoretical optimal fidelities ([\crule[Magenta]{0.2cm}{0.2cm}], magenta solid line) for this family of states, and the corresponding cost minimum ([\crule[Red]{0.2cm}{0.2cm}], red dashed line).}
    \label{fig:fixed_structure_phase_cov_cloning_swap_vs_no_swap}
\end{figure}

Let us begin with a problem-inspired option. Recall, the ideal circuit for performing phase-covariant cloning is given by \figref{fig:qubit_cloning_ideal_circ}. Here, we learn the parameters of this fixed circuit. This gives us the opportunity to illustrate the effect of measurement noise in using the $\SWAP$ test. The results of this can be seen in \figref{fig:fixed_structure_phase_cov_cloning_swap_vs_no_swap}. We compare the $\SWAP$ test in~\figref{fig:fixed_structure_phase_cov_cloning_swap_vs_no_swap}(a) to `direct simulation' in~\figref{fig:fixed_structure_phase_cov_cloning_swap_vs_no_swap} (b) to compute the fidelities. Direct simulation here refers to the explicit extraction of a classical description of the qubit density matrices. We do this using quantum state tomography~\cite{dariano_quantum_2003} with the \computerfont{forest-benchmarking} library~\cite{gulshen_forest_2019}. Note, that this intermediate step via tomography is not necessary in general as with sufficiently low hardware noise, the fidelity could be directly extracted via the an overlap test (i.e., the $\SWAP$ test). The effect of measurement noise can be clearly seen in the latter case. 

We note in the main text that we do not use the $\SWAP$ test when running the experiments on the \computerfont{Aspen} QPU. This is because the test fails to output the fidelity since both states to compare will be mixed due to device noise.
However, this essentially reproduces the findings of~\cite{jasek_experimental_2019} in a slightly different scenario. Furthermore, this was only possible because we had prior knowledge of an optimal circuit to implement the cloning transformation from~\cite{buzek_quantum_1997, fan_quantum_2014}. Of course, in generality this information is not available, which precludes the use of (problem-inspired) fixed-structure $\Ansatze$.

\subsubsection[\texorpdfstring{\color{black}}{} Phase-covariant cloning with a hardware-efficient fixed \texorpdfstring{$\Ansatz$}{}]{Phase-covariant cloning with a hardware-efficient fixed \texorpdfstring{$\Ansatz$}{}} \label{sssec:pc_cloning_fixed_hardware_efficient_ansatz}
\begin{figure}
    \centering
    \includegraphics[width=0.7\columnwidth]{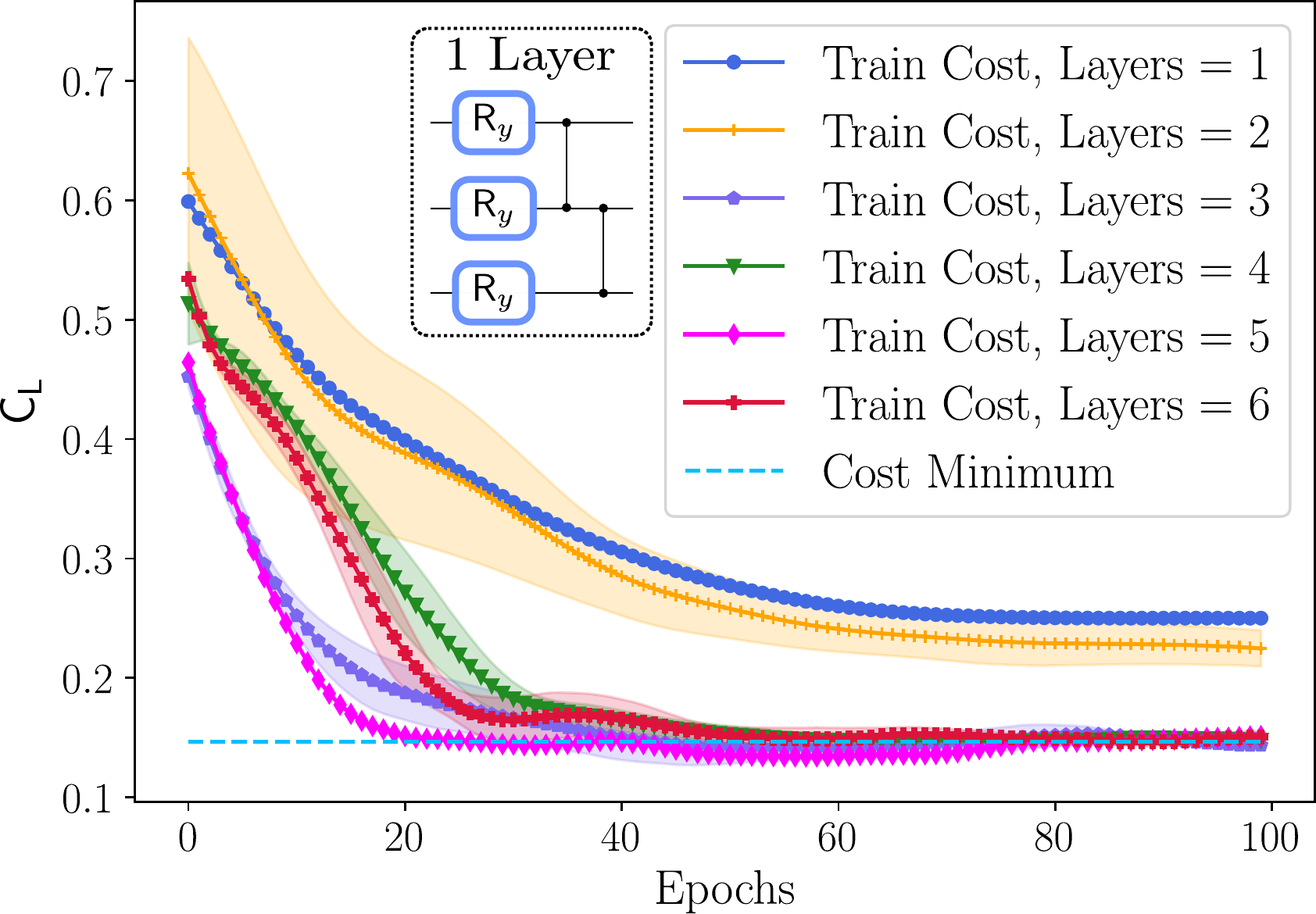}
    \caption[\color{black} Local $\VQC$ cost minimised on a training set of random phase-covariant states.]{\textbf{Local cost, $\Cbs_{\Lbs}$ minimised on a training set of $24$ random phase-covariant states.} We plot layers $K\in [1, \dots 6]$ of the hardware-efficient $\Ansatz$ shown in the inset. Fastest convergence is observed for $K=5$ but $K=3$ is sufficient to achieve a minimal cost value, which is the same number of entangling gates as in \figref{fig:qubit_cloning_ideal_circ}. Error bars shown mean and standard deviation over $5$ independent training runs.}
    \label{fig:hardware_efficient_ansatz_phase_cov_cloning}
\end{figure}

Let us now test the problem-agnostic hardware-efficient fixed-structure (HEFS) $\Ansatz$ for the sample problem as in the previous section. Here, we introduce a number of layers in the $\Ansatz$, $K$, in which each layer has a fixed-structure. For simplicity, we choose each layer to have parametrised single qubit rotations, $\RY(\theta)$, and nearest neighbour $\CZ$ gates. We deal again with $1\rightarrow 2$ cloning, so we use $3$ qubits and therefore we have $2$ $\CZ$ gates per layer. We show the results for $K=1$ layer to $K=6$ layers in \figref{fig:hardware_efficient_ansatz_phase_cov_cloning}. Not surprisingly, we observe convergence to the minimum as the number of layers increases, saturating at $K=3$.\\

\noindent \textbf{Barren Plateaus}\\

\noindent Furthermore, we can examine $\VQC$ for the existence of barren plateaus in this scenario. Recall from~\secref{ssec:vqa_cost_functions}, a local $\VQA$ cost can be written as:
\begin{align}
    \Cbs_{\Lbs} &= \kappa\Tr\left[\mathsf{O}_{\Lbs}U(\paramtheta)\rho U(\paramtheta)^{\dagger}\right],\\
    \mathsf{O}_{\Lbs} &= c_0\mathds{1} + \sum_{j}c_j\mathsf{O}_j
\end{align}
Note that taking $c_0 = 1, c_j = -1/N~\forall j$ and $\kappa = 1$ recovers the specific form of our cost, \eqref{eqn:local_cost_full}. We will prove that this cost does not exhibit barren plateaus for a sufficiently shallow alternating layered $\Ansatz$, i.e. $U(\paramtheta)$ contains blocks, $W$, acting on alternating pairs of qubits~\cite{cerezo_cost_2021}. To do so, we first recall the following Theorem from~\cite{cerezo_cost_2021}:
\begin{thmbox}
\begin{theorem}[Adapted from Theorem $2$ in~\cite{cerezo_cost_2021}]
Consider a trainable parameter, $\theta^l$ in a block, $W$ of an alternating layered $\Ansatz$ (denoted $U(\paramtheta)$). Let $\text{Var}[\partial_l \Cbs]$ be the variance of an $m$-local cost function, $\Cbs$ with respect to $\theta^l$. If each block in $U(\paramtheta)$ forms a local $2$-design, then $\text{Var}[\partial_l \Cbs]$ is lower bounded by:
\begin{align}
    G_N(K, k) &\leq \text{Var}[\partial_l \Cbs]   \\
     G_N(K, k) &= \frac{2^{m(k+1) - 1}}{(2^{2m} - 1)^2(2^m+1)^{K+k}} \sum\limits_{j \in j_{\mathcal{L}}} \sum\limits_{\substack{(p, p')\in p_{\mathcal{L}_B} \\ p' \geq p}}\chi_j(p, p', \mathsf{O}_j) \\
     \chi_j(p, p', \mathsf{O}_j) &:= c_j^2 \text{d}_{\HS}\left(\rho_{p, p'}, \frac{\Tr(\rho_{p, p'})\mathds{1}}{d_{\rho_{(p, p')}}}\right)\text{d}_{\HS}\left(\mathsf{O}_j, \frac{\Tr(\mathsf{O}_j)\mathds{1}}{d_{\mathsf{O}_j}}\right)
\end{align}
$j_\mathcal{L}$ are the set of $j$ indices in the forward light-cone\footnote{The forward light-cone is the subset of qubits which are causally connected (in the future) to the qubit(s) on which the block $W$ is applied.} $\mathcal{L}_B$ of the block $W$ and $\rho_{p, p'}$ is the partial trace of the input state, $\rho$, down to the subsystems $S_p, S_{p+1}, \dots, S_{p'}$. $d_M$ denotes the dimension of a matrix $M$ and $\text{d}_{\HS}$ denotes the Hilbert-Schmidt distance from~\eqref{eqn:hs_distance_defn}.
\end{theorem}
\end{thmbox}
$S_p$ in the above represents the qubit subsystem in which $W$ acts. Firstly, the operators $\mathsf{O}_j$ are all single qubit projectors ($m=1$ local), $\ketbra{\psi}{\psi}$, so we have:
\begin{align}
    d_{\HS}\left(\mathsf{O}_j, \Tr(\mathsf{O}_j)\frac{\mathds{1}}{d}\right) &=  \text{d}_{\HS}\left(\ketbra{\psi}{\psi}, \Tr\left[\ketbra{\psi}{\psi}\right]\frac{\mathds{1}}{2}\right) \\
    &=\sqrt{\Tr\left[\left(\ketbra{\psi}{\psi} - \frac{\mathds{1}}{2} \right)\left(\ketbra{\psi}{\psi} - \frac{\mathds{1}}{2}\right)^{\dagger}\right]}\\
    &= \sqrt{\Tr\left[\ketbra{\psi}{\psi} - \frac{\ketbra{\psi}{\psi}}{2} - \frac{\ketbra{\psi}{\psi}}{2} + \frac{\mathds{1}}{4}\right]} = \sqrt{\Tr\left(\frac{\mathds{1}}{4}\right)} = \frac{1}{\sqrt{2}}
\end{align}
So $G(K, k)$ simplifies to:
\begin{equation}
     G_N(K, k) = \frac{2^{k}}{3^{K+k+2}\sqrt{2}N^2} \sum\limits_{j \in j_{\mathcal{L}}} \sum\limits_{\substack{(p, p')\in p_{\mathcal{L}_B} \\ p' \geq p}} \text{d}_{\mathsf{HS}}\left(\rho_{p, p'}, \frac{\mathds{1}}{d_{\rho_{(p, p')}}}\right)
\end{equation}
If we now define $S_{P}$ to be the subsystems from $p$ to $p'$, the reduced state of $\rho$ in $S_P$ will be one of either $\ketbra{\psi}{\psi}^{\otimes |P|}$, $\ketbra{0}{0}^{\otimes |P|}$ or $\ketbra{\psi}{\psi}^{\otimes q}\ketbra{0}{0}^{\otimes |P|-q}$ for some $q<|P|$ where we denote $|P|$ to be the number of qubits in the reduced subsystem $S_P$. Since these are all pure states, we can compute $\text{d}_{\mathsf{HS}}(\rho_{p, p'}, \mathds{1}/\text{d}_{\rho_{(p, p')}}) = \sqrt{1-1/d_{\rho_{(p, p')}}}$. Lower bounding the sum over $j$ by $1$ and $\sqrt{1-1/d_{\rho_{(p, p')}}}$ by $1/\sqrt{2}$ ($d_{\rho_{(p, p')}}$ is at least $2$) gives:
\begin{equation}
     \frac{2^{k}}{3^{K+k+2}2N^2} \leq G_N(K, k)
\end{equation}
Finally, by choosing $K \in \mathcal{O}\left(\log(N)\right)$, we have that $k, K+k \in \mathcal{O}(\log(N))$ and so $G_n(K, k) \in \Omega\left(1/\poly(N)\right)$.
Since we have that if $G(K, k)$ vanishes no faster than $\Omega\left(1/\poly(N)\right)$, then so does the variance of the gradient and so will not require exponential resources to estimate. As a result, we can formalise the following corollary:
\begin{corrbox}
\begin{corollary}\label{corr:barren_plateau_vqc_local_cost}[Absence of barren plateaus in local cost]~ \\
Given the local $\VQC$ cost function, $\Cbs_{\Lbs}$ (\eqref{eqn:local_cost_full}) in $M\rightarrow N$ cloning, and a hardware-efficient fixed-structure $\Ansatz$, $U(\paramtheta)$, made up of alternating blocks, $W$, with a depth $\mathcal{O}(\log(N))$, where each block forms a local 2-design. Then the variance of the gradient of $\Cbs_{\Lbs}$ with respect to a parameter, $\theta_l$ can be lower bounded as:
 \begin{equation}
     G_N := \min(G_N(K, k)) \leq \text{Var}[\partial_l \Cbs],\qquad G_N(K, k) \in \Omega\left(\frac{1}{\text{poly}(N)}\right)
 \end{equation}
\end{corollary}
\end{corrbox}
One final thing to note, is that the $\Ansatz$ we choose in \figref{fig:hardware_efficient_ansatz_phase_cov_cloning}, does not form an exact local 2-design, but the same $\Ansatz$ is used in~\cite{cerezo_cost_2021}) and is sufficient to exhibit a cost function dependent barren plateau.

\newpage

\subsection[\texorpdfstring{\color{black}}{} Variable-structure \texorpdfstring{$\Ansatze$}{}]{Variable-structure \texorpdfstring{$\Ansatze$}{}} \label{app_ssec:variable_structure_ansatz}

For our variable-structure $\Ansatz$, we adopt the approach of~\cite{cincio_learning_2018} which fixes the length, $l$, of the circuit sequence to be used,contains parametrised single qubit gates, and un-parametrised entangling gates, which we chose to be $\CZ$ for simplicity. For example, with a three qubit chip, we have an example gatepool:
\begin{multline} \label{eqn:phase_covariant_cloning_gateset_appendix}
    \mathcal{G} = \left\{\right. \mathsf{R}^0_{z}(\theta), \mathsf{R}^1_{z}(\theta), \mathsf{R}^2_{z}(\theta),
    \mathsf{R}^0_{x}(\theta), \mathsf{R}^1_{x}(\theta), \mathsf{R}^2_{x}(\theta), \\
    \mathsf{R}^0_{y}(\theta), \mathsf{R}^1_{y}(\theta), \mathsf{R}^2_{y}(\theta), 
    \CZ_{0, 1}, \CZ_{1, 2}, \CZ_{0, 2}\left.\right\}
\end{multline}
We use the $\CZ$ gate as the entangler for two reasons. The first is that $\CZ$ is a native entangling gate on the Rigetti hardware. The second is that it simplifies our problem slightly, since it is symmetric on the control and target qubit, we do not need to worry about the ordering of the qubits: $\CZ_{i, j} = \CZ_{j, i}$.  The unitary to be learned is given by: 
\begin{align}\label{eqn:structure_learning_unitary}
U_{\boldsymbol{g}}(\paramtheta) = U_{g_1}(\theta_1)U_{g_2}(\theta_2) \dots U_{g_l}(\theta_l)
\end{align}
where each gate is from the above set $\mathcal{G}$. The sequence, $\boldsymbol{g} := [g_1, \dots, g_l]$, in~\eqref{eqn:variable_structure_ansatz_optimisation_problem} and~\eqref{eqn:structure_learning_unitary} above, corresponds to the indices of the gates in an ordered version of $\mathcal{G}$. So using $\mathcal{G}$ in \eqref{eqn:phase_covariant_cloning_gateset_appendix} as an example, $\boldsymbol{g} = [0, 6, 3, 2, 10]$ would give the unitary:
\begin{align}\label{eqn:structure_learning_unitary_example}
U_{\boldsymbol{g}}(\paramtheta) =  \mathsf{R}^0_{z}(\theta_1)\mathsf{R}^1_{y}(\theta_2)\mathsf{R}^0_{x}(\theta_3) \mathsf{R}^2_{z}(\theta_4) \CZ_{0, 1}
\end{align}
and $\paramtheta := [\theta_1, \theta_2, \theta_3, \theta_4, 0]$.
 
At the beginning of the procedure, the gate sequence is chosen randomly (, i.e. a random sequence for the gate indices, $\boldsymbol{g}$), and also the parameters ($\paramtheta$) therein\footnote{If some information is known about the problem beforehand, this could be used to initialise the sequence and improve performance.}.

The optimisation procedure proceeds over a number of \computerfont{epochs} and \computerfont{iterations}. In each \computerfont{iteration}, $\boldsymbol{g}$ is perturbed by altering $d$ gates, $\boldsymbol{g}^{\computerfont{\text{iter}}} \rightarrow \boldsymbol{g}^{\computerfont{\text{iter}} +1}$, so $\boldsymbol{g}^{\computerfont{\text{iter}} +1}$ has at most $d$ gates different from $\boldsymbol{g}^{\computerfont{\text{iter}}}$. The probability of changing $d$ gates is given by $1/2^d$, and the probability of doing nothing (i.e. $\boldsymbol{g}^{\computerfont{\text{iter}}} = \boldsymbol{g}^{\computerfont{\text{iter}} + 1} $) is:

\begin{equation}
    \Pr(d=0) = 1 - \sum_{d=1}^l\frac{1}{2^d} = 2 - \frac{1 - \frac{1}{2^l}}{ 1- \frac{1}{2} } - \frac{1}{2^l}
\end{equation}
The \computerfont{epochs} correspond to optimisation of the parameters $\paramtheta$ using the Adam optimiser (\eqref{eqn:adam_update_rule}). We typically set the maximum number of \computerfont{epochs} to be $100$ and \computerfont{iterations} to be $50$ in all this work. After each \computerfont{iteration}, the best cost, $\Cbs^{\text{best}}_t$ for a chosen cost - either the local, \eqref{eqn:local_cost_full} ($t = \Lbs$), the global, \eqref{eqn:global_cost_full} ($t = \Gbs$), squared, \eqref{eqn:squared_local_cost_mton} ($t = \sq$) or some other choice, is updated, if that \computerfont{iteration} has found a circuit with a lower cost. As in~\cite{cincio_learning_2018}, we repeatedly compress the sequence by removing redundant gates (e.g. combining $U_{g_i}(\theta_i)$ and $U_{g_{i+1}}(\theta_{i+1})$ if $g_i = g_i+1$), and adding random gates to keep the sequence length fixed at $g_l$.

\begin{figure}
    \centering
        \includegraphics[width=0.95\columnwidth]{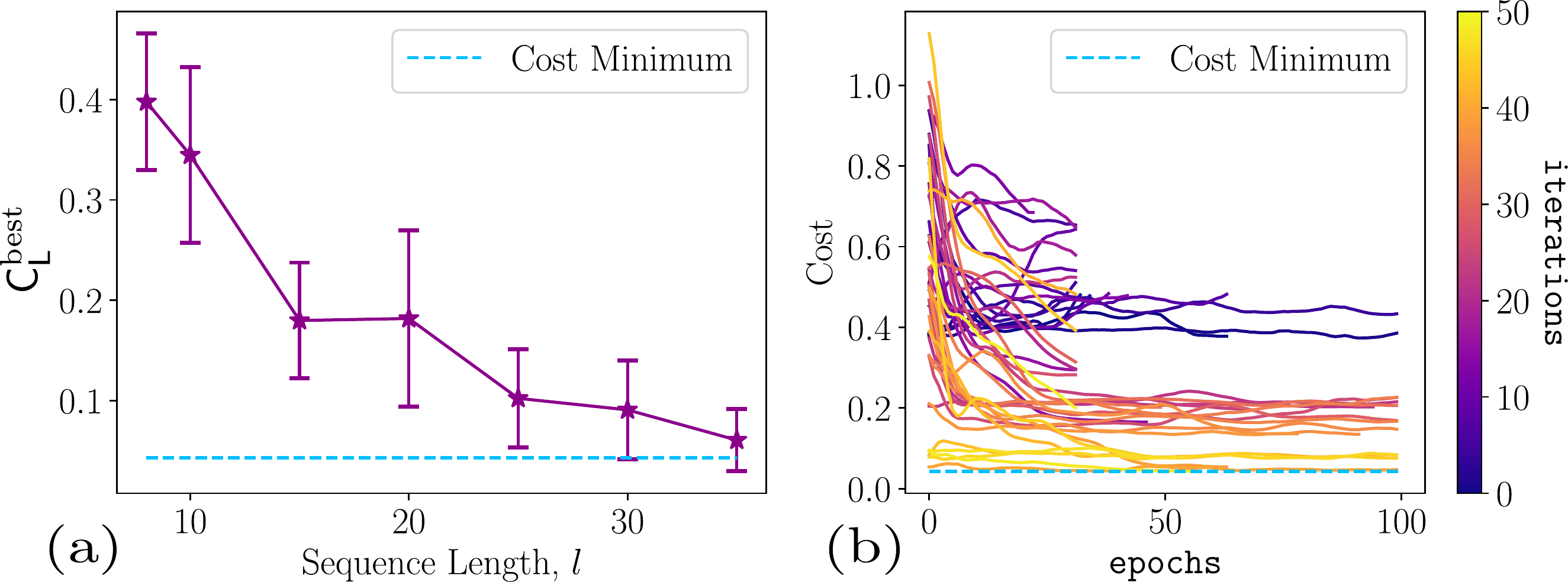}
    \caption[\color{black} Variable-structure $\Ansatz$ details for $\VQC$.]{\textbf{Variable-structure $\Ansatz$ details for $\VQC$.} (a) $\Cbs^{\text{best}}_{\Lbs}$ as a function of sequence length, $l$ in achieving the same task as \figref{fig:qubit_cloning_ideal_circ}, where the Bob and Eve's clones appear in qubits $1$ and $2$. As $l$ increases, the number of runs which successfully approach the theoretical minimum increases. Error bars shown mean and standard deviations for the minimum costs achieved over $20$ independent runs with each sequence length. (b) Cost achieved for 50 \computerfont{iterations} of the structure learning protocols, using a sequence length of $l=35$. Recall an \computerfont{iteration} is an update to the circuit \emph{structure}, whereas an \computerfont{epoch} is an update to the circuit \emph{parameters}. Each line (\computerfont{iteration}) corresponds to a slightly different circuit structure, $\boldsymbol{g}$. Early \computerfont{iterations} (darker lines) are not able to find the minimum, but eventually, a circuit is found which has this capacity. For each $\boldsymbol{g}$, $\paramtheta$ is trained for $100$ epochs of gradient descent, using the Adam optimiser. If an \computerfont{iteration} has not converged close enough to $\Cbs^{\text{best}}_{\Lbs}$ by $30$ epochs, the \computerfont{iteration} is ended.}
    \label{fig:all_runs_structure_learning}
\end{figure}

\figref{fig:all_runs_structure_learning} illustrates some results from this protocol. We find that with an increasing sequence length, the procedure is more likely to find circuits which achieve the minimum cost, and is able to first do so with a circuit with between $25$-$30$ gates from the above gateset in \eqref{eqn:phase_covariant_cloning_gateset}. We also plot the results achieved in a particular run of the protocol in \figref{fig:all_runs_structure_learning}(b). Here, an `\computerfont{iteration}' is an update to the circuit \emph{structure}, whereas an `\computerfont{epoch}' is an update to the circuit \emph{parameters} (as in previous chapters).  As the circuit learns, it is able to subsequently lower $\Cbs^{\text{best}}_t$, until it eventually finds a circuit capable of achieving the optimal cost for the problem.
%


\subsubsection[\texorpdfstring{\color{black}}{} Phase-covariant cloning]{Phase-covariant cloning}
\begin{figure}
\centering
    \includegraphics[width=\columnwidth, height=0.4\columnwidth]{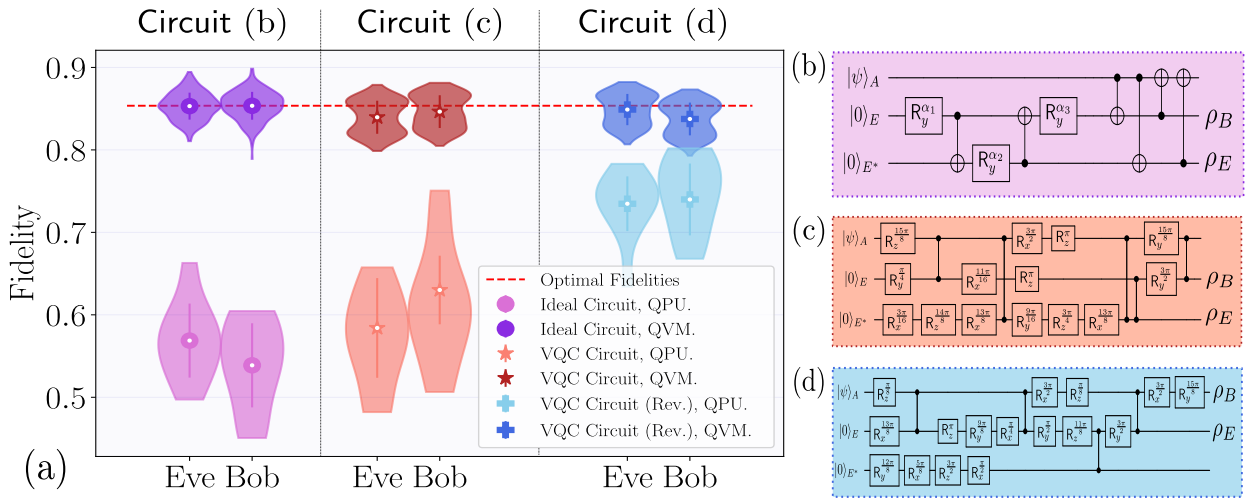}
    \caption[\color{black} Variational quantum cloning implemented on phase-covariant states using three qubits.]{\textbf{Variational quantum cloning implemented on phase-covariant states using three qubits of the Rigetti \computerfont{Aspen-8} chip (QPU), plus simulated results (QVM).} Violin plots in (a) show the cloning fidelities, for Bob and Eve, found using each of the circuits shown in (b)--(d) respectively. Shown in red is the maximal possible fidelity for this problem. (b) is the ideal circuit with clones appearing in registers $2$ and $3$ [\crule[Purple]{0.2cm}{0.2cm}, \crule[CarnationPink]{0.2cm}{0.2cm}]. (c) shows the structure-learned circuit for the same scenario, using one less entangling gate [\crule[BrickRed]{0.2cm}{0.2cm}, \crule[Peach]{0.2cm}{0.2cm}]. (d) demonstrates the effect of allowing clones to appear in registers $1$ and $2$ [\crule[Blue]{0.2cm}{0.2cm}, \crule[Cyan]{0.2cm}{0.2cm}]. In the latter case, only four (nearest-neighbour) entangling gates are used, demonstrating a significant boost in performance on the  QPU. In all cases the fidelity is extracted explicitly using quantum state tomography, as discussed in the text.}
    \label{fig:learned_vs_ideal_circ_on_hw_main_text}
\end{figure}

Now, we return to phase-covariant cloning, and implement the procedure detailed above\footnote{For all results shown using a variable-structure $\Ansatz$, we use the \computerfont{forest-benchmarking} library~\cite{gulshen_forest_2019} to reconstruct the output density matrix (and from this quantity then compute the overlaps and fidelities required) in order to mitigate the effect of quantum noise}. We consider this to be the proof of principle implementation of our methods. The results of this can be seen in 
\figref{fig:learned_vs_ideal_circ_on_hw_main_text}. Let us begin by describing some problem parameters. Firstly, we allow $3$ qubits ($2$ output clones plus $1$ ancilla) in the circuit. Next, we give the $\VarQlone$ the following fully connected (FC) gateset pool for this problem (indices represent qubits of the \computerfont{Aspen-8} sublattice):
\begin{multline} \label{eqn:phase_covariant_cloning_gateset}
    \mathcal{G}_{\text{PC}} = \left\{\right. \mathsf{R}^2_{z}(\theta), \mathsf{R}^3_{z}(\theta), \mathsf{R}^4_{z}(\theta),
    \mathsf{R}^2_{x}(\theta), \mathsf{R}^3_{x}(\theta), \mathsf{R}^4_{x}(\theta), \\
    \mathsf{R}^2_{y}(\theta), \mathsf{R}^3_{y}(\theta), \mathsf{R}^4_{y}(\theta), 
    \CZ_{2, 3}, \CZ_{3, 4}, \CZ_{2, 4}\left.\right\}
\end{multline}

Let us now discuss in greater detail the observations which can be drawn from \figref{fig:learned_vs_ideal_circ_on_hw_main_text}. Here, \figref{fig:learned_vs_ideal_circ_on_hw_main_text}(c, d) illustrates some candidate cloning circuits learned by $\VQC$, compared to the optimal `analytic' circuit (the same one from \figref{fig:qubit_cloning_ideal_circ}) given in~\cite{buzek_quantum_1997, fan_quantum_2001, fan_quantum_2014} (\figref{fig:learned_vs_ideal_circ_on_hw_main_text}(b)). 

We being by noting that all three circuits approximately saturate the optimal bound for phase-covariant cloning ($F_{\Lbs} = 0.85$, which is plotted as a red dotted line in \figref{fig:learned_vs_ideal_circ_on_hw_main_text}) when \emph{simulated} (i.e. without quantum noise). 

Next, we move to performance on the actual hardware, the \computerfont{Aspen-8} chip. We first notice that the ideal circuit in \figref{fig:learned_vs_ideal_circ_on_hw_main_text}(b) suffers a degradation in performance when implemented on the QPU since it requires $6$ entangling gates to transfer the information across the circuit. Furthermore, since the \computerfont{Aspen-8} chip does not have any $3$ qubit loops in its topology, it is necessary for the compiler to insert $\SWAP$ gates, further reducing the quality of the clones.

The two examples for our learned circuits \figref{fig:learned_vs_ideal_circ_on_hw_main_text}(c, d) are trained under two different restrictions. First, to ensure a fair comparison to the ideal circuit, we begin by forcing the qubit clones to appear in registers $2$ and $3$ of the circuit, (demonstrated in \figref{fig:learned_vs_ideal_circ_on_hw_main_text}(c)) exactly as in \figref{fig:learned_vs_ideal_circ_on_hw_main_text}(b). Secondly, we allow the clones to appear instead in registers $1$ and $2$ (demonstrated in \figref{fig:learned_vs_ideal_circ_on_hw_main_text}(d) - The circuit labelled `Rev.' (`Reverse').)  The ability to make such a subtle change clearly demonstrates the advantage of our flexible approach. We notice that the restriction imposed in \figref{fig:learned_vs_ideal_circ_on_hw_main_text}(c) results in only a modest performance improvement over the ideal. However, by allowing the clones to  appear in registers $1$ and $2$, $\VQC$ is able to find much more conservative circuits, having fewer entangling gates, and are directly implementable on a linear topology. This gives a significant improvement in the cloning fidelities, of about $15\%$ when the circuit is run on the \computerfont{Aspen-8} QPU.

Now, let us return to discussing phase-covariant cloning through the lens of an adversary eavesdropping on the BB84 protocol. We specifically analyse the performance of one of these $\VQC$-learned circuits (Circuit(c)) in such an attack. We will do so by computing the corresponding critical error rate, $D_\text{crit}$, via the expression for the key rate, $R$, in~\eqref{eqn:qkd-key-rate} (recall the discussion in~\secref{sec:qkd_and_cloning_attacks}) for the BB84 protocol protocol run in $\mathsf{X}$-$\mathsf{Y}$ Pauli basis.

We compute this for the circuit in \figref{fig:learned_vs_ideal_circ_on_hw_main_text}(c) only, since while the circuit in \figref{fig:learned_vs_ideal_circ_on_hw_main_text}(d) achieves higher fidelities on the \computerfont{Aspen} hardware, it does not actually make use of the ancillary qubit (the sequence of gates acting on it approximately resolve to the identity).

Now, we compute the resulting mixed states outputted over all input states to the cloning machine, for each basis state: $\{\ket{+}, \ket{-}, \ket{+\mathrm{i}}, \ket{-\mathrm{i}}\}$ so $\rho_{E}$ in \eqref{eqn:qkd-holevo-quantity} is given by:
\begin{equation}
    \rho_{E} := \frac{1}{4}(\rho_{E}^{+}+ \rho_{E}^{-} + \rho_{E}^{+\mathrm{i}} + \rho_{E}^{-\mathrm{i}})
\end{equation}
Similarly,  $\rho_{E}^0, \rho_{E}^1$ in \eqref{eqn:qkd-holevo-quantity} are the mixed states encoding the symbol $0$ (which have input $\ket{+}, \ket{+\mathrm{i}}$) and the symbol $1$ (which have input $\ket{-}, \ket{-\mathrm{i}}$), so are given by:
\begin{equation}
        \rho^0_{E} := \frac{1}{2}(\rho_{E}^{+}+  \rho_{E}^{+\mathrm{i}}) \qquad  \rho^1_{E} := \frac{1}{2}(\rho_{E}^{-}+  \rho_{E}^{-\mathrm{i}})
\end{equation}
Calculating the minimum Holevo quantity (denoted by $\chi_{\min}$) for the above density matrices outputted by the circuit in \figref{fig:learned_vs_ideal_circ_on_hw_main_text}(c) numerically gives the following:
\begin{equation} \label{eqn:qkd-ourcircuit-error-rate}
\begin{split}
          1 - H(D_{\text{crit}}) - \chi_{\min} &= 0 \\
          \Rightarrow  1 - \chi_{\min} + (D_{\text{crit}} \log_2{D_{\text{crit}}} + (1-D_{\text{crit}})\log_2{(1-D_{\text{crit}})}) &= 0 \\
         \Rightarrow D_{\text{crit}} &= 15.8 \%
\end{split}
\end{equation}
which is very close to the optimal bound for the individual attack. Nevertheless as pointed out in ~\cite{scarani_quantum_2005, ferenczi_symmetries_2012}, the same bound can be reached by a collective
attack (where Eve defers any measurements until the end of the reconciliation phase, and applies a general strategy to all collected states) where the individual quantum operations are still given by the optimal phase-covariant cloner. As such, the $\VQC$ learned attack can almost saturate the optimal collective bound as well.

Finally, one may observe that Circuit (d) in \figref{fig:learned_vs_ideal_circ_on_hw_main_text} achieves a higher still fidelity on hardware, but it does so by not using the ancilla to reduce the circuit depth. As such, it is a better circuit for purely performing cloning than either (c) or (d), but does not provide an optimal attack for BB84~\cite{scarani_quantum_2005}.\\

\noindent \textbf{Local versus Global Fidelities}

\vspace{2mm}
As a final remark on this experiment, we can investigate the difference between the global and local fidelities achieved by the circuits $\VQC$ (i.e. in \figref{fig:learned_vs_ideal_circ_on_hw_main_text}(c)) finds, versus the ideal one (shown in \figref{fig:learned_vs_ideal_circ_on_hw_main_text}(b)). Recall that in \secref{sssec:global_v_local_faithfulness}, we showed that the `ideal' circuit achieves both the optimal local and global fidelities for this problem:
\begin{equation} \label{eqn:ideal_circuit_local_global_fidelities}
    \figref{fig:learned_vs_ideal_circ_on_hw_main_text}(b) \implies
    \begin{cases}
    F_{\mathrm{B}}^{(b)}  = F_{\mathrm{E}}^{(b)}  = F_{\mathsf{L}}^{\opt} = \frac{1}{2}\left(1 + \frac{1}{\sqrt{2}}\right) \approx 0.853\\
    F_{\mathsf{G}}^{(b)} = F_{\mathsf{G}}^{\opt} = \frac{1}{8}\left(1 + \sqrt{2}\right)^2 \approx 0.72
    \end{cases}
\end{equation}
In contrast, our learned circuit (\figref{fig:learned_vs_ideal_circ_on_hw_main_text}(c)) maximises the local fidelity, but in order to gain an advantage in circuit depth, compromises with respect to the global fidelity:
\begin{equation} \label{eqn:learned_circuit_local_global_fidelities}
    \figref{fig:learned_vs_ideal_circ_on_hw_main_text}(c) \implies\begin{cases}
    F_{\mathrm{B}}^{(c)}  \approx F_{\mathrm{E}}^{(c)} \approx F_{\mathsf{L}}^{\opt} = 0.85\\
    F_{\mathsf{G}}^{(c)}  \approx 0.638 < F_{\mathsf{G}}^{\opt}
    \end{cases}
\end{equation}
%

\subsection[\texorpdfstring{\color{black}}{} State-dependent cloning]{State-dependent cloning}\label{ssec:results_state_dep_cloning}
Next, let us move onto the results of $\VQC$ when learning to clone the states used in the two coin flipping protocols described in \secref{sec:cloning/variational_cryptanalysis}. Firstly, we focus on the states used in the original protocol, $\mathcal{P}_1$ for $1\rightarrow 2$ cloning, and then move to the 4 state protocol, $\mathcal{P}_2$. In the latter we also extend from $1\rightarrow 2$ cloning to $1\rightarrow 3$ and $2\rightarrow 4$. These extensions will allow us to probe certain features of $\VQC$, in particular explicit symmetry in the cost functions. In all cases, we use the variable-structure Ansatz, and once a suitable candidate has been found, the solution is manually optimised further. The learned circuits used to produce the figures in this section are given in \secref{app_sec:vqc_learned_circuits}.

\subsubsection[\texorpdfstring{\color{black}}{} Cloning \texorpdfstring{$\mathcal{P}_1$}{} states]{Cloning \texorpdfstring{$\mathcal{P}_1$}{} states} \label{sssec:cloning/numerics/mayers_states}
As a reminder, the two states used in this protocol are:
\begin{align} \label{eqn:mayers_states_explicit_numerical}
    \ket{\phi_0} := \ket{\phi_{0, 0}} = \cos\left(\frac{\pi}{18}\right)\ket{0} + \sin\left(\frac{\pi}{18}\right)\ket{1}\\
    \ket{\phi_1} := \ket{\phi_{0, 1}} = \cos\left(\frac{\pi}{18}\right)\ket{0} - \sin\left(\frac{\pi}{18}\right)\ket{1}
\end{align}

\begin{figure}
\centering
    \includegraphics[width=\columnwidth,height=0.4\textwidth]{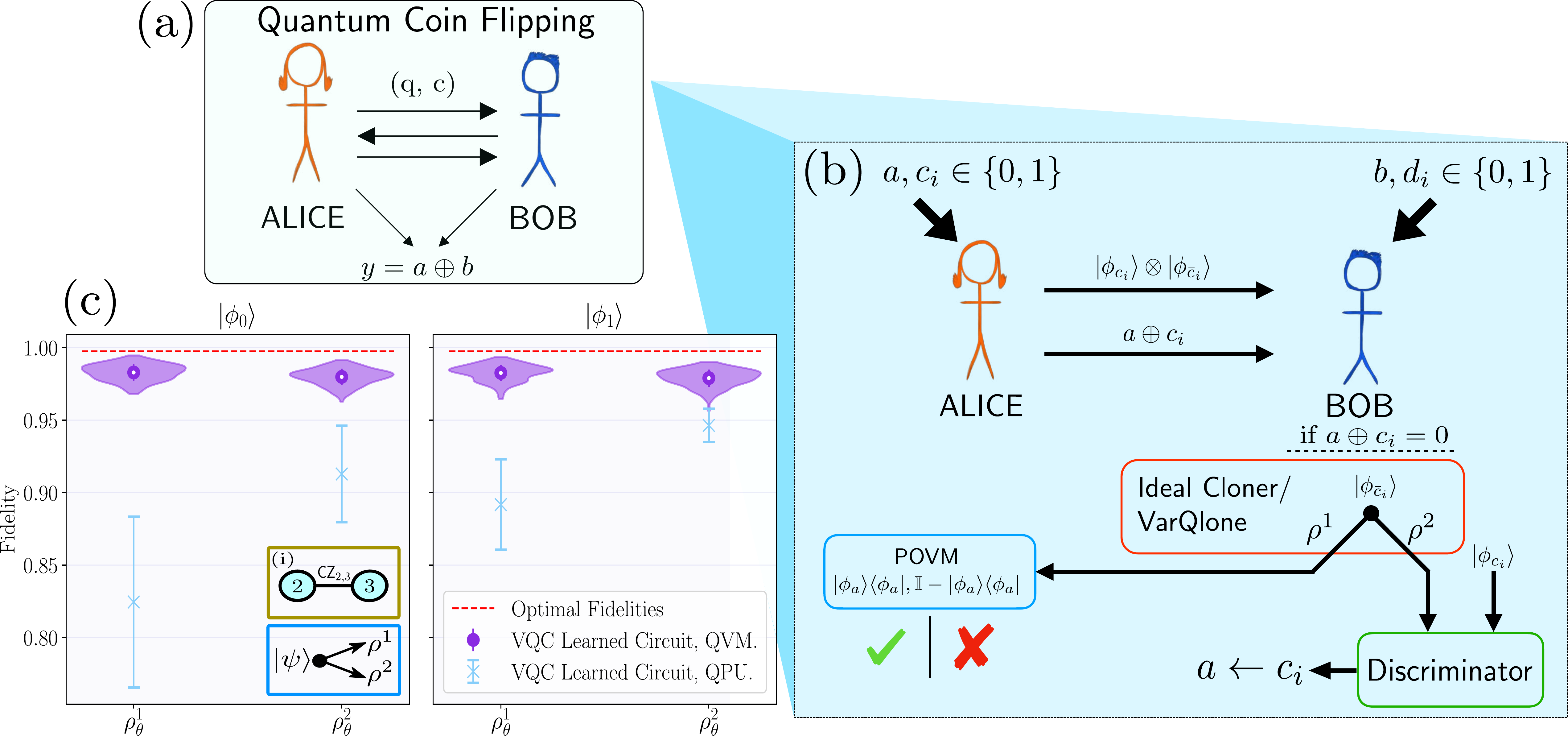}
    \caption[\color{black} Overview of cloning-based attack on the protocol of Mayers \textit{et. al.}, plus corresponding numerical results for $\VQC$.]{\textbf{Overview of cloning-based attack on the protocol of Mayers \textit{et. al.}~\cite{mayers_unconditionally_1999}, plus corresponding numerical results for $\VQC$.} (a) Cartoon of coin flipping protocols, Alice and Bob send quantum (q) and/or classical (c) information to agree on a final `coin flip' bit, $y$. (b) The relevant part of the protocol of Mayers \textit{et. al.}, $\mathcal{P}_1$, plus a cloning based attack on Bob's side. He builds a cloning machine using $\VQC$ to produce two clones of Alice's sent states, one of which he returns, and the other is used to guess Alice's input bit, $a$. (c) Fidelities of each output clone, $\rho^j_{\theta}$ achieved using $\VQC$ when ($1\rightarrow 2$) cloning the family of states used in, $\mathcal{P}_1$.  In the left (right) panel, $\ket{\phi_0}$ ($\ket{\phi_1}$) is. Figure shows both simulated (QVM -  [\crule[Purple]{0.2cm}{0.2cm}] purple circles) and on Rigetti hardware (QPU - [\crule[Cyan]{0.2cm}{0.2cm}] blue crosses). For the QVM (QPU) results, 256 (5) samples of each state are used to generate statistics. Violin plots show complete distribution of outcomes and error bars show the means and standard deviations. Inset (i) shows the two qubits of the \computerfont{Aspen-8} chip which were used, with the allowed connectivity of a $\CZ$ between them. Note an ancilla was also allowed, but $\VQC$ chose not to use it in this example. The corresponding learned circuit is given in \secref{app_sec:vqc_learned_circuits}.}
    \label{fig:mayers_1to2_cloning_fidelities_variational_main_text}
\end{figure}

The fidelities achieved by the $\VQC$ learned circuit can be seen in \figref{fig:mayers_1to2_cloning_fidelities_variational_main_text} using the gate pool \eqref{eqn:mayers_1to2_state_dependent_cloning_gateset_maintext} which allows a linear entangling connectivity:
\begin{equation} \label{eqn:mayers_1to2_state_dependent_cloning_gateset_maintext}
    \mathcal{G}_{\mathcal{P}_1^{1\rightarrow 2}} := \left\{
    \mathsf{R}^i_{j}(\theta), \CZ_{2, 3}, \CZ_{3, 4} \right\},  \forall i \in \{2, 3, 4\}, \forall j \in \{x, y, z\}
\end{equation}

A deviation from the optimal fidelity is observed in the simulated case, partly due to tomographic errors in reconstructing the cloned states. We note that the corresponding circuit for \figref{fig:mayers_1to2_cloning_fidelities_variational_main_text} only actually used $2$ qubits (see \secref{app_sec:vqc_learned_circuits}). This is because while $\VQC$ was \emph{allowed} to use the ancilla, it \emph{chose not} in this case by applying only identity gates to it. This mimics the behavior seen in the previous example of phase-covariant cloning. As such, we only use the two qubits shown in the inset (i) of the figure when running on the QPU to improve performance.

Now, returning to the attack on $\mathcal{P}_1$ above, we can compute the success probabilities using these fidelities. For illustration, let us return to the example in~\eqref{eqn:mayers_bob_pairs_discriminate_appendix}, where instead the cloned state is now produced from our $\VQC$ circuit, $\rho^0_c\rightarrow \rho^0_{\VQC}$.
\begin{thmbox}
\begin{theorem} \label{thm:vqc_bias_mayers_protocol}[$\VQC$ attack bias on $\mathcal{P}_1$]~ \\
Bob can achieve a bias of $\epsilon \approx 0.29$ using a state-dependent $\VQC$ attack on the protocol, $\mathcal{P}_1$, with a single copy of Alice's state.
\end{theorem}
\end{thmbox}
We can prove~\thmref{thm:vqc_bias_mayers_protocol} by computing the success probability in the same manner as was done in~\secref{app_ssec:mayers_protocol_and_attack}:
\begin{equation}\label{eqn:mayers_vqc_clone0}
\begin{split}
     P^{\VQC}_{\mathrm{succ}, \mathcal{P}_1} = \frac{1}{2}+\frac{1}{4}\Tr|\rho_1 - \ket{\phi_1}\bra{\phi_1}\otimes  \rho^0_{\VQC}| \approx 0.804
\end{split}
\end{equation}
The state $\rho_1= \ket{\phi_0}\bra{\phi_0}\otimes\ket{\phi_1}\bra{\phi_1}$ as in \eqref{eqn:mayers_bob_pairs_discriminate_appendix}. Here, we have a higher probability for Bob to correctly guess Alice's bit, $a$, but correspondingly the detection probability by Alice is higher than in the ideal case, due to a marginally lower local fidelity of $F^{\VQC}_{\Lbs} = 0.985$. 

\subsubsection[\texorpdfstring{\color{black}}{} Cloning \texorpdfstring{$\mathcal{P}_2$}{} states]{Cloning \texorpdfstring{$\mathcal{P}_2 $}{} states.} \label{sssec:cloning/numerics/aharonov_states}
Next, we turn to the family of states used in the $4$ states protocol, given by~\eqref{eqn:aharonov_coinflip_states}:\\

\noindent\textbf{$1 \rightarrow 2$ Cloning}

\begin{figure}
    \centering
        \includegraphics[width=0.9\columnwidth,height=0.35\textwidth]{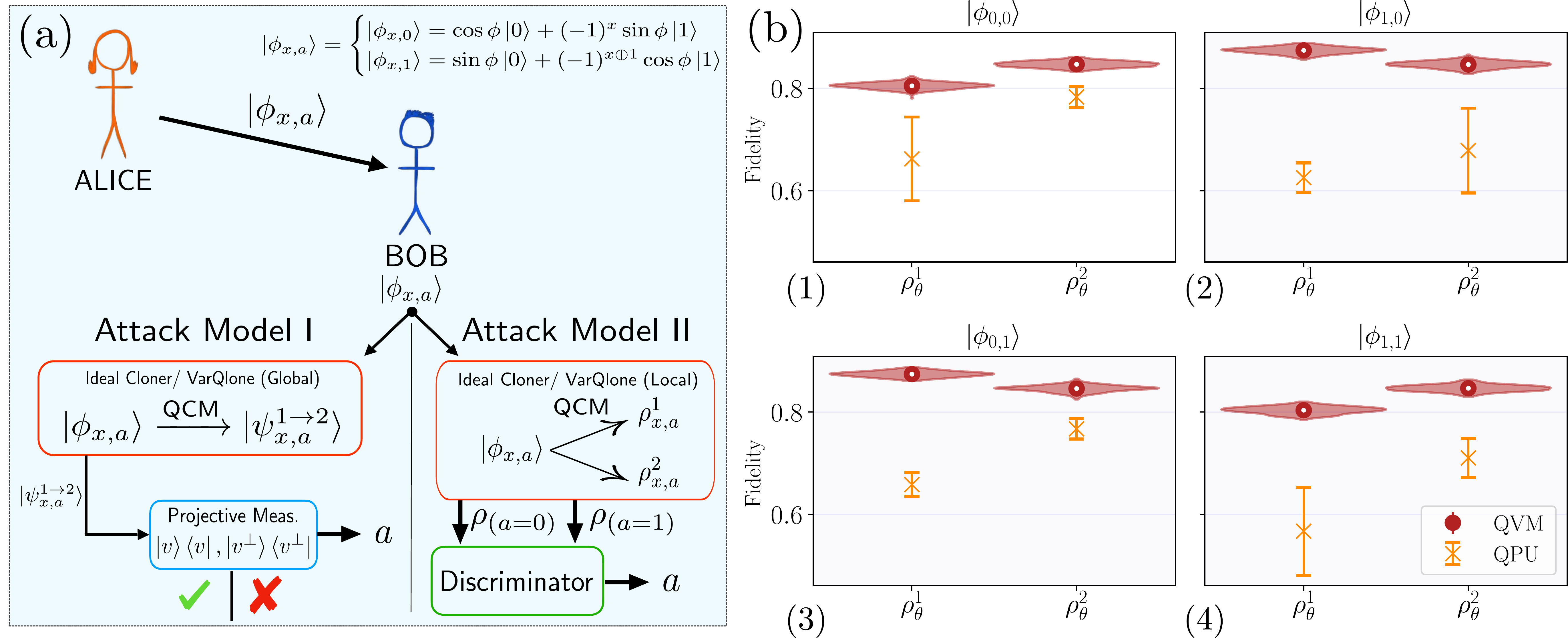}
    \caption[\color{black} Cloning attacks and numerical results for the protocol of Aharonov \textit{et. al.}.]{\textbf{Cloning attacks and numerical results for the protocol, $\mathcal{P}_2$.} (a) The two cloning based attacks we consider. In attack model I (left), Bob measures both output states with a set of fixed projective measurements, defined relative to the cloner output states, $\ket{\psi^{1\rightarrow 2}_{x, a}}$ and guesses Alice's bit, $a$. In attack model II, Bob keeps one clone for either testing Alice later or to send back the deposit qubit requested by Alice. He uses then the other local clone to discriminate and guess $a$. (b) The fidelities achieved cloning the each state, $\{\ket{\phi_{x, a}}\}$ used in $\mathcal{P}_2$ with $\VQC$. These numerics relate to scenario $1$ from attack model II (see \secref{app_sssec:aharonov_cloning_attack_II}). Each panel (1-4) shows both simulated (QVM - [\crule[BrickRed]{0.2cm}{0.2cm}] red circles) and on Rigetti hardware (QPU - [\crule[YellowOrange]{0.2cm}{0.2cm}] orange crosses). We indicate the fidelities of the each clone received by Alice and Bob. For the QVM (QPU) results, 256 (3) samples of each state are used to generate statistics. Violin plots show complete distribution of outcomes and error bars show the means and standard deviations. Inset (i) shows the connectivity we allow in $\VQC$ for this example. The corresponding learned circuit is shown in \secref{app_sec:vqc_learned_circuits}.}
    \label{fig:aharonov_1to2_cloning_fidelities_variational_plus_attack_models}
\end{figure}

\vspace{2mm}
Firstly, we repeat the exercise from above with the same scenario, using the same gateset and subset of the \computerfont{Aspen-8} lattice $(\mathcal{G}_{\mathcal{P}_2^{1\rightarrow 2}} = \mathcal{G}_{\mathcal{P}_1^{1\rightarrow 2}})$. We use the local cost, \eqref{eqn:local_cost_full}, to train the model, with a sequence length of $35$ gates. The results are seen in \figref{fig:aharonov_1to2_cloning_fidelities_variational_plus_attack_models}(b) both on the QVM and the QPU. We note that the solution exhibits some small degree of asymmetry in the output states, due to the form of the local cost function. This asymmetry is especially pronounced as we scale the problem size and try to produce $N$ output clones, which we discuss in the next section.

Now, we can relate the performance of the $\VQC$ cloner to the attacks discussed in \secref{ssec:aharonov_protocol_attack}. We do this by explicitly analysing the output states produced in the circuits used to achieve fidelities shown in \figref{fig:aharonov_1to2_cloning_fidelities_variational_plus_attack_models}(b) and following the derivation in \secref{app_sec:coin_flipping_cloning_attacks} for \thmref{thm:aharonov_attack_I_bias_probability_maintext_vqc} and \thmref{thm:aharonov_4state_attack_II_bias_probability_maintext_vqc}:
\begin{thmbox}
\begin{theorem}\label{thm:aharonov_attack_I_bias_probability_maintext_vqc}[$\VQC$ cloning attack (I) bias on $\mathcal{P}_2$]~ \\
Using a cloning attack on the protocol, $\mathcal{P}_2$, (in attack model I) Bob can achieve a bias:
\begin{equation}
    \epsilon^{\mathrm{I}}_{\mathcal{P}_2, \VQC} \approx 0.345
\end{equation}
\end{theorem}
\end{thmbox}
Similarly, we have the bias which can be achieved with attack II:
\begin{thmbox}
\begin{theorem}\label{thm:aharonov_4state_attack_II_bias_probability_maintext_vqc}[$\VQC$ Cloning Attack (II) Bias on $\mathcal{P}_2$]~ \\
Using a cloning attack on the protocol, $\mathcal{P}_2$, (in attack model II) Bob can achieve a bias:
\begin{equation}\label{eqn:attack_2_aharonov_success_probability_bound_maintext_real}
    \epsilon^{\mathrm{II}}_{\mathcal{P}_2, \VQC} = 0.241
\end{equation}
\end{theorem}
\end{thmbox}
The discrepancy between these results and the ideal biases are primarily due to the small degree of asymmetry induced by the heuristics of $\VQC$. However, we emphasise that these biases can now be achieved constructively.\\

\noindent \textbf{$1 \rightarrow 3$ and $2\rightarrow 4$ Cloning}

\vspace{2mm}
Finally, we extend the above to the more general scenario of $M\rightarrow N$ cloning, taking $M=1, 2$ and $N=3, 4$. These examples are illustrative since they demonstrate strengths of the squared local cost function  (\eqref{eqn:squared_local_cost_mton}) over the local cost function (\eqref{eqn:local_cost_full}). In particular, we find the local cost function does not enforce symmetry strongly enough in the output clones, and using only the local cost function, suboptimal solutions are found. This was particularly observed in the example of $2\rightarrow 4$ cloning, where $\VQC$ tended to take a shortcut by allowing one of the input states to fly through the circuit (resulting in nearly $100\%$ fidelity for that clone), and then attempt to perform $1\rightarrow 3$ cloning with the remaining input state. By strongly enforcing symmetry in the output clones using the squared cost, this can be avoided as we demonstrate explicitly in \secref{fig:local_vs_squared_cost_aharanoov_2to4}.
\begin{figure*}[ht]
    \begin{center}
    \includegraphics[width=0.9\columnwidth, height=0.55\columnwidth]{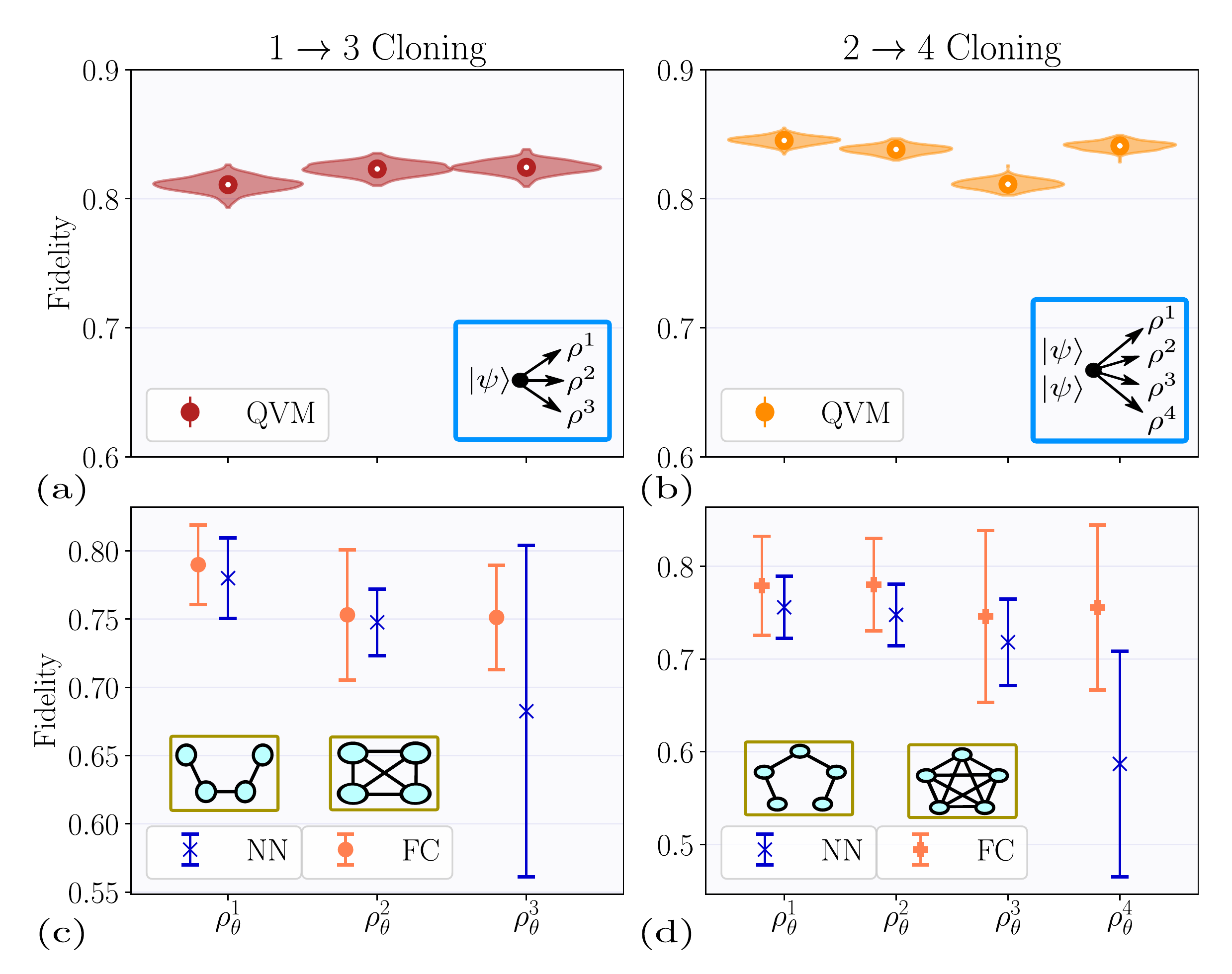}
    \caption[\color{black} $\VQC$ for $1\rightarrow 3$ and $2 \rightarrow 4$ cloning with the states of the in the coin flipping protocol of Aharonov \textit{et al.}, and the effect of gate pool connectivity.]{\textbf{Clone fidelities for optimal circuits learned by $\VQC$ for (a) $1\rightarrow 3$ [\crule[BrickRed]{0.2cm}{0.2cm}] and (b) $2 \rightarrow 4$ [\crule[YellowOrange]{0.2cm}{0.2cm}] cloning of the states used in the coin flipping protocol of \cite{aharonov_quantum_2000} \textit{et al.}\@.} Mean and standard deviations of $256$ samples are shown (violin plots show full distribution of fidelities), where the fidelities are computed using tomography only on the Rigetti QVM. In both cases, $\VQC$ is able to achieve average fidelities $> 80\%$. (c-d) shows the mean and standard deviation of the optimal fidelities found by $\VQC$ over 15 independent runs ($15$ random initial structures, $\boldsymbol{g}$) for a nearest neighbour (NN - [\crule[BlueViolet]{0.2cm}{0.2cm}] purple) versus (d) fully connected (FC - [\crule[Salmon]{0.2cm}{0.2cm}] pink) entanglement connectivity allowed in the variable-structure $\Ansatz$ for $1\rightarrow 3$ and $2 \rightarrow 4$ cloning of $\mathcal{P}_2$ states. Insets of (c-d) shown corresponding allowed $\CZ$ gates in each example. 
    }\label{fig:1to3_2to4_aharonov_optimal_fidelities_plus_nn_vs_fc}
        \end{center}
\end{figure*}
We also test two connectivities in these examples, a fully connected (FC) and a nearest neighbour (NN) architecture as allowed by the following gatesets:
\begin{align} 
    \mathcal{G}^{\textnormal{NN}}_{\mathcal{P}_2^{1\rightarrow 3}} = 
    \left\{\right. \mathsf{R}^i_{z}(\theta),  \mathsf{R}^i_{x}(\theta), \mathsf{R}^i_{y}(\theta), 
    \CZ_{2, 3}, \CZ_{3, 4}, \CZ_{4, 5}\left.\right\} &\forall i \in \{2, 3, 4, 5\}
    \label{eqn:aharonov_1to3_state_dependent_cloning_gateset_NN}\\
    \mathcal{G}^{\textnormal{FC}}_{\mathcal{P}_2^{1\rightarrow 3}} =
    \left\{\right. \mathsf{R}^i_{z}(\theta),  \mathsf{R}^i_{x}(\theta), \mathsf{R}^i_{y}(\theta), \CZ_{2, 3},  \CZ_{2, 4},  \CZ_{2, 5}, 
    &\CZ_{3, 4}, \CZ_{3, 5},  \CZ_{4, 5}\left.\right\} \\
    \forall i \in \{2, 3, 4, 5\}  \label{eqn:aharonov_1to3_state_dependent_cloning_gateset_FC}
\end{align}
Note, that for $1\rightarrow 3$ ($2\rightarrow 4$) cloning, we actually use $4$ ($5$) qubits, with one being an ancilla. The results of these experiments are given in \figref{fig:1to3_2to4_aharonov_optimal_fidelities_plus_nn_vs_fc}. We use the following hyperparameters for this experiment: 1) a sequence length of $l=35$ for $1\rightarrow 3$, and $l=40$ for $1\rightarrow 4$ with $50$ iterations over $\boldsymbol{g}$ in both cases,  2) the Adam optimiser with an initial learning rate of $\eta_{\text{init}} = 0.05$, 3) $50$ training samples. In all cases, we use the squared cost function, $\Cbs_{\sq}$ to train, and its gradients.

\subsection[\texorpdfstring{\color{black}}{} Training sample complexity]{Training sample complexity} \label{app_ssec:training_sample_complexity}

Here we study the sample complexity of the training procedure by retraining the continuous parameters of the learned circuit (\figref{fig:learned_vs_ideal_circ_on_hw_main_text}(c)) starting from a random initialisation of the parameters, $\paramtheta$. As expected, as the number of training samples increases (i.e. the number of random choices of the phase parameter, $\eta$, in~\eqref{eqn:x_y_plane_states}), the generalisation error (difference between training and test error) approaches zero. This is not surprising, since the training set will eventually cover all states on the equator of the Bloch sphere.

\begin{figure}[ht]
    \begin{center}
        \includegraphics[width=0.95\columnwidth]{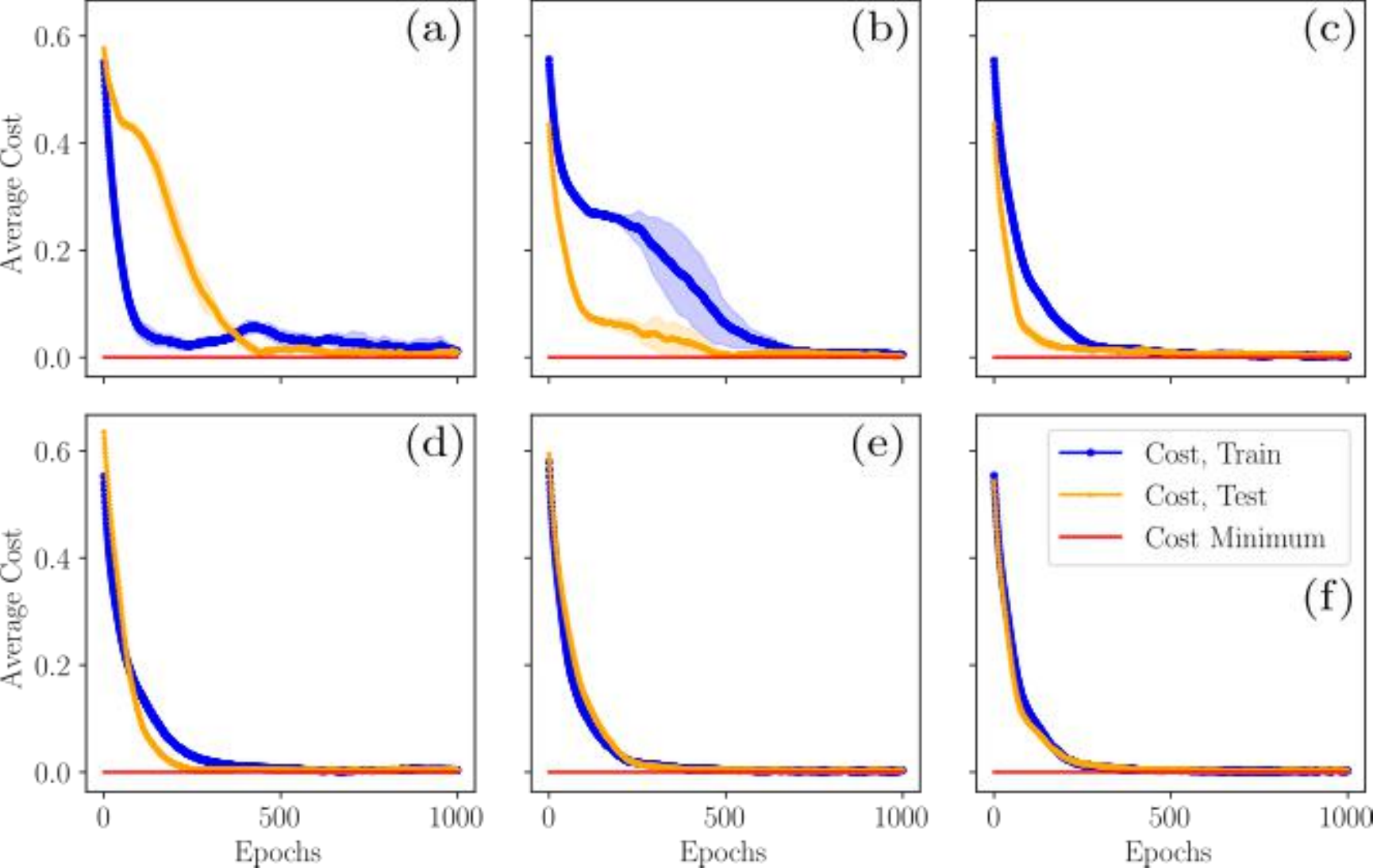}
    \caption[\color{black} Numerical sample complexity of $\VQC$.]{\textbf{Numerical sample complexity of $\VQC$ using the squared cost.} We begin with a random initialisation of the structure learned circuit in \figref{fig:learned_vs_ideal_circ_on_hw_main_text}(c) and optimise the parameters using different sizes in the training/text set, and different mini-batch sizes. All of the following using a train/test split of $20\%$ and we denote the tuple $(i, j, k)$ as $i = $ number of training samples, $j = $  number of test samples, $k = $ batch size. (a) $(1, 1, 1)$, (b) $(4, 1, 2)$, (c) $(8, 2, 5)$, (d) $(16, 4, 8)$ (e) $(40, 10, 15)$, (f) $(80, 20, 20)$. }
    \label{fig:phase_cov_learned_circuit_sample_complexity}
        \end{center}
\end{figure}

\subsection[\texorpdfstring{\color{black}}{} Local cost function comparison]{Local cost function comparison} \label{app_ssec:connectivity_and_cost_function_comparison_aharanov}

\begin{figure}[ht]
    \begin{center}
        \includegraphics[width=0.85\columnwidth]{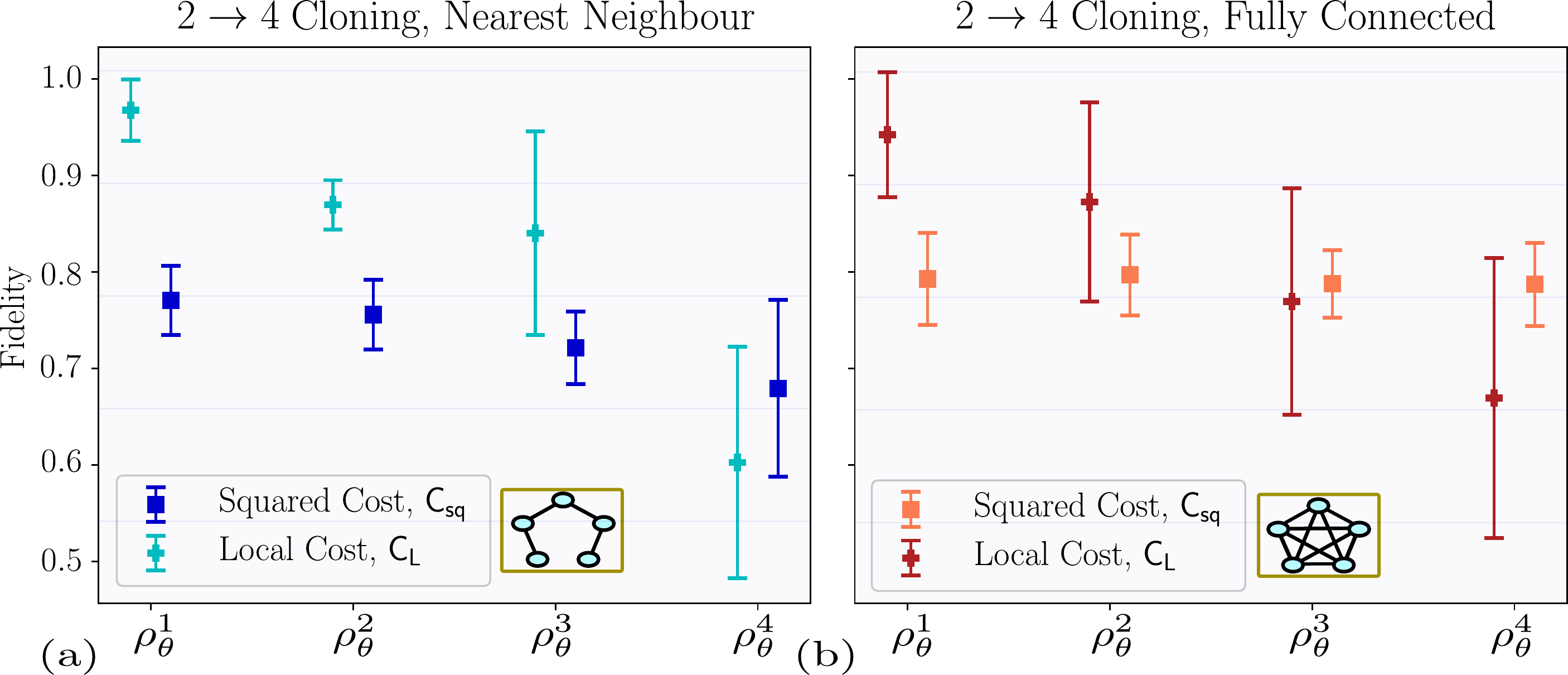}
    \caption[\color{black} Comparison between the local and squared cost functions for $2\rightarrow 4$ cloning.]{Comparison between the local (\eqref{eqn:local_cost_full}) [\crule[Cyan]{0.2cm}{0.2cm}, \crule[BrickRed]{0.2cm}{0.2cm}] and squared (\eqref{eqn:squared_local_cost_mton}) [\crule[Blue]{0.2cm}{0.2cm}, \crule[Salmon]{0.2cm}{0.2cm}] cost functions for $2\rightarrow 4$ cloning. (a) shows nearest neighbour (NN) and (b) has a fully connected (FC) entanglement connectivity allowed in the variable-structure $\Ansatz$. Again, we use the family of states in the protocol $\mathcal{P}_2$. Plots show the mean and standard deviation of the optimal fidelities found by $\VQC$ over 10 independent runs ($10$ random initial circuit structures). A sequence length of $35$ is used for $1\rightarrow 3$ and $40$ for $2\rightarrow 4$, with $50$ iterations of the variable-structure Ansatz search in both cases. Here we use the same experiment hyperparameters as in \figref{fig:1to3_2to4_aharonov_optimal_fidelities_plus_nn_vs_fc}.
    }
    \label{fig:local_vs_squared_cost_aharanoov_2to4}
        \end{center}
\end{figure}

In \figref{fig:local_vs_squared_cost_aharanoov_2to4}, we revisit the weakness of the local cost function, $\mathsf{C}_{\Lbs}$, in not enforcing symmetry strongly enough in the problem output, and how the squared cost function, $\mathsf{C}_{\mathsf{sq}}$ can alleviate this, for $2\rightarrow 4$ cloning specifically. Here we show the optimal fidelities found by $\VQC$ with a variable-structure $\Ansatz$, starting from a random structure. The local cost tends towards local minima, where one of the initial states ($\rho^1_\theta$) ends up with high fidelity, while the last qubit ($\rho^4_\theta$) has a low fidelity. This is alleviated with the squared cost function which is clearly more symmetric, on average, in the output fidelities. This is observed for both circuit connectivities we try (although a NN architecture is less able to transfer information across the circuit for a fixed depth).

\section[\texorpdfstring{\color{black}}{} \texorpdfstring{$\VQC$}{} learned circuits]{\texorpdfstring{$\VQC$}{} learned circuits} \label{app_sec:vqc_learned_circuits}

To round off this chapter, we give the explicit circuits we use to generate the results above. We mention as above, that these are only representative examples, and many alternatives were also found in each case.

\subsection[\texorpdfstring{\color{black}}{} Ancilla-free phase-covariant cloning]{Ancilla-free phase-covariant cloning} \label{app_ssec:ancialla_free_phase_covariant}

\begin{figure}
    \centering
    \includegraphics[width=\columnwidth]{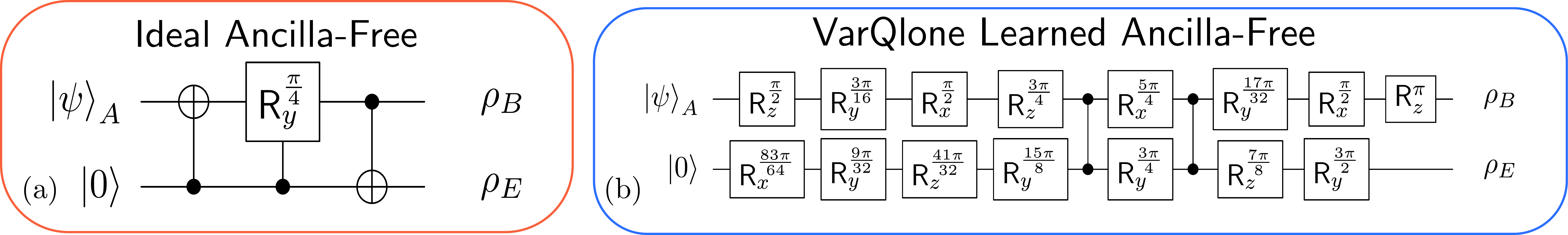}
    \caption[\color{black} Circuits to clone phase-covariant states, without ancilla for $2$ qubits.]{\textbf{Circuits to clone phase-covariant states, without ancilla for $2$ qubits} (a) Optimal circuit from~\cite{du_experimental_2005}. (b) Circuit learned by $\VQC$. It can be checked that the ideal circuit in (a) can be compiled to \emph{also} use $2$ $\CZ$ plus single qubit gates, so $\VQC$ has achieved close to optimality. The average fidelities for $B, E$ for the circuit in (b) is $F_{\mathsf{L},  \VQC}^{B, \text{PC}}  \approx 0.854$ and  $F_{\mathsf{L},  \VQC}^{E, \text{PC}}  \approx 0.851$ respectively, over $256$ input samples, $\ket{\psi}_A$ (comparing to the ideal fidelity of $F_{\mathsf{L},  \opt}^{\text{PC}} = 0.853$).}
    \label{fig:phase_covariant_cloning_circuits_2_qubits}
\end{figure}

The circuits found in \figref{fig:learned_vs_ideal_circ_on_hw_main_text} to clone phase-covariant states are slightly more general than we may wish to use. In particular, the circuit \figref{fig:qubit_cloning_ideal_circ} also has the ability to clone \emph{universal} states, due to the addition of the ancilla, which can be used as a resource. However, it is known that phase-covariant cloning can be implemented economically, i.e. \emph{without} the ancilla~\cite{niu_two-qubit_1999, scarani_quantum_2005} as mentioned above\footnote{Although as discussed above in \secref{sec:qkd_and_cloning_attacks}, this the economical version does not provide the optimal attack on related protocols.}. As such, we could compare against a shorter depth circuit which also does not use the ancilla. For example, the circuit from~~\cite{du_experimental_2005} shown in \figref{fig:phase_covariant_cloning_circuits_2_qubits}(a) is also able to achieve the optimal cloning fidelities ($\sim 0.85$). An example $\VQC$ learned circuit for this task can be seen in \figref{fig:phase_covariant_cloning_circuits_2_qubits}(b) which has $2$ $\CZ$ gates. We note that this ideal circuit can be compiled to \emph{also} use $2$ $\CZ$ gates, so in this case $\VQC$ finds a circuit which is approximately comparable up to single qubit rotations.

\subsection[\texorpdfstring{\color{black}}{} State-dependent cloning circuits]{State-dependent cloning circuits} \label{app_ssec:state-dependent-circuits}

\begin{figure}
    \centering
    \includegraphics[width=0.8\columnwidth]{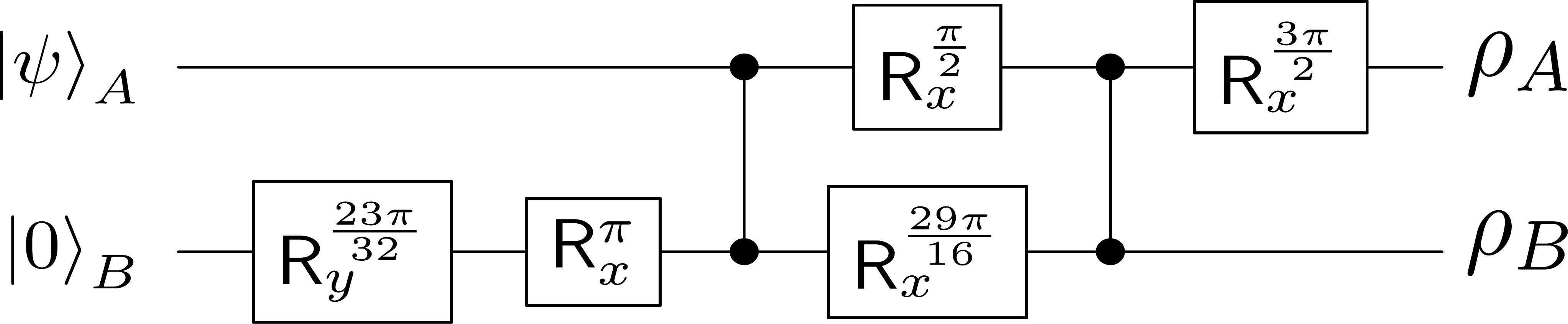}
    \caption[\color{black} Circuit learned by $\VQC$ in to clone states in the protocol of Mayers \textit{et. al.}.]{\textbf{Circuit learned by $\VQC$ in to clone states, $\ket{\phi_0}, \ket{\phi_1}$, with an overlap $s=\cos\left(\pi/9\right)$ in the protocol, $\mathcal{P}_1$.} For example, $\rho_A$ is the clone sent back to Alice, while $\rho_B$ is kept by Bob.}
    \label{fig:mayers_1to2_cloning_vqc_circuit}
\end{figure}

\figref{fig:mayers_1to2_cloning_vqc_circuit} shows the circuit used to achieve the fidelities in the attack on $\mathcal{P}_1$. In training, we still allow an ancilla to aid the cloning, but the example in \figref{fig:mayers_1to2_cloning_vqc_circuit} did not make use of it (in other words, $\VQC$ only applied gates which resolved to the identity on the ancilla), so we remove it to improve hardware performance. This repeats the behaviour seen for the circuits learned in phase-covariant cloning. We mention again, that some of the learned circuits did make use of the ancilla with similar performance.

\begin{figure}
    \centering
    \includegraphics[width=0.7\columnwidth]{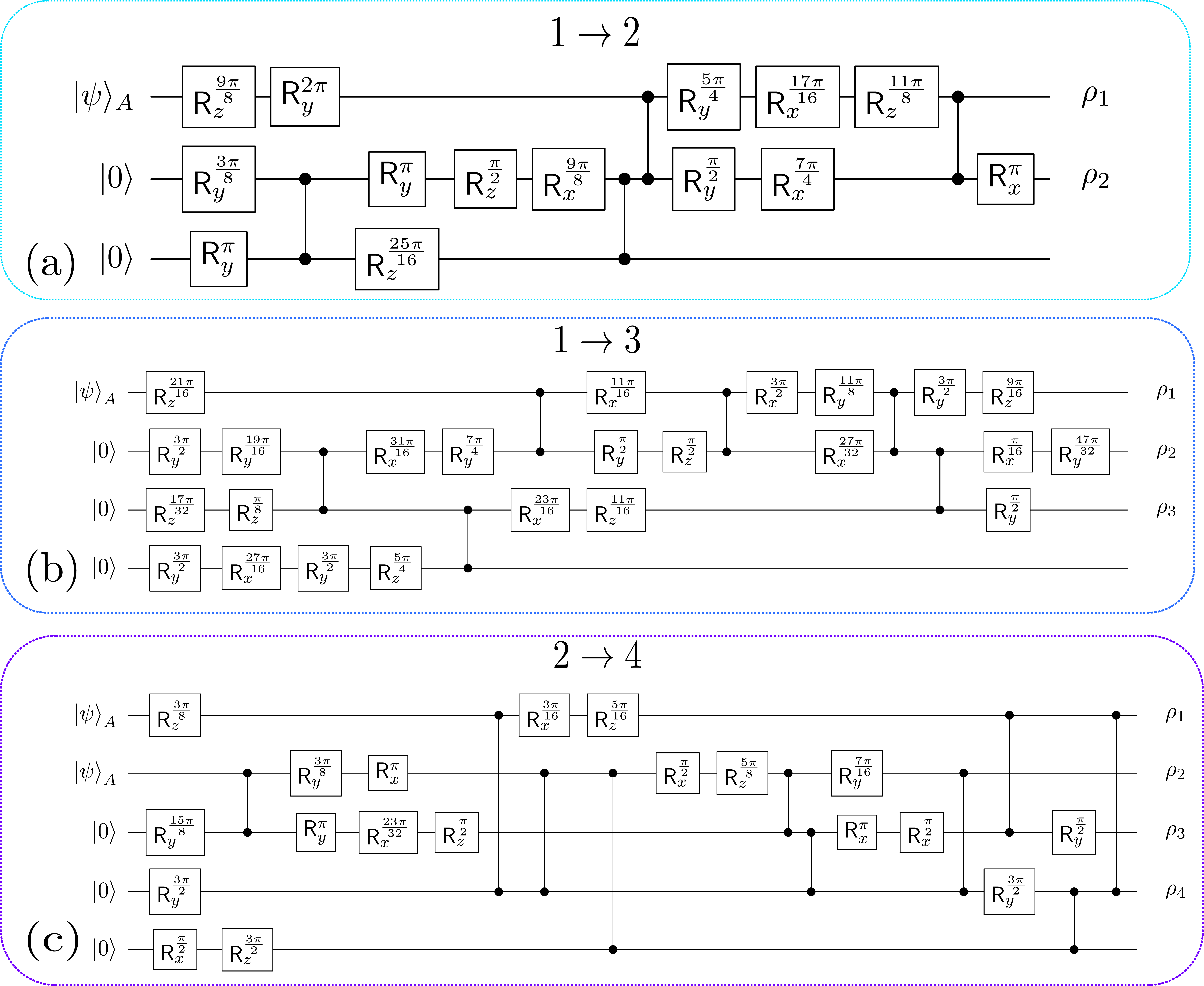}
    \caption[\color{black} Circuits learned by $\VQC$ to clone states from the protocol of Aharonov \textit{et. al.} for $1\rightarrow 2$, $1\rightarrow 3$ and $2 \rightarrow 4$ cloning.]{\textbf{Circuits learned by $\VQC$ to clone states from the protocol, $\mathcal{P}_2$ for (a) $1\rightarrow 2$, (b) $1\rightarrow 3$ and (c) $2 \rightarrow 4$ cloning.} These specific circuits produce the fidelities in \figref{fig:aharonov_1to2_cloning_fidelities_variational_plus_attack_models}(b) for $1\rightarrow 2$, (using the local cost function), and in \figref{fig:1to3_2to4_aharonov_optimal_fidelities_plus_nn_vs_fc} for $1\rightarrow 3$ and $2 \rightarrow 4$ (using the squared cost function). We allow an ancilla for all circuits, and $\rho_k$ indicates the qubit which will be the $k^{th}$ output clone.}
    \label{fig:aharonov__1to2_1to3_2to4_circuits}
\end{figure}

\figref{fig:aharonov__1to2_1to3_2to4_circuits} shows the circuits learned by $\VQC$ and approximately clone all four states in \eqref{eqn:aharonov_coinflip_states} in the protocol, $\mathcal{P}_2$, for $1\rightarrow 2, 1 \rightarrow 3$ and $2 \rightarrow 4$ cloning. These are the specific circuits used to produce the fidelities in \figref{fig:aharonov_1to2_cloning_fidelities_variational_plus_attack_models}(b) and \figref{fig:1to3_2to4_aharonov_optimal_fidelities_plus_nn_vs_fc}.

\section[\texorpdfstring{\texorpdfstring{\color{black}}{}}{} Discussion and conclusions]{Discussion and conclusions} \label{sec:cloning/conclusions}

We have now concluded the discussion of our third and final application for near-term quantum computers; the use of quantum machine learning in quantum information and quantum cryptography. We saw how quantum cloning is one of the most important ingredients not just as a tool in quantum cryptanalysis, but also with roots in foundational questions of quantum mechanics. However, given the amount of attention this field has received, a fundamental question remained elusive: how do we construct efficient, flexible, and noise-tolerant circuits to actually perform approximate or probabilistic cloning? We demonstrated how $\VQC$ provided a novel toolkit to such an exercise. We further believe that $\VQC$ is unique among variational/quantum machine learning algorithms for a number of reasons. The first, and most obvious reason for this is that it bridges the fields of quantum machine learning and quantum cryptography in an interesting manner. We demonstrated how $\VQC$ brings into view a whole new domain of performing realistic implementation of attacks on quantum cryptographic systems.  Secondly, since it is an example using quantum data directly, there is no obvious method of dequantisation, or indeed no obvious argument why dequantisation would even make sense in such a context. Finally, unlike many other variational algorithms, which aim to mimic performance of fault-tolerant algorithms in asymptotic scaling regimes, it is useful in both the small scale, \emph{and} the large scale. The former is more obvious; we have already demonstrated example use cases ion cryptography. At the larger scale, one could imagine studying connections to quantum state estimation, or simply the question of cloning by itself.

In future work, one of the most exciting possible directions is in actually using $\VQC$ to aid the attack of a quantum protocol in a physical experiment. Other directions of study include improving the robustness and performance of $\VQC$. One could utilise noise tolerant methods for quantum architecture search, or incorporating error mitigation methods. The latter presents an interesting question in itself, since many error mitigation methods aim to correct expectation values of computations by classical post-processing. It is not obvious how such techniques may be useful when the desired output is not classical.

\chapter{Conclusion} \label{chap:conclusion}

This Thesis began with the goal of finding, and developing, applications and models for near-term quantum computers. Focusing on problems and techniques from quantum machine learning, we aimed to attach theoretical rigour, where possible, to an experimental plug-and-play mindset in the development of such applications. We focused on a particular primitive from modern quantum machine learning; the variational quantum algorithm (VQA) equipped with the parameterised quantum circuit (PQC). Let us now conclude the Thesis by briefly summarising the contributions of each chapter, before discussing a more general outlook and future directions.

As a reminder, this Thesis was structured in accordance with the increasing complexity of the data available to each application. 

We began in~\chapref{chap:classifier} with labelled vectors as data and focused on the use of a PQC in supervised learning for the problem of classification. In order to build such models to be implementable across multiple types of near-term quantum hardware, it is important to build them in a robust manner, to account for the inevitable quantum noise which will plague them for at least the next several years. We did this by focusing attention on the means of encoding data into the PQC for subsequent classification. We provided several definitions for `robust' data encodings, and we proved that such encodings must always exist, for a given noise channel. We also studied the design of the PQC outside of the data encoding, and found that by incorporating simple alternative design principles, the model could be naturally robust. However, we noted that this robustness came at a price; namely in the expressiveness or learnability of the model. Finally, we performed several numerical experiments demonstrating that an `optimal' classifier for near-term devices involves a trade-off between robustness and learnability.

Next, in~\chapref{chap:born_machine} we turned to the more complicated unsupervised learning problem of generative modelling. Here, the data becomes probability distributions and sample access to them. To attack this problem, we turn to a PQC in the form of a Born machine, an implicit generative model suitable for near-term devices. We began by discussing the possibility of a (provable) quantum advantage in generative modelling and natural reasons why the Born machine may demonstrate such a thing. We then developed new gradient-based training methods for the Born machine by changing the cost function used. Finally, we performed an extensive numerical and experimental study for the Born machine with three different datasets. Of particular interest, was that one of these datasets originated with a quantum process, which one of the others was relevant to a real world use case in finances. In this latter example, we also performed a comparison to a comparable classical generative model, the restricted Boltzmann machine, and found that the quantum model could outperform its classical counterpart (a \emph{numerical} quantum advantage). We also surprisingly found some evidence that entanglement may be partially responsible for such an advantage, by adding correlations which are not classically accessible.

Finally, in~\chapref{chap:cloning} we changed course slightly away from `machine learning' problems, into the domain of quantum foundations. However, we still used very similar ML tools in the study of quantum cloning. Here, we designed a variational quantum algorithm ($\VQC$) to learn to approximately clone quantum states. For the algorithm, we provided cost functions with faithfulness guarantees, derived suitable gradients for gradient-based training, and incorporated quantum architecture search to find high quality cloning circuits. In this chapter, we also made an interesting bridge between quantum machine learning and quantum cryptography by demonstrating how QML methods could be used as a subroutine in tools for the cryptanalysis of quantum protocols. Specifically for this latter use case, we focused on the common examples of quantum key distribution, and quantum coin flipping. We illustrated how $\VQC$ could be used to find novel attacks on such protocols, which could subsequently be employed to improve the protocols.

Returning to an abridged version of the question posed in the Introduction (``How can we use the quantum computers we have now . . . ''), it is clear that there are a multitude of potential applications for such devices. Whether any of these will prove to be the `killer application', only time (and bigger quantum computers) will tell. This breakthrough may come from a dedicated research team in a multinational conglomerate, or even from a lone `quantum hacker'. Both of these are now possible due to the abundance of resources and learning material. Regardless, the sheer scope of possibility is one of the main reasons why there is excitement and hype surrounding the field. While such excitement is extremely beneficial for a number of reasons, we must also be careful to temper unfounded claims of quantum computers revolutionising all aspects of science and technology. Throughout this thesis, I hope I have been able to provide a balanced view. 

My PhD has been fortunately situated in the era of the first commercially available quantum computers. As such, it was perfectly positioned in the QML transition period to take advantage of these computers. With this in mind, I would summarise my experience of performing research in quantum machine learning as the perfectly epitomised quote by the Doctor: ``\emph{. . . It's a bit dodgy, this process. You never know what you're going to end up with.}'', which applies equally well to both its parent fields.

\appendix

\chapter{Proofs and derivations}\label{chap:proofs}

\section[\texorpdfstring{\color{black}}{} Proofs for \texorpdfstring{\chapref{chap:classifier}}{}]{Proofs for \texorpdfstring{\chapref{chap:classifier}}{}} \label{app:appendices/classifier}

\subsection[\texorpdfstring{\color{black}}{} Proofs of \texorpdfstring{\thmref{thm:factorizable_noise_before_meas}}{} and \texorpdfstring{\thmref{thm:measurement_noise_robustness}}{}]{Proofs of \texorpdfstring{\thmref{thm:factorizable_noise_before_meas}}{} and \texorpdfstring{\thmref{thm:measurement_noise_robustness}}{}}\label{app_ssec:noise_robustness}

Here we give the proofs of the above two Theorems, which broadly have the same structure.
\begin{thmbox}
\begin{reptheorem}{thm:factorizable_noise_before_meas}
If $\E$ is any noise channel which factorises into a single qubit channel, and a multiqubit channel as follows:
\begin{equation*}
    \E(\rho) = \E_{\Bar{c}}(\rhotildex^{\Bar{c}}) \otimes \E_c(\rhotildex^c)
\end{equation*}
where WLOG $\E_c$ acts only on the classification qubit $(\rhotildex^c = \Tr_{\Bar{c}}(\rhotildex))$ after encoding and unitary evolution, and $\E_{\Bar{c}}$ acts on all other qubits arbitrarily, $(\rhotildex^{\Bar{c}} = \Tr_c(\rhotildex))$. Further assume the state meets the robust classification requirements for the single qubit error channel $\E_c$. Then the classifier will be robust to $\E$.
\end{reptheorem}
\end{thmbox}

\begin{proof}
The correct classification again depends on the classification qubit measurement probabilities:
$\Tr(\ketbra{0}{0}^{c}\otimes \mathds{1}^{\otimes n-1}\rhotildex), \Tr(\ketbra{1}{1}^{c}\otimes \mathds{1}^{\otimes n-1} \rhotildex)$. If $\rhotildex$ is robust to the single qubit error channel $\E_c$, this means 
\begin{align*}
    \Tr(\Pi_0^c\rhotildex) \geq 1/2 &\implies \Tr(\Pi_0^c \E_c(\rhotildex)) \geq 1/2 \\
    \Tr(\Pi_1^c \rhotildex) < 1/2  &\implies \Tr(\Pi_1^c \E_c(\rhotildex)) < 1/2
\end{align*}
Then WLOG, assume the point $\xbs$ classified as $y(\rhotildex) = 0$ before the noise, then:
\begin{align*}
    \Tr\left(\Pi_0^c\E(\rhotildex)\right)&= \Tr\left(\Pi_0^c \left[\E_{\Bar{c}}(\rhotildex^{\Bar{c}} \otimes \E_c(\rhotildex^c)\right]\right) \\
    &=\Tr_{\Bar{c}}\left(\E_{\Bar{c}}(\rhotildex^{\Bar{c}})\right) \Tr_{c}\left(\ketbra{0}{0} \E_c(\rhotildex^c)\right) \\
    &=\Tr \left(\ketbra{0}{0} \E_c(\rhotildex^c)\right) \geq 1/2 \qedhere
\end{align*}
\end{proof}

\begin{thmbox}
\begin{reptheorem}{thm:measurement_noise_robustness}
Let $\mathcal{E}_{\vec{p}}^{\textnormal{meas}}$ define measurement noise acting on the classification qubit and consider a quantum classifier on data from the set $\mathcal{X}$. 
Then, for any encoding $E: \mathcal{X} \rightarrow \mathcal{S}_2$, we have complete robustness
    \begin{equation*}
        \mathcal{R} (\mathcal{E}_{\vec{p}}^{\textnormal{meas}}, E, \yhat) = \mathcal{X}
    \end{equation*}
     if the measurement assignment probabilities satisfy $p_{00} > p_{01}, p_{11} > p_{10}$.
\end{reptheorem}
\end{thmbox}

\begin{proof}
We can write the measurement noise channel acting on the POVM elements as:
\begin{equation*}
    \mathcal{E}_{\vec{p}}^{\textnormal{meas}}(\Pi^{c}_0\otimes \mathds{1}^{\otimes n-1})  = (p_{00}\ketbra{0}{0} + p_{01}\ketbra{1}{1}) \otimes \mathds{1}^{\otimes n-1}
\end{equation*}
Again, if we had correct classification before the noise, $\Tr(\ketbra{0}{0}\rhotildex) \geq 1/2$, then:
\begin{align*}
    &\Tr[\mathcal{E}_{\vec{p}}^{\textnormal{meas}}(\Pi^{c}_0\otimes \mathds{1}^{\otimes n-1})\rhotildex]= \Tr[(\{p_{00}\ketbra{0}{0} + p_{01}\ketbra{1}{1}\} \otimes \mathds{1}^{\otimes n-1})\rhotildex]\\
    &= p_{00}\Tr[(\ketbra{0}{0}\otimes \mathds{1}^{\otimes n-1})\rhotildex] + p_{01}\Tr[(\ketbra{1}{1} \otimes \mathds{1}^{\otimes n-1})\rhotildex]\\
    &= p_{00}\Tr[(\ketbra{0}{0}\otimes \mathds{1}^{\otimes n-1})\rhotildex] + p_{01}(1- \Tr[(\ketbra{0}{0}\otimes \mathds{1}^{\otimes n-1}) \rhotildex])\\
    &= (p_{00} - p_{01})\Tr[(\ketbra{0}{0}\otimes \mathds{1}^{\otimes n-1})\rhotildex] + p_{01}\\
    &\geq (p_{00} - p_{01})\frac{1}{2} + p_{01} = \frac{1}{2}(p_{00} + p_{01}) = \frac{1}{2}
\end{align*}
\end{proof}

\subsection[\texorpdfstring{\color{black}}{} Proof of \texorpdfstring{\thmref{thm:partial_upper_bound}}{}]{Proof of \texorpdfstring{\thmref{thm:partial_upper_bound}}{}} \label{app_ssec:fidelity_bound}

Here we prove the above Theorem for a bound on the size of the robust set.
\begin{thmbox}
\begin{reptheorem}{thm:partial_upper_bound}
    Consider a quantum classifier using the mixed state encoding~\eqref{eqn:mixed-state-encoding-def} and indicator cost function~\eqref{eqn:indicator_cost_over_dataset}. Assuming that a noise channel $\E$ acts only on the encoded feature vectors $\rhoxi$ (and not the encoded labels $|y_i\rangle \langle y_i |$), then
    \begin{equation} \label{eqn:partially_robust_set_bound_appendix}
         |\mathcal{R}(\mathcal{E}, E, \hat{y})| \geq M\left[1 -  \sqrt{1 - F( \E(\tilde{\sigma}), \tilde{\sigma})}\right] 
    \end{equation}
    where $F$ is the fidelity between quantum states, \eqref{eqn:fidelity_defn}.
\end{reptheorem}
\end{thmbox}

This Theorem is useful since, if one has an expression for the output fidelity of a quantum circuit after a noise channel is applied, one can directly determine a bound on how the classification results will be affected. In order to prove it, we relate the difference between the noisy cost and the noiseless cost to the size of the robust set.

\begin{proof}
    Let the indicator cost~\eqref{eqn:indicator_cost_over_dataset} of the noisy classifier be $\Cbs_\E$ and the noiseless classifier be $\Cbs$. Define
    \begin{equation} \label{eqn:change_in_cost}
        \Delta_\E \Cbs := | \Cbs_\E - \Cbs |
    \end{equation}
    to be the magnitude of the difference between noisy and noiseless costs. Due to our choice of cost function~\eqref{eqn:indicator_cost_over_dataset}, we can relate $\Delta_\E \Cbs$ to the $\delta$-robustness of the model in \defref{def:partial_noise_robustness} as follows.
     
    First, it is easy to see that the quantity $1 - \Delta_\E \Cbs$ is the the fraction $\delta$ of robust points~\eqref{eqn:delta_robust_encoding_condition} so that:
    \begin{equation} \label{eqn:connecting-robust-set-to-delta-cost}
        |\mathcal{R}(\mathcal{E}, E, \hat{y})| = M \left(1 - \Delta_\E \Cbs \right)
    \end{equation}

    Now, the PQC acts only on the ``data subsystem'' of the mixed state encoding~\eqref{eqn:mixed-state-encoding-def} so that the evolved state before measurement is:
    \begin{equation}\label{eqn:mixed_state_over_data}
        \tilde{\sigma} = \frac{1}{M} \sum_{i = 1}^{M} \rhotildexi \otimes \ketbra{y_i}{y_i}
    \end{equation}
    By assumption, the channel $\E$ leaves the ``label subsystem'' $|y_i\rangle \langle y_i |$ invariant so that:
    \begin{equation*} 
        \E (\tilde{\sigma} ) = \frac{1}{M} \sum_{i=i}^M \E \left( \rhotildexi \right) \otimes \ketbra{y_i}{y_i}
    \end{equation*}
    %
    Now, a related cost $\Cbs'$ can be evaluated by measuring the expectation of:
    \begin{equation*} 
        D := \mathds{1}^{\otimes n-1} \otimes \ZG_c \otimes \ZG_l 
    \end{equation*}
    where $c$ and $l$ denote classification and label qubits, respectively~\cite{cao_cost_2019}. That is, the noiseless cost is given by
    \begin{equation*}
        \Cbs' = \Tr(D \tilde{\sigma}) 
    \end{equation*}
    and the noisy cost is given by
    \begin{equation*}
        \Cbs'_\E = \Tr(D \E ( \tilde{\sigma}) ) .
    \end{equation*}
    Now, it is easy to see that $0 \leq\Delta_\E \Cbs' \leq2$. When all points are robust, $\Delta_\E \Cbs' = 0$, and when a single point goes from robust to not robust, $\Delta_\E \Cbs'$ increases by at most $2 / M$. As such, we can write down a corresponding inequality for the cost, $\Cbs'$, as in \eqref{eqn:connecting-robust-set-to-delta-cost} so that
    \begin{equation} \label{eqn:connecting-robust-set-to-delta-cost-prime}
        |\mathcal{R}(\mathcal{E}, E, \hat{y})| \geq M \left(1 - \frac{\Delta_\E \Cbs'}{2} \right).
    \end{equation}
    The difference in cost~\eqref{eqn:change_in_cost} due to noise is thus
    \begin{align}
        \Delta_\E \Cbs' 
            := \left| \Cbs'_\E - \Cbs' \right| &= \left|  \Tr [ D ( \E (\tilde{\sigma} ) - \tilde{\sigma}) ] \right| \nonumber \\[1.0ex]
            &\leq ||D||_\infty || \E (\tilde{\sigma}) - \tilde{\sigma} ||_1 \nonumber \\[1.0ex]
            &\leq 2 \sqrt{ 1 - F(\E (\tilde{\sigma}), \tilde{\sigma}) } . \label{eqn:fidelity_bound_mixed_state}
    \end{align}
    The third line in this derivation follows from H\"{o}lders inequality and the last line from the Fuchs-van de Graaf inequality~\cite{gentini_noise-resilient_2020, fuchs_cryptographic_1999}. We also used the fact that $||D||_\infty := \max_j | \lambda_j(D) | = 1$.
    
    Substituting the inequality~\eqref{eqn:connecting-robust-set-to-delta-cost-prime} in~\eqref{eqn:fidelity_bound_mixed_state} completes the proof.

\end{proof}

\section[\texorpdfstring{\color{black}}{} Proofs for \texorpdfstring{\chapref{chap:cloning}}{}]{Proofs for \texorpdfstring{\chapref{chap:cloning}}{}} \label{app:appendices/cloning}

\subsection[\texorpdfstring{\color{black}}{} Proof of gradients for \texorpdfstring{$\VQC$}{}]{Proof of gradients for \texorpdfstring{$\VQC$}{}}\label{app_ssec:varqlone_gradients}

We do this explicitly for the squared local cost,~\eqref{eqn:squared_local_cost_mton}, and for brevity neglect the gradient derivations for the others, since they are simpler and follow analogously. For this cost function we have the gradient with respect to a parameter, $\theta_l$ as:
\begin{multline}  \label{eqn:analytic_squared_grad_mton_2}
    \frac{\partial \Cbs_{\sq}(\boldsymbol{\theta})}{\partial \theta_l}  = 
    2{\mathbb{E}}\left[\sum\limits_{i<j} (F^i_{\Lbs}- F^j_{\Lbs}) \left(F^{i, l+ \frac{\pi}{2} }_{\Lbs} - F^{i,  l- \frac{\pi}{2} }_{\Lbs} -F^{j,  l+ \frac{\pi}{2}}_{\Lbs} + F^{j, l-\frac{\pi}{2}}_{\Lbs} \right) \right.\\
    \left. - \sum\limits_{i} (1- F^i_{\Lbs})(F^{i, l+\frac{\pi}{2}}_{\Lbs}
    - F^{i, l- \frac{\pi}{2}}_{\Lbs})  \right]
\end{multline}
where $F^{j, l\pm\pi/2}_{\Lbs}(\boldsymbol{\theta})$ denotes the fidelity of a particular state, $\ket{\psi}$, with $j^{th}$ reduced state of the $\VQC$ circuit which has the $l^{th}$ parameter shifted by $\pm \pi/2$. We suppress the $\paramtheta$ dependence in the above.

\begin{proof}
We begin as before, by taking the derivative of \eqref{eqn:squared_local_cost_mton} with respect to a single parameter, $\theta_l$:
\begin{multline*}
    \frac{\partial \Cbs_{\sq}(\boldsymbol{\theta})}{\partial \theta_l} =
    2\mathop{\mathbb{E}}_{\substack{\ket{\psi} \in \mathcal{S}}}\left[ \sum\limits_{i=1}^N (1-F^i_{\Lbs}(\paramtheta)) \left[-\frac{\partial  F^i_{\Lbs}(\boldsymbol{\theta})}{\partial \theta_l}\right]\right.\\
    \left.+ \sum\limits_{i<j}^N (F^i_{\Lbs}(\paramtheta)-F^j_{\Lbs}(\paramtheta)) \left[\frac{\partial  F^i_{\Lbs}(\boldsymbol{\theta})}{\partial \theta_l}  - \frac{\partial  F^j_{\Lbs}(\boldsymbol{\theta})}{\partial \theta_l} \right]\right] 
\end{multline*}
We can rewrite the expression for the fidelity of the $j^{th}$ clone as:
\begin{align}
    F^{j}_{\Lbs}(\boldsymbol{\theta}) = \bra{\psi}\rho_j(\boldsymbol{\theta})\ket{\psi} = \tr\left[\ket{\psi}\bra{\psi}\rho_{j}\right] =  \tr_j\left[\ket{\psi}\bra{\psi}\tr_{\bar{j}}\left(U(\boldsymbol{\theta}) \rho_{\text{init}}U(\boldsymbol{\theta})^{\dagger} \right)\right] 
\end{align}

Using the linearity of the trace, the derivative of the fidelities with respect to the parameters, $\theta_l$, can be computed:
\begin{align}
    \frac{\partial   F^{j}_{\Lbs}(\boldsymbol{\theta})}{\partial \theta_l} =   \tr_j\left[\ket{\psi}\bra{\psi}\tr_{\bar{j}}\left(\frac{\partial  U(\boldsymbol{\theta})\rho_{\text{init}}U(\boldsymbol{\theta})^{\dagger}}{\partial \theta_l} \right)\right] 
\end{align}
Now plugging in the parameter shift rule:
\begin{align} \label{eqn:cloning_gradient_proof_1}
    \frac{\partial  U(\boldsymbol{\theta})\rho_{\text{init}}U(\boldsymbol{\theta})^{\dagger}}{\partial \theta_l} = U^{l+\frac{\pi}{2}}(\boldsymbol{\theta})\rho_{\text{init}}(U(\boldsymbol{\theta})^{l+\frac{\pi}{2}})^{\dagger} -  U^{l-\frac{\pi}{2}}(\boldsymbol{\theta})\rho_{\text{init}}(U(\boldsymbol{\theta})^{l-\frac{\pi}{2}})^{\dagger}
\end{align}
Where the notation, $U^{l\pm\frac{\pi}{2}}$, indicates the $l^{th}$ parameter has been shifted by $\pm \frac{\pi}{2}$, i.e. $U^{l\pm\frac{\pi}{2}} := U(\theta_d) \dots U(\theta_l \pm  \sfrac{\pi}{2}) \dots  U(\theta_{1})$. Now:
\begin{align}
    \frac{\partial  F^{j}_{\Lbs}(\boldsymbol{\theta})}{\partial \theta_l} &=   \tr_j\left[\ket{\psi}\bra{\psi}\tr_{\bar{j}}\left(U^{l+\frac{\pi}{2}}(\boldsymbol{\theta})\rho_{\text{init}}(U(\boldsymbol{\theta})^{l+\frac{\pi}{2}})^{\dagger} \right)\right] \nonumber  \\
    &\qquad -   \tr_j\left[\ket{\psi}\bra{\psi}\tr_{\bar{j}}\left(U^{l-\frac{\pi}{2}}(\paramtheta)\rho_{\text{init}}(U(\boldsymbol{\theta})^{l-\frac{\pi}{2}})^{\dagger} \right)\right] \nonumber \\
        \implies \frac{\partial   F^{ j}_{\Lbs}(\boldsymbol{\theta})}{\partial \theta_l} &=   \tr\left[\ket{\psi}\bra{\psi}\rho_j^{l+\frac{\pi}{2}}(\boldsymbol{\theta})\right]
    -   \tr\left[\ket{\psi}\bra{\psi}\rho_j^{l-\frac{\pi}{2}}(\boldsymbol{\theta})\right]\\
    &= F^{(j, l+\frac{\pi}{2})}_{\Lbs}(\boldsymbol{\theta}) - F^{(j, l-\frac{\pi}{2})}_{\Lbs}(\boldsymbol{\theta}) \label{eqn:cloning_gradient_proof_2}
\end{align}
where we define $F^{(l\pm \frac{\pi}{2})}_j(\boldsymbol{\theta}) \coloneqq \bra{\psi}\rho_j^{l\pm\frac{\pi}{2}}(\boldsymbol{\theta})\ket{\psi}$ the fidelity of the $j^{th}$ clone, when prepared using a unitary whose $l^{th}$ parameter is shifted by $\pm \frac{\pi}{2}$, with respect to a target input state, $\ket{\psi}$. 

Plugging \eqref{eqn:cloning_gradient_proof_2} into \eqref{eqn:cloning_gradient_proof_1}, we recover the expression in \eqref{eqn:analytic_squared_grad_mton_2}.
\end{proof}

\subsection[\texorpdfstring{\color{black}}{} Proof of~\texorpdfstring{\thmref{thm:squared_local_cost_squared_FS_weak_faithful_appendix}}{}]{Proof of~\texorpdfstring{\thmref{thm:squared_local_cost_squared_FS_weak_faithful_appendix}}{}} \label{app_ssec:proof_faithfulness}

\begin{thmbox}
\begin{reptheorem}{thm:squared_local_cost_squared_FS_weak_faithful_appendix}
The squared cost function as defined~\eqref{eqn:squared_local_cost_mton_supp_redo}, is $\epsilon$-weakly faithful with respect to the Bures angle distance measure $\text{d}_{\BA}$.
In other words, if we have:
\begin{equation}\label{eqn:squared_cost_to_epsilon_appendix_2}
    \Cbs_{\sq}(\paramtheta) - \Cbs^{\opt}_{\sq} \leq \epsilon
\end{equation}
where $\Cbs^{\opt}_{\sq} := (F_i(\paramtheta)-F_j(\paramtheta))^2 = N(1-F_{\opt})^2$ is the optimal cost, then the following fact holds:

\begin{equation}     \label{eqn:fubini_study_bound_squared_appendix_2}
    \text{d}_{\BA}(\rho^{\psi, j}_{\paramtheta}, \rho_{\opt}^{\psi, j}) \leq \frac{\mathcal{N}\epsilon}{2(1 - F_{\opt})\sin(F_{\opt})} := f_1(\epsilon),   \qquad \forall \ket{\psi} \in \mathcal{S}, \forall j \in [N]
\end{equation}
\end{reptheorem}
\end{thmbox}

\begin{proof}
To prove \thmref{thm:squared_local_cost_squared_FS_weak_faithful_appendix}, we revisit and rewrite the Bures angle from \eqref{eqn:bures_angle_definition} as~\cite{nielsen_quantum_2010}:
\begin{equation} \label{eqn:fubini_study_defn_appendix_2}
    \text{d}_{\BA}(\rho,\sigma) = \arccos{\sqrt{F(\rho, \sigma)}} = \arccos \bra{\phi}\tau\rangle
\end{equation}
where $\ket{\phi}$ and $\ket{\tau}$ are the purifications\footnote{A purification of a mixed state, $\rho\in \hilb_A$, is a pure state, $\ket{\psi}$, in a larger Hilbert space ($\hilb_A\otimes \hilb_B$) such that tracing out $B$ leaves the original state, $\rho$, $\rho=\tr_B(\ketbra{\psi}{\psi})$.} of $\rho$ and $\sigma$ respectively which maximise the overlap. We note that $\text{d}_{\BA}(\rho,\sigma)$ lies between $[0, \pi/2]$, with $\text{d}_{\BA}(\rho,\sigma) = \pi/2$ corresponding to $\rho = \sigma$. Since this distance is a metric,  it obeys the triangle inequality (\defref{defn:metric}), i.e., for any three states $\rho, \sigma$ and $\delta$,
\begin{equation}
     \text{d}_{\BA}(\rho,\sigma) \leq  \text{d}_{\BA}(\rho,\delta) +  \text{d}_{\BA}(\sigma,\delta)
     \label{eqn:triangle-inequality_2}
\end{equation}

Rewriting the result of~\lemref{lemma:trace_bound_squared_cost} in terms of fidelity for each $\ket{\psi} \in \mathcal{S}$ and correspondingly in terms of Bures angle using \eqref{eqn:fubini_study_defn_appendix_2} is, 
\begin{equation}
    \begin{split}
    F(\rho_{\opt}^{\psi, j}, \ket{\psi}) - F(\rho^{\psi,j}_{\paramtheta}, \ket{\psi}) &\leq \epsilon' \\
    \implies  \cos^2(\text{d}_{\BA}(\rho_{\opt}^{\psi, j}, \ket{\psi})) -  \cos^2(\text{d}_{\BA}(\rho^{\psi,j}_{\paramtheta}, \ket{\psi})) &\leq \epsilon' 
    \end{split}
    \label{eqn:fidelityFS_squared_local_2}
\end{equation}
where $\epsilon' = \frac{\mathcal{N}\epsilon}{2(1 - F_{\opt})}$. 
Let us denote $D_{\pm}^{\psi} = \text{d}_{\BA}(\rho_{\opt}^{\psi, j}, \ket{\psi}) \pm \text{d}_{\BA}(\rho^{\psi,j}_{\paramtheta}, \ket{\psi})$
This inequality in \eqref{eqn:fidelityFS_squared_local_2} can be further rewritten as,
\begin{small}
    \begin{equation}
    \begin{split}
        \cos(\text{d}_{\BA}(\rho_{\opt}^{\psi, j}, \ket{\psi})) -  \cos(\text{d}_{\BA}(\rho^{\psi,j}_{\paramtheta}, \ket{\psi})) &\leq \frac{\epsilon'}{\cos(\text{d}_{\BA}(\rho_{\opt}^{\psi, j}, \ket{\psi})) +  \cos(\text{d}_{\BA}(\rho^{\psi,j}_{\paramtheta}, \ket{\psi}))} \\
        \cos(\text{d}_{\BA}(\rho_{\opt}^{\psi, j}, \ket{\psi})) -  \cos(\text{d}_{\BA}(\rho^{\psi,j}_{\paramtheta}, \ket{\psi})) &\lessapprox \frac{\epsilon'}{2\cos(\text{d}_{\BA}(\rho_{\opt}^{\psi, j}, \ket{\psi}))} \\
        2\sin\left(\frac{D^{\psi}_{+}}{2}\right)\sin\left(\frac{\text{d}^{\psi}_{-}}{2}\right) &\leq \frac{\epsilon'}{2\cos(\text{d}_{\BA}(\rho_{\opt}^{\psi, j}, \ket{\psi}))} \\
        \implies \text{d}^{\psi}_{-} &\leq \frac{\epsilon'}{\sin(\text{d}_{\BA}(\rho_{\opt}^{\psi, j}, \ket{\psi}))}  = \frac{\mathcal{N}\epsilon}{2(1 - F_{\opt})\sin(F_{\opt})}
    \end{split}
    \label{eqn:cosinerelation_2}
    \end{equation}
\end{small}
In the above, we use the approximations that as $\epsilon \rightarrow 0$, $\text{d}_{\BA}(\rho_{\opt}^{\psi, j}, \ket{\psi}) \approx \text{d}_{\BA}(\rho^{\psi,j}_{\paramtheta}, \ket{\psi})$ and the trigonometric identities $\cos\left( x - y \right) = 2\sin \left(\frac{x+y}{2}\right)\sin \left(\frac{x-y}{2}\right)$, and $\sin 2x = 2\sin x \cos x$. 

Now, using the triangle inequality on the states $\{\rho_{\opt}^{\psi, j}, \rho^{\psi,j}_{\paramtheta}, \ket{\psi}\}$, we have:
\begin{equation}
    \text{d}_{\BA}(\rho^{\psi,j}_{\paramtheta}, \ket{\psi}) \leq \text{d}_{\BA}(\rho_{\opt}^{\psi, j}, \ket{\psi}) + \text{d}_{\BA}(\rho^{\psi,j}_{\paramtheta}, \rho_{\opt}^{\psi, j}) 
\end{equation}
Combining the above inequality and \eqref{eqn:cosinerelation_2} completes the proof:
\begin{equation} \label{eqn:squared_cost_fubini_study_bound_appendix_2}
    \text{d}_{\BA}(\rho^{\psi,j}_{\paramtheta}, \rho_{\opt}^{\psi, j}) \leq \frac{\mathcal{N}\epsilon}{2(1 - F_{\opt})\sin(F_{\opt})}, \hspace{3mm} \forall \ket{\psi} \in \mathcal{S}
\end{equation}
\end{proof}

\subsection[\texorpdfstring{\color{black}}{} Proof of~\texorpdfstring{\thmref{thm:squared_local_cost_squared_trace_weak_faithful_appendix}}{}]{Proof of~\texorpdfstring{\thmref{thm:squared_local_cost_squared_trace_weak_faithful_appendix}}{}}\label{app_ssec:proof_faithfulness_trace}

Here we give the proof of~\thmref{thm:squared_local_cost_squared_trace_weak_faithful_appendix}. We begin by repeating the Theorem:
\begin{thmbox}
\begin{reptheorem}{thm:squared_local_cost_squared_trace_weak_faithful_appendix}
The squared cost function, \eqref{eqn:squared_local_cost_mton_supp_redo}, is $\epsilon$-weakly faithful with respect to the trace distance $\text{d}_{\tr}$.
\begin{equation}     \label{eqn:trace_distance_bound_squared_appendix_2}
        \text{d}_{\tr}(\rho_{\opt}^{\psi, j},  \rho^{\psi, j}_{\paramtheta})  \leq g_1(\epsilon), \qquad \forall j \in [N]
\end{equation}
where:
\begin{equation} \label{eqn:trace_distance_squared_cost_bound_function_2}
    g_1(\epsilon) \approx \frac{1}{2}\sqrt{4F_{\opt}(1 - F_{\opt}) + \epsilon\frac{\mathcal{N}(1 - 2F_{\opt})}{2(1 - F_{\opt})}}
\end{equation}
\end{reptheorem}
\end{thmbox}

\begin{proof}

Firstly, we note that $F_{\opt} = \bra{\psi}\rho_{\opt}^{\psi, j}\ket{\psi}$ is the same value for all input states $\ket\psi \in \mathcal{S}$. We apply the change of basis from $\ket{\psi} \rightarrow \ket{0}$ by applying the unitary $V\ket{\psi} = \ket{0}$. Then the effective change on the state $\rho_{\opt}^{\psi, j}$ to have a fidelity $F_{\opt}$ with the state $\ket{0}$ is, $\rho_{\opt}^{\psi, j} \rightarrow V\rho_{\opt}^{\psi, j}V^{\dagger}$, written as:
\begin{equation}
    V\rho_{\opt}^{\psi, j}V^{\dagger} = \begin{pmatrix}
F_{\opt} & a^*\\ 
a & 1- F_{\opt}
\end{pmatrix}
\end{equation}
where we use the usual properties of a density matrix and $a \in \mathbb{C}$. The upper bound condition in \eqref{eqn:tracecloseness_squared_local_appendix} states that:
\begin{equation*}
    \bra{\psi}\rho_{\opt}^{\psi, j} - \rho^{\psi, j}_{\paramtheta}\ket{\psi} = \bra{0}V(\rho_{\opt}^{\psi, j} - \rho^{\psi, j}_{\paramtheta})V^{\dagger}\ket{0}
= \frac{\mathcal{N}\epsilon}{2(1 - F_{\opt})} =: \epsilon'
\end{equation*}
$V$ has the following effect on $\rho^{\psi, j}_{\paramtheta}$:
\begin{equation}
    V\rho^{\psi, j}_{\paramtheta}V^{\dagger} = \begin{pmatrix}
F_{\opt} + \epsilon' & b^*\\ 
b & 1- (F_{\opt} + \epsilon')
\end{pmatrix}
\end{equation}
for some $b \in \mathbb{C}$. The condition that $V\rho_{\opt}^{\psi, j}V^{\dagger}, V\rho^{\psi, j}_{\paramtheta}V^{\dagger} \geqslant 0$ i.e. they are positive, implies that,
\begin{equation}
 |a|^2 \leq F_{\opt}(1 - F_{\opt}) =: r_{F_{\opt}}^2, \hspace{3mm} |b|^2 \leq (F_{\opt} + \epsilon')(1 - [F_{\opt} + \epsilon']) =: r_{F_{\opt} + \epsilon'}
\end{equation}
Using the relationship between the trace distance and the positive eigenvalue of the difference between the states, we consider the eigenvalues of $V\rho_{\opt}^{\psi, j}V^{\dagger} -  V\rho^{\psi, j}_{\paramtheta}V^{\dagger}$. The two eigenvalues of this matrix is $\lambda_{\pm} = \pm \sqrt{\epsilon'^2 + |a - b|^2}$. From this, the trace distance between the two states is,
\begin{equation}
    d_{\tr}(V\rho_{\opt}^{\psi, j}V^{\dagger}, V\rho^{\psi, j}_{\paramtheta}V^{\dagger}) = \frac{1}{2}\left|\left|V\rho_{\opt}^{\psi, j}V^{\dagger} - V\rho^{\psi, j}_{\paramtheta}V^{\dagger}\right|\right| = \frac{1}{2}|\lambda_{+}| = \frac{1}{2}\sqrt{\epsilon'^2 + |a - b|^2}
\end{equation}
We note that the trace distance is unitary invariant. Thus, 
\begin{equation}
    \begin{split}
        d_{\tr}(\rho_{\opt}^{\psi, j},  \rho^{\psi, j}_{\paramtheta})  &= d_{\tr}(V\rho_{\opt}^{\psi, j}V^{\dagger}, V\rho^{\psi, j}_{\paramtheta}V^{\dagger}) \\
        &= \frac{1}{2}\sqrt{\epsilon'^2 + |a - b|^2} \\
        &\leq \frac{1}{2}\sqrt{\epsilon'^2 + (r_{F_{\opt}} + r_{F_{\opt} + \epsilon'})^2} \\
        &\approx \frac{1}{2}\sqrt{4F_{\opt}(1 - F_{\opt}) + \epsilon'(1 - 2F_{\opt})} \\
        &=  \frac{1}{2}\sqrt{4F_{\opt}(1 - F_{\opt}) + \epsilon\frac{\mathcal{N}(1 - 2F_{\opt})}{2(1 - F_{\opt})}}
    \end{split}
    \label{eqn:trace-distance-closeness}
\end{equation}
where we have used the inequality $|a - b|^2 \leq \left||a| + |b|\right|^2$ for all $a,b \in \mathbb{C}$.

\end{proof}

\subsection[\texorpdfstring{\color{black}}{} Proof of~\texorpdfstring{\thmref{thm:relationship-local-global-appendix}}{}]{Proof of~\texorpdfstring{\thmref{thm:relationship-local-global-appendix}}{}}\label{app_ssec:proof_local_gloabl_relationship}

Here we give the proof of~\thmref{thm:relationship-local-global-appendix} which is repeated below:
\begin{thmbox}
\begin{reptheorem}{thm:relationship-local-global-appendix}
For the general case of $M \rightarrow N$ cloning, the global cost function $\Cbs_{\Gbs}(\paramtheta)$ and the local cost function $\Cbs_{\Lbs}(\paramtheta)$ satisfy the inequality,
\begin{equation}
 \Cbs_{\Lbs}(\paramtheta) \leq \Cbs_{\Gbs}(\paramtheta) \leq N\cdot\Cbs_{\Lbs}(\paramtheta)  
\end{equation}
\end{reptheorem}
\end{thmbox}
\begin{proof}
We first prove the first part of the inequality,
\begin{small}
    \begin{equation}
    \begin{split}
        \Cbs_{\Gbs}(\paramtheta) - \Cbs_{\Lbs}(\paramtheta) &= \frac{1}{\mathcal{N}}\int_{\mathcal{S}}\tr((\mathsf{O}_{\Gbs}^{\psi} - \mathsf{O}_{\Lbs}^{\psi})\rho_{\paramtheta}^{\psi})d\psi \\
        &= \frac{1}{\mathcal{N}N}\int_{\mathcal{S}}\tr\left(\left(
        \sum\limits_{j=1}^N\left(\ketbra{\psi}{\psi}_j \otimes \mathds{1}_{\Bar{j}} - \ketbra{\psi}{\psi}_1 \otimes \cdots \ketbra{\psi}{\psi}_N \right)\right)\rho^{\psi}_{\paramtheta}\right) \geq 0\\
        &\qquad\implies \Cbs_{\Gbs}(\paramtheta) \geq \Cbs_{\Lbs}(\paramtheta)
    \end{split}   
    \label{eqn:global-local}
    \end{equation}
\end{small}
where $\mathsf{O}_{\Lbs}^{\psi}$ is defined in \eqref{eqn:local_cost_observable}, and the inequality in the second line holds because
\begin{multline}
 \sum\limits_{j=1}^N\left(\ketbra{\psi}{\psi}_j \otimes \mathds{1}_{\Bar{j}} - \ketbra{\psi}{\psi}_1 \otimes \cdots \ketbra{\psi}{\psi}_N \right) \\
 =  \sum\limits_{j=1}^N\ketbra{\psi}{\psi}_j \otimes (\mathds{1}_{\Bar{j}} - \ketbra{\psi}{\psi}_{\Bar{j}}) \geq 0, \qquad \forall \ket{\psi} \in \mathcal{S}   
\end{multline}

For the second part of the inequality, we consider the operator $N \mathsf{O}_{\Lbs}^{\psi} - \mathsf{O}_{\Gbs}^{\psi}$,
\begin{equation}
     \begin{split}
         N \mathsf{O}_{\Lbs}^{\psi} - \mathsf{O}_{\Gbs}^{\psi} &= (N - 1)\mathds{1} - \sum_{j=1}^{N}\left(\ketbra{\psi}{\psi}_j \otimes \mathds{1}_{\Bar{j}}\right) + \ketbra{\psi}{\psi}_1 \otimes \cdots \otimes \ketbra{\psi}{\psi}_N \\
         &= \sum_{j = 1}^{N-1}\left(\mathds{1}_j \otimes \mathds{1}_{\Bar{j}} - \ketbra{\psi}{\psi}_j \otimes \mathds{1}_{\Bar{j}}\right) - \ketbra{\psi}{\psi}_{N} \otimes \mathds{1}_{\Bar{N}} + \ketbra{\psi}{\psi}_1 \otimes \cdots \otimes \ketbra{\psi}{\psi}_N \\
         &= \sum_{j = 1}^{N-1}\left((\mathds{1} - \ketbra{\psi}{\psi})_{j} \otimes \mathds{1}_{\Bar{j}}\right) - \bigotimes_{j=1}^{N-1}(\mathds{1} - \ketbra{\psi}{\psi})_j \otimes \ketbra{\psi}{\psi}_N \\
         &= (\mathds{1} - \ketbra{\psi}{\psi})_{1}\otimes \left(\mathds{1}_{\Bar{1}} - \bigotimes_{j=2}^{N-1}(\mathds{1} -  \ketbra{\psi}{\psi}_{j}) \otimes \ketbra{\psi}{\psi}_N\right) \\
         &\qquad + \sum_{j = 2}^{N-1}\left((\mathds{1} - \ketbra{\psi}{\psi})_{j}) \otimes \mathds{1}_{\Bar{j}}\right) \\
         &\geq 0
     \end{split}
\end{equation}   
where the second last line is positive because each individual operator is positive for all $\ket{\psi} \in \mathcal{S}$.
\end{proof}

\subsection[\texorpdfstring{\color{black}}{} Proof of~\texorpdfstring{\thmref{thm:asymmetric_cost_FS_weak_faithful_appendix}}{}]{Proof of~\texorpdfstring{\thmref{thm:asymmetric_cost_FS_weak_faithful_appendix}}{}}\label{app_ssec:asymetric_faithful}

Here we give the proof of faithfulness for the asymmetric cost function,~\thmref{thm:asymmetric_cost_FS_weak_faithful_appendix}.
\begin{thmbox}
\begin{reptheorem}{thm:asymmetric_cost_FS_weak_faithful_appendix}
The asymmetric cost function, \eqref{eqn:asymmetric_cost_full_appendix}, is $\epsilon$-weakly faithful with respect to $\text{d}_{\BA}$:
\begin{equation}  \label{eqn:asymmetric_cost_function_epsilon_guarantee_2}
 \Cbs_{\Lbs, \mathrm{asym}}(\paramtheta) - \Cbs^{\opt}_{\Lbs, \mathrm{asym}} \leq \epsilon   
\end{equation}
where $\Cbs^{\opt}_{\Lbs, \mathrm{asym}} = 0$. then the following fact holds for Bob and Eve's reduced states::
%
\begin{equation} \label{eqn:asymmetric_cost_function_FS_bound_appendix_2}
     \text{d}_{\BA}(\rho^{\psi,B}_{\paramtheta}, \rho^{\psi,B}_{\opt}) \leq \frac{\sqrt{\mathcal{N}\epsilon}}{\sin(1 - p^2/2)}, \hspace{3mm} \text{d}_{\BA}(\rho^{\psi,E}_{\paramtheta}, \rho^{\psi,E}_{\opt}) \leq \frac{\sqrt{\mathcal{N} \epsilon}}{\sin(1 - q^2/2)}
\end{equation}
Furthermore, we also have the following trace distance bounds:
\begin{align}    \label{eqn:asymmmetric_trace_distance_bound_Bob_Eve_2}
  \text{d}_{\tr}(\rho^{\psi, B},  \rho^{\psi, B}_{\paramtheta})  
        &\leq \frac{1}{2}\sqrt{p^2(2 - p^2) - \sqrt{\mathcal{N}\epsilon}(1 - p^2)}, \\
        \text{d}_{\tr}(\rho^{\psi, E},  \rho^{\psi, E}_{\paramtheta})  
        &\leq \frac{1}{2}\sqrt{q^2(2 - q^2) - \sqrt{\mathcal{N}\epsilon}(1 - q^2)} 
\end{align}

\end{reptheorem}
\end{thmbox}

\begin{proof}
Firstly, we derive a similar result to \lemref{lemma:trace_bound_squared_cost} and \lemref{lemma:trace_bound_local_cost}. By expanding the term $|\Cbs_{\Lbs, \text{asym}}(\paramtheta) - \Cbs_{\Lbs, \opt}|$ in terms of the corresponding output states, we obtain,
\begin{equation} \label{eqn:asymmetric_optimal_cost_expanded_2}
\begin{split}
     &|\Cbs_{\Lbs,\text{asym}}(\paramtheta) - \Cbs_{\Lbs, \opt}| = 
     \left|\frac{1}{\mathcal{N}}\int_{\mathcal{S}}\left((F_{\Lbs}^{p, B}- F_{\Lbs}^{B}(\paramtheta))^2 + (F_{\Lbs}^{p, E} - F_{\Lbs}^{ E}(\paramtheta))^2\right)d\psi \right| \\
     &= \frac{1}{\mathcal{N}}\int_{\mathcal{S}}\left[\left(\tr[(\rho^{\psi, B} - \rho^{\psi, B}_{\paramtheta})\ket{\psi}\bra{\psi}] \right)^2 + \left(\tr[(\rho^{\psi, E} - \rho^{\psi, E}_{\paramtheta})\ket{\psi}\bra{\psi}] \right)^2 \right]d\psi
\end{split}
\end{equation}
Using the inequality \eqref{eqn:asymmetric_cost_function_epsilon_guarantee_2} and \eqref{eqn:asymmetric_optimal_cost_expanded_2}, we get,
\begin{equation}\label{eqn:aymmmetric_cost_local_trace_bound_2}
     \frac{1}{\mathcal{N}}\int_{\mathcal{S}}\left(\text{Tr}[(\rho^{\psi, j}_{\opt} - \rho^{\psi, j}_{\paramtheta})\ket{\psi}\bra{\psi}] \right)^2 d\psi \leq \epsilon 
     \implies \text{Tr}[(\rho^{\psi, j}_{\opt} - \rho^{j}_{\paramtheta})\ket{\psi}\bra{\psi}]\leq \sqrt{\mathcal{N}\epsilon}
\end{equation}
where $j \in \{B,E\}$.

Next, to derive \eqref{eqn:asymmetric_cost_function_FS_bound_appendix_2} we rewrite the \eqref{eqn:aymmmetric_cost_local_trace_bound_2} in terms of the Bures angle:
\begin{equation}    \label{eqn:fidelityFS_asymmetric_appendix_2}
\begin{split}
    &F(\rho^{\psi, j}_{\opt}, \ket{\psi}) - F(\rho^{\psi, j}_{\paramtheta}, \ket{\psi}) \leq \sqrt{\mathcal{N}\epsilon}\\ \implies  &\cos^2(\text{d}_{\BA}(\rho^{\psi, j}_{\opt}, \ket{\psi})) -  \cos^2(\text{d}_{\BA}(\rho^{\psi, j}_{\paramtheta}, \ket{\psi})) \leq \sqrt{\mathcal{N}\epsilon} 
    \end{split}
\end{equation}
Following the derivations in the previous sections, we have:
\begin{equation}     \label{eqn:FSbound-asym-standard-local_2}
    \text{d}_{\BA}(\rho^{\psi,j}_{\paramtheta}, \rho^{\psi, j}_{\opt}) \leq \frac{\sqrt{\mathcal{N}\epsilon}}{\sin(F_{\Lbs}^{r, j})}, \hspace{3mm} \forall \ket{\psi} \in \mathcal{S}
\end{equation}
where $F_{\Lbs}^{r, j}$ is the optimal cloning fidelity corresponding to $j \in \{B,E\}$ with $r \in \{p, q\}$. Finally, plugging in the optimal asymmetric fidelities, $F_{\Lbs}^{p, B} = 1-p^2/2$, and similarly for $F_{\Lbs}^{q, E}$ we arrive at:
\begin{equation}
     \text{d}_{\BA}(\rho^{\psi,B}_{\paramtheta}, \rho^{\psi,B}_{\opt}) \leq \frac{\sqrt{\mathcal{N} \epsilon}}{\sin(1 - p^2/2)}, \qquad \text{d}_{\BA}(\rho^{\psi,E}_{\paramtheta}, \rho^{\psi,E}_{\opt}) \leq \frac{\sqrt{\mathcal{N}\epsilon}}{\sin(1 - q^2/2)}
\end{equation}
Finally, to prove \eqref{eqn:asymmmetric_trace_distance_bound_Bob_Eve_2}, we follow the trace distance derivation bounds as in previous sections and obtain:
\begin{equation}     \label{eqn:asym-trace-distance-closeness_2}
        \text{d}_{\tr}(\rho^{\psi, j},  \rho^{\psi,j}_{\paramtheta})
        \leq \frac{1}{2}\sqrt{4F_{\Lbs}^{r, j}(1 - F_{\Lbs}^{r, j}) + \sqrt{\mathcal{N}\epsilon}(1 - 2F_{\Lbs}^{r, j} ) }
\end{equation}
Again, plugging in the optimal fidelities for Bob and Eve completes the proof.

\end{proof}

\subsection[\texorpdfstring{\color{black}}{} Proof of~\texorpdfstring{\thmref{thm:sample-complexity-appendix}}{}]{Proof of~\texorpdfstring{\thmref{thm:sample-complexity-appendix}}{}}\label{app_ssec:vqc_sample_complexity}

Here we prove~\thmref{thm:sample-complexity-appendix} in the main text.
\begin{thmbox}
\begin{reptheorem}{thm:sample-complexity-appendix}[Sample complexity of $\VQC$]
The number of samples $L \times K$ required to estimate the $\VQC$ cost function $\Cbs(\paramtheta)$ up to $\epsilon'$-additive error with a success probability $\delta$ is:
\begin{equation}\label{eqn:number-of-samples-appendix_2}
    L \times K =  \mathcal{O}\left(\frac{1}{\epsilon'^2}\log \frac{2}{\delta}\right)
\end{equation}
where $K$ is the number of distinct states $\ket{\psi}$ sampled uniformly at random from the distribution $\mathcal{S}$, and $L$ is the number of copies of each input state. 
\end{reptheorem}
\end{thmbox}
\begin{proof}

We provide the proof for the cost function $\Cbs_{\Gbs}(\paramtheta)$. However, this proof extends in a straightforward manner to other cost functions. As a reminder, the global cost function is defined as,
\begin{equation}
    \Cbs_{\Gbs}^{\psi}(\paramtheta) = 1 - \bra{\phi}\rho_{\paramtheta}^{\psi}\ket{\phi} \qquad \implies \qquad \Cbs_{\Gbs}(\paramtheta) = 1 - \frac{1}{\mathcal{N}}\int_{\mathcal{S}}\bra{\phi}\rho^{\psi}_{\paramtheta}\ket{\phi}d\psi
\end{equation}
The estimation of $\Cbs_{\Gbs}^{\psi}(\paramtheta)$ requires the computation of the overlap $\bra{\phi}\rho_{\paramtheta}^{\psi}\ket{\phi}$. Let us denote $\ket{\phi} := \ket{\psi}^{\otimes N}$ as an $N$-fold tensor product of the input state. 

This test inputs the states $\ket{\phi}$ and $\rho^{\psi}_{\paramtheta}$ with an additional ancilla qubit $\ket{0}$, and measures the ancilla in the end in the computational basis. The probability of obtaining an outcome `1' in the measurement is (recall $1-\text{Pr}[\ket{\text{anc}} = \ket{0}]$ from \eqref{eqn:swap_test_probability}):
\begin{equation}
 \text{Pr}[\ket{\text{anc}} = \ket{1}] = p^{\psi} = \frac{1}{2}(1 -  \bra{\phi}\rho_{\paramtheta}^{\psi}\ket{\phi})   
\end{equation}
From this, we see that $\Cbs^{\psi}_{\Gbs}(\paramtheta) = 2p^{\psi}$. To estimate this cost function value, we run the $\SWAP$ test $L$ times and estimate the averaged number of `1' outcomes. Let us the define the estimator for cost function with $L$ samples to be,
\begin{equation}
 \widehat{\Cbs}^{\psi}_{\Gbs, \text{avg}}(\paramtheta) = \frac{1}{L}\sum_{i=1}^{L} \widehat{\Cbs}^{\psi}_{\Gbs, i}(\paramtheta)    
\end{equation}
where $\widehat{\Cbs}^{\psi}_{\Gbs, i}(\paramtheta)$ is equal to 2 if the $\SWAP$ test outcome at the $i$-th run is `1' and is 0 otherwise. From this we can see that the expected value of $ \widehat{\Cbs}^{\psi}_{\Gbs, \text{avg}}(\paramtheta)$ is,
\begin{equation}
    \mathbb{E}\left[ \widehat{\Cbs}^{\psi}_{\Gbs, \text{avg}}(\paramtheta)\right] = \Cbs^{\psi}_{\Gbs}(\paramtheta) = 2p^{\psi}
\end{equation}
Now consider $K$ different states $\{\ket{\psi_1},\cdots\ket{\psi_K}\}$ are chosen uniformly at random from the distribution $\mathcal{S}$. The average value of the cost over $K$ is:
\begin{equation}
 \widehat{\Cbs}_{\Gbs, \text{avg}}(\paramtheta) = \frac{1}{K}\sum_{j=1}^{K}\widehat{\Cbs}^{\psi_j}_{\Gbs, \text{avg}}(\paramtheta) = \frac{1}{LK}\sum_{j=1}^{K}\sum_{i=1}^{L} \widehat{\Cbs}^{\psi_j}_{\Gbs, i}(\paramtheta)    
\end{equation}
with,
\begin{equation}
    \mathbb{E}\left[\widehat{\Cbs}_{\Gbs, \text{avg}}(\paramtheta)\right] = \frac{1}{K}\sum_{j=1}^{K}\frac{1}{\mathcal{N}}\int_{\mathcal{S}} \widehat{\Cbs}^{\psi}_{\Gbs, \text{avg}}(\paramtheta) d\psi = \Cbs_{\Gbs}(\paramtheta)
\end{equation}
Using H\"{o}effding's inequality~\cite{hoeffding_probability_1963}, we obtain a probabilistic bound on $| \widehat{\Cbs}_{\Gbs, \text{avg}}(\paramtheta) - \Cbs_{\Gbs}(\paramtheta)|$,
\begin{equation}\label{eqn:Hoeffeding1-appendix}
    \text{Pr}\left[| \widehat{\Cbs}_{\Gbs, \text{avg}}(\paramtheta) - \Cbs_{\Gbs}(\paramtheta)| \geq \epsilon'\right] \leqslant 2e^{-2KL\epsilon'^2}
\end{equation}
Now, setting $2e^{-2KL\epsilon'^2} = \delta$ and solving for $L\times K$ gives:
\begin{equation}
    L \times K = \mathcal{O}\left(\frac{1}{\epsilon'^2}\log \frac{2}{\delta}\right)
\end{equation}
\end{proof}

\subsection[\texorpdfstring{\color{black}}{} Proof of~\texorpdfstring{\thmref{thm:aharonov_4state_attack_II_bias_probability_appendix}}{}]{Proof of~\texorpdfstring{\thmref{thm:aharonov_4state_attack_II_bias_probability_appendix}}{}}\label{app_ssec:proof_aharonov_attack_II}

Here we prove~\thmref{thm:aharonov_4state_attack_II_bias_probability_appendix}. We repeat the Theorem first:

\begin{thmbox}
\begin{reptheorem}{thm:aharonov_4state_attack_II_bias_probability_appendix}[Ideal cloning attack (II) bias on $\mathcal{P}_2$ in scenario $1$.]~ \\
Using a cloning attack on the protocol, $\mathcal{P}_2$, (in attack model II with $4$ states) Bob can achieve a bias:
\begin{equation}\label{eqn:attack_4_state_aharonov_success_probability_exact_appendix_2}
    \epsilon^{\mathrm{II}}_{\mathcal{P}_2, \mathrm{ideal}} = 0.25
\end{equation}
\end{reptheorem}
\end{thmbox}
\begin{proof}
Considering the 4 states to be in the $\mathsf{X} - \mathsf{Z}$ plane of the Bloch sphere, the density matrices of each state can be represented as:
\begin{equation} \label{eqn:4state-density-bloch}
\rho_{ij} = \frac{1}{2}(\mathds{1} + m^{ij}_x\XG + m^{ij}_z\ZG)
\end{equation}
where $\XG$ and $\ZG$ are Pauli matrices and each $m^{ij}$ is a three dimensional vector given by:
\begin{equation} \label{eqn:4state-cloner-vectors-bloch}
\begin{split}
    & m^{00} := [\sin(2\phi),~ 0,~ \cos(2\phi)]\\
    & m^{01} := [-\sin(2\phi),~ 0,~ -\cos(2\phi)]\\
    & m^{10} := [-\sin(2\phi),~ 0,~ \cos(2\phi)]\\
    & m^{11} := [\sin(2\phi),~ 0,~ -\cos(2\phi)]
\end{split}
\end{equation}
After the cloning (in the ideal case), the density matrix of each clone will become:
\begin{equation} \label{eqn:4state-cloner-density-bloch}
\rho_{ij}^c = \frac{1}{2}(\mathds{1} + \eta_x m^{ij}_x\XG + \eta_z m^{ij}_z\ZG)
\end{equation}
where $\eta_x$ and $\eta_z$ are the shrinking factors in each direction given as follows:
\begin{equation} \label{eqn:4state-cloner-shrinking-factors}
\eta_x = \sin^2(2\phi)\sqrt{\frac{1}{\sin^4(2\phi) + \cos^4(2\phi)}}, \quad
\eta_z = \cos^2(2\phi)\sqrt{\frac{1}{\sin^4(2\phi) + \cos^4(2\phi)}}
\end{equation}
For the states used in $\mathcal{P}_2$, we have $\phi = \frac{\pi}{8}$ and hence $\eta_x = \eta_z := \eta =  \frac{1}{\sqrt{2}}$. Again, we can return to the discrimination probability between the two ensembles encoding $a=0$ and $a=1$ in \eqref{eqn:ensembles}. Here, we have (let us define $\rho^c$ to be the output clone that Bob chooses to use ($c\in \{1, 2\}$):
\begin{align*} \label{eqn:4state-cloner-final-probability}
     &P^{\opt, \mathrm{II}}_{\mathrm{disc}, \mathcal{P}_2} = \frac{1}{2} + \frac{1}{4}\left|\left|\rho_{(a=0)} - \rho_{(a=1)}\right|\right|_{\Tr}\\
     &= \frac{1}{2} + \frac{1}{4}\left|\left|\frac{1}{2}\left[(\rho_{00}^c - \rho_{11}^c) + (\rho_{10}^c - \rho_{01}^c)\right]\right|\right|_{\Tr} \\
     &= \frac{1}{2} + \frac{1}{4}\left|\left|\frac{\eta}{4}(m_{00}^x - m_{11}^x + m_{10}^x - m_{01}^x)\XG + (m_{00}^z - m_{11}^z + m_{10}^z - m_{01}^z)\ZG \right|\right|_{\Tr} \\
      &= \frac{1}{2} + \frac{\eta\cos(2\phi)}{4}\left|\left|\ZG \right|\right|_{\Tr} \\
     &= \frac{1}{2} + \frac{\eta\cos(2\phi)}{2} = \frac{3}{4}
\end{align*}
Computing the bias in the same way as done previously completes the proof.
\end{proof}

\bibliographystyle{alpha}

\singlespace



\begin{thebibliography}{PSCLGFL20}

\bibitem[AAB{\etalchar{+}}19]{arute_quantum_2019}
Frank Arute, Kunal Arya, Ryan Babbush, Dave Bacon, Joseph~C. Bardin, Rami
  Barends, Rupak Biswas, Sergio Boixo, Fernando G. S.~L. Brandao, David~A.
  Buell, Brian Burkett, Yu~Chen, Zijun Chen, Ben Chiaro, Roberto Collins,
  William Courtney, Andrew Dunsworth, Edward Farhi, Brooks Foxen, Austin
  Fowler, Craig Gidney, Marissa Giustina, Rob Graff, Keith Guerin, Steve
  Habegger, Matthew~P. Harrigan, Michael~J. Hartmann, Alan Ho, Markus Hoffmann,
  Trent Huang, Travis~S. Humble, Sergei~V. Isakov, Evan Jeffrey, Zhang Jiang,
  Dvir Kafri, Kostyantyn Kechedzhi, Julian Kelly, Paul~V. Klimov, Sergey Knysh,
  Alexander Korotkov, Fedor Kostritsa, David Landhuis, Mike Lindmark, Erik
  Lucero, Dmitry Lyakh, Salvatore Mandr{\`a}, Jarrod~R. McClean, Matthew
  McEwen, Anthony Megrant, Xiao Mi, Kristel Michielsen, Masoud Mohseni, Josh
  Mutus, Ofer Naaman, Matthew Neeley, Charles Neill, Murphy~Yuezhen Niu, Eric
  Ostby, Andre Petukhov, John~C. Platt, Chris Quintana, Eleanor~G. Rieffel,
  Pedram Roushan, Nicholas~C. Rubin, Daniel Sank, Kevin~J. Satzinger, Vadim
  Smelyanskiy, Kevin~J. Sung, Matthew~D. Trevithick, Amit Vainsencher, Benjamin
  Villalonga, Theodore White, Z.~Jamie Yao, Ping Yeh, Adam Zalcman, Hartmut
  Neven, and John~M. Martinis.
\newblock Quantum supremacy using a programmable superconducting processor.
\newblock {\em Nature}, 574(7779):505--510, October 2019.
\newblock \href{https://www.nature.com/articles/s41586-019-1666-5}{Nature,
  574(7779):505–510}.

\bibitem[Aar09]{aaronson_quantum_2009}
Scott Aaronson.
\newblock Quantum {Copy}-{Protection} and {Quantum} {Money}.
\newblock In {\em Proceedings of the 2009 24th {Annual} {IEEE} {Conference} on
  {Computational} {Complexity}}, {CCC} '09, pages 229--242, USA, July 2009.
  IEEE Computer Society.
\newblock \href{https://doi.org/10.1109/CCC.2009.42}{CCC ’09, pages
  229–242}.

\bibitem[Aar15]{aaronson_read_2015}
Scott Aaronson.
\newblock Read the fine print.
\newblock {\em Nature Physics}, 11(4):291--293, April 2015.
\newblock \href{https://www.nature.com/articles/nphys3272}{Nature Physics,
  11(4):291–293}.

\bibitem[AAR{\etalchar{+}}18]{amin_quantum_2018}
Mohammad~H. Amin, Evgeny Andriyash, Jason Rolfe, Bohdan Kulchytskyy, and Roger
  Melko.
\newblock Quantum {Boltzmann} {Machine}.
\newblock {\em Physical Review X}, 8(2):21050, May 2018.
\newblock \href{https://link.aps.org/doi/10.1103/PhysRevX.8.021050}{PRX,
  8(2):21050}.

\bibitem[ABB{\etalchar{+}}21]{arrazola_quantum_2021}
J.~M. Arrazola, V.~Bergholm, K.~Br{\'a}dler, T.~R. Bromley, M.~J. Collins,
  I.~Dhand, A.~Fumagalli, T.~Gerrits, A.~Goussev, L.~G. Helt, J.~Hundal,
  T.~Isacsson, R.~B. Israel, J.~Izaac, S.~Jahangiri, R.~Janik, N.~Killoran,
  S.~P. Kumar, J.~Lavoie, A.~E. Lita, D.~H. Mahler, M.~Menotti, B.~Morrison,
  S.~W. Nam, L.~Neuhaus, H.~Y. Qi, N.~Quesada, A.~Repingon, K.~K. Sabapathy,
  M.~Schuld, D.~Su, J.~Swinarton, A.~Sz{\'a}va, K.~Tan, P.~Tan, V.~D. Vaidya,
  Z.~Vernon, Z.~Zabaneh, and Y.~Zhang.
\newblock Quantum circuits with many photons on a programmable nanophotonic
  chip.
\newblock {\em Nature}, 591(7848):54--60, March 2021.
\newblock \href{https://www.nature.com/articles/s41586-021-03202-1}{Nature,
  591(7848):54–60}.

\bibitem[ABC{\etalchar{+}}16]{abadi_tensorflow_2016}
Mart{\'i}n Abadi, Paul Barham, Jianmin Chen, Zhifeng Chen, Andy Davis, Jeffrey
  Dean, Matthieu Devin, Sanjay Ghemawat, Geoffrey Irving, Michael Isard,
  Manjunath Kudlur, Josh Levenberg, Rajat Monga, Sherry Moore, Derek~G. Murray,
  Benoit Steiner, Paul Tucker, Vijay Vasudevan, Pete Warden, Martin Wicke, Yuan
  Yu, and Xiaoqiang Zheng.
\newblock {TensorFlow}: {A} system for large-scale machine learning.
\newblock {\em arXiv:1605.08695 [cs]}, May 2016.
\newblock \href{http://arxiv.org/abs/1605.08695}{ArXiv: 1605.08695}.

\bibitem[ABG06]{aimeur_machine_2006}
Esma A{\"i}meur, Gilles Brassard, and S{\'e}bastien Gambs.
\newblock Machine {Learning} in a {Quantum} {World}.
\newblock In Luc Lamontagne and Mario Marchand, editors, {\em Advances in
  {Artificial} {Intelligence}}, Lecture {Notes} in {Computer} {Science}, pages
  431--442, Berlin, Heidelberg, 2006. Springer.
\newblock \href{https://link.springer.com/chapter/10.1007/11766247_37}{ LNCS,23
  vol 4013}.

\bibitem[ABO08]{aharonov_fault-tolerant_2008}
Dorit Aharonov and Michael Ben-Or.
\newblock Fault-{Tolerant} {Quantum} {Computation} with {Constant} {Error}
  {Rate}.
\newblock {\em SIAM Journal on Computing}, 38(4):1207--1282, January 2008.
\newblock \href{https://epubs.siam.org/doi/10.1137/S0097539799359385}{SIAM JoC,
  38(4):1207–1282}.

\bibitem[ACC{\etalchar{+}}20]{arrasmith_effect_2020}
Andrew Arrasmith, M.~Cerezo, Piotr Czarnik, Lukasz Cincio, and Patrick~J.
  Coles.
\newblock Effect of barren plateaus on gradient-free optimization.
\newblock {\em arXiv:2011.12245 [quant-ph, stat]}, November 2020.
\newblock \href{http://arxiv.org/abs/2011.12245}{ArXiv: 2011.12245}.

\bibitem[ACL{\etalchar{+}}20]{arunachalam_two_2020}
Srinivasan Arunachalam, Sourav Chakraborty, Troy Lee, Manaswi Paraashar, and
  Ronald de~Wolf.
\newblock Two new results about quantum exact learning.
\newblock {\em arXiv:1810.00481 [quant-ph]}, April 2020.
\newblock \href{http://arxiv.org/abs/1810.00481}{ArXiv: 1810.00481}.

\bibitem[ADBL20]{arrazola_quantum-inspired_2020}
Juan~Miguel Arrazola, Alain Delgado, Bhaskar~Roy Bardhan, and Seth Lloyd.
\newblock Quantum-inspired algorithms in practice.
\newblock {\em Quantum}, 4:307, August 2020.
\newblock \href{https://quantum-journal.org/papers/q-2020-08-13-307/}{Quantum,
  4:307}.

\bibitem[ADJ{\etalchar{+}}20]{abrams_implementation_2020}
Deanna~M. Abrams, Nicolas Didier, Blake~R. Johnson, Marcus P.~da Silva, and
  Colm~A. Ryan.
\newblock Implementation of {XY} entangling gates with a single calibrated
  pulse.
\newblock {\em Nature Electronics}, 3(12):744--750, December 2020.
\newblock \href{https://www.nature.com/articles/s41928-020-00498-1}{Nature
  Electronics, 3(12):744–750}.

\bibitem[AdW17a]{arunachalam_guest_2017}
Srinivasan Arunachalam and Ronald de~Wolf.
\newblock Guest {Column}: {A} {Survey} of {Quantum} {Learning} {Theory}.
\newblock {\em ACM SIGACT News}, 48(2):41--67, June 2017.
\newblock \href{https://doi.org/10.1145/3106700.3106710}{ACM SIGACT News,
  48(2):41–67}.

\bibitem[AdW17b]{arunachalam_optimal_2017}
Srinivasan Arunachalam and Ronald de~Wolf.
\newblock Optimal quantum sample complexity of learning algorithms.
\newblock In {\em Proceedings of the 32nd {Computational} {Complexity}
  {Conference}}, {CCC} '17, pages 1--31, Riga, Latvia, July 2017. Schloss
  Dagstuhl{\textendash}Leibniz-Zentrum fuer Informatik.
\newblock \href{https://dl.acm.org/doi/10.5555/3135595.3135620}{CCC ’17,
  1–31}.

\bibitem[AGS19]{arunachalam_quantum_2019}
Srinivasan Arunachalam, Alex~B. Grilo, and Aarthi Sundaram.
\newblock Quantum hardness of learning shallow classical circuits.
\newblock {\em arXiv:1903.02840 [quant-ph]}, September 2019.
\newblock \href{http://arxiv.org/abs/1903.02840}{ArXiv: 1903.02840}.

\bibitem[AGY20]{arunachalam_quantum_2020}
Srinivasan Arunachalam, Alex~B. Grilo, and Henry Yuen.
\newblock Quantum statistical query learning.
\newblock {\em arXiv:2002.08240 [quant-ph]}, February 2020.
\newblock \href{http://arxiv.org/abs/2002.08240}{ArXiv: 2002.08240}.

\bibitem[AHCC21]{arrasmith_equivalence_2021}
Andrew Arrasmith, Zo{\"e} Holmes, M.~Cerezo, and Patrick~J. Coles.
\newblock Equivalence of quantum barren plateaus to cost concentration and
  narrow gorges.
\newblock {\em arXiv:2104.05868 [quant-ph]}, April 2021.
\newblock \href{http://arxiv.org/abs/2104.05868}{ArXiv: 2104.05868}.

\bibitem[AKE{\etalchar{+}}20]{alexander_qiskit_2020}
Thomas Alexander, Naoki Kanazawa, Daniel~J. Egger, Lauren Capelluto,
  Christopher~J. Wood, Ali Javadi-Abhari, and David~C. McKay.
\newblock Qiskit pulse: programming quantum computers through the cloud with
  pulses.
\newblock {\em Quantum Science and Technology}, 5(4):044006, August 2020.
\newblock \href{https://doi.org/10.1088/2058-9565/aba404}{QST, 5(4):044006}.

\bibitem[{Ale}03]{alexei_kitaev_quantum_2003}
{Alexei Kitaev}.
\newblock Quantum coin-flipping. {Unpublished} result.
\newblock {\em 6th Annual workshop on Quantum Information Processing (QIP)},
  2003.
\newblock
  \href{http://web.archive.org/web/20151016194300/https://www.msri.org/workshops/204}{QIP
  2003}.

\bibitem[ALOPO20]{alcazar_classical_2020}
Javier Alcazar, Vicente Leyton-Ortega, and Alejandro Perdomo-Ortiz.
\newblock Classical versus quantum models in machine learning: insights from a
  finance application.
\newblock {\em Machine Learning: Science and Technology}, 1(3):035003, July
  2020.
\newblock \href{https://doi.org/10.1088\%2F2632-2153\%2Fab9009}{MLST,
  1(3):035003}.

\bibitem[AM19]{amy_t-count_2019}
Matthew Amy and Michele Mosca.
\newblock T-{Count} {Optimization} and {Reed}{\textendash}{Muller} {Codes}.
\newblock {\em IEEE Transactions on Information Theory}, 65(8):4771--4784,
  August 2019.
\newblock \href{https://ieeexplore.ieee.org/document/8672175/}{IEEE TIT,
  65(8):4771–4784}.

\bibitem[Ang92]{angluin_computational_1992}
Dana Angluin.
\newblock Computational learning theory: survey and selected bibliography.
\newblock In {\em Proceedings of the twenty-fourth annual {ACM} symposium on
  {Theory} of {Computing}}, {STOC} '92, pages 351--369, New York, NY, USA, July
  1992. Association for Computing Machinery.
\newblock \href{https://doi.org/10.1145/129712.129746}{STOC ’92, pages
  351–369}.

\bibitem[ARDAG20]{anand_experimental_2020}
Abhinav Anand, Jonathan Romero, Matthias Degroote, and Al{\'a}n Aspuru-Guzik.
\newblock Experimental demonstration of a quantum generative adversarial
  network for continuous distributions.
\newblock {\em arXiv:2006.01976 [quant-ph]}, June 2020.
\newblock \href{http://arxiv.org/abs/2006.01976}{ArXiv: 2006.01976}.

\bibitem[AS66]{ali_general_1966}
S.~M. Ali and S.~D. Silvey.
\newblock A {General} {Class} of {Coefficients} of {Divergence} of {One}
  {Distribution} from {Another}.
\newblock {\em Journal of the Royal Statistical Society. Series B
  (Methodological)}, 28(1):131--142, 1966.
\newblock \href{http://www.jstor.org/stable/2984279}{JRSS, Series B
  (Methodological), 28(1):131–142}.

\bibitem[AS05]{atici_improved_2005}
Alp Atici and Rocco~A. Servedio.
\newblock Improved {Bounds} on {Quantum} {Learning} {Algorithms}.
\newblock {\em Quantum Information Processing}, 4(5):355--386, November 2005.
\newblock \href{https://doi.org/10.1007/s11128-005-0001-2}{QIP, 4(5):355-386}.

\bibitem[ASZ{\etalchar{+}}21]{abbas_power_2021}
Amira Abbas, David Sutter, Christa Zoufal, Aurelien Lucchi, Alessio Figalli,
  and Stefan Woerner.
\newblock The power of quantum neural networks.
\newblock {\em Nature Computational Science}, 1(6):403--409, June 2021.
\newblock \href{https://www.nature.com/articles/s43588-021-00084-1}{Nature
  Computational Science, 1(6):403–409}.

\bibitem[ATSVY00]{aharonov_quantum_2000}
Dorit Aharonov, Amnon Ta-Shma, Umesh~V. Vazirani, and Andrew~C. Yao.
\newblock Quantum bit escrow.
\newblock In {\em Proceedings of the thirty-second annual {ACM} symposium on
  {Theory} of computing}, {STOC} '00, pages 705--714, Portland, Oregon, USA,
  May 2000. Association for Computing Machinery.
\newblock \href{https://doi.org/10.1145/335305.335404}{SIAM JoC,
  38(4):1207–1282}.

\bibitem[BB14]{bennett_quantum_2014}
Charles~H. Bennett and Gilles Brassard.
\newblock Quantum cryptography: {Public} key distribution and coin tossing.
\newblock {\em Theoretical Computer Science}, 560:7--11, December 2014.
\newblock
  \href{http://www.sciencedirect.com/science/article/pii/S0304397514004241}{TCS,
  560:7–11}.

\bibitem[BBBG09]{berlin_fair_2009}
Guido Berl{\'i}n, Gilles Brassard, F{\'e}lix Bussi{\`e}res, and Nicolas
  Godbout.
\newblock Fair loss-tolerant quantum coin flipping.
\newblock {\em Phys. Rev. A}, 80(6):062321, December 2009.
\newblock \href{https://link.aps.org/doi/10.1103/PhysRevA.80.062321}{PRA,
  80(6):062321}.

\bibitem[BBHB97]{buzek_quantum_1997}
V.~Bu{\v z}ek, S.~L. Braunstein, M.~Hillery, and D.~Bru{\ss}.
\newblock Quantum copying: {A} network.
\newblock {\em Physical Review A}, 56(5):3446--3452, November 1997.
\newblock \href{https://link.aps.org/doi/10.1103/PhysRevA.56.3446}{PRA,
  56(5):3446–3452}.

\bibitem[BC21]{banchi_measuring_2021}
Leonardo Banchi and Gavin~E. Crooks.
\newblock Measuring {Analytic} {Gradients} of {General} {Quantum} {Evolution}
  with the {Stochastic} {Parameter} {Shift} {Rule}.
\newblock {\em Quantum}, 5:386, January 2021.
\newblock \href{https://quantum-journal.org/papers/q-2021-01-25-386/}{Quantum,
  5:386}.

\bibitem[BCF{\etalchar{+}}96]{barnum_noncommuting_1996}
Howard Barnum, Carlton~M. Caves, Christopher~A. Fuchs, Richard Jozsa, and
  Benjamin Schumacher.
\newblock Noncommuting {Mixed} {States} {Cannot} {Be} {Broadcast}.
\newblock {\em Physical Review Letters}, 76(15):2818--2821, April 1996.
\newblock \href{https://link.aps.org/doi/10.1103/PhysRevLett.76.2818}{PRL,
  76(15):2818–2821}.

\bibitem[BCF{\etalchar{+}}21]{benedetti_variational_2021}
Marcello Benedetti, Brian Coyle, Mattia Fiorentini, Michael Lubasch, and
  Matthias Rosenkranz.
\newblock Variational inference with a quantum computer.
\newblock {\em Phys. Rev. Applied}, 16:044057, Oct 2021.
\newblock
  \href{https://link.aps.org/doi/10.1103/PhysRevApplied.16.044057}{Phys. Rev.
  Applied 16, 044057}.

\bibitem[BCLK{\etalchar{+}}21]{bharti_noisy_2021}
Kishor Bharti, Alba Cervera-Lierta, Thi~Ha Kyaw, Tobias Haug, Sumner
  Alperin-Lea, Abhinav Anand, Matthias Degroote, Hermanni Heimonen, Jakob~S.
  Kottmann, Tim Menke, Wai-Keong Mok, Sukin Sim, Leong-Chuan Kwek, and Al{\'a}n
  Aspuru-Guzik.
\newblock Noisy intermediate-scale quantum ({NISQ}) algorithms.
\newblock {\em arXiv:2101.08448 [cond-mat, physics:quant-ph]}, January 2021.
\newblock \href{http://arxiv.org/abs/2101.08448}{ArXiv: 2101.08448}.

\bibitem[BCMDM00]{brus_phase-covariant_2000}
Dagmar Bru{\ss}, Mirko Cinchetti, G.~Mauro~D'Ariano, and Chiara Macchiavello.
\newblock Phase-covariant quantum cloning.
\newblock {\em Physical Review A}, 62(1):012302, June 2000.
\newblock \href{https://link.aps.org/doi/10.1103/PhysRevA.62.012302}{PRA,
  62(1):012302}.

\bibitem[BCWdW01]{buhrman_quantum_2001}
Harry Buhrman, Richard Cleve, John Watrous, and Ronald de~Wolf.
\newblock Quantum {Fingerprinting}.
\newblock {\em Physical Review Letters}, 87(16):167902, September 2001.
\newblock \href{https://link.aps.org/doi/10.1103/PhysRevLett.87.167902}{PRL,
  87(16):167902}.

\bibitem[BDE{\etalchar{+}}98]{brus_optimal_1998}
Dagmar Bru{\ss}, David~P. DiVincenzo, Artur Ekert, Christopher~A. Fuchs, Chiara
  Macchiavello, and John~A. Smolin.
\newblock Optimal universal and state-dependent quantum cloning.
\newblock {\em Physical Review A}, 57(4):2368--2378, April 1998.
\newblock \href{https://link.aps.org/doi/10.1103/PhysRevA.57.2368}{PRA,
  57(4):2368–2378}.

\bibitem[BEHW89]{blumer_learnability_1989}
Anselm Blumer, A.~Ehrenfeucht, David Haussler, and Manfred~K. Warmuth.
\newblock Learnability and the {Vapnik}-{Chervonenkis} dimension.
\newblock {\em Journal of the ACM}, 36(4):929--965, October 1989.
\newblock \href{https://doi.org/10.1145/76359.76371}{JACM, 36(4):929–965}.

\bibitem[Bel64]{bell_einstein_1964}
J.~S. Bell.
\newblock On the {Einstein} {Podolsky} {Rosen} paradox.
\newblock {\em Physics Physique Fizika}, 1(3):195--200, November 1964.
\newblock
  \href{https://link.aps.org/doi/10.1103/PhysicsPhysiqueFizika.1.195}{PPF,
  1(3):195–200}.

\bibitem[BFK09]{broadbent_universal_2009}
Anne Broadbent, Joseph Fitzsimons, and Elham Kashefi.
\newblock Universal blind quantum computation.
\newblock In {\em 2009 50th {Annual} {IEEE} {Symposium} on {Foundations} of
  {Computer} {Science}}, pages 517--526. IEEE, 2009.
\newblock \href{https://ieeexplore.ieee.org/document/5438603}{FOCS `09,
  517–526}.

\bibitem[BFNV19]{bouland_complexity_2019}
Adam Bouland, Bill Fefferman, Chinmay Nirkhe, and Umesh Vazirani.
\newblock On the complexity and verification of quantum random circuit
  sampling.
\newblock {\em Nature Physics}, 15(2):159--163, February 2019.
\newblock \href{https://www.nature.com/articles/s41567-018-0318-2}{Nat. Phys.,
  15(2):159–163}.

\bibitem[BGMT17]{bravyi_tapering_2017}
Sergey Bravyi, Jay~M. Gambetta, Antonio Mezzacapo, and Kristan Temme.
\newblock Tapering off qubits to simulate fermionic {Hamiltonians}.
\newblock {\em arXiv:1701.08213 [quant-ph]}, January 2017.
\newblock \href{http://arxiv.org/abs/1701.08213}{ArXiv: 1701.08213}.

\bibitem[BGPP{\etalchar{+}}19]{benedetti_generative_2019}
Marcello Benedetti, Delfina Garcia-Pintos, Oscar Perdomo, Vicente
  Leyton-Ortega, Yunseong Nam, and Alejandro Perdomo-Ortiz.
\newblock A generative modeling approach for benchmarking and training shallow
  quantum circuits.
\newblock {\em npj Quantum Information}, 5(1):1--9, May 2019.
\newblock \href{https://www.nature.com/articles/s41534-019-0157-8}{npj QI,
  5(1):1–9}.

\bibitem[BGR{\etalchar{+}}06]{borgwardt_integrating_2006}
Karsten~M Borgwardt, Arthur Gretton, Malte~J Rasch, Hans-Peter Kriegel,
  Bernhard Sch{\"o}lkopf, and Alex~J Smola.
\newblock Integrating structured biological data by {Kernel} {Maximum} {Mean}
  {Discrepancy}.
\newblock {\em Bioinformatics}, 22(14):e49--e57, 2006.
\newblock
  \href{http://dx.doi.org/10.1093/bioinformatics/btl242}{Bioinformatics,
  22(14):e49–e57}.

\bibitem[BGW09]{bot_introduction_2009}
Radu~Ioan Bo{\c t}, Sorin-Mihai Grad, and Gert Wanka.
\newblock Introduction.
\newblock In Radu~Ioan Bot, Sorin-Mihai Grad, and Gert Wanka, editors, {\em
  Duality in {Vector} {Optimization}}, Vector {Optimization}, pages 1--7.
  Springer, Berlin, Heidelberg, 2009.
\newblock \href{https://doi.org/10.1007/978-3-642-02886-1_1}{Duality in Vector
  Optimization, Vector Optimization, pages 1–7.}

\bibitem[BH96]{buzek_quantum_1996}
V.~Bu{\v z}ek and M.~Hillery.
\newblock Quantum copying: {Beyond} the no-cloning theorem.
\newblock {\em Physical Review A}, 54(3):1844--1852, September 1996.
\newblock \href{https://link.aps.org/doi/10.1103/PhysRevA.54.1844}{PRA,
  54(3):1844–1852}.

\bibitem[BH98]{buzek_universal_1998}
V.~Bu{\v z}ek and Mark Hillery.
\newblock Universal {Optimal} {Cloning} of {Arbitrary} {Quantum} {States}:
  {From} {Qubits} to {Quantum} {Registers}.
\newblock {\em Physical Review Letters}, 81(22):5003--5006, November 1998.
\newblock \href{https://link.aps.org/doi/10.1103/PhysRevLett.81.5003}{PRL,
  81(22):5003–5006}.

\bibitem[Bia19]{biamonte_universal_2019}
Jacob Biamonte.
\newblock Universal {Variational} {Quantum} {Computation}.
\newblock {\em arXiv:1903.04500 [quant-ph]}, March 2019.
\newblock \href{http://arxiv.org/abs/1903.04500}{ArXiv: 1903.04500}.

\bibitem[Bia20]{biamonte_theory_2020}
Jacob Biamonte.
\newblock On the {Theory} of {Modern} {Quantum} {Algorithms}.
\newblock {\em arXiv:2009.10088 [math-ph, physics:quant-ph]}, September 2020.
\newblock \href{http://arxiv.org/abs/2009.10088}{ArXiv: 2009.10088}.

\bibitem[BIS{\etalchar{+}}20]{bergholm_pennylane_2020}
Ville Bergholm, Josh Izaac, Maria Schuld, Christian Gogolin, M.~Sohaib Alam,
  Shahnawaz Ahmed, Juan~Miguel Arrazola, Carsten Blank, Alain Delgado, Soran
  Jahangiri, Keri McKiernan, Johannes~Jakob Meyer, Zeyue Niu, Antal Sz{\'a}va,
  and Nathan Killoran.
\newblock {PennyLane}: {Automatic} differentiation of hybrid quantum-classical
  computations.
\newblock {\em arXiv:1811.04968 [physics, physics:quant-ph]}, February 2020.
\newblock \href{http://arxiv.org/abs/1811.04968}{ArXiv: 1811.04968}.

\bibitem[BJ95]{bshouty_learning_1995}
Nader~H. Bshouty and Jeffrey~C. Jackson.
\newblock Learning {DNF} over the uniform distribution using a quantum example
  oracle.
\newblock In {\em Proceedings of the eighth annual conference on
  {Computational} learning theory}, {COLT} '95, pages 118--127, New York, NY,
  USA, July 1995. Association for Computing Machinery.
\newblock \href{https://doi.org/10.1145/225298.225312}{COLT ’95, pages
  118–127}.

\bibitem[BJ12]{booth_jr_quantum_2012}
Jeffrey Booth~Jr.
\newblock Quantum {Compiler} {Optimizations}.
\newblock {\em arXiv:1206.3348 [quant-ph]}, June 2012.
\newblock \href{http://arxiv.org/abs/1206.3348}{ArXiv: 1206.3348}.

\bibitem[BJS11]{bremner_classical_2011}
Michael~J Bremner, Richard Jozsa, and Dan~J Shepherd.
\newblock Classical simulation of commuting quantum computations implies
  collapse of the polynomial hierarchy.
\newblock {\em Proc. R. Soc. London A Math. Phys. Eng. Sci.},
  467(2126):459--472, 2011.
\newblock
  \href{http://rspa.royalsocietypublishing.org/content/467/2126/459}{PRS London
  AMPES, 467(2126):459–472}.

\bibitem[BK10]{blume-kohout_optimal_2010}
Robin Blume-Kohout.
\newblock Optimal, reliable estimation of quantum states.
\newblock {\em New Journal of Physics}, 12(4):043034, April 2010.
\newblock \href{https://doi.org/10.1088/1367-2630/12/4/043034}{NJP,
  12(4):043034}.

\bibitem[BK15]{bae_quantum_2015}
Joonwoo Bae and Leong-Chuan Kwek.
\newblock Quantum state discrimination and its applications.
\newblock {\em Journal of Physics A: Mathematical and Theoretical},
  48(8):083001, January 2015.
\newblock \href{https://doi.org/10.1088/1751-8113/48/8/083001}{JPA: M\&T,
  48(8):083001}.

\bibitem[BKL{\etalchar{+}}21a]{bombin_interleaving_2021}
Hector Bombin, Isaac~H. Kim, Daniel Litinski, Naomi Nickerson, Mihir Pant,
  Fernando Pastawski, Sam Roberts, and Terry Rudolph.
\newblock Interleaving: {Modular} architectures for fault-tolerant photonic
  quantum computing.
\newblock {\em arXiv:2103.08612 [quant-ph]}, March 2021.
\newblock \href{http://arxiv.org/abs/2103.08612}{ArXiv: 2103.08612}.

\bibitem[BKL{\etalchar{+}}21b]{bu_effects_2021}
Kaifeng Bu, Dax~Enshan Koh, Lu~Li, Qingxian Luo, and Yaobo Zhang.
\newblock Effects of quantum resources on the statistical complexity of quantum
  circuits.
\newblock {\em arXiv:2102.03282 [quant-ph, stat]}, February 2021.
\newblock \href{http://arxiv.org/abs/2102.03282}{ArXiv: 2102.03282}.

\bibitem[BKL{\etalchar{+}}21c]{bu_statistical_2021}
Kaifeng Bu, Dax~Enshan Koh, Lu~Li, Qingxian Luo, and Yaobo Zhang.
\newblock On the statistical complexity of quantum circuits.
\newblock {\em arXiv:2101.06154 [quant-ph, stat]}, January 2021.
\newblock \href{http://arxiv.org/abs/2101.06154}{ArXiv: 2102.03282}.

\bibitem[BKL{\etalchar{+}}21d]{bu_rademacher_2021}
Kaifeng Bu, Dax~Enshan Koh, Lu~Li, Qingxian Luo, and Yaobo Zhang.
\newblock Rademacher complexity of noisy quantum circuits.
\newblock {\em arXiv:2103.03139 [quant-ph]}, March 2021.
\newblock \href{http://arxiv.org/abs/2103.03139}{ArXiv: 2103.03139}.

\bibitem[BL17]{bernstein_post-quantum_2017}
Daniel~J. Bernstein and Tanja Lange.
\newblock Post-quantum cryptography.
\newblock {\em Nature}, 549(7671):188--194, September 2017.
\newblock \href{https://www.nature.com/articles/nature23461}{Nature,
  549(7671):188–194}.

\bibitem[BLSF19]{benedetti_parameterized_2019}
Marcello Benedetti, Erika Lloyd, Stefan Sack, and Mattia Fiorentini.
\newblock Parameterized quantum circuits as machine learning models.
\newblock {\em Quantum Science \& Technology}, 4(4):043001, November 2019.
\newblock \href{https://doi.org/10.1088\%2F2058-9565\%2Fab4eb5}{QST,
  4(4):043001}.

\bibitem[Blu83]{blum_coin_1983}
Manuel Blum.
\newblock {\em Coin flipping by telephone a protocol for solving impossible
  problems}.
\newblock Association for Computing Machinery, January 1983.
\newblock \href{https://doi.org/10.1145/1008908.1008911}{ACM January 1983}.

\bibitem[BM08]{brus_approximate_2008}
Dagmar Bru{\ss} and Chiara Macchiavello.
\newblock Approximate {Quantum} {Cloning}.
\newblock In {\em Lectures on {Quantum} {Information}}, pages 53--71. John
  Wiley \& Sons, Ltd, 2008.
\newblock
  \href{https://onlinelibrary.wiley.com/doi/abs/10.1002/9783527618637.ch4}{Link}.

\bibitem[BMS16]{bremner_average-case_2016}
Michael~J. Bremner, Ashley Montanaro, and Dan~J. Shepherd.
\newblock Average-{Case} {Complexity} {Versus} {Approximate} {Simulation} of
  {Commuting} {Quantum} {Computations}.
\newblock {\em Physical Review Letters}, 117(8):080501, August 2016.
\newblock \href{https://link.aps.org/doi/10.1103/PhysRevLett.117.080501}{PRL,
  117(8):080501}.

\bibitem[BNSS96]{behrman_quantum_1996}
E.~C. Behrman, J.~Niemel, J.~E. Steck, and S.~R. Skinner.
\newblock A {Quantum} {Dot} {Neural} {Network}, 1996.
\newblock
  \href{http://citeseerx.ist.psu.edu/viewdoc/summary?doi=10.1.1.56.1507}{Link}.

\bibitem[BPLC{\etalchar{+}}19]{bravo-prieto_variational_2019}
Carlos Bravo-Prieto, Ryan LaRose, M.~Cerezo, Yigit Subasi, Lukasz Cincio, and
  Patrick~J. Coles.
\newblock Variational {Quantum} {Linear} {Solver}: {A} {Hybrid} {Algorithm} for
  {Linear} {Systems}.
\newblock {\em arXiv:1909.05820 [quant-ph]}, September 2019.
\newblock \href{http://arxiv.org/abs/1909.05820}{ArXiv: 1909.05820}.

\bibitem[Bre03]{brennen_observable_2003}
Gavin~K. Brennen.
\newblock An observable measure of entanglement for pure states of multi-qubit
  systems.
\newblock {\em arXiv:quant-ph/0305094}, November 2003.
\newblock \href{http://arxiv.org/abs/quant-ph/0305094}{ArXiv: 0305094}.

\bibitem[BRGPO18]{benedetti_quantum-assisted_2018}
Marcello Benedetti, John Realpe-G{\'o}mez, and Alejandro Perdomo-Ortiz.
\newblock Quantum-assisted {Helmholtz} machines: {A} quantum-classical deep
  learning framework for industrial datasets in near-term devices.
\newblock {\em Quantum Science \& Technology}, 3(3):034007, July 2018.
\newblock
  \href{http://stacks.iop.org/2058-9565/3/i=3/a=034007?key=crossref.b76591ee4bc7a7df6153c5cfaf284353}{QST,
  3(3):034007}.

\bibitem[Bru19]{brun_quantum_2019}
Todd~A. Brun.
\newblock Quantum {Error} {Correction}.
\newblock {\em arXiv:1910.03672 [quant-ph]}, October 2019.
\newblock \href{http://arxiv.org/abs/1910.03672}{ArXiv: 1910.03672}.

\bibitem[BS02]{beyer_evolution_2002}
Hans-Georg Beyer and Hans-Paul Schwefel.
\newblock Evolution strategies: {A} comprehensive introduction.
\newblock {\em Natural Computing}, 1(1):3--52, March 2002.
\newblock \href{https://doi.org/10.1023/A:1015059928466}{Natural Computing,
  1(1):3–52}.

\bibitem[BVM{\etalchar{+}}20]{broughton_tensorflow_2020}
Michael Broughton, Guillaume Verdon, Trevor McCourt, Antonio~J. Martinez,
  Jae~Hyeon Yoo, Sergei~V. Isakov, Philip Massey, Murphy~Yuezhen Niu, Ramin
  Halavati, Evan Peters, Martin Leib, Andrea Skolik, Michael Streif, David
  Von~Dollen, Jarrod~R. McClean, Sergio Boixo, Dave Bacon, Alan~K. Ho, Hartmut
  Neven, and Masoud Mohseni.
\newblock {TensorFlow} {Quantum}: {A} {Software} {Framework} for {Quantum}
  {Machine} {Learning}.
\newblock {\em arXiv:2003.02989 [cond-mat, physics:quant-ph]}, March 2020.
\newblock \href{http://arxiv.org/abs/2003.02989}{ArXiv: 2003.02989}.

\bibitem[BWP{\etalchar{+}}17]{biamonte_quantum_2017}
Jacob Biamonte, Peter Wittek, Nicola Pancotti, Patrick Rebentrost, Nathan
  Wiebe, and Seth Lloyd.
\newblock Quantum {Machine} {Learning}.
\newblock {\em Nature}, 549(7671):195--202, September 2017.
\newblock \href{https://www.nature.com/articles/nature23474}{Nature,
  549(7671):195–202}.

\bibitem[CAB{\etalchar{+}}21]{cerezo_variational_2021}
M.~Cerezo, Andrew Arrasmith, Ryan Babbush, Simon~C. Benjamin, Suguru Endo,
  Keisuke Fujii, Jarrod~R. McClean, Kosuke Mitarai, Xiao Yuan, Lukasz Cincio,
  and Patrick~J. Coles.
\newblock Variational quantum algorithms.
\newblock {\em Nature Reviews Physics}, pages 1--20, August 2021.
\newblock \href{https://www.nature.com/articles/s42254-021-00348-9}{Nat. Rev.
  Phys., 1–20}.

\bibitem[CCC19]{coles_strong_2019}
Patrick~J. Coles, M.~Cerezo, and Lukasz Cincio.
\newblock Strong bound between trace distance and {Hilbert}-{Schmidt} distance
  for low-rank states.
\newblock {\em Physical Review A}, 100(2):022103, August 2019.
\newblock \href{https://link.aps.org/doi/10.1103/PhysRevA.100.022103}{PRA,
  100(2):022103}.

\bibitem[{\v C}CK20]{cepaite_continuous_2020}
Ieva {\v C}epait{\.e}, Brian Coyle, and Elham Kashefi.
\newblock A {Continuous} {Variable} {Born} {Machine}.
\newblock {\em arXiv:2011.00904 [quant-ph]}, November 2020.
\newblock \href{http://arxiv.org/abs/2011.00904}{ArXiv: 2011.00904}.

\bibitem[CCL19]{cong_quantum_2019}
Iris Cong, Soonwon Choi, and Mikhail~D. Lukin.
\newblock Quantum convolutional neural networks.
\newblock {\em Nature Physics}, 15(12):1273--1278, December 2019.
\newblock \href{https://www.nature.com/articles/s41567-019-0648-8}{Nat. Phys.,
  15(12):1273–1278}.

\bibitem[CCW18]{cheng_information_2018}
Song Cheng, Jing Chen, and Lei Wang.
\newblock Information {Perspective} to {Probabilistic} {Modeling}: {Boltzmann}
  {Machines} versus {Born} {Machines}.
\newblock {\em Entropy}, 20(8):583, August 2018.
\newblock \href{http://www.mdpi.com/1099-4300/20/8/583}{Entropy, 20(8):583}.

\bibitem[CD20]{caro_pseudo-dimension_2020}
Matthias~C. Caro and Ishaun Datta.
\newblock Pseudo-dimension of quantum circuits.
\newblock {\em Quantum Machine Intelligence}, 2(2):14, November 2020.
\newblock \href{https://doi.org/10.1007/s42484-020-00027-5}{QUMI, 2(2):14}.

\bibitem[CDKK20]{coyle_variational_2020}
Brian Coyle, Mina Doosti, Elham Kashefi, and Niraj Kumar.
\newblock Variational {Quantum} {Cloning}: {Improving} {Practicality} for
  {Quantum} {Cryptanalysis}.
\newblock {\em arXiv:2012.11424 [quant-ph]}, December 2020.
\newblock \href{http://arxiv.org/abs/2012.11424}{ArXiv: 2012.11424}.

\bibitem[CDKK22]{coyle_progress_2022}
Brian Coyle, Mina Doosti, Elham Kashefi, and Niraj Kumar.
\newblock Progress toward practical quantum cryptanalysis by variational
  quantum cloning.
\newblock {\em Phys. Rev. A}, 105:042604, Apr 2022.
\newblock \href{https://link.aps.org/doi/10.1103/PhysRevA.105.042604}{PRA,
  99(3):03233}.

\bibitem[CFUZ02]{chen_alternative_2002}
Jing-Ling Chen, Libin Fu, Abraham~A. Ungar, and Xian-Geng Zhao.
\newblock Alternative fidelity measure between two states of an {N}-state
  quantum system.
\newblock {\em Physical Review A}, 65(5):054304, May 2002.
\newblock \href{https://link.aps.org/doi/10.1103/PhysRevA.65.054304}{PRA,
  65(5):054304}.

\bibitem[CGL{\etalchar{+}}20]{chia_sampling-based_2020}
Nai-Hui Chia, Andr{\'a}s Gily{\'e}n, Tongyang Li, Han-Hsuan Lin, Ewin Tang, and
  Chunhao Wang.
\newblock Sampling-based sublinear low-rank matrix arithmetic framework for
  dequantizing quantum machine learning.
\newblock In {\em Proceedings of the 52nd {Annual} {ACM} {SIGACT} {Symposium}
  on {Theory} of {Computing}}, {STOC} 2020, pages 387--400, New York, NY, USA,
  June 2020. Association for Computing Machinery.
\newblock \href{https://doi.org/10.1145/3357713.3384314}{STOC '20, 387–400}.

\bibitem[CH17]{campbell_unified_2017}
Earl~T. Campbell and Mark Howard.
\newblock Unified framework for magic state distillation and multiqubit gate
  synthesis with reduced resource cost.
\newblock {\em Physical Review A}, 95(2):022316, February 2017.
\newblock \href{https://link.aps.org/doi/10.1103/PhysRevA.95.022316}{PRA,
  95(2):022316}.

\bibitem[CHL{\etalchar{+}}21]{coyle_quantum_2021}
Brian Coyle, Maxwell Henderson, Justin Chan~Jin Le, Niraj Kumar, Marco Paini,
  and Elham Kashefi.
\newblock Quantum versus classical generative modelling in finance.
\newblock {\em Quantum Science and Technology}, 6(2):024013, April 2021.
\newblock \href{https://doi.org/10.1088/2058-9565/abd3db}{QST: 6(2):024013}.

\bibitem[Cho19]{cho_ibm_2019}
Adrian Cho.
\newblock {IBM} casts doubt on {Google}{\textquoteright}s claims of quantum
  supremacy, October 2019.
\newblock
  \href{https://www.sciencemag.org/news/2019/10/ibm-casts-doubt-googles-claims-quantum-supremacy}{https://www.sciencemag.org/news/2019/10/ibm-casts-doubt-googles-claims-quantum-supremacy}.

\bibitem[CIVA02]{cerf_cloning_2002}
N.J. Cerf, S.~Iblisdir, and G.~Van~Assche.
\newblock Cloning and cryptography with quantum continuous variables.
\newblock {\em Eur. Phys. J. D}, 18(2):211--218, February 2002.
\newblock \href{https://doi.org/10.1140/epjd/e20020025}{EPJD, 18(2):211–218}.

\bibitem[CKH19]{coyle_certified_2019}
Brian Coyle, Elham Kashefi, and Matty~J. Hoban.
\newblock Certified {Randomness} {From} {Steering} {Using} {Sequential}
  {Measurements}.
\newblock {\em Cryptography}, 3(4):27, December 2019.
\newblock \href{https://www.mdpi.com/2410-387X/3/4/27}{Cryptography, 3(4):27}.

\bibitem[CLW18]{chia_quantum-inspired_2018}
Nai-Hui Chia, Han-Hsuan Lin, and Chunhao Wang.
\newblock Quantum-inspired sublinear classical algorithms for solving low-rank
  linear systems.
\newblock {\em arXiv:1811.04852 [quant-ph]}, November 2018.
\newblock \href{http://arxiv.org/abs/1811.04852}{ArXiv: 1811.04852}.

\bibitem[CMDK20]{coyle_born_2020}
Brian Coyle, Daniel Mills, Vincent Danos, and Elham Kashefi.
\newblock The {Born} supremacy: quantum advantage and training of an {Ising}
  {Born} machine.
\newblock {\em npj Quantum Information}, 6(1):1--11, July 2020.
\newblock \href{https://www.nature.com/articles/s41534-020-00288-9}{npj QI,
  6(1):1-11}.

\bibitem[CMMS20]{cade_strategies_2020}
Chris Cade, Lana Mineh, Ashley Montanaro, and Stasja Stanisic.
\newblock Strategies for solving the {Fermi}-{Hubbard} model on near-term
  quantum computers.
\newblock {\em Physical Review B}, 102(23):235122, December 2020.
\newblock \href{https://link.aps.org/doi/10.1103/PhysRevB.102.235122}{PRB,
  102(23):235122}.

\bibitem[CMO{\etalchar{+}}20]{carolan_variational_2020}
Jacques Carolan, Masoud Mohseni, Jonathan~P. Olson, Mihika Prabhu, Changchen
  Chen, Darius Bunandar, Murphy~Yuezhen Niu, Nicholas~C. Harris, Franco N.~C.
  Wong, Michael Hochberg, Seth Lloyd, and Dirk Englund.
\newblock Variational quantum unsampling on a quantum photonic processor.
\newblock {\em Nature Physics}, 16(3):322--327, March 2020.
\newblock \href{https://www.nature.com/articles/s41567-019-0747-6}{Nat. Phys.
  16, 322–327}.

\bibitem[Co15]{chollet_keras_2015}
Francois Chollet and {others}.
\newblock Keras, 2015.
\newblock
  \href{https://github.com/fchollet/keras}{https://github.com/fchollet/keras}.

\bibitem[Coy18]{coyle_project_2018}
Brian Coyle.
\newblock {\em Project / {Dissertation} {Submission} {Index}}.
\newblock {PhD} {Thesis}, University of Edinburgh, August 2018.
\newblock
  \href{https://project-archive.inf.ed.ac.uk/msc/2018-outstanding.html}{Link}.

\bibitem[Coy20]{coyle_noiserobustclassifier_2020}
Brian Coyle.
\newblock {\em {NoiseRobustClassifier}: {Noise} {Robust} {Data} {Encodings} for
  {Quantum} {Classifiers}}.
\newblock Zenodo, February 2020.
\newblock
  \href{https://github.com/BrianCoyle/NoiseRobustClassifier/}{https://github.com/BrianCoyle/NoiseRobustClassifier/}.

\bibitem[CPCC20]{cerezo_variational_2020-2}
Marco Cerezo, Alexander Poremba, Lukasz Cincio, and Patrick~J. Coles.
\newblock Variational {Quantum} {Fidelity} {Estimation}.
\newblock {\em Quantum}, 4:248, March 2020.
\newblock \href{https://quantum-journal.org/papers/q-2020-03-26-248/}{Quantum,
  4:248}.

\bibitem[CPH05]{carreira-perpin_contrastive_2005}
Miguel~\'A. Carreira-Perpi{\~n}\'an and Geoffrey Hinton.
\newblock On contrastive divergence learning.
\newblock In Robert~G. Cowell and Zoubin Ghahramani, editors, {\em Proceedings
  of the Tenth International Workshop on Artificial Intelligence and
  Statistics}, volume~R5 of {\em Proceedings of Machine Learning Research},
  pages 33--40. PMLR, 06--08 Jan 2005.
\newblock
  \href{http://proceedings.mlr.press/r5/carreira-perpinan05a.html}{AISTATS '05,
  33-40}.

\bibitem[Cro19]{crooks_gradients_2019}
Gavin~E. Crooks.
\newblock Gradients of parameterized quantum gates using the parameter-shift
  rule and gate decomposition.
\newblock {\em arXiv:1905.13311 [quant-ph]}, May 2019.
\newblock \href{http://arxiv.org/abs/1905.13311}{ArXiv: 1905.13311}.

\bibitem[CRSC21]{cincio_machine_2021}
Lukasz Cincio, Kenneth Rudinger, Mohan Sarovar, and Patrick~J. Coles.
\newblock Machine {Learning} of {Noise}-{Resilient} {Quantum} {Circuits}.
\newblock {\em PRX Quantum}, 2(1):010324, February 2021.
\newblock \href{https://link.aps.org/doi/10.1103/PRXQuantum.2.010324}{PRX
  Quantum, 2(1):010324}.

\bibitem[CSAC20]{cerezo_variational_2020}
M.~Cerezo, Kunal Sharma, Andrew Arrasmith, and Patrick~J. Coles.
\newblock Variational {Quantum} {State} {Eigensolver}.
\newblock {\em arXiv:2004.01372 [quant-ph]}, April 2020.
\newblock \href{http://arxiv.org/abs/2004.01372}{ArXiv: 2004.01372}.

\bibitem[Csi67]{csiszar_information-type_1967}
I.~Csiszar.
\newblock Information-type measures of difference of probability distributions
  and indirect observation.
\newblock {\em Studia Scientiarum Mathematicarum Hungarica}, 2:229--318, 1967.
\newblock \href{https://ci.nii.ac.jp/naid/10028997448/en/}{SSMH, 2:229–318}.

\bibitem[CSSC18]{cincio_learning_2018}
Lukasz Cincio, Yi{\u g}it Suba{\c s}{\i}, Andrew~T. Sornborger, and Patrick~J.
  Coles.
\newblock Learning the quantum algorithm for state overlap.
\newblock {\em New Journal of Physics}, 20(11):113022, November 2018.
\newblock
  \href{https://iopscience.iop.org/article/10.1088/1367-2630/aae94a/meta}{NJP,
  20(11):113022}.

\bibitem[CSU{\etalchar{+}}20]{chivilikhin_mog-vqe_2020}
D.~Chivilikhin, A.~Samarin, V.~Ulyantsev, I.~Iorsh, A.~R. Oganov, and
  O.~Kyriienko.
\newblock {MoG}-{VQE}: {Multiobjective} genetic variational quantum
  eigensolver.
\newblock {\em arXiv:2007.04424 [cond-mat, physics:quant-ph]}, July 2020.
\newblock \href{http://arxiv.org/abs/2007.04424}{ArXiv: 2007.04424}.

\bibitem[CSV{\etalchar{+}}21]{cerezo_cost_2021}
M.~Cerezo, Akira Sone, Tyler Volkoff, Lukasz Cincio, and Patrick~J. Coles.
\newblock Cost function dependent barren plateaus in shallow parametrized
  quantum circuits.
\newblock {\em Nature Communications}, 12(1):1791, March 2021.
\newblock \href{https://www.nature.com/articles/s41467-021-21728-w}{Nature
  Communications, 12(1):1791}.

\bibitem[CT17]{carleo_solving_2017}
Giuseppe Carleo and Matthias Troyer.
\newblock Solving the {Quantum} {Many}-{Body} {Problem} with {Artificial}
  {Neural} {Networks}.
\newblock {\em Science}, 355(6325):602--606, February 2017.
\newblock \href{https://science.sciencemag.org/content/355/6325/602}{Science,
  355(6325):602–60}.

\bibitem[Cut13]{cuturi_sinkhorn_2013}
Marco Cuturi.
\newblock Sinkhorn distances: Lightspeed computation of optimal transport.
\newblock In C.~J.~C. Burges, L.~Bottou, M.~Welling, Z.~Ghahramani, and K.~Q.
  Weinberger, editors, {\em Advances in Neural Information Processing Systems},
  volume~26. Curran Associates, Inc., 2013.
\newblock
  \href{https://proceedings.neurips.cc/paper/2013/hash/af21d0c97db2e27e13572cbf59eb343d-Abstract.html}{NeurIPS
  '13}.

\bibitem[CWV{\etalchar{+}}19]{cao_cost_2019}
Shuxiang Cao, Leonard Wossnig, Brian Vlastakis, Peter Leek, and Edward Grant.
\newblock Cost function embedding and dataset encoding for machine learning
  with parameterized quantum circuits.
\newblock {\em arXiv:1910.03902 [quant-ph]}, October 2019.
\newblock \href{http://arxiv.org/abs/1910.03902}{ArXiv: 1910.03902}.

\bibitem[DB18]{dunjko_machine_2018}
Vedran Dunjko and Hans~J Briegel.
\newblock Machine learning \& artificial intelligence in the quantum domain: a
  review of recent progress.
\newblock {\em Reports Prog. Phys.}, 81(7):074001, July 2018.
\newblock
  \href{http://stacks.iop.org/0034-4885/81/i=7/a=074001?key=crossref.484b39e1cdde454de1cbc5aba8d6de34}{Reports
  Prog. Phys., 81(7):074001}.

\bibitem[DCEL09]{dankert_exact_2009}
Christoph Dankert, Richard Cleve, Joseph Emerson, and Etera Livine.
\newblock Exact and approximate unitary 2-designs and their application to
  fidelity estimation.
\newblock {\em Physical Review A}, 80(1):012304, July 2009.
\newblock \href{https://link.aps.org/doi/10.1103/PhysRevA.80.012304}{PRA,
  80(1):012304}.

\bibitem[DCLT08]{dong_quantum_2008}
Daoyi Dong, Chunlin Chen, Hanxiong Li, and Tzyh-Jong Tarn.
\newblock Quantum {Reinforcement} {Learning}.
\newblock {\em IEEE Transactions on Systems, Man, and Cybernetics, Part B
  (Cybernetics)}, 38(5):1207--1220, October 2008.
\newblock \href{10.1109/TSMCB.2008.925743}{IEEE TSMC, 38(5):1207-1220}.

\bibitem[DDZ{\etalchar{+}}05]{du_experimental_2005}
Jiangfeng Du, Thomas Durt, Ping Zou, Hui Li, L.~C. Kwek, C.~H. Lai, C.~H. Oh,
  and Artur Ekert.
\newblock Experimental {Quantum} {Cloning} with {Prior} {Partial}
  {Information}.
\newblock {\em Physical Review Letters}, 94(4):040505, February 2005.
\newblock \href{https://link.aps.org/doi/10.1103/PhysRevLett.94.040505}{PRL,
  94(4):040505}.

\bibitem[DG84]{diggle_monte_1984}
Peter~J. Diggle and Richard~J. Gratton.
\newblock Monte {Carlo} {Methods} of {Inference} for {Implicit} {Statistical}
  {Models}.
\newblock {\em Journal of the Royal Statistical Society. Series B
  (Methodological)}, 46(2):193--227, 1984.
\newblock \href{https://link.aps.org/doi/10.1103/PhysRevLett.94.040505}{JRSS,
  46(2):193–227}.

\bibitem[DG97]{duan_two_1997}
Lu-Ming Duan and Guang-Can Guo.
\newblock Two non-orthogonal states can be cloned by a unitary-reduction
  process.
\newblock {\em arXiv:quant-ph/9704020}, April 1997.
\newblock \href{http://arxiv.org/abs/quant-ph/9704020}{ArXiv: 9704020}.

\bibitem[DG98]{duan_probabilistic_1998}
Lu-Ming Duan and Guang-Can Guo.
\newblock Probabilistic {Cloning} and {Identification} of {Linearly}
  {Independent} {Quantum} {States}.
\newblock {\em Physical Review Letters}, 80(22):4999--5002, June 1998.
\newblock \href{https://link.aps.org/doi/10.1103/PhysRevLett.80.4999}{PRL,
  80(22):4999–5002}.

\bibitem[DHL{\etalchar{+}}21]{du_quantum_2021}
Yuxuan Du, Min-Hsiu Hsieh, Tongliang Liu, Dacheng Tao, and Nana Liu.
\newblock Quantum noise protects quantum classifiers against adversaries.
\newblock {\em Phys. Rev. Research}, 3(2):023153, May 2021.
\newblock \href{https://link.aps.org/doi/10.1103/PhysRevResearch.3.023153}{PRR,
  3(2):023153}.

\bibitem[DHM{\etalchar{+}}18]{dervovic_quantum_2018}
Danial Dervovic, Mark Herbster, Peter Mountney, Simone Severini, Na{\"i}ri
  Usher, and Leonard Wossnig.
\newblock Quantum linear systems algorithms: a primer.
\newblock {\em arXiv:1802.08227 [quant-ph]}, February 2018.
\newblock \href{http://arxiv.org/abs/1802.08227}{ArXiv: 1802.08227}.

\bibitem[DHS11]{duchi_adaptive_2011}
John Duchi, Elad Hazan, and Yoram Singer.
\newblock Adaptive {Subgradient} {Methods} for {Online} {Learning} and
  {Stochastic} {Optimization}.
\newblock {\em Journal of Machine Learning Research}, 12(61):2121--2159, 2011.
\newblock \href{http://jmlr.org/papers/v12/duchi11a.html}{JMLR,
  12(61):2121–2159}.

\bibitem[DI00]{divincenzo_physical_2000}
David~P. DiVincenzo and {IBM}.
\newblock The {Physical} {Implementation} of {Quantum} {Computation}.
\newblock {\em Fortschritte der Physik}, 48(9):771--783, September 2000.
\newblock
  \href{10.1002/1521-3978(200009)48:9/11<771::AID-PROP771>3.0.CO;2-E}{FdP,48(9):771-783}.

\bibitem[Die82]{dieks_communication_1982}
D.~Dieks.
\newblock Communication by {EPR} devices.
\newblock {\em Physics Letters A}, 92(6):271--272, November 1982.
\newblock
  \href{https://www.sciencedirect.com/science/article/pii/0375960182900846}{PLA,
  92(6):271–272}.

\bibitem[DLWT18]{dunjko_exponential_2018}
Vedran Dunjko, Yi-Kai Liu, Xingyao Wu, and Jacob~M. Taylor.
\newblock Exponential improvements for quantum-accessible reinforcement
  learning.
\newblock {\em arXiv:1710.11160 [quant-ph]}, August 2018.
\newblock \href{http://arxiv.org/abs/1710.11160}{ArXiv: 1710.11160}.

\bibitem[Doz16]{dozat_incorporating_2016}
Timothy Dozat.
\newblock Incorporating {Nesterov} {Momentum} into {Adam}.
\newblock {\em -}, February 2016.
\newblock
  \href{https://openreview.net/forum?id=OM0jvwB8jIp57ZJjtNEZ}{OpenReview}.

\bibitem[DPMTL20]{de_palma_quantum_2020}
Giacomo De~Palma, Milad Marvian, Dario Trevisan, and Seth Lloyd.
\newblock The quantum {Wasserstein} distance of order 1.
\newblock {\em arXiv:2009.04469 [math-ph, physics:quant-ph]}, September 2020.
\newblock \href{http://arxiv.org/abs/2009.04469}{ArXiv: 2009.04469}.

\bibitem[DPS03]{dariano_quantum_2003}
G.~Mauro D'Ariano, Matteo G.~A. Paris, and Massimiliano~F. Sacchi.
\newblock Quantum {Tomography}.
\newblock {\em arXiv:quant-ph/0302028}, February 2003.
\newblock \href{http://arxiv.org/abs/quant-ph/0302028}{ArXiv: 0302028}.

\bibitem[DTB17]{dunjko_advances_2017}
Vedran Dunjko, Jacob~M. Taylor, and Hans~J. Briegel.
\newblock Advances in quantum reinforcement learning.
\newblock In {\em 2017 {IEEE} {International} {Conference} on {Systems}, {Man},
  and {Cybernetics} ({SMC})}, pages 282--287, October 2017.
\newblock \href{https://ieeexplore.ieee.org/document/8122616/}{IEEE ICSMC,
  282–287}.

\bibitem[Dud02]{dudley_real_2002}
R.~M. Dudley.
\newblock {\em Real {Analysis} and {Probability}}.
\newblock Cambridge {Studies} in {Advanced} {Mathematics}. Cambridge University
  Press, 2 edition, 2002.
\newblock
  \href{https://www.cambridge.org/core/books/real-analysis-and-probability/26DDF2D09E526185F2347AA5658B96F6}{Link}.

\bibitem[ECBY21]{endo_hybrid_2021}
Suguru Endo, Zhenyu Cai, Simon~C. Benjamin, and Xiao Yuan.
\newblock Hybrid {Quantum}-{Classical} {Algorithms} and {Quantum} {Error}
  {Mitigation}.
\newblock {\em Journal of the Physical Society of Japan}, 90(3):032001, March
  2021.
\newblock \href{https://journals.jps.jp/doi/full/10.7566/JPSJ.90.032001}{JPSJ,
  90(3):032001}.

\bibitem[FG99]{fuchs_cryptographic_1999}
C.~A. Fuchs and J.~van~de Graaf.
\newblock Cryptographic distinguishability measures for quantum-mechanical
  states.
\newblock {\em IEEE Transactions on Information Theory}, 45(4):1216--1227, May
  1999.
\newblock \href{https://arxiv.org/abs/quant-ph/9712042}{IEEE TIT,
  45(4):1216–1227}.

\bibitem[FGG{\etalchar{+}}97]{fuchs_optimal_1997}
Christopher~A. Fuchs, Nicolas Gisin, Robert~B. Griffiths, Chi-Sheng Niu, and
  Asher Peres.
\newblock Optimal eavesdropping in quantum cryptography. {I}. {Information}
  bound and optimal strategy.
\newblock {\em Physical Review A}, 56(2):1163--1172, August 1997.
\newblock \href{https://link.aps.org/doi/10.1103/PhysRevA.56.1163}{PRA,
  56(2):1163–1172}.

\bibitem[FGG14]{farhi_quantum_2014}
Edward Farhi, Jeffrey Goldstone, and Sam Gutmann.
\newblock A {Quantum} {Approximate} {Optimization} {Algorithm}.
\newblock {\em arXiv:1411.4028 [quant-ph]}, November 2014.
\newblock \href{http://arxiv.org/abs/1411.4028}{ArXiv: 1411.4028}.

\bibitem[FGGS00]{farhi_quantum_2000}
Edward Farhi, Jeffrey Goldstone, Sam Gutmann, and Michael Sipser.
\newblock Quantum {Computation} by {Adiabatic} {Evolution}.
\newblock {\em arXiv:quant-ph/0001106}, January 2000.
\newblock \href{http://arxiv.org/abs/quant-ph/0001106}{ArXiv: 0001106}.

\bibitem[FGP20]{franca_limitations_2020}
Daniel~Stilck Franca and Raul Garcia-Patron.
\newblock Limitations of optimization algorithms on noisy quantum devices.
\newblock {\em arXiv:2009.05532 [quant-ph]}, September 2020.
\newblock \href{http://arxiv.org/abs/2009.05532}{ArXiv: 2009.05532}.

\bibitem[FGSS07]{fukumizu_kernel_2007}
Kenji Fukumizu, Arthur Gretton, Xiaohai Sun, and Bernhard Sch{\"o}lkopf.
\newblock Kernel {Measures} of {Conditional} {Dependence}.
\newblock In {\em {NIPS}}, 2007.
\newblock
  \href{https://papers.nips.cc/paper/2007/hash/3a0772443a0739141292a5429b952fe6-Abstract.html}{NIPS
  '07}.

\bibitem[FH16]{farhi_quantum_2016}
Edward Farhi and Aram~W. Harrow.
\newblock Quantum {Supremacy} through the {Quantum} {Approximate}
  {Optimization} {Algorithm}.
\newblock {\em arXiv:1602.07674 [quant-ph]}, February 2016.
\newblock \href{http://arxiv.org/abs/1602.07674}{ArXiv: 1602.07674}.

\bibitem[Fis36]{fisher_use_1936}
R.~A. Fisher.
\newblock The {Use} of {Multiple} {Measurements} in {Taxonomic} {Problems}.
\newblock {\em Annals of Eugenics}, 7(2):179--188, 1936.
\newblock
  \href{https://onlinelibrary.wiley.com/doi/abs/10.1111/j.1469-1809.1936.tb02137.x}{AoE,
  7(2):179–188}.

\bibitem[FL12]{ferenczi_symmetries_2012}
Agnes Ferenczi and Norbert L{\"u}tkenhaus.
\newblock Symmetries in quantum key distribution and the connection between
  optimal attacks and optimal cloning.
\newblock {\em Physical Review A}, 85(5):052310, May 2012.
\newblock \href{https://link.aps.org/doi/10.1103/PhysRevA.85.052310}{PRA,
  85(5):052310}.

\bibitem[FM17]{fujii_commuting_2017}
Keisuke Fujii and Tomoyuki Morimae.
\newblock Commuting quantum circuits and complexity of {Ising} partition
  functions.
\newblock {\em New J. Phys.}, 19(3):33003, March 2017.
\newblock
  \href{http://stacks.iop.org/1367-2630/19/i=3/a=033003?key=crossref.cefbe34cf11242886552ceea447a4526}{NJP,
  19(3):33003}.

\bibitem[FMWW01]{fan_quantum_2001}
Heng Fan, Keiji Matsumoto, Xiang-Bin Wang, and Miki Wadati.
\newblock Quantum cloning machines for equatorial qubits.
\newblock {\em Physical Review A}, 65(1):012304, December 2001.
\newblock \href{https://link.aps.org/doi/10.1103/PhysRevA.65.012304}{PRA,
  65(1):012304}.

\bibitem[FN18]{farhi_classification_2018}
Edward Farhi and Hartmut Neven.
\newblock Classification with {Quantum} {Neural} {Networks} on {Near} {Term}
  {Processors}.
\newblock {\em arXiv:1802.06002 [quant-ph]}, February 2018.
\newblock \href{http://arxiv.org/abs/1802.06002}{ArXiv: 1802.06002}.

\bibitem[Fow11]{fowler_constructing_2011}
Austin~G. Fowler.
\newblock Constructing arbitrary steane code single logical qubit
  fault-tolerant gates.
\newblock {\em Quantum Information \& Computation}, 11(9-10):867--873,
  September 2011.
\newblock \href{https://dl.acm.org/doi/10.5555/2230936.2230946}{QIC,
  11(9-10):867–873}.

\bibitem[FSV{\etalchar{+}}19]{feydy_interpolating_2019}
Jean Feydy, Thibault S{\'e}journ{\'e}, Fran{\c c}ois-Xavier Vialard, Shun-ichi
  Amari, Alain Trouve, and Gabriel Peyr{\'e}.
\newblock Interpolating between {Optimal} {Transport} and {MMD} using
  {Sinkhorn} {Divergences}.
\newblock In Kamalika Chaudhuri and Masashi Sugiyama, editors, {\em Proc.
  {Mach}. {Learn}. {Res}.}, volume~89 of {\em Proceedings of {Machine}
  {Learning} {Research}}, pages 2681--2690. PMLR, April 2019.
\newblock \href{http://proceedings.mlr.press/v89/feydy19a.html}{PMLR,
  (89)2681-2690}.

\bibitem[Fub04]{fubini_sulle_1904}
Guido Fubini.
\newblock Sulle metriche definite da una forma hermitiana: nota.
\newblock {\em Atti del Reale Istituto Veneto di Scienze, Lettere ed Arti},
  63:502--513, 1904.
\newblock \href{https://link.springer.com/article/10.1007/BF02420184}{LeA,
  63:502–513}.

\bibitem[Fuc96]{fuchs_information_1996}
Christopher~A. Fuchs.
\newblock Information {Gain} vs. {State} {Disturbance} in {Quantum} {Theory}.
\newblock {\em arXiv:quant-ph/9611010}, November 1996.
\newblock \href{http://arxiv.org/abs/quant-ph/9611010}{ArXiv: 9611010}.

\bibitem[FWJ{\etalchar{+}}14]{fan_quantum_2014}
Heng Fan, Yi-Nan Wang, Li~Jing, Jie-Dong Yue, Han-Duo Shi, Yong-Liang Zhang,
  and Liang-Zhu Mu.
\newblock Quantum {Cloning} {Machines} and the {Applications}.
\newblock {\em Physics Reports}, 544(3):241--322, November 2014.
\newblock
  \href{https://www.sciencedirect.com/science/article/abs/pii/S0370157314002099}{Phys.
  Rep., 544(3):241–322}.

\bibitem[GAW{\etalchar{+}}21]{gao_enhancing_2021}
Xun Gao, Eric~R. Anschuetz, Sheng-Tao Wang, J.~Ignacio Cirac, and Mikhail~D.
  Lukin.
\newblock Enhancing {Generative} {Models} via {Quantum} {Correlations}.
\newblock {\em arXiv:2101.08354 [cond-mat, physics:quant-ph, stat]}, January
  2021.
\newblock \href{http://arxiv.org/abs/2101.08354}{ArXiv: 2101.08354}.

\bibitem[GBC16]{goodfellow_deep_2016}
Ian Goodfellow, Yoshua Bengio, and Aaron Courville.
\newblock {\em Deep {Learning}}.
\newblock MIT Press, 2016.
\newblock
  \href{https://www.deeplearningbook.org/}{https://www.deeplearningbook.org/}.

\bibitem[GBC{\etalchar{+}}18]{grant_hierarchical_2018}
Edward Grant, Marcello Benedetti, Shuxiang Cao, Andrew Hallam, Joshua Lockhart,
  Vid Stojevic, Andrew~G. Green, and Simone Severini.
\newblock Hierarchical quantum classifiers.
\newblock {\em npj Quantum Information}, 4(1):1--8, December 2018.
\newblock \href{https://www.nature.com/articles/s41534-018-0116-9}{npj QI,
  4(1):1–8}.

\bibitem[GBR{\etalchar{+}}07]{gretton_kernel_2007}
Arthur Gretton, Karsten~M. Borgwardt, Malte Rasch, Bernhard Sch{\"o}lkopf, and
  Alex~J. Smola.
\newblock A {Kernel} {Method} for the {Two}-{Sample}-{Problem}.
\newblock In B.~Sch{\"o}lkopf, J.~C. Platt, and T.~Hoffman, editors, {\em
  Advances in {Neural} {Information} {Processing} {Systems} 19}, pages
  513--520. MIT Press, 2007.
\newblock
  \href{https://proceedings.neurips.cc/paper/2006/hash/e9fb2eda3d9c55a0d89c98d6c54b5b3e-Abstract.html}{NIPS
  '19, 513-520}.

\bibitem[GC01]{gottesman_quantum_2001}
Daniel Gottesman and Isaac Chuang.
\newblock Quantum {Digital} {Signatures}.
\newblock {\em arXiv:quant-ph/0105032}, November 2001.
\newblock \href{http://arxiv.org/abs/quant-ph/0105032}{ArXiv: 0105032}.

\bibitem[GCB{\etalchar{+}}19]{genevay_sample_2019}
Aude Genevay, L{\'e}na{\"i}c Chizat, Francis Bach, Marco Cuturi, and Gabriel
  Peyr{\'e}.
\newblock Sample {Complexity} of {Sinkhorn} {Divergences}.
\newblock In Kamalika Chaudhuri and Masashi Sugiyama, editors, {\em Proc.
  {Mach}. {Learn}. {Res}.}, volume~89 of {\em Proceedings of {Machine}
  {Learning} {Research}}, pages 1574--1583. PMLR, 2019.
\newblock \href{http://proceedings.mlr.press/v89/genevay19a.html}{PMLR,
  89:1574-1583}.

\bibitem[GCH{\etalchar{+}}19]{gulshen_forest_2019}
Kyle Gulshen, Joshua Combes, Matthew~P. Harrigan, Peter~J. Karalekas, Marcus
  P.~da Silva, M.~Sohaib Alam, Amy Brown, Shane Caldwell, Lauren Capelluto,
  Gavin Crooks, Daniel Girshovich, Blake~R. Johnson, Eric~C. Peterson, Anthony
  Polloreno, Nicholas~C. Rubin, Colm~A. Ryan, Alexa Staley, Nikolas~A. Tezak,
  and Joseph Valery.
\newblock Forest {Benchmarking}: {QCVV} using {PyQuil}, 2019.
\newblock
  \href{https://doi.org/10.5281/zenodo.3455847}{https://github.com/rigetti/forest-benchmarking/}.

\bibitem[GCP{\etalchar{+}}20]{gentini_noise-resilient_2020}
Laura Gentini, Alessandro Cuccoli, Stefano Pirandola, Paola Verrucchi, and
  Leonardo Banchi.
\newblock Noise-resilient variational hybrid quantum-classical optimization.
\newblock {\em Physical Review A}, 102(5):052414, November 2020.
\newblock \href{https://link.aps.org/doi/10.1103/PhysRevA.102.052414}{PRA,
  02(5):052414}.

\bibitem[GE21]{gidney_how_2019}
Craig Gidney and Martin Eker{\aa{}}.
\newblock How to factor 2048 bit {RSA} integers in 8 hours using 20 million
  noisy qubits.
\newblock {\em {Quantum}}, 5:433, April 2021.
\newblock \href{https://doi.org/10.22331/q-2021-04-15-433}{Quantum, 5:433}.

\bibitem[GEBM19]{grimsley_adaptive_2019}
Harper~R. Grimsley, Sophia~E. Economou, Edwin Barnes, and Nicholas~J. Mayhall.
\newblock An adaptive variational algorithm for exact molecular simulations on
  a quantum computer.
\newblock {\em Nature Communications}, 10(1):3007, July 2019.
\newblock \href{https://www.nature.com/articles/s41467-019-10988-2}{Nat. Comm.,
  10(1):3007}.

\bibitem[GF19]{gidney_efficient_2019}
Craig Gidney and Austin~G. Fowler.
\newblock Efficient magic state factories with a catalyzed {CCZ} to {2T}
  transformation.
\newblock {\em Quantum}, 3:135, April 2019.
\newblock \href{https://quantum-journal.org/papers/q-2019-04-30-135/}{Quantum,
  3:135}.

\bibitem[GFY21]{guan_robustness_2021}
Ji~Guan, Wang Fang, and Mingsheng Ying.
\newblock Robustness {Verification} of {Quantum} {Classifiers}.
\newblock {\em arXiv:2008.07230 [quant-ph]}, May 2021.
\newblock \href{http://arxiv.org/abs/2008.07230}{ArXiv: 2008.07230}.

\bibitem[GKK19]{gheorghiu_verification_2019}
Alexandru Gheorghiu, Theodoros Kapourniotis, and Elham Kashefi.
\newblock Verification of {Quantum} {Computation}: {An} {Overview} of
  {Existing} {Approaches}.
\newblock {\em Theory of Computing Systems}, 63(4):715--808, May 2019.
\newblock \href{https://doi.org/10.1007/s00224-018-9872-3}{TCS,
  63(4):715–808}.

\bibitem[GLF{\etalchar{+}}10]{gross_quantum_2010}
David Gross, Yi-Kai Liu, Steven~T. Flammia, Stephen Becker, and Jens Eisert.
\newblock Quantum {State} {Tomography} via {Compressed} {Sensing}.
\newblock {\em Physical Review Letters}, 105(15):150401, October 2010.
\newblock \href{https://link.aps.org/doi/10.1103/PhysRevLett.105.150401}{PRL,
  105(15):150401}.

\bibitem[GLM08]{giovannetti_quantum_2008}
Vittorio Giovannetti, Seth Lloyd, and Lorenzo Maccone.
\newblock Quantum {Random} {Access} {Memory}.
\newblock {\em Phys. Rev. Letters}, 100(16), April 2008.
\newblock \href{https://link.aps.org/doi/10.1103/PhysRevLett.100.160501}{PRL,
  100(16)}.

\bibitem[GLT18]{gilyen_quantum-inspired_2018}
Andr{\'a}s Gily{\'e}n, Seth Lloyd, and Ewin Tang.
\newblock Quantum-inspired low-rank stochastic regression with logarithmic
  dependence on the dimension.
\newblock {\em arXiv:1811.04909 [quant-ph]}, November 2018.
\newblock \href{http://arxiv.org/abs/1811.04909}{ArXiv: 1811.04909}.

\bibitem[GM97]{gisin_optimal_1997}
N.~Gisin and S.~Massar.
\newblock Optimal {Quantum} {Cloning} {Machines}.
\newblock {\em Physical Review Letters}, 79(11):2153--2156, September 1997.
\newblock \href{https://link.aps.org/doi/10.1103/PhysRevLett.79.2153}{PRL,
  79(11):2153–2156}.

\bibitem[GPAM{\etalchar{+}}14]{goodfellow_generative_2014}
Ian Goodfellow, Jean Pouget-Abadie, Mehdi Mirza, Bing Xu, David Warde-Farley,
  Sherjil Ozair, Aaron Courville, and Yoshua Bengio.
\newblock Generative {Adversarial} {Nets}.
\newblock In Z.~Ghahramani, M.~Welling, C.~Cortes, N.~Lawrence, and K.~Q.
  Weinberger, editors, {\em Advances in {Neural} {Information} {Processing}
  {Systems}}, volume~27. Curran Associates, Inc., 2014.
\newblock
  \href{https://proceedings.neurips.cc/paper/2014/hash/5ca3e9b122f61f8f06494c97b1afccf3-Abstract.html}{NIPS
  '14, 27}.

\bibitem[GPC18]{genevay_learning_2018}
Aude Genevay, Gabriel Peyre, and Marco Cuturi.
\newblock Learning {Generative} {Models} with {Sinkhorn} {Divergences}.
\newblock In {\em Proc. {Twenty}-{First} {Int}. {Conf}. {Artif}. {Intell}.
  {Stat}.}, volume~84 of {\em Proceedings of {Machine} {Learning} {Research}},
  pages 1608--1617, Playa Blanca, Lanzarote, Canary Islands, April 2018. PMLR.
\newblock \href{http://proceedings.mlr.press/v84/genevay18a.html}{PMLR,
  84:1608-1617}.

\bibitem[Gro96]{grover_fast_1996}
Lov~K. Grover.
\newblock A fast quantum mechanical algorithm for database search.
\newblock In {\em Proceedings of the twenty-eighth annual {ACM} symposium on
  {Theory} of {Computing}}, {STOC} '96, pages 212--219, Philadelphia,
  Pennsylvania, USA, July 1996. Association for Computing Machinery.
\newblock \href{https://doi.org/10.1145/237814.237866}{STOC ’96, 212–219}.

\bibitem[GS02]{gibbs_choosing_2002}
Alison~L. Gibbs and Francis~Edward Su.
\newblock On {Choosing} and {Bounding} {Probability} {Metrics}.
\newblock {\em International Statistical Review / Revue Internationale de
  Statistique}, 70(3):419--435, 2002.
\newblock \href{http://www.jstor.org/stable/1403865}{ISV, 70(3):419–435}.

\bibitem[GVT20]{gil_vidal_input_2020}
Francisco~Javier Gil~Vidal and Dirk~Oliver Theis.
\newblock Input {Redundancy} for {Parameterized} {Quantum} {Circuits}.
\newblock {\em Frontiers in Physics}, 8, 2020.
\newblock
  \href{https://www.frontiersin.org/articles/10.3389/fphy.2020.00297/full}{FiP,
  8}.

\bibitem[GvVD21]{gyurik_structural_2021}
Casper Gyurik, Dyon van Vreumingen, and Vedran Dunjko.
\newblock Structural risk minimization for quantum linear classifiers.
\newblock {\em arXiv:2105.05566 [quant-ph]}, May 2021.
\newblock \href{http://arxiv.org/abs/2105.05566}{ArXiv: 2105.05566}.

\bibitem[GWOB19]{grant_initialization_2019}
Edward Grant, Leonard Wossnig, Mateusz Ostaszewski, and Marcello Benedetti.
\newblock An initialization strategy for addressing barren plateaus in
  parametrized quantum circuits.
\newblock {\em Quantum}, 3:214, December 2019.
\newblock \href{https://quantum-journal.org/papers/q-2019-12-09-214/}{Quantum,
  3:214}.

\bibitem[GZCW21]{gacon_simultaneous_2021}
Julien Gacon, Christa Zoufal, Giuseppe Carleo, and Stefan Woerner.
\newblock Simultaneous {Perturbation} {Stochastic} {Approximation} of the
  {Quantum} {Fisher} {Information}.
\newblock {\em arXiv:2103.09232 [quant-ph]}, March 2021.
\newblock \href{http://arxiv.org/abs/2103.09232}{ArXiv: 2103.09232}.

\bibitem[GZD18]{gao_quantum_2018}
X.~Gao, Z.-Y. Zhang, and L.-M. Duan.
\newblock A quantum machine learning algorithm based on generative models.
\newblock {\em Science Advances}, 4(12):eaat9004, December 2018.
\newblock \href{https://advances.sciencemag.org/content/4/12/eaat9004}{Sci.
  Adv., 4(12):eaat9004}.

\bibitem[Han16]{hanneke_optimal_2016}
Steve Hanneke.
\newblock The {Optimal} {Sample} {Complexity} of {PAC} {Learning}.
\newblock {\em Journal of Machine Learning Research}, 17(38):1--15, 2016.
\newblock \href{http://jmlr.org/papers/v17/15-389.html}{JMLR, 17(38):1–15}.

\bibitem[HBM{\etalchar{+}}21]{huang_power_2021}
Hsin-Yuan Huang, Michael Broughton, Masoud Mohseni, Ryan Babbush, Sergio Boixo,
  Hartmut Neven, and Jarrod~R. McClean.
\newblock Power of data in quantum machine learning.
\newblock {\em Nature Communications}, 12(1):2631, May 2021.
\newblock \href{https://www.nature.com/articles/s41467-021-22539-9}{ArXiv:
  2105.02276}.

\bibitem[HC18]{heyfron_efficient_2018}
Luke Heyfron and Earl~T. Campbell.
\newblock An {Efficient} {Quantum} {Compiler} that reduces {T} count.
\newblock {\em arXiv:1712.01557 [quant-ph]}, June 2018.
\newblock \href{http://arxiv.org/abs/1712.01557}{ArXiv: 1712.01557}.

\bibitem[HCT{\etalchar{+}}19]{havlicek_supervised_2019}
Vojt{\v e}ch Havl{\'i}{\v c}ek, Antonio~D. C{\'o}rcoles, Kristan Temme, Aram~W.
  Harrow, Abhinav Kandala, Jerry~M. Chow, and Jay~M. Gambetta.
\newblock Supervised learning with quantum-enhanced feature spaces.
\newblock {\em Nature}, 567(7747):209--212, March 2019.
\newblock \href{https://www.nature.com/articles/s41586-019-0980-2}{Nature,
  567(7747):209–212}.

\bibitem[Hel69]{helstrom_quantum_1969}
Carl~W. Helstrom.
\newblock Quantum detection and estimation theory.
\newblock {\em Journal of Statistical Physics}, 1(2):231--252, June 1969.
\newblock \href{https://doi.org/10.1007/BF01007479}{J. Stat. Phys.,
  1(2):231–252}.

\bibitem[HHL09]{harrow_quantum_2009}
Aram~W. Harrow, Avinatan Hassidim, and Seth Lloyd.
\newblock Quantum {Algorithm} for {Linear} {Systems} of {Equations}.
\newblock {\em Physical Review Letters}, 103(15):150502, October 2009.
\newblock \href{https://link.aps.org/doi/10.1103/PhysRevLett.103.150502}{PRL,
  103(15):150502}.

\bibitem[Hin02]{hinton_training_2002}
Geoffrey~E. Hinton.
\newblock Training {Products} of {Experts} by {Minimizing} {Contrastive}
  {Divergence}.
\newblock {\em Neural Computation}, 14(8):1771--1800, August 2002.
\newblock \href{https://doi.org/10.1162/089976602760128018}{Neural Comp.,
  14(8):1771–1800}.

\bibitem[Hin12]{hinton_practical_2012}
Geoffrey~E Hinton.
\newblock A {Practical} {Guide} to {Training} {Restricted} {Boltzmann}
  {Machines}.
\newblock In Gr{\'e}goire Montavon, Genevi{\`e}ve~B Orr, and Klaus-Robert
  M{\"u}ller, editors, {\em Neural {Networks}: {Tricks} of the {Trade}}, volume
  7700, pages 599--619. Springer Berlin Heidelberg, Berlin, Heidelberg, 2012.
\newblock \href{http://link.springer.com/10.1007/978-3-642-35289-8_32}{NNs:
  TotT, 7700: 599–619}.

\bibitem[HKEG19]{hangleiter_sample_2019}
Dominik Hangleiter, Martin Kliesch, Jens Eisert, and Christian Gogolin.
\newblock Sample {Complexity} of {Device}-{Independently} {Certified}
  ``{Quantum} {Supremacy}".
\newblock {\em Physical Review Letters}, 122(21):210502, May 2019.
\newblock \href{https://link.aps.org/doi/10.1103/PhysRevLett.122.210502}{PRL,
  122(21):210502}.

\bibitem[HM17]{harrow_quantum_2017}
Aram~W. Harrow and Ashley Montanaro.
\newblock Quantum computational supremacy.
\newblock {\em Nature}, 549(7671):203--209, September 2017.
\newblock \href{https://www.nature.com/articles/nature23458}{Nature,
  549(7671):203–209}.

\bibitem[HN21]{harrow_low-depth_2021}
Aram~W. Harrow and John~C. Napp.
\newblock Low-{Depth} {Gradient} {Measurements} {Can} {Improve} {Convergence}
  in {Variational} {Hybrid} {Quantum}-{Classical} {Algorithms}.
\newblock {\em Physical Review Letters}, 126(14):140502, April 2021.
\newblock \href{https://link.aps.org/doi/10.1103/PhysRevLett.126.140502}{PRL,
  126(14):140502}.

\bibitem[Ho95]{ho_random_1995}
Tin~Kam Ho.
\newblock Random decision forests.
\newblock In {\em Proceedings of 3rd {International} {Conference} on {Document}
  {Analysis} and {Recognition}}, volume~1, pages 278--282 vol.1, August 1995.
\newblock \href{https://ieeexplore.ieee.org/document/598994}{ICDAR '95,
  1:278-282}.

\bibitem[Hoe63]{hoeffding_probability_1963}
Wassily Hoeffding.
\newblock Probability {Inequalities} for {Sums} of {Bounded} {Random}
  {Variables}.
\newblock {\em Journal of the American Statistical Association},
  58(301):13--30, 1963.
\newblock
  \href{https://www.tandfonline.com/doi/abs/10.1080/01621459.1963.10500830}{JASA,
  58(301):13–3}.

\bibitem[Hol73]{holevo_statistical_1973}
A.~S Holevo.
\newblock Statistical decision theory for quantum systems.
\newblock {\em Journal of Multivariate Analysis}, 3(4):337--394, December 1973.
\newblock
  \href{http://www.sciencedirect.com/science/article/pii/0047259X73900286}{JMA,
  3(4):337–394}.

\bibitem[Hop82]{hopfield_neural_1982}
J.~J. Hopfield.
\newblock Neural networks and physical systems with emergent collective
  computational abilities.
\newblock {\em Proceedings of the National Academy of Sciences},
  79(8):2554--2558, April 1982.
\newblock \href{https://www.pnas.org/content/79/8/2554}{PNAS,
  79(8):2554–2558}.

\bibitem[HP20]{hamilton_error-mitigated_2020}
Kathleen~E. Hamilton and Raphael~C. Pooser.
\newblock Error-mitigated data-driven circuit learning on noisy quantum
  hardware.
\newblock {\em Quantum Machine Intelligence}, 2(1):10, July 2020.
\newblock \href{https://doi.org/10.1007/s42484-020-00021-x}{QUMI, (1):10}.

\bibitem[HPSB21]{hubregtsen_evaluation_2021}
Thomas Hubregtsen, Josef Pichlmeier, Patrick Stecher, and Koen Bertels.
\newblock Evaluation of parameterized quantum circuits: on the relation between
  classification accuracy, expressibility, and entangling capability.
\newblock {\em Quantum Machine Intelligence}, 3(1):9, March 2021.
\newblock \href{https://doi.org/10.1007/s42484-021-00038-w}{QUMI, 3(1):9}.

\bibitem[HSCC21]{holmes_connecting_2021}
Zo{\"e} Holmes, Kunal Sharma, M.~Cerezo, and Patrick~J. Coles.
\newblock Connecting ansatz expressibility to gradient magnitudes and barren
  plateaus.
\newblock {\em arXiv:2101.02138 [quant-ph, stat]}, January 2021.
\newblock \href{http://arxiv.org/abs/2101.02138}{ArXiv: 2101.02138}.

\bibitem[HSNF18]{heya_variational_2018}
Kentaro Heya, Yasunari Suzuki, Yasunobu Nakamura, and Keisuke Fujii.
\newblock Variational {Quantum} {Gate} {Optimization}.
\newblock {\em arXiv:1810.12745 [quant-ph]}, October 2018.
\newblock \href{http://arxiv.org/abs/1810.12745}{ArXiv:1810.12745}.

\bibitem[HSPC20]{henderson_quanvolutional_2020}
Maxwell Henderson, Samriddhi Shakya, Shashindra Pradhan, and Tristan Cook.
\newblock Quanvolutional neural networks: powering image recognition with
  quantum circuits.
\newblock {\em Quantum Machine Intelligence}, 2(1):1--9, June 2020.
\newblock \href{https://doi.org/10.1007/s42484-020-00012-y}{QUMI, 2(1):1–9}.

\bibitem[HSST18]{haner_software_2018}
Thomas H{\"a}ner, Damian~S. Steiger, Krysta Svore, and Matthias Troyer.
\newblock A software methodology for compiling quantum programs.
\newblock {\em Quantum Science and Technology}, 3(2):020501, February 2018.
\newblock \href{https://doi.org/10.1088/2058-9565/aaa5cc}{QST, 3(2):020501}.

\bibitem[HWFZ20]{huang_superconducting_2020}
He-Liang Huang, Dachao Wu, Daojin Fan, and Xiaobo Zhu.
\newblock Superconducting quantum computing: a review.
\newblock {\em Science China Information Sciences}, 63(8):180501, July 2020.
\newblock \href{https://doi.org/10.1007/s11432-020-2881-9}{SCIS, 63(8):180501}.

\bibitem[HWGF{\etalchar{+}}21]{hubregtsen_training_2021}
Thomas Hubregtsen, David Wierichs, Elies Gil-Fuster, Peter-Jan H.~S. Derks,
  Paul~K. Faehrmann, and Johannes~Jakob Meyer.
\newblock Training {Quantum} {Embedding} {Kernels} on {Near}-{Term} {Quantum}
  {Computers}.
\newblock {\em arXiv:2105.02276 [quant-ph]}, May 2021.
\newblock \href{http://arxiv.org/abs/2105.02276}{ArXiv: 2105.02276}.

\bibitem[HWO{\etalchar{+}}19]{hadfield_quantum_2019}
Stuart Hadfield, Zhihui Wang, Bryan O{\textquoteright}Gorman, Eleanor~G.
  Rieffel, Davide Venturelli, and Rupak Biswas.
\newblock From the {Quantum} {Approximate} {Optimization} {Algorithm} to a
  {Quantum} {Alternating} {Operator} {Ansatz}.
\newblock {\em Algorithms}, 12(2):34, February 2019.
\newblock \href{https://www.mdpi.com/1999-4893/12/2/34}{Algorithms, 12(2):34}.

\bibitem[JB18]{jones_quantum_2018}
Tyson Jones and Simon~C. Benjamin.
\newblock Quantum compilation and circuit optimisation via energy dissipation.
\newblock {\em arXiv:1811.03147 [quant-ph]}, December 2018.
\newblock \href{http://arxiv.org/abs/1811.03147}{ArXiv:1811.03147}.

\bibitem[JCKK21]{jain_graph_2021}
Nishant Jain, Brian Coyle, Elham Kashefi, and Niraj Kumar.
\newblock Graph neural network initialisation of quantum approximate
  optimisation.
\newblock {\em arXiv:2111.03016 [quant-ph]}, November 2021.
\newblock \href{http://arxiv.org/abs/2111.03016}{ArXiv: 2111.03016}.

\bibitem[JDM{\etalchar{+}}21]{johri_nearest_2021}
Sonika Johri, Shantanu Debnath, Avinash Mocherla, Alexandros Singk, Anupam
  Prakash, Jungsang Kim, and Iordanis Kerenidis.
\newblock Nearest centroid classification on a trapped ion quantum computer.
\newblock {\em npj Quantum Information}, 7(1):1--11, August 2021.
\newblock \href{https://www.nature.com/articles/s41534-021-00456-5}{npj QI,
  7(1):1–11}.

\bibitem[{Jea}18]{jeanfeydy_mmd_2018}
{Jeanfeydy}.
\newblock {MMD}, {Hausdorff} and {Sinkhorn} divergences scaled up to 1,000,000
  samples., November 2018.
\newblock
  \href{https://github.com/jeanfeydy/global-divergences}{https://github.com/jeanfeydy/global-divergences}.

\bibitem[JGM{\etalchar{+}}21]{jerbi_variational_2021}
Sofiene Jerbi, Casper Gyurik, Simon Marshall, Hans~J. Briegel, and Vedran
  Dunjko.
\newblock Variational quantum policies for reinforcement learning.
\newblock {\em arXiv:2103.05577 [quant-ph, stat]}, March 2021.
\newblock \href{http://arxiv.org/abs/2103.05577}{ArXiv: 2103.05577}.

\bibitem[JJB{\etalchar{+}}19]{jasek_experimental_2019}
Jan Ja{\v s}ek, Kate{\v r}ina Jir{\'a}kov{\'a}, Karol Bartkiewicz, Karol
  Bartkiewicz, Anton{\'i}n {\v C}ernoch, Tom{\'a}{\v s} F{\"u}rst, and Karel
  Lemr.
\newblock Experimental hybrid quantum-classical reinforcement learning by boson
  sampling: how to train a quantum cloner.
\newblock {\em Optics Express}, 27(22):32454--32464, October 2019.
\newblock
  \href{https://www.osapublishing.org/oe/abstract.cfm?uri=oe-27-22-32454}{OE,
  27(22):32454–32464}.

\bibitem[Joz94]{jozsa_fidelity_1994}
Richard Jozsa.
\newblock Fidelity for {Mixed} {Quantum} {States}.
\newblock {\em Journal of Modern Optics}, 41(12):2315--2323, December 1994.
\newblock \href{https://doi.org/10.1080/09500349414552171}{JMO,
  41(12):2315–2323}.

\bibitem[JRO{\etalchar{+}}17]{johnson_qvector_2017}
Peter~D. Johnson, Jonathan Romero, Jonathan Olson, Yudong Cao, and Al{\'a}n
  Aspuru-Guzik.
\newblock {QVECTOR}: an algorithm for device-tailored quantum error correction.
\newblock {\em arXiv:1711.02249 [quant-ph]}, November 2017.
\newblock \href{http://arxiv.org/abs/1711.02249}{ArXiv: 1711.02249}.

\bibitem[JTPN{\etalchar{+}}21]{jerbi_quantum_2021}
Sofiene Jerbi, Lea~M. Trenkwalder, Hendrik Poulsen~Nautrup, Hans~J. Briegel,
  and Vedran Dunjko.
\newblock Quantum {Enhancements} for {Deep} {Reinforcement} {Learning} in
  {Large} {Spaces}.
\newblock {\em PRX Quantum}, 2(1):010328, February 2021.
\newblock \href{https://link.aps.org/doi/10.1103/PRXQuantum.2.010328}{PRX
  Quantum, 2(1):010328}.

\bibitem[JWHT13a]{james_introduction_2013}
Gareth James, Daniela Witten, Trevor Hastie, and Robert Tibshirani.
\newblock Introduction.
\newblock In {\em An {Introduction} to {Statistical} {Learning}: with
  {Applications} in {R}}, Springer {Texts} in {Statistics}, pages 1--14.
  Springer, New York, NY, 2013.
\newblock \href{https://doi.org/10.1007/978-1-4614-7138-7_2}{STS, 1-14}.

\bibitem[JWHT13b]{james_statistical_2013}
Gareth James, Daniela Witten, Trevor Hastie, and Robert Tibshirani.
\newblock Statistical {Learning}.
\newblock In {\em An {Introduction} to {Statistical} {Learning}: with
  {Applications} in {R}}, Springer {Texts} in {Statistics}, pages 15--57.
  Springer, New York, NY, 2013.
\newblock \href{https://doi.org/10.1007/978-1-4614-7138-7_2}{STS, 15-57}.

\bibitem[KACC20]{kubler_adaptive_2020}
Jonas~M. K{\"u}bler, Andrew Arrasmith, Lukasz Cincio, and Patrick~J. Coles.
\newblock An {Adaptive} {Optimizer} for {Measurement}-{Frugal} {Variational}
  {Algorithms}.
\newblock {\em Quantum}, 4:263, May 2020.
\newblock \href{https://quantum-journal.org/papers/q-2020-05-11-263/}{Quantum,
  4:263}.

\bibitem[KB15]{kingma_adam_2015}
Diederik~P Kingma and Jimmy Ba.
\newblock Adam: {A} {Method} for {Stochastic} {Optimization}.
\newblock In Yoshua Bengio and Yann LeCun, editors, {\em 3rd {Int}. {Conf}.
  {Learn}. {Represent}. {ICLR} 2015, {San} {Diego}, {CA}, {USA}, {May} 7-9,
  2015, {Conf}. {Track} {Proc}.}, 2015.
\newblock \href{http://arxiv.org/abs/1412.6980}{ICLR '15}.

\bibitem[KB20a]{koczor_quantum_2020-1}
B{\'a}lint Koczor and Simon~C. Benjamin.
\newblock Quantum {Analytic} {Descent}.
\newblock {\em arXiv:2008.13774 [quant-ph]}, December 2020.
\newblock \href{http://arxiv.org/abs/2008.13774}{ArXiv: 2008.13774}.

\bibitem[KB20b]{koczor_quantum_2020}
B{\'a}lint Koczor and Simon~C. Benjamin.
\newblock Quantum natural gradient generalised to non-unitary circuits.
\newblock {\em arXiv:1912.08660 [quant-ph]}, December 2020.
\newblock \href{http://arxiv.org/abs/1912.08660}{ArXiv: 1912.08660}.

\bibitem[KBDW83]{kraus_states_1983}
Karl Kraus, A.~B{\"o}hm, J.~D. Dollard, and W.~H. Wootters, editors.
\newblock {\em States, {Effects}, and {Operations} {Fundamental} {Notions} of
  {Quantum} {Theory}}.
\newblock Lecture {Notes} in {Physics}. Springer, Berlin, Heidelberg, 1983.
\newblock \href{https://doi.org/10.1007/3540127321_22}{Link}.

\bibitem[KCW21]{kieferova_quantum_2021}
Maria Kieferova, Ortiz~Marrero Carlos, and Nathan Wiebe.
\newblock Quantum {Generative} {Training} {Using} {R{\'e}nyi} {Divergences}.
\newblock {\em arXiv:2106.09567 [quant-ph]}, June 2021.
\newblock \href{http://arxiv.org/abs/2106.09567}{ArXiv: 2106.09567}.

\bibitem[KGV83]{kirkpatrick_optimization_1983}
S.~Kirkpatrick, C.~D. Gelatt, and M.~P. Vecchi.
\newblock Optimization by {Simulated} {Annealing}.
\newblock {\em Science}, 220(4598):671--680, May 1983.
\newblock \href{http://science.sciencemag.org/content/220/4598/671}{Science,
  220(4598):671–680}.

\bibitem[Kit97]{kitaev_quantum_1997}
A.~Yu Kitaev.
\newblock Quantum computations: algorithms and error correction.
\newblock {\em Russian Mathematical Surveys}, 52(6):1191, December 1997.
\newblock
  \href{https://iopscience.iop.org/article/10.1070/RM1997v052n06ABEH002155/meta}{RMS,
  2(6):1191}.

\bibitem[Kit03]{kitaev_fault-tolerant_2003}
A.~Yu. Kitaev.
\newblock Fault-tolerant quantum computation by anyons.
\newblock {\em Annals of Physics}, 303(1):2--30, January 2003.
\newblock
  \href{https://www.sciencedirect.com/science/article/pii/S0003491602000180}{AoP,
  303(1):2–30}.

\bibitem[KKR06]{kempe_complexity_2006}
Julia Kempe, Alexei Kitaev, and Oded Regev.
\newblock The {Complexity} of the {Local} {Hamiltonian} {Problem}.
\newblock {\em SIAM Journal on Computing}, 35(5):1070--1097, January 2006.
\newblock \href{https://epubs.siam.org/doi/10.1137/S0097539704445226}{SIAM JoC,
  35(5):1070–1097}.

\bibitem[KL51]{kullback_information_1951}
S~Kullback and R~A Leibler.
\newblock On {Information} and {Sufficiency}.
\newblock {\em Annals of Mathematical Statistics}, 22(1):79--86, 1951.
\newblock \href{https://doi.org/10.1214/aoms/1177729694}{AMS, 22(1):79–86}.

\bibitem[KL20]{kerenidis_classification_2020}
Iordanis Kerenidis and Alessandro Luongo.
\newblock Classification of the {MNIST} data set with quantum slow feature
  analysis.
\newblock {\em Physical Review A}, 101(6):062327, June 2020.
\newblock \href{https://link.aps.org/doi/10.1103/PhysRevA.101.062327}{PRA,
  101(6):062327}.

\bibitem[KLP{\etalchar{+}}19]{khatri_quantum-assisted_2019}
Sumeet Khatri, Ryan LaRose, Alexander Poremba, Lukasz Cincio, Andrew~T.
  Sornborger, and Patrick~J. Coles.
\newblock Quantum-assisted quantum compiling.
\newblock {\em Quantum}, 3:140, May 2019.
\newblock \href{https://quantum-journal.org/papers/q-2019-05-13-140/}{Quantum,
  3:140}.

\bibitem[KLZ98]{knill_resilient_1998}
Emanuel Knill, Raymond Laflamme, and Wojciech~H. Zurek.
\newblock Resilient {Quantum} {Computation}.
\newblock {\em Science}, 279(5349):342--345, January 1998.
\newblock \href{https://science.sciencemag.org/content/279/5349/342}{Science,
  279(5349):342-345}.

\bibitem[KMF{\etalchar{+}}16]{krenn_automated_2016}
Mario Krenn, Mehul Malik, Robert Fickler, Radek Lapkiewicz, and Anton
  Zeilinger.
\newblock Automated {Search} for new {Quantum} {Experiments}.
\newblock {\em Physical Review Letters}, 116(9):090405, March 2016.
\newblock \href{https://link.aps.org/doi/10.1103/PhysRevLett.116.090405}{PRL,
  116(9):090405}.

\bibitem[KMR{\etalchar{+}}94]{kearns_learnability_1994}
Michael Kearns, Yishay Mansour, Dana Ron, Ronitt Rubinfeld, Robert~E Schapire,
  and Linda Sellie.
\newblock On the {Learnability} of {Discrete} {Distributions}.
\newblock In {\em Proc. {Twenty}-sixth {Annual} {ACM} {Symposium} {Theory}
  {Computing}}, {STOC} '94, pages 273--282, New York, NY, USA, 1994. ACM.
\newblock \href{http://doi.acm.org/10.1145/195058.195155}{STOC '94, 273-282}.

\bibitem[KMS19]{kubler_quantum_2019}
Jonas~M. K{\"u}bler, Krikamol Muandet, and Bernhard Sch{\"o}lkopf.
\newblock Quantum mean embedding of probability distributions.
\newblock {\em Phys. Rev. Research}, 1(3):033159, December 2019.
\newblock \href{https://link.aps.org/doi/10.1103/PhysRevResearch.1.033159}{PRR,
  1(3):033159}.

\bibitem[Kon21]{kondratyev_non-differentiable_2021}
Alex Kondratyev.
\newblock Non-{Differentiable} {Leaning} of {Quantum} {Circuit} {Born}
  {Machine} with {Genetic} {Algorithm}.
\newblock {\em Wilmott}, 2021(114):50--61, 2021.
\newblock
  \href{https://onlinelibrary.wiley.com/doi/pdf/10.1002/wilm.10943}{Wilmott,
  2021(114):50–6}.

\bibitem[Kot14]{kothari_optimal_2014}
Robin Kothari.
\newblock An optimal quantum algorithm for the oracle identification problem.
\newblock In {\em 31st {International} {Symposium} on {Theoretical} {Aspects}
  of {Computer} {Science} ({STACS} 2014)}, volume~25 of {\em Leibniz
  {International} {Proceedings} in {Informatics} ({LIPIcs})}, pages 482--493,
  Dagstuhl, Germany, 2014. Schloss Dagstuhl{\textendash}Leibniz-Zentrum fuer
  Informatik.
\newblock \href{http://drops.dagstuhl.de/opus/volltexte/2014/4481}{STACS '14,
  482-493}.

\bibitem[KP16]{kerenidis_quantum_2016}
Iordanis Kerenidis and Anupam Prakash.
\newblock Quantum {Recommendation} {Systems}.
\newblock {\em arXiv:1603.08675 [quant-ph]}, September 2016.
\newblock \href{http://arxiv.org/abs/1603.08675}{ArXiv: 1603.08675}.

\bibitem[KP20]{kerenidis_quantum_2020}
Iordanis Kerenidis and Anupam Prakash.
\newblock Quantum gradient descent for linear systems and least squares.
\newblock {\em Physical Review A}, 101(2):022316, February 2020.
\newblock \href{https://link.aps.org/doi/10.1103/PhysRevA.101.022316}{PRA,
  101(2):022316}.

\bibitem[KS19]{kondratyev_market_2019}
Alexei Kondratyev and Christian Schwarz.
\newblock The {Market} {Generator}.
\newblock {\em Available SSRN 3384948}, 2019.
\newblock
  \href{https://papers.ssrn.com/sol3/papers.cfm?abstract_id=3384948}{SSRN
  3384948}.

\bibitem[KSB{\etalchar{+}}20]{kjaergaard_superconducting_2020}
Morten Kjaergaard, Mollie~E. Schwartz, Jochen Braum{\"u}ller, Philip Krantz,
  Joel I.-J. Wang, Simon Gustavsson, and William~D. Oliver.
\newblock Superconducting {Qubits}: {Current} {State} of {Play}.
\newblock {\em Annual Review of Condensed Matter Physics}, 11(1):369--395,
  2020.
\newblock
  \href{https://doi.org/10.1146/annurev-conmatphys-031119-050605}{ARCMP,
  1(1):369–395}.

\bibitem[KTP{\etalchar{+}}20]{karalekas_quantum-classical_2020}
Peter~J Karalekas, Nikolas~A Tezak, Eric~C Peterson, Colm~A Ryan, Marcus~P
  da~Silva, and Robert~S Smith.
\newblock A quantum-classical cloud platform optimized for variational hybrid
  algorithms.
\newblock {\em Quantum Science \& Technology}, 5(2):24003, April 2020.
\newblock
  \href{https://onlinelibrary.wiley.com/doi/pdf/10.1002/wilm.10943}{QST,
  5(2):24003}.

\bibitem[KV94]{kearns_introduction_1994}
Michael~J. Kearns and Umesh~V. Vazirani.
\newblock {\em An introduction to computational learning theory}.
\newblock MIT Press, Cambridge, MA, USA, 1994.
\newblock
  \href{https://mitpress.mit.edu/books/introduction-computational-learning-theory}{Link}.

\bibitem[KW99]{keyl_optimal_1999}
M.~Keyl and R.~F. Werner.
\newblock Optimal cloning of pure states, testing single clones.
\newblock {\em Journal of Mathematical Physics}, 40(7):3283--3299, June 1999.
\newblock \href{https://aip.scitation.org/doi/abs/10.1063/1.532887}{JMP,
  40(7):3283–3299}.

\bibitem[KW17]{kieferova_tomography_2017}
Maria Kieferova and Nathan Wiebe.
\newblock Tomography and {Generative} {Data} {Modeling} via {Quantum}
  {Boltzmann} {Training}.
\newblock {\em Physical Review A}, 96(6), December 2017.
\newblock \href{https://link.aps.org/doi/10.1103/PhysRevA.96.062327}{PRA,
  96(6):062327}.

\bibitem[LA87]{laarhoven_simulated_1987}
P.~J.~van Laarhoven and E.~H. Aarts.
\newblock {\em Simulated {Annealing}: {Theory} and {Applications}}.
\newblock Mathematics and {Its} {Applications}. Springer Netherlands, 1987.
\newblock \href{https://www.springer.com/gp/book/9789027725134}{Link}.

\bibitem[LaR19]{larose_overview_2019}
Ryan LaRose.
\newblock Overview and {Comparison} of {Gate} {Level} {Quantum} {Software}
  {Platforms}.
\newblock {\em Quantum}, 3:130, March 2019.
\newblock \href{https://quantum-journal.org/papers/q-2019-03-25-130/}{Quantum,
  3:130}.

\bibitem[LAT21]{liu_rigorous_2021}
Yunchao Liu, Srinivasan Arunachalam, and Kristan Temme.
\newblock A rigorous and robust quantum speed-up in supervised machine
  learning.
\newblock {\em Nature Physics}, pages 1--5, July 2021.
\newblock \href{https://www.nature.com/articles/s41567-021-01287-z}{Nat. Phys.,
  1–5}.

\bibitem[LBBH98]{lecun_gradient-based_1998}
Y.~Lecun, L.~Bottou, Y.~Bengio, and P.~Haffner.
\newblock Gradient-based learning applied to document recognition.
\newblock {\em Proceedings of the IEEE}, 86(11):2278--2324, November 1998.
\newblock \href{https://ieeexplore.ieee.org/document/726791}{PIEEE,
  86(11):2278-2324}.

\bibitem[LC98]{lo_why_1998}
Hoi-Kwong Lo and H.~F. Chau.
\newblock Why quantum bit commitment and ideal quantum coin tossing are
  impossible.
\newblock {\em Physica D: Nonlinear Phenomena}, 120(1):177--187, September
  1998.
\newblock
  \href{https://www.sciencedirect.com/science/article/pii/S0167278998000530}{PD:NP,
  120(1):177–187}.

\bibitem[LC20]{larose_robust_2020}
Ryan LaRose and Brian Coyle.
\newblock Robust data encodings for quantum classifiers.
\newblock {\em Physical Review A}, 102(3):032420, September 2020.
\newblock \href{https://link.aps.org/doi/10.1103/PhysRevA.102.032420}{PRA,
  102(3):032420}.

\bibitem[LCW98]{lidar_decoherence_1998}
D.~A. Lidar, I.~L. Chuang, and K.~B. Whaley.
\newblock Decoherence {Free} {Subspaces} for {Quantum} {Computation}.
\newblock {\em Phys. Rev. Letters}, 81(12):2594--2597, September 1998.
\newblock \href{http://arxiv.org/abs/quant-ph/9807004}{ArXiv: 9807004}.

\bibitem[LFC{\etalchar{+}}20]{li_quantum_2020}
Li~Li, Minjie Fan, Marc Coram, Patrick Riley, and Stefan Leichenauer.
\newblock Quantum optimization with a novel {Gibbs} objective function and
  ansatz architecture search.
\newblock {\em Phys. Rev. Research}, 2(2):023074, April 2020.
\newblock \href{https://link.aps.org/doi/10.1103/PhysRevResearch.2.023074}{PRR,
  2(2):023074}.

\bibitem[LHF21]{liao_quantum_2021}
Yidong Liao, Min-Hsiu Hsieh, and Chris Ferrie.
\newblock Quantum {Optimization} for {Training} {Quantum} {Neural} {Networks}.
\newblock {\em arXiv:2103.17047 [quant-ph]}, March 2021.
\newblock \href{http://arxiv.org/abs/2103.17047}{ArXiv: 2103.17047}.

\bibitem[Lid14]{lidar_review_2014}
Daniel~A. Lidar.
\newblock Review of {Decoherence}-{Free} {Subspaces}, {Noiseless} {Subsystems},
  and {Dynamical} {Decoupling}.
\newblock In {\em Quantum {Information} and {Computation} for {Chemistry}},
  pages 295--354. John Wiley \& Sons, Ltd, 2014.
\newblock
  \href{https://onlinelibrary.wiley.com/doi/abs/10.1002/9781118742631.ch11}{QICC,
  295–354}.

\bibitem[LJL{\etalchar{+}}10]{ladd_quantum_2010}
T.~D. Ladd, F.~Jelezko, R.~Laflamme, Y.~Nakamura, C.~Monroe, and J.~L.
  O{\textquoteright}Brien.
\newblock Quantum computers.
\newblock {\em Nature}, 464(7285):45--53, March 2010.
\newblock \href{https://www.nature.com/articles/nature08812}{Science,
  279(5349):342–345}.

\bibitem[LLJ16]{liu_kernelized_2016}
Qiang Liu, Jason~D. Lee, and Michael Jordan.
\newblock A {Kernelized} {Stein} {Discrepancy} for {Goodness}-of-fit {Tests}.
\newblock In {\em Proceedings of the 33rd {International} {Conference} on
  {International} {Conference} on {Machine} {Learning} - {Volume} 48},
  {ICML}'16, pages 276--284, New York, NY, USA, 2016. JMLR.org.
\newblock \href{http://dl.acm.org/citation.cfm?id=3045390.3045421}{ICML '16,
  276-284}.

\bibitem[LMR13]{lloyd_quantum_2013}
Seth Lloyd, Masoud Mohseni, and Patrick Rebentrost.
\newblock Quantum algorithms for supervised and unsupervised machine learning.
\newblock {\em arXiv:1307.0411 [quant-ph]}, July 2013.
\newblock \href{http://arxiv.org/abs/1307.0411}{ArXiv: 1307.0411}.

\bibitem[LMR14]{lloyd_quantum_2014}
Seth Lloyd, Masoud Mohseni, and Patrick Rebentrost.
\newblock Quantum principal component analysis.
\newblock {\em Nature Physics}, 10(9):631--633, September 2014.
\newblock \href{https://www.nature.com/articles/nphys3029}{Nat. Phys.,
  10(9):631–633}.

\bibitem[LS20a]{lockwood_reinforcement_2020}
Owen Lockwood and Mei Si.
\newblock Reinforcement {Learning} with {Quantum} {Variational} {Circuit}.
\newblock {\em Proceedings of the AAAI Conference on Artificial Intelligence
  and Interactive Digital Entertainment}, 16(1):245--251, October 2020.
\newblock \href{https://ojs.aaai.org/index.php/AIIDE/article/view/7437}{CAIIDE,
  16(1):245–251}.

\bibitem[LS20b]{lostaglio_contextual_2020}
Matteo Lostaglio and Gabriel Senno.
\newblock Contextual advantage for state-dependent cloning.
\newblock {\em Quantum}, 4:258, April 2020.
\newblock \href{https://quantum-journal.org/papers/q-2020-04-27-258/}{Quantum,
  4:258}.

\bibitem[LSCB21]{leadbeater_f-divergences_2021}
Chiara Leadbeater, Louis Sharrock, Brian Coyle, and Marcello Benedetti.
\newblock F-divergences and cost function locality in generative modelling with
  quantum circuits.
\newblock {\em Entropy}, 23(10), 2021.
\newblock \href{https://www.mdpi.com/1099-4300/23/10/1281}{Entropy 2021,
  23(10), 1281}.

\bibitem[LSI{\etalchar{+}}20]{lloyd_quantum_2020}
Seth Lloyd, Maria Schuld, Aroosa Ijaz, Josh Izaac, and Nathan Killoran.
\newblock Quantum embeddings for machine learning.
\newblock {\em arXiv:2001.03622 [quant-ph]}, February 2020.
\newblock \href{http://arxiv.org/abs/2001.03622}{ArXiv: 2001.03622}.

\bibitem[LSY19]{liu_darts_2019}
Hanxiao Liu, Karen Simonyan, and Yiming Yang.
\newblock {DARTS}: {Differentiable} {Architecture} {Search}.
\newblock In {\em 7th {International} {Conference} on {Learning}
  {Representations}, {ICLR} 2019, {New} {Orleans}, {LA}, {USA}, {May} 6-9,
  2019}. OpenReview.net, 2019.
\newblock \href{https://openreview.net/forum?id=S1eYHoC5FX}{OpenReview}.

\bibitem[LT18]{li_gradient_2018}
Yingzhen Li and Richard~E. Turner.
\newblock Gradient estimators for implicit models.
\newblock In {\em 6th International Conference on Learning Representations,
  {ICLR} 2018, Vancouver, BC, Canada, April 30 - May 3, 2018, Conference Track
  Proceedings}. OpenReview.net, 2018.
\newblock \href{https://openreview.net/forum?id=SJi9WOeRb}{ICLR '18}.

\bibitem[LTOJ{\etalchar{+}}19]{larose_variational_2019}
Ryan LaRose, Arkin Tikku, {\'E}tude O{\textquoteright}Neel-Judy, Lukasz Cincio,
  and Patrick~J. Coles.
\newblock Variational quantum state diagonalization.
\newblock {\em npj Quantum Information}, 5(1):1--10, June 2019.
\newblock \href{https://www.nature.com/articles/s41534-019-0167-6}{npj QI,
  5(1):1–10}.

\bibitem[LW18a]{liu_differentiable_2018}
Jin-Guo Liu and Lei Wang.
\newblock Differentiable learning of quantum circuit {Born} machines.
\newblock {\em Physical Review A}, 98(6):062324, December 2018.
\newblock \href{https://link.aps.org/doi/10.1103/PhysRevA.98.062324}{PRA,
  98(6):062324}.

\bibitem[LW18b]{lloyd_quantum_2018}
Seth Lloyd and Christian Weedbrook.
\newblock Quantum {Generative} {Adversarial} {Learning}.
\newblock {\em Physical Review Letters}, 121(4):040502, July 2018.
\newblock \href{https://link.aps.org/doi/10.1103/PhysRevLett.121.040502}{PRL,
  121(4):040502}.

\bibitem[LW20]{liu_vulnerability_2019}
Nana Liu and Peter Wittek.
\newblock Vulnerability of quantum classification to adversarial perturbations.
\newblock {\em Physical Review A}, 101:062331, Jun 2020.
\newblock \href{https://link.aps.org/doi/10.1103/PhysRevA.101.062331}{PRA,
  101(6):062331}.

\bibitem[LWF{\etalchar{+}}17]{lekitsch_blueprint_2017}
Bjoern Lekitsch, Sebastian Weidt, Austin~G. Fowler, Klaus M{\o}lmer, Simon~J.
  Devitt, Christof Wunderlich, and Winfried~K. Hensinger.
\newblock Blueprint for a microwave trapped ion quantum computer.
\newblock {\em Science Advances}, 3(2):e1601540, February 2017.
\newblock \href{https://advances.sciencemag.org/content/3/2/e1601540}{Sci.
  Adv., 3(2):e1601540}.

\bibitem[MBK21]{mari_estimating_2021}
Andrea Mari, Thomas~R. Bromley, and Nathan Killoran.
\newblock Estimating the gradient and higher-order derivatives on quantum
  hardware.
\newblock {\em Physical Review A}, 103(1):012405, January 2021.
\newblock \href{https://link.aps.org/doi/10.1103/PhysRevA.103.012405}{PRA,
  103(1):012405}.

\bibitem[MBS{\etalchar{+}}18]{mcclean_barren_2018}
Jarrod~R. McClean, Sergio Boixo, Vadim~N. Smelyanskiy, Ryan Babbush, and
  Hartmut Neven.
\newblock Barren plateaus in quantum neural network training landscapes.
\newblock {\em Nature Communications}, 9(1):1--6, November 2018.
\newblock \href{https://www.nature.com/articles/s41467-018-07090-4}{Nat. Comm.,
  9(1):1-6}.

\bibitem[MEAG{\etalchar{+}}20]{mcardle_quantum_2020}
Sam McArdle, Suguru Endo, Al{\'a}n Aspuru-Guzik, Simon~C. Benjamin, and Xiao
  Yuan.
\newblock Quantum computational chemistry.
\newblock {\em Reviews of Modern Physics}, 92(1):015003, March 2020.
\newblock \href{https://link.aps.org/doi/10.1103/RevModPhys.92.015003}{Rev.Mod.
  Phys., 92(1):015003}.

\bibitem[MFSS17]{muandet_kernel_2017}
Krikamol Muandet, Kenji Fukumizu, Bharath Sriperumbudur, and Bernhard
  Sch{\"o}lkopf.
\newblock Kernel {Mean} {Embedding} of {Distributions}: {A} {Review} and
  {Beyond}.
\newblock {\em Found. Trends{\textregistered}in Mach. Learn.}, 10(1-2):1--141,
  2017.
\newblock \href{https://www.nowpublishers.com/article/Details/MAL-060}{FTML,
  10(1-2):1–141}.

\bibitem[MJE{\etalchar{+}}19]{mcardle_variational_2019}
Sam McArdle, Tyson Jones, Suguru Endo, Ying Li, Simon~C. Benjamin, and Xiao
  Yuan.
\newblock Variational ansatz-based quantum simulation of imaginary time
  evolution.
\newblock {\em npj Quantum Information}, 5(1):1--6, September 2019.
\newblock \href{https://www.nature.com/articles/s41534-019-0187-2}{npj QI,
  5(1): 1-6}.

\bibitem[MKW21]{marrero_entanglement_2021}
Carlos~Ortiz Marrero, M{\'a}ria Kieferov{\'a}, and Nathan Wiebe.
\newblock Entanglement {Induced} {Barren} {Plateaus}.
\newblock {\em arXiv:2010.15968 [quant-ph]}, March 2021.
\newblock \href{http://arxiv.org/abs/2010.15968}{ArXiv: 2010.15968}.

\bibitem[ML17]{mohamed_learning_2017}
Shakir Mohamed and Balaji Lakshminarayanan.
\newblock Learning in {Implicit} {Generative} {Models}.
\newblock {\em arXiv:1610.03483 [cs, stat]}, February 2017.
\newblock \href{http://arxiv.org/abs/1610.03483}{ArXiv: 1610.03483}.

\bibitem[MNK{\etalchar{+}}18]{melnikov_active_2018}
Alexey~A. Melnikov, Hendrik~Poulsen Nautrup, Mario Krenn, Vedran Dunjko, Markus
  Tiersch, Anton Zeilinger, and Hans~J. Briegel.
\newblock Active learning machine learns to create new quantum experiments.
\newblock {\em Proceedings of the National Academy of Sciences},
  115(6):1221--1226, February 2018.
\newblock \href{https://www.pnas.org/content/115/6/1221}{PNAS,
  115(6):1221–1226}.

\bibitem[MNKF18]{mitarai_quantum_2018}
K.~Mitarai, M.~Negoro, M.~Kitagawa, and K.~Fujii.
\newblock Quantum circuit learning.
\newblock {\em Physical Review A}, 98(3):032309, September 2018.
\newblock \href{https://link.aps.org/doi/10.1103/PhysRevA.98.032309}{PRA,
  98(3):032309}.

\bibitem[MNM{\etalchar{+}}08]{mendonca_alternative_2008}
Paulo E. M.~F. Mendon{\c c}a, Reginaldo d.~J. Napolitano, Marcelo~A.
  Marchiolli, Christopher~J. Foster, and Yeong-Cherng Liang.
\newblock Alternative fidelity measure between quantum states.
\newblock {\em Physical Review A}, 78(5):052330, November 2008.
\newblock \href{https://link.aps.org/doi/10.1103/PhysRevA.78.052330}{PRA,
  78(5):052330}.

\bibitem[Moc89]{mockus_global_1989}
Jonas Mockus.
\newblock Global {Optimization} and the {Bayesian} {Approach}.
\newblock In Jonas Mockus, editor, {\em Bayesian {Approach} to {Global}
  {Optimization}: {Theory} and {Applications}}, Mathematics and {Its}
  {Applications}, pages 1--3. Springer Netherlands, Dordrecht, 1989.
\newblock \href{https://doi.org/10.1007/978-94-009-0909-0_1}{Link}.

\bibitem[Mon16]{montanaro_quantum_2016}
Ashley Montanaro.
\newblock Quantum algorithms: an overview.
\newblock {\em npj Quantum Information}, 2:15023, January 2016.
\newblock \href{https://www.nature.com/articles/npjqi201523}{npj QI, 2:15023}.

\bibitem[MP43]{mcculloch_logical_1943}
Warren~S. McCulloch and Walter Pitts.
\newblock A logical calculus of the ideas immanent in nervous activity.
\newblock {\em The bulletin of mathematical biophysics}, 5(4):115--133,
  December 1943.
\newblock \href{https://doi.org/10.1007/BF02478259}{BMB, 5(4):115–133}.

\bibitem[MRBAG16]{mcclean_theory_2016}
Jarrod~R. McClean, Jonathan Romero, Ryan Babbush, and Al{\'a}n Aspuru-Guzik.
\newblock The theory of variational hybrid quantum-classical algorithms.
\newblock {\em New Journal of Physics}, 18(2):023023, February 2016.
\newblock \href{https://doi.org/10.1088\%2F1367-2630\%2F18\%2F2\%2F023023}{NJP,
  18(2):023023}.

\bibitem[MSCK99]{mayers_unconditionally_1999}
Dominic Mayers, Louis Salvail, and Yoshie Chiba-Kohno.
\newblock Unconditionally secure quantum coin tossing, 1999.
\newblock \href{https://core.ac.uk/display/2669299}{ArXiv: 9904078}.

\bibitem[MSSD21]{mills_application-motivated_2021}
Daniel Mills, Seyon Sivarajah, Travis~L. Scholten, and Ross Duncan.
\newblock Application-{Motivated}, {Holistic} {Benchmarking} of a {Full}
  {Quantum} {Computing} {Stack}.
\newblock {\em Quantum}, 5:415, March 2021.
\newblock \href{https://quantum-journal.org/papers/q-2021-03-22-415/}{Quantum,
  5:415}.

\bibitem[MW02]{meyer_global_2002}
David~A. Meyer and Nolan~R. Wallach.
\newblock Global entanglement in multiparticle systems.
\newblock {\em Journal of Mathematical Physics}, 43(9):4273--4278, September
  2002.
\newblock \href{http://aip.scitation.org/doi/10.1063/1.1497700}{JMP,
  43(9):4273–4278}.

\bibitem[NC10]{nielsen_quantum_2010}
Michael~A. Nielsen and Isaac~L. Chuang.
\newblock {\em Quantum computation and quantum information}.
\newblock Cambridge University Press, Cambridge ; New York, 10th anniversary ed
  edition, 2010.
\newblock \href{https://dl.acm.org/doi/10.5555/1972505}{Link}.

\bibitem[NG99]{niu_two-qubit_1999}
Chi-Sheng Niu and Robert~B. Griffiths.
\newblock Two-qubit copying machine for economical quantum eavesdropping.
\newblock {\em Physical Review A}, 60(4):2764--2776, October 1999.
\newblock \href{https://link.aps.org/doi/10.1103/PhysRevA.60.2764}{PRA,
  60(4):2764–2776}.

\bibitem[NM65]{nelder_simplex_1965}
J.~A. Nelder and R.~Mead.
\newblock A {Simplex} {Method} for {Function} {Minimization}.
\newblock {\em The Computer Journal}, 7(4):308--313, January 1965.
\newblock \href{https://academic.oup.com/comjnl/article/7/4/308/354237}{ArXiv:
  1811.04909}.

\bibitem[NRS{\etalchar{+}}18]{nam_automated_2018}
Yunseong Nam, Neil~J. Ross, Yuan Su, Andrew~M. Childs, and Dmitri Maslov.
\newblock Automated optimization of large quantum circuits with continuous
  parameters.
\newblock {\em npj Quantum Information}, 4(1):1--12, May 2018.
\newblock \href{https://www.nature.com/articles/s41534-018-0072-4}{npj QI,
  4(1):1–12}.

\bibitem[Nys30]{nystrom_uber_1930}
E.~J. Nystr{\"o}m.
\newblock {\"U}ber {Die} {Praktische} {Aufl{\"o}sung} von {Integralgleichungen}
  mit {Anwendungen} auf {Randwertaufgaben}.
\newblock {\em Acta Math.}, 54:185--204, 1930.
\newblock \href{https://doi.org/10.1007/BF02547521}{Am, 54:185-204}.

\bibitem[OBK{\etalchar{+}}16]{omalley_scalable_2016}
P.~J.~J. O'Malley, R.~Babbush, I.~D. Kivlichan, J.~Romero, J.~R. McClean,
  R.~Barends, J.~Kelly, P.~Roushan, A.~Tranter, N.~Ding, B.~Campbell, Y.~Chen,
  Z.~Chen, B.~Chiaro, A.~Dunsworth, A.~G. Fowler, E.~Jeffrey, E.~Lucero,
  A.~Megrant, J.~Y. Mutus, M.~Neeley, C.~Neill, C.~Quintana, D.~Sank,
  A.~Vainsencher, J.~Wenner, T.~C. White, P.~V. Coveney, P.~J. Love, H.~Neven,
  A.~Aspuru-Guzik, and J.~M. Martinis.
\newblock Scalable {Quantum} {Simulation} of {Molecular} {Energies}.
\newblock {\em Physical Review X}, 6(3):031007, July 2016.
\newblock \href{https://link.aps.org/doi/10.1103/PhysRevX.6.031007}{PRX,
  6(3):031007}.

\bibitem[OGB21]{ostaszewski_structure_2021}
Mateusz Ostaszewski, Edward Grant, and Marcello Benedetti.
\newblock Structure optimization for parameterized quantum circuits.
\newblock {\em Quantum}, 5:391, January 2021.
\newblock \href{https://quantum-journal.org/papers/q-2021-01-28-391/}{Quantum,
  5:391}.

\bibitem[Par70]{park_concept_1970}
James~L. Park.
\newblock The concept of transition in quantum mechanics.
\newblock {\em Foundations of Physics}, 1(1):23--33, March 1970.
\newblock \href{https://doi.org/10.1007/BF00708652}{FoP, 1(1):23–33}.

\bibitem[PBO20]{poland_no_2020}
Kyle Poland, Kerstin Beer, and Tobias~J. Osborne.
\newblock No {Free} {Lunch} for {Quantum} {Machine} {Learning}.
\newblock {\em arXiv:2003.14103 [quant-ph]}, March 2020.
\newblock \href{http://arxiv.org/abs/2003.14103}{ArXiv: 2003.14103}.

\bibitem[PC19]{peyre_computational_2019}
Gabriel Peyr{\'e} and Marco Cuturi.
\newblock Computational {Optimal} {Transport}: {With} {Applications} to {Data}
  {Science}.
\newblock {\em Foundations and Trends{\textregistered} in Machine Learning},
  11(5-6):355--607, February 2019.
\newblock \href{https://www.nowpublishers.com/article/Details/MAL-073}{FTML,
  1(5-6):355–607}.

\bibitem[PCW{\etalchar{+}}20]{pesah_absence_2020}
Arthur Pesah, M.~Cerezo, Samson Wang, Tyler Volkoff, Andrew~T. Sornborger, and
  Patrick~J. Coles.
\newblock Absence of {Barren} {Plateaus} in {Quantum} {Convolutional} {Neural}
  {Networks}.
\newblock {\em arXiv:2011.02966 [quant-ph, stat]}, November 2020.
\newblock \href{http://arxiv.org/abs/2011.02966}{ArXiv: 2011.02966}.

\bibitem[PDF{\etalchar{+}}21]{pino_demonstration_2021}
J.~M. Pino, J.~M. Dreiling, C.~Figgatt, J.~P. Gaebler, S.~A. Moses, M.~S.
  Allman, C.~H. Baldwin, M.~Foss-Feig, D.~Hayes, K.~Mayer, C.~Ryan-Anderson,
  and B.~Neyenhuis.
\newblock Demonstration of the trapped-ion quantum {CCD} computer architecture.
\newblock {\em Nature}, 592(7853):209--213, April 2021.
\newblock \href{https://www.nature.com/articles/s41586-021-03318-4}{Nature,
  592(7853):209–213}.

\bibitem[PGM{\etalchar{+}}19]{paszke_pytorch_2019}
Adam Paszke, Sam Gross, Francisco Massa, Adam Lerer, James Bradbury, Gregory
  Chanan, Trevor Killeen, Zeming Lin, Natalia Gimelshein, Luca Antiga, Alban
  Desmaison, Andreas Kopf, Edward Yang, Zachary DeVito, Martin Raison, Alykhan
  Tejani, Sasank Chilamkurthy, Benoit Steiner, Lu~Fang, Junjie Bai, and Soumith
  Chintala.
\newblock {PyTorch}: {An} {Imperative} {Style}, {High}-{Performance} {Deep}
  {Learning} {Library}.
\newblock In {\em Advances in {Neural} {Information} {Processing} {Systems}
  32}, pages 8024--8035. Curran Associates, Inc., 2019.
\newblock
  \href{https://papers.nips.cc/paper/2019/hash/bdbca288fee7f92f2bfa9f7012727740-Abstract.html}{NeurIPS,
  32:8024–8035}.

\bibitem[PM09]{puchala_bound_2009}
Zbigniew Pucha{\l }a and Jaros{\l }aw~Adam Miszczak.
\newblock Bound on trace distance based on superfidelity.
\newblock {\em Physical Review A}, 79(2):024302, February 2009.
\newblock \href{https://link.aps.org/doi/10.1103/PhysRevA.79.024302}{PRA,
  79(2):024302}.

\bibitem[PMS{\etalchar{+}}14]{peruzzo_variational_2014}
Alberto Peruzzo, Jarrod McClean, Peter Shadbolt, Man-Hong Yung, Xiao-Qi Zhou,
  Peter~J. Love, Al{\'a}n Aspuru-Guzik, and Jeremy~L. O{\textquoteright}Brien.
\newblock A variational eigenvalue solver on a photonic quantum processor.
\newblock {\em Nature Communications}, 5(1):1--7, July 2014.
\newblock \href{https://www.nature.com/articles/ncomms5213}{Nat. Comm.,
  5(1):1–7}.

\bibitem[PNGY20]{patti_entanglement_2020}
Taylor~L. Patti, Khadijeh Najafi, Xun Gao, and Susanne~F. Yelin.
\newblock Entanglement {Devised} {Barren} {Plateau} {Mitigation}.
\newblock {\em arXiv:2012.12658 [quant-ph]}, December 2020.
\newblock \href{http://arxiv.org/abs/2012.12658}{ArXiv: 2012.12658}.

\bibitem[PPA{\etalchar{+}}20]{pirandola_advances_2020}
S.~Pirandola, S.~Pirandola, U.~L. Andersen, L.~Banchi, M.~Berta, D.~Bunandar,
  R.~Colbeck, D.~Englund, T.~Gehring, C.~Lupo, C.~Ottaviani, J.~L. Pereira,
  M.~Razavi, J.~Shamsul Shaari, J.~Shamsul Shaari, M.~Tomamichel,
  M.~Tomamichel, V.~C. Usenko, G.~Vallone, P.~Villoresi, and P.~Wallden.
\newblock Advances in quantum cryptography.
\newblock {\em Advances in Optics and Photonics}, 12(4):1012--1236, December
  2020.
\newblock
  \href{https://www.osapublishing.org/aop/abstract.cfm?uri=aop-12-4-1012}{AOP,
  12(4):1012–1236}.

\bibitem[Pre18]{preskill_quantum_2018}
John Preskill.
\newblock Quantum {Computing} in the {NISQ} era and beyond.
\newblock {\em Quantum}, 2:79, August 2018.
\newblock \href{https://quantum-journal.org/papers/q-2018-08-06-79/}{Quantum,
  2:79}.

\bibitem[PSCLGFL20]{perez-salinas_data_2020}
Adri{\'a}n P{\'e}rez-Salinas, Alba Cervera-Lierta, Elies Gil-Fuster, and
  Jos{\'e}~I. Latorre.
\newblock Data re-uploading for a universal quantum classifier.
\newblock {\em Quantum}, 4:226, February 2020.
\newblock \href{https://quantum-journal.org/papers/q-2020-02-06-226/}{Quantum,
  4:226}.

\bibitem[PT21]{pirhooshyaran_quantum_2021}
Mohammad Pirhooshyaran and Tamas Terlaky.
\newblock Quantum {Circuit} {Design} {Search}.
\newblock {\em arXiv:2012.04046 [quant-ph]}, January 2021.
\newblock \href{http://arxiv.org/abs/2012.04046}{ArXiv: 2012.04046}.

\bibitem[PVG{\etalchar{+}}11]{pedregosa_scikit-learn_2011}
F.~Pedregosa, G.~Varoquaux, A.~Gramfort, V.~Michel, B.~Thirion, O.~Grisel,
  M.~Blondel, P.~Prettenhofer, R.~Weiss, V.~Dubourg, J.~Vanderplas, A.~Passos,
  D.~Cournapeau, M.~Brucher, M.~Perrot, and E.~Duchesnay.
\newblock Scikit-learn: {Machine} {Learning} in {Python}.
\newblock {\em Journal of Machine Learning Research}, 12:2825--2830, 2011.
\newblock \href{https://www.jmlr.org/papers/v12/pedregosa11a.html}{JMLR,
  12:2825–2830}.

\bibitem[PWH19]{padilha_qxsqa_2019}
D.~Padilha, S.~Weinstock, and M.~Hodson.
\newblock {QxSQA}: {GPGPU}-{Accelerated} {Simulated} {Quantum} {Annealer}
  within a {Non}-{Linear} {Optimization} and {Boltzmann} {Sampling}
  {Framework}.
\newblock In {\em 2019 {IEEE} {High} {Performance} {Extreme} {Computing}
  {Conference} ({HPEC})}, pages 1--8, 2019.
\newblock \href{https://ieeexplore.ieee.org/document/8916450}{HPEC '19, 1-8 }.

\bibitem[RAG21]{romero_variational_2021}
Jonathan Romero and Al{\'a}n Aspuru-Guzik.
\newblock Variational {Quantum} {Generators}: {Generative} {Adversarial}
  {Quantum} {Machine} {Learning} for {Continuous} {Distributions}.
\newblock {\em Advanced Quantum Technologies}, 4(1):2000003, 2021.
\newblock
  \href{https://onlinelibrary.wiley.com/doi/abs/10.1002/qute.202000003}{AQT,
  4(1):2000003}.

\bibitem[RB01]{raussendorf_one-way_2001}
Robert Raussendorf and Hans~J. Briegel.
\newblock A {One}-{Way} {Quantum} {Computer}.
\newblock {\em Physical Review Letters}, 86(22):5188--5191, May 2001.
\newblock \href{https://link.aps.org/doi/10.1103/PhysRevLett.86.5188}{PRL,
  86(22):5188–5191}.

\bibitem[RGS{\etalchar{+}}18]{rocchetto_learning_2018}
Andrea Rocchetto, Edward Grant, Sergii Strelchuk, Giuseppe Carleo, and Simone
  Severini.
\newblock Learning hard quantum distributions with variational autoencoders.
\newblock {\em npj Quantum Inf.}, 4(1):28, June 2018.
\newblock \href{https://doi.org/10.1038/s41534-018-0077-z}{npj QI, 4(1):28}.

\bibitem[RHP{\etalchar{+}}20]{rattew_domain-agnostic_2020}
Arthur~G. Rattew, Shaohan Hu, Marco Pistoia, Richard Chen, and Steve Wood.
\newblock A {Domain}-agnostic, {Noise}-resistant, {Hardware}-efficient
  {Evolutionary} {Variational} {Quantum} {Eigensolver}.
\newblock {\em arXiv:1910.09694 [quant-ph]}, January 2020.
\newblock \href{http://arxiv.org/abs/1910.09694}{ArXiv: 1910.09694}.

\bibitem[RM87]{rumelhart_learning_1987}
David~E. Rumelhart and James~L. McClelland.
\newblock Learning {Internal} {Representations} by {Error} {Propagation}.
\newblock In {\em Parallel {Distributed} {Processing}: {Explorations} in the
  {Microstructure} of {Cognition}: {Foundations}}, pages 318--362. MITP, 1987.
\newblock \href{https://ieeexplore.ieee.org/document/6302929}{PDP, EMC:Found.:,
  318–362}.

\bibitem[RML14]{rebentrost_quantum_2014}
Patrick Rebentrost, Masoud Mohseni, and Seth Lloyd.
\newblock Quantum support vector machine for big data classification.
\newblock {\em Phys. Rev. Letters}, 113(13), September 2014.
\newblock \href{https://link.aps.org/doi/10.1103/PhysRevLett.113.130503}{PRL,
  113(13)}.

\bibitem[ROAG17]{romero_quantum_2017}
Jonathan Romero, Jonathan~P. Olson, and Alan Aspuru-Guzik.
\newblock Quantum autoencoders for efficient compression of quantum data.
\newblock {\em Quantum Science and Technology}, 2(4):045001, December 2017.
\newblock
  \href{https://iopscience.iop.org/article/10.1088/2058-9565/aa8072}{QST,
  2(4):045001}.

\bibitem[RPK{\etalchar{+}}17]{raghu_expressive_2017}
Maithra Raghu, Ben Poole, Jon Kleinberg, Surya Ganguli, and Jascha
  Sohl-Dickstein.
\newblock On the {Expressive} {Power} of {Deep} {Neural} {Networks}.
\newblock In {\em Proceedings of the 34th {International} {Conference} on
  {Machine} {Learning}}, volume~70 of {\em Proceedings of {Machine} {Learning}
  {Research}}, pages 2847--2854, August 2017.
\newblock \href{http://proceedings.mlr.press/v70/raghu17a.html}{ICLR '17}.

\bibitem[RS13]{rios_derivative-free_2013}
Luis~Miguel Rios and Nikolaos~V. Sahinidis.
\newblock Derivative-free optimization: a review of algorithms and comparison
  of software implementations.
\newblock {\em J Glob Optim}, 56(3):1247--1293, July 2013.
\newblock
  \href{https://link.springer.com/article/10.1007/s10898-012-9951-y}{JGO,
  56(3):1247–1293}.

\bibitem[RSA78]{rivest_method_1978}
R.~L. Rivest, A.~Shamir, and L.~Adleman.
\newblock A method for obtaining digital signatures and public-key
  cryptosystems.
\newblock {\em Communications of the ACM}, 21(2):120--126, February 1978.
\newblock \href{https://doi.org/10.1145/359340.359342}{Comm. ACM,
  21(2):120–126}.

\bibitem[RTC17]{ramdas_wasserstein_2017}
Aaditya Ramdas, Nicol{\'a}s~Garc{\'i}a Trillos, and Marco Cuturi.
\newblock On {Wasserstein} {Two}-{Sample} {Testing} and {Related} {Families} of
  {Nonparametric} {Tests}.
\newblock {\em Entropy}, 19(2):47, February 2017.
\newblock \href{http://creativecommons.org/licenses/by/3.0/}{Entropy,
  19(2):47}.

\bibitem[RTK{\etalchar{+}}20]{rudolph_generation_2020}
Manuel~S. Rudolph, Ntwali~Bashige Toussaint, Amara Katabarwa, Sonika Johri,
  Borja Peropadre, and Alejandro Perdomo-Ortiz.
\newblock Generation of {High}-{Resolution} {Handwritten} {Digits} with an
  {Ion}-{Trap} {Quantum} {Computer}.
\newblock {\em arXiv:2012.03924 [quant-ph]}, December 2020.
\newblock \href{http://arxiv.org/abs/2012.03924}{ArXiv: 2012.03924}.

\bibitem[RWJ{\etalchar{+}}14]{ronnow_defining_2014}
Troels~F. R{\o}nnow, Zhihui Wang, Joshua Job, Sergio Boixo, Sergei~V. Isakov,
  David Wecker, John~M. Martinis, Daniel~A. Lidar, and Matthias Troyer.
\newblock Defining and detecting quantum speedup.
\newblock {\em Science}, 345(6195):420--424, July 2014.
\newblock \href{https://science.sciencemag.org/content/345/6195/420}{Science,
  345(6195):420–424}.

\bibitem[RXY{\etalchar{+}}17]{ren_ground--satellite_2017}
Ji-Gang Ren, Ping Xu, Hai-Lin Yong, Liang Zhang, Sheng-Kai Liao, Juan Yin,
  Wei-Yue Liu, Wen-Qi Cai, Meng Yang, Li~Li, Kui-Xing Yang, Xuan Han,
  Yong-Qiang Yao, Ji~Li, Hai-Yan Wu, Song Wan, Lei Liu, Ding-Quan Liu, Yao-Wu
  Kuang, Zhi-Ping He, Peng Shang, Cheng Guo, Ru-Hua Zheng, Kai Tian, Zhen-Cai
  Zhu, Nai-Le Liu, Chao-Yang Lu, Rong Shu, Yu-Ao Chen, Cheng-Zhi Peng, Jian-Yu
  Wang, and Jian-Wei Pan.
\newblock Ground-to-satellite quantum teleportation.
\newblock {\em Nature}, 549(7670):70--73, September 2017.
\newblock \href{https://www.nature.com/articles/nature23675}{Nature,
  549(7670):70–73}.

\bibitem[SB09]{shepherd_temporally_2009}
Dan Shepherd and Michael~J Bremner.
\newblock Temporally unstructured quantum computation.
\newblock {\em Proceedings of the Royal Society A}, 465:1413–1439, January
  2009.
\newblock
  \href{http://rspa.royalsocietypublishing.org/content/early/2009/02/18/rspa.2008.0443.abstract}{Proc.
  R. Soc. A. 465:1413–1439}.

\bibitem[SBG{\etalchar{+}}19]{schuld_evaluating_2019}
Maria Schuld, Ville Bergholm, Christian Gogolin, Josh Izaac, and Nathan
  Killoran.
\newblock Evaluating analytic gradients on quantum hardware.
\newblock {\em Physical Review A}, 99(3):032331, March 2019.
\newblock \href{https://link.aps.org/doi/10.1103/PhysRevA.99.032331}{PRA,
  99(3):032331}.

\bibitem[SBSW20]{schuld_circuit-centric_2020}
Maria Schuld, Alex Bocharov, Krysta~M. Svore, and Nathan Wiebe.
\newblock Circuit-centric quantum classifiers.
\newblock {\em Physical Review A}, 101(3):032308, March 2020.
\newblock \href{https://link.aps.org/doi/10.1103/PhysRevA.101.032308}{PRA,
  99(3):032331}.

\bibitem[SCD{\etalchar{+}}20]{selvaraju_grad-cam_2020}
Ramprasaath~R. Selvaraju, Michael Cogswell, Abhishek Das, Ramakrishna Vedantam,
  Devi Parikh, and Dhruv Batra.
\newblock Grad-{CAM}: {Visual} {Explanations} from {Deep} {Networks} via
  {Gradient}-{Based} {Localization}.
\newblock {\em International Journal of Computer Vision}, 128(2):336--359,
  February 2020.
\newblock \href{https://doi.org/10.1007/s11263-019-01228-7}{IJCV,
  128(2):336-359}.

\bibitem[Sch05]{schlosshauer_decoherence_2005}
Maximilian Schlosshauer.
\newblock Decoherence, the measurement problem, and interpretations of quantum
  mechanics.
\newblock {\em Rev. Mod. Phys.}, 76(4):1267--1305, February 2005.
\newblock \href{https://link.aps.org/doi/10.1103/RevModPhys.76.1267}{Rev. Mod.
  Phys., 76(4):1267–1305}.

\bibitem[Sch15]{schmidhuber_deep_2015}
Juergen Schmidhuber.
\newblock Deep {Learning} in {Neural} {Networks}: {An} {Overview}.
\newblock {\em Neural Networks}, 61:85--117, January 2015.
\newblock
  \href{https://www.sciencedirect.com/science/article/pii/S0893608014002135}{Neural
  Networks, 61:85–117.}

\bibitem[Sch16]{schmied_quantum_2016}
Roman Schmied.
\newblock Quantum state tomography of a single qubit: comparison of methods.
\newblock {\em Journal of Modern Optics}, 63(18):1744--1758, October 2016.
\newblock \href{https://doi.org/10.1080/09500340.2016.1142018}{JMO,
  63(18):1744–1758}.

\bibitem[SCH{\etalchar{+}}20]{sharma_reformulation_2020}
Kunal Sharma, M.~Cerezo, Zo{\"e} Holmes, Lukasz Cincio, Andrew Sornborger, and
  Patrick~J. Coles.
\newblock Reformulation of the {No}-{Free}-{Lunch} {Theorem} for {Entangled}
  {Data} {Sets}.
\newblock {\em arXiv:2007.04900 [quant-ph]}, July 2020.
\newblock \href{http://arxiv.org/abs/2007.04900}{ArXiv: 2007.04900}.

\bibitem[Sch21]{schuld_supervised_2021}
Maria Schuld.
\newblock Supervised quantum machine learning models are kernel methods.
\newblock {\em arXiv:2101.11020 [quant-ph, stat]}, April 2021.
\newblock \href{http://arxiv.org/abs/2101.11020}{ArXiv: 2101.11020}.

\bibitem[SCZ17]{smith_practical_2017}
Robert~S. Smith, Michael~J. Curtis, and William~J. Zeng.
\newblock A {Practical} {Quantum} {Instruction} {Set} {Architecture}.
\newblock {\em arXiv:1608.03355 [quant-ph]}, February 2017.
\newblock \href{http://arxiv.org/abs/2004.01372}{ArXiv: 2004.01372}.

\bibitem[SFG{\etalchar{+}}09]{sriperumbudur_integral_2009}
Bharath~K. Sriperumbudur, Kenji Fukumizu, Arthur Gretton, Bernhard
  Sch{\"o}lkopf, and Gert R.~G. Lanckriet.
\newblock On integral probability metrics, phi-divergences and binary
  classification.
\newblock {\em arXiv:0901.2698 [cs, math]}, October 2009.
\newblock \href{http://arxiv.org/abs/0901.2698}{ArXiv: 0901.2698}.

\bibitem[SFL11]{sriperumbudur_universality_2011}
Bharath~K Sriperumbudur, Kenji Fukumizu, and Gert R~G Lanckriet.
\newblock Universality, {Characteristic} {Kernels} and {RKHS} {Embedding} of
  {Measures}.
\newblock {\em Journal of Machine Learning Research}, 12(Jul):2389--2410, 2011.
\newblock \href{http://www.jmlr.org/papers/v12/sriperumbudur11a.html}{JMLR,
  12(Jul):2389–2410}.

\bibitem[SG01]{scarani_quantum_2001}
Valerio Scarani and Nicolas Gisin.
\newblock Quantum key distribution between {N} partners: {Optimal}
  eavesdropping and {Bell}'s inequalities.
\newblock {\em Physical Review A}, 65(1):012311, December 2001.
\newblock \href{https://link.aps.org/doi/10.1103/PhysRevA.65.012311}{PRA,
  65(1):012311}.

\bibitem[SG04]{servedio_equivalences_2004}
Rocco~A. Servedio and Steven~J. Gortler.
\newblock Equivalences and {Separations} {Between} {Quantum} and {Classical}
  {Learnability}.
\newblock {\em SIAM J. Comput.}, 33(5):1067--1092, May 2004.
\newblock \href{https://doi.org/10.1137/S0097539704412910}{QIP,
  4(5):355–386}.

\bibitem[SGF{\etalchar{+}}08]{sriperumbudur_injective_2008}
BK. Sriperumbudur, A.~Gretton, K.~Fukumizu, G.~Lanckriet, and B.~Sch{\"o}lkopf.
\newblock Injective {Hilbert} {Space} {Embeddings} of {Probability} {Measures}.
\newblock In {\em Proceedings of the 21st {Annual} {Conference} on {Learning}
  {Theory}}, pages 111--122, Madison, WI, USA, July 2008. Biologische
  Kybernetik.
\newblock \href{https://www.is.mpg.de/publications/5122}{COLT '08, 111-122}.

\bibitem[SHLZ18]{situ_quantum_2018}
Haozhen Situ, Zhimin He, Lvzhou Li, and Shenggen Zheng.
\newblock Quantum generative adversarial network for generating discrete data.
\newblock {\em arXiv:1807.01235 [quant-ph]}, July 2018.
\newblock \href{http://arxiv.org/abs/1807.01235}{ArXiv: 1807.01235}.

\bibitem[Sho94]{shor_algorithms_1994}
P.W. Shor.
\newblock Algorithms for quantum computation: discrete logarithms and
  factoring.
\newblock In {\em Proceedings 35th {Annual} {Symposium} on {Foundations} of
  {Computer} {Science}}, pages 124--134, November 1994.
\newblock \href{https://ieeexplore.ieee.org/document/365700}{FOCS '94,
  124-134}.

\bibitem[SIGA05]{scarani_quantum_2005}
Valerio Scarani, Sofyan Iblisdir, Nicolas Gisin, and Antonio Ac{\'i}n.
\newblock Quantum cloning.
\newblock {\em Rev. Mod. Phys.}, 77(4):1225--1256, November 2005.
\newblock \href{https://link.aps.org/doi/10.1103/RevModPhys.77.1225}{Rev. Mod.
  Phys., 77(4):1225–1256}.

\bibitem[SIKC20]{stokes_quantum_2020}
James Stokes, Josh Izaac, Nathan Killoran, and Giuseppe Carleo.
\newblock Quantum {Natural} {Gradient}.
\newblock {\em Quantum}, 4:269, May 2020.
\newblock \href{https://quantum-journal.org/papers/q-2020-05-25-269/}{Quantum,
  4:269}.

\bibitem[Sin64]{sinkhorn_relationship_1964}
Richard Sinkhorn.
\newblock A {Relationship} {Between} {Arbitrary} {Positive} {Matrices} and
  {Doubly} {Stochastic} {Matrices}.
\newblock {\em Annals of Mathematical Statistics}, 35(2):876--879, June 1964.
\newblock \href{https://projecteuclid.org/euclid.aoms/1177703591}{AMS,
  35(2):876–879}.

\bibitem[SJAG19]{sim_expressibility_2019}
Sukin Sim, Peter~D. Johnson, and Al{\'a}n Aspuru-Guzik.
\newblock Expressibility and {Entangling} {Capability} of {Parameterized}
  {Quantum} {Circuits} for {Hybrid} {Quantum}-{Classical} {Algorithms}.
\newblock {\em Advanced Quantum Technologies}, 2(12):1900070, December 2019.
\newblock
  \href{https://onlinelibrary.wiley.com/doi/abs/10.1002/qute.201900070}{AQT,
  2(12):1900070}.

\bibitem[SJD21]{skolik_quantum_2021}
Andrea Skolik, Sofiene Jerbi, and Vedran Dunjko.
\newblock Quantum agents in the {Gym}: a variational quantum algorithm for deep
  {Q}-learning.
\newblock {\em arXiv:2103.15084 [quant-ph]}, March 2021.
\newblock \href{http://arxiv.org/abs/2103.15084}{ArXiv: 2103.15084}.

\bibitem[SK19]{schuld_quantum_2019}
Maria Schuld and Nathan Killoran.
\newblock Quantum {Machine} {Learning} in {Feature} {Hilbert} {Spaces}.
\newblock {\em Physical Review Letters}, 122(4):040504, February 2019.
\newblock \href{https://link.aps.org/doi/10.1103/PhysRevLett.122.040504}{PRL,
  122(4):040504}.

\bibitem[SKCC20]{sharma_noise_2020}
Kunal Sharma, Sumeet Khatri, Marco Cerezo, and Patrick Coles.
\newblock Noise {Resilience} of {Variational} {Quantum} {Compiling}.
\newblock {\em New J. Phys.}, 22(4):043006, 2020.
\newblock
  \href{https://iopscience.iop.org/article/10.1088/1367-2630/ab784c}{NJP, 22
  043006}.

\bibitem[SM21]{shao_faster_2021}
Changpeng Shao and Ashley Montanaro.
\newblock Faster quantum-inspired algorithms for solving linear systems.
\newblock {\em arXiv:2103.10309 [quant-ph]}, March 2021.
\newblock \href{http://arxiv.org/abs/2103.10309}{ArXiv: 2103.10309}.

\bibitem[SP18]{schuld_supervised_2018}
Maria Schuld and Francesco Petruccione.
\newblock {\em Supervised {Learning} with {Quantum} {Computers}}.
\newblock Quantum {Science} and {Technology}. Springer International
  Publishing, 2018.
\newblock \href{https://www.springer.com/us/book/9783319964232}{Link}.

\bibitem[SPR{\etalchar{+}}20]{sarovar_detecting_2020}
Mohan Sarovar, Timothy Proctor, Kenneth Rudinger, Kevin Young, Erik Nielsen,
  and Robin Blume-Kohout.
\newblock Detecting crosstalk errors in quantum information processors.
\newblock {\em Quantum}, 4:321, September 2020.
\newblock \href{https://quantum-journal.org/papers/q-2020-09-11-321/}{Quantum,
  4:321}.

\bibitem[SS16]{stoudenmire_supervised_2016}
Edwin Stoudenmire and David~J Schwab.
\newblock Supervised {Learning} with {Tensor} {Networks}.
\newblock In D.~D. Lee, M.~Sugiyama, U.~V. Luxburg, I.~Guyon, and R.~Garnett,
  editors, {\em Advances in {Neural} {Information} {Processing} {Systems} 29},
  pages 4799--4807. Curran Associates, Inc., 2016.
\newblock
  \href{https://papers.nips.cc/paper/2016/hash/5314b9674c86e3f9d1ba25ef9bb32895-Abstract.html}{NIPS
  '16: 4799-4807}.

\bibitem[SSHE21]{sweke_quantum_2021}
Ryan Sweke, Jean-Pierre Seifert, Dominik Hangleiter, and Jens Eisert.
\newblock On the {Quantum} versus {Classical} {Learnability} of {Discrete}
  {Distributions}.
\newblock {\em Quantum}, 5:417, March 2021.
\newblock \href{https://quantum-journal.org/papers/q-2021-03-23-417/}{Quantum,
  5:417}.

\bibitem[SSM21]{schuld_effect_2021}
Maria Schuld, Ryan Sweke, and Johannes~Jakob Meyer.
\newblock Effect of data encoding on the expressive power of variational
  quantum-machine-learning models.
\newblock {\em Physical Review A}, 103(3):032430, March 2021.
\newblock \href{https://link.aps.org/doi/10.1103/PhysRevA.103.032430}{PRA,
  103(3):032430}.

\bibitem[SSP14]{schuld_quest_2014}
Maria Schuld, Ilya Sinayskiy, and Francesco Petruccione.
\newblock The quest for a {Quantum} {Neural} {Network}.
\newblock {\em Quantum Information Processing}, 13(11):2567--2586, November
  2014.
\newblock \href{https://doi.org/10.1007/s11128-014-0809-8}{QIP,
  13(11):2567–2586}.

\bibitem[SSP15]{schuld_introduction_2015}
Maria Schuld, Ilya Sinayskiy, and Francesco Petruccione.
\newblock An introduction to quantum machine learning.
\newblock {\em Contemporary Physics}, 56(2):172--185, April 2015.
\newblock \href{https://doi.org/10.1080/00107514.2014.964942}{CP,
  56(2):172–185}.

\bibitem[SSZ18]{shi_spectral_2018}
Jiaxin Shi, Shengyang Sun, and Jun Zhu.
\newblock A {Spectral} {Approach} to {Gradient} {Estimation} for {Implicit}
  {Distributions}.
\newblock {\em arXiv:1806.02925 [cs, stat]}, June 2018.
\newblock \href{http://arxiv.org/abs/1806.02925}{ArXiv: 1806.02925}.

\bibitem[Ste72]{stein_bound_1972}
Charles Stein.
\newblock A bound for the error in the normal approximation to the distribution
  of a sum of dependent random variables.
\newblock In {\em Proceedings of the {Sixth} {Berkeley} {Symposium} on
  {Mathematical} {Statistics} and {Probability}, {Volume} 2: {Probability}
  {Theory}}, pages 583--602, Berkeley, Calif., 1972. University of California
  Press.
\newblock \href{https://projecteuclid.org/euclid.bsmsp/1200514239}{SMSP '72,
  2:583-602}.

\bibitem[Sto76]{stockmeyer_polynomial-time_1976}
Larry~J Stockmeyer.
\newblock The polynomial-time hierarchy.
\newblock {\em Theor. Comput. Sci.}, 3(1):1--22, 1976.
\newblock
  \href{http://www.sciencedirect.com/science/article/pii/030439757690061X}{TCS,
  3(1):1–22}.

\bibitem[Stu05]{study_kurzeste_1905}
E.~Study.
\newblock K{\"u}rzeste {Wege} im komplexen {Gebiet}.
\newblock {\em Mathematische Annalen}, 60(3):321--378, September 1905.
\newblock \href{https://doi.org/10.1007/BF01457616}{MA, 60(3):321–378}.

\bibitem[SWM{\etalchar{+}}20]{sweke_stochastic_2020}
Ryan Sweke, Frederik Wilde, Johannes Meyer, Maria Schuld, Paul~K. Faehrmann,
  Barth{\'e}l{\'e}my Meynard-Piganeau, and Jens Eisert.
\newblock Stochastic gradient descent for hybrid quantum-classical
  optimization.
\newblock {\em Quantum}, 4:314, August 2020.
\newblock \href{https://quantum-journal.org/papers/q-2020-08-31-314/}{Quantum,
  4:314}.

\bibitem[Tan18a]{tang_quantum-inspired_2018}
Ewin Tang.
\newblock A quantum-inspired classical algorithm for recommendation systems.
\newblock {\em arXiv:1807.04271 [quant-ph]}, July 2018.
\newblock \href{http://arxiv.org/abs/1807.04271}{ArXiv: 1807.04271}.

\bibitem[Tan18b]{tang_quantum-inspired_2018-1}
Ewin Tang.
\newblock Quantum-inspired classical algorithms for principal component
  analysis and supervised clustering.
\newblock {\em arXiv:1811.00414 [quant-ph]}, October 2018.
\newblock \href{http://arxiv.org/abs/1811.00414}{ArXiv: 1811.00414}.

\bibitem[TGR15]{trask_modeling_2015}
Andrew Trask, David Gilmore, and Matthew Russell.
\newblock Modeling {Order} in {Neural} {Word} {Embeddings} at {Scale}.
\newblock {\em arXiv:1506.02338 [cs]}, June 2015.
\newblock \href{http://arxiv.org/abs/1506.02338}{ArXiv: 1506.02338}.

\bibitem[VBB17]{verdon_quantum_2017}
Guillaume Verdon, Michael Broughton, and Jacob Biamonte.
\newblock A quantum algorithm to train neural networks using low-depth
  circuits.
\newblock {\em arXiv:1712.05304 [cond-mat, physics:quant-ph]}, December 2017.
\newblock \href{http://arxiv.org/abs/1712.05304}{ArXiv: 1712.05304}.

\bibitem[VC15]{vapnik_uniform_2015}
V.~N. Vapnik and A.~Ya. Chervonenkis.
\newblock On the {Uniform} {Convergence} of {Relative} {Frequencies} of
  {Events} to {Their} {Probabilities}.
\newblock In Vladimir Vovk, Harris Papadopoulos, and Alexander Gammerman,
  editors, {\em Measures of {Complexity}: {Festschrift} for {Alexey}
  {Chervonenkis}}, pages 11--30. Springer International Publishing, Cham, 2015.
\newblock \href{https://doi.org/10.1007/978-3-319-21852-6_3}{TPA XVI, 2,
  264-280}.

\bibitem[VC21]{volkoff_large_2021}
Tyler Volkoff and Patrick~J. Coles.
\newblock Large gradients via correlation in random parameterized quantum
  circuits.
\newblock {\em Quantum Science and Technology}, 6(2):025008, January 2021.
\newblock \href{https://doi.org/10.1088/2058-9565/abd891}{QST, 6(2):025008}.

\bibitem[VD04]{vidal_universal_2004}
G.~Vidal and C.~M. Dawson.
\newblock Universal quantum circuit for two-qubit transformations with three
  controlled-{NOT} gates.
\newblock {\em Physical Review A}, 69(1):010301, January 2004.
\newblock \href{https://link.aps.org/doi/10.1103/PhysRevA.69.010301}{PRA,
  69(1):010301}.

\bibitem[VDRF18]{venturelli_compiling_2018}
Davide Venturelli, Minh Do, Eleanor Rieffel, and Jeremy Frank.
\newblock Compiling quantum circuits to realistic hardware architectures using
  temporal planners.
\newblock {\em Quantum Science and Technology}, 3(2):025004, February 2018.
\newblock \href{https://doi.org/10.1088/2058-9565/aaa331}{QST, 3(2):025004}.

\bibitem[Vil09]{villani_optimal_2009}
C{\'e}dric Villani.
\newblock {\em Optimal {Transport}: {Old} and {New}}.
\newblock Grundlehren der mathematischen {Wissenschaften}. Springer-Verlag,
  Berlin Heidelberg, 2009.
\newblock \href{https://www.springer.com/gp/book/9783540710493}{Link}.

\bibitem[VMN{\etalchar{+}}19]{verdon_quantum_2019}
Guillaume Verdon, Jacob Marks, Sasha Nanda, Stefan Leichenauer, and Jack
  Hidary.
\newblock Quantum {Hamiltonian}-{Based} {Models} and the {Variational}
  {Quantum} {Thermalizer} {Algorithm}.
\newblock {\em arXiv:1910.02071 [quant-ph]}, October 2019.
\newblock \href{http://arxiv.org/abs/1910.02071}{ArXiv: 1910.02071}.

\bibitem[VPB18]{verdon_universal_2018}
Guillaume Verdon, Jason Pye, and Michael Broughton.
\newblock A {Universal} {Training} {Algorithm} for {Quantum} {Deep} {Learning}.
\newblock {\em arXiv:1806.09729 [quant-ph]}, June 2018.
\newblock \href{http://arxiv.org/abs/1806.09729}{ArXiv: 1806.09729}.

\bibitem[VSC07]{vaccaro_quantum_2007}
J.~A. Vaccaro, Joseph Spring, and Anthony Chefles.
\newblock Quantum protocols for anonymous voting and surveying.
\newblock {\em Physical Review A}, 75:012333, Jan 2007.
\newblock \href{https://link.aps.org/doi/10.1103/PhysRevA.75.012333}{PRA,
  75(1):012333}.

\bibitem[Wat02]{watrous_quantum_2002}
John Watrous.
\newblock Quantum statistical zero-knowledge.
\newblock {\em arXiv:quant-ph/0202111}, February 2002.
\newblock \href{http://arxiv.org/abs/quant-ph/0202111}{ArXiv: 0202111}.

\bibitem[WBD{\etalchar{+}}19]{wright_benchmarking_2019}
K.~Wright, K.~M. Beck, S.~Debnath, J.~M. Amini, Y.~Nam, N.~Grzesiak, J.-S.
  Chen, N.~C. Pisenti, M.~Chmielewski, C.~Collins, K.~M. Hudek, J.~Mizrahi,
  J.~D. Wong-Campos, S.~Allen, J.~Apisdorf, P.~Solomon, M.~Williams, A.~M.
  Ducore, A.~Blinov, S.~M. Kreikemeier, V.~Chaplin, M.~Keesan, C.~Monroe, and
  J.~Kim.
\newblock Benchmarking an 11-qubit quantum computer.
\newblock {\em Nature Communications}, 10(1):5464, November 2019.
\newblock \href{https://www.nature.com/articles/s41467-019-13534-2}{Nat. Comm.,
  10(1):5464}.

\bibitem[WBL12]{wiebe_quantum_2012}
Nathan Wiebe, Daniel Braun, and Seth Lloyd.
\newblock Quantum {Algorithm} for {Data} {Fitting}.
\newblock {\em Physical Review Letters}, 109(5):050505, August 2012.
\newblock \href{https://link.aps.org/doi/10.1103/PhysRevLett.109.050505}{PRL,
  109(5):050505}.

\bibitem[Wer98]{werner_optimal_1998}
R.~F. Werner.
\newblock Optimal cloning of pure states.
\newblock {\em Physical Review A}, 58(3):1827--1832, September 1998.
\newblock \href{https://link.aps.org/doi/10.1103/PhysRevA.58.1827}{PRA,
  58(3):1827–1832}.

\bibitem[WFC{\etalchar{+}}21]{wang_noise-induced_2021}
Samson Wang, Enrico Fontana, M.~Cerezo, Kunal Sharma, Akira Sone, Lukasz
  Cincio, and Patrick~J. Coles.
\newblock Noise-{Induced} {Barren} {Plateaus} in {Variational} {Quantum}
  {Algorithms}.
\newblock {\em arXiv:2007.14384 [quant-ph]}, February 2021.
\newblock \href{http://arxiv.org/abs/2007.14384}{ArXiv: 2007.14384}.

\bibitem[Wie83]{wiesner_conjugate_1983}
Stephen Wiesner.
\newblock Conjugate coding.
\newblock {\em ACM SIGACT News}, 15(1):78--88, January 1983.
\newblock \href{https://doi.org/10.1145/1008908.1008920}{ACM SIGACT News,
  15(1):78–88}.

\bibitem[Wit14]{wittek_quantum_2014}
Peter Wittek.
\newblock {\em Quantum {Machine} {Learning}: {What} {Quantum} {Computing}
  {Means} to {Data} {Mining}}.
\newblock Academic Press, August 2014.
\newblock
  \href{https://www-sciencedirect-com.ezproxy.is.ed.ac.uk/book/9780128009536/quantum-machine-learning}{Link}.

\bibitem[WKGS15]{wiebe_quantum_2015}
Nathan Wiebe, Ashish Kapoor, Christopher Granade, and Krysta~M Svore.
\newblock Quantum {Inspired} {Training} for {Boltzmann} {Machines}.
\newblock {\em arXiv:1507.02642 [quant-ph]}, July 2015.
\newblock \href{http://arxiv.org/abs/1507.02642}{ArXiv: 1507.02642}.

\bibitem[WLL{\etalchar{+}}21]{weber_optimal_2021}
Maurice Weber, Nana Liu, Bo~Li, Ce~Zhang, and Zhikuan Zhao.
\newblock Optimal provable robustness of quantum classification via quantum
  hypothesis testing.
\newblock {\em npj Quantum Information}, 7(1):1--12, May 2021.
\newblock \href{https://www.nature.com/articles/s41534-021-00410-5}{npj QI,
  7(1):1–12}.

\bibitem[WM97]{wolpert_no_1997}
D.H. Wolpert and W.G. Macready.
\newblock No free lunch theorems for optimization.
\newblock {\em IEEE Transactions on Evolutionary Computation}, 1(1):67--82,
  April 1997.
\newblock \href{https://ieeexplore.ieee.org/document/585893}{IEEE TEC,
  1(1):67-82}.

\bibitem[WM20]{wright_capacity_2020}
Logan~G. Wright and Peter~L. McMahon.
\newblock The {Capacity} of {Quantum} {Neural} {Networks}.
\newblock In {\em Conference on {Lasers} and {Electro}-{Optics} (2020), paper
  {JM4G}.5}, page JM4G.5. Optical Society of America, May 2020.
\newblock
  \href{https://www.osapublishing.org/abstract.cfm?uri=CLEO_SI-2020-JM4G.5}{CLEO,
  JM4G.5}.

\bibitem[WMDB20]{wallnofer_machine_2019}
Julius Walln\"ofer, Alexey~A. Melnikov, Wolfgang D\"ur, and Hans~J. Briegel.
\newblock Machine learning for long-distance quantum communication.
\newblock {\em PRX Quantum}, 1:010301, Sep 2020.
\newblock \href{https://link.aps.org/doi/10.1103/PRXQuantum.1.010301}{PRX
  Quantum 1(1):010301}.

\bibitem[WVHR19]{wilson_quantum-assisted_2019}
Max Wilson, Thomas Vandal, Tad Hogg, and Eleanor Rieffel.
\newblock Quantum-assisted associative adversarial network: {Applying} quantum
  annealing in deep learning.
\newblock {\em arXiv e-prints}, page arXiv:1904.10573, April 2019.
\newblock \href{http://arxiv.org/abs/1904.10573}{ArXiv: 1904.10573}.

\bibitem[WW19]{wiebe_generative_2019}
Nathan Wiebe and Leonard Wossnig.
\newblock Generative training of quantum {Boltzmann} machines with hidden
  units.
\newblock {\em arXiv:1905.09902 [quant-ph]}, May 2019.
\newblock \href{http://arxiv.org/abs/1905.09902}{ArXiv: 1905.09902}.

\bibitem[WZ82]{wootters_single_1982}
W.~K. Wootters and W.~H. Zurek.
\newblock A single quantum cannot be cloned.
\newblock {\em Nature}, 299(5886):802--803, October 1982.
\newblock \href{https://www.nature.com/articles/299802a0}{Nature,
  299(5886):802–803}.

\bibitem[WZdS{\etalchar{+}}20]{wiersema_exploring_2020}
Roeland Wiersema, Cunlu Zhou, Yvette de~Sereville, Juan~Felipe Carrasquilla,
  Yong~Baek Kim, and Henry Yuen.
\newblock Exploring {Entanglement} and {Optimization} within the {Hamiltonian}
  {Variational} {Ansatz}.
\newblock {\em PRX Quantum}, 1(2):020319, December 2020.
\newblock \href{https://link.aps.org/doi/10.1103/PRXQuantum.1.020319}{PRX
  Quantum, 1(2):020319}.

\bibitem[YCL{\etalchar{+}}17]{yin_satellite-based_2017}
Juan Yin, Yuan Cao, Yu-Huai Li, Sheng-Kai Liao, Liang Zhang, Ji-Gang Ren,
  Wen-Qi Cai, Wei-Yue Liu, Bo~Li, Hui Dai, Guang-Bing Li, Qi-Ming Lu, Yun-Hong
  Gong, Yu~Xu, Shuang-Lin Li, Feng-Zhi Li, Ya-Yun Yin, Zi-Qing Jiang, Ming Li,
  Jian-Jun Jia, Ge~Ren, Dong He, Yi-Lin Zhou, Xiao-Xiang Zhang, Na~Wang, Xiang
  Chang, Zhen-Cai Zhu, Nai-Le Liu, Yu-Ao Chen, Chao-Yang Lu, Rong Shu,
  Cheng-Zhi Peng, Jian-Yu Wang, and Jian-Wei Pan.
\newblock Satellite-based entanglement distribution over 1200 kilometers.
\newblock {\em Science}, 356(6343):1140--1144, June 2017.
\newblock \href{https://science.sciencemag.org/content/356/6343/1140}{Science,
  356(6343):1140–1144}.

\bibitem[YLRN18]{yang_goodness--fit_2018}
Jiasen Yang, Qiang Liu, Vinayak Rao, and Jennifer Neville.
\newblock Goodness-of-{Fit} {Testing} for {Discrete} {Distributions} via
  {Stein} {Discrepancy}.
\newblock In Jennifer Dy and Andreas Krause, editors, {\em Proceedings of the
  35th {International} {Conference} on {Machine} {Learning}}, volume~80 of {\em
  Proceedings of {Machine} {Learning} {Research}}, pages 5561--5570,
  Stockholmsm{\"a}ssan, Stockholm Sweden, July 2018. PMLR.
\newblock \href{http://proceedings.mlr.press/v80/yang18c.html}{ICML, PMLR, 80,
  5561–5570}.

\bibitem[YWC{\etalchar{+}}19]{yao_taking_2019}
Quanming Yao, Mengshuo Wang, Yuqiang Chen, Wenyuan Dai, Yu-Feng Li, Wei-Wei Tu,
  Qiang Yang, and Yang Yu.
\newblock Taking {Human} out of {Learning} {Applications}: {A} {Survey} on
  {Automated} {Machine} {Learning}.
\newblock {\em arXiv:1810.13306 [cs, stat]}, December 2019.
\newblock \href{http://arxiv.org/abs/1810.13306}{ArXiv: 1810.13306}.

\bibitem[Zei12]{zeiler_adadelta_2012}
Matthew~D. Zeiler.
\newblock {ADADELTA}: {An} {Adaptive} {Learning} {Rate} {Method}.
\newblock {\em arXiv:1212.5701 [cs]}, December 2012.
\newblock \href{http://arxiv.org/abs/1212.5701}{ArXiv: 1212.5701}.

\bibitem[ZFF19]{zhao_quantum-assisted_2019}
Zhikuan Zhao, Jack~K Fitzsimons, and Joseph~F Fitzsimons.
\newblock Quantum-assisted {Gaussian} process regression.
\newblock {\em Physical Review A}, 99(5):52331, May 2019.
\newblock \href{https://link.aps.org/doi/10.1103/PhysRevA.99.052331}{PRA,
  99(5):52331}.

\bibitem[ZFR{\etalchar{+}}21]{zhao_smooth_2021}
Zhikuan Zhao, Jack~K. Fitzsimons, Patrick Rebentrost, Vedran Dunjko, and
  Joseph~F. Fitzsimons.
\newblock Smooth input preparation for quantum and quantum-inspired machine
  learning.
\newblock {\em Quantum Machine Intelligence}, 3(1):14, April 2021.
\newblock \href{https://doi.org/10.1007/s42484-021-00045-x}{QUMI, 3(1):14}.

\bibitem[ZG21]{zhao_analyzing_2021}
Chen Zhao and Xiao-Shan Gao.
\newblock Analyzing the barren plateau phenomenon in training quantum neural
  networks with the {ZX}-calculus.
\newblock {\em Quantum}, 5:466, June 2021.
\newblock \href{https://quantum-journal.org/papers/q-2021-06-04-466/}{Quantum,
  5:466}.

\bibitem[ZHLT20]{zhang_toward_2020}
Kaining Zhang, Min-Hsiu Hsieh, Liu Liu, and Dacheng Tao.
\newblock Toward {Trainability} of {Quantum} {Neural} {Networks}.
\newblock {\em arXiv:2011.06258 [quant-ph]}, December 2020.
\newblock \href{http://arxiv.org/abs/2011.06258}{ArXiv: 2011.06258}.

\bibitem[ZHZY20]{zhang_differentiable_2020}
Shi-Xin Zhang, Chang-Yu Hsieh, Shengyu Zhang, and Hong Yao.
\newblock Differentiable {Quantum} {Architecture} {Search}.
\newblock {\em arXiv:2010.08561 [quant-ph]}, October 2020.
\newblock \href{http://arxiv.org/abs/2010.08561}{ArXiv: 2010.08561}.

\bibitem[ZS05]{zyczkowski_average_2005}
Karol Zyczkowski, Karol and Hans-J{\"u}rgen Sommers.
\newblock Average fidelity between random quantum states.
\newblock {\em Physical Review A}, 71(3):032313, March 2005.
\newblock \href{https://link.aps.org/doi/10.1103/PhysRevA.71.032313}{PRA,
  71(3):032313}.

\bibitem[ZTB{\etalchar{+}}20]{zhu_adaptive_2020}
Linghua Zhu, Ho~Lun Tang, George~S. Barron, F.~A. Calderon-Vargas, Nicholas~J.
  Mayhall, Edwin Barnes, and Sophia~E. Economou.
\newblock An adaptive quantum approximate optimization algorithm for solving
  combinatorial problems on a quantum computer.
\newblock {\em arXiv:2005.10258 [quant-ph]}, December 2020.
\newblock \href{http://arxiv.org/abs/2005.10258}{ArXiv: 2005.10258}.

\bibitem[ZWD{\etalchar{+}}20]{zhong_quantum_2020}
Han-Sen Zhong, Hui Wang, Yu-Hao Deng, Ming-Cheng Chen, Li-Chao Peng, Yi-Han
  Luo, Jian Qin, Dian Wu, Xing Ding, Yi~Hu, Peng Hu, Xiao-Yan Yang, Wei-Jun
  Zhang, Hao Li, Yuxuan Li, Xiao Jiang, Lin Gan, Guangwen Yang, Lixing You,
  Zhen Wang, Li~Li, Nai-Le Liu, Chao-Yang Lu, and Jian-Wei Pan.
\newblock Quantum computational advantage using photons.
\newblock {\em Science}, 370(6523):1460--1463, December 2020.
\newblock \href{https://science.sciencemag.org/content/370/6523/1460}{Science,
  370(6523):1460–1463}.

\bibitem[ZWL{\etalchar{+}}19]{zeng_learning_2019}
Jinfeng Zeng, Yufeng Wu, Jin-Guo Liu, Lei Wang, and Jiangping Hu.
\newblock Learning and inference on generative adversarial quantum circuits.
\newblock {\em Physical Review A}, 99(5):052306, May 2019.
\newblock \href{https://link.aps.org/doi/10.1103/PhysRevA.99.052306}{PRA,
  99(5):052306}.

\end{thebibliography}

\newcommand{\etalchar}[1]{$^{#1}$}


\end{document}